\documentclass[10pt]{article}

\usepackage[numbers]{natbib}
\usepackage[english]{babel}
\usepackage[figuresright]{rotating}
\usepackage{array}
\usepackage{graphicx}
\usepackage{pdflscape}
\usepackage{colortbl}
\usepackage{xcolor}
\usepackage{tikz}
\usepackage{chngcntr}
\usepackage{caption}
\usetikzlibrary{arrows}

\usepackage[english]{babel}
\usepackage[figuresright]{rotating}
\usepackage{array}
\usepackage{graphicx}
\usepackage{pdflscape}
\usepackage{colortbl}
\usepackage{xcolor}
\usepackage{algorithm}
\usepackage{algorithmicx}
\RequirePackage{amsthm,amsmath,amsfonts,amssymb}
\usepackage{bbm}
\usepackage{algpseudocode}
\usepackage{algorithm}
\usepackage{algorithmicx}
\usepackage{caption}
\usepackage{amsmath,amsfonts}
\usepackage{amsbsy}
\usepackage{verbatim}
\usepackage{setspace}
\usepackage{rotating}
\usepackage{array}
\usepackage{graphicx}
\usepackage{pdflscape}
\usepackage{colortbl}

\usepackage{tabularx,ragged2e,booktabs,caption}

\usepackage{amsmath,amsfonts}
\usepackage{amsbsy}
\usepackage{color,soul}
\usepackage{verbatim}
\usepackage{setspace}
\usepackage{rotating}
\usepackage{array}
\usepackage{graphicx}
\usepackage{pdflscape}
\usepackage{booktabs}
\usepackage{colortbl}
\usepackage{xcolor}
\usepackage{multirow}
\definecolor{lightgray}{gray}{0.9}
\usepackage{xcolor}

\usepackage{enumerate}

\def\bc{\begin{center}}
\def\ec{\end{center}}
\def\mR{\mathbb{R}}

\def\tx{{\tilde x}}

\def\definedas{\stackrel{\Delta}{=}}

\newcommand{\cX}{{\mathcal X}}

\newcommand{\cN}{{\mathcal N}}

\newcommand{\cL}{{\mathcal L}}

\newcommand{\cZ}{{\mathcal Z}}

\newcommand{\red}[1]{\textcolor{red}{\textsf{#1}}}

\def\tZ{{\tilde Z}}

\def\real{\mR}
\newcommand{\mc}{\multicolumn}
\def\bc{\begin{center}}
\def\ec{\end{center}}
\newcommand{\beq}{\begin{eqnarray}}
\newcommand{\eeq}{\end{eqnarray}}
\newcommand{\beqq}{\begin{eqnarray*}}
\newcommand{\eeqq}{\end{eqnarray*}}%\pagestyle{plain}
 \theoremstyle{definition}
 
 \theoremstyle{remark}

 \numberwithin{equation}{section}
\title{Improved Modeling of Persistence Diagram}
\author{Sarit Agami\\
Department of Economics\\
Hebrew University, Mount Scopus, Jerusalem, Israel\\
email:sarit.agami@mail.huji.ac.il}
\begin{document}
\maketitle

%\begin{document}
%\maketitle

\begin{abstract}
High-dimensional reduction methods are powerful tools for describing the main patterns in big data. One of these methods is the topological data analysis (TDA), which modeling the shape of the data in terms of topological properties. This method specifically translates the original data into two-dimensional system, which is graphically represented via  the 'persistence diagram'. The outliers points on this diagram present the data pattern, whereas the other points behave as a random noise. In order to determine which points are significant outliers, replications of the original data set are needed. Once only one original data is available, replications can be created by fitting a model  for the points on the persistence diagram, and then using the MCMC methods.
One of such model is the RST (Replicating Statistical Topology). In this paper we suggest a modification of the RST model. Using a simulation study, we show that the modified RST improves the performance of the RST in terms of goodness of fit. We use the MCMC Metropolis-Hastings algorithm for sampling according to the fitted model.
\end{abstract}

% Keywords
%\keyword{RST, Persistence diagram, Metropolis-Hastings, Gibbs sampler.}

\section{Introduction}
Topological data analysis (TDA) is an emerging field in which topological
properties of data are analyzed (\cite{Munch}). This should provide useful information about the structure and geometry of the data. The idea is to reduce high dimensional data sets to lower dimensions without sacrificing their most relevant topological properties. It is done by four steps (\cite{Kim}): First, the given data set (which contains data samples in the rows and multiple attributes in the columns) is converted into a 'point cloud' by calculating the similarity value using some distance metric. That is, every row in the given data is extracted into a single data point in the point cloud. Next, the point cloud is converted into a simplicial complex, and based on it, homology groups, which are algebraic analogues of certain properties of the manifold, are constructed. Specifically, persistent homology computes topological features of the manifold at different spatial resolutions. More persistent features are detected over a wide range of spatial scales and are deemed more likely to represent true features of the underlying space rather than artifacts of sampling, noise, or particular choice of parameters. Persistent homology techniques reveal topological features such as connected components, holes, and voids. Finally, these topological features are summarized in a 'persistence diagram' (PD), a multiset of points in $R^2$ that tracks the information about the "birth" and "death" scale of each topological feature. The difference between the birth and death scales is called the persistence of a feature and in some sense indicates its prominence.
Given the topological features, the question of statistical inference is arise, and there exist a literature on this issue: \cite{DIVOL} studied the persistence diagram with deterministic measure on $R^2$ ('Expected Persistence Diagram' (\cite{DivolLacombe})), and discussed the density of such diagrams, and a kernel based estimation of this density; \cite{MILEYKO} showed that the space of persistence diagrams has properties that allow to define on it expectation, variance, percentile and conditional probability; \cite{Kusano} studied the expectation of a persistence diagram by the persistence weighted kernel; \cite{Turner} suggested the Fréchet Means for Distributions of Persistence Diagrams.
\\
\indent For generating multiple instances of persistence diagrams when only one such original diagram is available, there exist two approaches in the literature:
One approach is the bootstrap approach as in \cite{chazal} and \cite{fasy}; this approach produces replicates of persistence diagram by subsampling either the data or the diagram. Another approach is the RST that was suggested by \cite{adleragamprat}, and was improved in \cite{adleragam2}. The RST is a parametric modeling of the points on a single diagram having the same rank of homology. It is based on a Gibbs model that involves the distances of the $K$ nearest neighbours of each point of the persistence diagram, multiplying by the kernel density estimator (KDE). The model's parameters are estimated via the maximum likelihood method. However, sometimes, no maximal solution exists, i.e., the estimation diverges. Such cases arises, for example, when the number of points on the persistence diagram for a given homology rank is relative large, and the points are dense. Then the distances of the $K$ nearest neighbors are relative small, which lead to increase more and more the value of the optimization solution, and divergence is obtained. By this, the log likelihood becomes undefined when no constrains are taken on the parameters values. One possible solution is to put some weight on each closeness level of the nearest neighbors. A reasonable weight is the KDE. That is, instead of weighting all levels of distance closeness levels of the nearest neighbours together, the weighting will be on each level of distance closeness of the nearest neighbours. In this paper we examine the goodness of fit of this modification, and compare it with the performance of the original RST.
The outline of the paper is as follows. Section 2 presents the notation and background, and gives a short description of the RST method along with the suggested modified model. Section 3 examined the performance and goodness of fit of the suggested modified model via a simulation study. Section 4 describes the results, and Section 5 presents a brief summary and conclusions.

\section{Background and setting}
\subsection{Notation}
\indent Let $\cZ$ be a compact subset of $\mathbb R^D$, typically a sub-manifold or stratified sub-manifold, and suppose that we observe a sample $\tZ_n=\{Z_1,\dots,Z_n\}$ drawn from a distribution $P$ supported on $\cZ$.
For defining the persistence diagram of a dataset in computational topology, one can use for example the usual distance function, or a smooth function such as the kernel density estimator.
The points on the persistence diagram are 'birth' and 'death' and are denoted by ${(b_i,d_i)}_{i=1}^{N}$, where $N$ is the number of points that have the same rank of homology $k$.
\subsection{The RST Model}
Define a new set of $N$ points $\tx_N =\{x_i\}_{i=1}^N$, with $x^{(1)}_i=b_i$ and
 $x^{(2)}_i=d_i-b_i$. That is, $\tx_N$ a set of $N$ points in $\cX=\real\times\real_+$.
This (invertible) transformation has the effect of moving the points in the original persistence diagram downwards, so that the diagonal line projects onto the horizontal axis, but still leaves a visually informative diagram, which \cite{adleragamprat} call the projected persistence diagram, or PPD. The goal is a parametric model for $\tx_N$.
The description of the suggested model of \cite{adleragam2} is as follows.
 %While the usual distance function is the most commonly used method for defining the persistence diagram of a dataset in computational topology, but it is  highly influenced by outliers, even few outliers, and it is highly non-robust. One option is to replace this distance function with
Define a  kernel density estimator (KDE), $\hat f_n$, given by
\beq
\label{eq:kernel}
\hat f_n(p) \ = \  \frac{1}{n(\sqrt{2\pi}\eta)^D}  \sum_{i=1}^{n} e^{{-\|p-z_i\|^2}/{2\eta^2}},\qquad p\in\mathbb R^D,
\eeq
 where $\eta >0$ is a bandwidth parameter for the Gaussian kernel defining $\hat f_n$.
In addition, for  $x\in\cX$ and for $k\geq 1$ let
$x^{nn}(k) \in \cX$ be the $k$-th nearest neighbour to $x$, and set
\beq
\label{eq:distance}
%\mathcal L_{\delta,k}     \equiv
\mathcal L_{k}(\tx_N) =  \sum_{x\in\tx_N} \|x-x^{nn}(k) \|.
\eeq
Also define
\beq
\label{eq:hamiltonian1}
\tilde H_{\Theta}^K(\tx_N)
=\sum_{k=1}^K \theta_k {\mathcal L}_{k}(\tx_N),
\eeq
where $\Theta=(\theta_1,\dots,\theta_K)$, and $K$ is the cluster size.
Then, the likelihood (pseudolikelihood \cite{Besag,chalmond}) is
\beq
\label{eq:pseudo1}
\tilde{L}^ K_{\alpha,\Theta}(\tx_N)  \definedas
\prod _{x\in\tx_N}  f_{\alpha,\Theta} \left(x\rvert\    \cN_{K}(x) \right),
\eeq
 where
$\cN_{K}(x)$ denotes the $K$ nearest neighbours of $x$ in $\tx_N$, and
%$f_\Theta\left(x\big|    \cN_{\delta,K}(x) \right)$
% are given by
 \beq
  f_{\alpha,\Theta}\left(x\rvert\    \cN_{K}(x) \right)=\frac{
 (KDE(x))^{\alpha}\times \exp \left(-\tilde{H}^K_{\Theta}\left(x\rvert\   \cN_{K}(x) \right)\right)
  }{
\int _{\real}\int _{\real_+} (KDE(z))^{\alpha} \times \exp \left(-\tilde{H}^K_{\Theta}\left(z\rvert\   \cN_{K}(x) \right)\right)\,dz^{(1)}dz^{(2)},
}
\label{eq:conditionalham1}
\eeq
with
\beqq
\tilde{H}^K_{\Theta}\left(x\rvert\   \cN_{K}(x)\right)=\sum_{k=1}^K \theta_k{\cL}_{k}\left(\cN_{K}(x)\right).
\eeqq
For considering some values of $K$, the best model can be chosen by the automated statistical procedures such as AIC, BIC, etc. The nuisance parameter $\alpha$ is estimated by the bisection method, where after considerable experimentation, \cite{adleragam2} found that it is enough to take the search (non-negative) range to be $[0,3]$. Given the value of $\alpha$ that maximizes the log likelihood, the next step is searching for $\Theta$ that maximizes the log likelihood.

\subsection{The Modified RST Model}
The density function (\ref{eq:conditionalham1}) is weighting the KDE over all the closeness levels of the nearest neighbors. The suggested modification is to weight the KDE separately for each closeness level of the nearest neighbors. That is, based on (\ref{eq:distance}), define
\beq
\label{eq:hamiltonian2}
\tilde H_{\alpha,\Theta}^K(\tx_N)
=\sum_{k=1}^K \theta_k {\mathcal L}_{k}(\tx_N)\times (KDE(\tx_N))^{\alpha}
\eeq
Then, the likelihood is
\beq
\label{eq:pseudo2}
\tilde{L}^ K_{\alpha,\Theta}(\tx_N)  \definedas
\prod _{x\in\tx_N}  f_{\alpha,\Theta} \left(x\rvert\    \cN_{K}(x) \right),
\eeq
 where
$\cN_{K}(x)$ denotes the $K$ nearest neighbours of $x$ in $\tx_N$, and
%$f_\Theta\left(x\big|    \cN_{\delta,K}(x) \right)$
% are given by
 \beq
  f_{\alpha,\Theta}\left(x\rvert\    \cN_{K}(x) \right)=\frac{
 \exp \left(-\tilde{H}^K_{\alpha,\Theta}\left(x\rvert\   \cN_{K}(x) \right)\right)
  }{
\int _{\real}\int _{\real_+} \exp \left(-\tilde{H}^K_{\alpha,\Theta}\left(z\rvert\   \cN_{K}(x) \right)\right)\,dz^{(1)}dz^{(2)}
}
\label{eq:conditionalham3}
\eeq
with
\beqq
\tilde{H}^K_{\alpha,\Theta}\left(x\rvert\   \cN_{K}(x)\right)=\sum_{k=1}^K \theta_k{\cL}_{k}\left(\cN_{K}(x)\right)\times (KDE(x))^{\alpha}.
\eeqq

\subsection{Algorithm for replicated persistence diagrams}
Based on the RST model, \cite{adleragam2} used the Metropolis-Hastings MCMC \cite{RobertCasella,Handbook} to generate simulated replications of the points on the original persistence diagram, as follows.
Firstly, given a $\tx_N$,  define a `proposal distribution' $q(\cdot \rvert\ \tx_N)$ to be the KDE by using the inverse transform method \cite{RobertCasella,Handbook}.
%Then set $q(X|\tX)=2\phi(X|\tX)$ if $X\in\real\times\real_+$, $\tX^*$, and zero otherwise.
% in which, for $X\in \tX$, $X^*$ a proposed replacement for $X$ chosen with (conditional, on $\tX$) the same as $\tX$ but with the point $X$ replaced by the point $X^*$, and
Next, for two points $x,x^*\in\real\times\real_+$  define an `acceptance probability', according to which $x\in\tx_N$ is replaced by $x^*$, leading to the updated PPD $\tx_N^*$, as
\beqq
\rho \left(x, x^*\right)
= \min \left\{1,\frac{f_\Theta\left(x^*\rvert\ N_{\delta,K}(x)\right) \cdot q(x\rvert\ \tx^*_N)}{f_\Theta \left(x\rvert\  N_{\delta,K}(x)\right)  \cdot q(x^*\rvert\ \tx_N)}\right\}.
\eeqq

\noindent Then the algorithm is Algorithm \ref{MCMC:algorithm}.

%Note that when the PD is obtained by the KDE, the condition of $w^{(2)}>0$ in step 6 should be replaced with $W>0$.
\begin{algorithm}
\caption{MCMC step updating diagram for $\tx_N$}
\label{MCMC:algorithm}
\begin{algorithmic}[1]
%\Procedure{CH\textendash Election}{}
\State  $k =0$
 \State $k \gets k+1$
\State Choose $x^*$ according to $q(\cdot \rvert\ \tx_N)$
\State Compute $\rho(x_k,x^*)$
\State Choose $U$ a standard uniform variable on $[0,1]$

\If{$U<\rho(x_k,x^*)$} set $x_k =x^*$
\EndIf
\If{$k<N$}  go to Step 2
\EndIf
%\EndProcedure
\end{algorithmic}
\end{algorithm}

To obtain $B$ approximately independent PPD's, the procedure dependents on a burn in period, see \cite{adleragamprat} SI Appendix (Sec.\ 2.1) for more details.
Given the collection of $B$ simulated PPDs, each PPD is converted back to a regular persistence diagram with the mapping {$x_m\to (x_m^{(1)}+x_m^{(2)}, x_m^{(1)})=(b_m,d_m)$ }of its component points.
%and write  $\mathcal S({\widetilde B}) = \{\widehat B_1,\dots,\widehat B_n\}$ for the resulting collection of simulated PDs generated from $\widetilde B$.

\subsection{Goodness of Fit}
The goodness of fit of each model versions is the degree of closeness between the resulted simulated PD by each model version with the real PD.
In order to evaluate the goodness of fit of the modified RST relative to the performance of the original RST, a simulation study was used, and it is presented below in Section 3. The general idea is as follows. We calculated 100 real PDs corresponded to 100 samples from some geometrical object, one PD for each sample. For each PD, we fitted both the original and the modified models. Then we calculated the simulated PD using the Metropolis-Hastings algorithm based on each of the two fitted models. As the next step, we examined two criteria of goodness of fit over the 100 PDs. Criterion 1 is the distance between the real PD and its corresponded simulated PD, using the Bottleneck and the Wasserstein distances. Smaller distances indicate on a better fitting. The bottleneck distance is the cruder of the two distances, and the Wasserstein distance is more sensitive to details in the persistence diagram \cite{Edels}. Criterion 2 is a comparison of  distributional properties of the real PDs with those of the simulated PDs: We used as the distributional properties the averaged distances of the first, second, and third nearest neighbors.
\\
Using the distributions of these criteria over the 100 PDs made the comparison of goodness of fit of the modified model vs. the original model.

\section{Simulation Study}
In the simulation study we took the various data to be of two, three, and forth dimensions. For the two dimensional data we examined the one unit circle, two concentric circles, and two distinct circles. These examples behave as one, two, and separated geometrical objects, respectively. For the three and four dimensional data sets we consider the unit 2-sphere ($S^2$) and the unit 3-sphere ($S^3$), respectively. In each example, the persistence diagram was generated by the upper level sets of a smoothed empirical density of the data, as defined in (2.1), with $\eta=0.1$. The grid for the calculation of this density was based on 100 points over the range of each coordinate. We used this grid in the all considered examples except the example of $S^3$ which has 4 dimensions, and due to computer's memory we took the grid to be based on 15 points over the range of each coordinate. For the calculations of the model likelihood, we used the plug-in bandwidth for the KDE as obtained by the function Hpi.diag in R software ("ks" package).
In addition, we took the search for $\alpha$ estimator over the range [0,4].
\\
Some notes regarding the MCMC algorithm that we used: (i) The KDE as the 'proposal distribution' had sometimes a negligible value, which should be ignored. Therefore, for each example in the simulation study, we dropped the proposal values that had KDE$<10^{-4}$. (ii) The proposal distribution is based on a two-dimensional grid to sample from it. In
\cite{adleragam2}, the optimal grid was considered, but it had resulted in a low acceptance rate in the MCMC. Generally, the acceptance rate is a one measure for the goodness of the MCMC algorithm performance. It depends largely on the proposal distribution, where distribution with smaller variance is resulted in a higher acceptance rate, and vice versa. Usually, the standard rate of acceptance is supposed to be around 0.2-0.25. But, using the original grid in the proposal distribution yields a smaller rate relative to that obtained by the standard grid. Increasing the grid size yields a better acceptance rate. More of that, increasing the grid size is itself better since the aim of the proposal distribution is to approximate the distribution of the points on the PD, therefore a finer grid may obtain better results. For these two reasons, we examined the performance of the MCMC under the grids of 25x25, 50x50, 100x100. (iii) For the burn-in parameter (which we call 'step' in the following results), we examined the values of 25, 50, 100, and we took the PD at that step to be the simulated PD.
%This yields a general influence of the KDE. But, after using the MCMC algorithm, it generates points from the main mass of the distribution, but does not generate "outliers" points, that is, the points that are far from the points mass. This can be solved by considering a more detailed influence of the KDE, which can be obtained by using it for each level of the nearest neighbours. One measure for the goodness of the MCMC algorithm performance is its acceptance rate.

\subsection{One geometrical object}
\begin{figure}[h!]
\bc
\includegraphics[width=2.8in, height=2.8in]{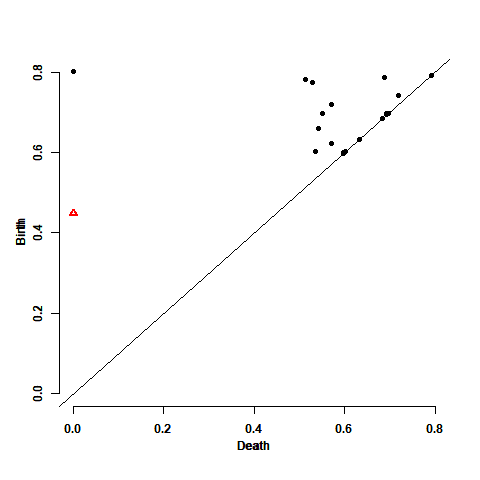}
\ec
%\caption{\footnotesize
% A random sample from two circles, 500 points from the larger circle and 300 from the smaller one,  with a kernel density
\caption{\footnotesize The persistence diagram of a sample of $n=1,000$ points from the unit circle, for its upper level sets. Black circles are connected components ($H_0$ persistence points), red triangles are holes ($H_1$ points). Birth times are on the vertical axis.}
\label{fig:circle}
\end{figure}

As one geometrical object data we took a sample of $n=1000$ points drawn from a circle with radius $r=1$ (the unit circle).
The typical corresponded persistence diagram is presented in Figure\ \ref{fig:circle}. The black circles indicating connected components ($H_0$ persistence), and the red triangles corresponding to holes ($H_1$).

%\begin{landscape}
\begin{landscape}
\begin{figure}[h!]
\bc
\includegraphics[width=1.2in, height=1.4in]{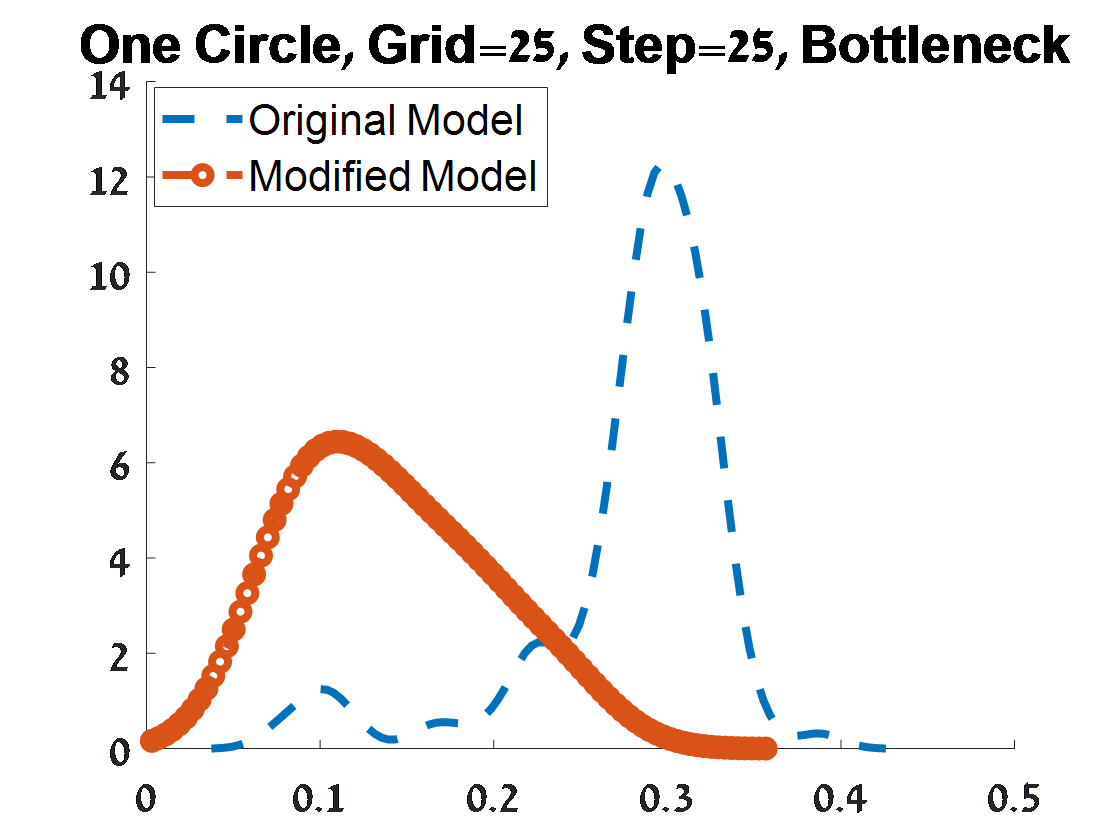}
\includegraphics[width=1.2in, height=1.4in]{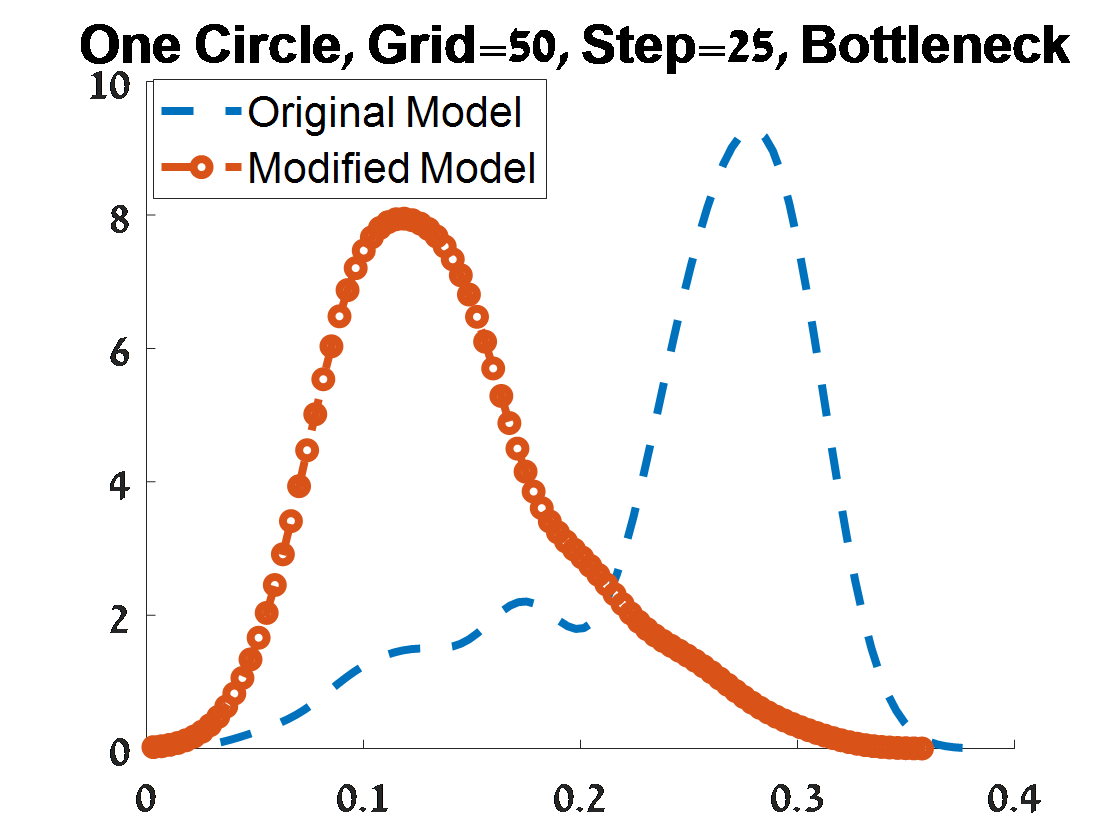}
\includegraphics[width=1.2in, height=1.4in]{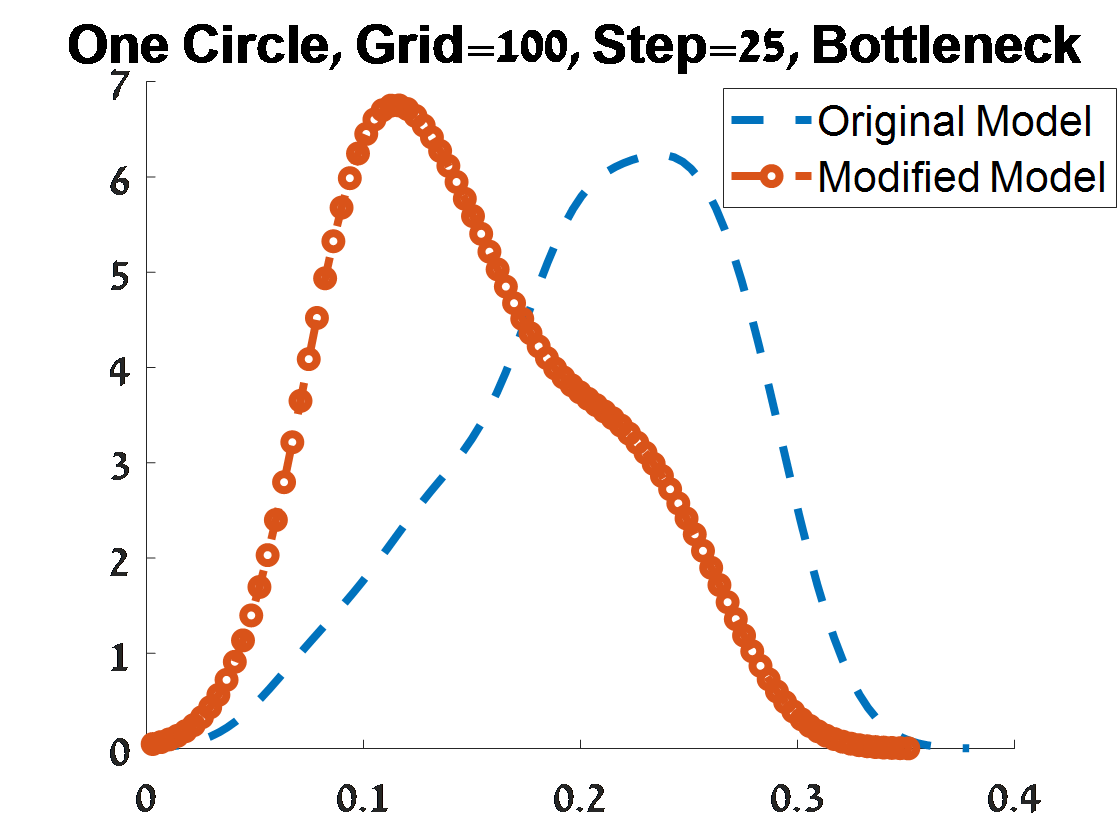}
\includegraphics[width=1.2in, height=1.4in]{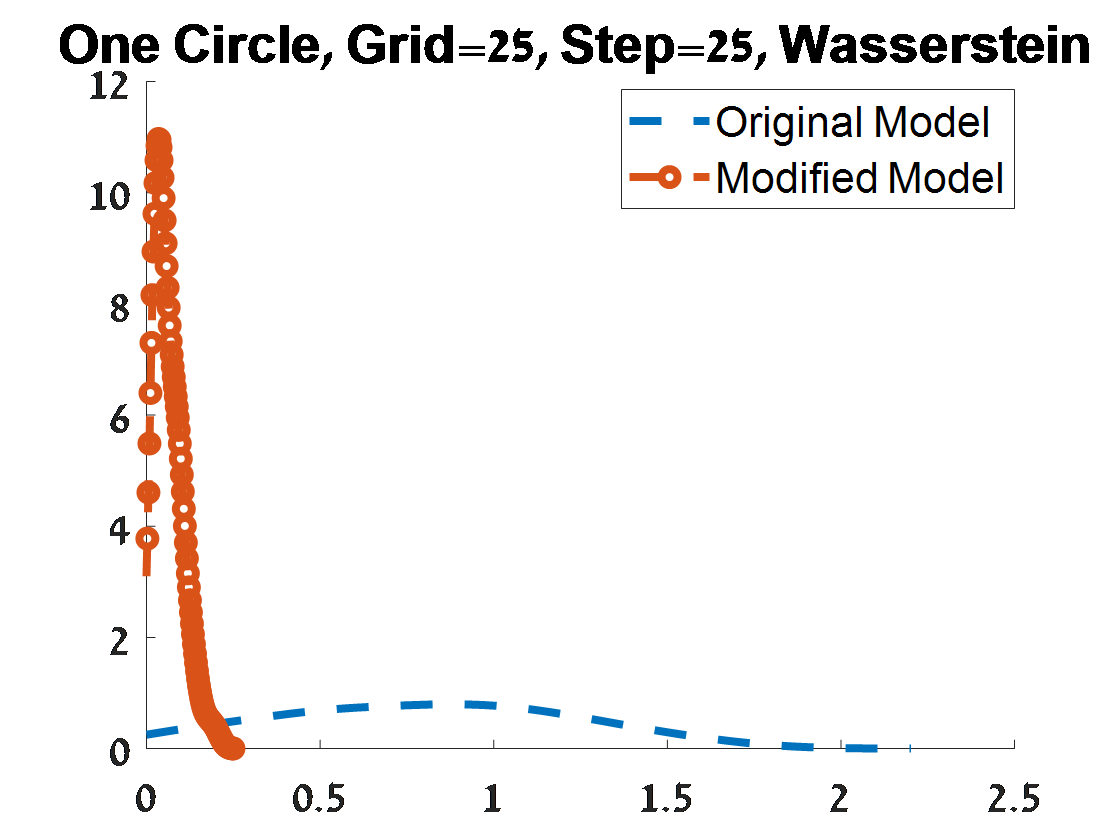}
\includegraphics[width=1.2in, height=1.4in]{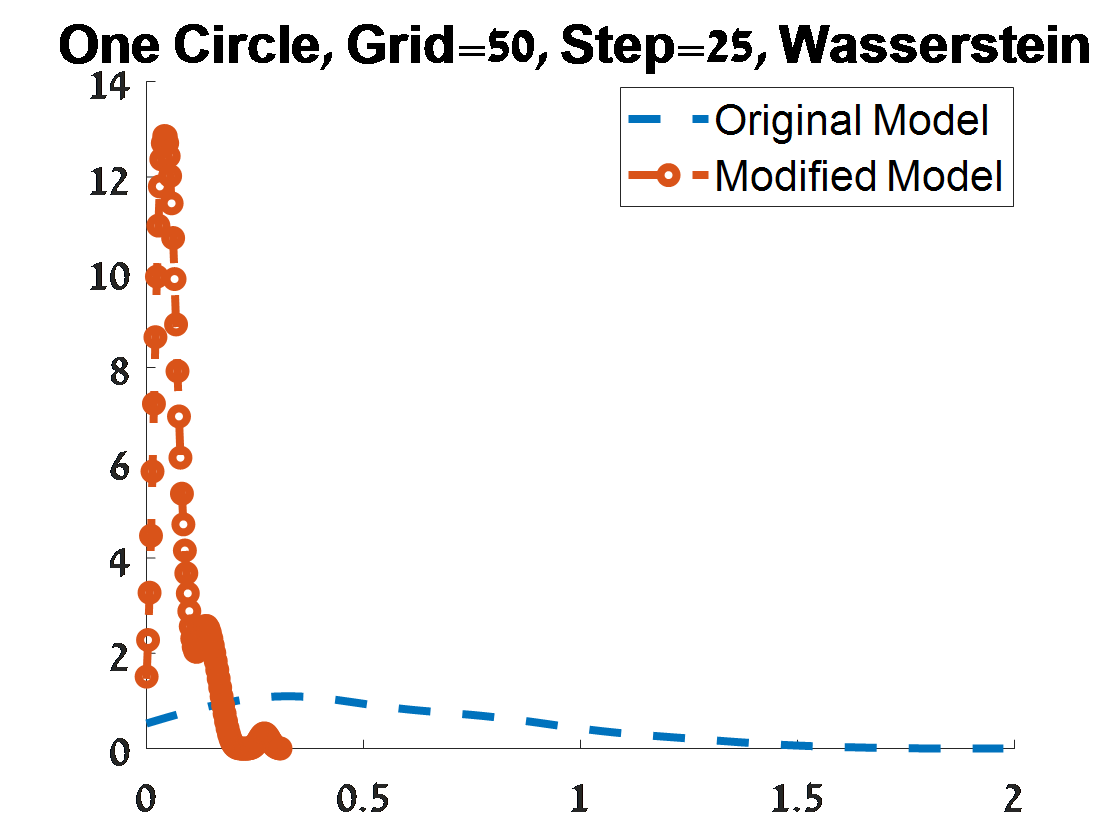}
\includegraphics[width=1.2in, height=1.4in]{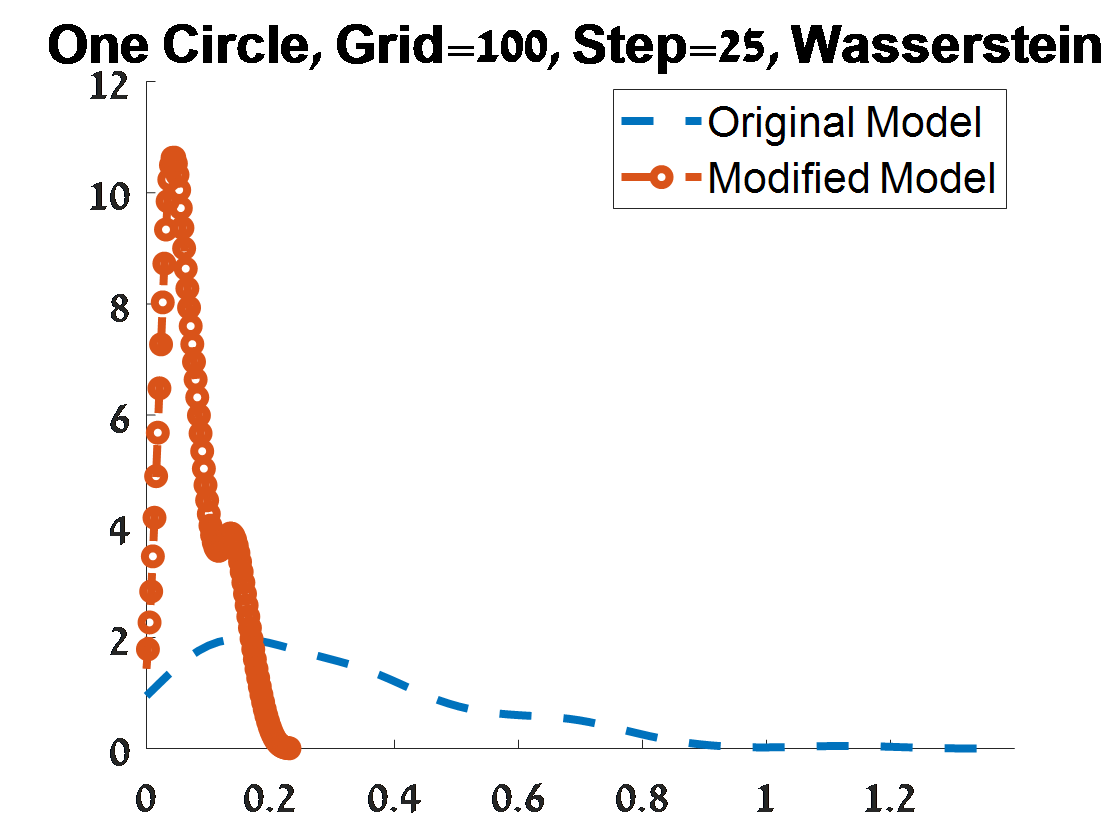}
\\
\includegraphics[width=1.2in, height=1.4in]{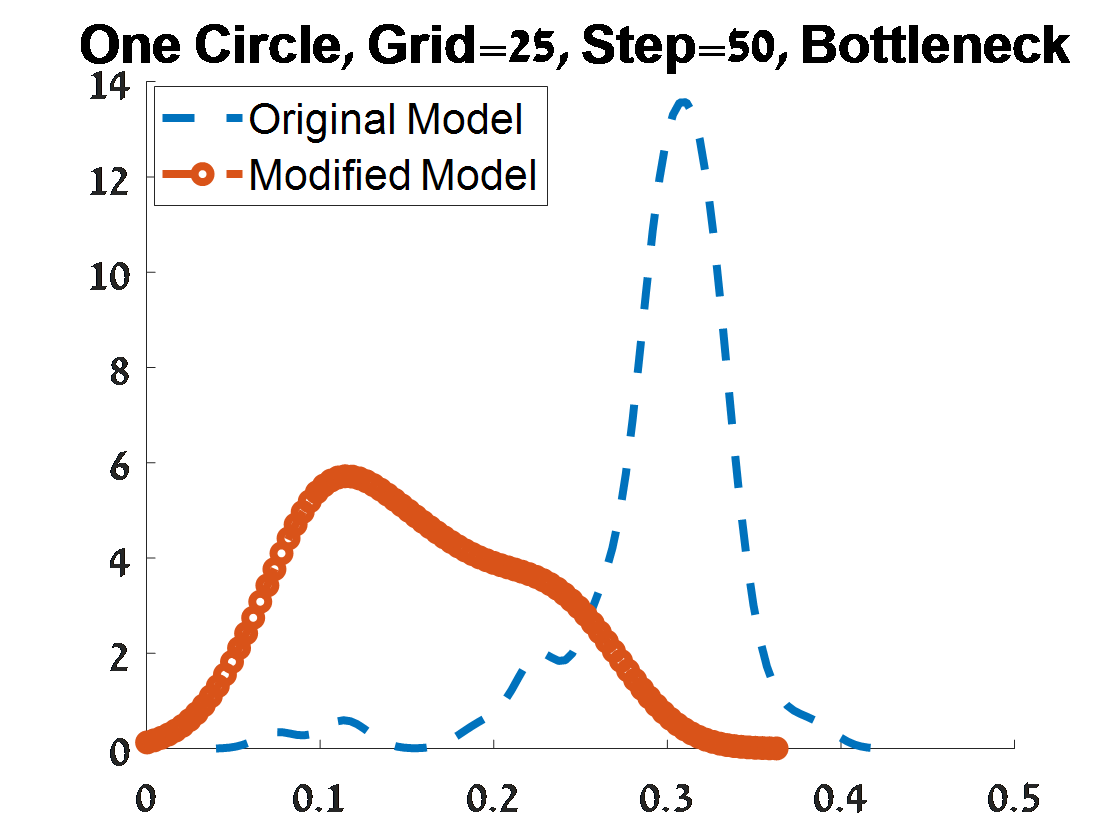}
\includegraphics[width=1.2in, height=1.4in]{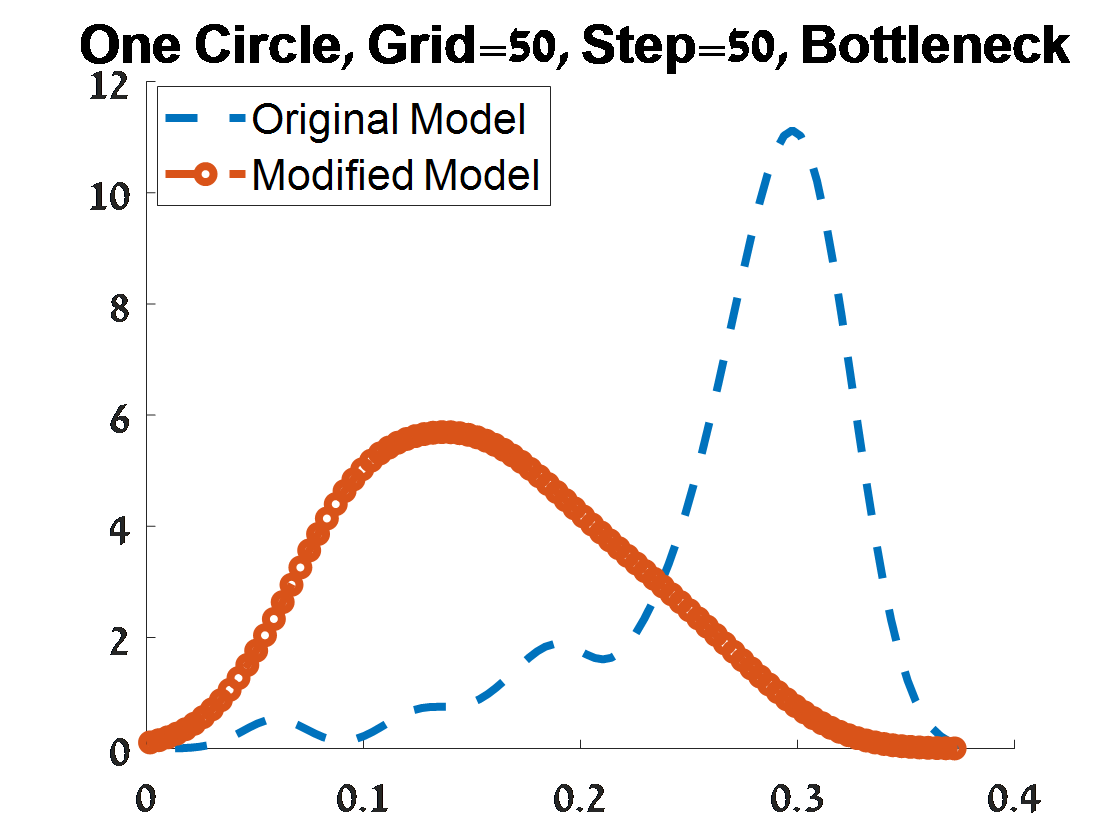}
\includegraphics[width=1.2in, height=1.4in]{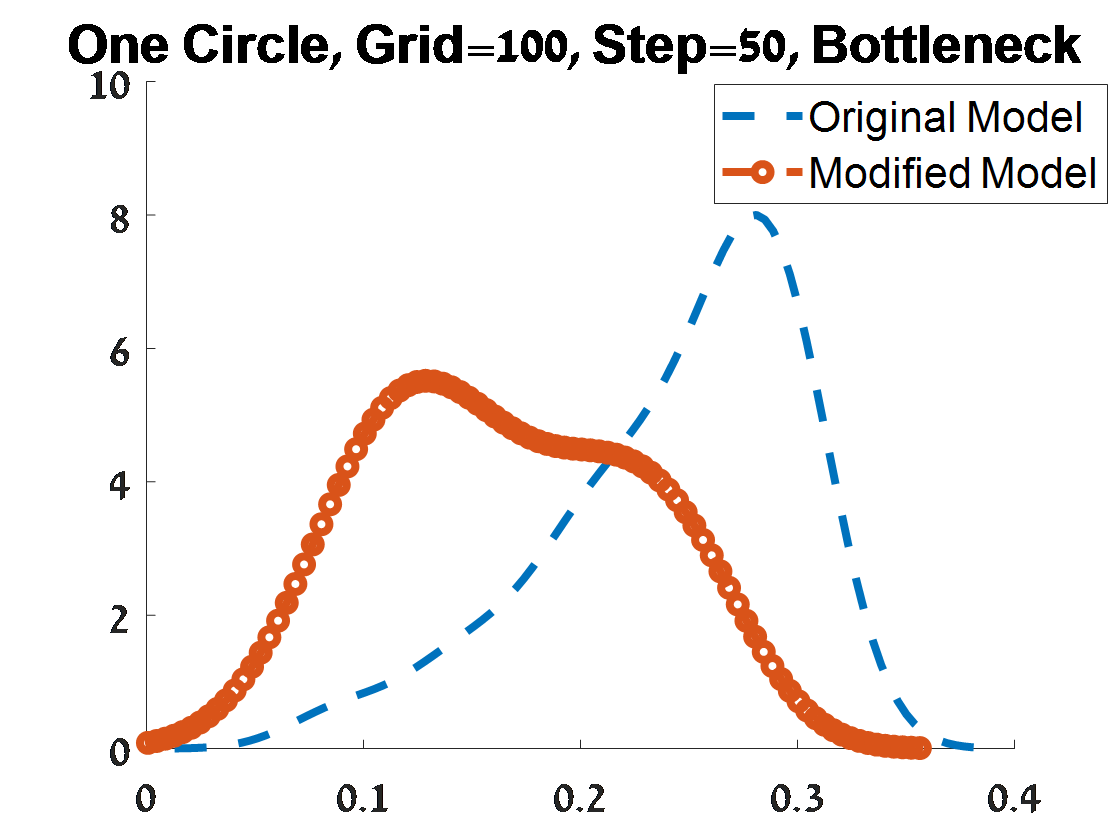}
\includegraphics[width=1.2in, height=1.4in]{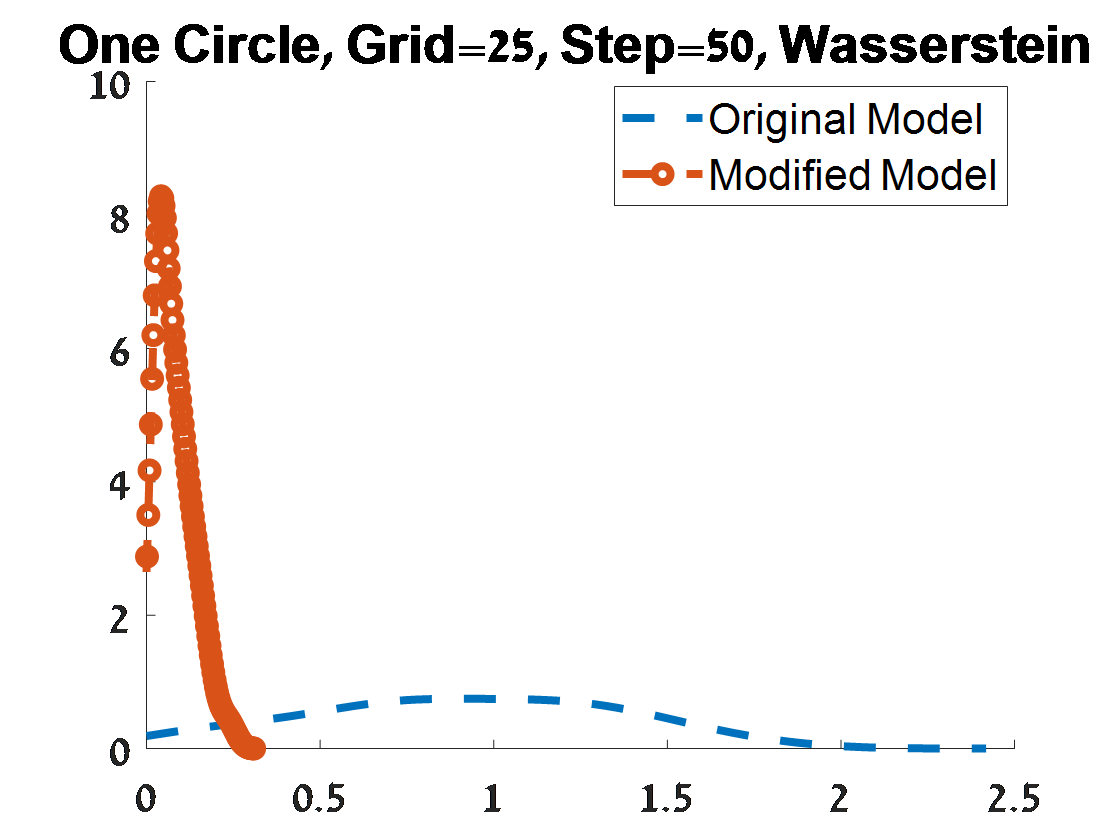}
\includegraphics[width=1.2in, height=1.4in]{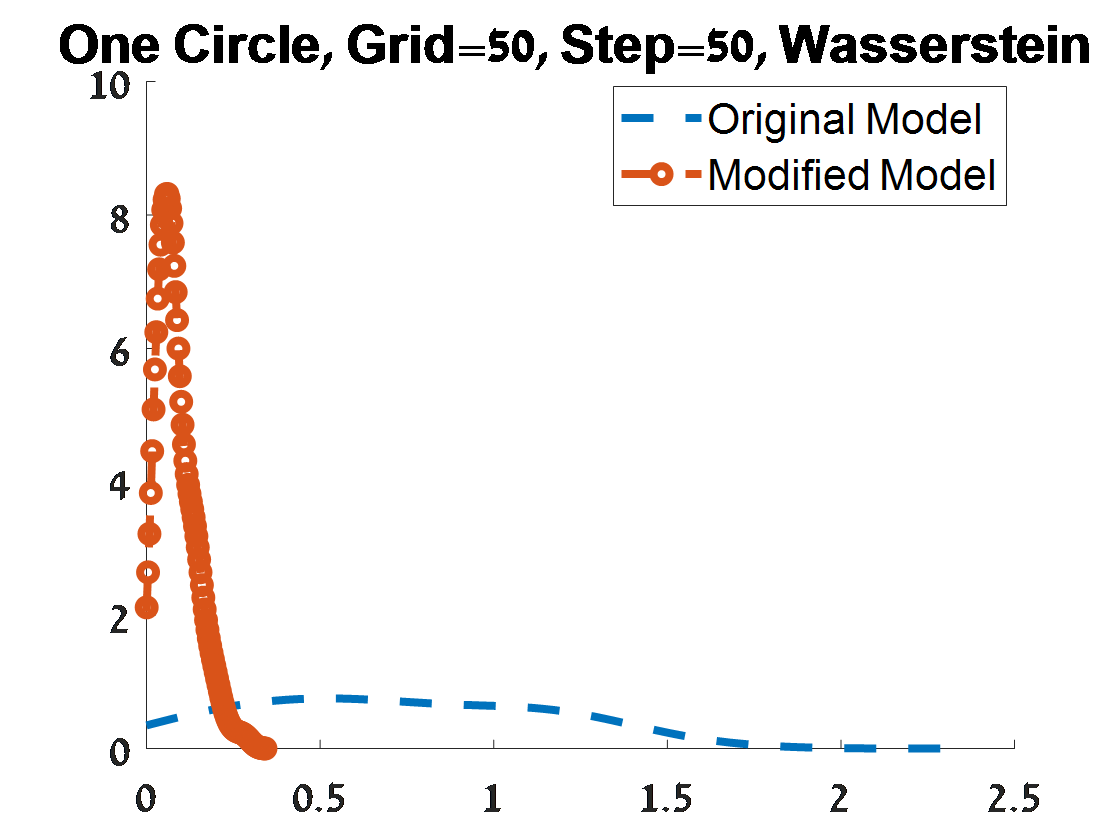}
\includegraphics[width=1.2in, height=1.4in]{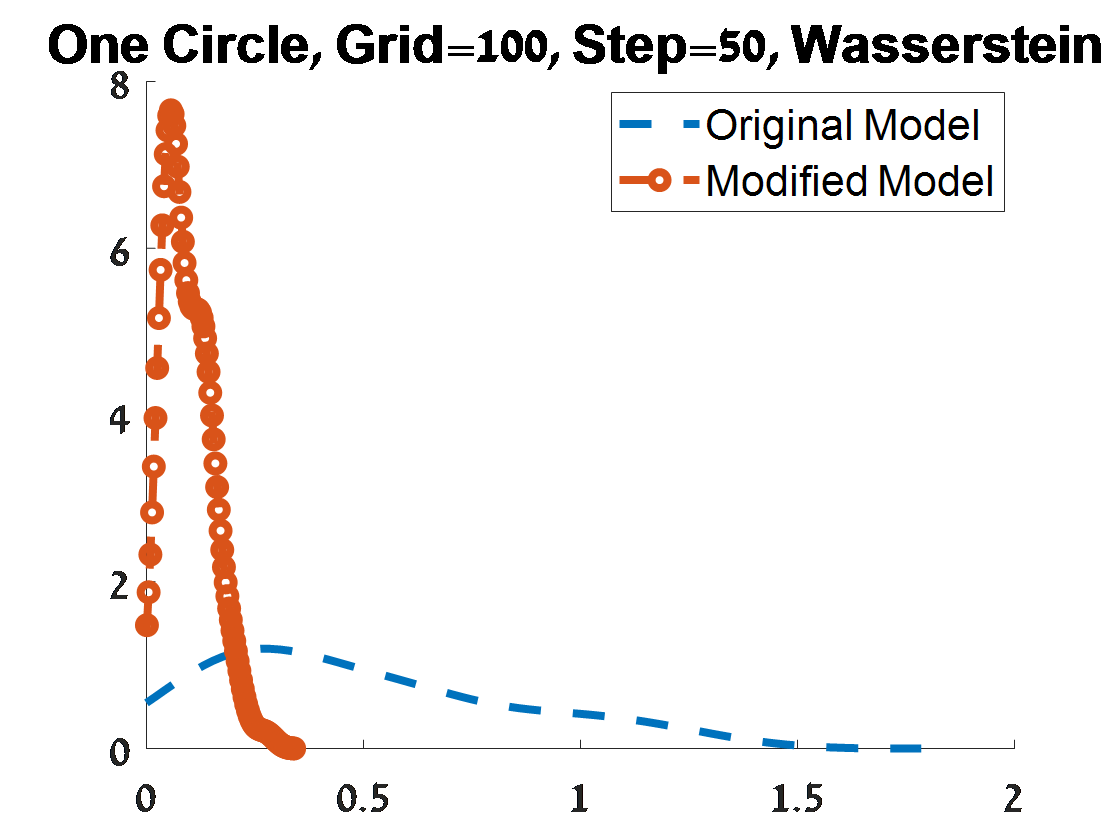}
\\
\includegraphics[width=1.2in, height=1.4in]{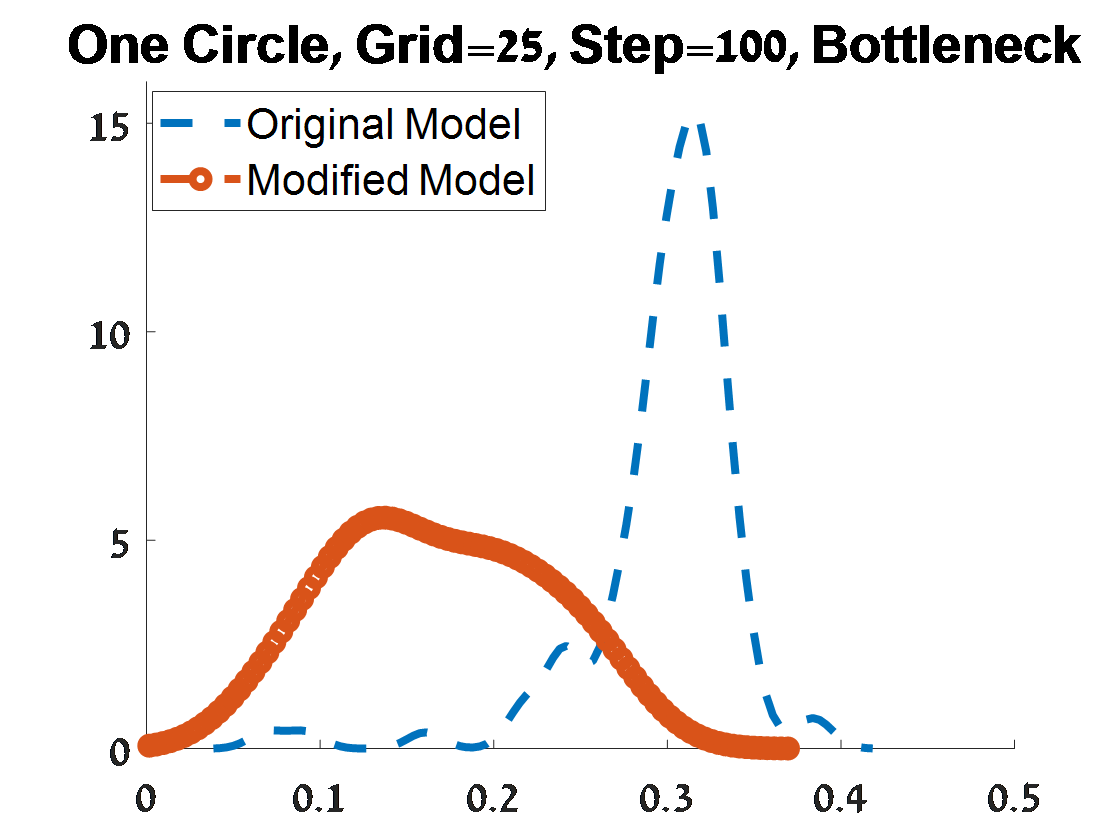}
\includegraphics[width=1.2in, height=1.4in]{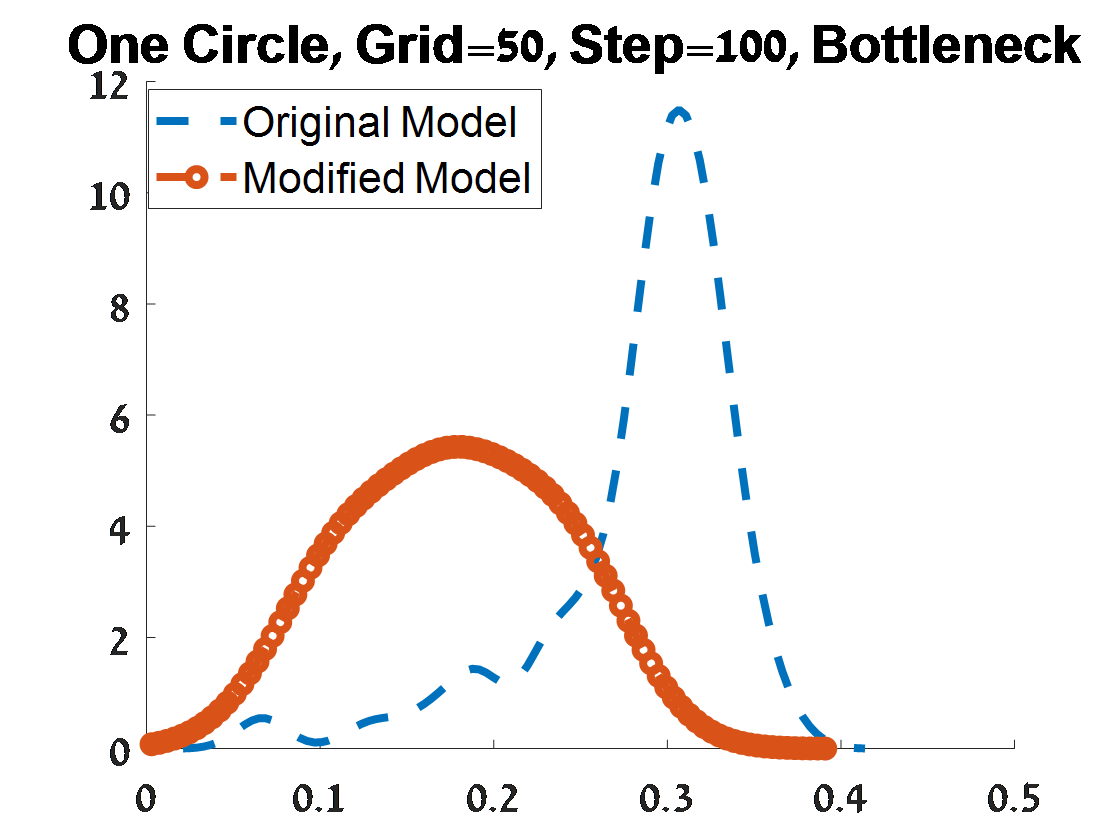}
\includegraphics[width=1.2in, height=1.4in]{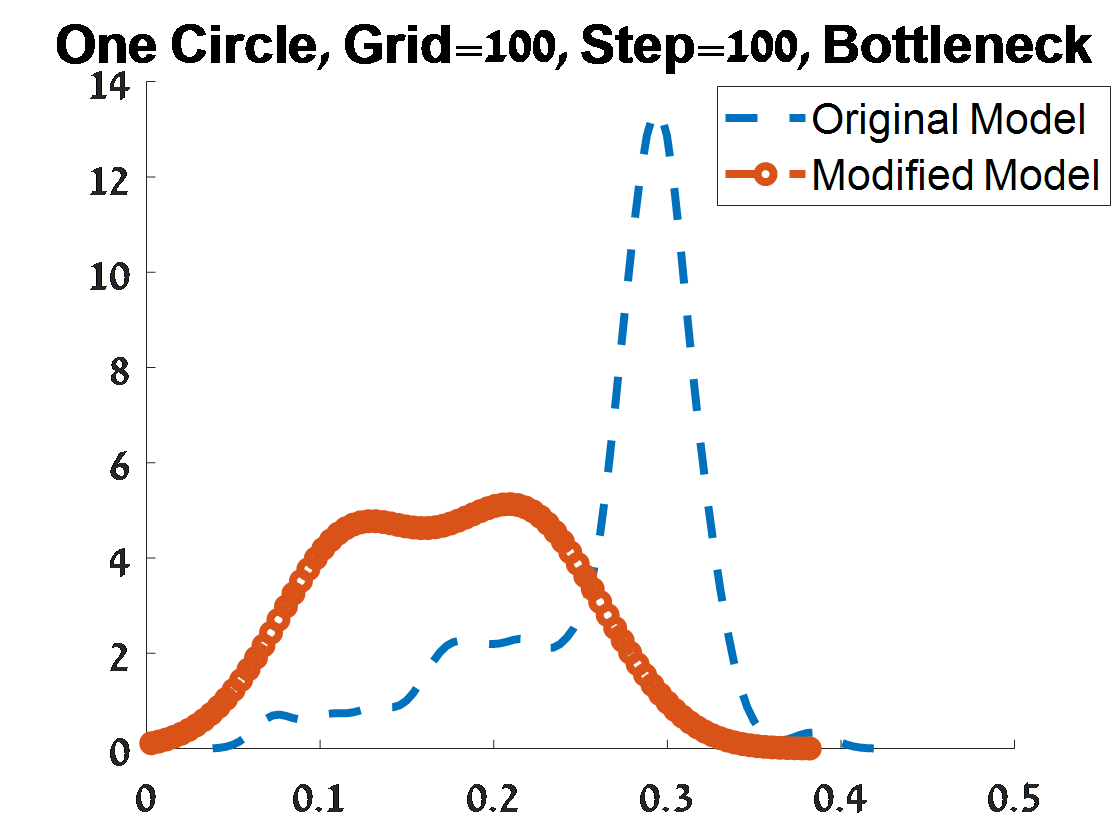}
\includegraphics[width=1.2in, height=1.4in]{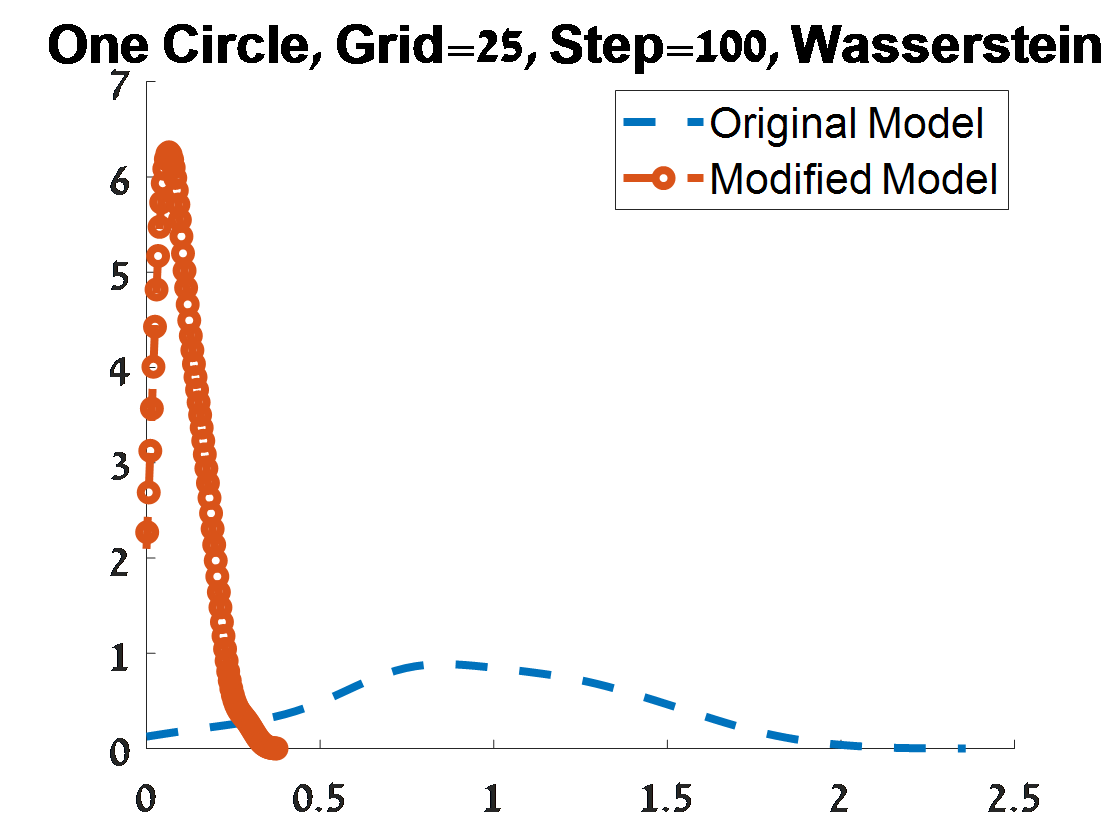}
\includegraphics[width=1.2in, height=1.4in]{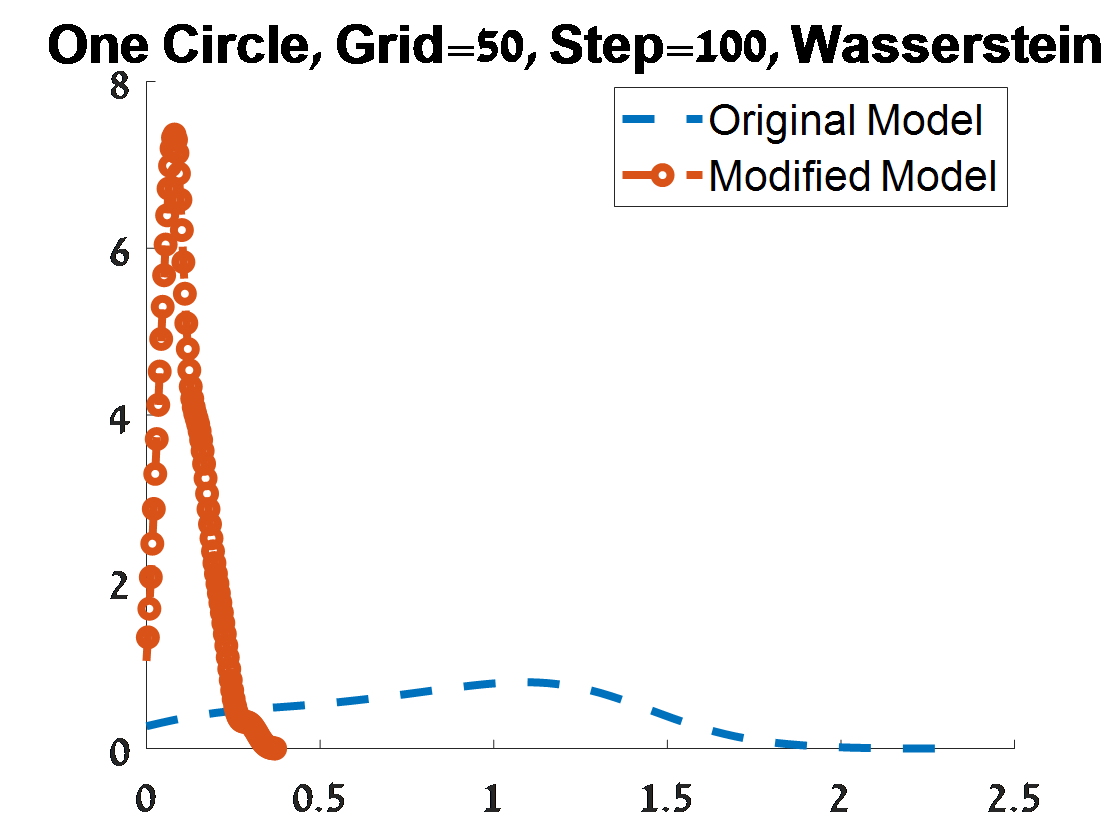}
\includegraphics[width=1.2in, height=1.4in]{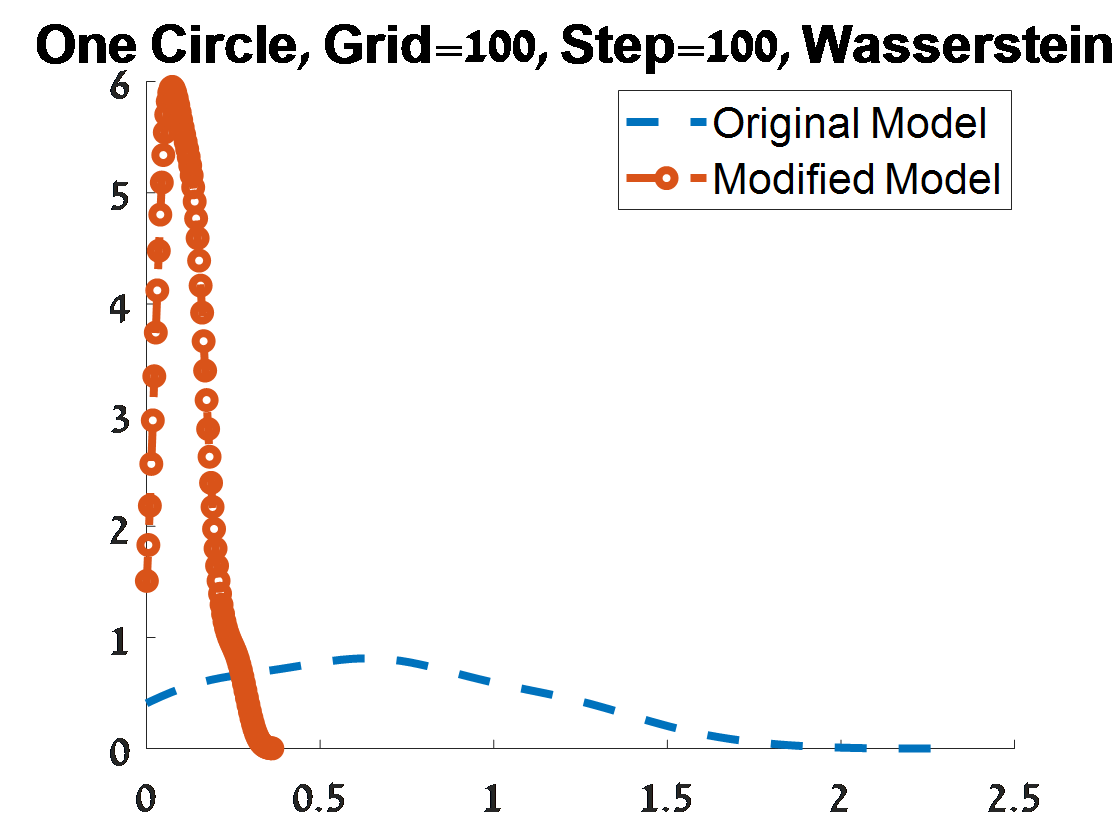}
\ec
%\caption{\footnotesize
% A random sample from two circles, 500 points from the larger circle and 300 from the smaller one,  with a kernel density
\caption{\footnotesize
Criterion 1 of goodness of fit for 100 PDs corresponded to 100 samples from a unit circle. The plots depend on the grid size of the proposal distribution ("Grid"), and the burn-in ("Step") of the MCMC algorithm. }
\label{fig:circle_a}
\end{figure}
\end{landscape}

%\end{landscape}
We generated 100 such samples, and calculated their corresponded PDs. For each PD we fitted both the original and modified models for the $H_0$ points, according to the steps that were mentioned above in Section 2.5. Figure\ \ref{fig:circle_a} describes the distributions over the 100 PDs of the first criterion of goodness of fit, and Figures 3-4 describe the distributions of the second criterion of goodness of fit.
In criterion 1, we have that the distance of the simulated PD from the real PD is smaller under the modified model relative to the distance under the original model. This is prominent in the Wasserstein distance, as expected due to its sensitivity to details in the PD, as was mentioned in Section 2.5. For a given grid, the burn-in value over the considered values has a negligible influence on both distances. But the larger grid size (for a given burn-in) yields smaller distances for both model's versions.
For criterion 2, the distributional properties of the modified model are close to those of the real PDs rather those of the original model, for all considered values of the grid size and the burn-in. More of that, the grid sizes of 50x50 and 100x100 are better in terms of the first, second and third nearest neighbors, and step of 25 is the best for each of them.
%This closeness is better in grid sizes of 50x50 and 100x100 for a given burn-in value, and in burn-in of 25 for a given grid size.
%In Figure\ \ref{fig:circle_d} we present two examples, each of them contains a real PD and its simulated PD based on the two model versions, only for the best scenarios we found, that is grid size of 50x50 and 100x100, and burn-in of 25. Also here we see that the best fitted MCMC is under the modified model rather than the original one.
That is, based on criteria 1-2 we conclude for this example that the modified RST is better than the original RST, where the best fitting is under grid sizes of 50x50 and 100x100, and burn-in of 25.
\\

\begin{landscape}
\begin{figure}[h!]
\bc
\includegraphics[width=1.2in, height=1.25in]{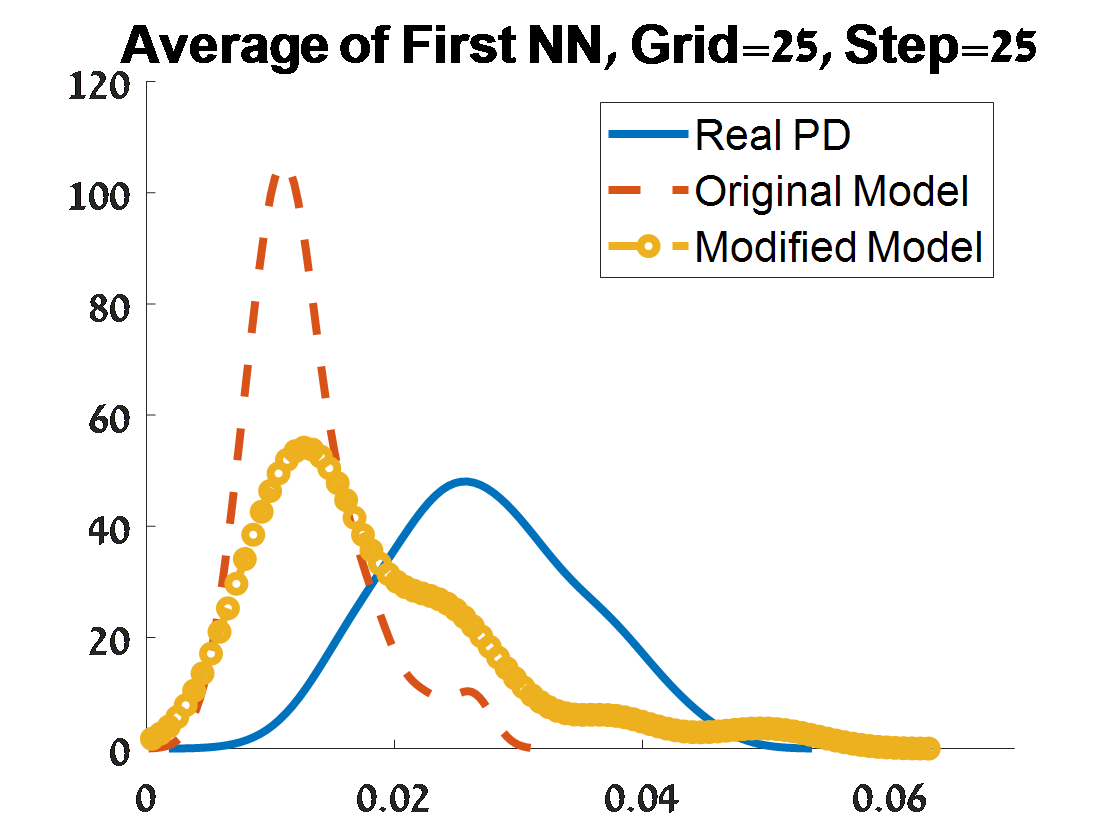}
\includegraphics[width=1.2in, height=1.25in]{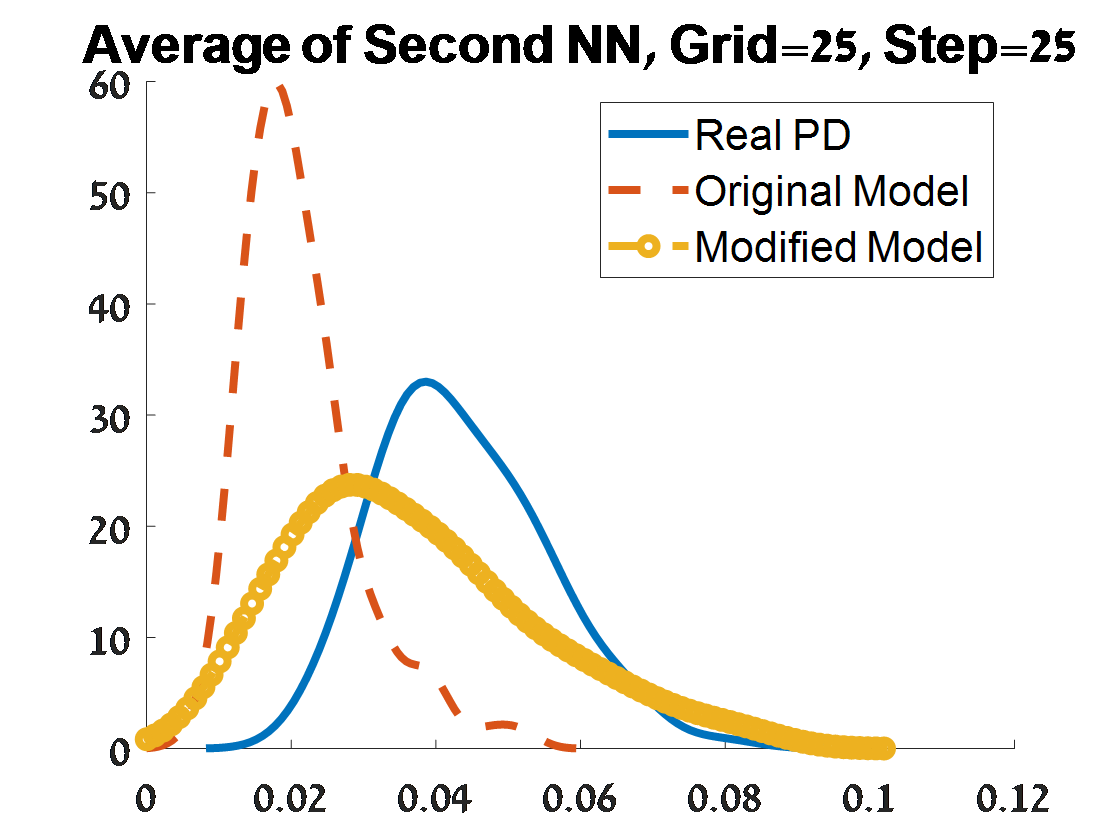}
\includegraphics[width=1.2in, height=1.25in]{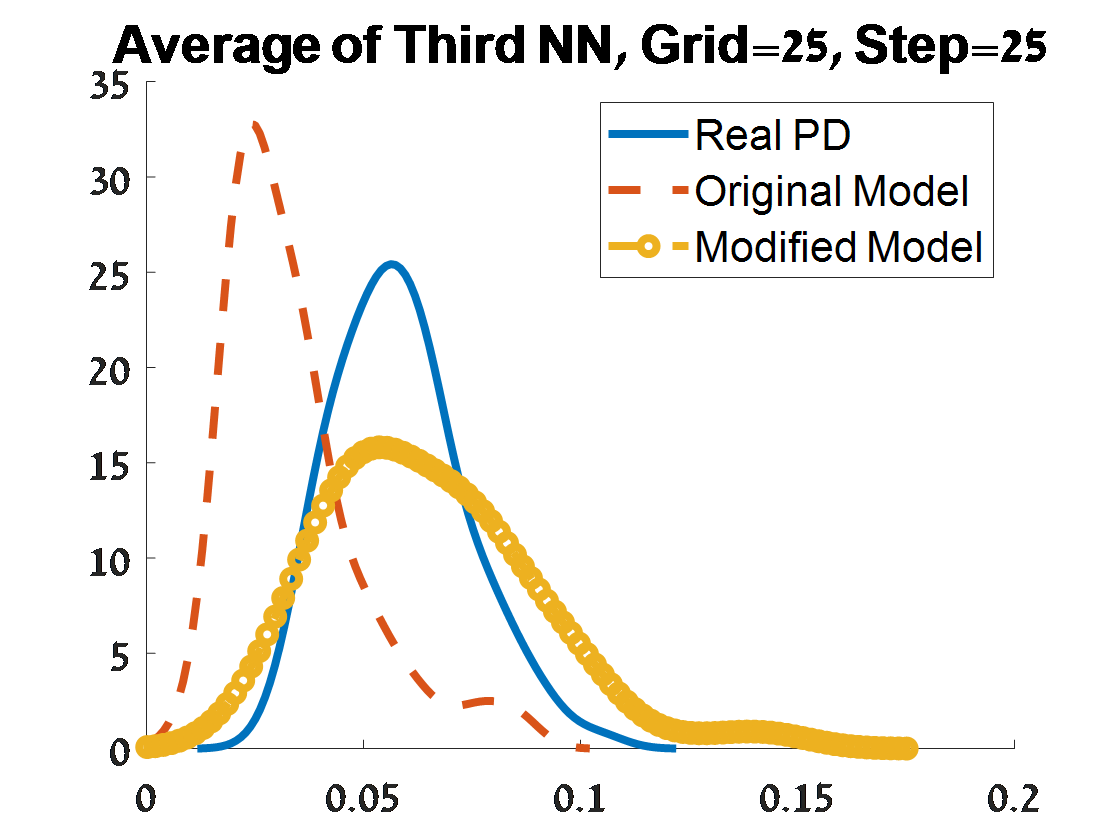}
\includegraphics[width=1.2in, height=1.25in]{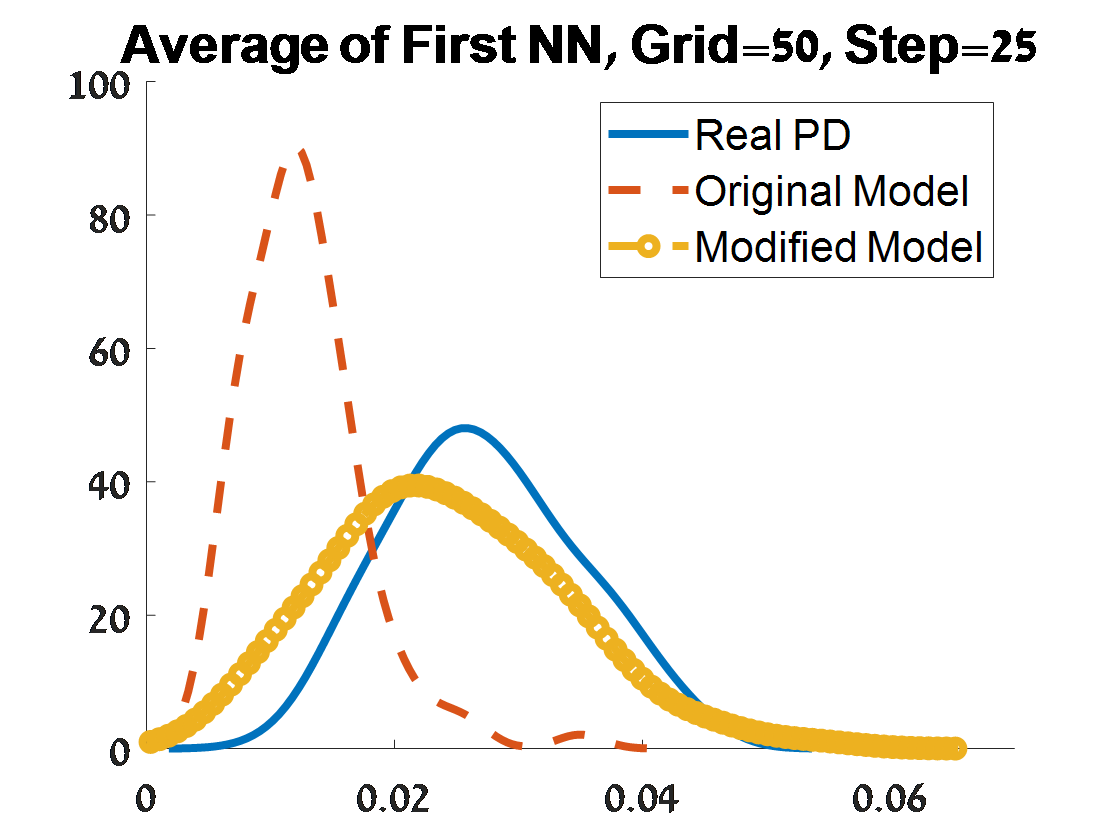}
\includegraphics[width=1.2in, height=1.25in]{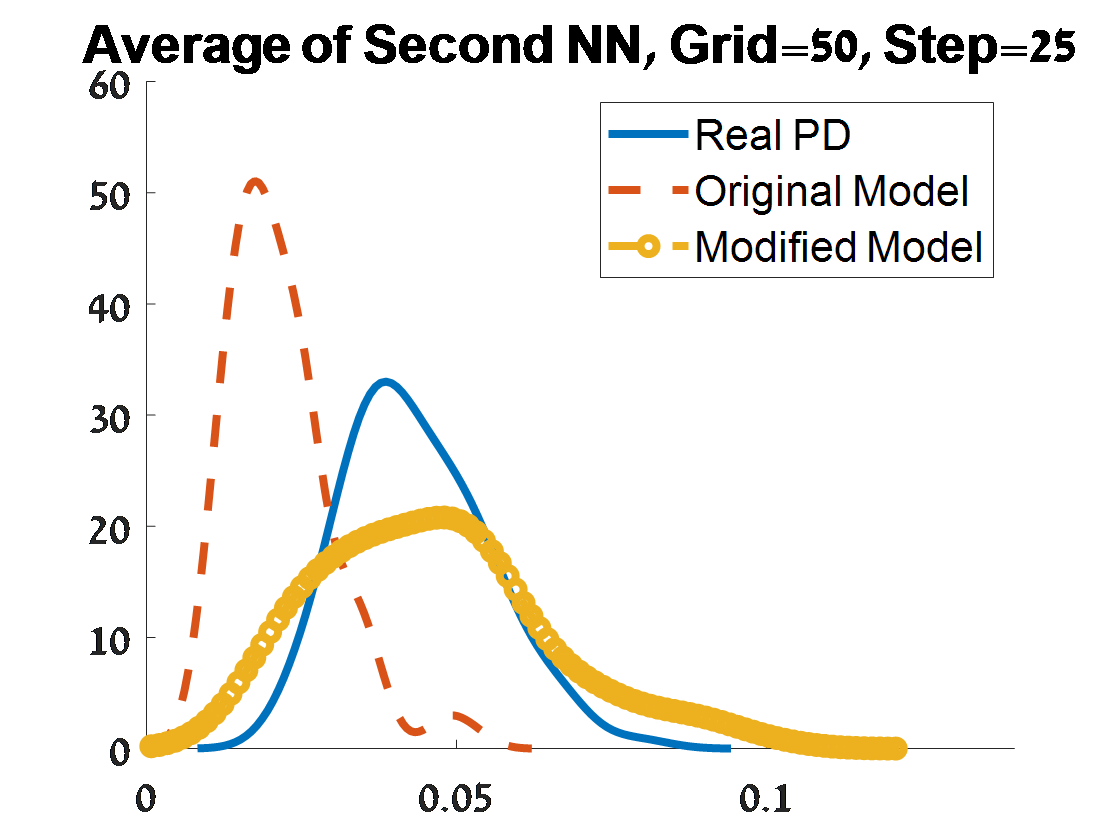}
\includegraphics[width=1.2in, height=1.25in]{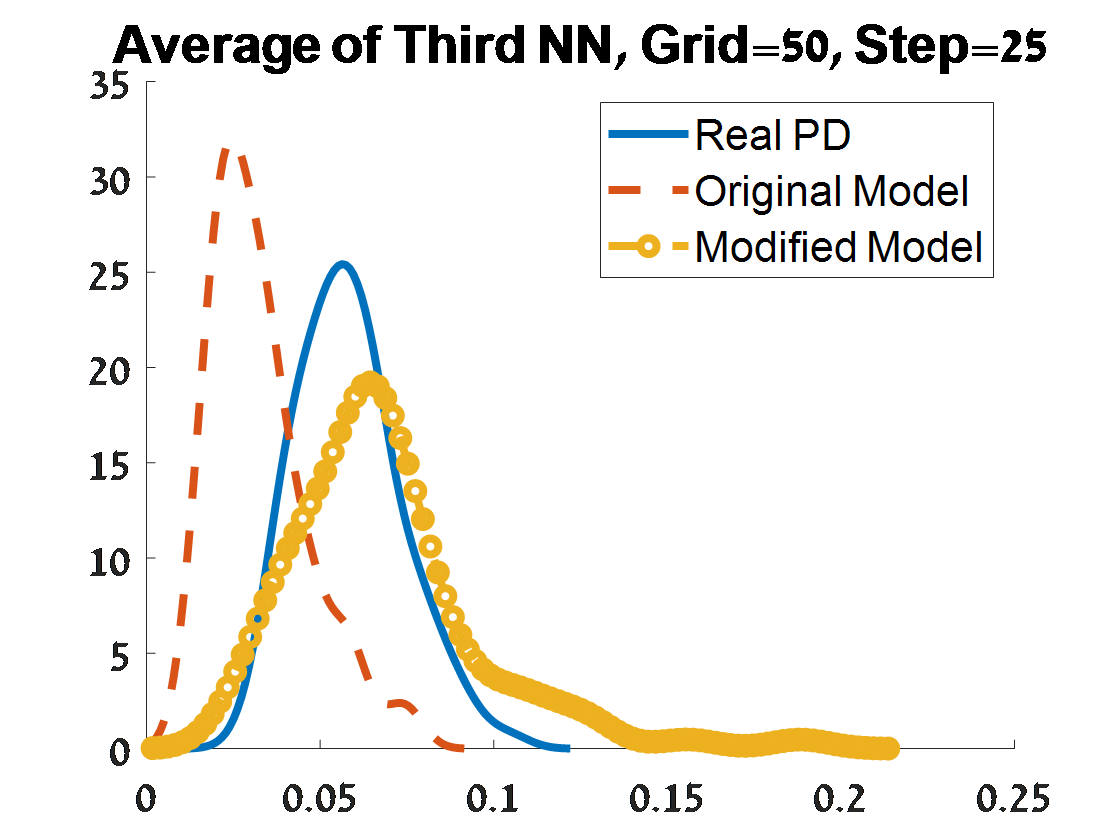}
\includegraphics[width=1.2in, height=1.25in]{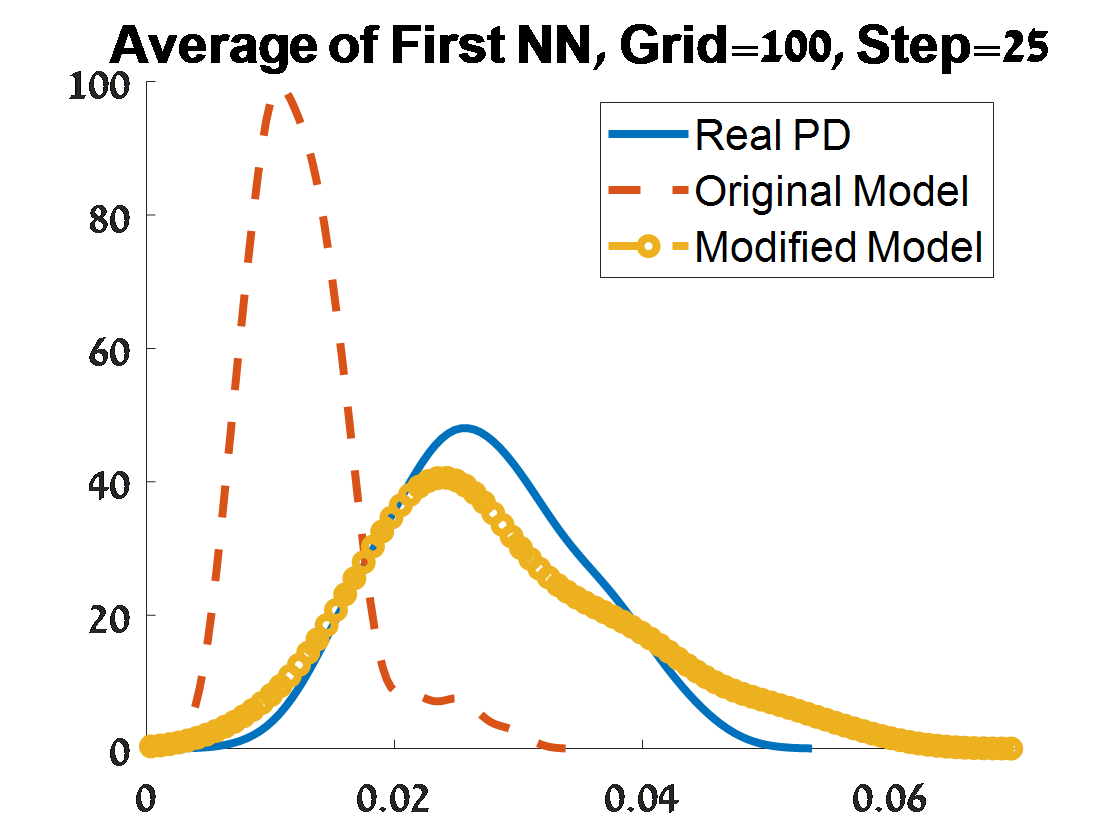}
\includegraphics[width=1.2in, height=1.25in]{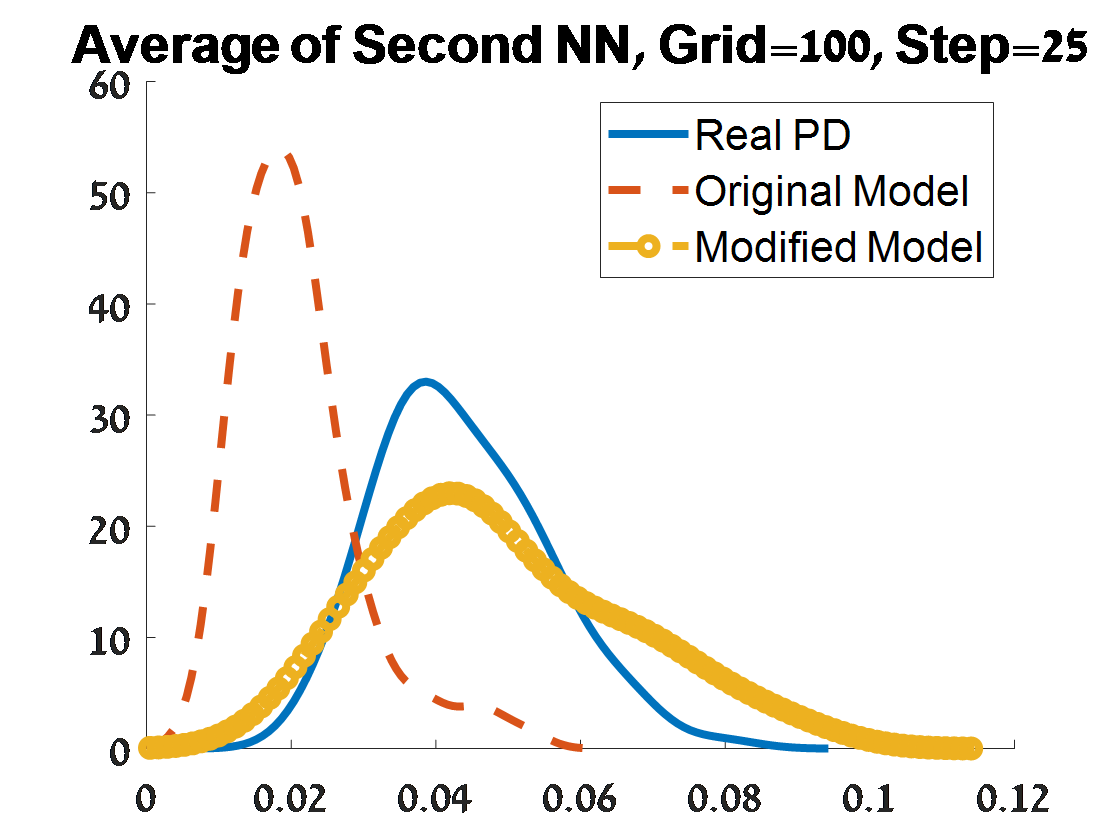}
\includegraphics[width=1.2in, height=1.25in]{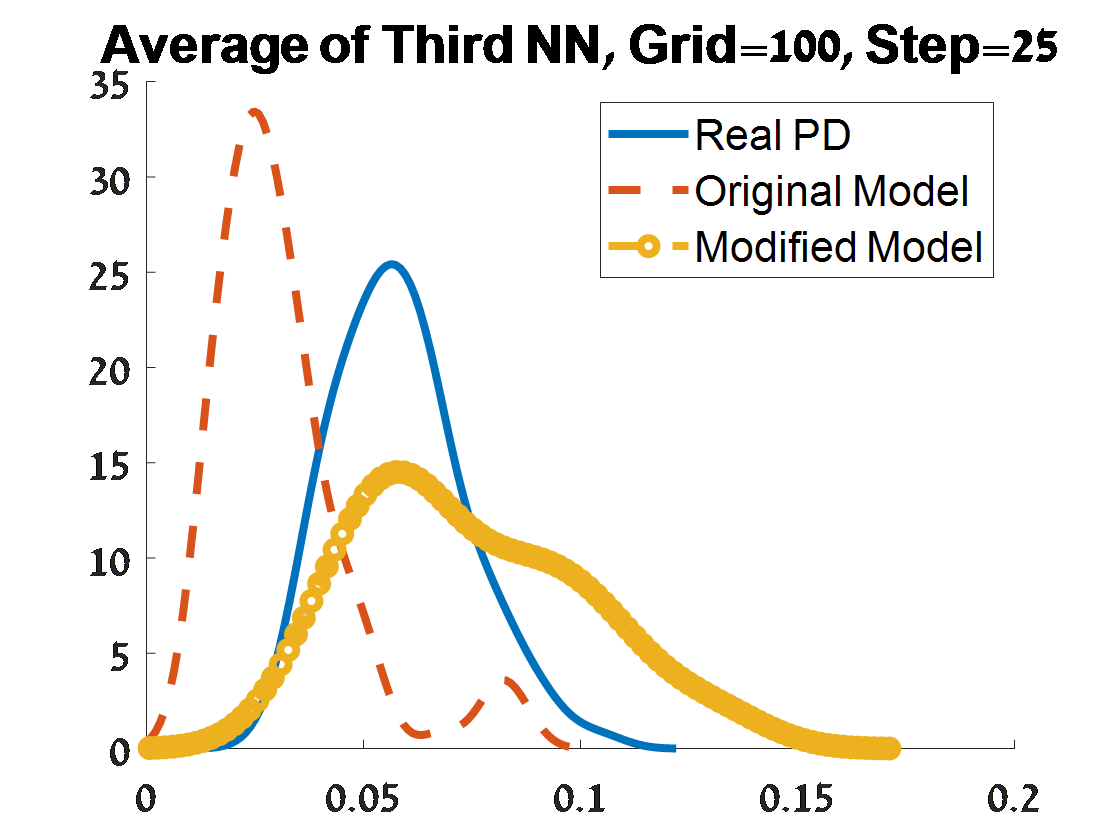}
\\
\includegraphics[width=1.2in, height=1.25in]{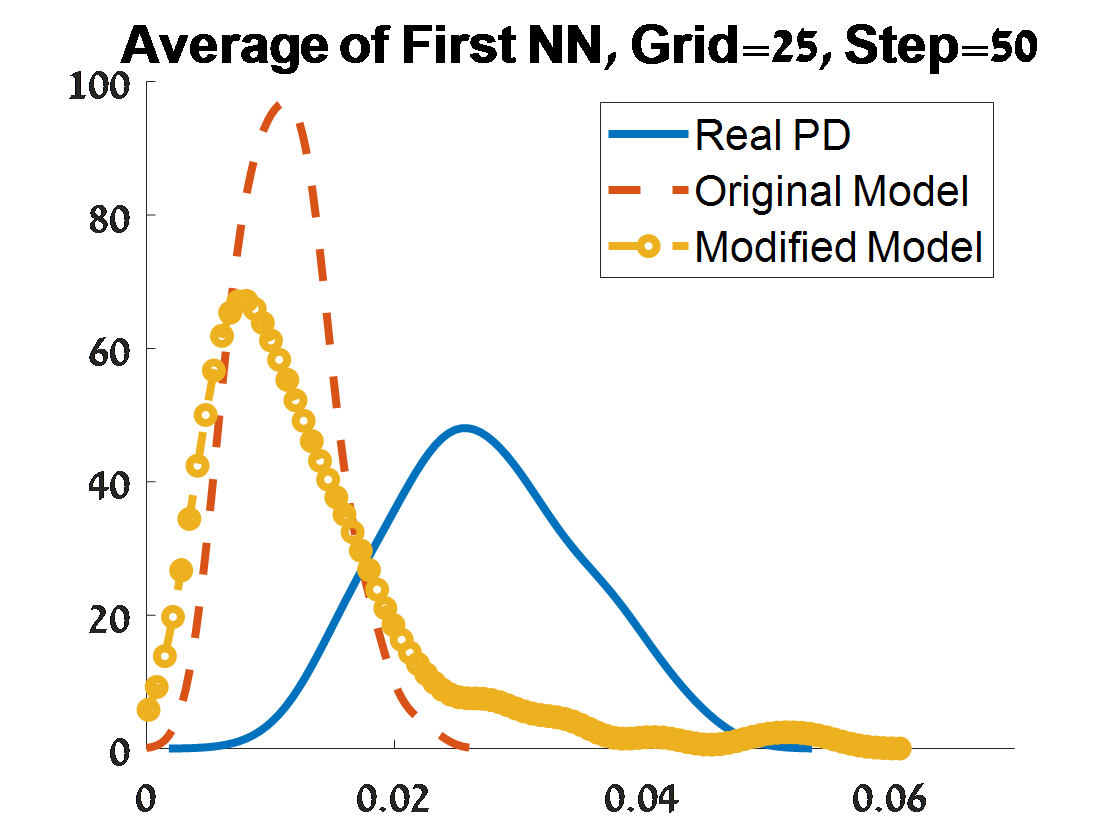}
\includegraphics[width=1.2in, height=1.25in]{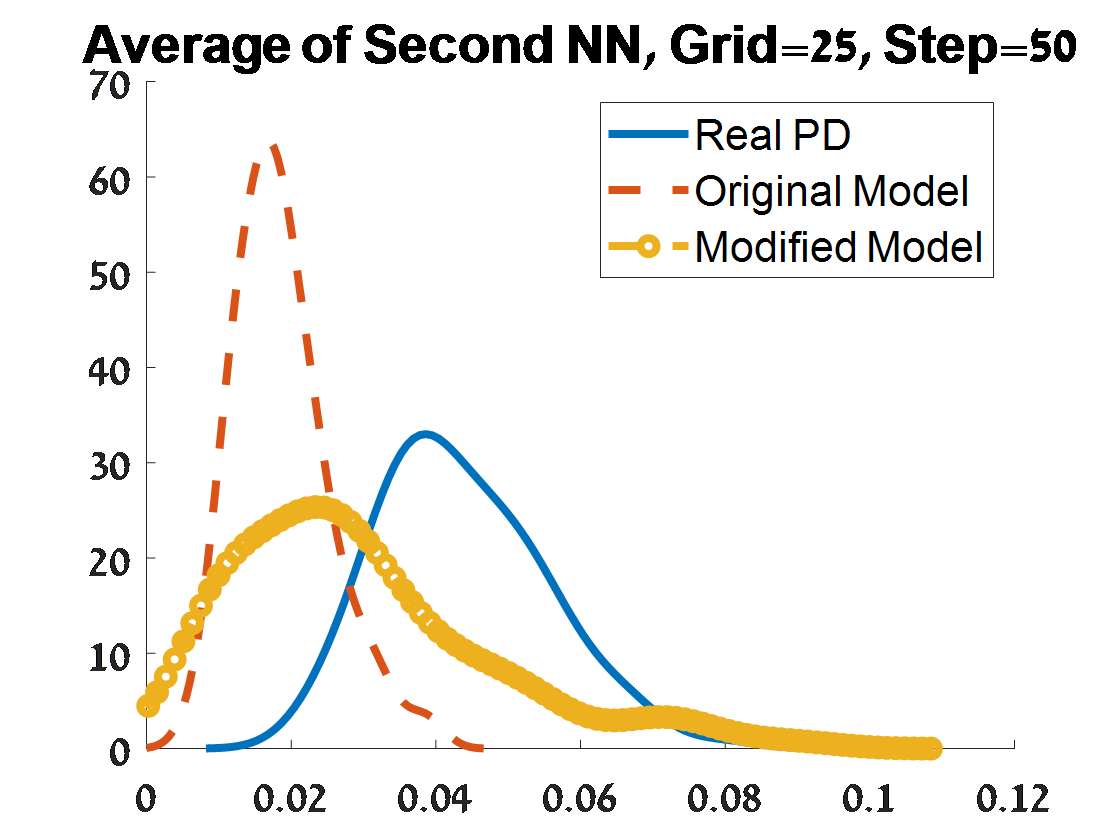}
\includegraphics[width=1.2in, height=1.25in]{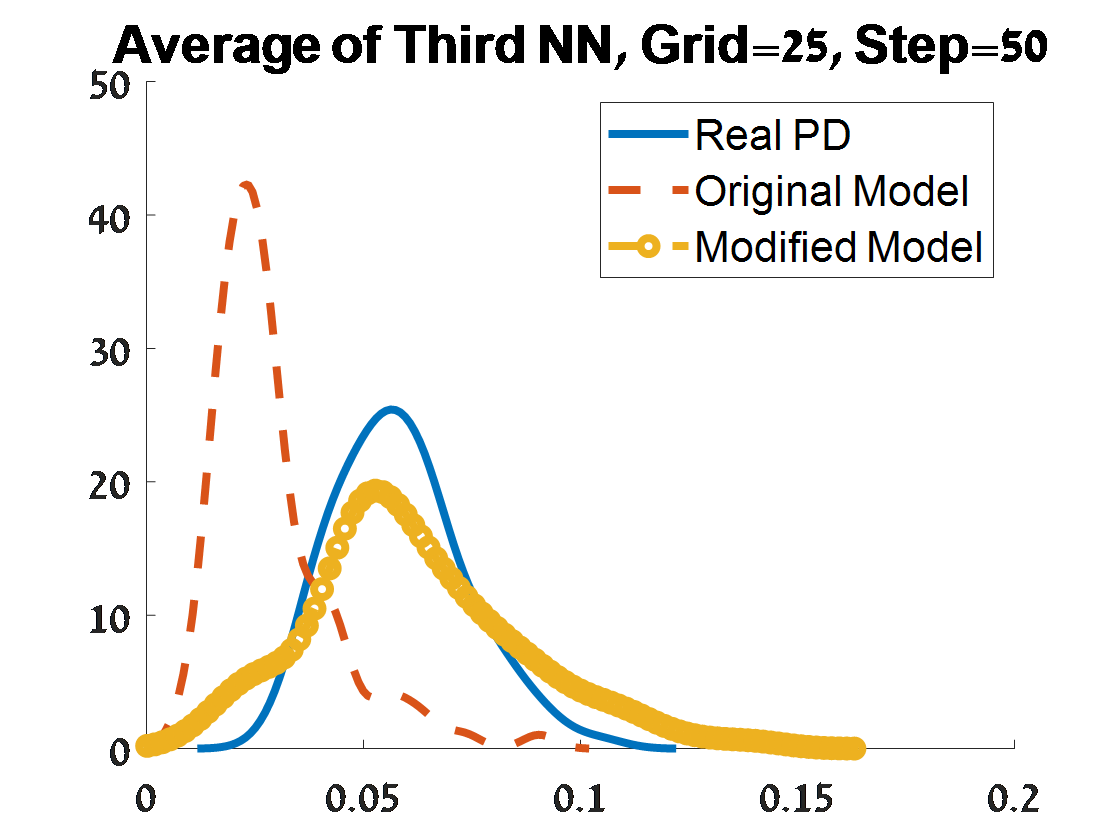}
\includegraphics[width=1.2in, height=1.25in]{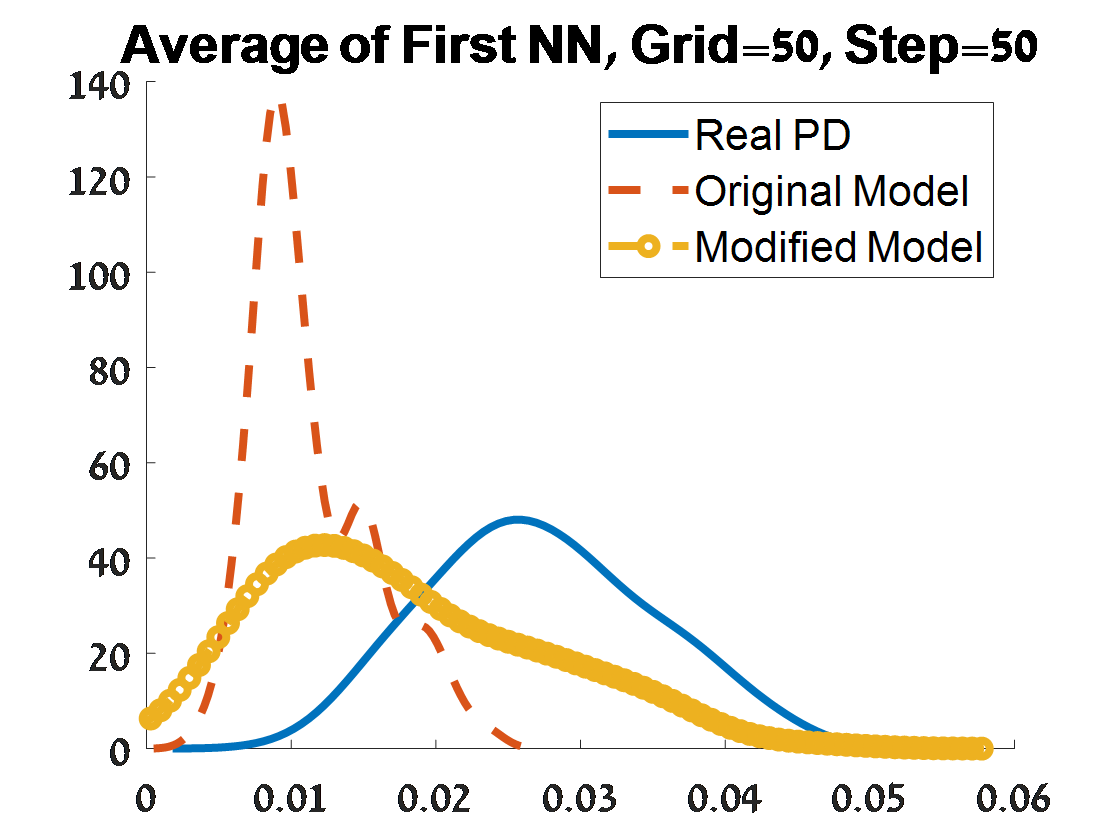}
\includegraphics[width=1.2in, height=1.25in]{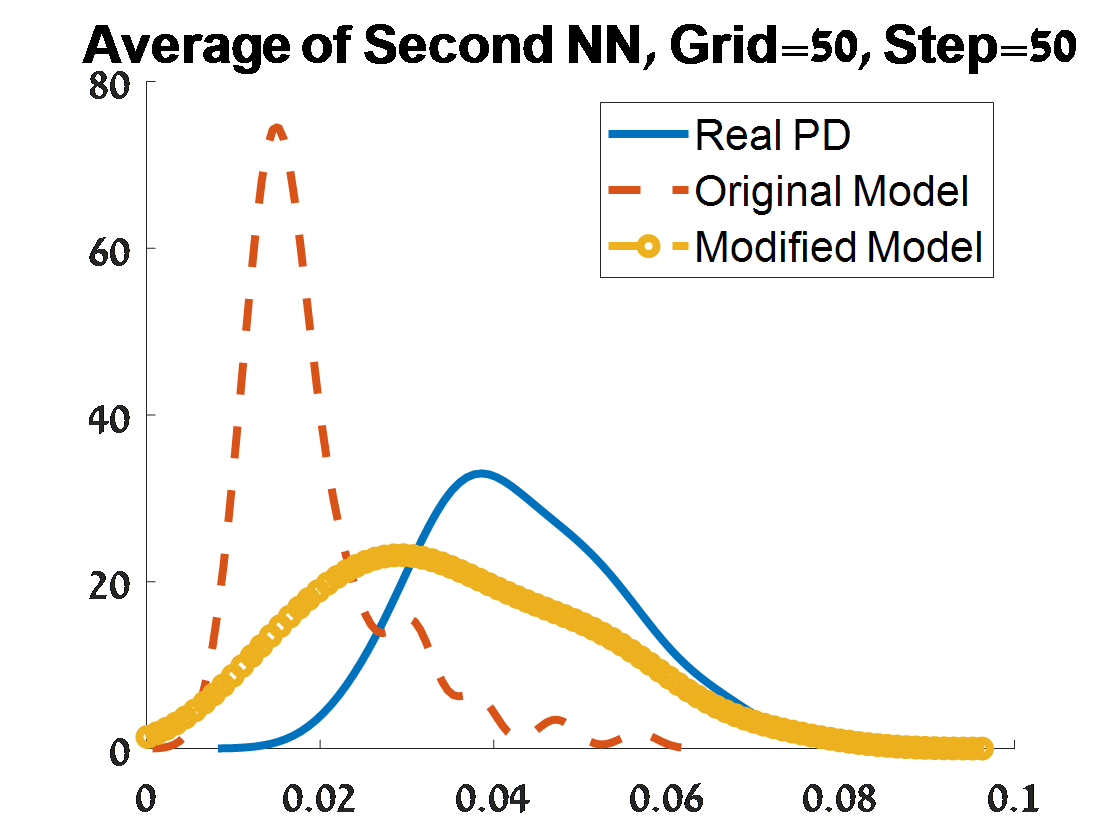}
\includegraphics[width=1.2in, height=1.25in]{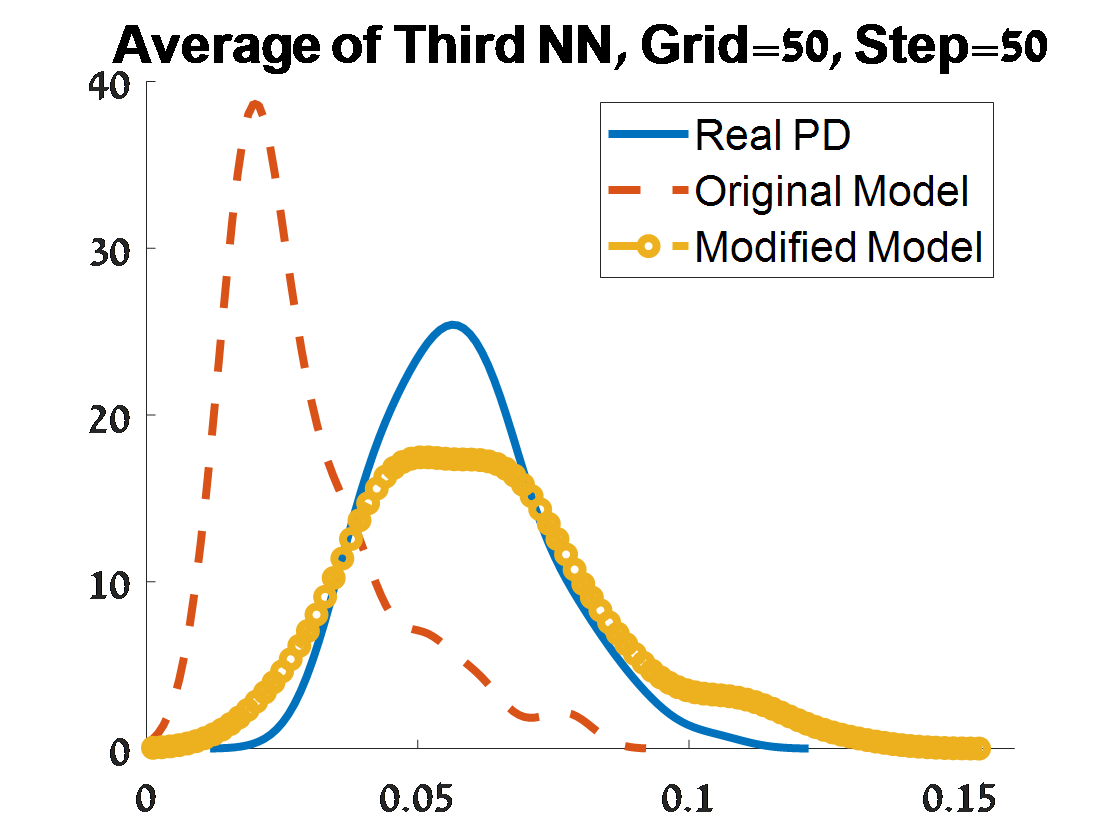}
\includegraphics[width=1.2in, height=1.25in]{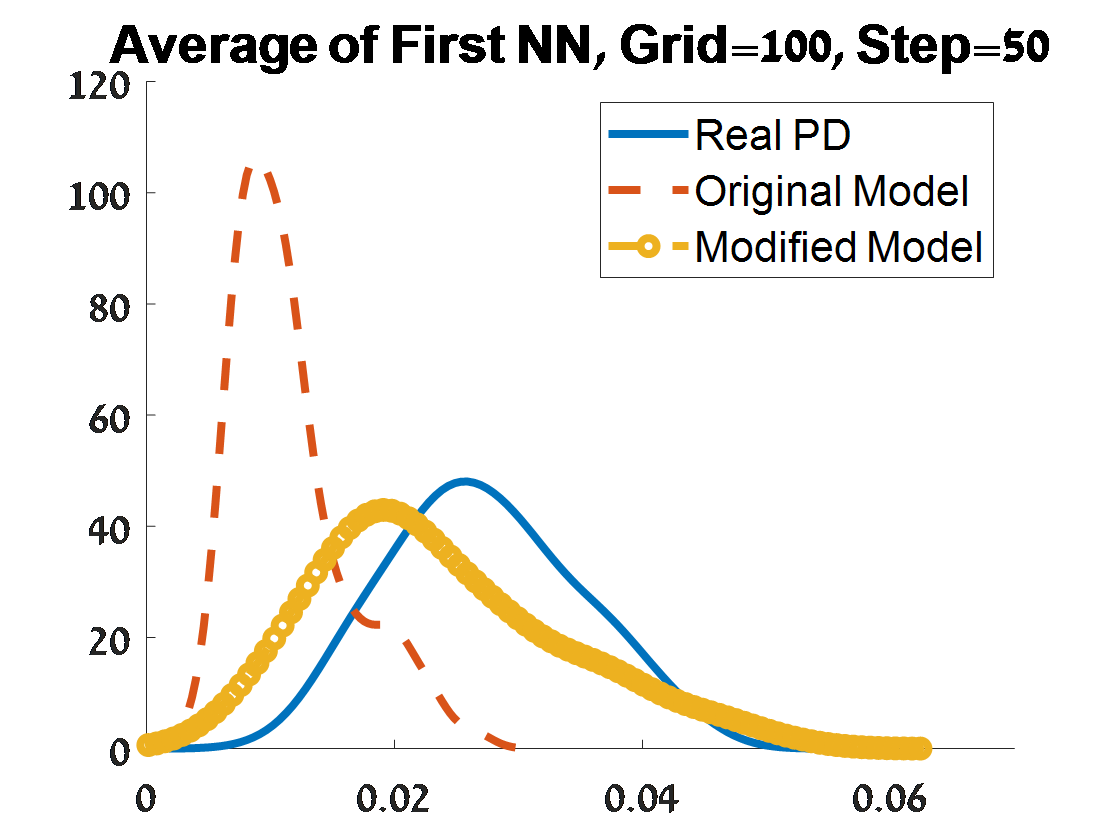}
\includegraphics[width=1.2in, height=1.25in]{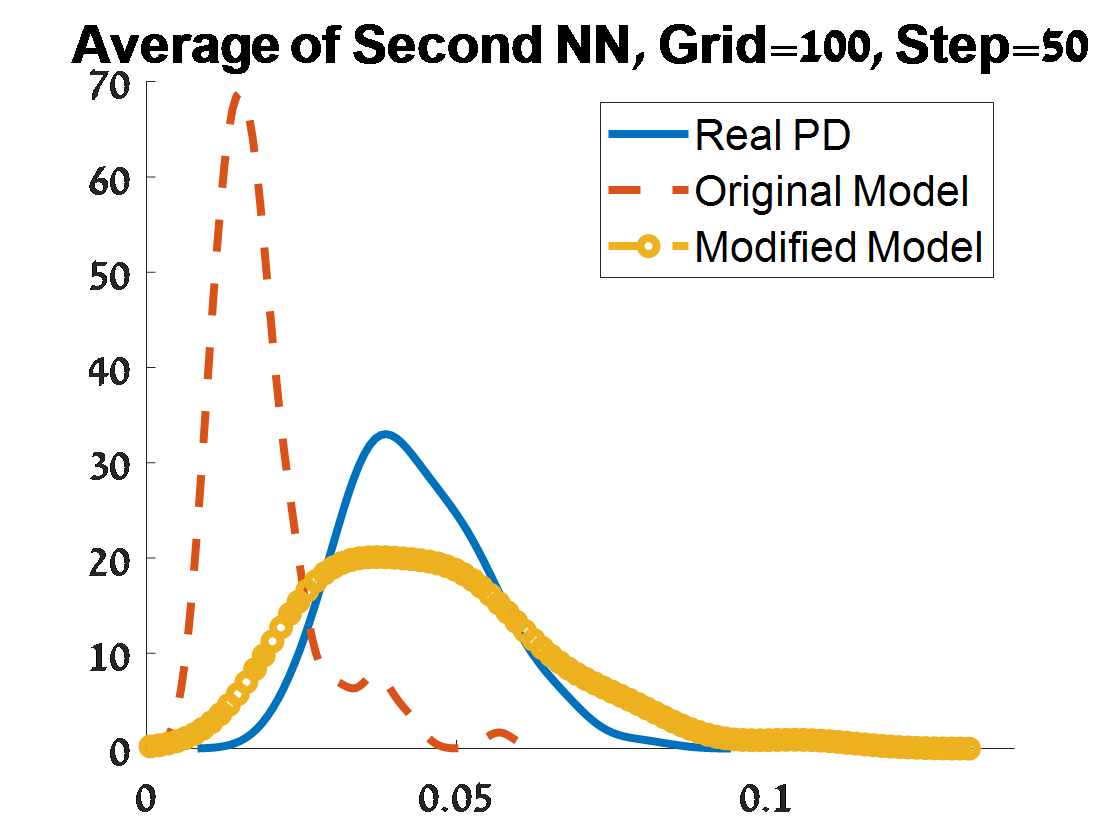}
\includegraphics[width=1.2in, height=1.25in]{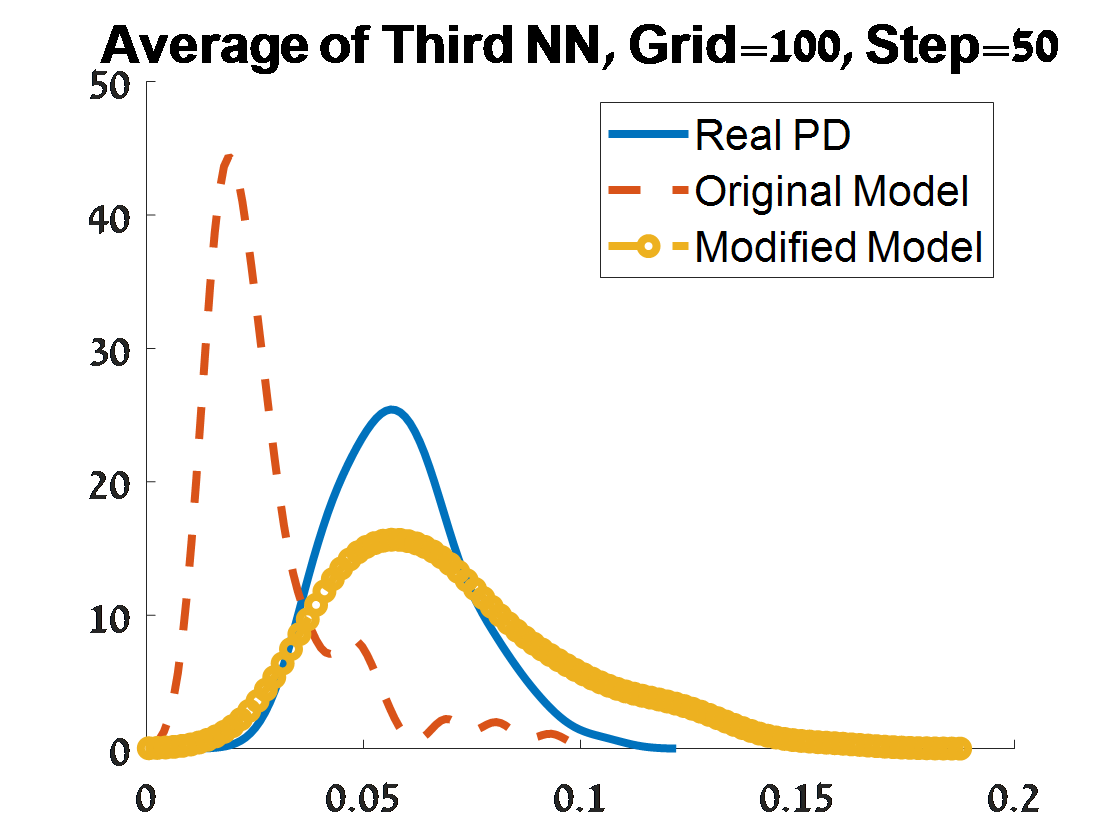}
\ec
%\caption{\footnotesize
% A random sample from two circles, 500 points from the larger circle and 300 from the smaller one,  with a kernel density
\caption{\footnotesize
 Criterion 2 of goodness of fit for 100 PDs corresponded to 100 samples from a unit circle. The plots depend on the grid size of the proposal distribution ("Grid"), and the burn-in ("Step") of the MCMC algorithm. }
\label{fig:circle_b}
\end{figure}
\end{landscape}

\begin{landscape}
\begin{figure}[h!]
\bc
\includegraphics[width=1.2in, height=1.25in]{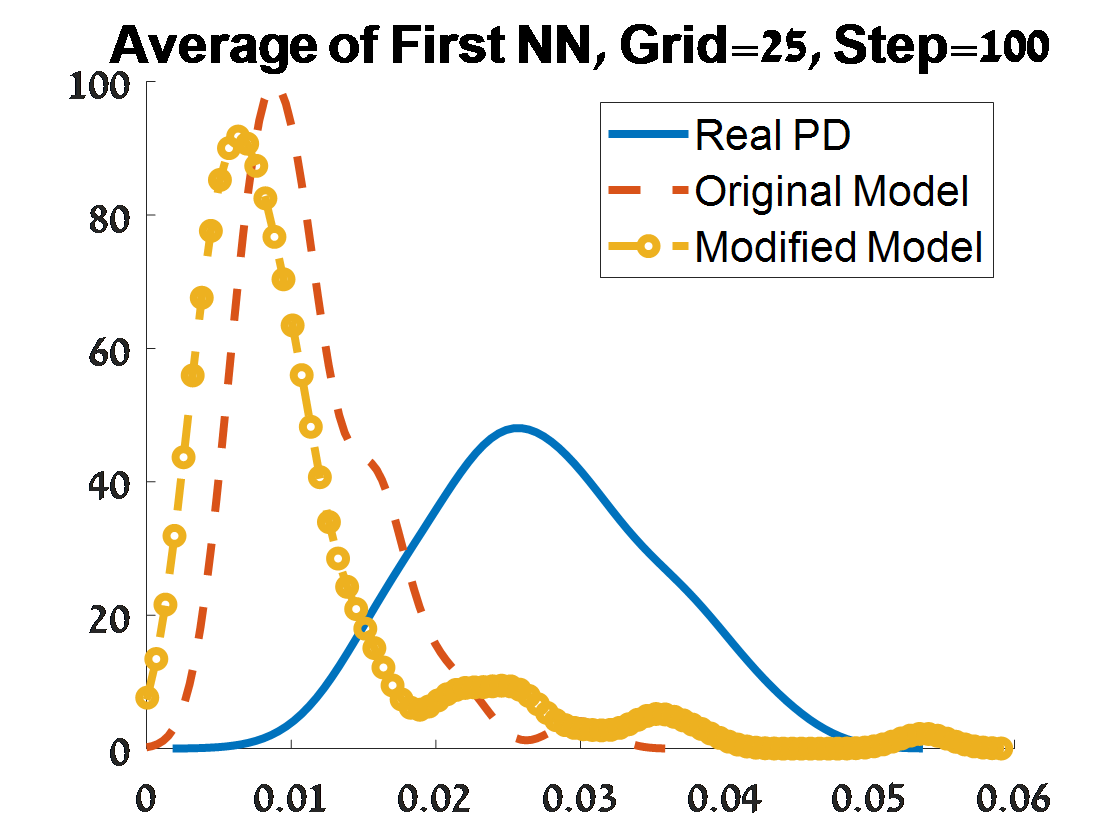}
\includegraphics[width=1.2in, height=1.25in]{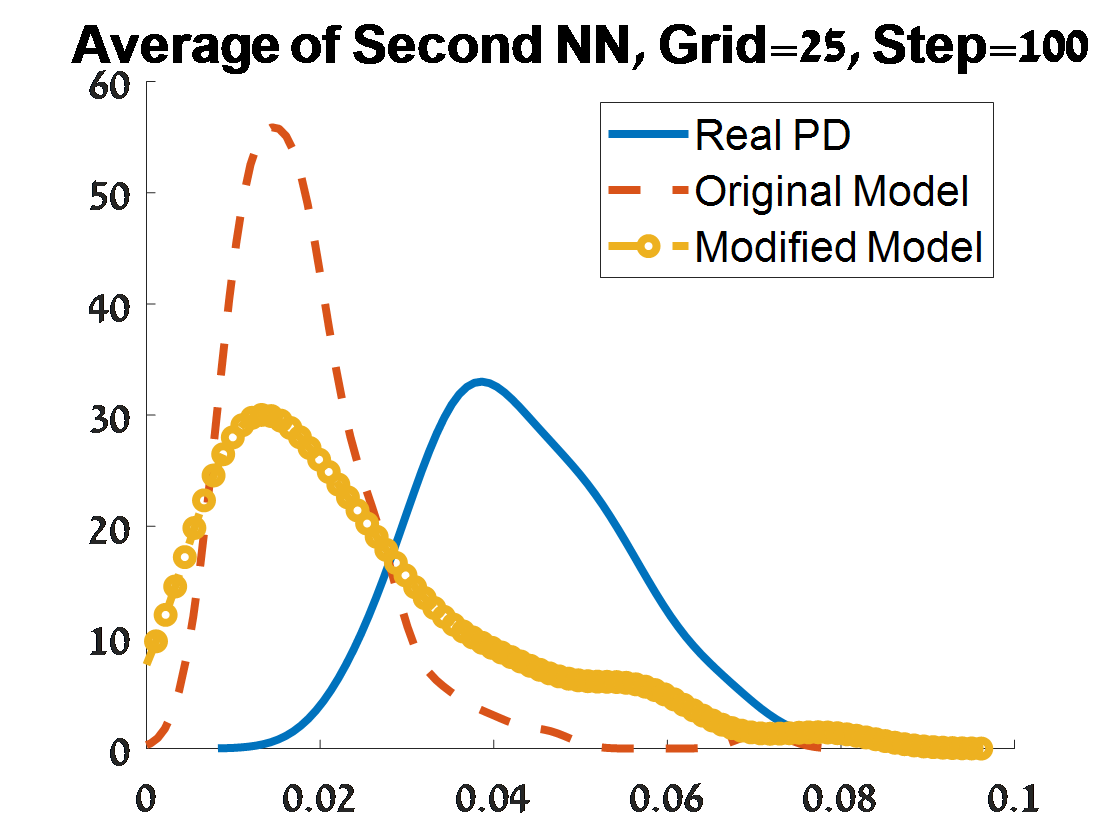}
\includegraphics[width=1.2in, height=1.25in]{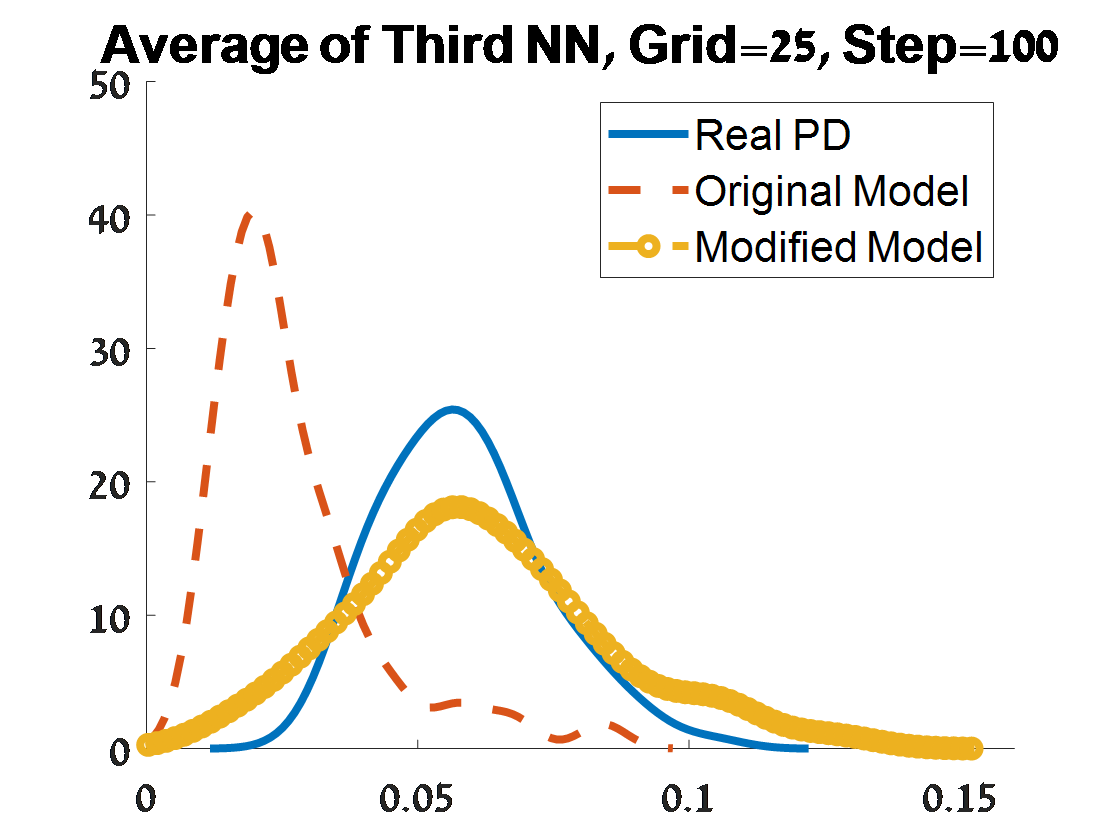}
\includegraphics[width=1.2in, height=1.25in]{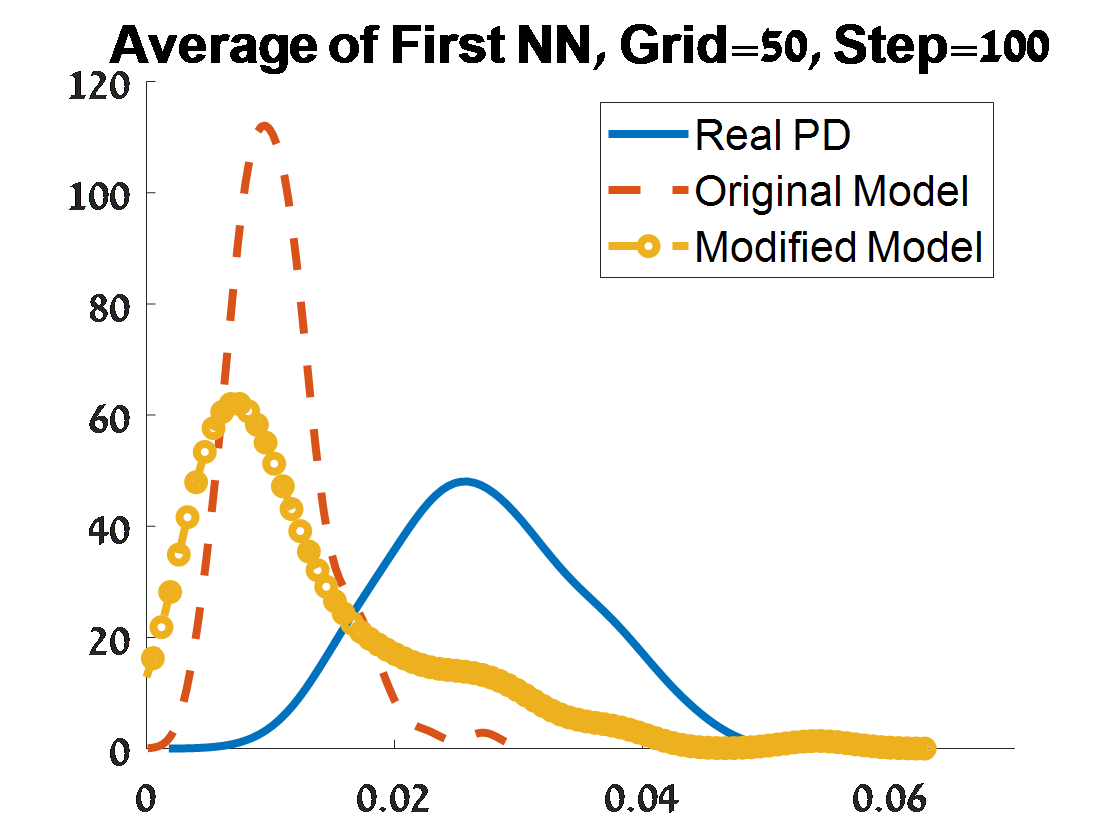}
\includegraphics[width=1.2in, height=1.25in]{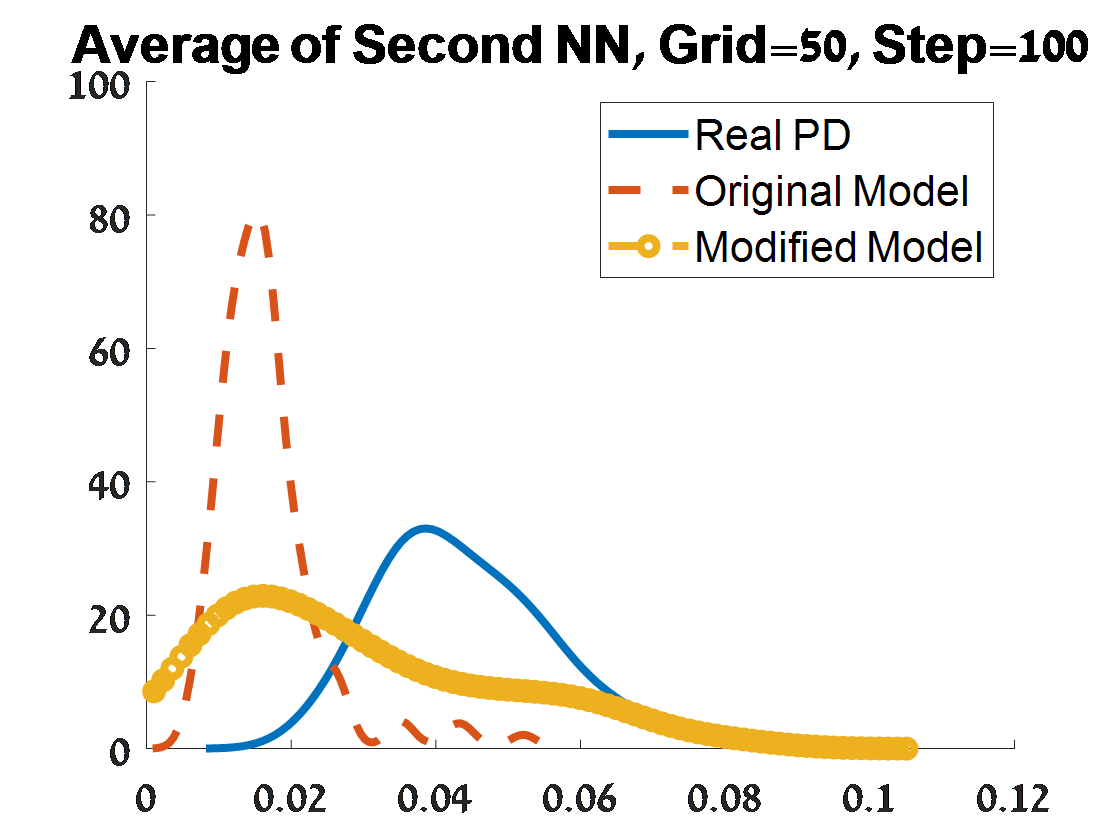}
\includegraphics[width=1.2in, height=1.25in]{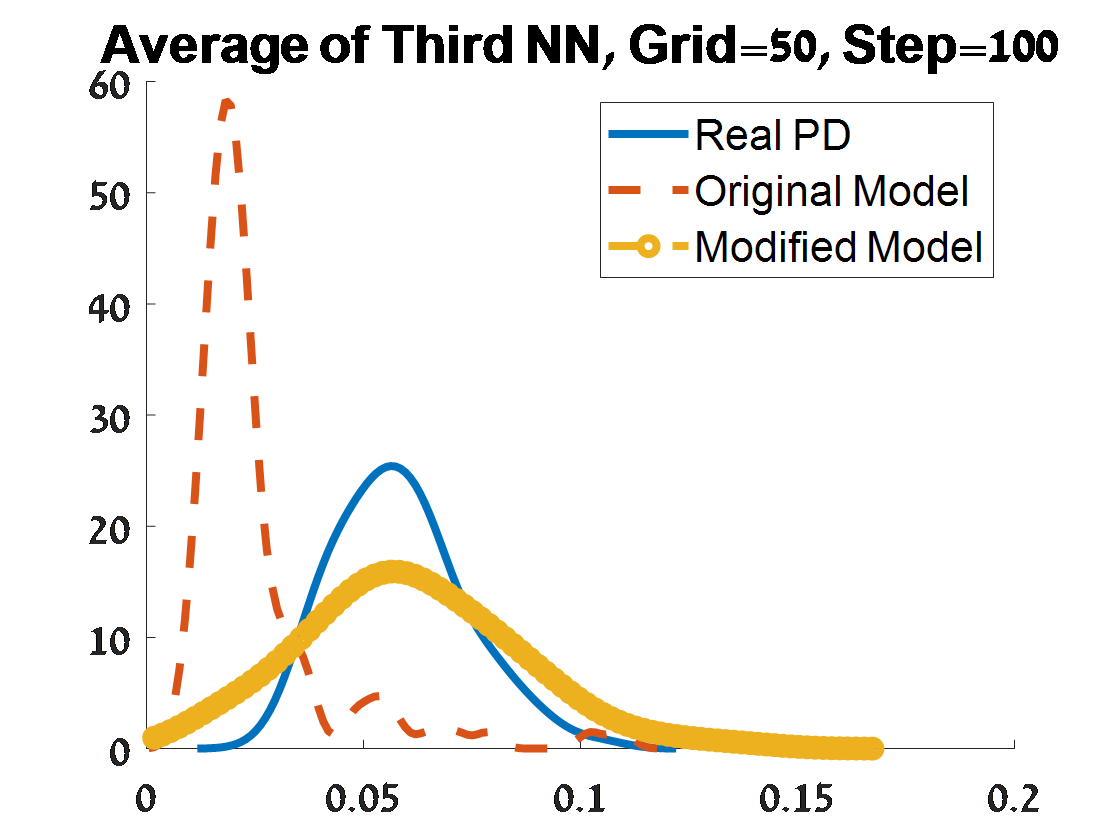}
\includegraphics[width=1.2in, height=1.25in]{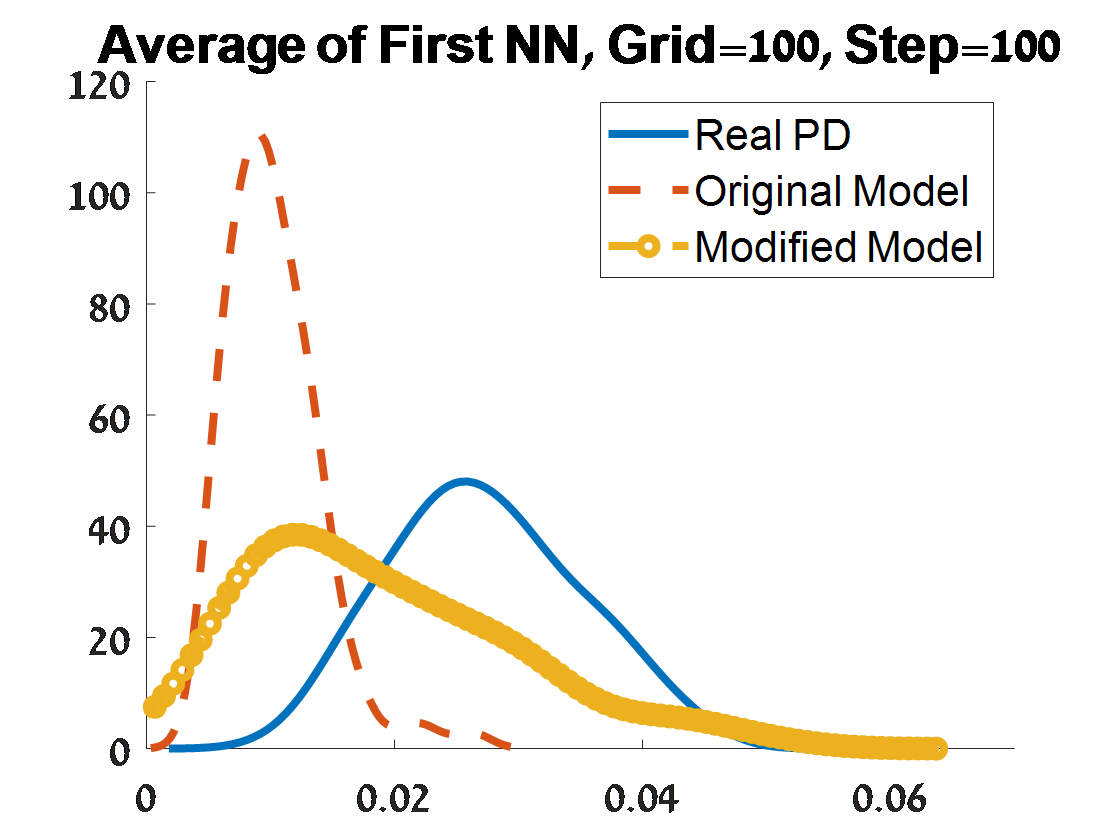}
\includegraphics[width=1.2in, height=1.25in]{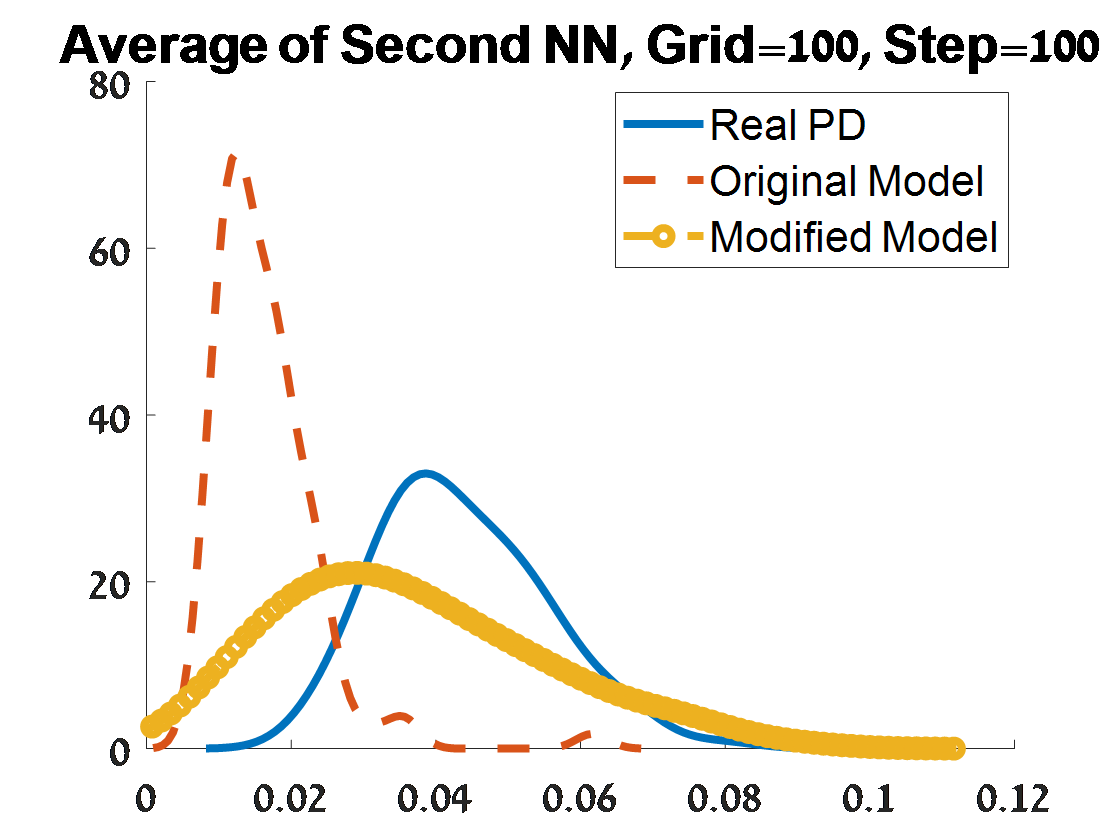}
\includegraphics[width=1.2in, height=1.25in]{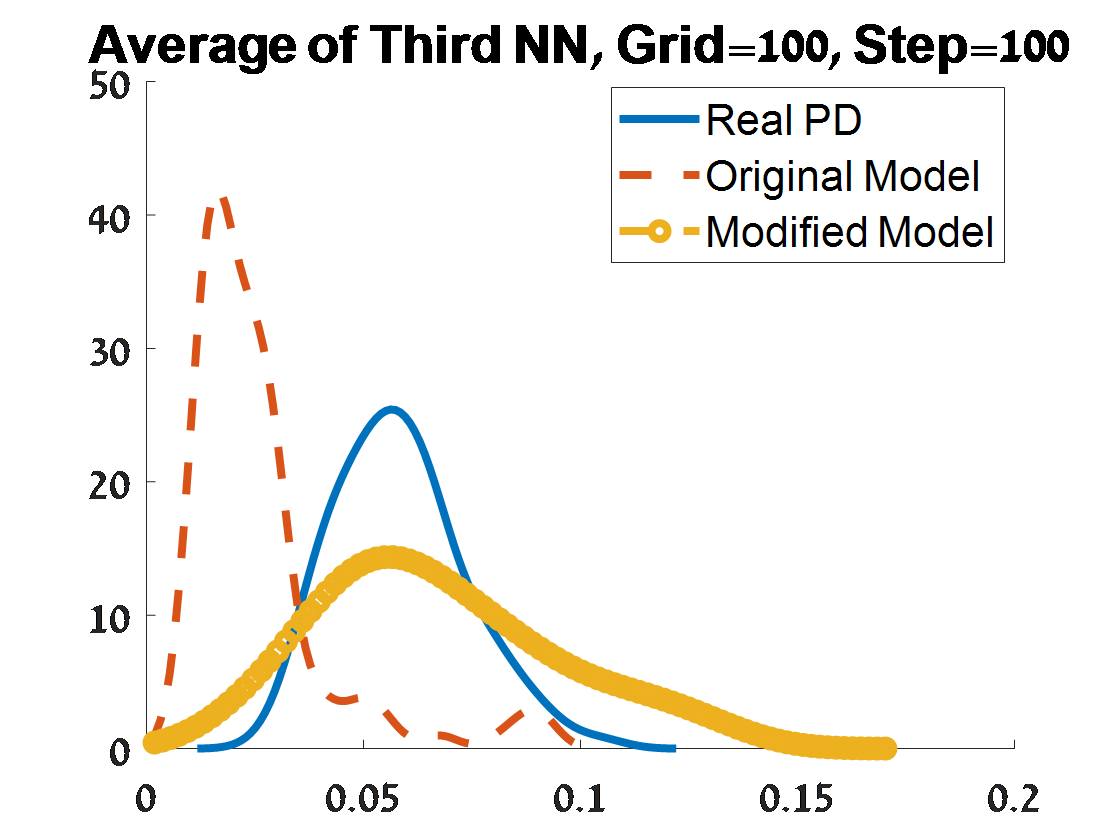}
\ec
%\caption{\footnotesize
% A random sample from two circles, 500 points from the larger circle and 300 from the smaller one,  with a kernel density
\caption{\footnotesize
 Continue of Criterion 2 of goodness of fit for 100 PDs corresponded to 100 samples from a unit circle. The plots depend on the grid size of the proposal distribution ("Grid"), and the burn-in ("Step") of the MCMC algorithm.}
\label{fig:circle_c}
\end{figure}
\end{landscape}

%\begin{landscape}
%\begin{figure}[h!]
%\bc
%\includegraphics[width=1.8in, height=1.8in]{OneCircle_pd30_grid50_step25}
%\includegraphics[width=1.8in, height=1.8in]{OneCircle_pd30_grid100_step25}
%\includegraphics[width=1.8in, height=1.8in]{OneCircle_pd60_grid50_step25}
%\includegraphics[width=1.8in, height=1.8in]{OneCircle_pd60_grid100_step25}
%\ec
%\caption{\footnotesize
% Examples of two PDs (from the 100  PDs we used), each one is corresponded to a sample from a unit circle. For each PD, the simulated PD are based on the two model's versions. The plots depend on the grid size of the proposal distribution ("Grid"), and the burn-in ("Step") of the MCMC algorithm.}
%\label{fig:circle_d}
%\end{figure}
%\end{landscape}
\subsection{Two geometrical objects}
\begin{figure}[h!]
\bc
\includegraphics[width=2in, height=2in]{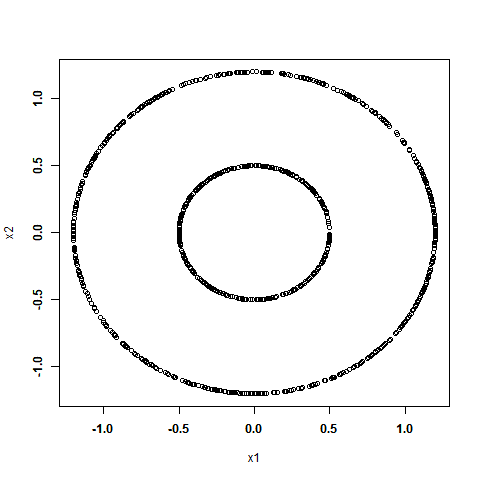}
\includegraphics[width=2in, height=2in]{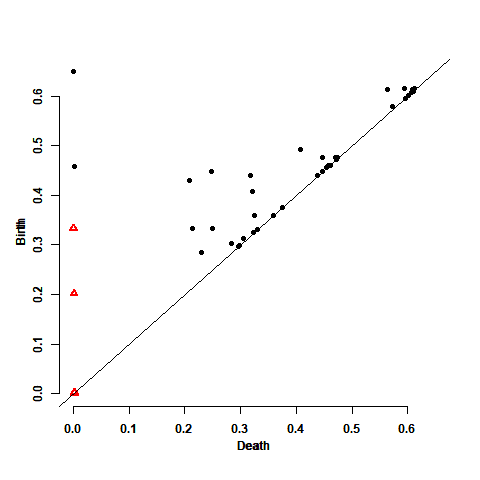}
\ec
%\caption{\footnotesize
% A random sample from two circles, 500 points from the larger circle and 300 from the smaller one,  with a kernel density
\caption{\footnotesize Left: A sample of $n=1,000$ points from two concentric circles. Right: The corresponded persistence diagram for its upper level sets. Black circles are connected components ($H_0$ persistence points), red triangles are holes ($H_1$ points). Birth times are on the vertical axis.}
\label{fig:concentric}
\end{figure}

In the contrary to the previous example that included one geometrical object, the following example describes two geometrical objects, and specifically two concentric circles: one circle has a radius $r_1=0.5$, and the second circle has a radius $r_2=1.2$. For a sample size of $n$ points from this geometrical object, the number of points of the smaller circle and the larger circle is $0.4n$ and $0.6n$, respectively. The both circles together obtain a smaller circle inside a larger one.
We consider a sample of $n=1,000$ points from this object. The typical object is presented in the left side of Figure\ \ref{fig:concentric}. Its corresponded persistence diagram is presented in the right side of Figure\ \ref{fig:concentric}.

\begin{landscape}
\begin{figure}[h!]
\bc
\includegraphics[width=1.2in, height=1.4in]{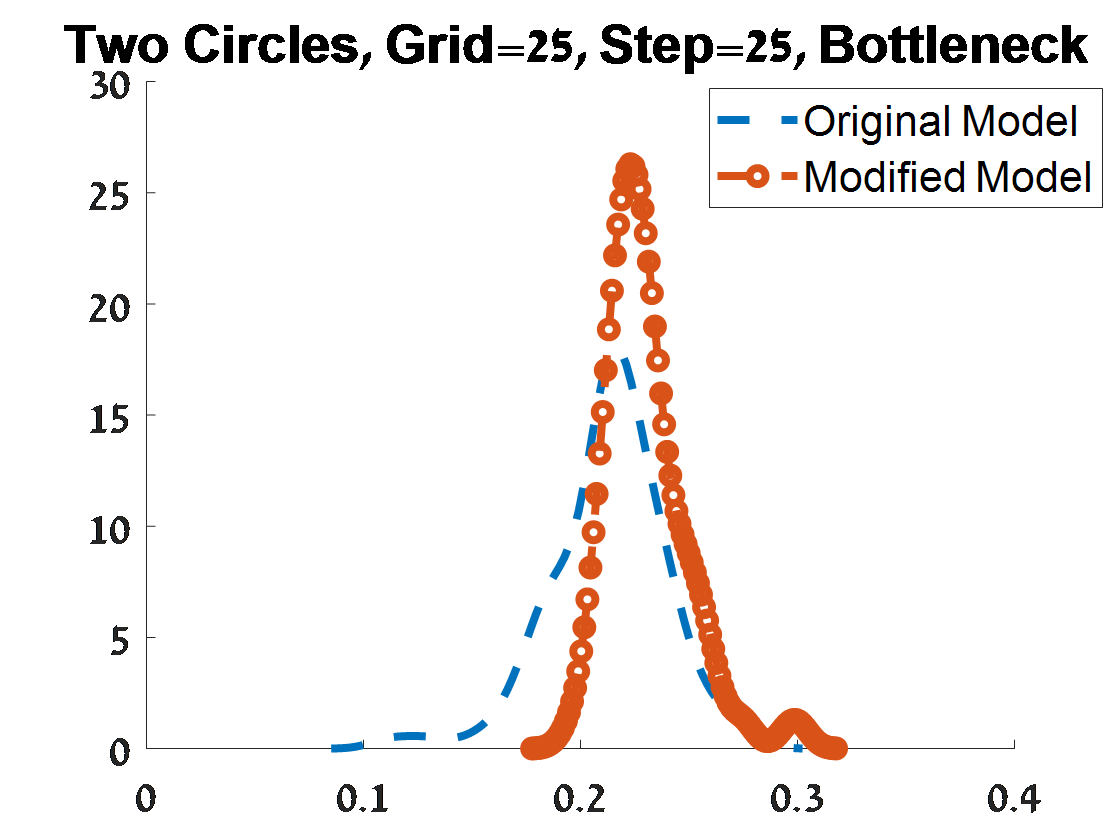}
\includegraphics[width=1.2in, height=1.4in]{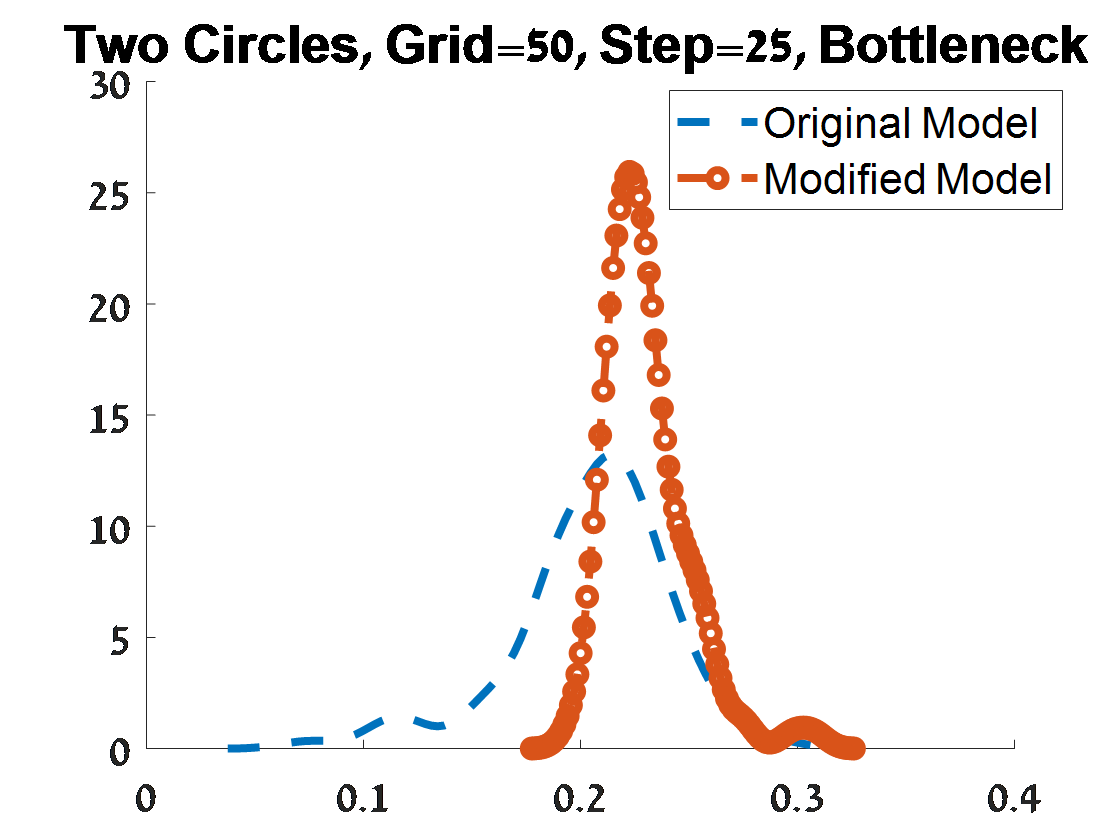}
\includegraphics[width=1.2in, height=1.4in]{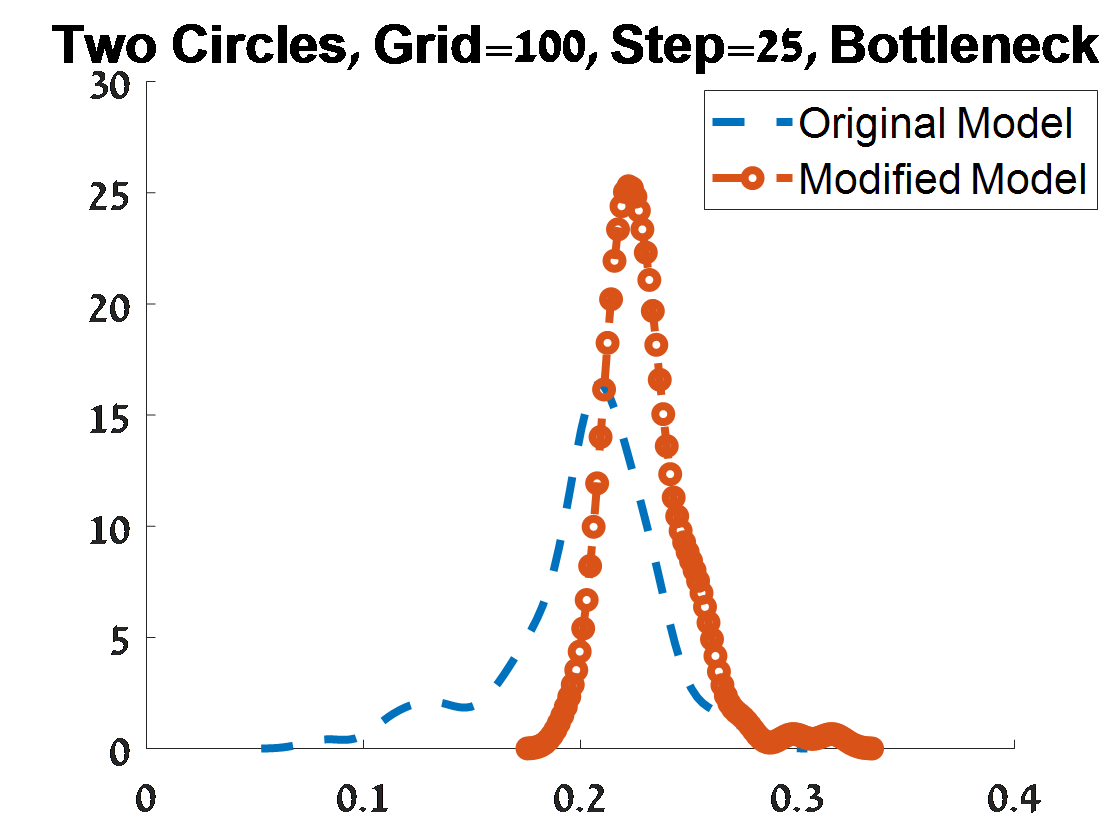}
\includegraphics[width=1.2in, height=1.4in]{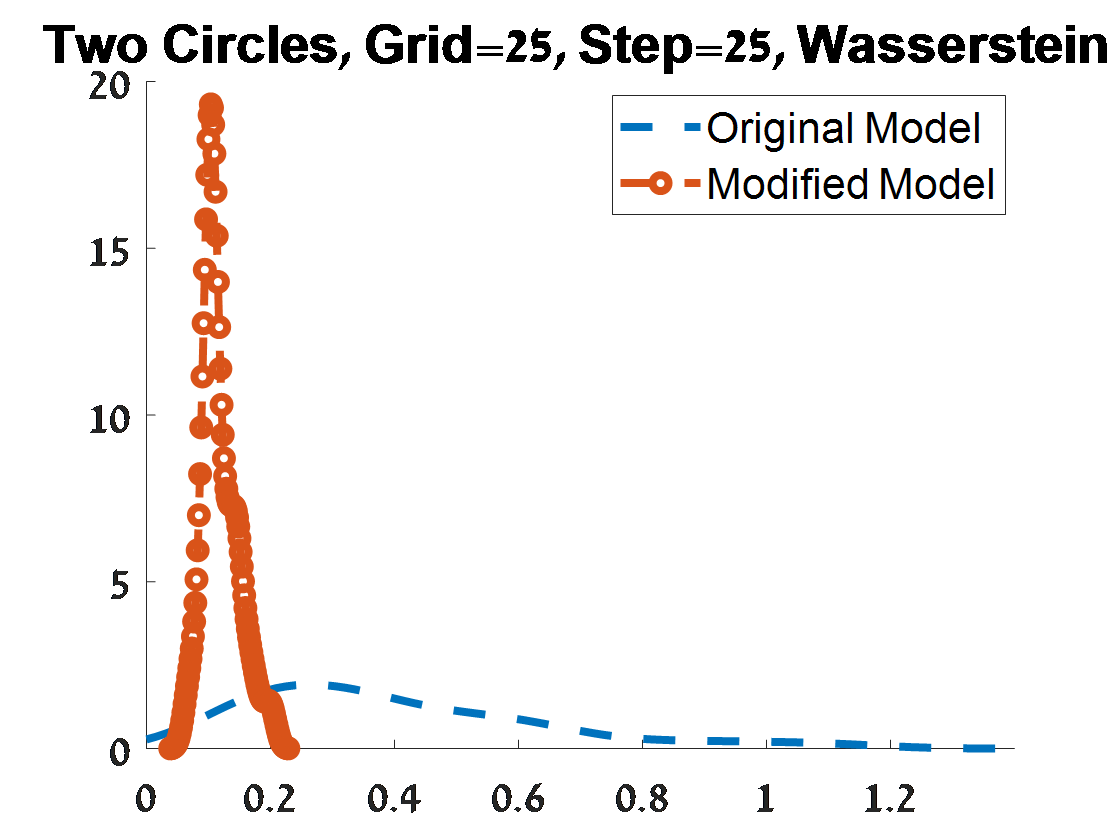}
\includegraphics[width=1.2in, height=1.4in]{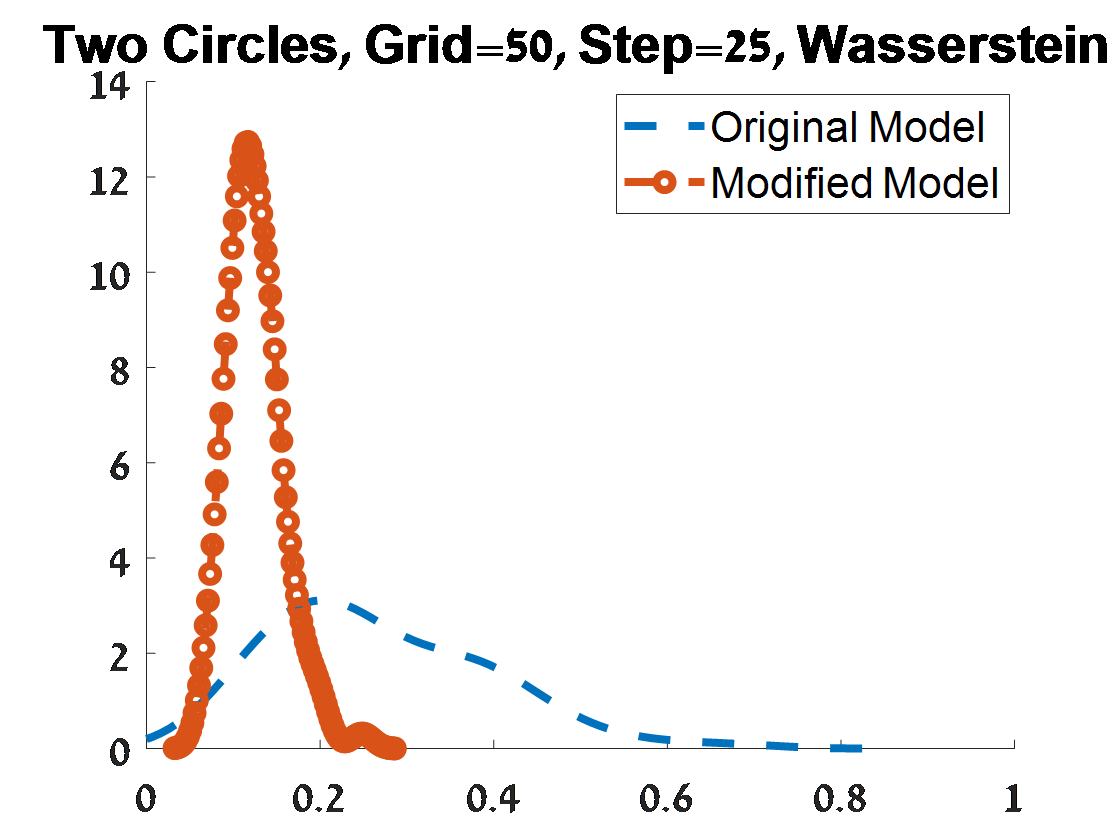}
\includegraphics[width=1.2in, height=1.4in]{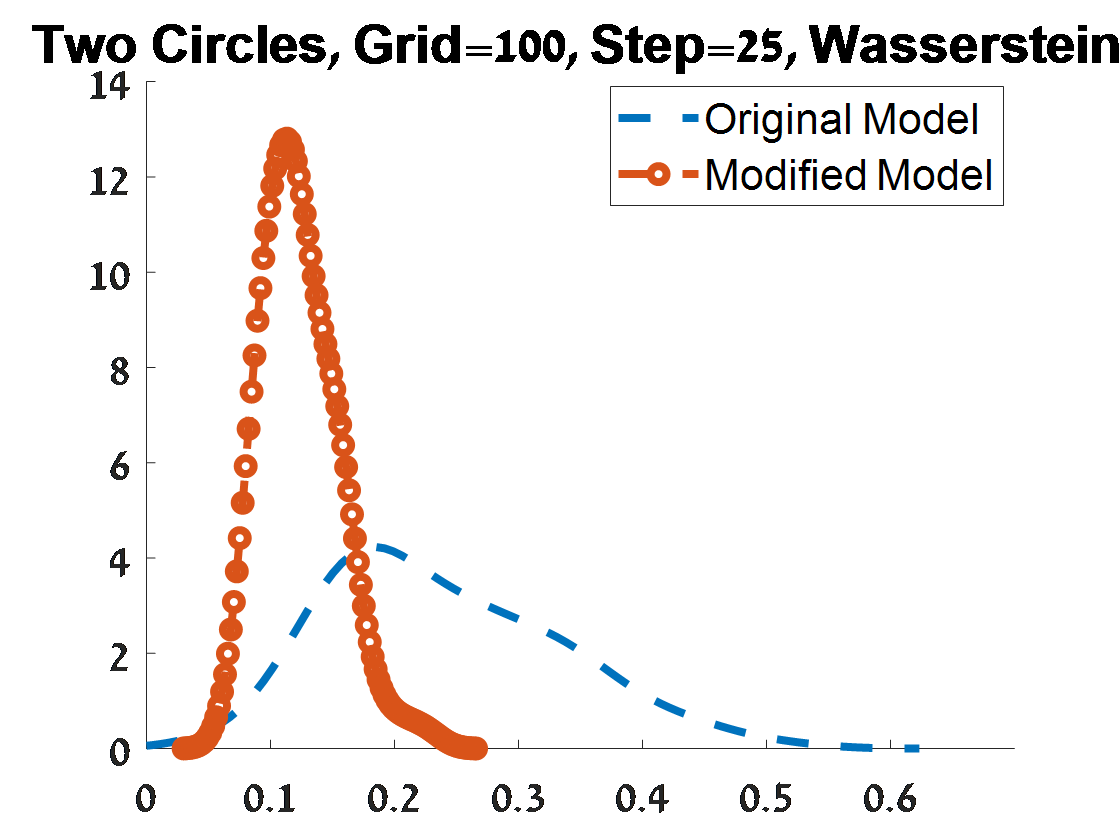}
\\
\includegraphics[width=1.2in, height=1.4in]{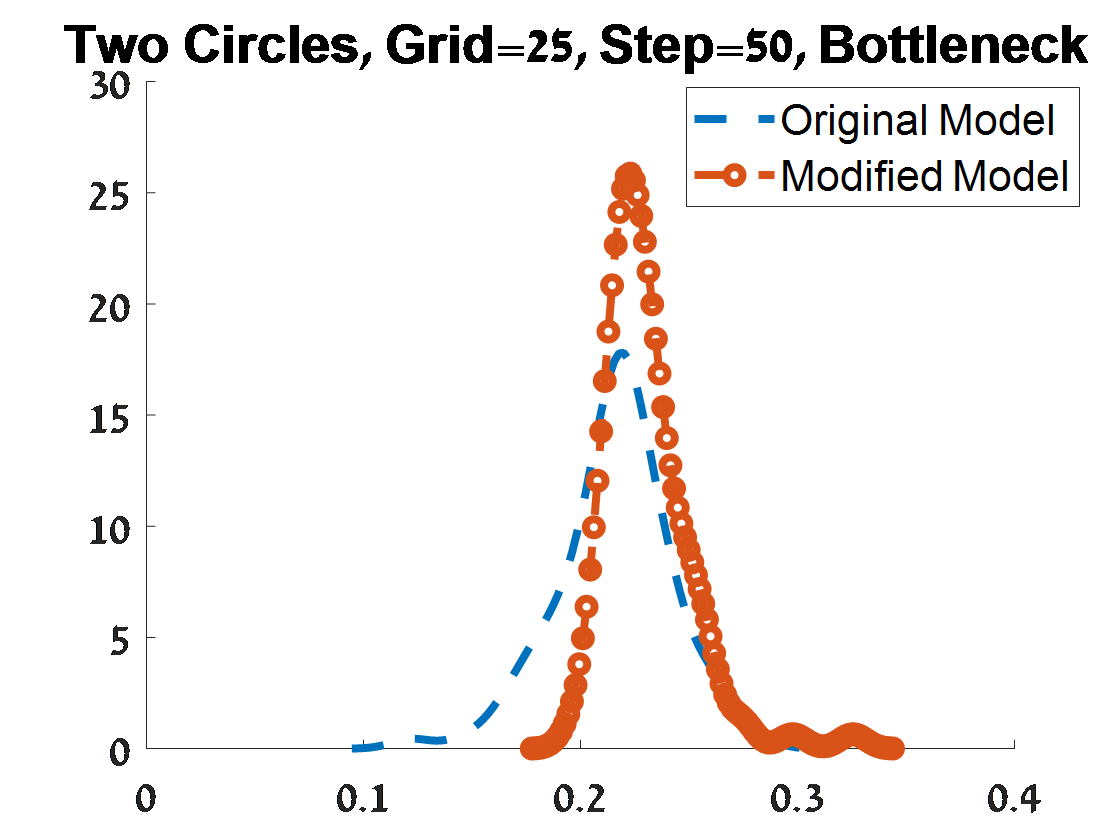}
\includegraphics[width=1.2in, height=1.4in]{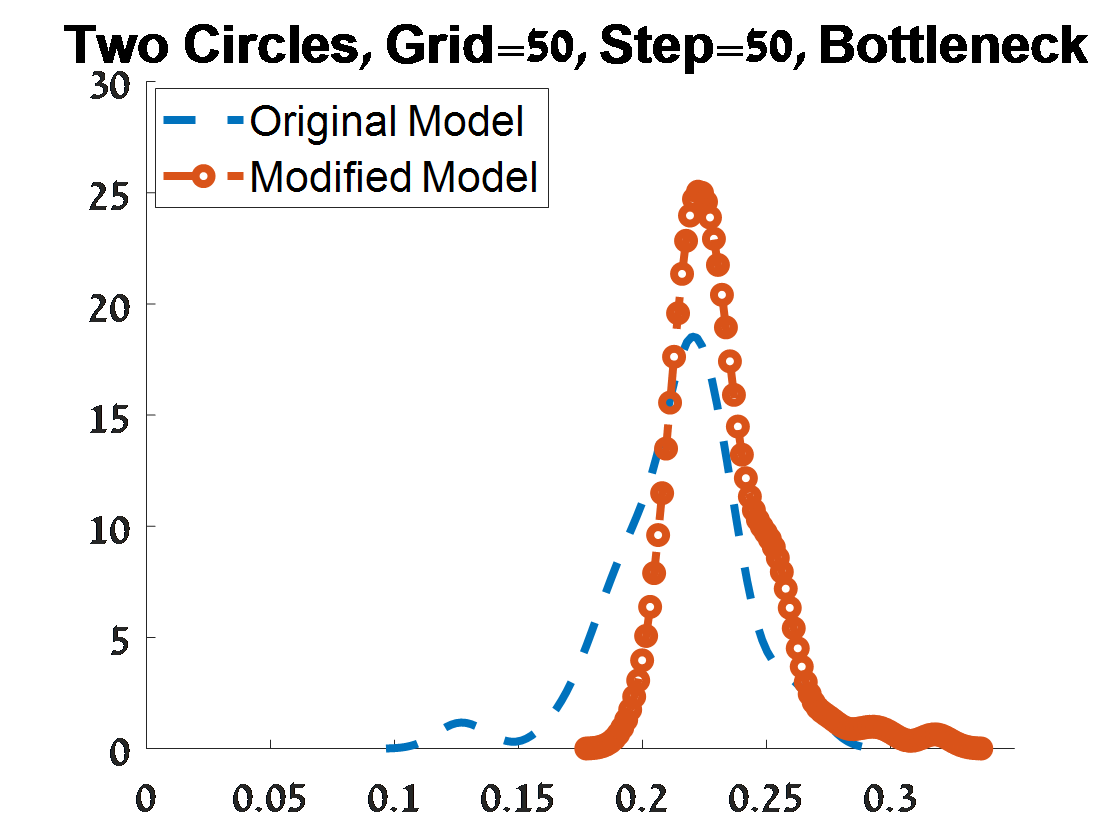}
\includegraphics[width=1.2in, height=1.4in]{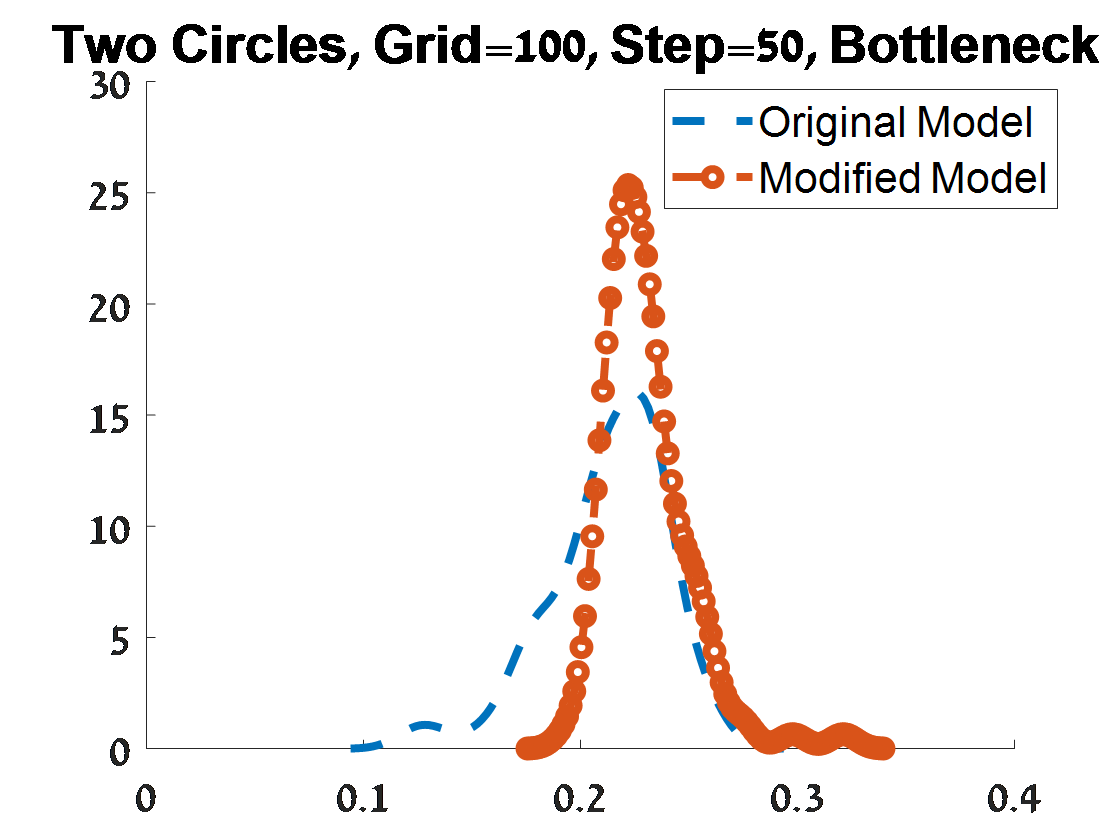}
\includegraphics[width=1.2in, height=1.4in]{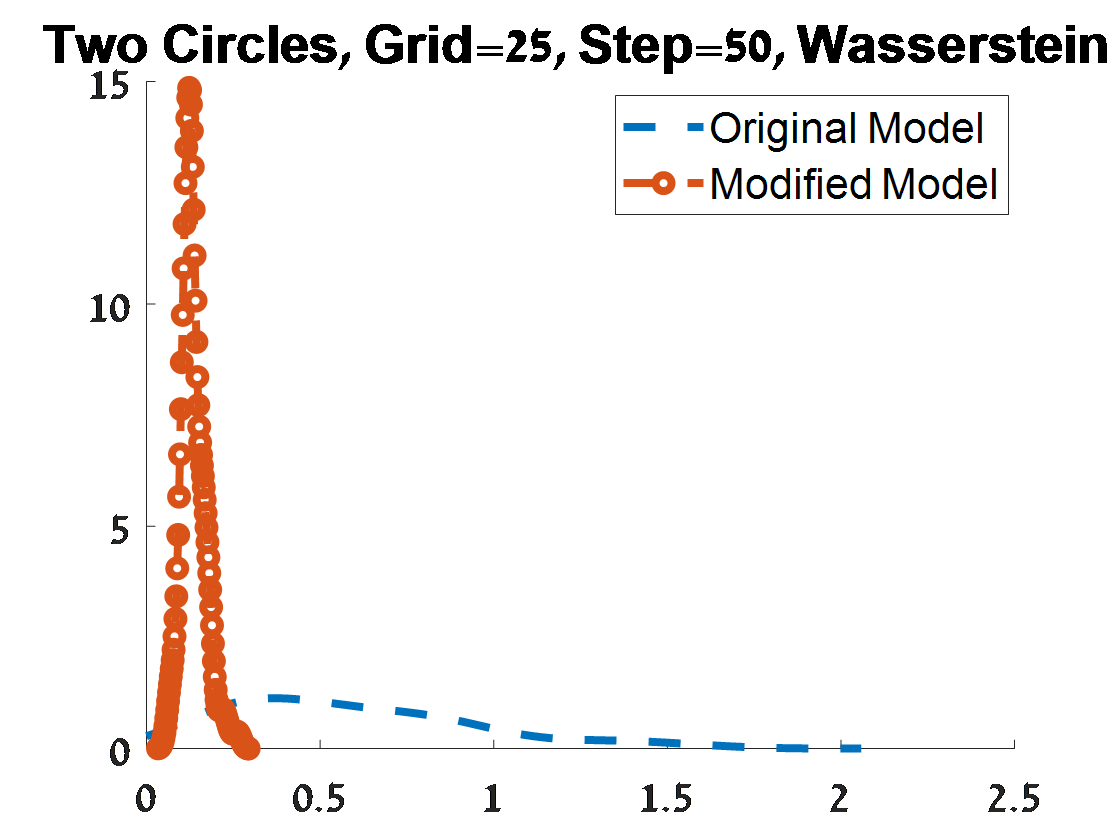}
\includegraphics[width=1.2in, height=1.4in]{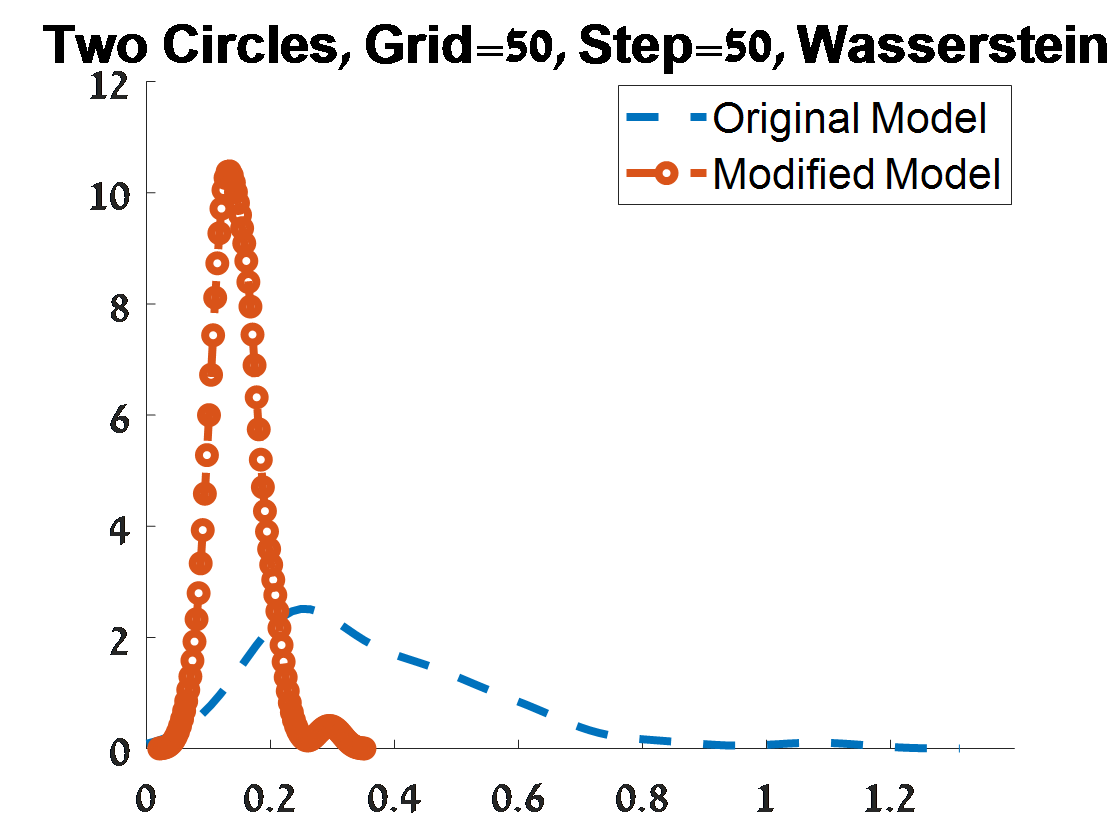}
\includegraphics[width=1.2in, height=1.4in]{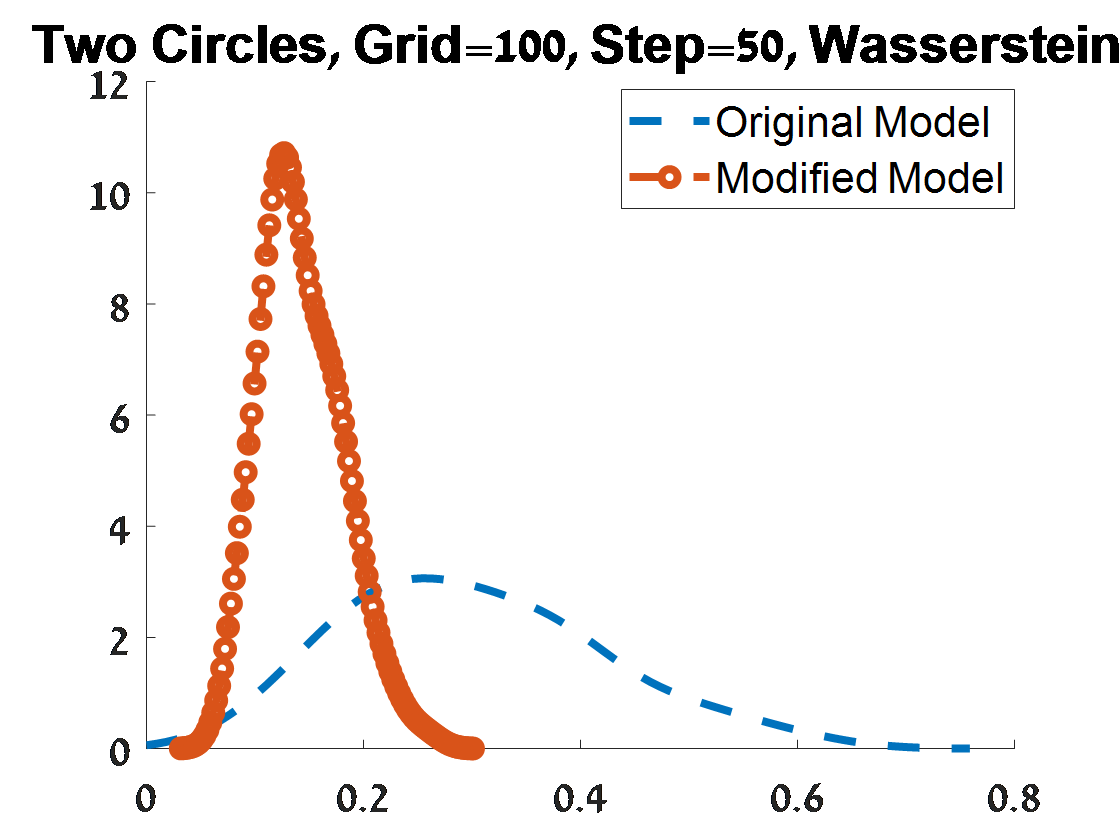}
\\
\includegraphics[width=1.2in, height=1.4in]{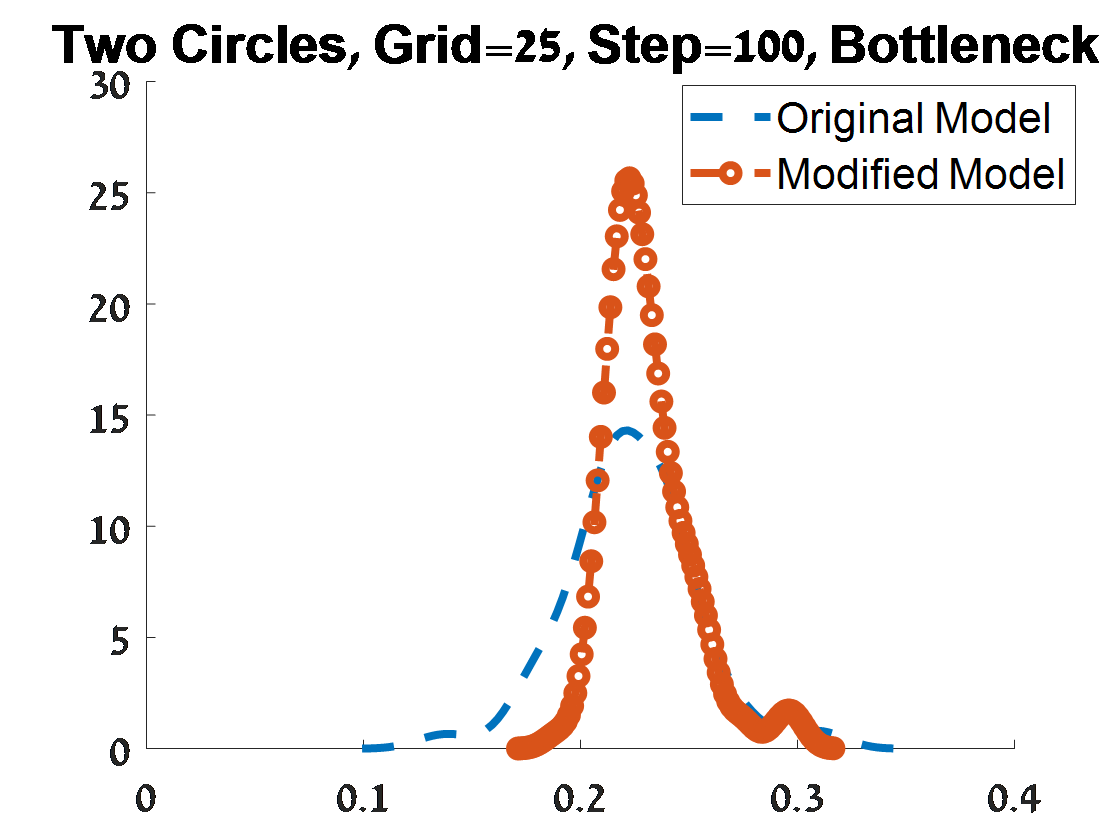}
\includegraphics[width=1.2in, height=1.4in]{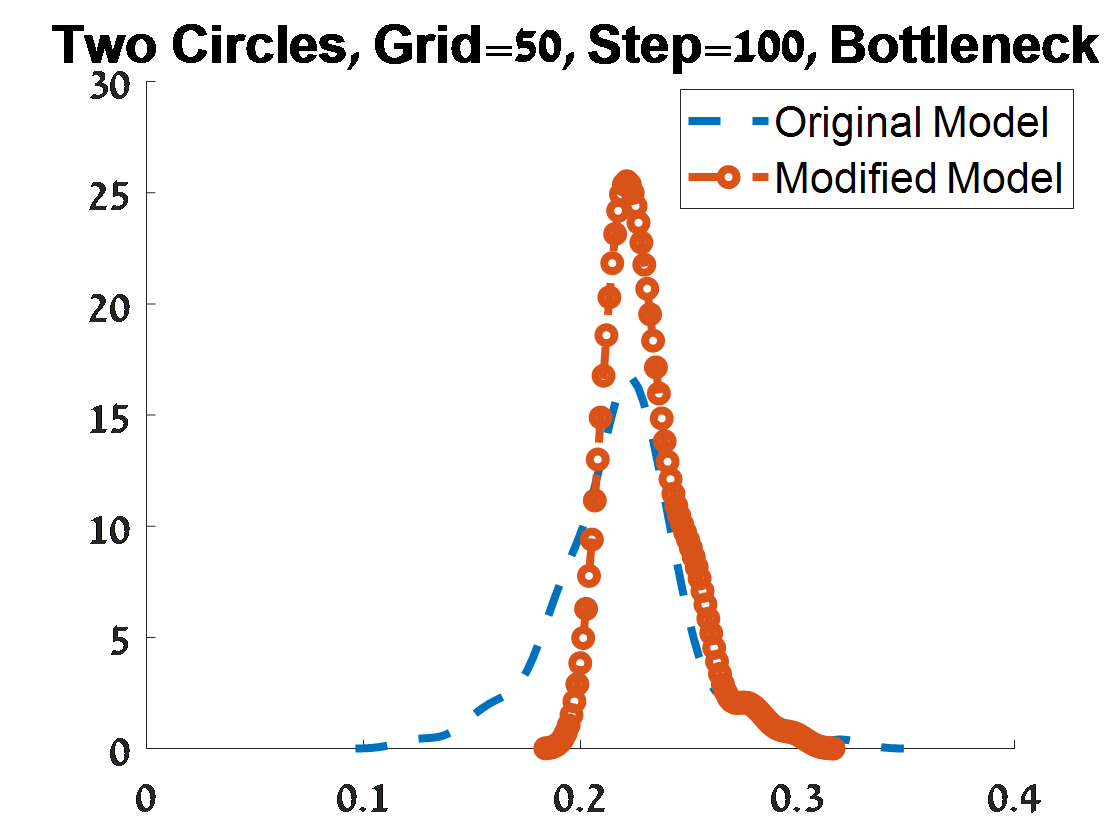}
\includegraphics[width=1.2in, height=1.4in]{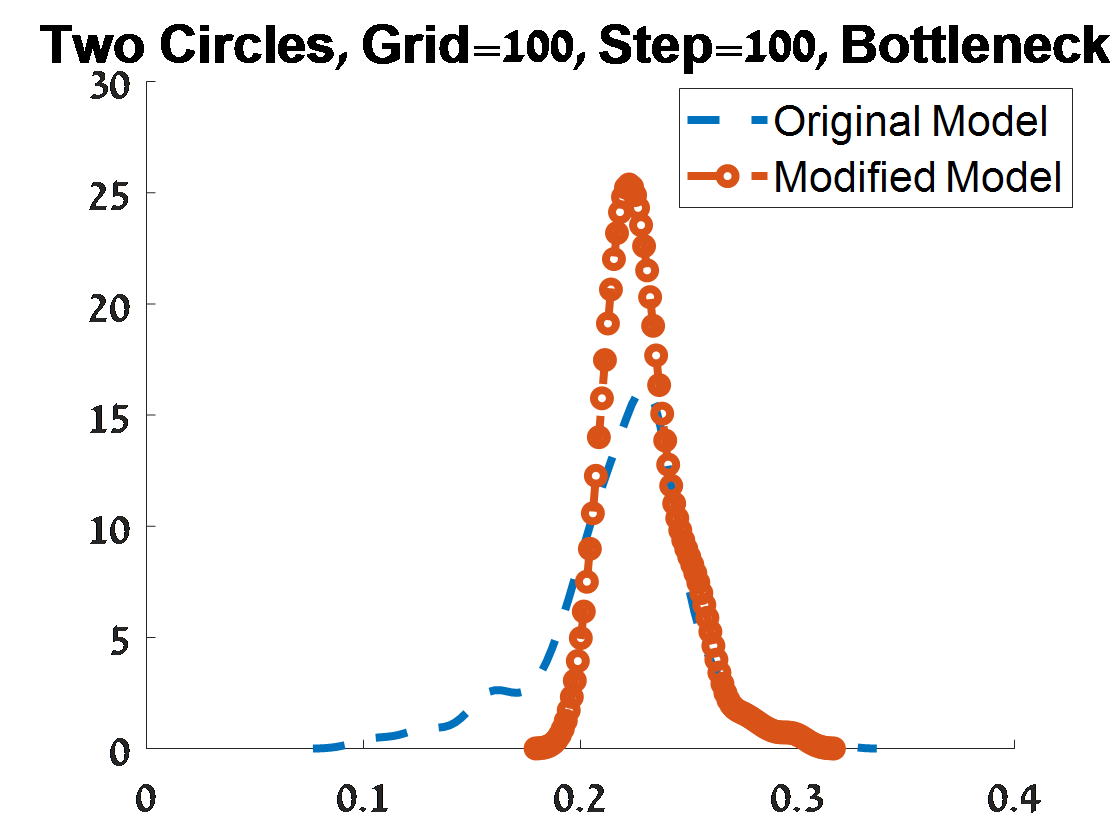}
\includegraphics[width=1.2in, height=1.4in]{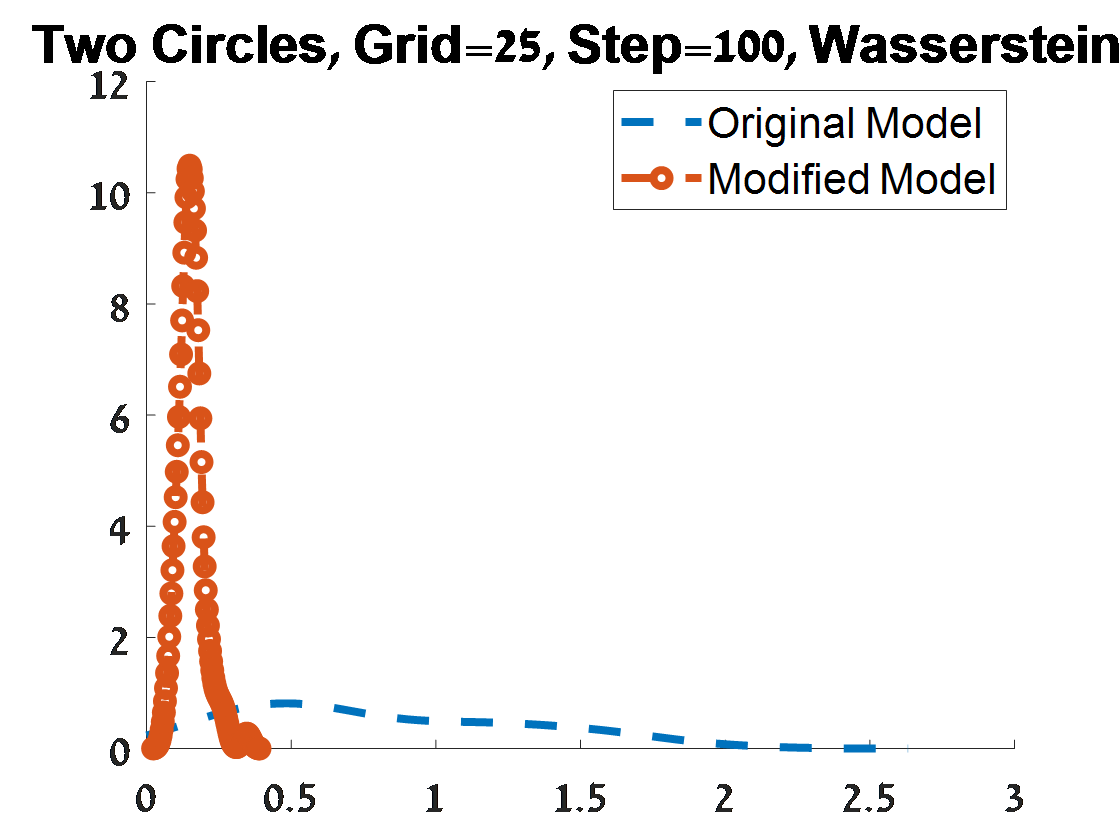}
\includegraphics[width=1.2in, height=1.4in]{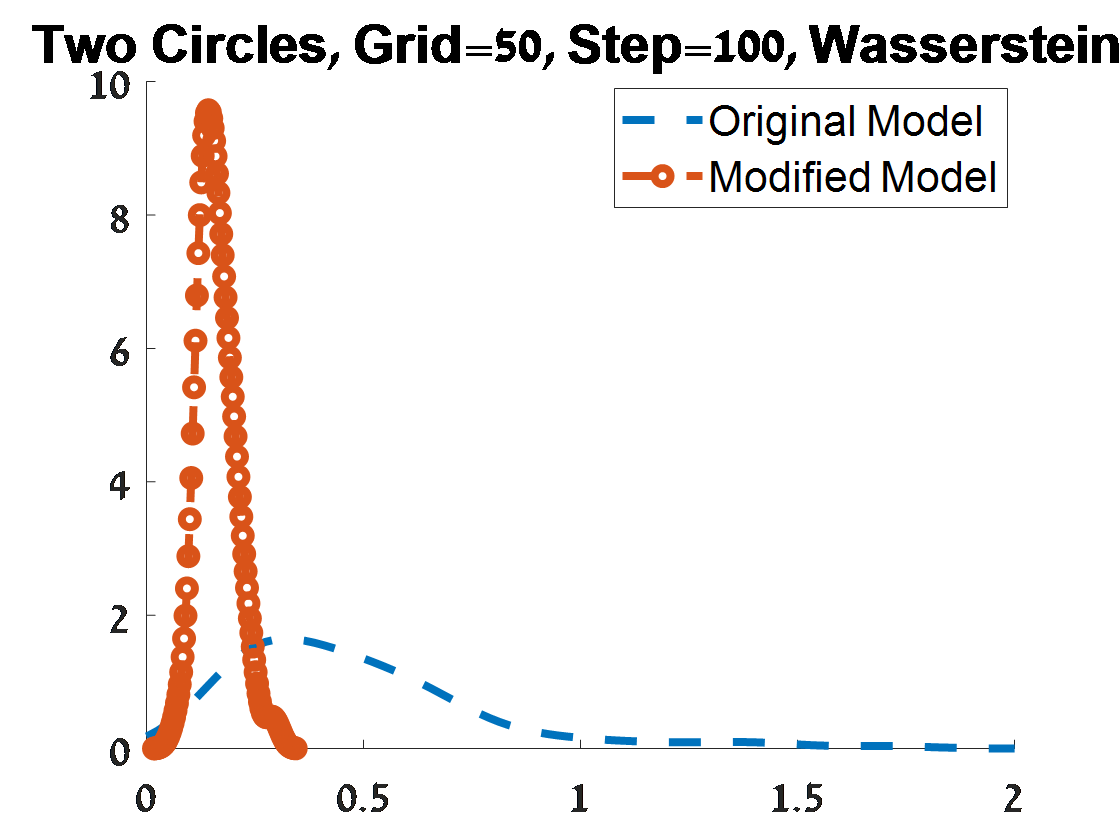}
\includegraphics[width=1.2in, height=1.4in]{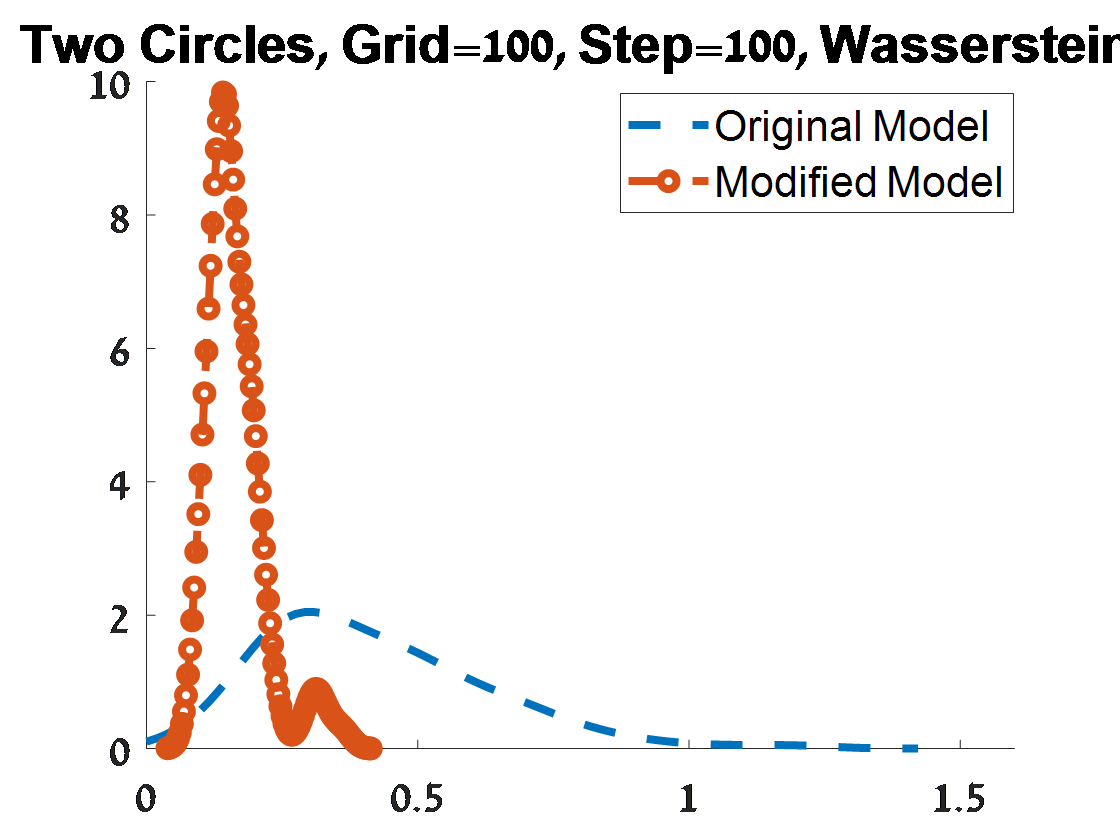}
\ec
%\caption{\footnotesize
% A random sample from two circles, 500 points from the larger circle and 300 from the smaller one,  with a kernel density
\caption{\footnotesize
 Criterion 1 of goodness of fit for 100 PDs corresponded to 100 samples from an object of two concentric circles. The plots depend on the grid of the proposal distribution ("Grid"), and the burn-in ("Step") of the MCMC algorithm.}
\label{fig:concentric_a}
\end{figure}
\end{landscape}

We generated 100 such samples, calculated their corresponded PDs, and fitted the both model's versions for the $H_0$ points of each PD.
Figure\ \ref{fig:concentric_a} describes the distributions over the 100 PDs of the first criterion of goodness of fit, and Figures 7-8 describe the distributions of the second criterion of goodness of fit.
In criterion 1, the Wasserstein distance between the simulated PD and the real PD is smaller under the modified model relative to this distance under the original model. For this distance, the best fitting is in burn-in of 25, and the distance decreases as the grid size increases. The Bottleneck distances are relative similar for both model's versions, with a similar impact of the grid size and the burn-in value.

For criterion 2,  the distributional properties of the modified model are close to those of the real PDs rather than the distributional properties of the original model. This is true for all considered values of grid size and burn-in. Specifically, given a burn-in of 25, the fitting is better in grid size of 100x100 for the three distributional properties, whereas given a burn-in of 50, 100, the fitting is better in grid of 25x25 for the second and third properties, and is better in grid of 100x100 for the first property.

%In Figure\ \ref{fig:concentric_d} we present two examples, each of them contains a real PD and its simulated PD based on the two model versions, only for the best scenarios we found, that is grid size of 50x50 and 100x100, and burn-in of 25. Also here we see that the best fitted MCMC is under the modified model rather than the original one. In addition, the modified model is better and succeed in creating the far points from the diagonal, where the original model is concentrating in creating more the points that are close to the diagonal and almost cannot succeed in generating the far points from the diagonal.
That is, here as in the previous example, that the modified RST is better than the original RST, where the best fitting is under grid sizes of 50x50 and 100x100, and burn-in of 25.

\begin{landscape}
\begin{figure}[h!]
\bc
\includegraphics[width=1.2in, height=1.25in]{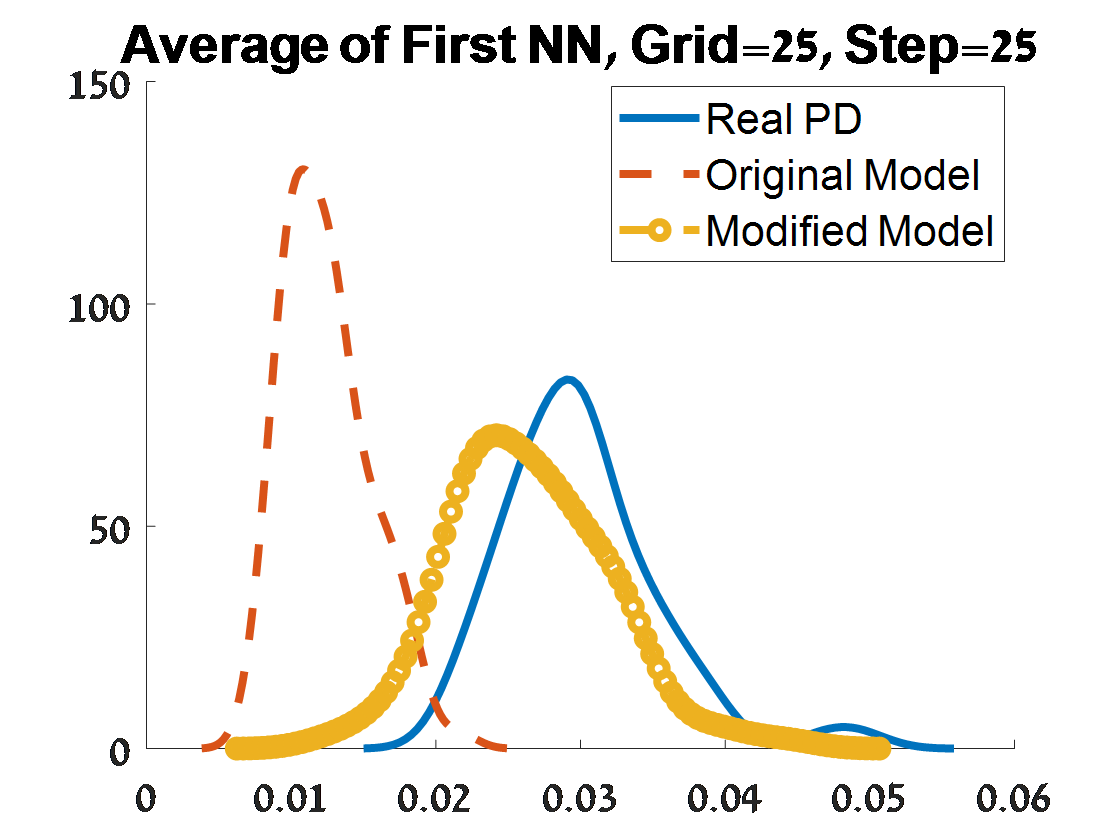}
\includegraphics[width=1.2in, height=1.25in]{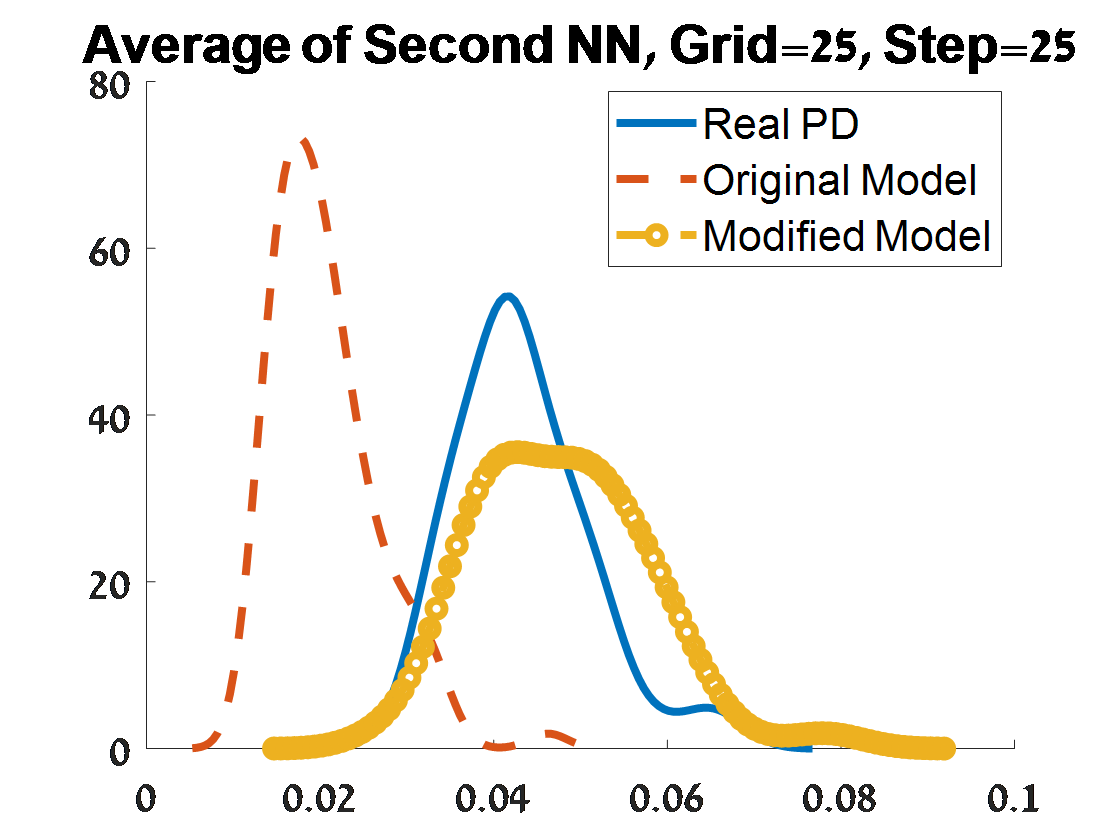}
\includegraphics[width=1.2in, height=1.25in]{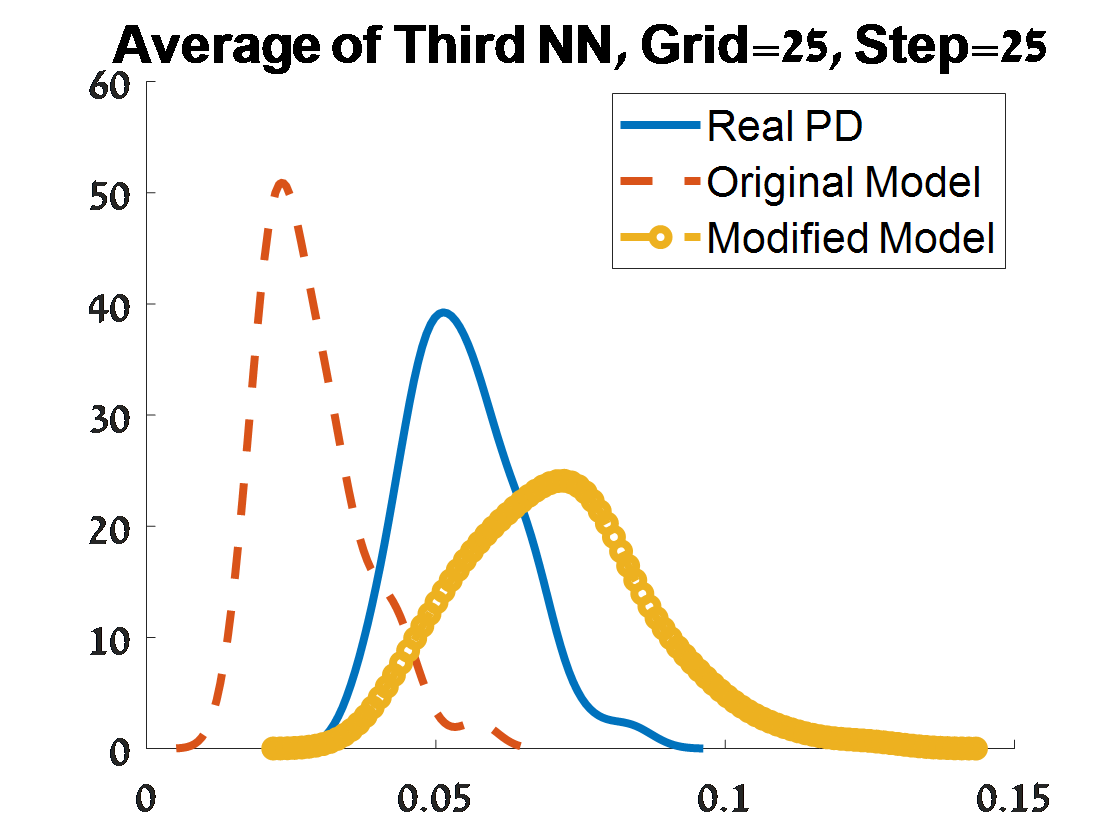}
\includegraphics[width=1.2in, height=1.25in]{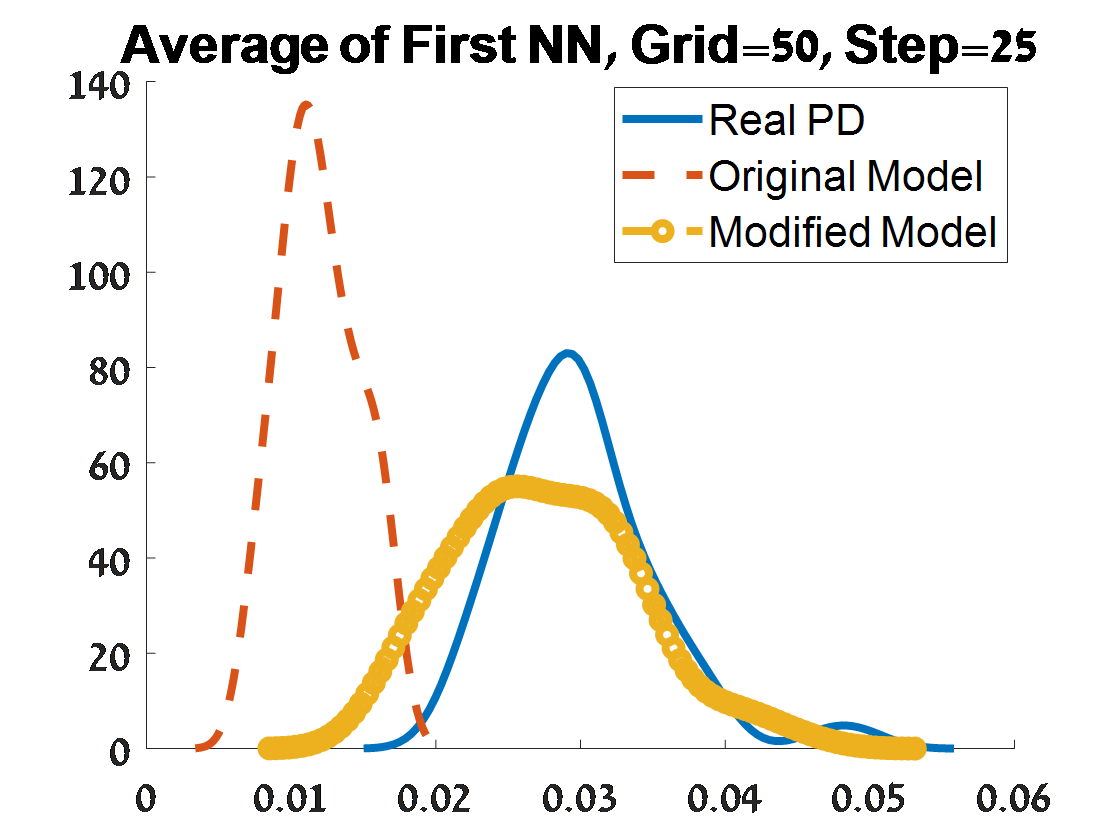}
\includegraphics[width=1.2in, height=1.25in]{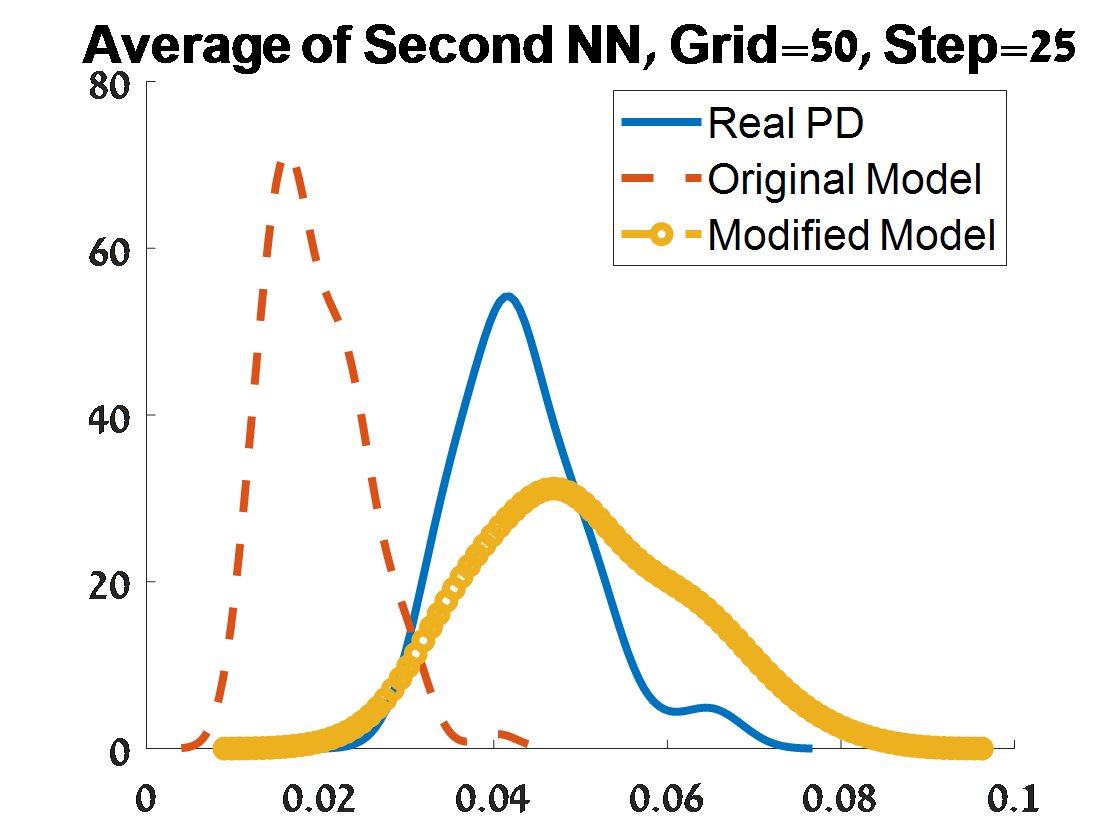}
\includegraphics[width=1.2in, height=1.25in]{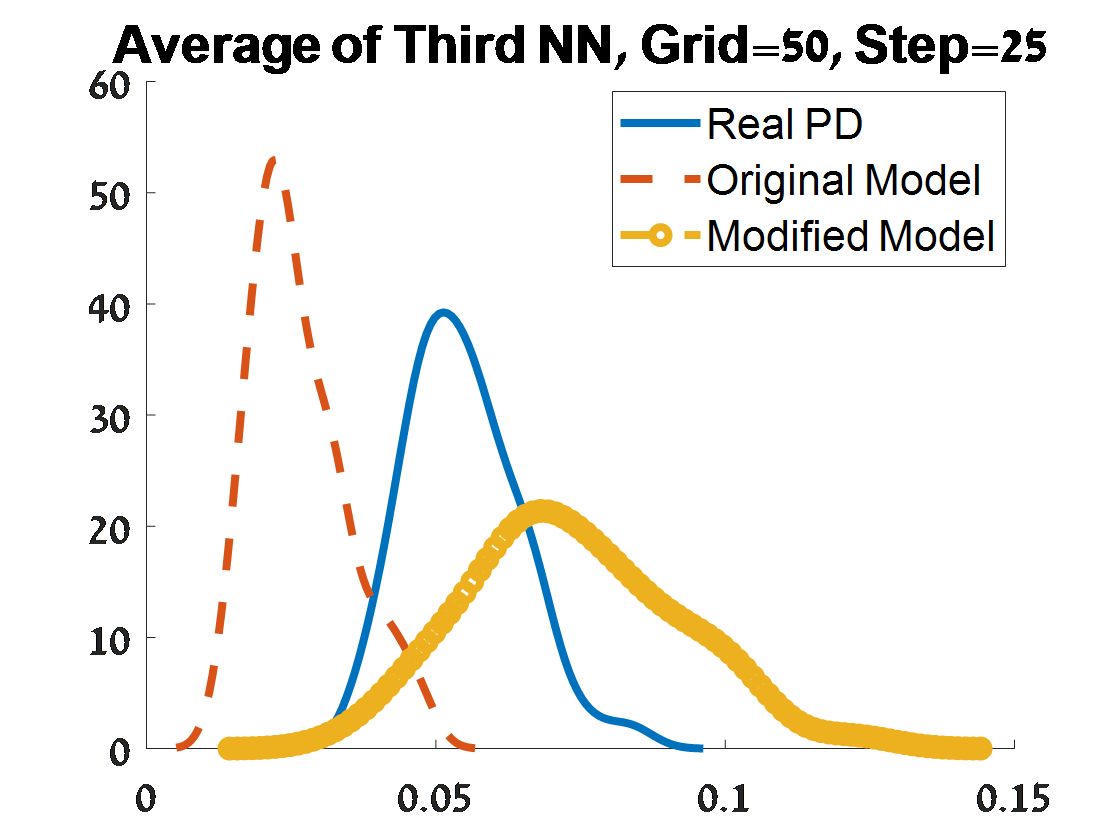}
\includegraphics[width=1.2in, height=1.25in]{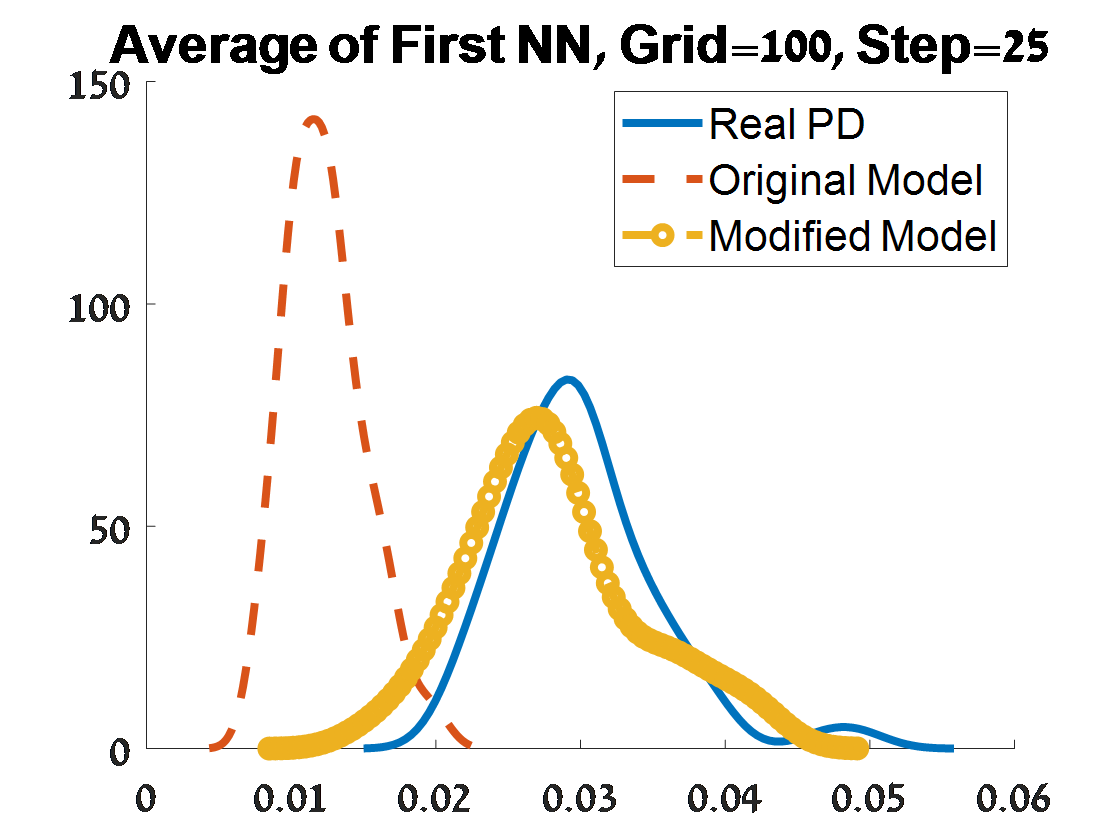}
\includegraphics[width=1.2in, height=1.25in]{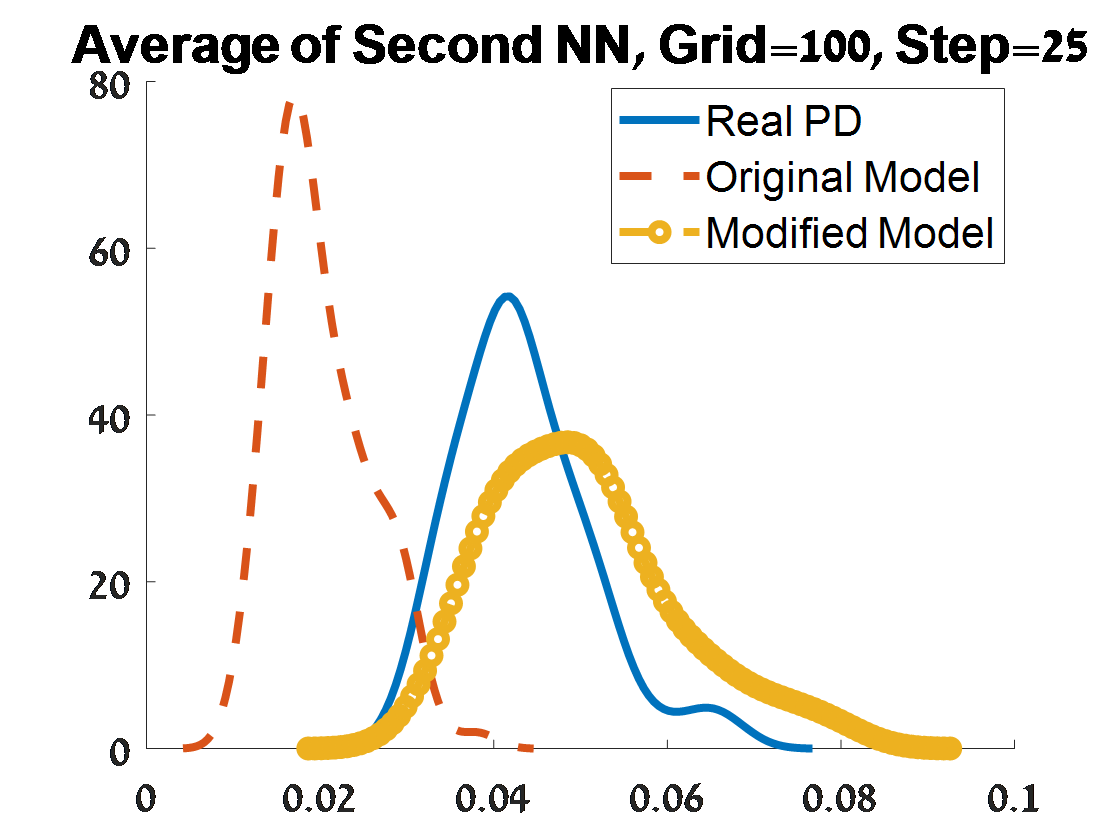}
\includegraphics[width=1.2in, height=1.25in]{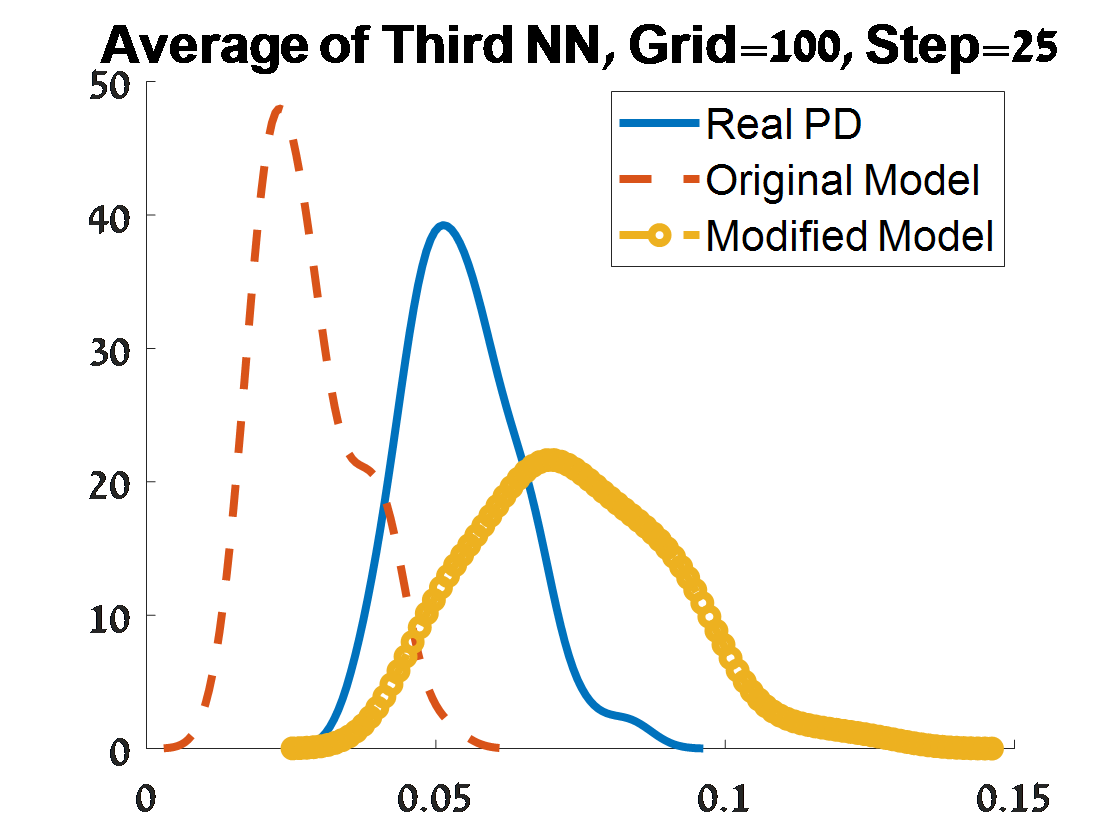}
\\
\includegraphics[width=1.2in, height=1.25in]{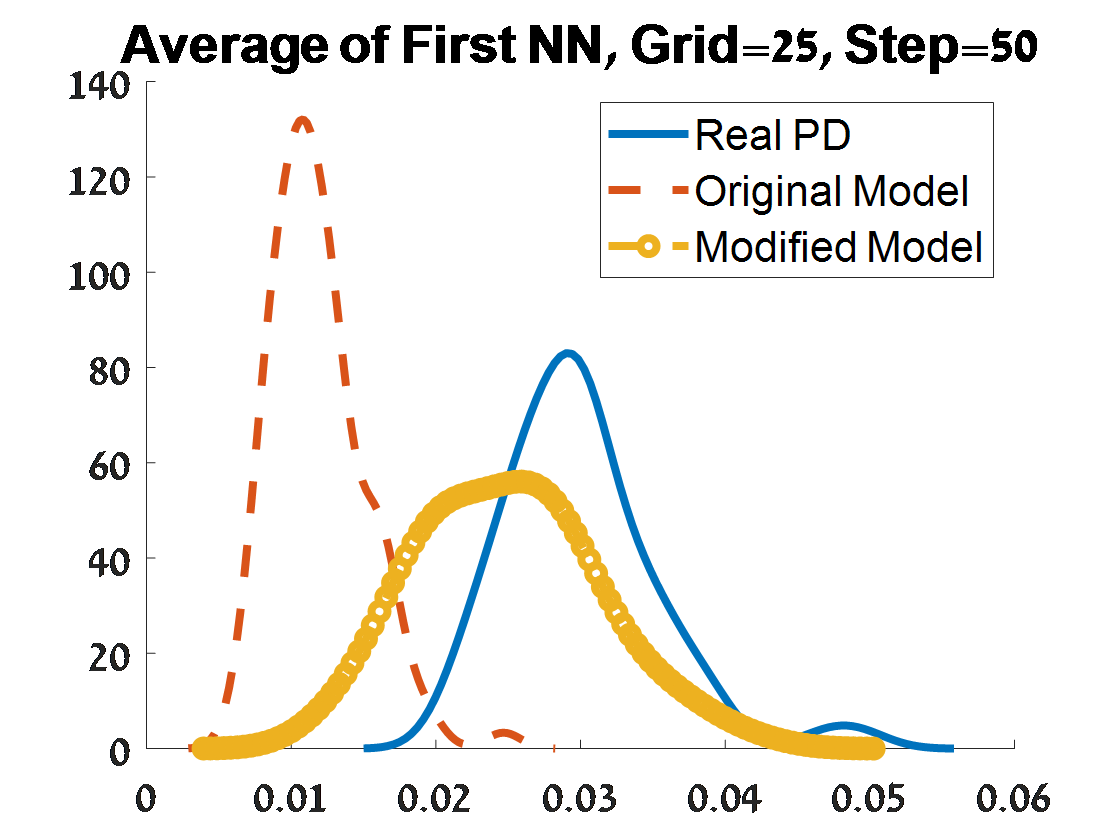}
\includegraphics[width=1.2in, height=1.25in]{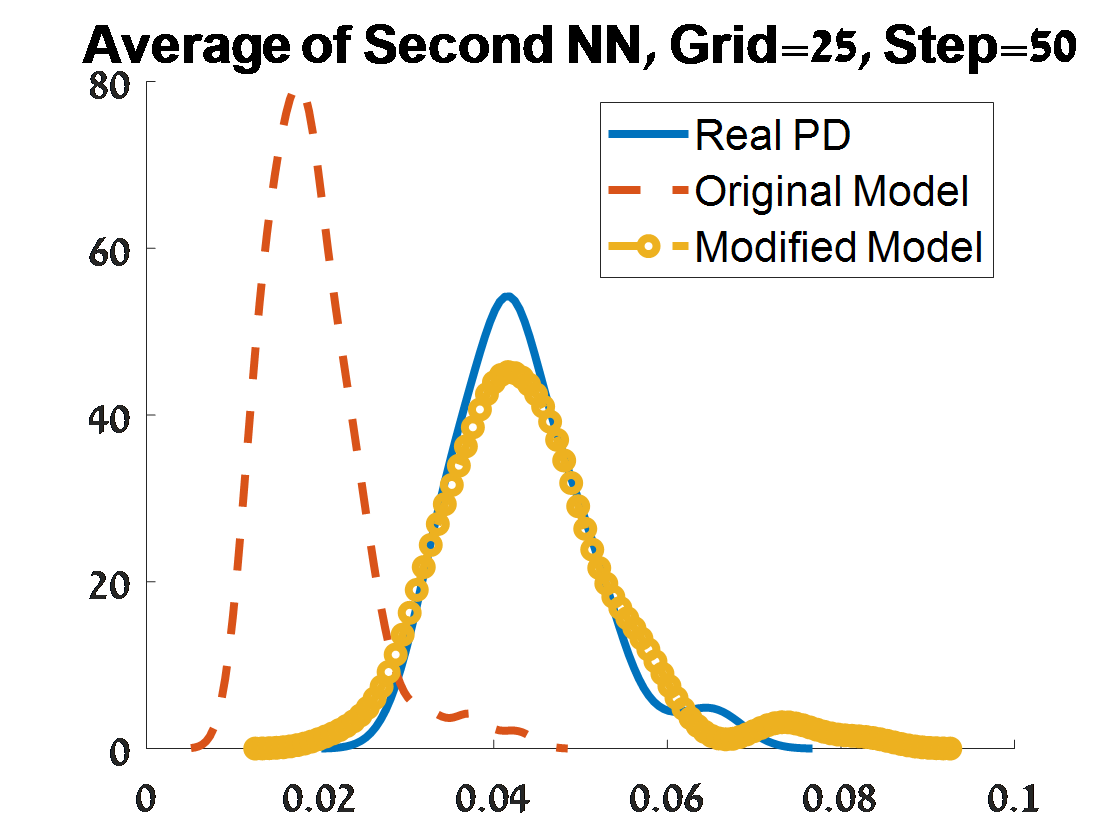}
\includegraphics[width=1.2in, height=1.25in]{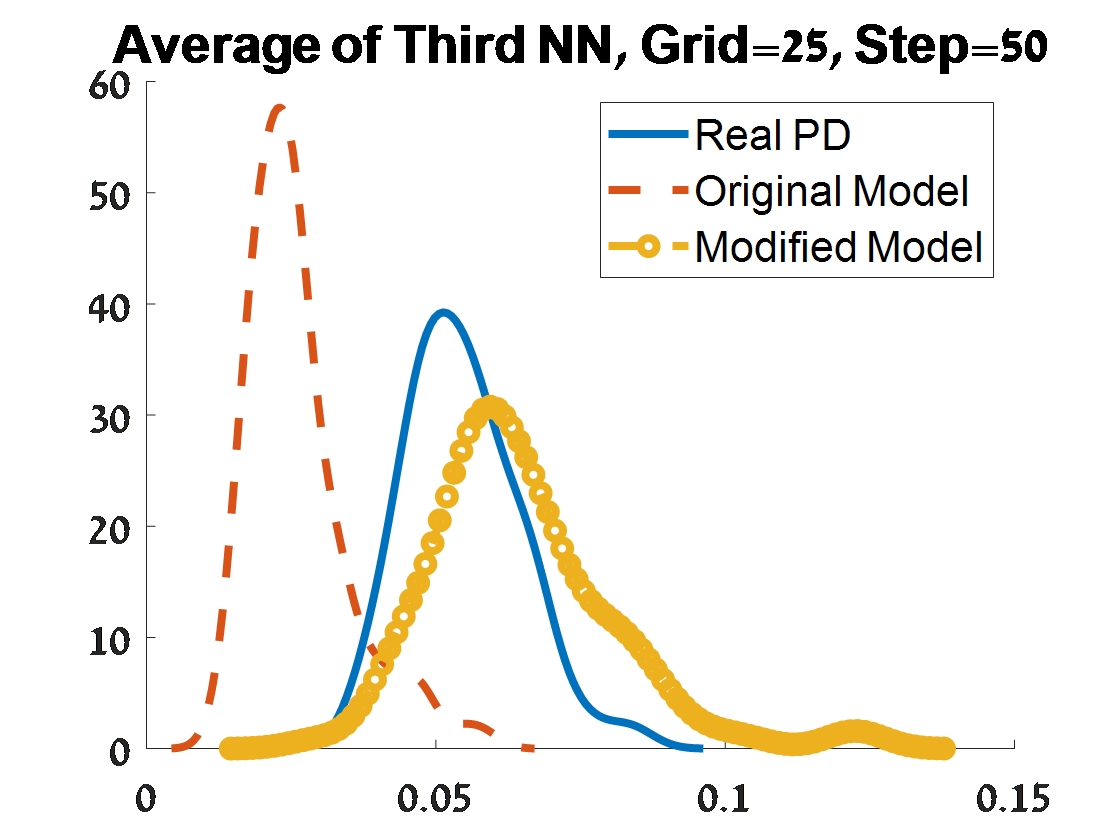}
\includegraphics[width=1.2in, height=1.25in]{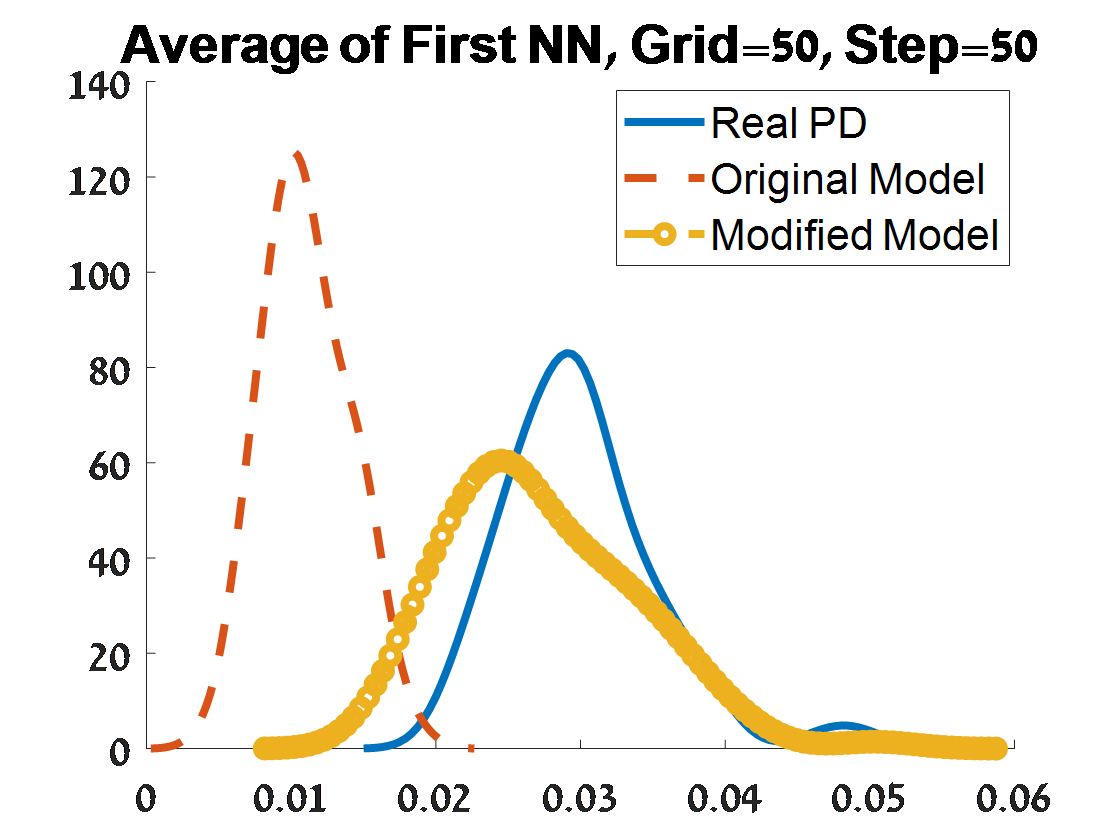}
\includegraphics[width=1.2in, height=1.25in]{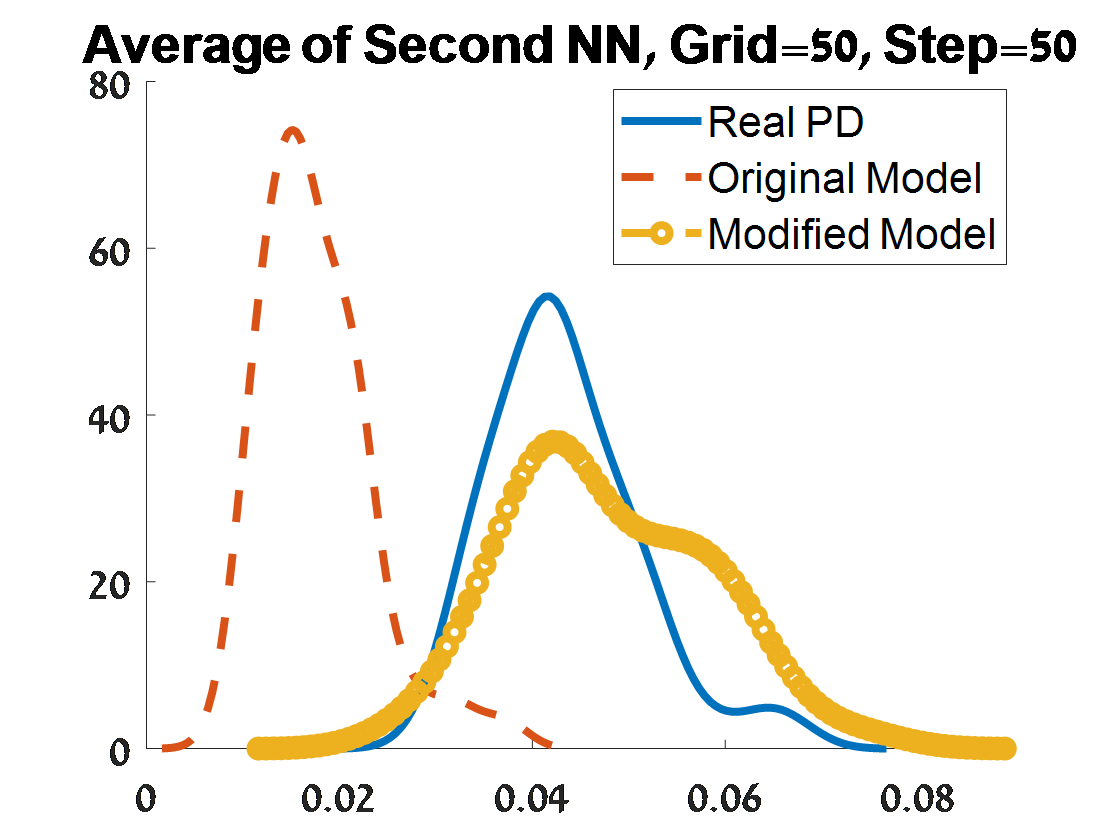}
\includegraphics[width=1.2in, height=1.25in]{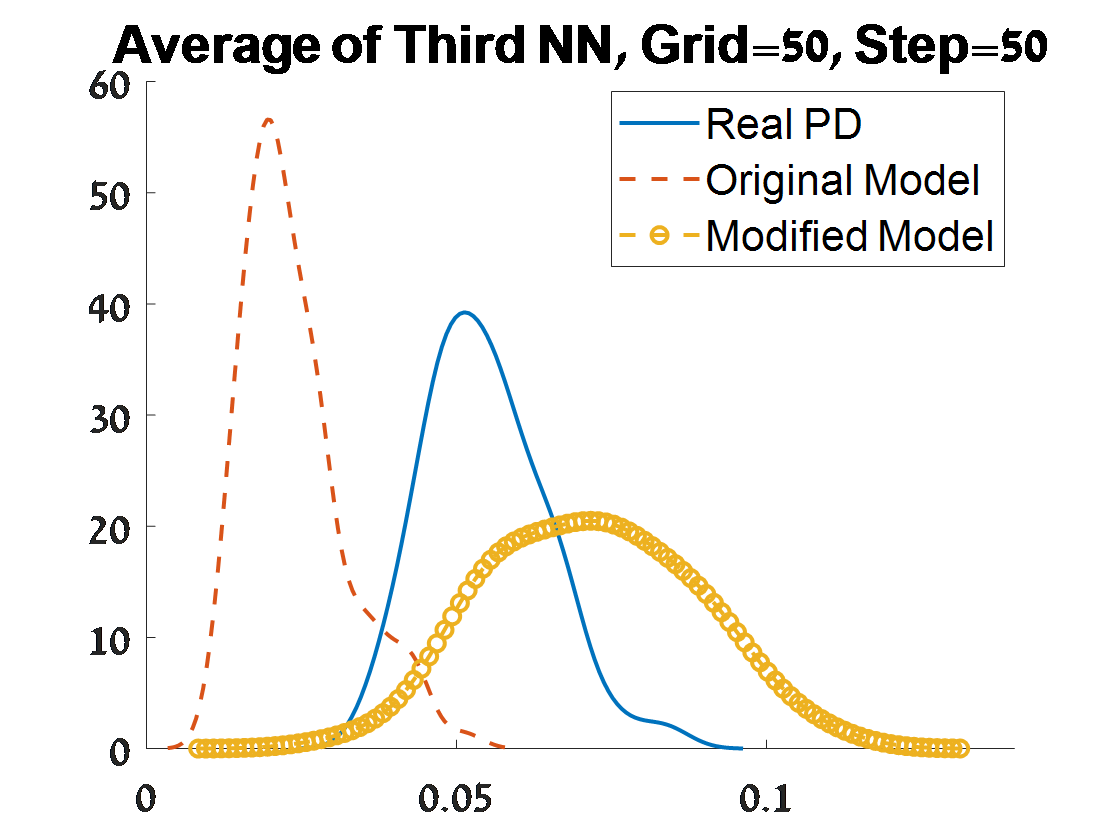}
\includegraphics[width=1.2in, height=1.25in]{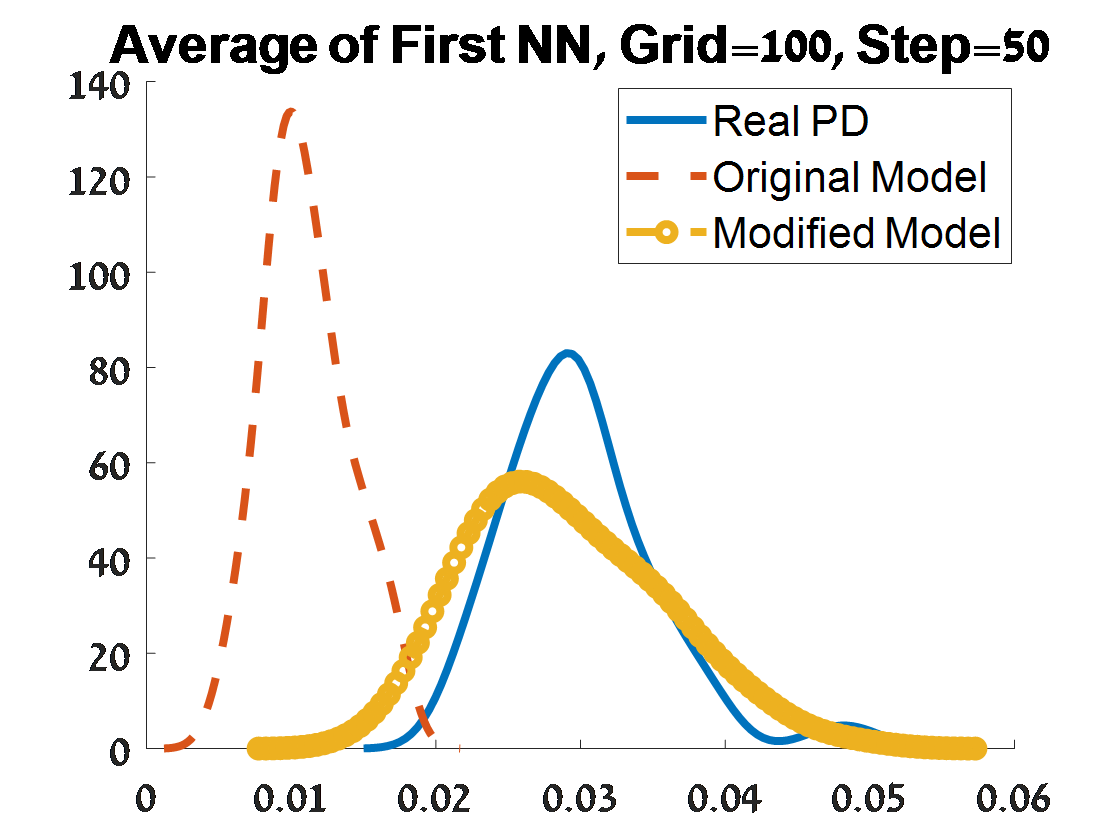}
\includegraphics[width=1.2in, height=1.25in]{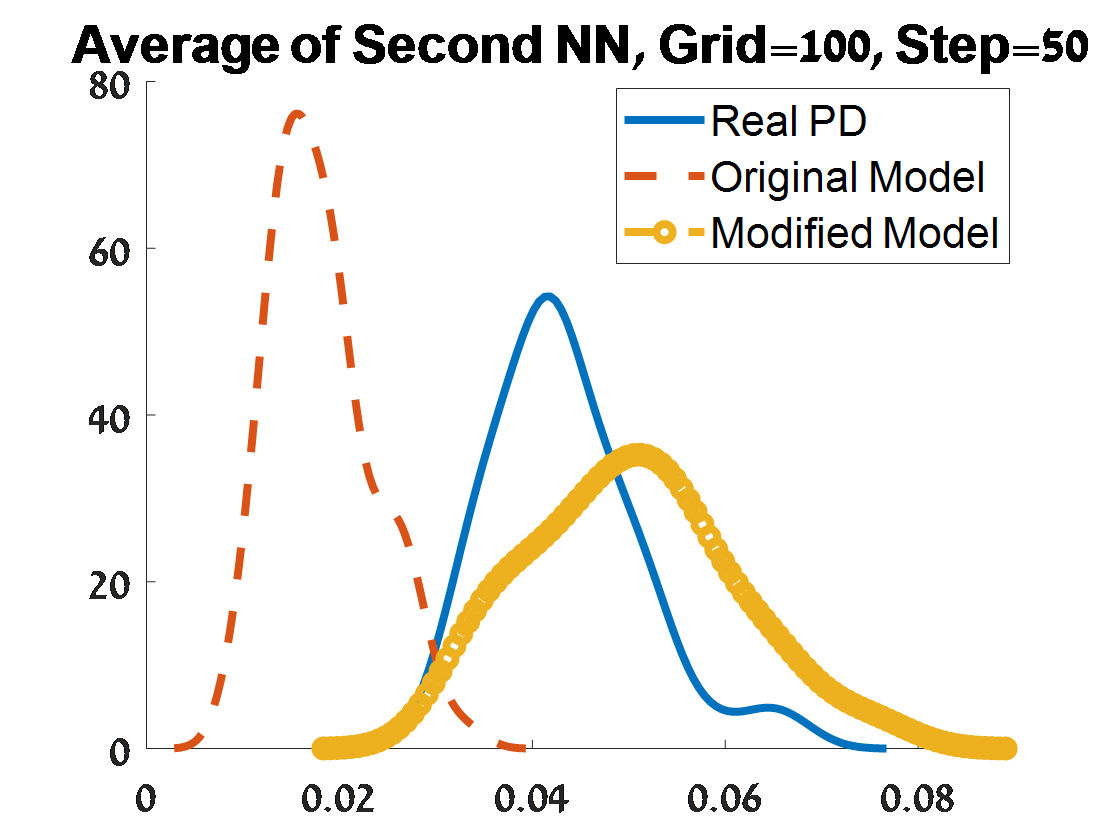}
\includegraphics[width=1.2in, height=1.25in]{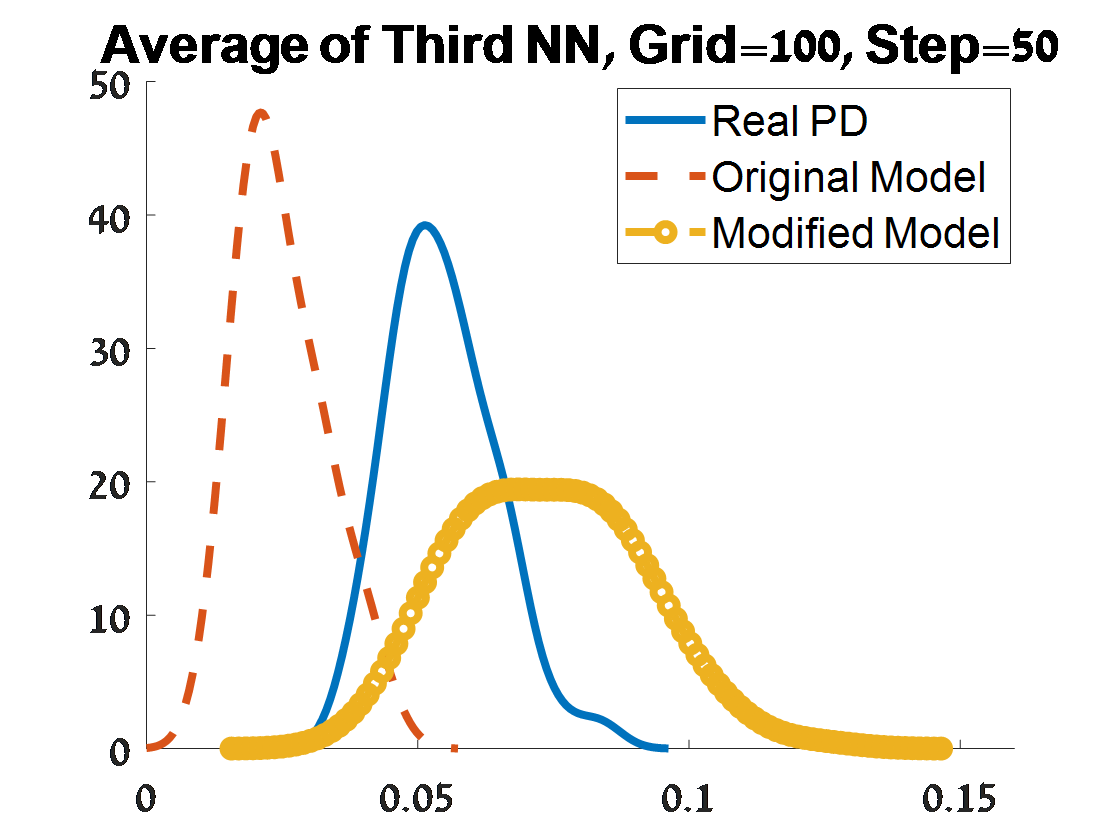}
\ec
%\caption{\footnotesize
% A random sample from two circles, 500 points from the larger circle and 300 from the smaller one,  with a kernel density
\caption{\footnotesize
Criterion 2 of goodness of fit for 100 PDs corresponded to 100 samples from an object of two concentric circles. The plots depend on the grid of the proposal distribution ("Grid"), and the burn-in ("Step") of the MCMC algorithm.}
\label{fig:concentric_b}
\end{figure}
\end{landscape}

\begin{landscape}
\begin{figure}[h!]
\bc
\includegraphics[width=1.2in, height=1.25in]{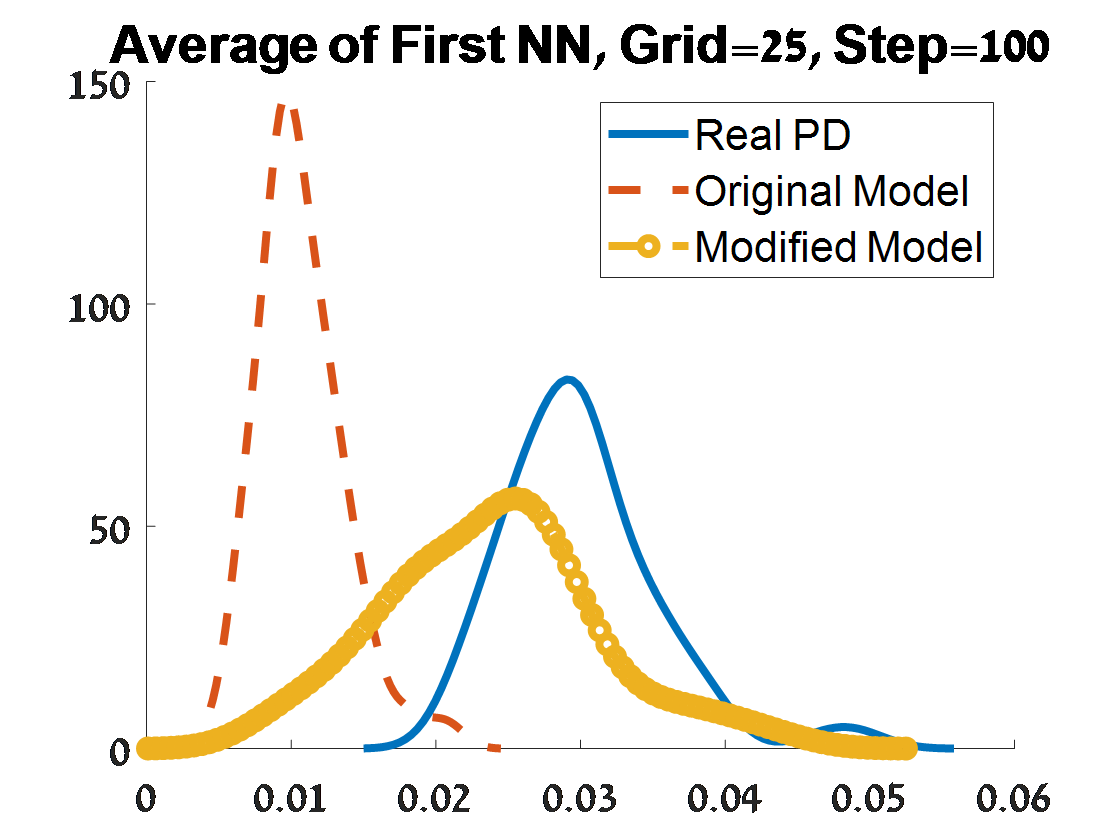}
\includegraphics[width=1.2in, height=1.25in]{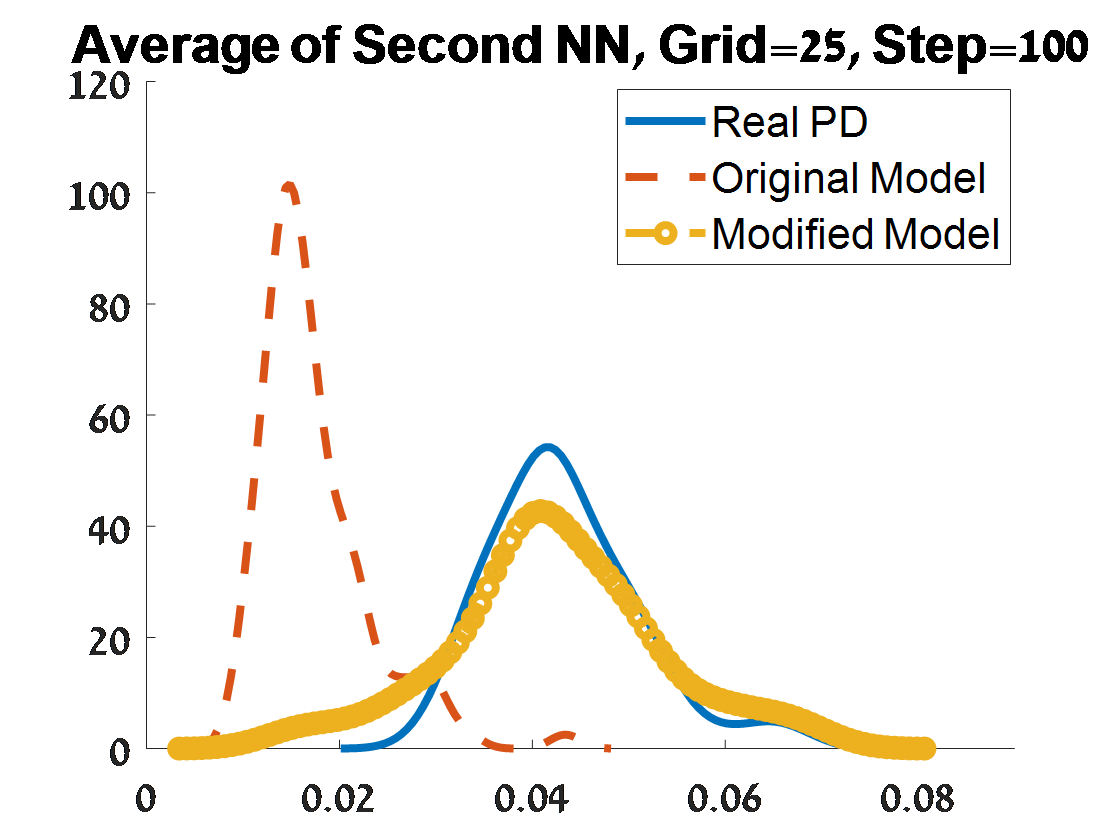}
\includegraphics[width=1.2in, height=1.25in]{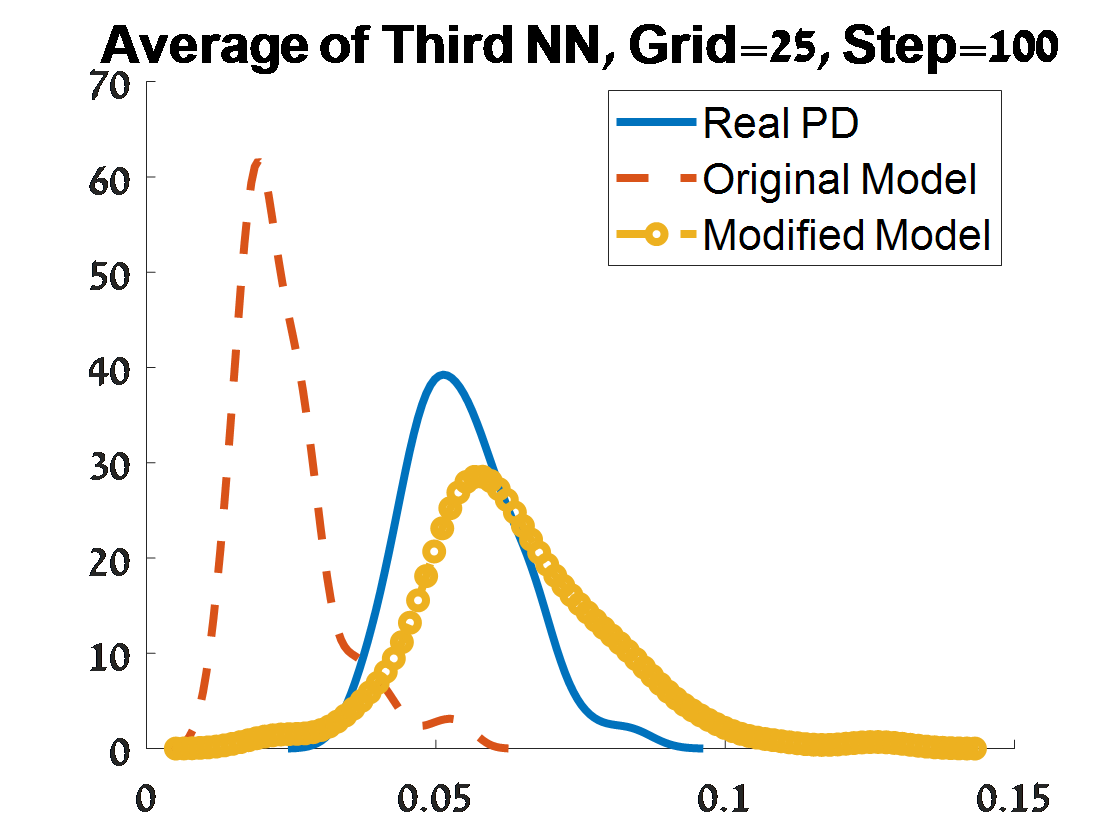}
\includegraphics[width=1.2in, height=1.25in]{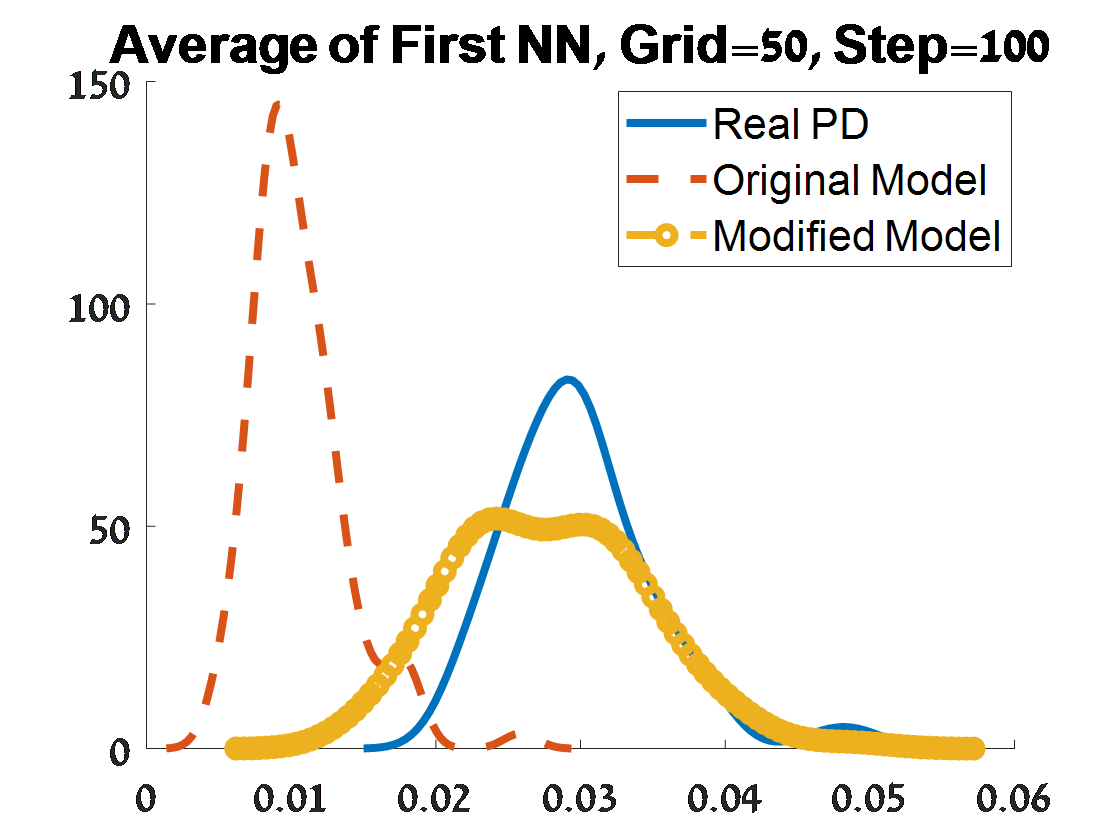}
\includegraphics[width=1.2in, height=1.25in]{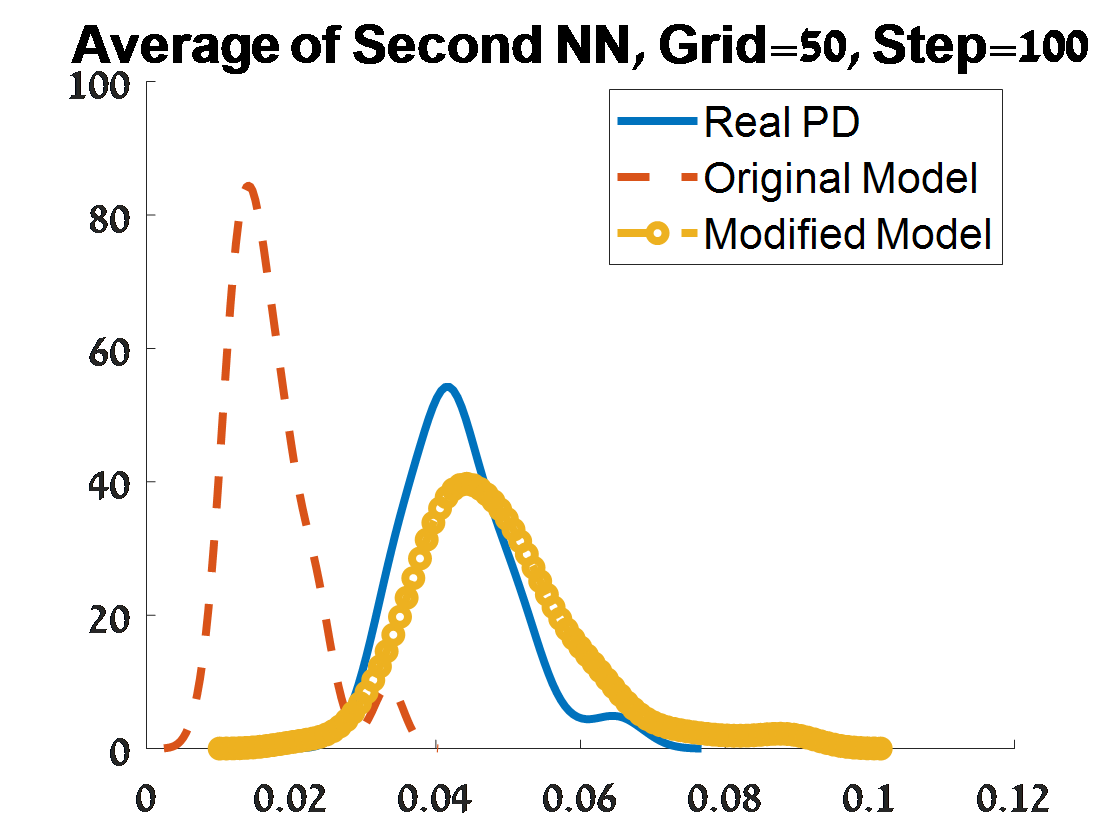}
\includegraphics[width=1.2in, height=1.25in]{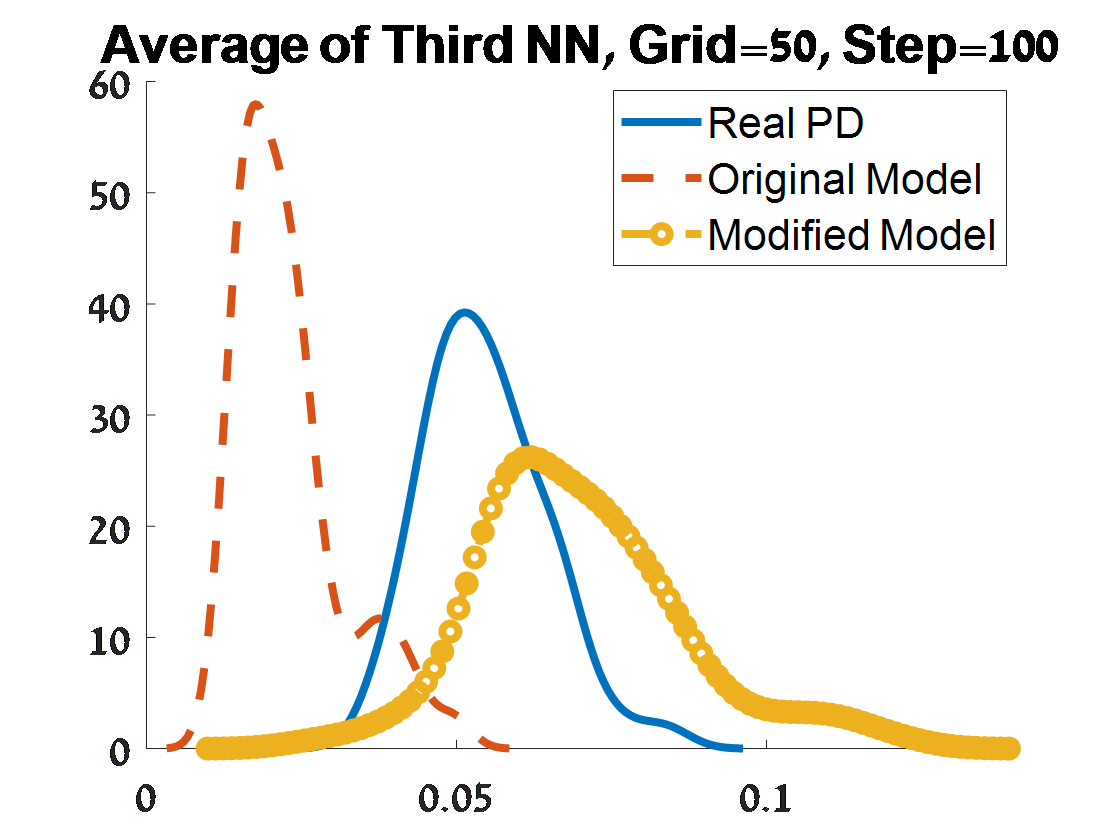}
\includegraphics[width=1.2in, height=1.25in]{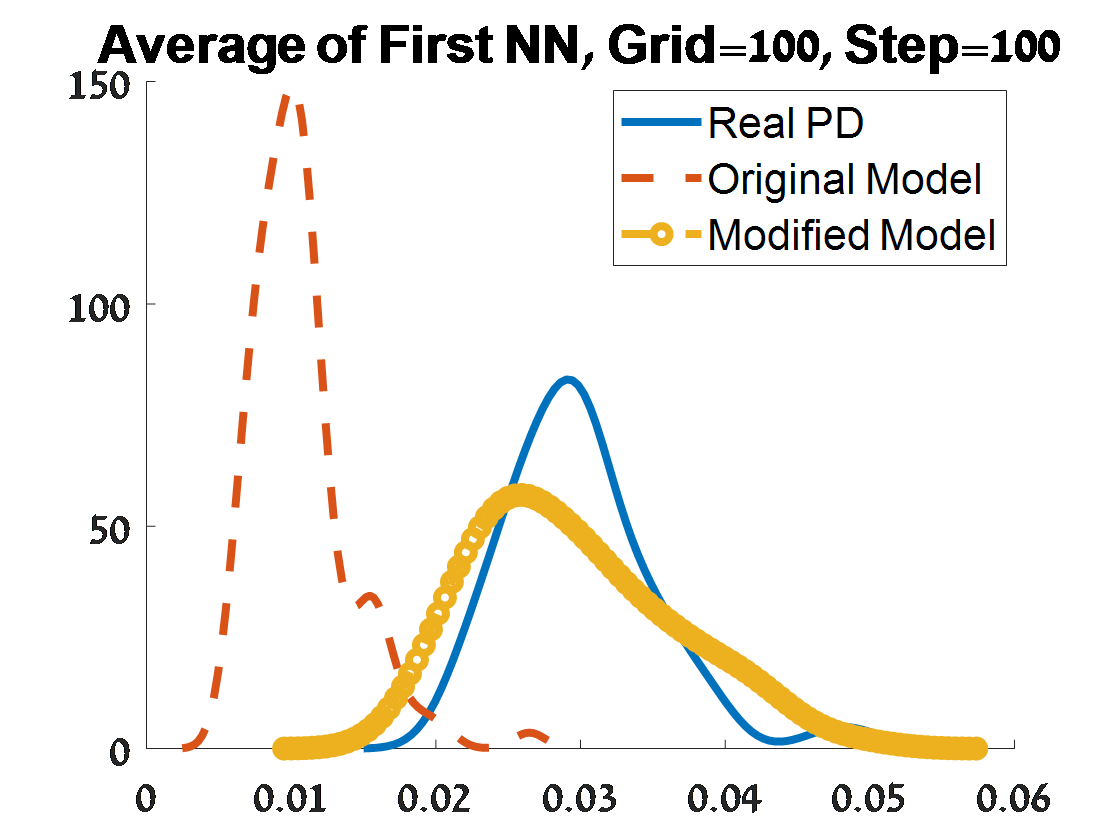}
\includegraphics[width=1.2in, height=1.25in]{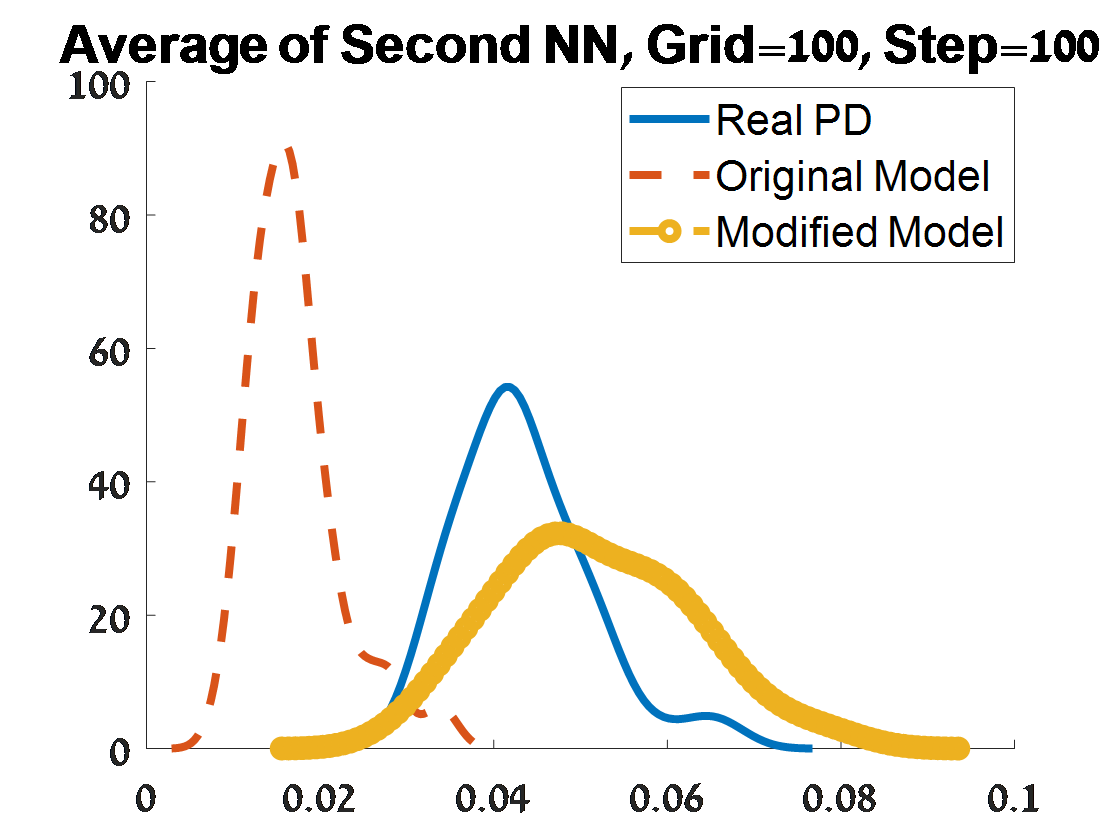}
\includegraphics[width=1.2in, height=1.25in]{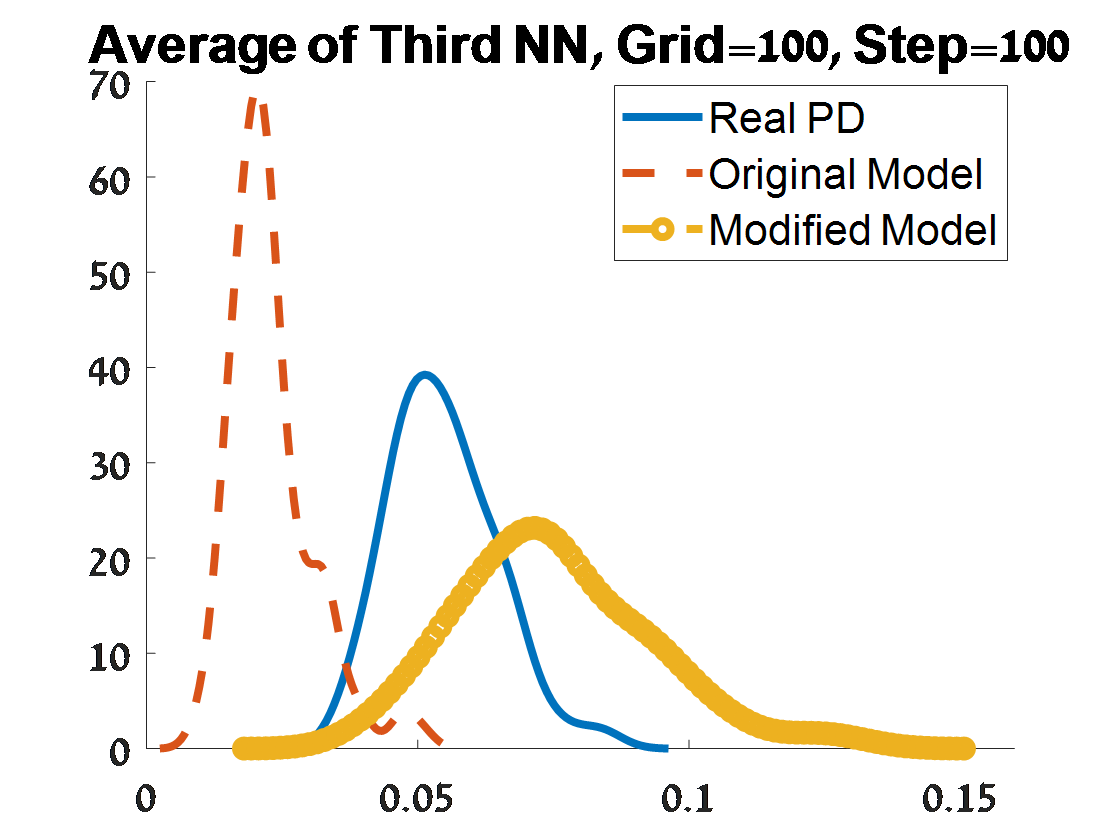}
\ec
%\caption{\footnotesize
% A random sample from two circles, 500 points from the larger circle and 300 from the smaller one,  with a kernel density
\caption{\footnotesize
 Continue of Criterion 2 of goodness of fit for 100 PDs corresponded to 100 samples from an object of two concentric circles. The plots depend on the grid of the proposal distribution ("Grid"), and the burn-in ("Step") of the MCMC algorithm.}
\label{fig:concentric_c}
\end{figure}
\end{landscape}

%\begin{landscape}
%\begin{figure}[h!]
%\bc
%\includegraphics[width=1.8in, height=1.8in]{TwoCircles_pd30_grid50_step25}
%\includegraphics[width=1.8in, height=1.8in]{TwoCircles_pd30_grid100_step25}
%\includegraphics[width=1.8in, height=1.8in]{TwoCircles_pd60_grid50_step25}
%\includegraphics[width=1.8in, height=1.8in]{TwoCircles_pd60_grid100_step25}
%\ec
%%\caption{\footnotesize
%% A random sample from two circles, 500 points from the larger circle and 300 from the smaller one,  with a kernel density
%\caption{\footnotesize
% Examples of two PDs (from the 100  PDs we used), each one is corresponded to a sample from an object of two concentric circles. For each PD, the simulated PD based on the two model's versions. The plots depend on the grid size of the proposal distribution ("Grid"), and the burn-in ("Step") of the MCMC algorithm. }
%\label{fig:concentric_d}
%\end{figure}
%\end{landscape}
\subsection{Two separated geometrical objects}
\begin{figure}[h!]
\bc
\includegraphics[width=2in, height=2in]{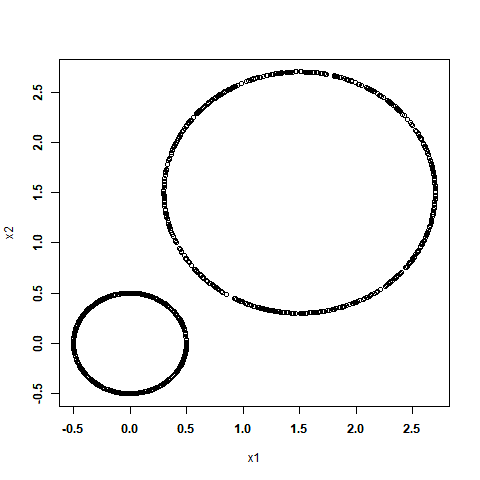}
\includegraphics[width=2in, height=2in]{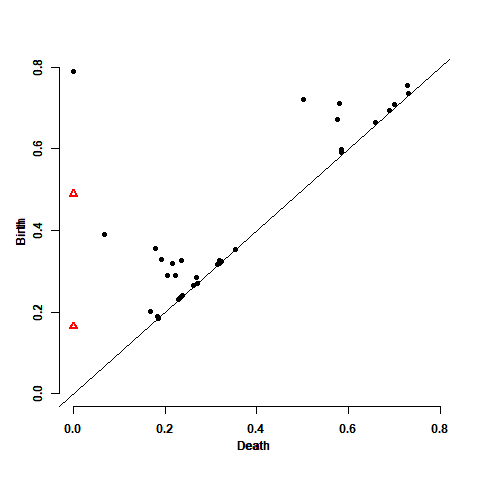}
\ec
%\caption{\footnotesize
% A random sample from two circles, 500 points from the larger circle and 300 from the smaller one,  with a kernel density
\caption{\footnotesize Left: A sample of $n=1,300$ points from a two distinct circles object, each circle has 650 points. Right: The corresponded persistence diagram for its upper level sets. Black circles are connected components ($H_0$ persistence points), red triangles are holes ($H_1$ points). Birth times are on the vertical axis.}
\label{fig:distinct}
\end{figure}

\begin{landscape}
\begin{figure}[h!]
\bc
\includegraphics[width=1.2in, height=1.4in]{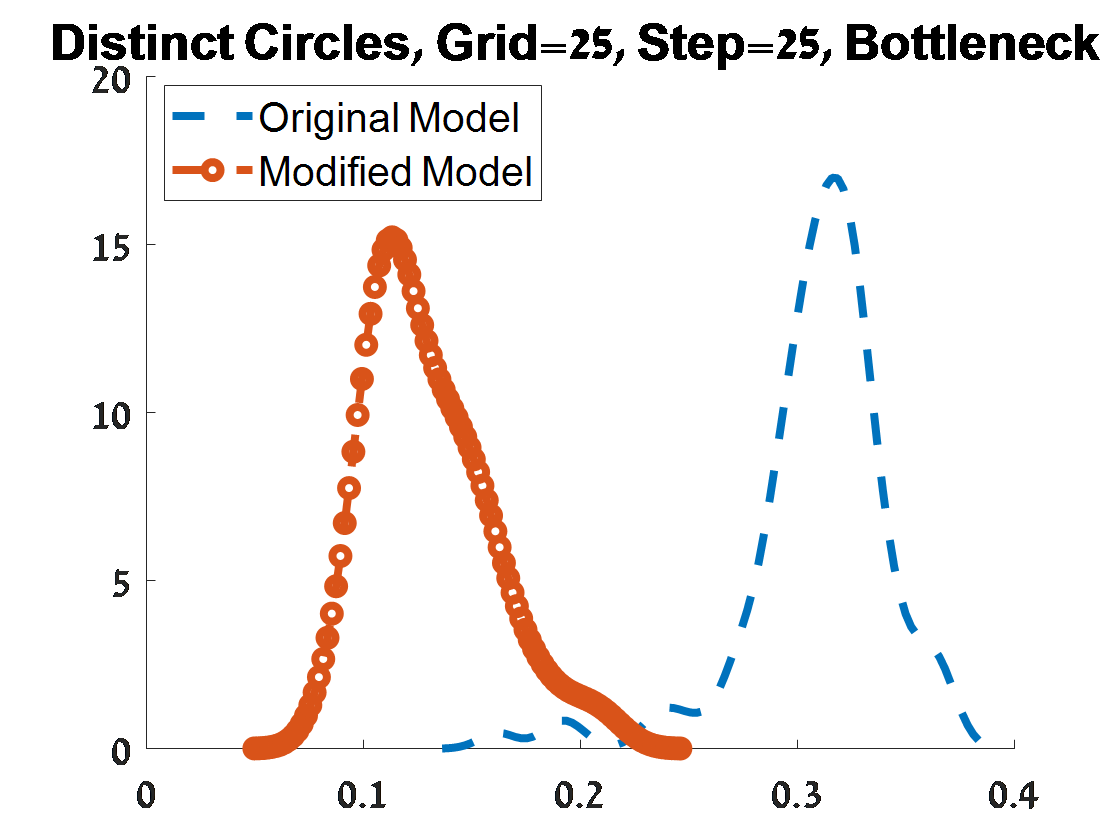}
\includegraphics[width=1.2in, height=1.4in]{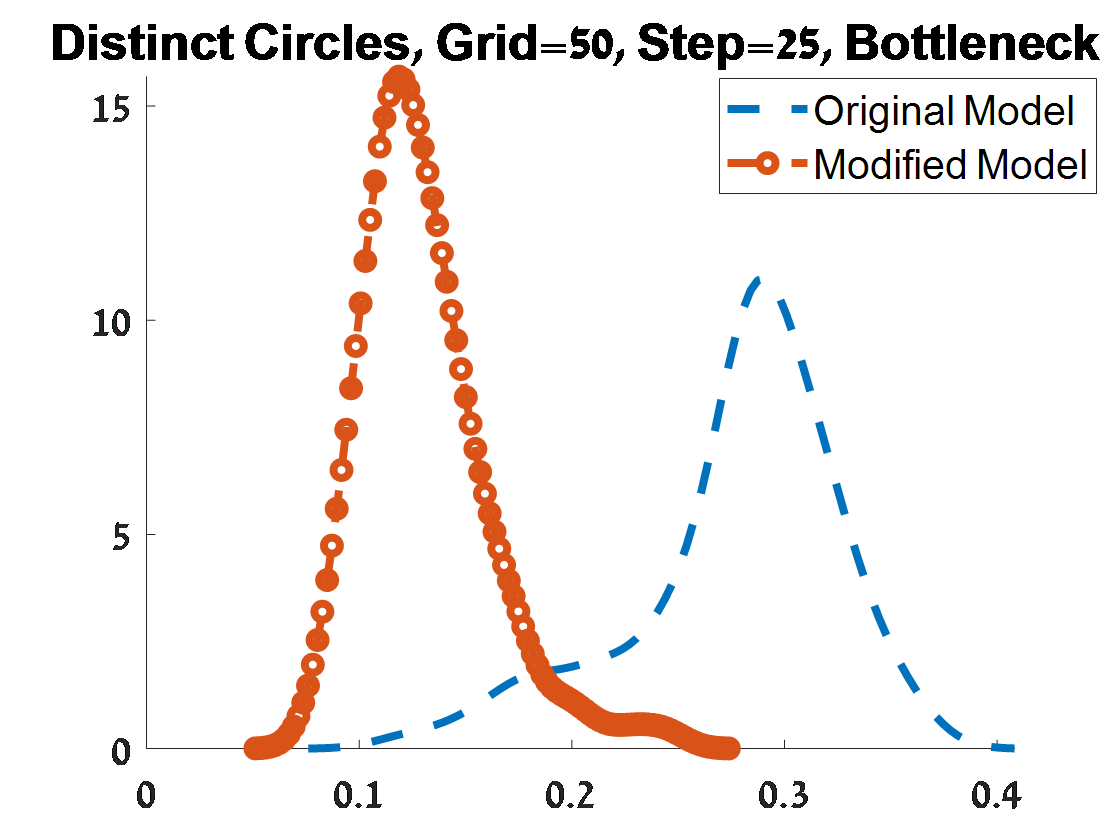}
\includegraphics[width=1.2in, height=1.4in]{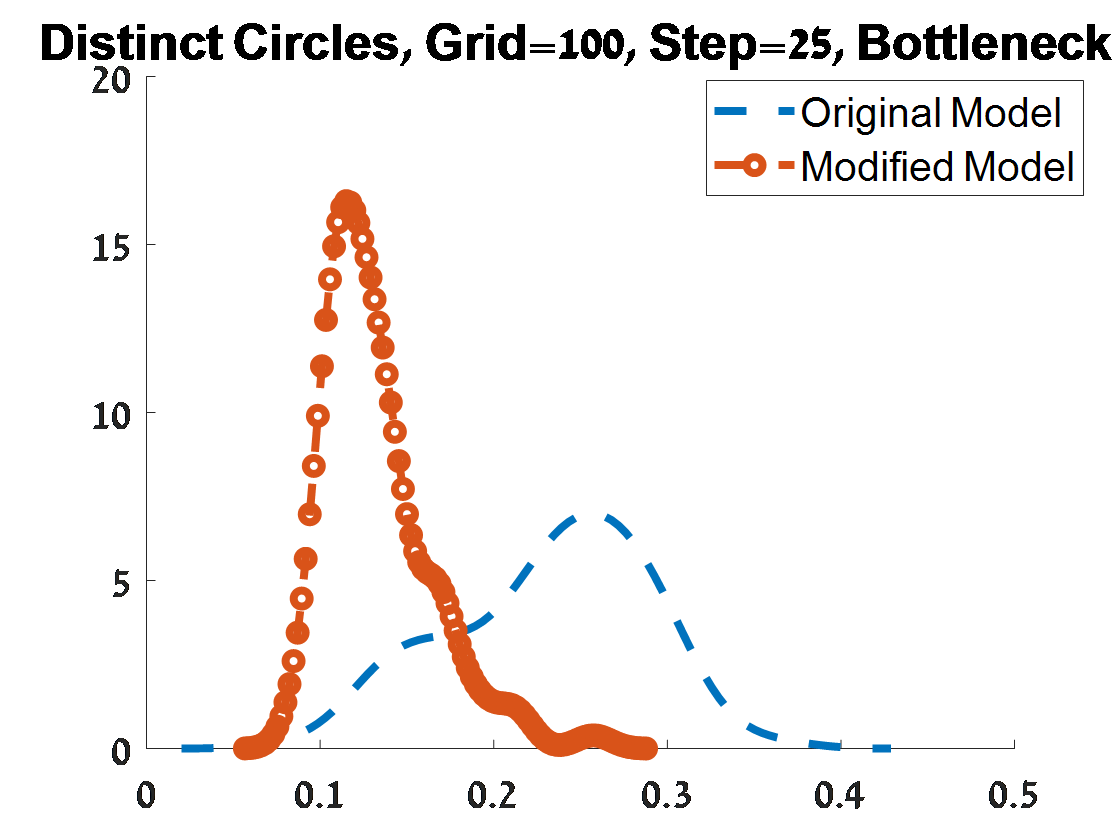}
\includegraphics[width=1.2in, height=1.4in]{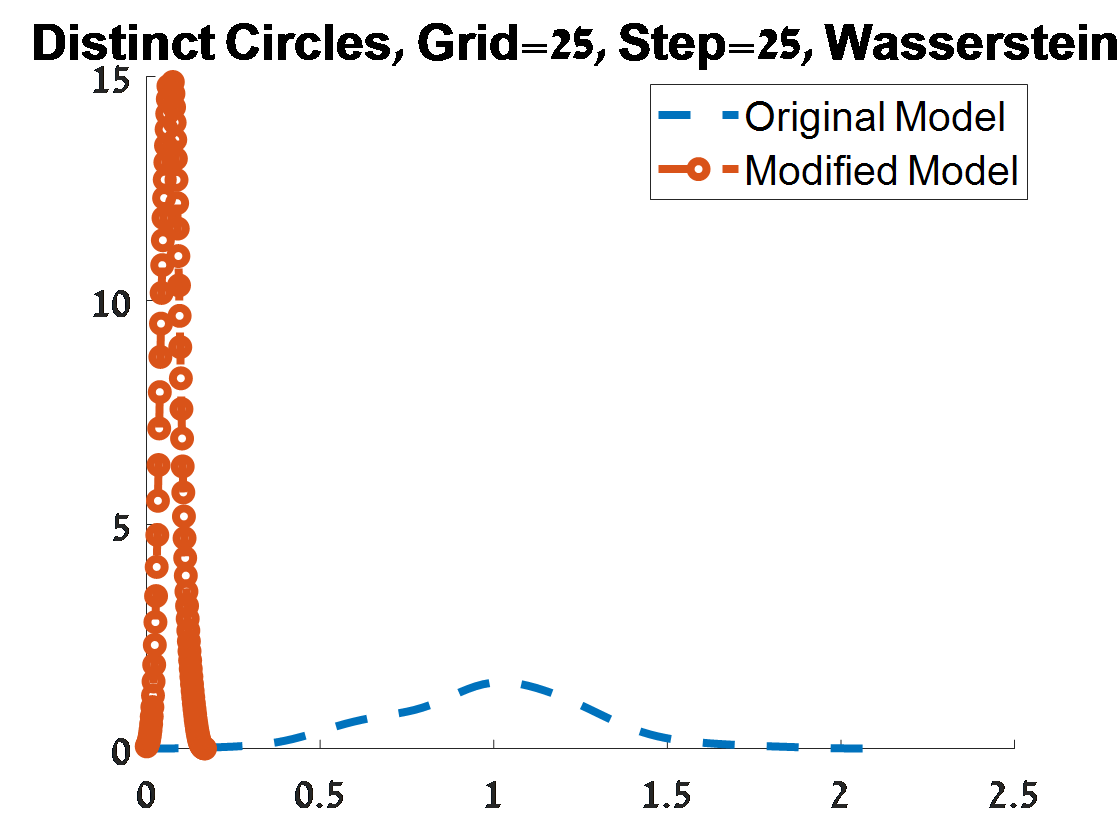}
\includegraphics[width=1.2in, height=1.4in]{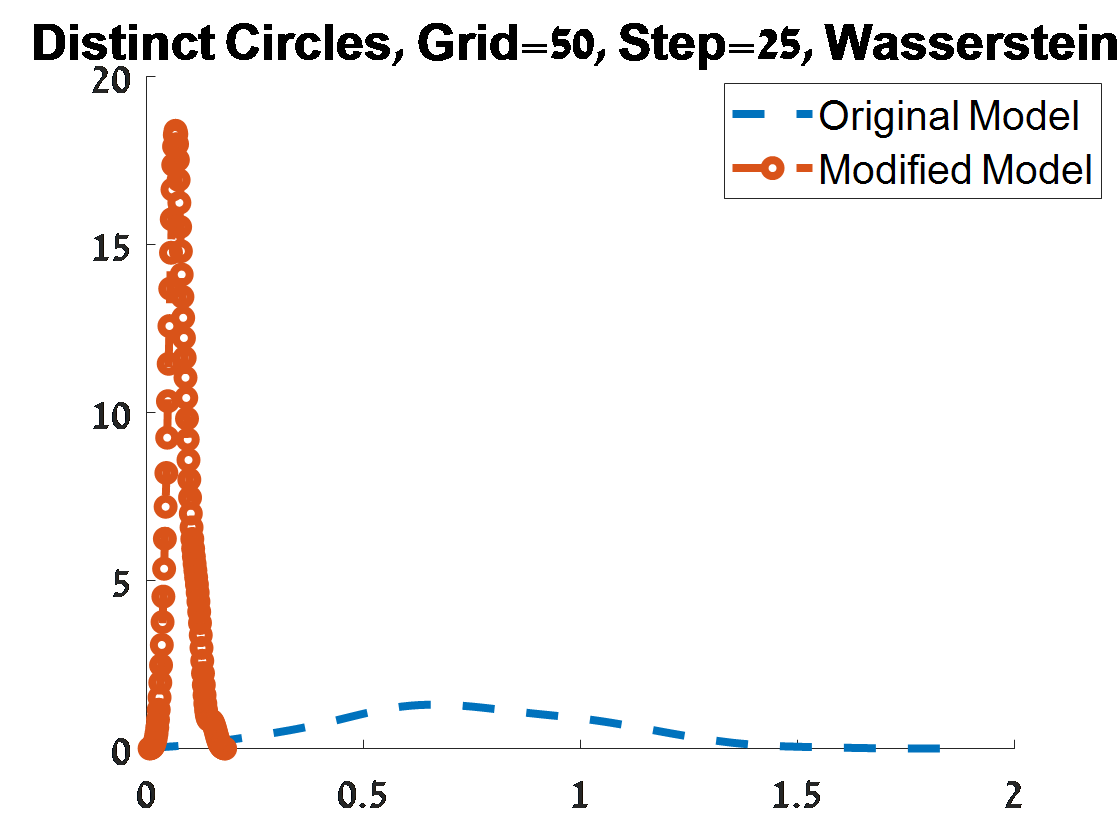}
\includegraphics[width=1.2in, height=1.4in]{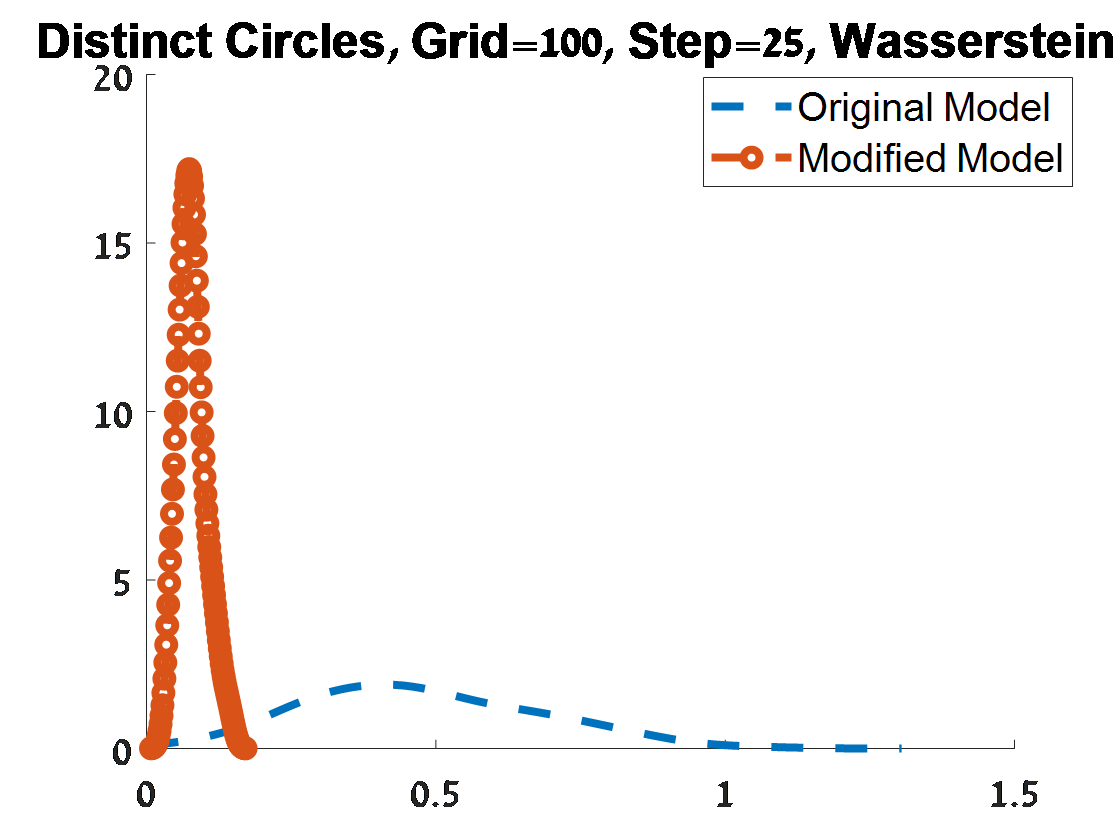}
%\\
\includegraphics[width=1.2in, height=1.4in]{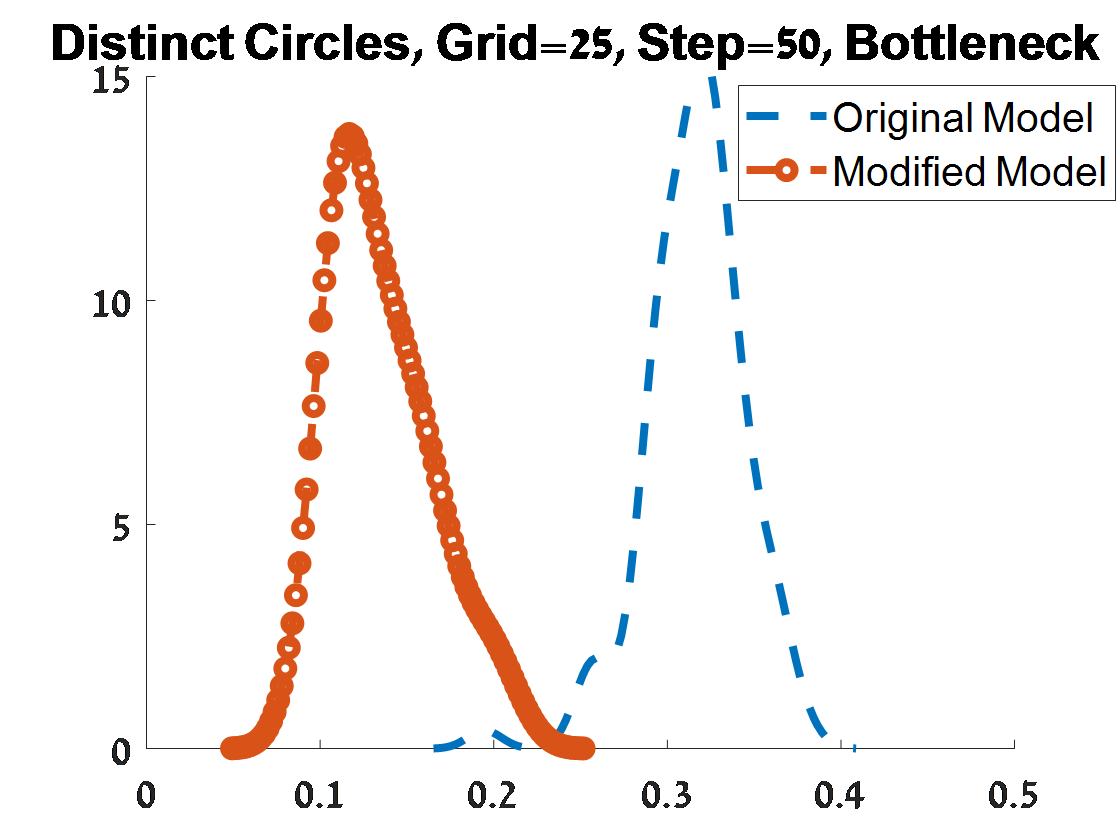}
\includegraphics[width=1.2in, height=1.4in]{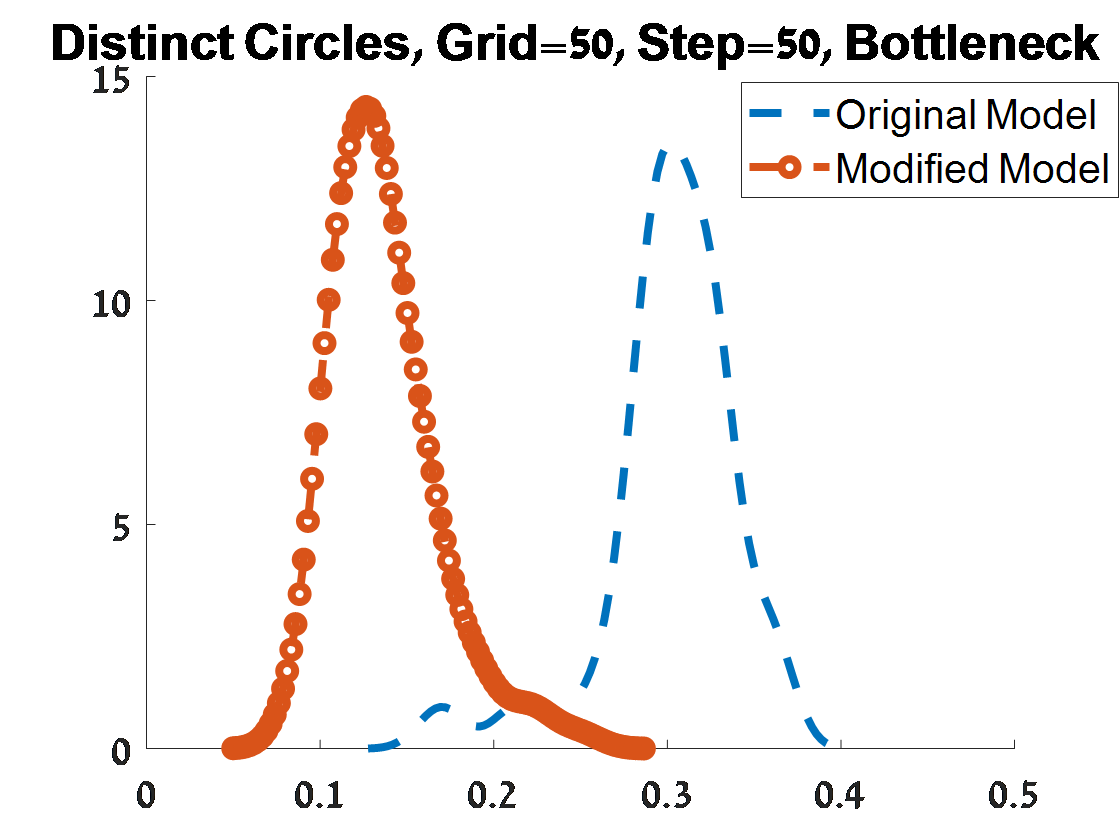}
\includegraphics[width=1.2in, height=1.4in]{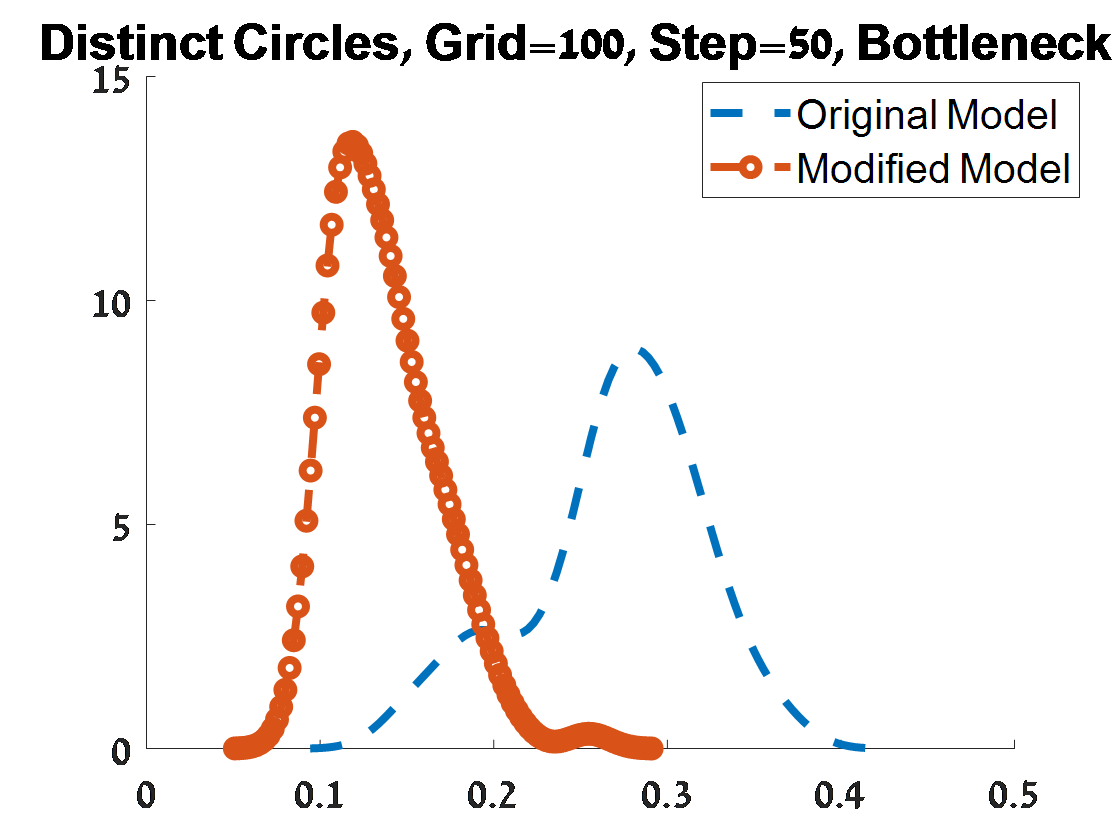}
\includegraphics[width=1.2in, height=1.4in]{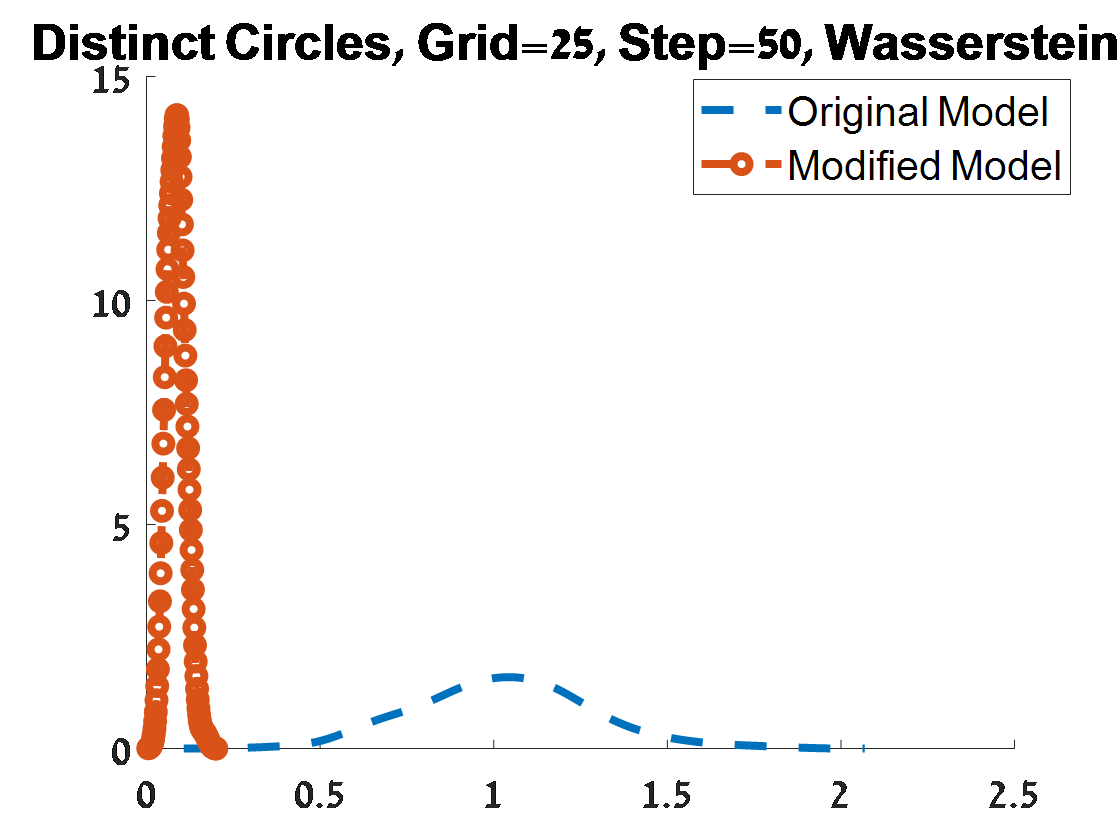}
\includegraphics[width=1.2in, height=1.4in]{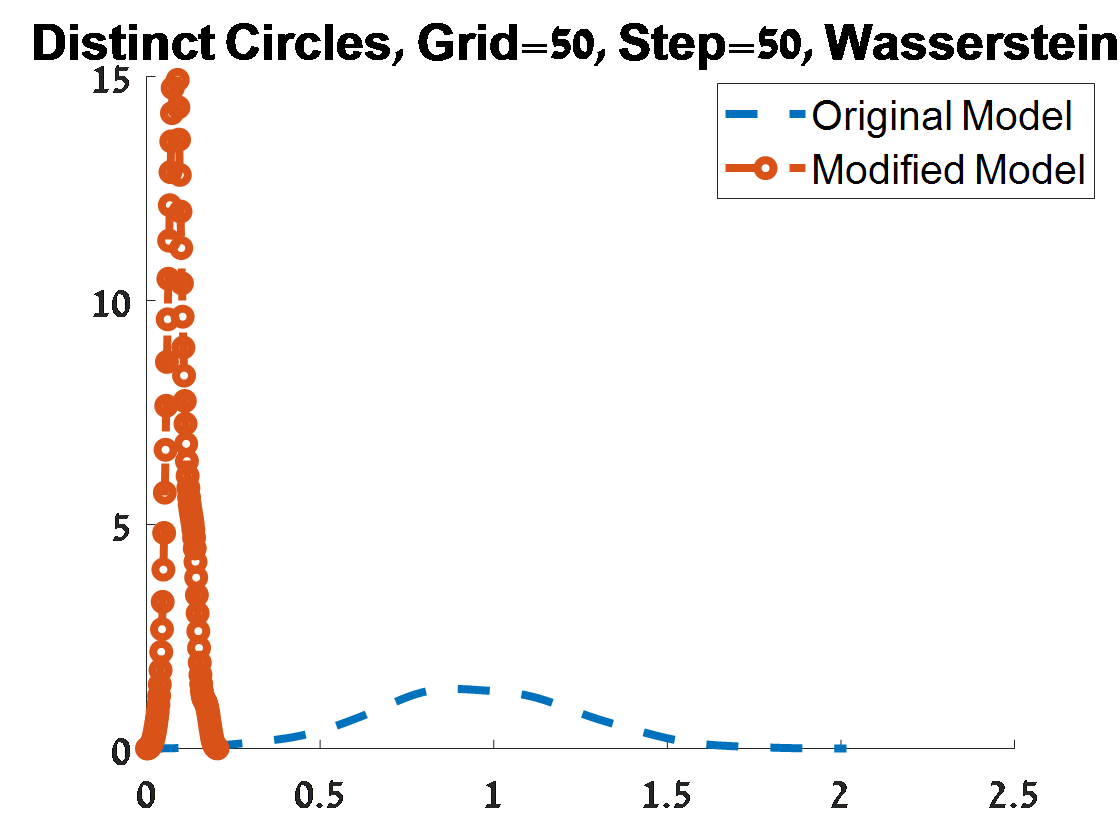}
\includegraphics[width=1.2in, height=1.4in]{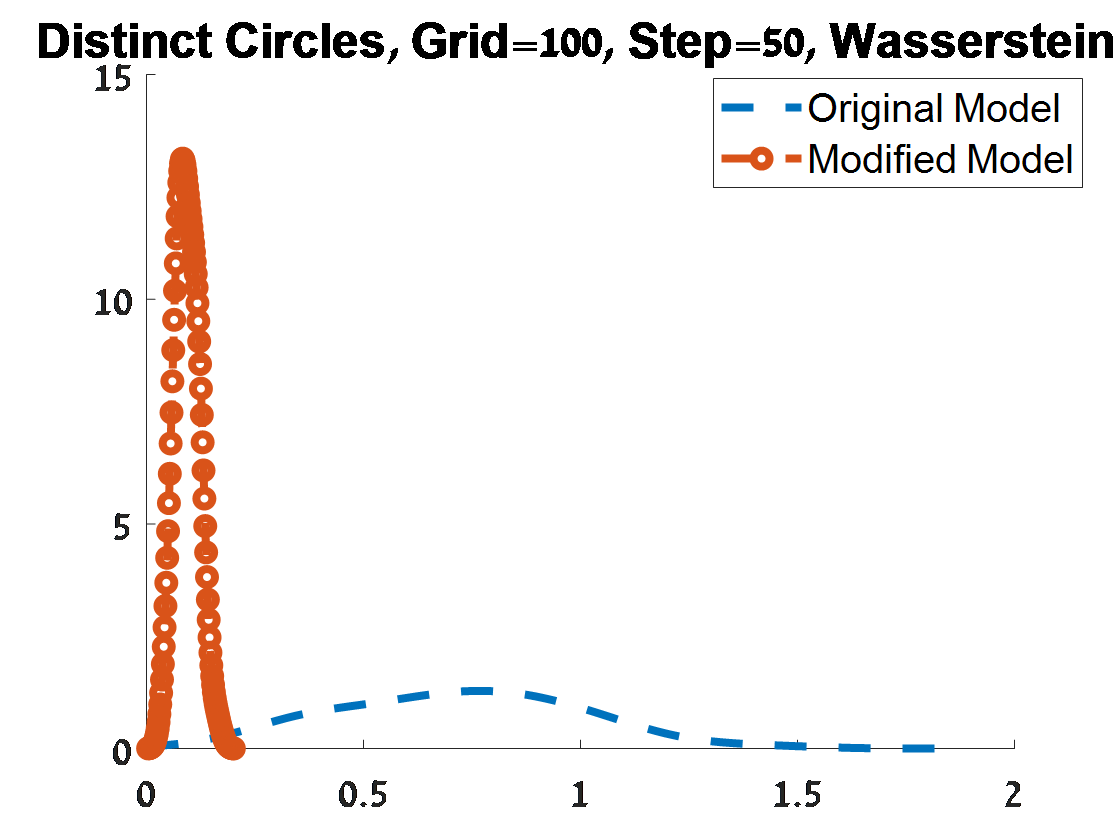}
%\\
\includegraphics[width=1.2in, height=1.3in]{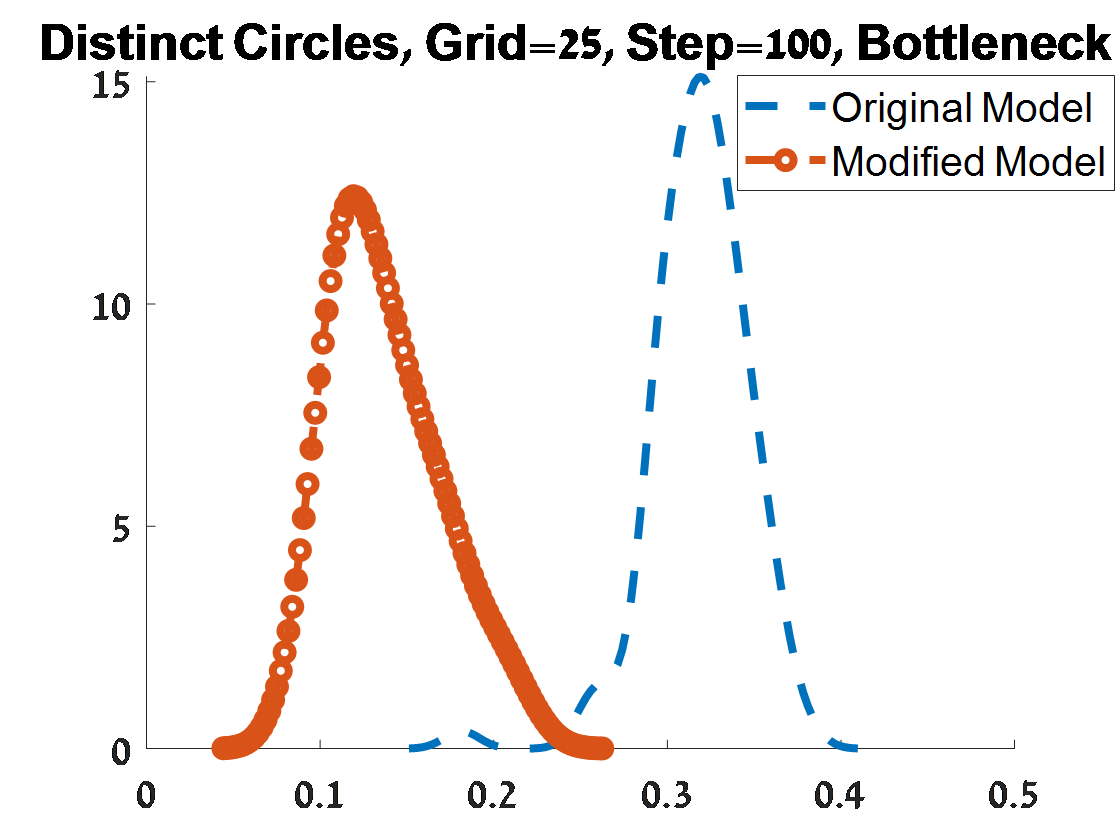}
\includegraphics[width=1.2in, height=1.3in]{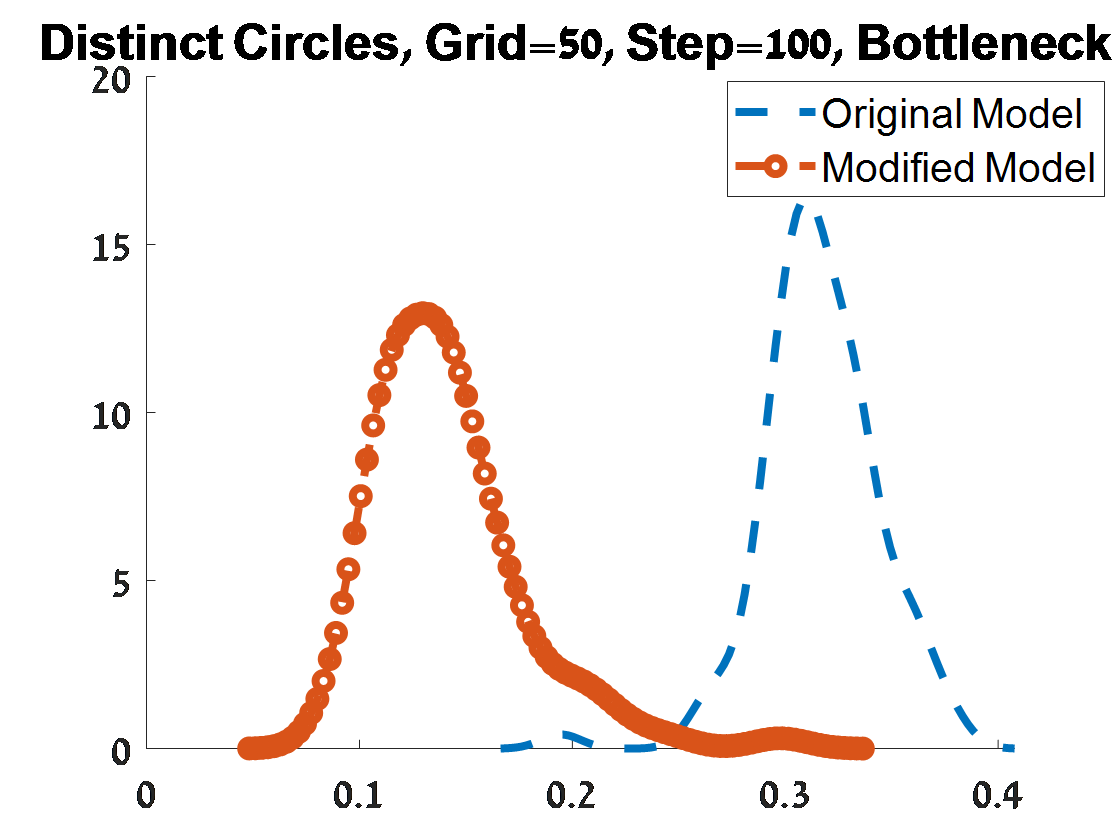}
\includegraphics[width=1.2in, height=1.3in]{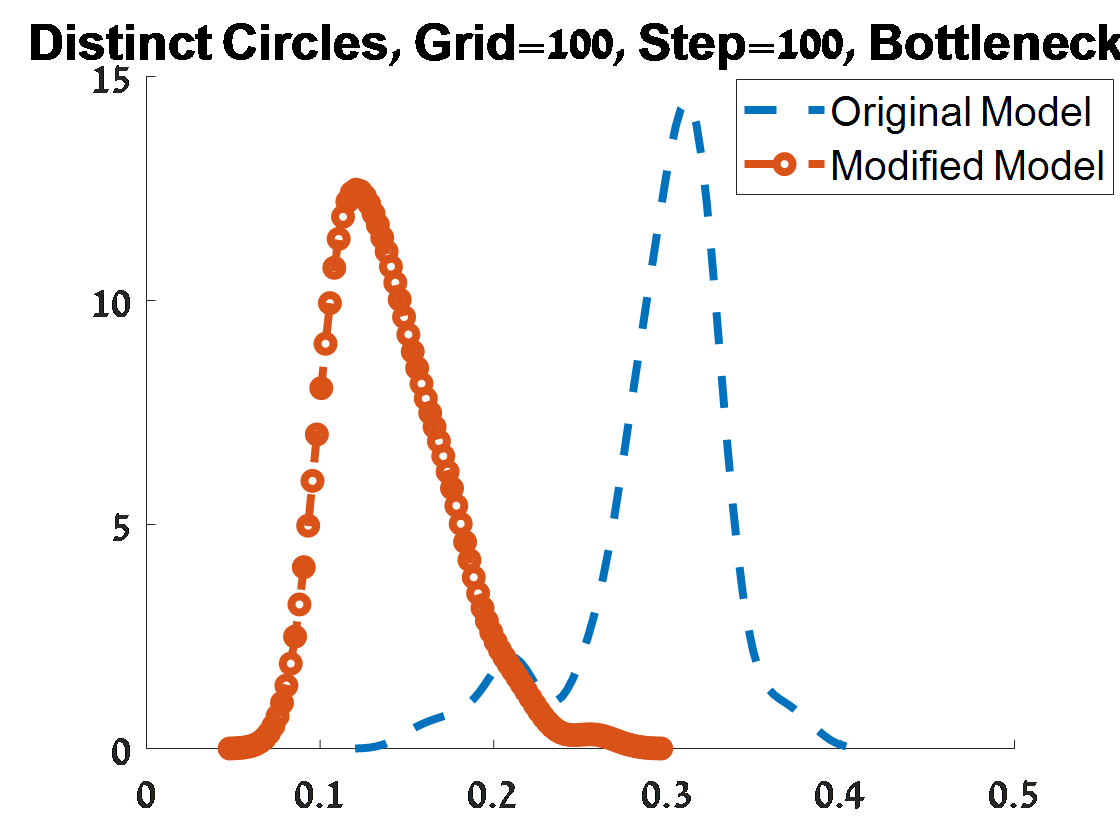}
\includegraphics[width=1.2in, height=1.3in]{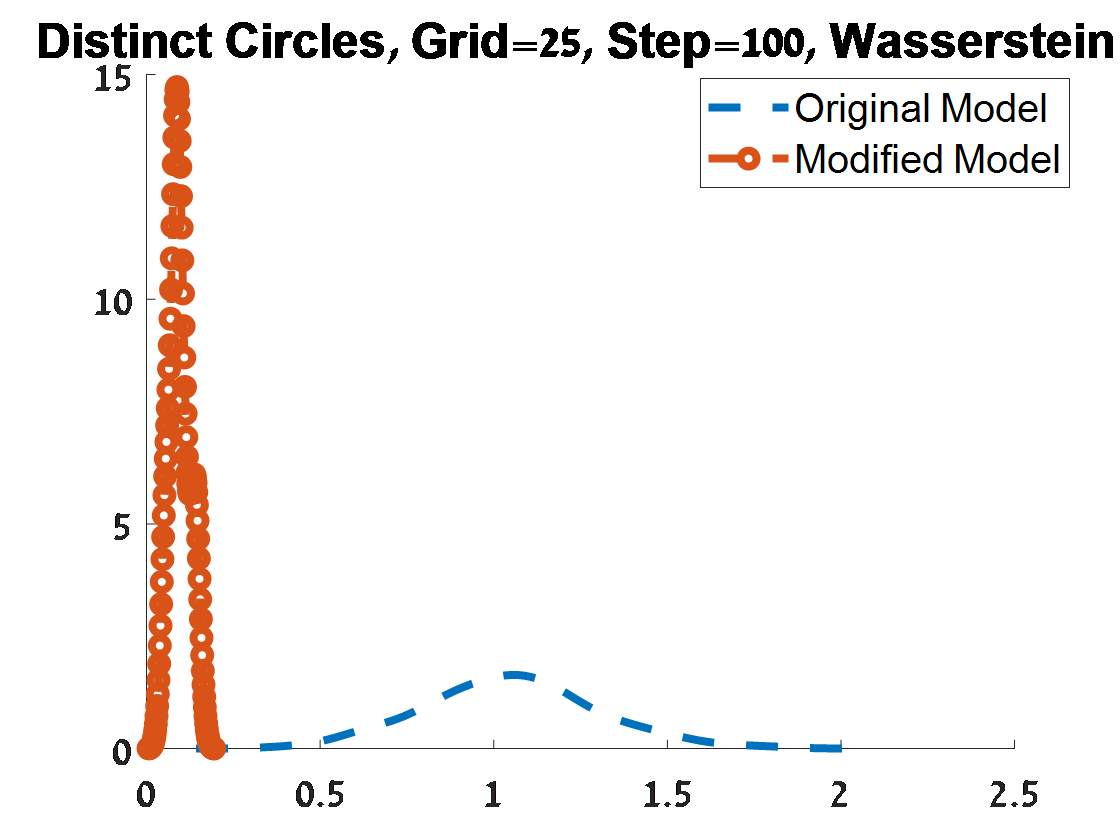}
\includegraphics[width=1.2in, height=1.3in]{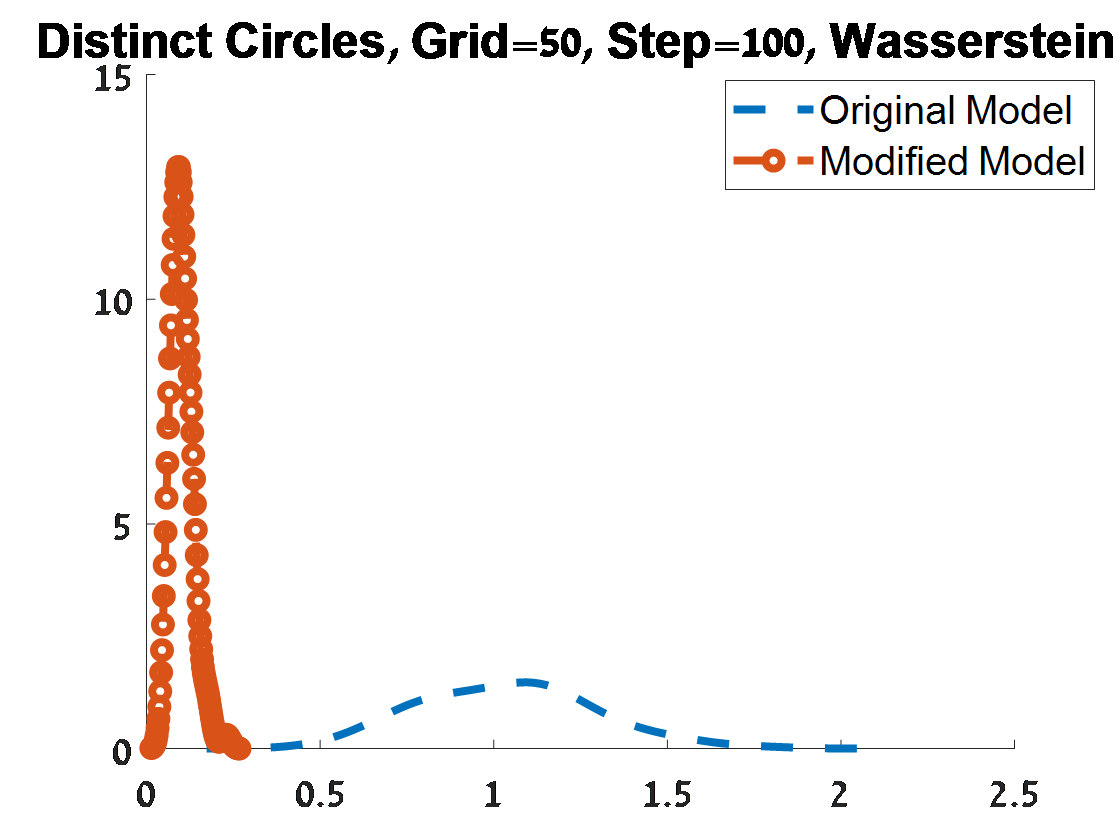}
\includegraphics[width=1.2in, height=1.3in]{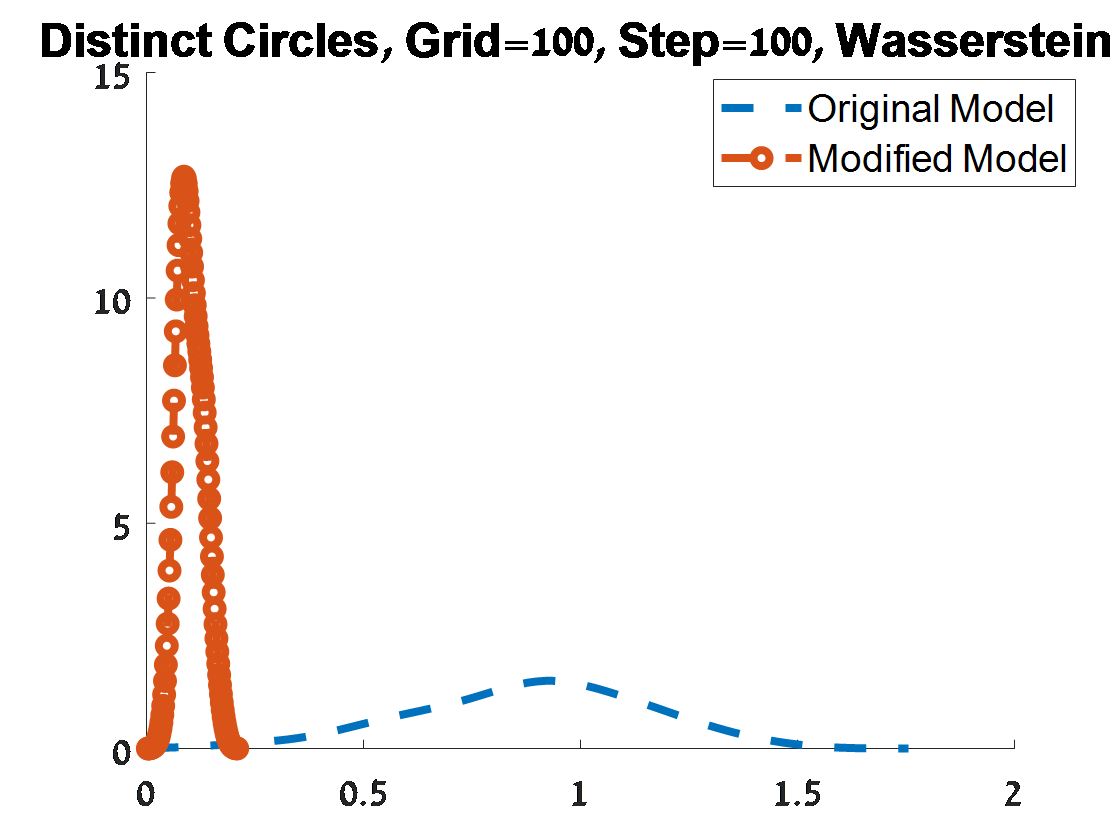}
\ec
%\caption{\footnotesize
% A random sample from two circles, 500 points from the larger circle and 300 from the smaller one,  with a kernel density
\caption{\footnotesize
 Criterion 1 of goodness of fit for 100 PDs corresponded to 100 samples from an object of two distinct circles. The figures depend on the grid of the proposal distribution ("Grid"), and the burn-in ("Step") of the MCMC algorithm.}
\label{fig:distinct_a}
\end{figure}
\end{landscape}

While the previous example contained data of two circles, this example contains data of two distinct circles. One circle has a radius $r_1=0.5$, the second circle has a radius $r_2=1.2$, and the distance between these two circles is 1.5 for each point. We consider a sample of $n=1,300$ points, where the number of points on each circle is 650.
\\
The typical sample is presented in the left side of Figure\ \ref{fig:distinct}. To its right, we have its corresponded PD.
We generated 100 such samples, calculated their corresponded PDs, and fitted the both model's versions for the $H_0$ points of each PD.
Figure\ \ref{fig:distinct_a} describes the distributions over the 100 PD of the first criterion of goodness of fit, and Figures 11-12 describe the distributions of the second criterion of goodness of fit.
In criterion 1, the Bottleneck distance between the simulated PD and the real PD is better under the modified model relative to that distance under the original model. This distance's distribution is similar for all considered values of grid size and burn-in. In Wasserstein distance, the best fitting is under the modified model relative to the fitting under the original model. Moreover, the fitting in Wasserstein distance is better (that is, a smaller distance) as the grid size increases for a given step, and this fitting is better for step of 25 given a specific considered value of grid size.
For criterion 2,  the distributional properties of the modified model are better than those of the real PDs for each considered value of grid size and the burn-in.
%In Figure\ \ref{fig:distinct_d} we present two examples, each of them contains a real PD and its simulated PD based on the two model versions, only for the best scenarios we found, that is grid size of 50x50 and 100x100, and burn-in of 25. Also here we see that the best fitted MCMC is under the modified model rather than the original one. In addition, the modified model is better and succeed in creating the far points from the diagonal, where the original model is concentrating in creating more the points that are close to the diagonal and almost cannot succeed in generating the far points from the diagonal.
That is, for this example we have that the modified RST is better than the original RST, where the best fitting is under grid sizes of 50x50 and 100x100, and burn-in of 25.

\begin{landscape}
\begin{figure}[h!]
\bc
\includegraphics[width=1.2in, height=1.25in]{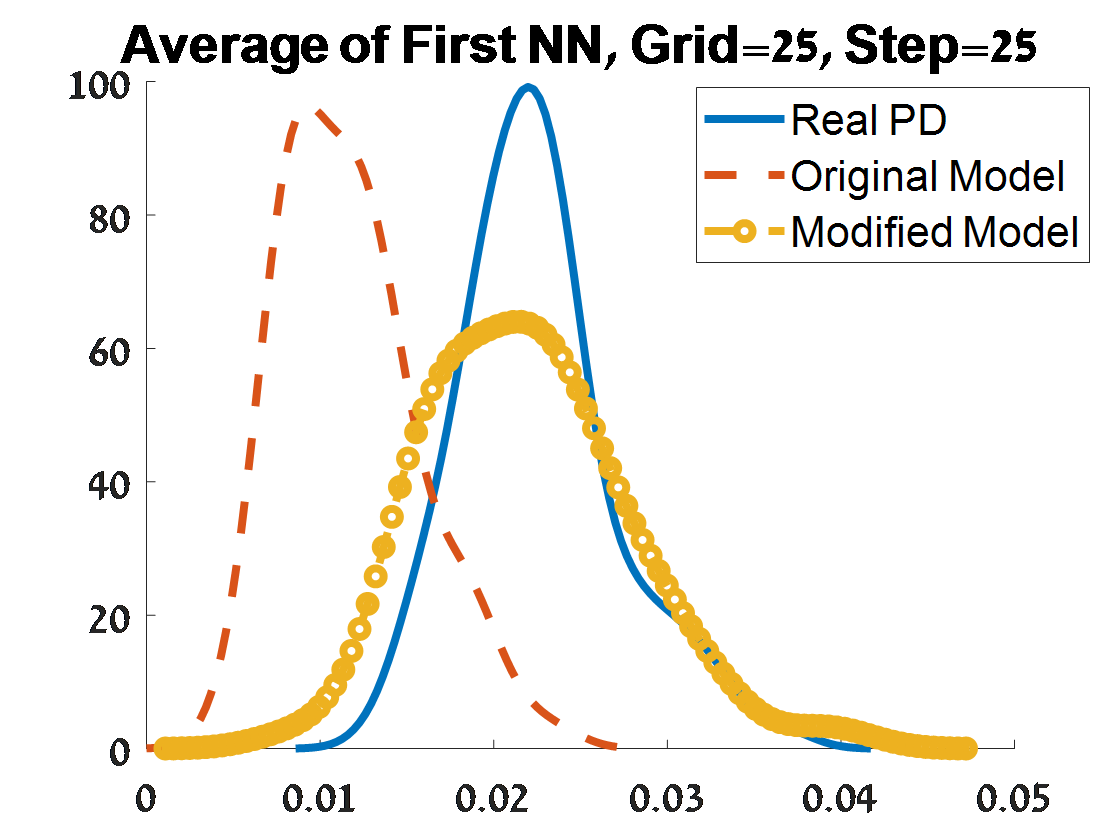}
\includegraphics[width=1.2in, height=1.25in]{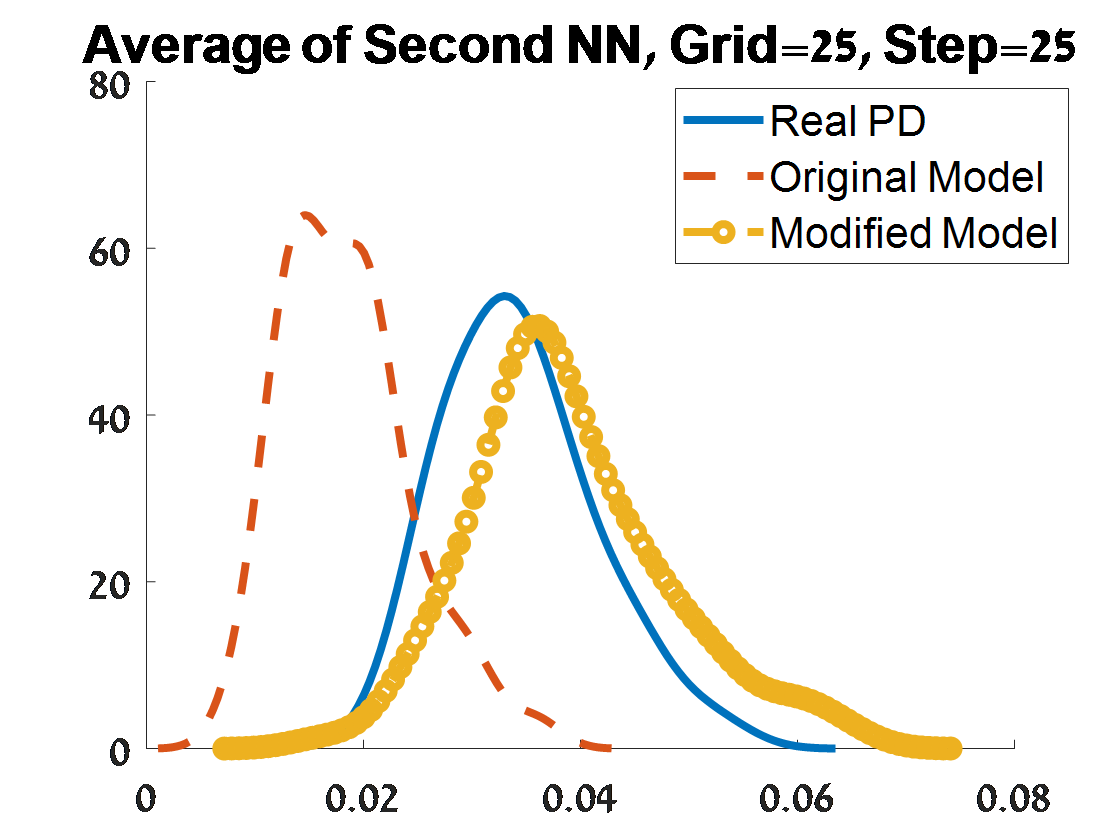}
\includegraphics[width=1.2in, height=1.25in]{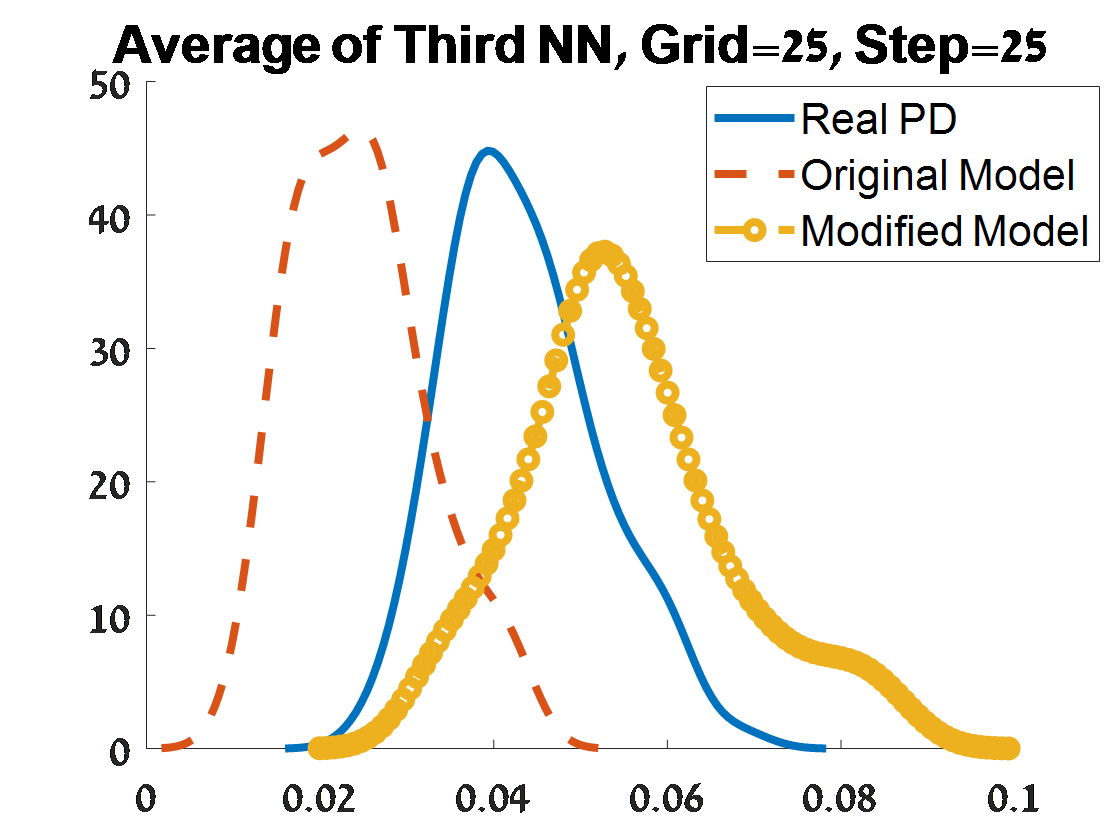}
\includegraphics[width=1.2in, height=1.25in]{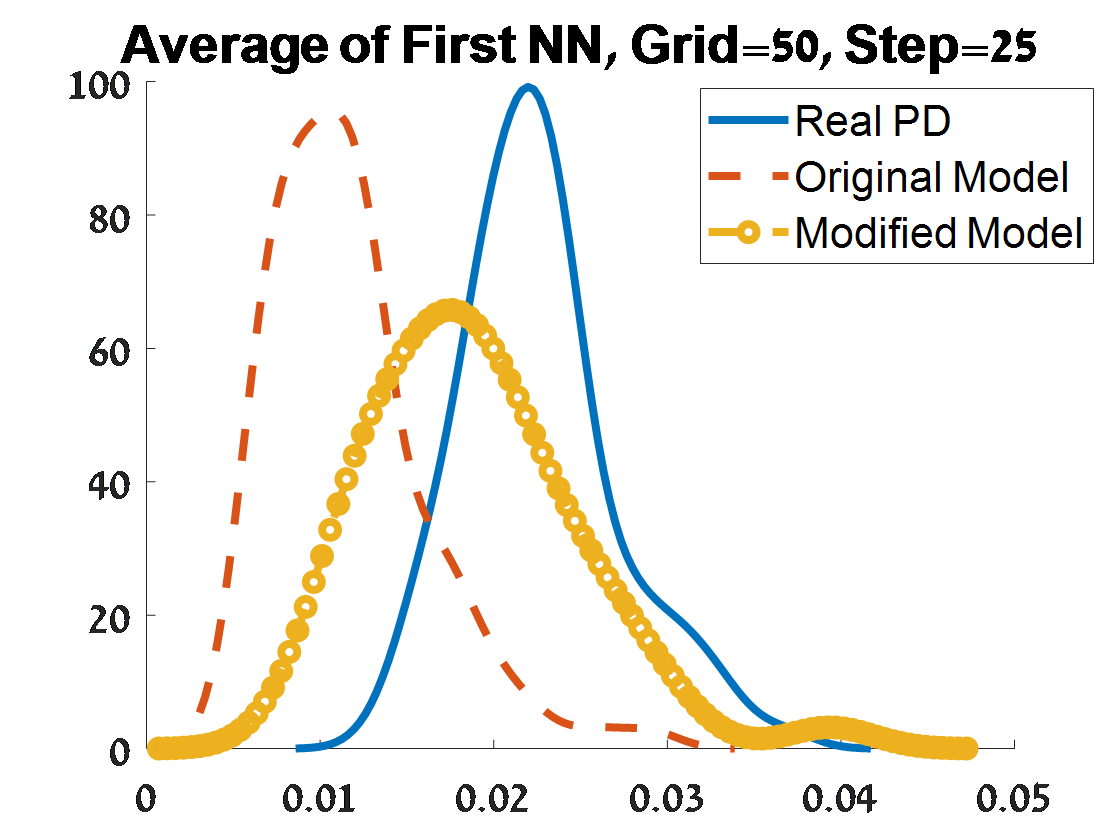}
\includegraphics[width=1.2in, height=1.25in]{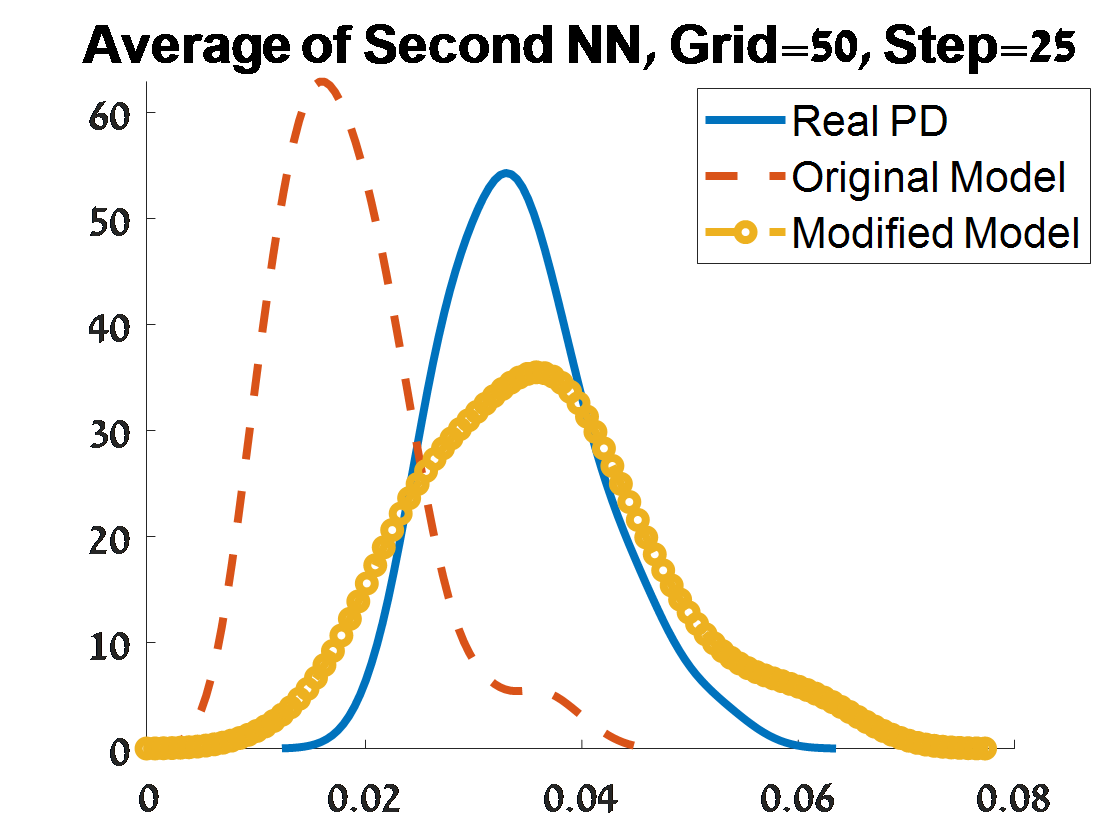}
\includegraphics[width=1.2in, height=1.25in]{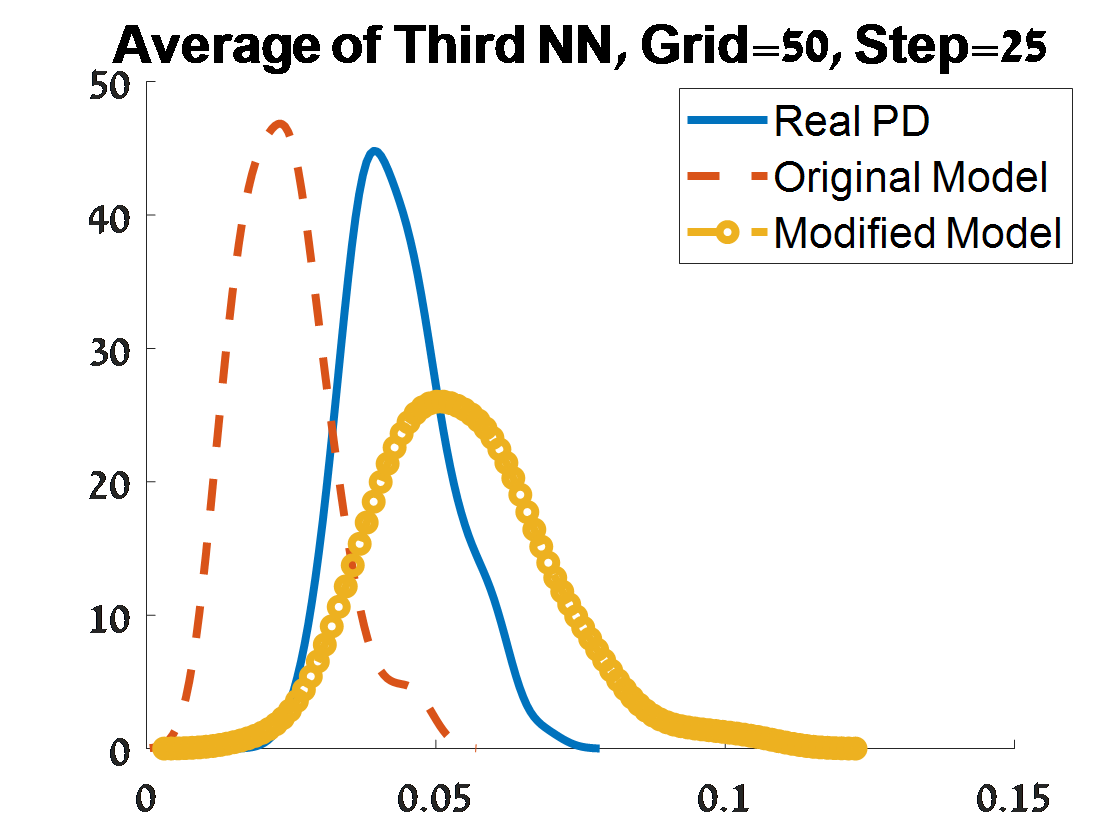}
\includegraphics[width=1.2in, height=1.25in]{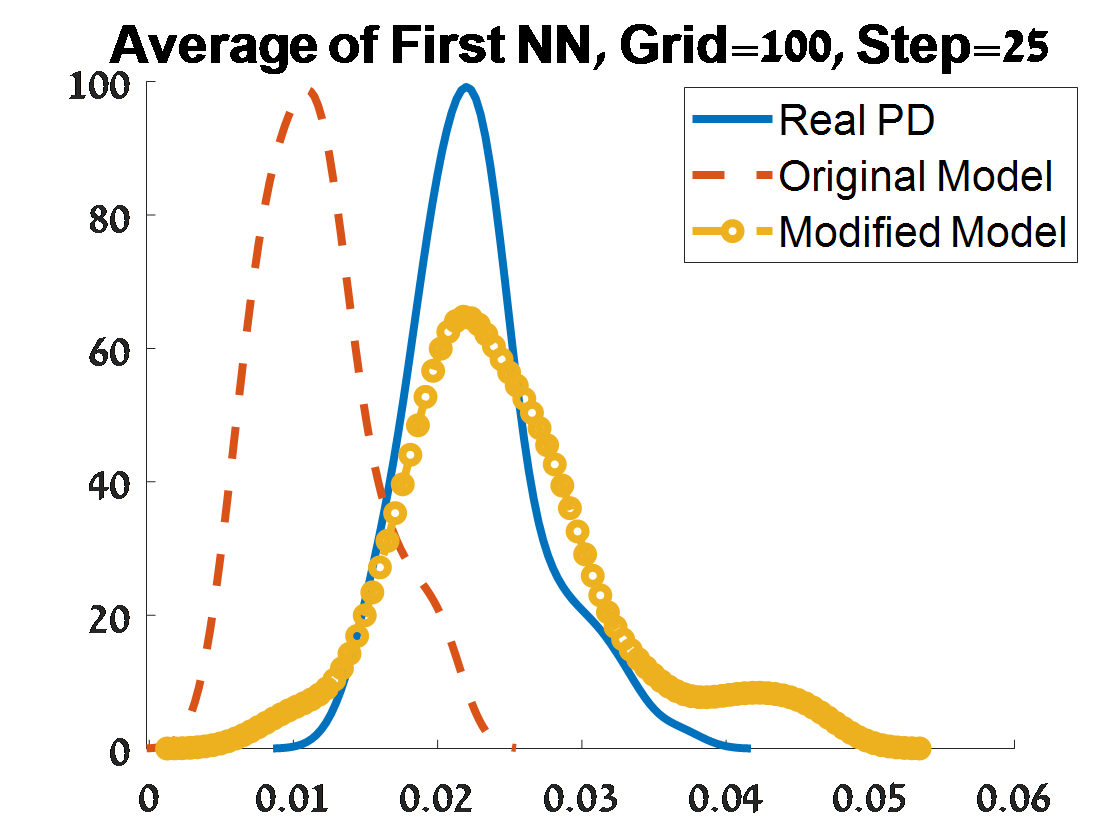}
\includegraphics[width=1.2in, height=1.25in]{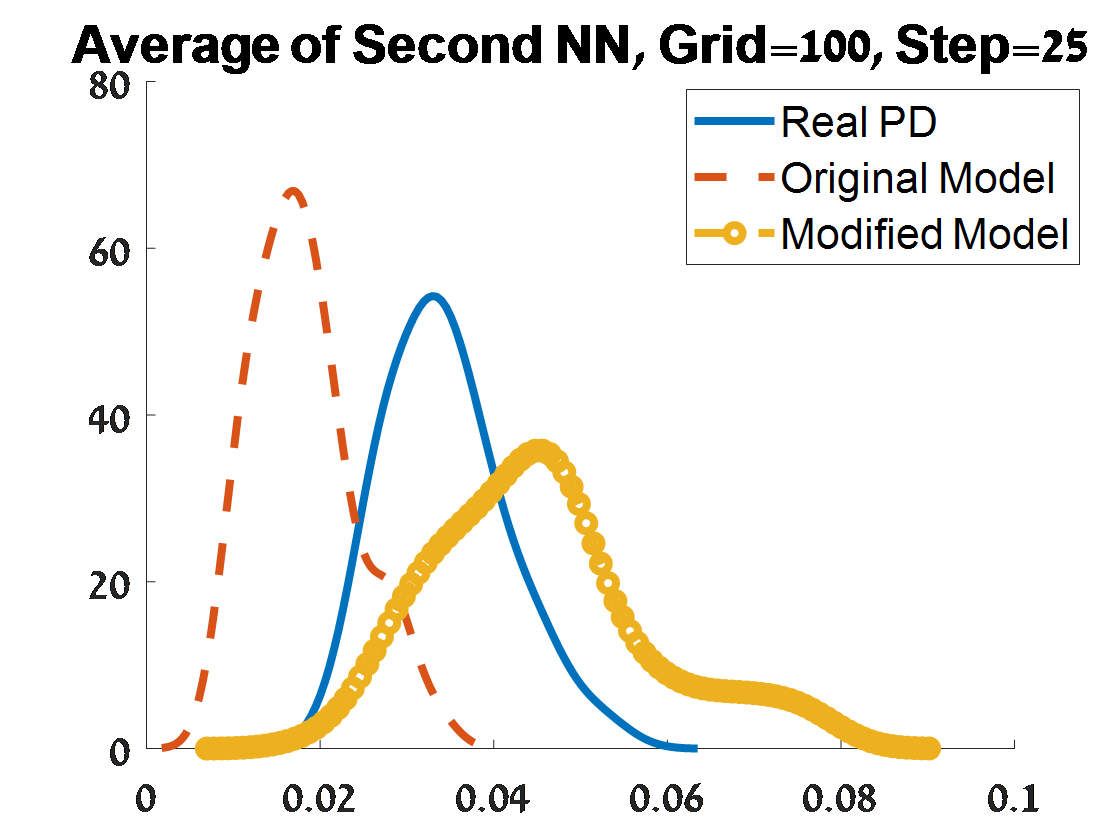}
\includegraphics[width=1.2in, height=1.25in]{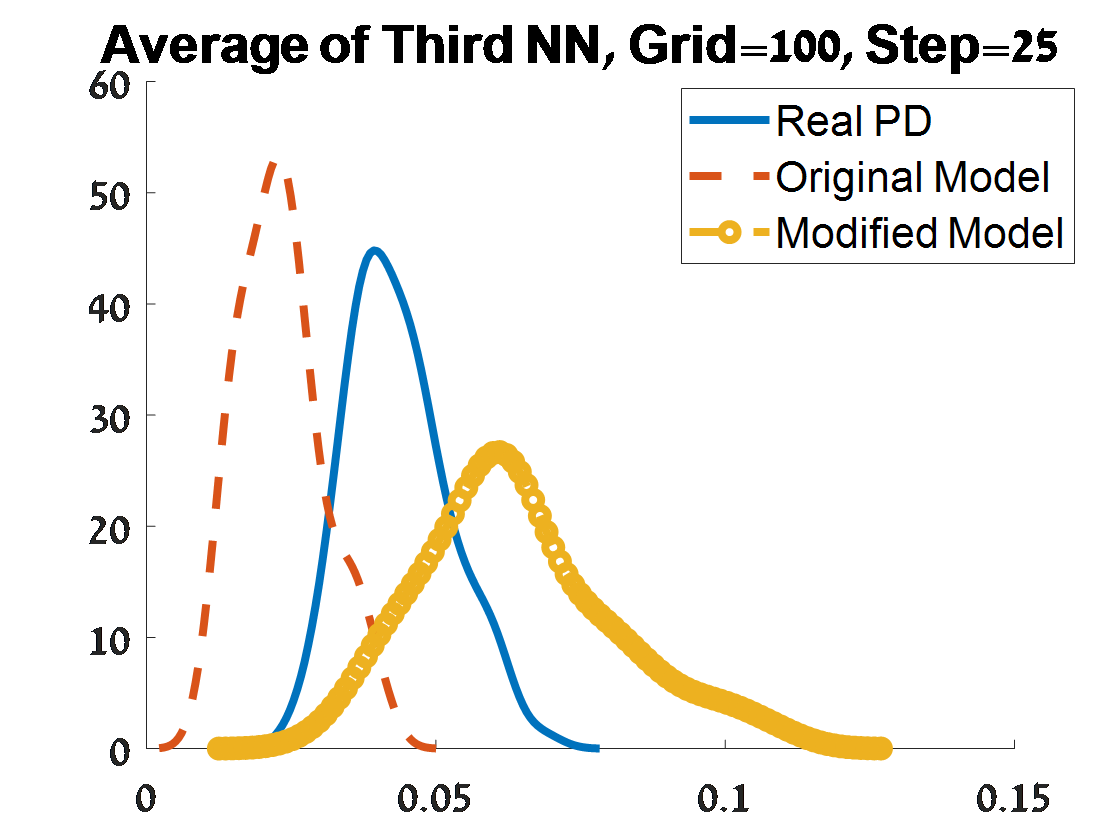}
\\
\includegraphics[width=1.2in, height=1.25in]{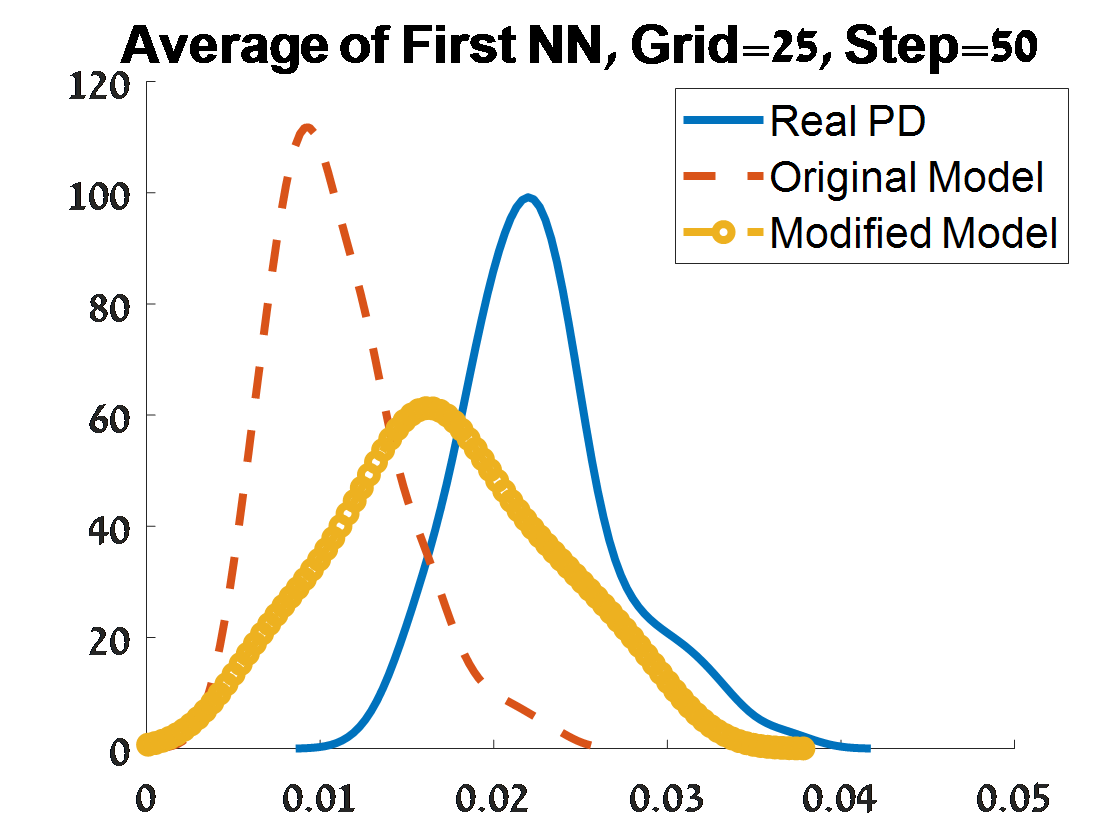}
\includegraphics[width=1.2in, height=1.25in]{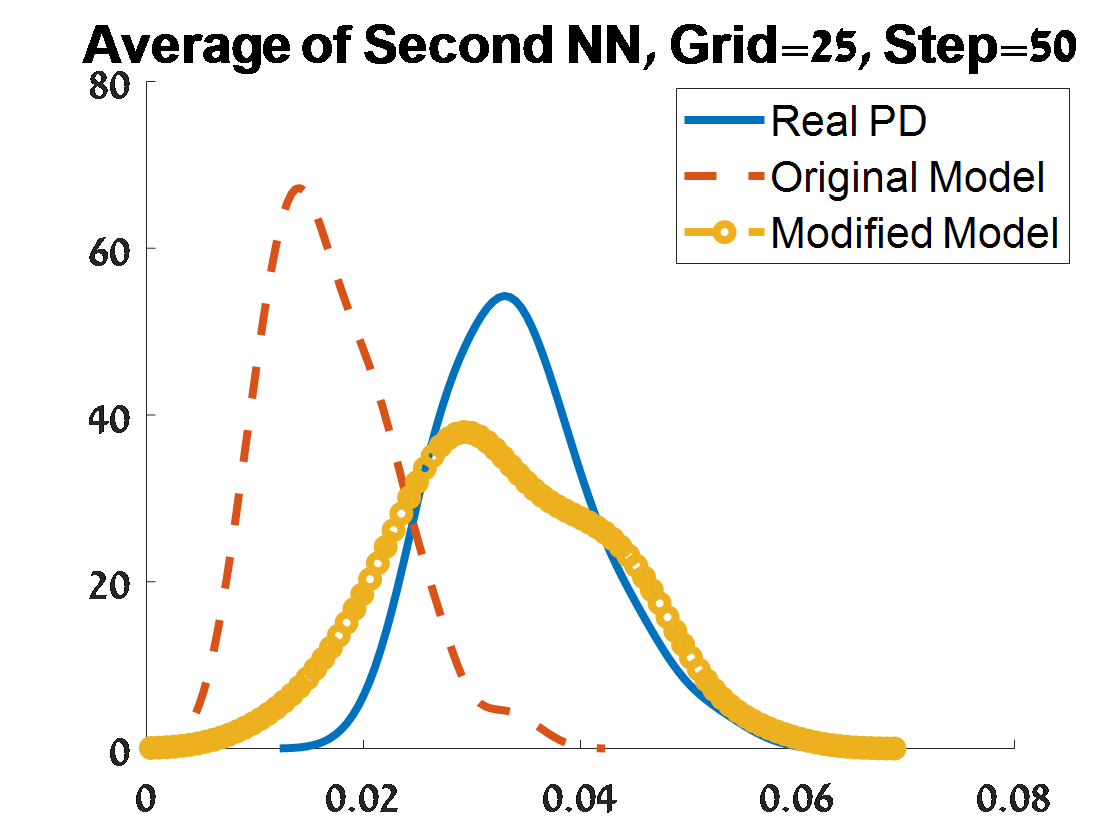}
\includegraphics[width=1.2in, height=1.25in]{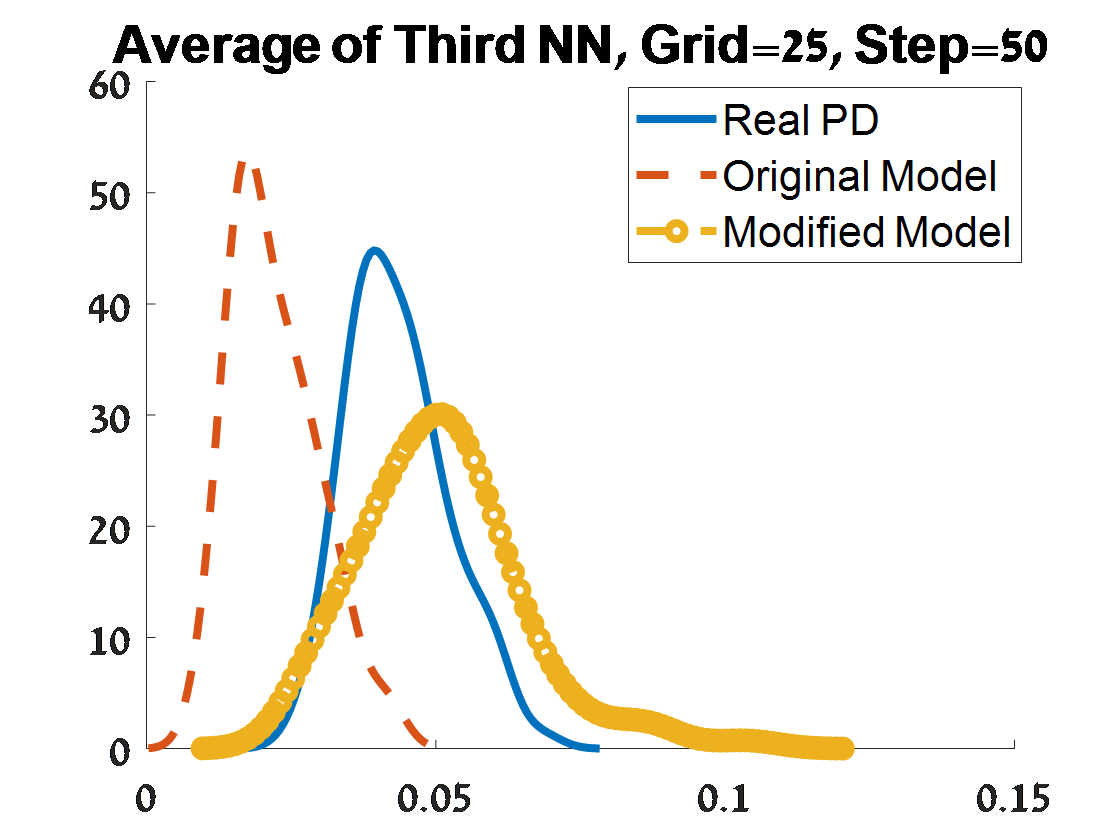}
\includegraphics[width=1.2in, height=1.25in]{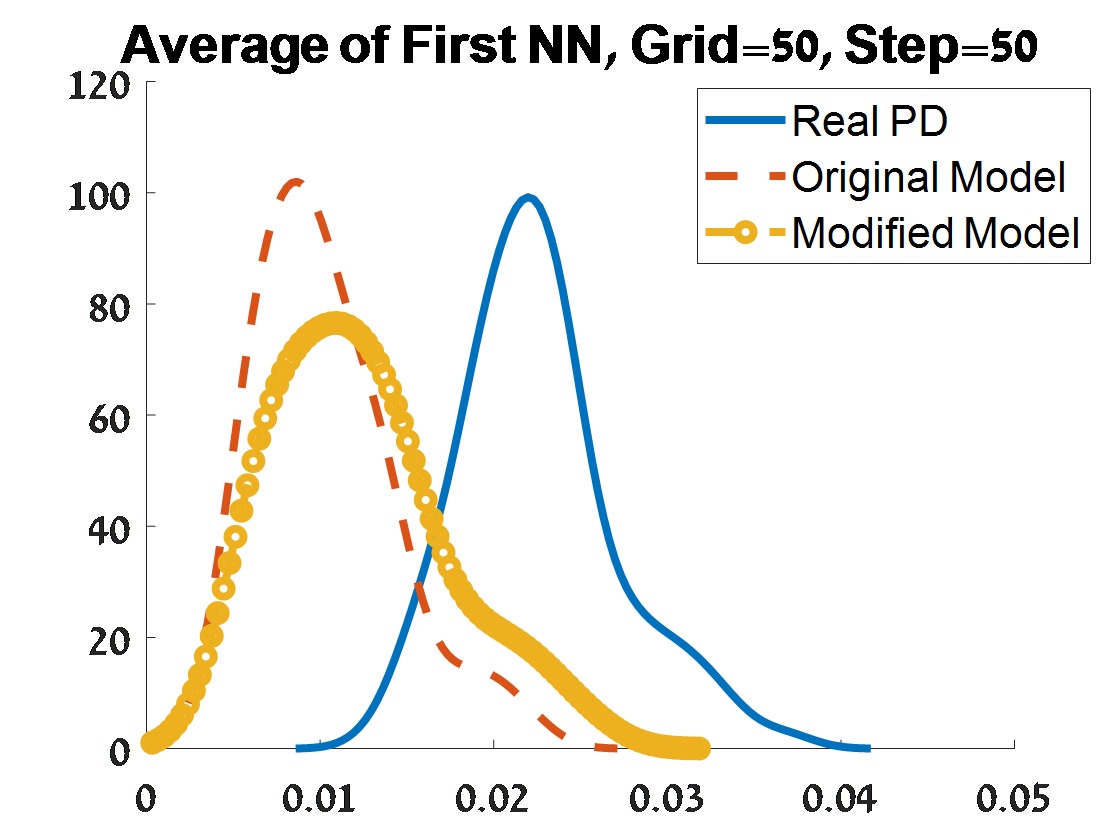}
\includegraphics[width=1.2in, height=1.25in]{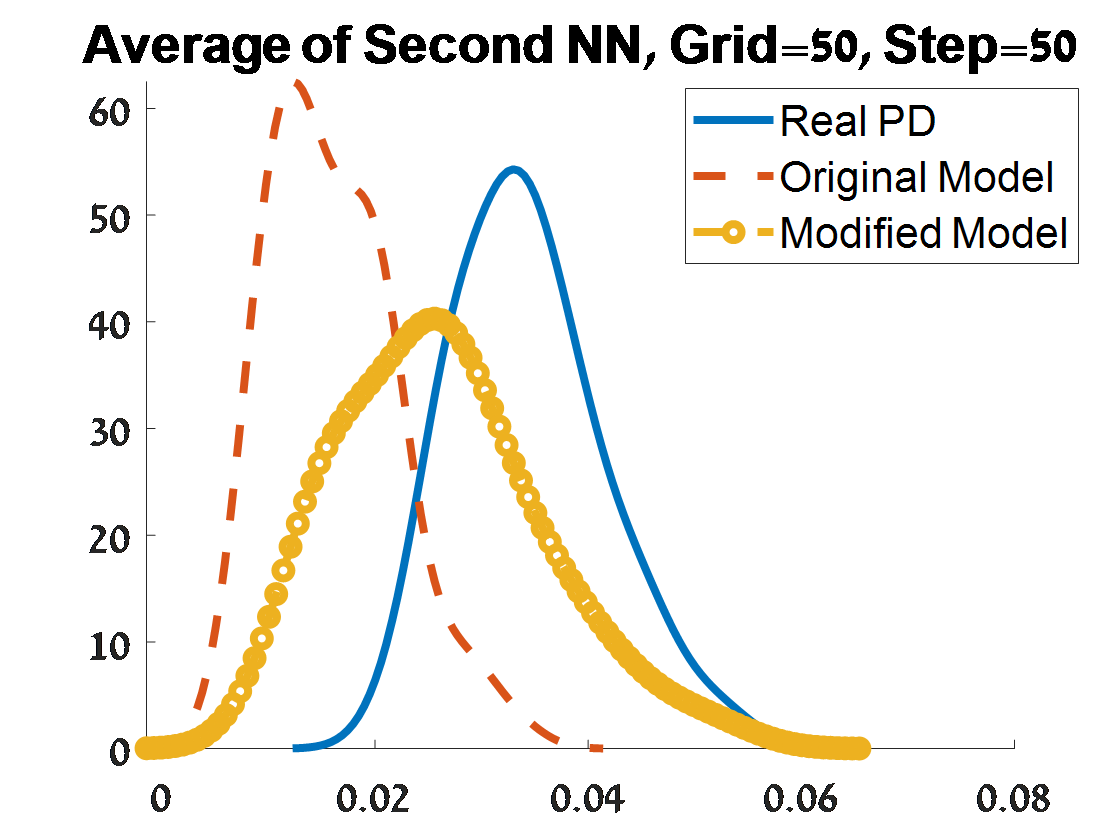}
\includegraphics[width=1.2in, height=1.25in]{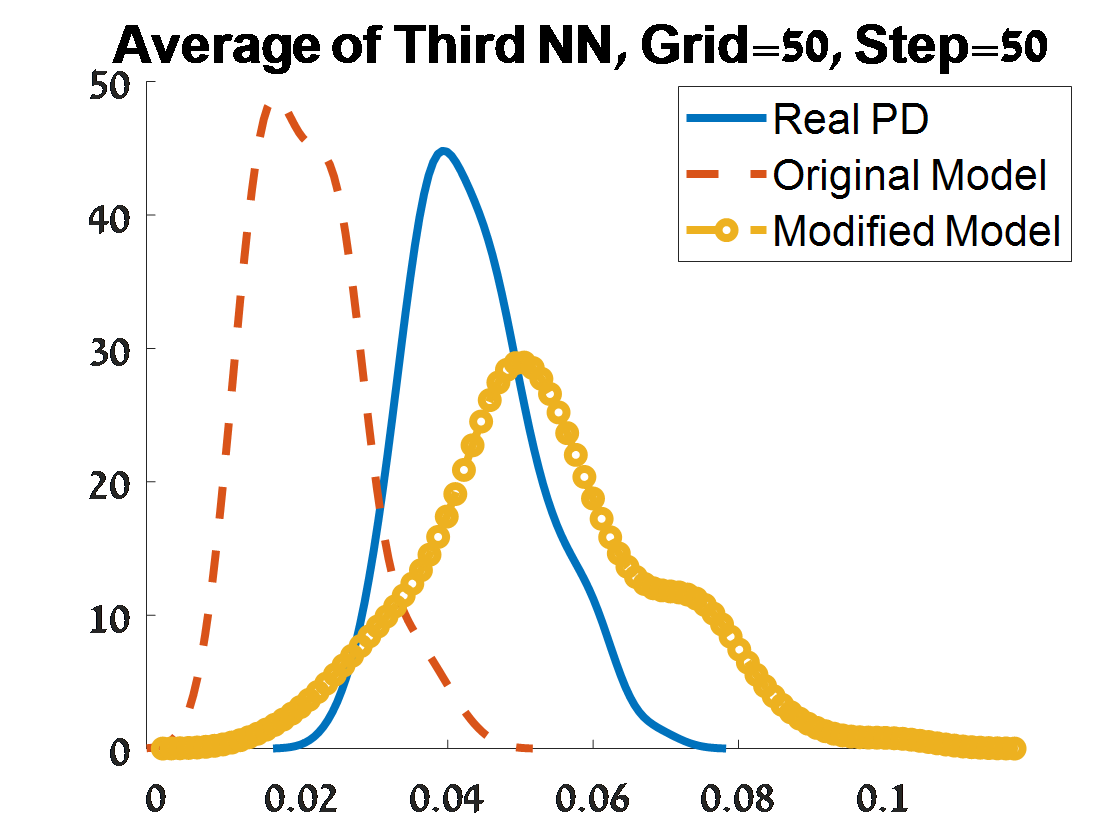}
\includegraphics[width=1.2in, height=1.25in]{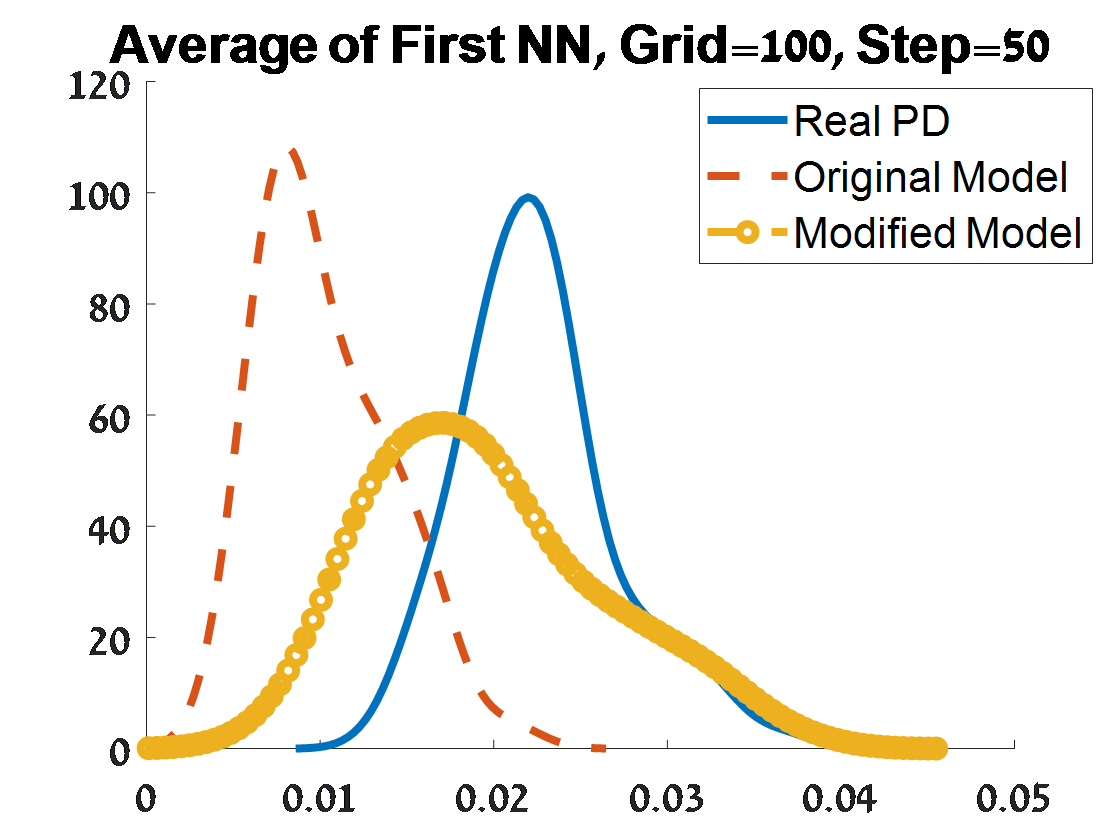}
\includegraphics[width=1.2in, height=1.25in]{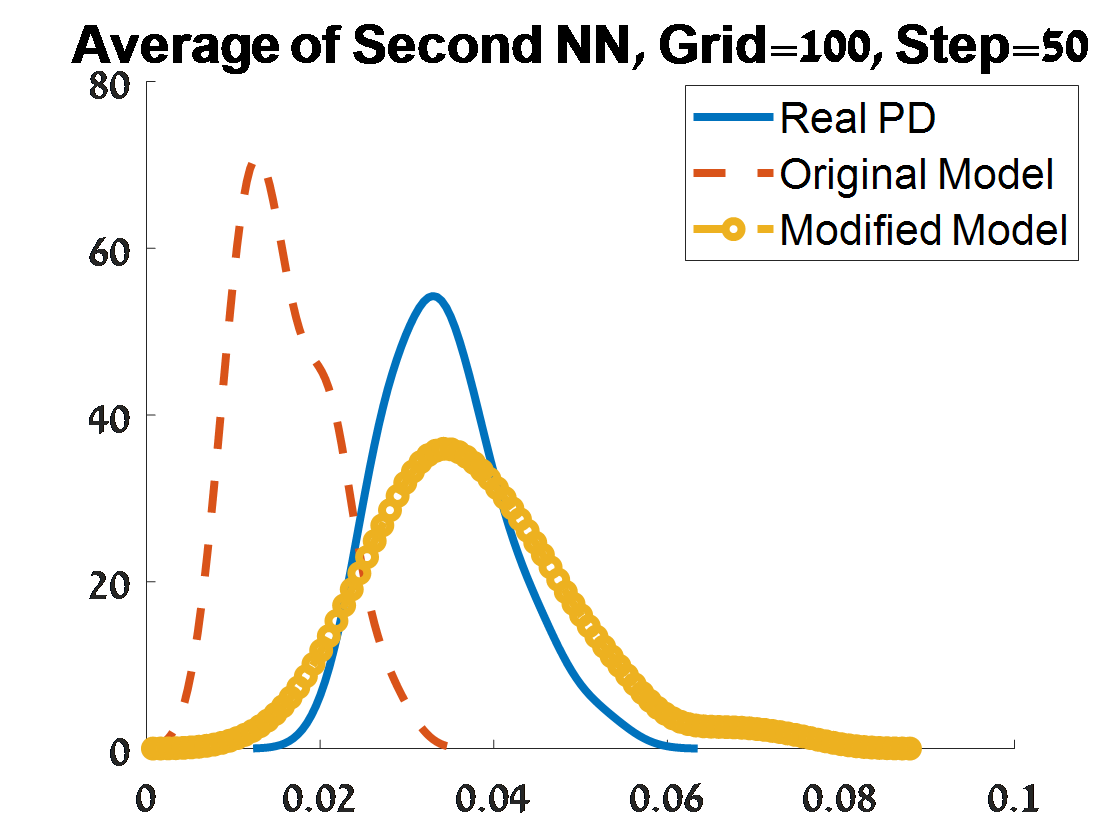}
\includegraphics[width=1.2in, height=1.25in]{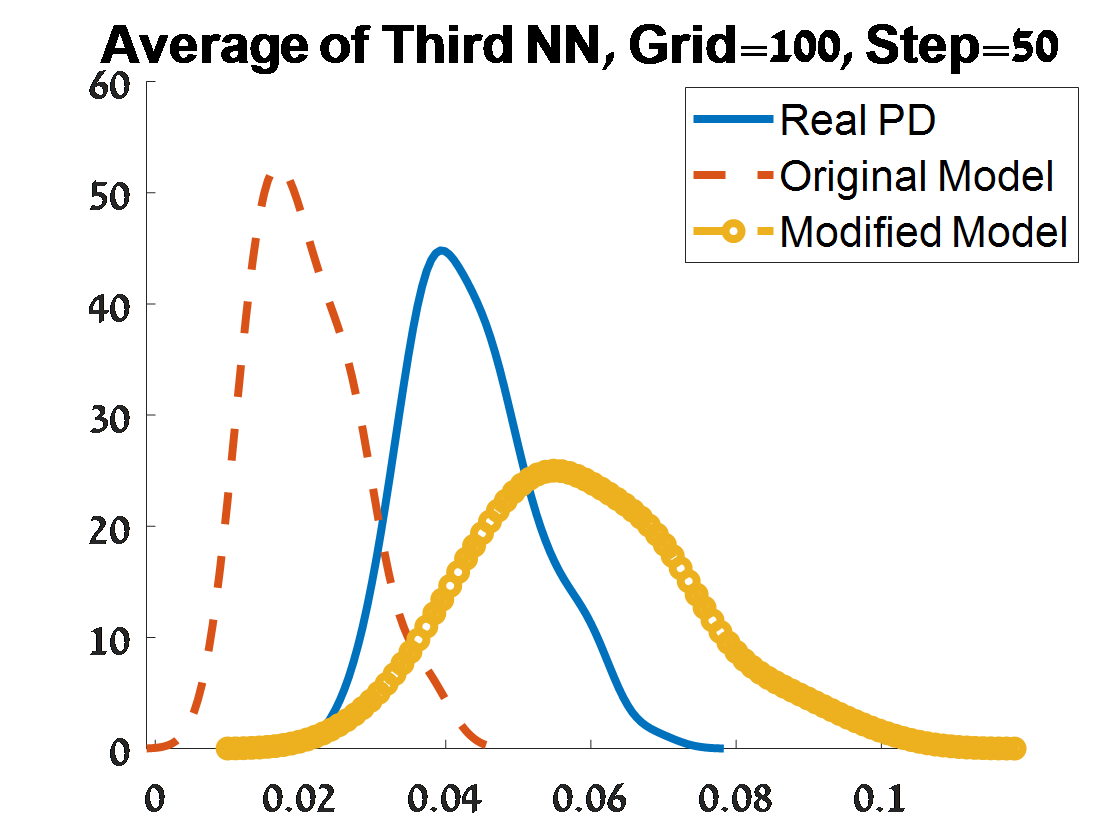}
\ec
%\caption{\footnotesize
% A random sample from two circles, 500 points from the larger circle and 300 from the smaller one,  with a kernel density
\caption{\footnotesize
 Criterion 2 of goodness of fit for 100 PDs corresponded to 100 samples from an object of two distinct circles. The figures depend on the grid of the proposal distribution ("Grid"), and the burn-in ("Step") of the MCMC algorithm.}
\label{fig:distinct_b}
\end{figure}
\end{landscape}

\begin{landscape}
\begin{figure}[h!]
\bc
\includegraphics[width=1.2in, height=1.25in]{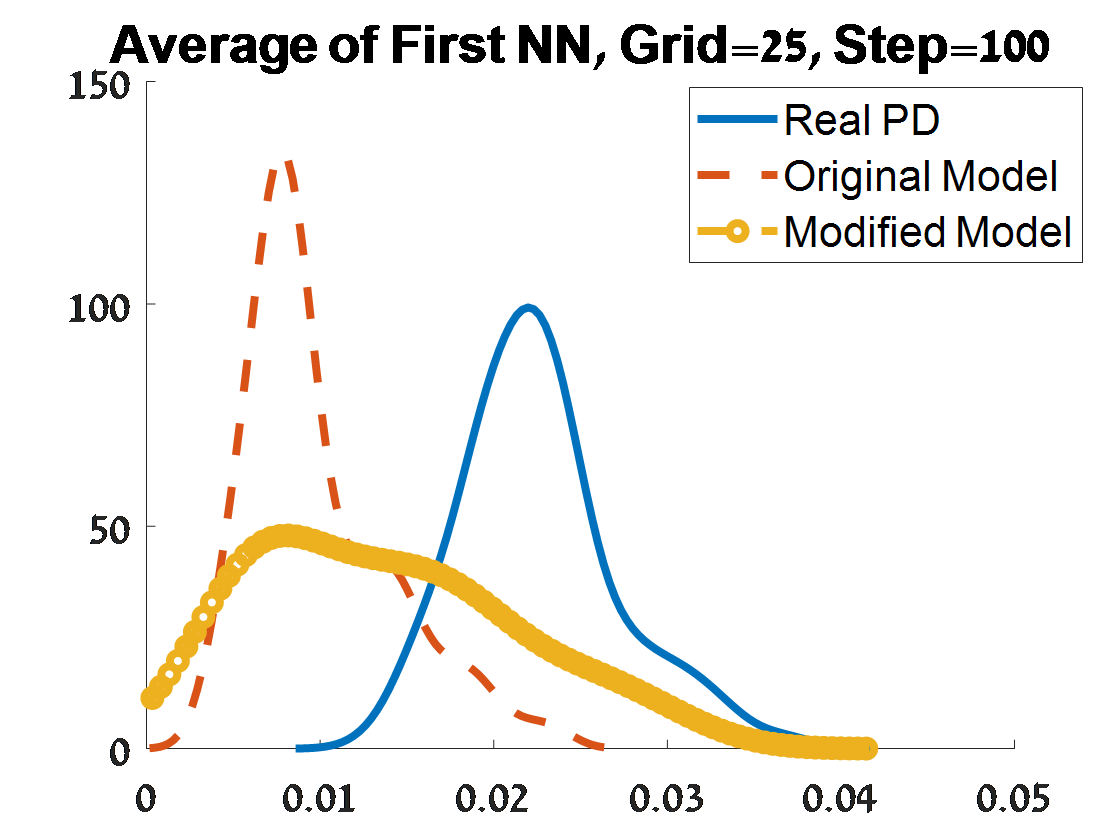}
\includegraphics[width=1.2in, height=1.25in]{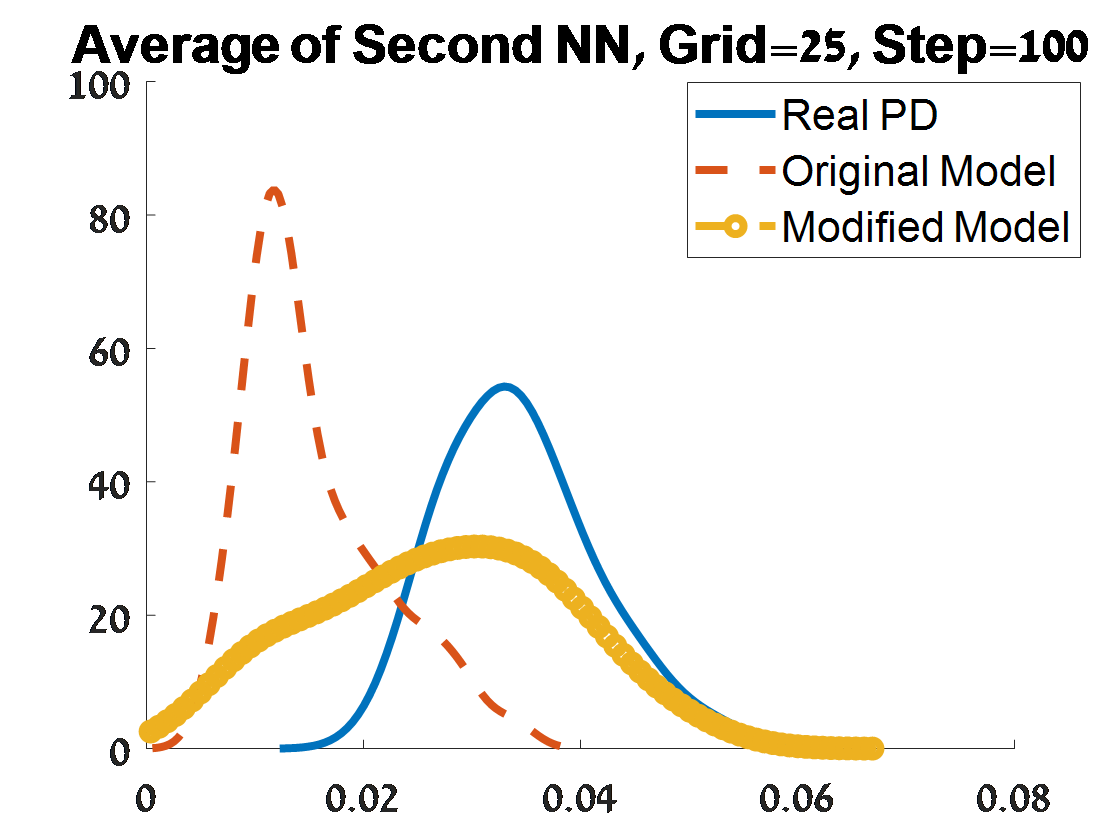}
\includegraphics[width=1.2in, height=1.25in]{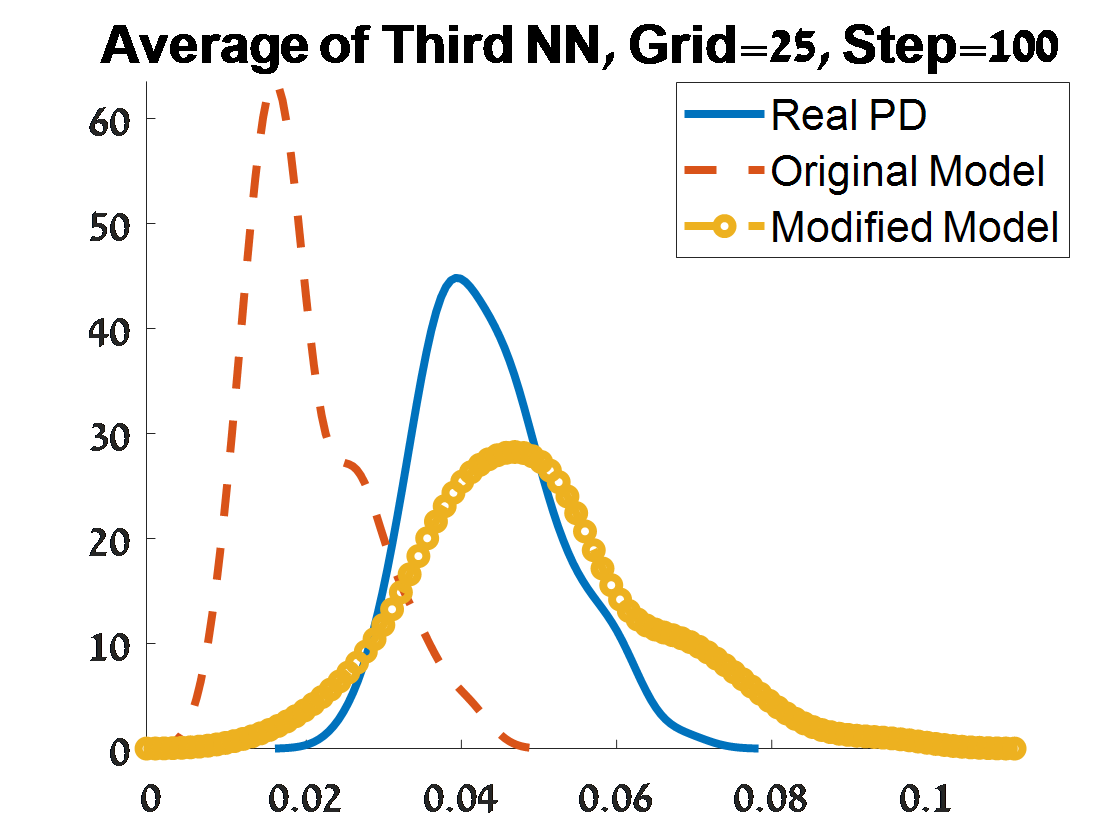}
\includegraphics[width=1.2in, height=1.25in]{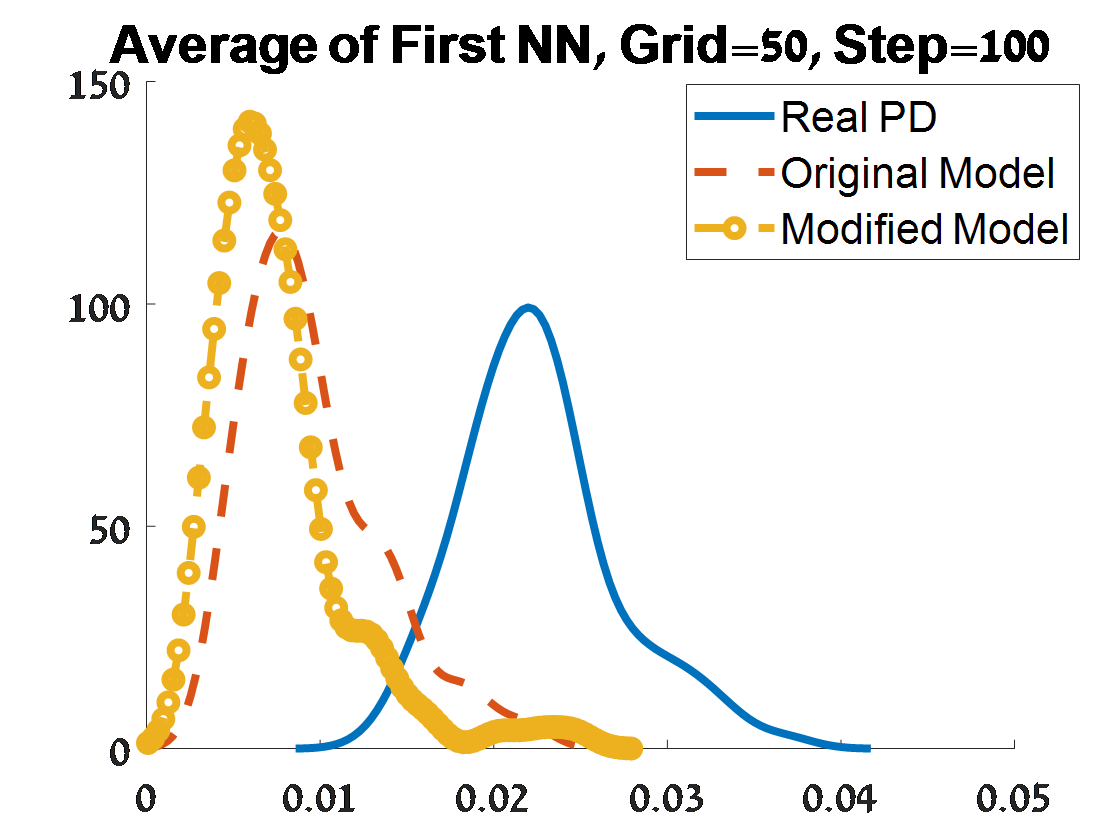}
\includegraphics[width=1.2in, height=1.25in]{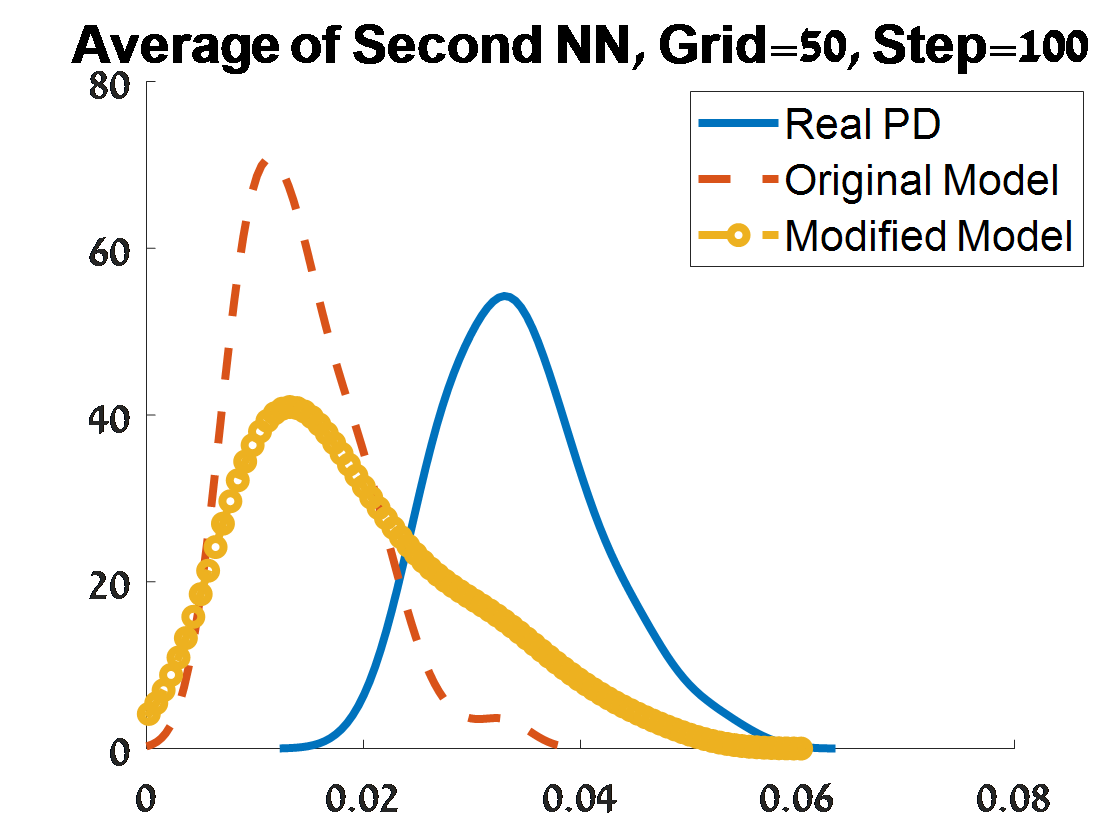}
\includegraphics[width=1.2in, height=1.25in]{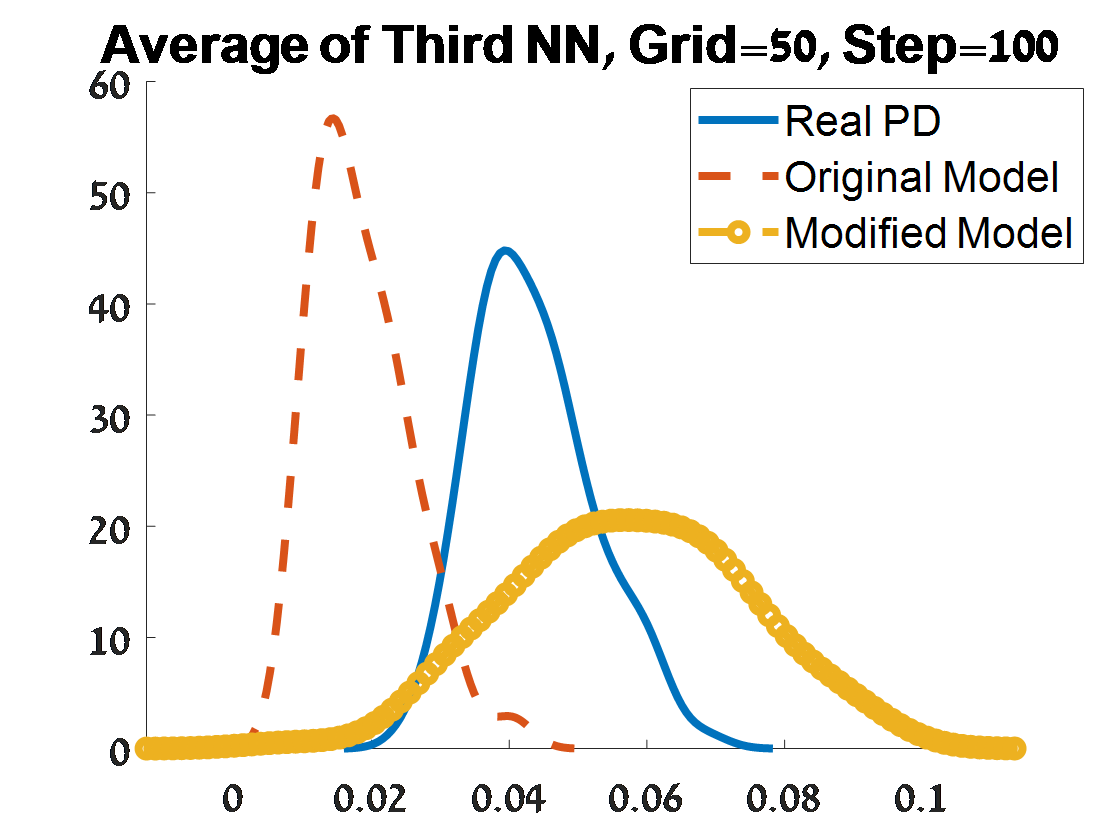}
\includegraphics[width=1.2in, height=1.25in]{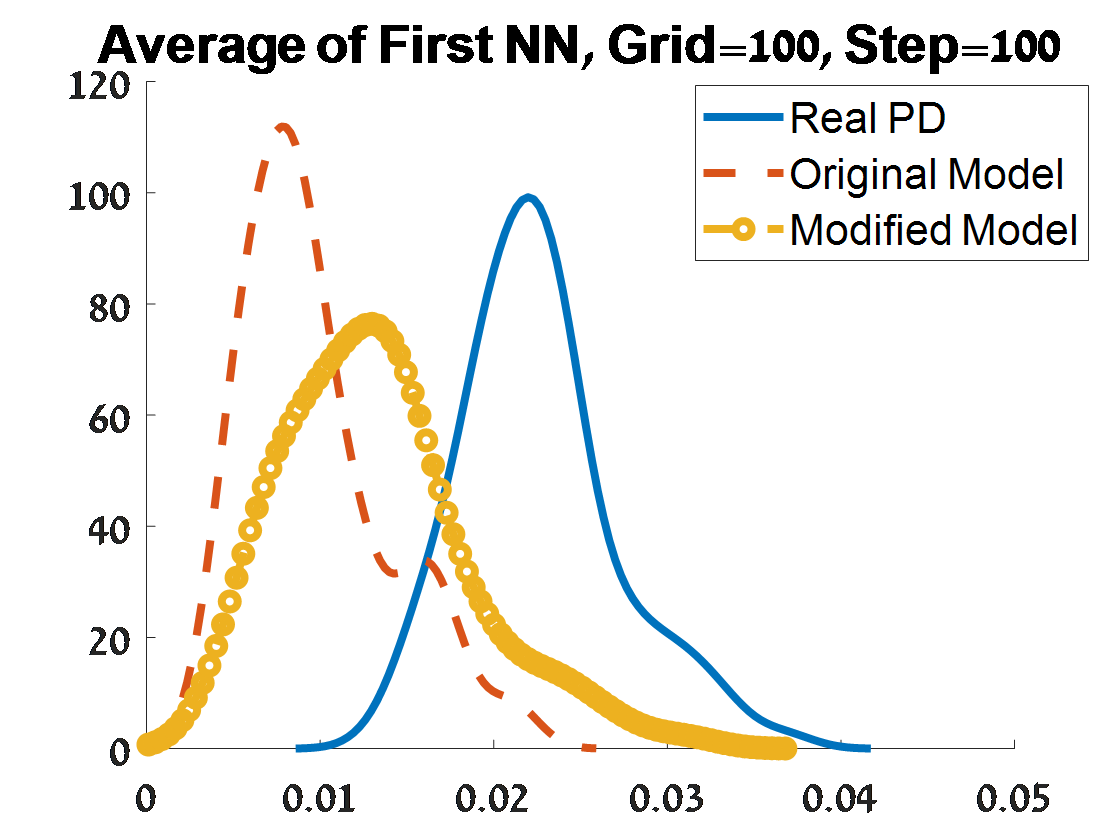}
\includegraphics[width=1.2in, height=1.25in]{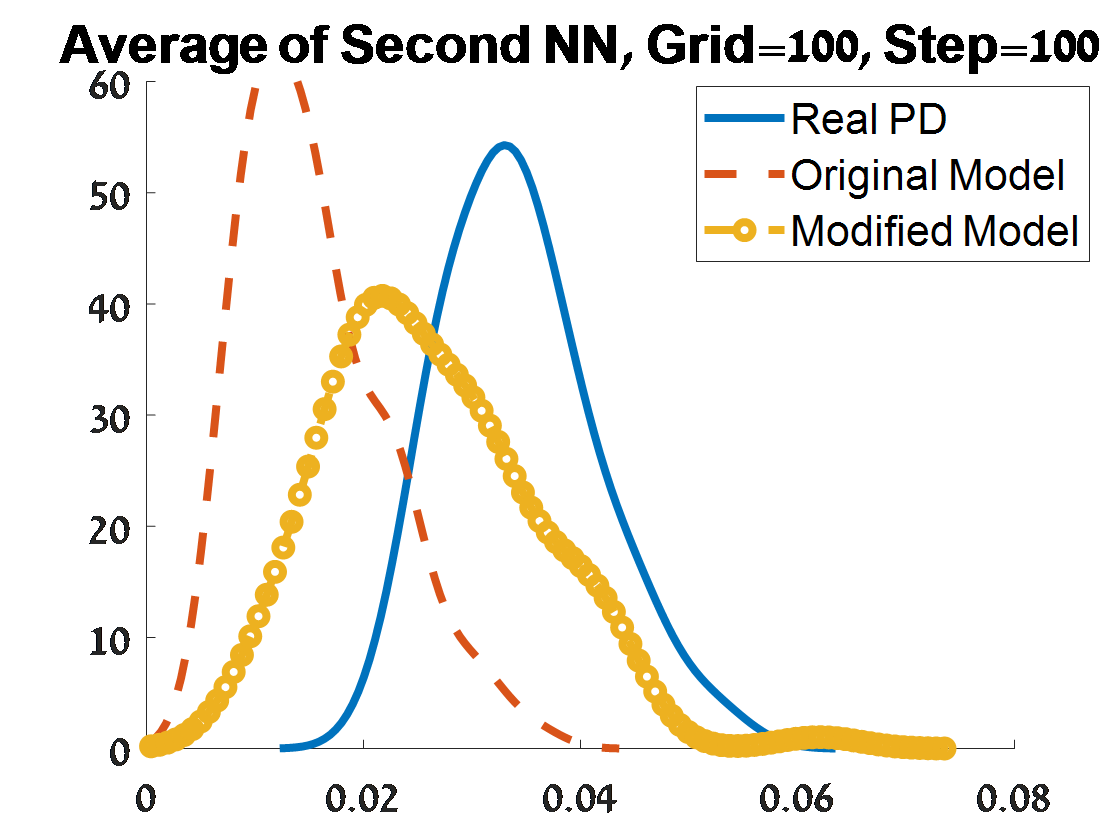}
\includegraphics[width=1.2in, height=1.25in]{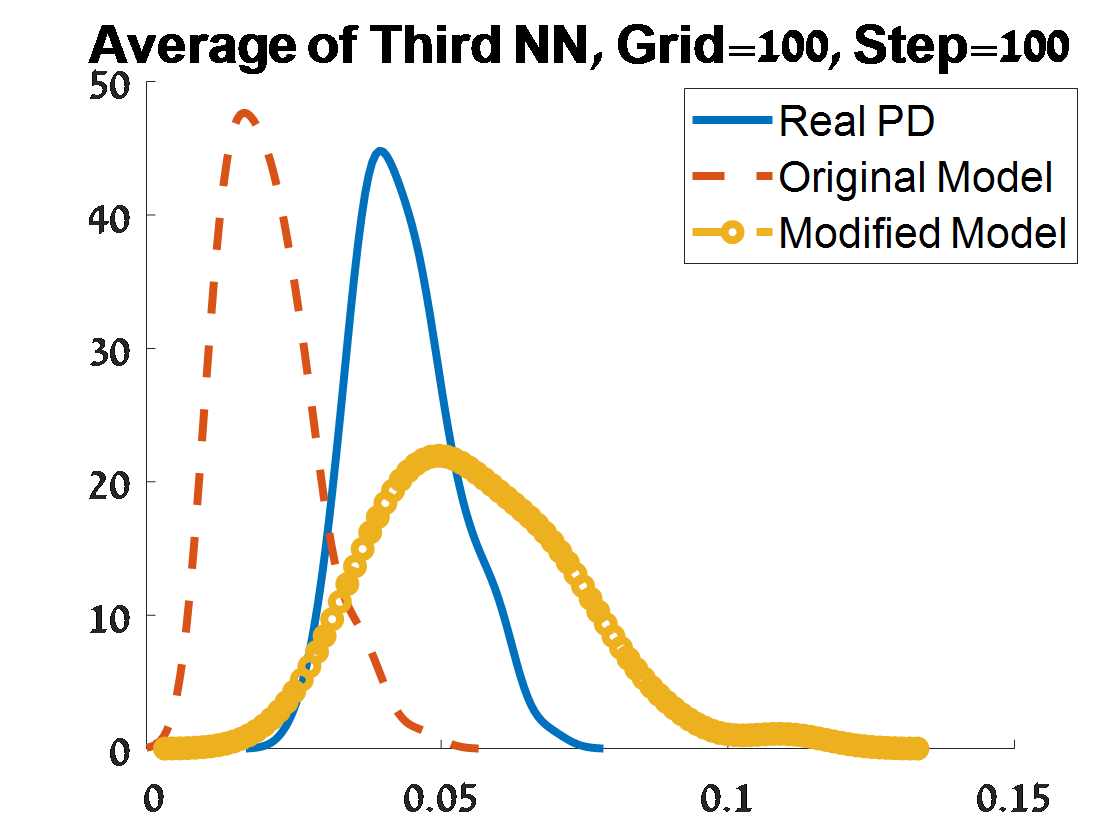}
\ec
%\caption{\footnotesize
% A random sample from two circles, 500 points from the larger circle and 300 from the smaller one,  with a kernel density
\caption{\footnotesize
 Continue of Criterion 2 of goodness of fit for 100 PDs corresponded to 100 samples from an object of two distinct circles. The figures depend on the grid of the proposal distribution ("Grid"), and the burn-in ("Step") of the MCMC algorithm.}
\label{fig:distinct_c}
\end{figure}
\end{landscape}

%\begin{landscape}
%\begin{figure}[h!]
%\bc
%\includegraphics[width=1.8in, height=1.8in]{DistinctCircles_pd30_grid50_step25}
%\includegraphics[width=1.8in, height=1.8in]{DistinctCircles_pd30_grid100_step25}
%\includegraphics[width=1.8in, height=1.8in]{DistinctCircles_pd60_grid50_step25}
%\includegraphics[width=1.8in, height=1.8in]{DistinctCircles_pd60_grid100_step25}
%\ec
%%\caption{\footnotesize
%% A random sample from two circles, 500 points from the larger circle and 300 from the smaller one,  with a kernel density
%\caption{\footnotesize
% Examples of two PDs, each one is corresponded to a sample from an object of two distinct circles. For each PD, the simulated PD based on the two model versions is described. The figures depend on the grid of the proposal distribution ("Grid"), and the burn-in ("Step") of the MCMC algorithm. }
%\label{fig:distinct_d}
%\end{figure}
%\end{landscape}
\subsection{2-Sphere ($S^2$)}
This example includes a random sample of $n=1,000$ points from the uniform distribution on the sphere $S^2$ in $R^3$ with radius $r=1$.

\begin{figure}[h!]
\bc
\includegraphics[width=2.1in, height=2.1in]{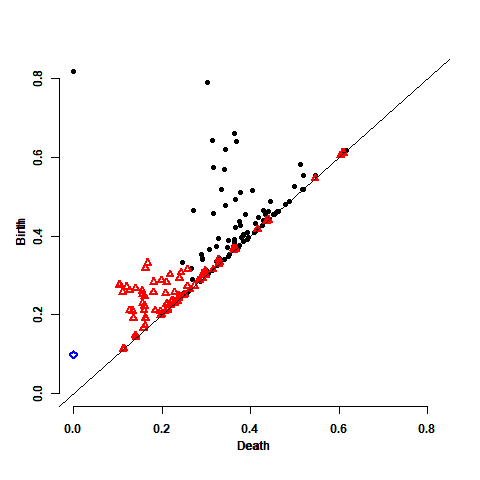}
\\
\includegraphics[width=2.1in, height=2.1in]{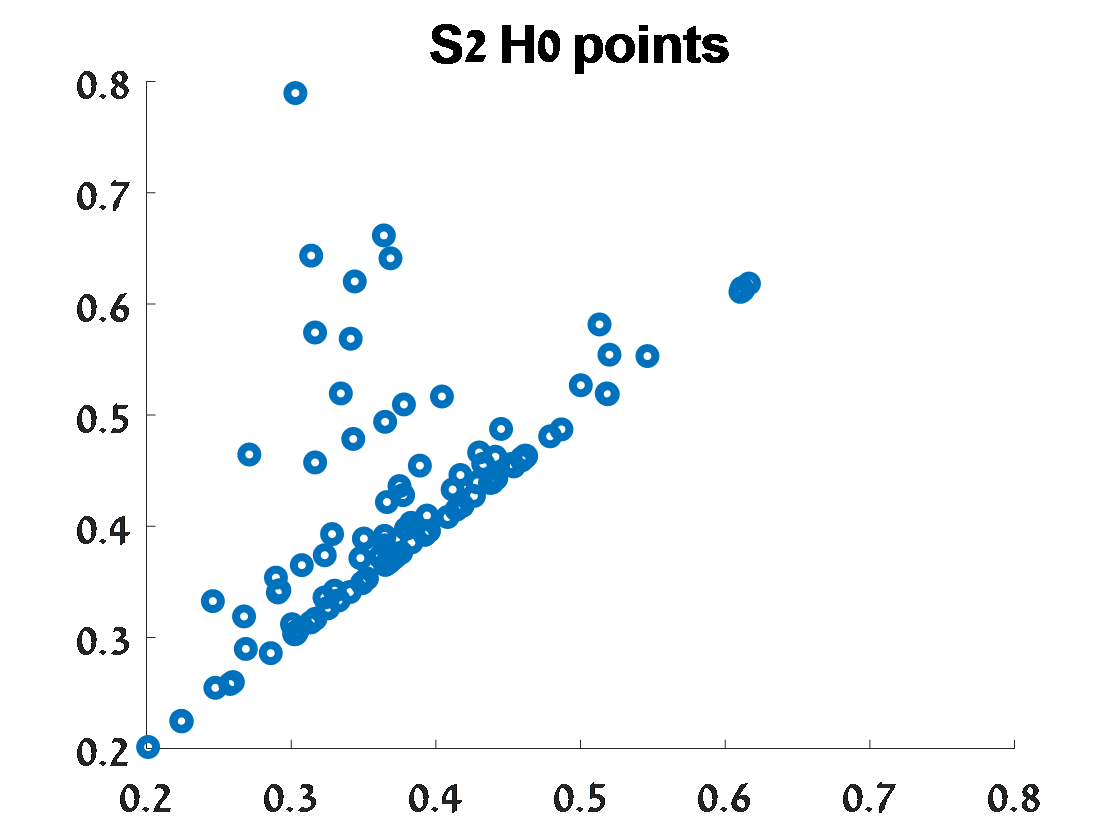}
\includegraphics[width=2.1in, height=2.1in]{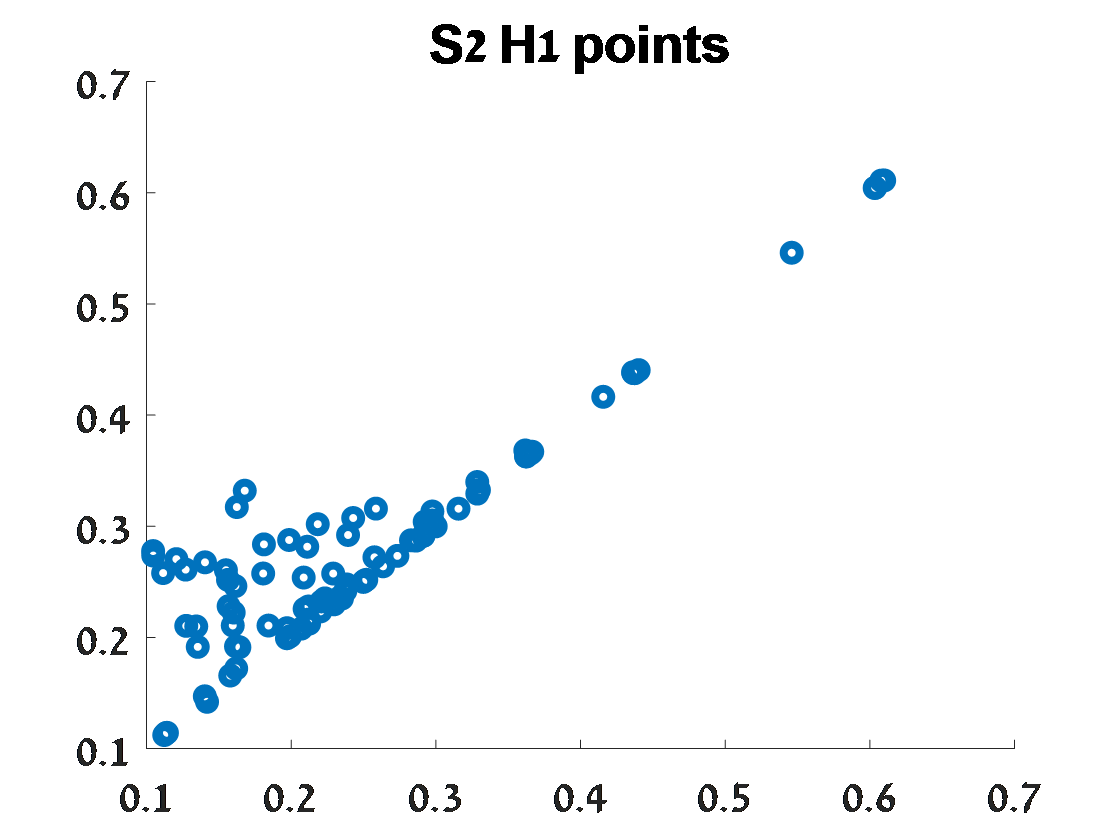}
\ec
%\caption{\footnotesize
% A random sample from two circles, 500 points from the larger circle and 300 from the smaller one,  with a kernel density
\caption{\footnotesize Top: The persistence diagram of a sample of $n=1,000$ points from the unit $S^2$, for its upper level sets. Black circles are connected components ($H_0$ persistence points), red triangles are holes ($H_1$ points), and the blue diamond are voids ($H_2$ points). Birth times are on the vertical axis. Bottom: The corresponded persistence diagram separately for each homology, except $H_2$.}
\label{fig:sphere_s2}
\end{figure}

The typical corresponded persistence diagram is presented in Figure\ \ref{fig:sphere_s2}. The black circles indicating connected components ($H_0$ persistence), the red triangles corresponding to holes ($H_1$), and the blue diamond corresponding to void ($H_2$). The next plots to its right present the persistence diagram for each homology separately, except $H_2$ which has only 1 point.

%\begin{landscape}

In this example, in the contrary to the setting of the previous examples, there are enough points in $H_1$, so we could fitted the model for $H_1$ points in addition to the model's fitting for the $H_0$ points.

\subsubsection{The fitted model for $H_0$}
\begin{landscape}
\begin{figure}[h!]
\bc
\includegraphics[width=1.2in, height=1.4in]{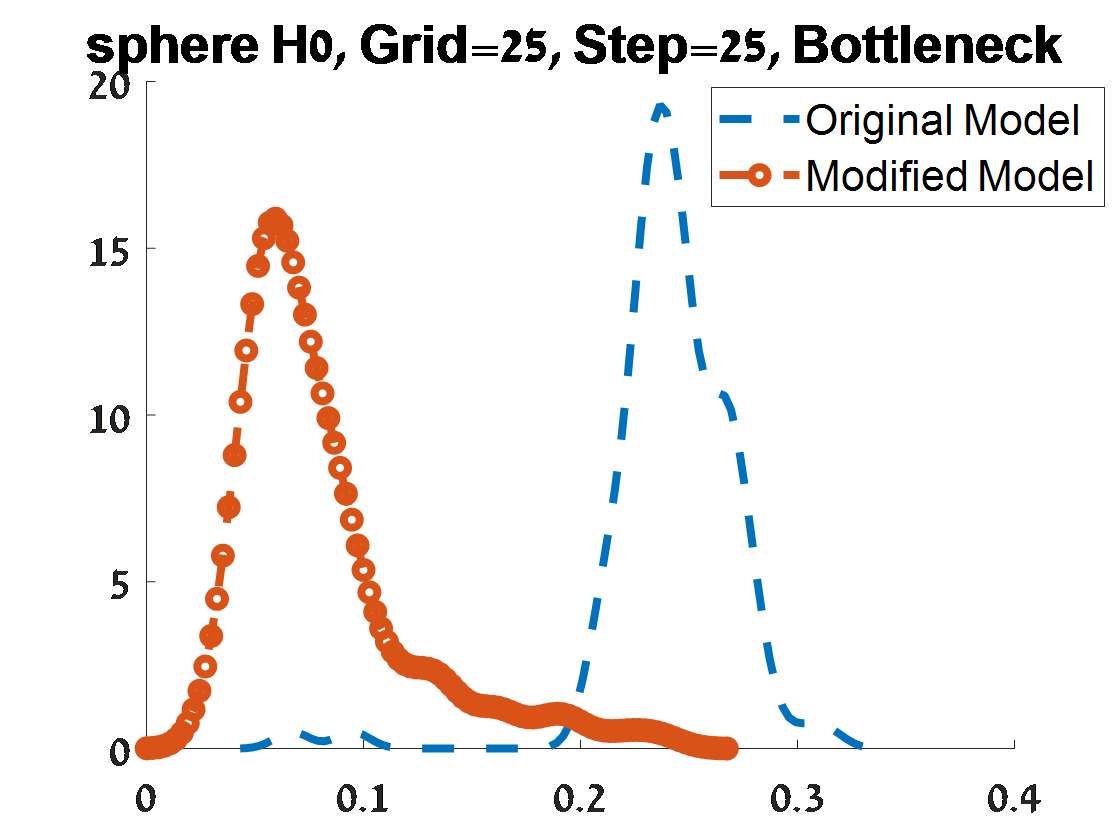}
\includegraphics[width=1.2in, height=1.4in]{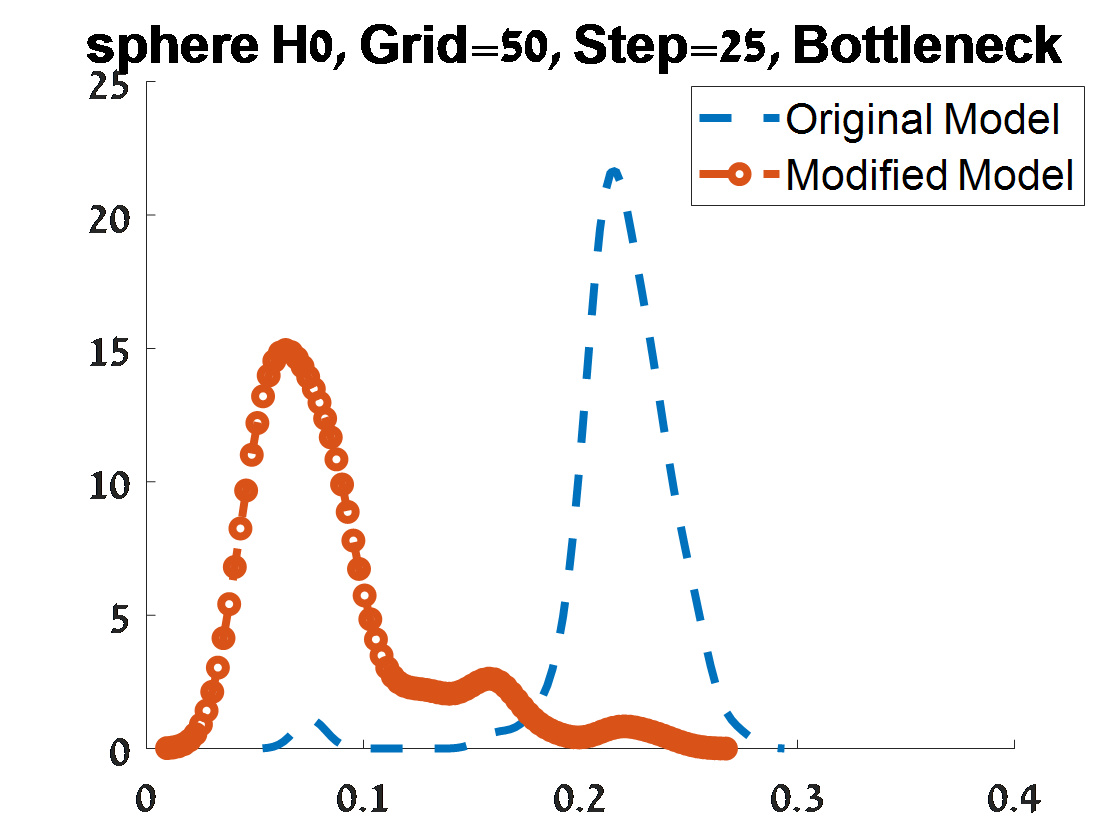}
\includegraphics[width=1.2in, height=1.4in]{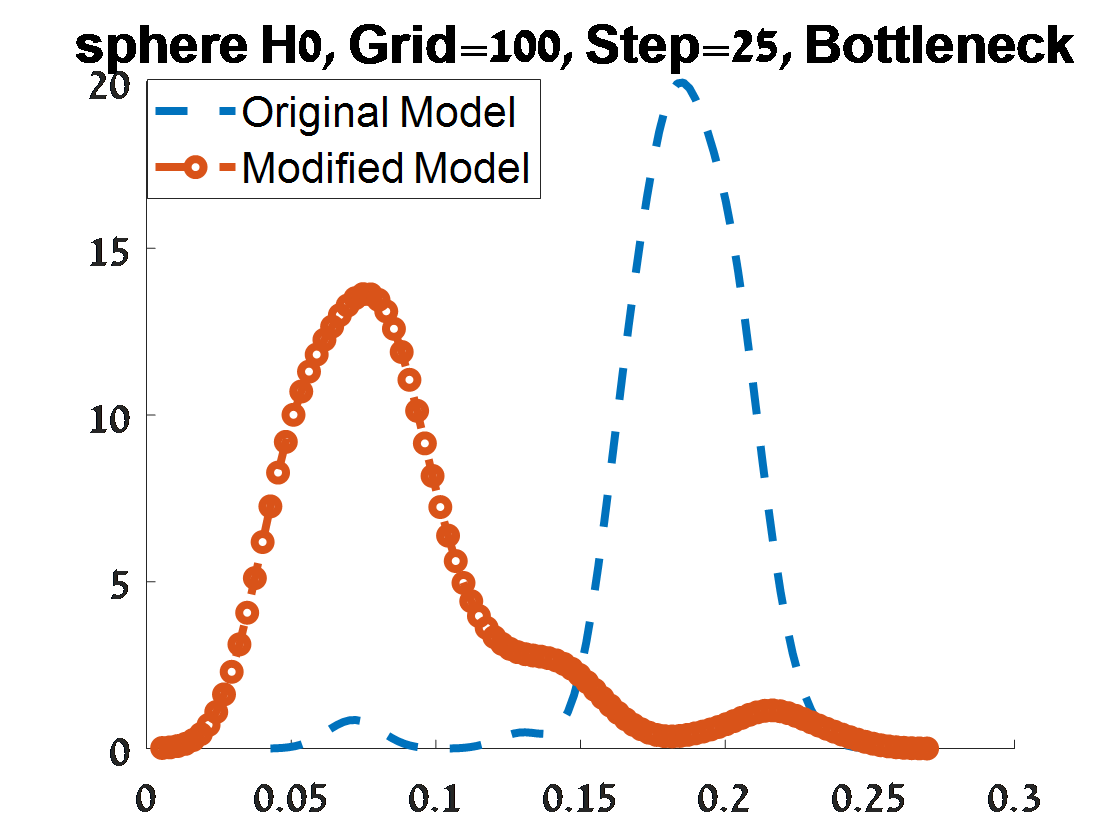}
\includegraphics[width=1.2in, height=1.4in]{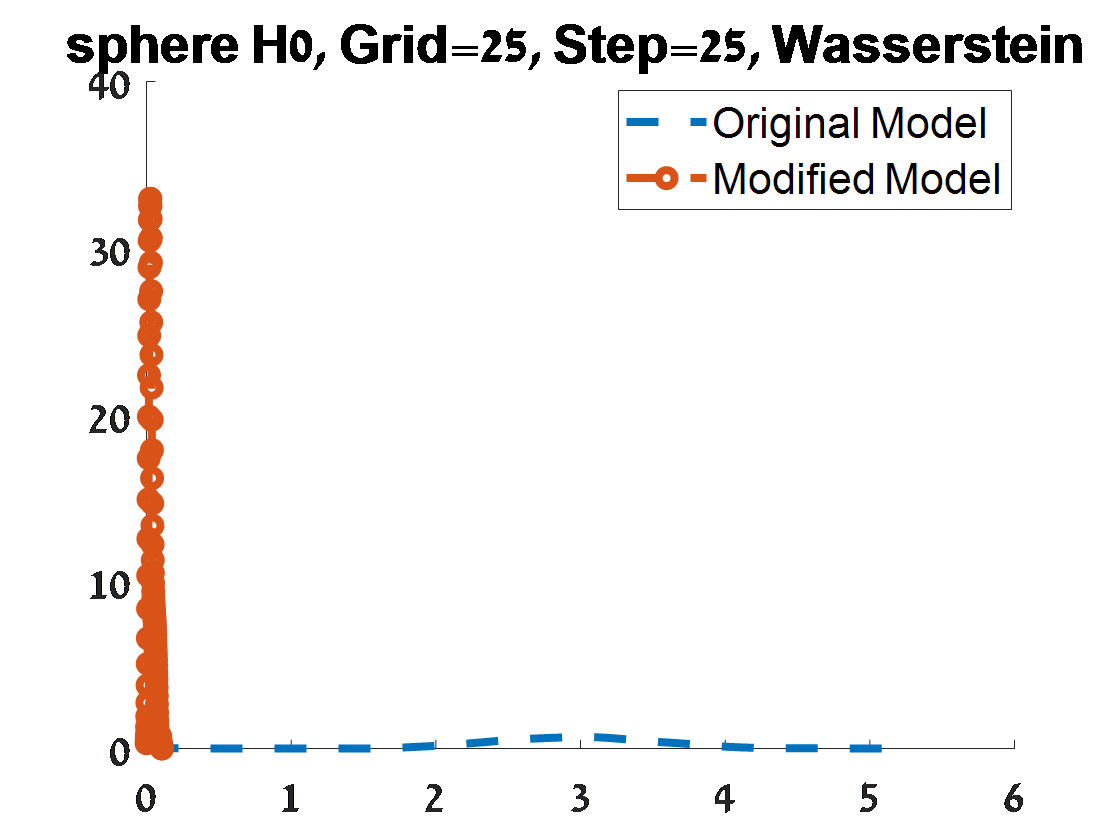}
\includegraphics[width=1.2in, height=1.4in]{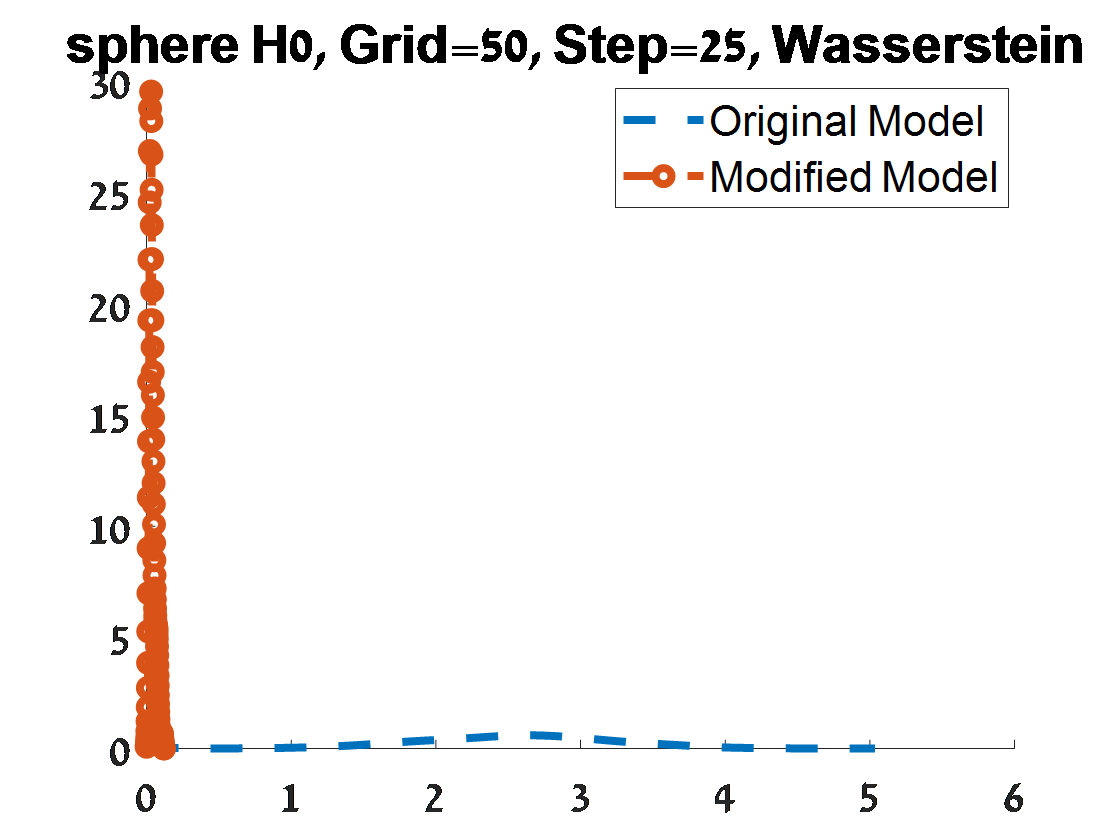}
\includegraphics[width=1.2in, height=1.4in]{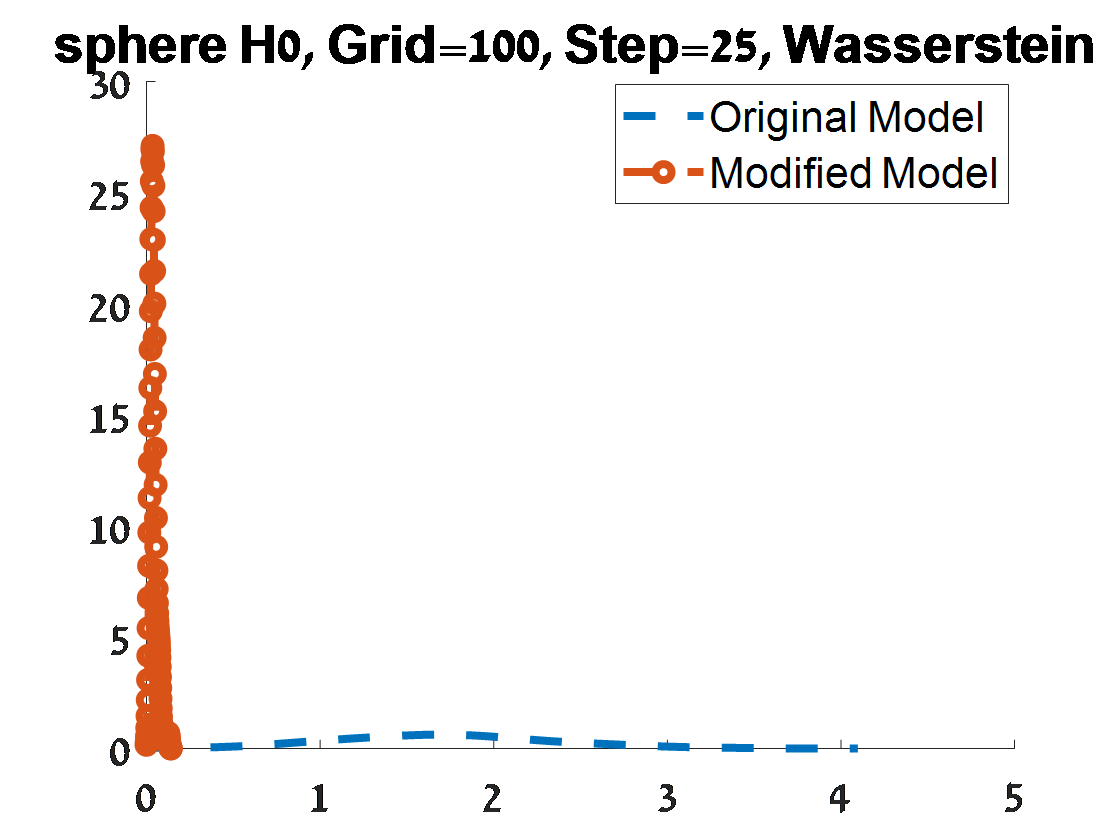}
\\
\includegraphics[width=1.2in, height=1.4in]{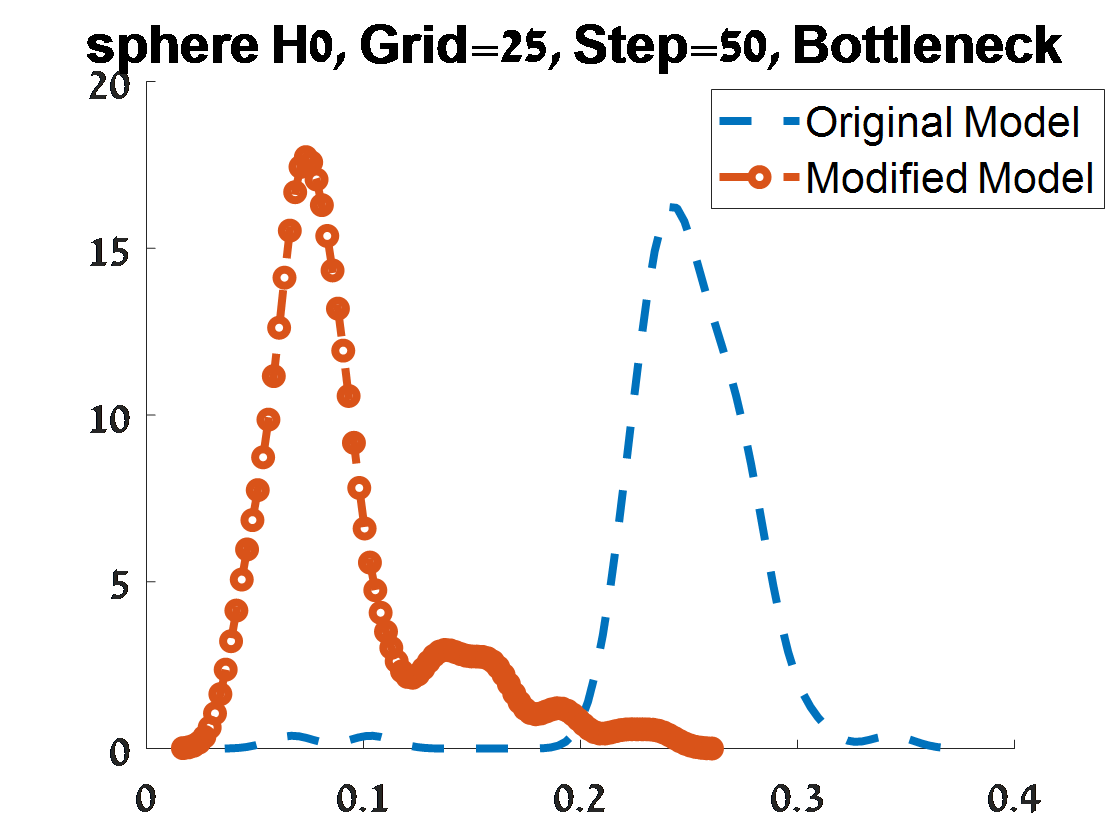}
\includegraphics[width=1.2in, height=1.4in]{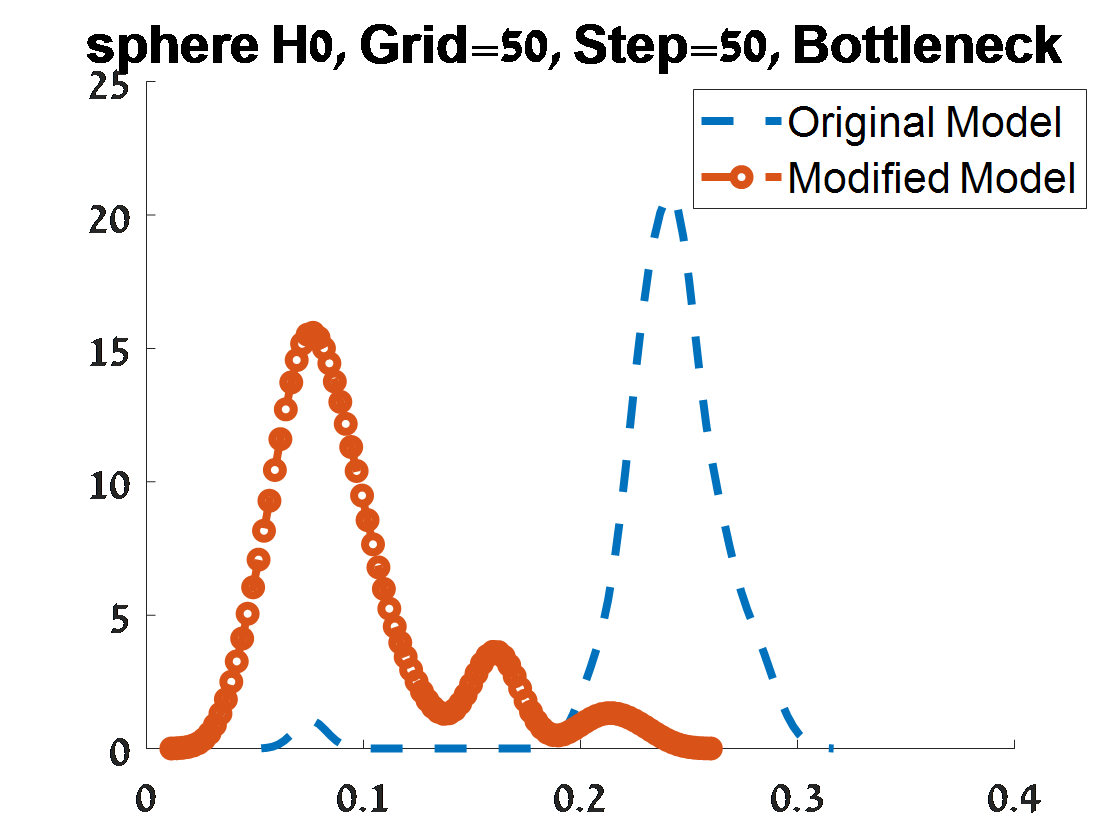}
\includegraphics[width=1.2in, height=1.4in]{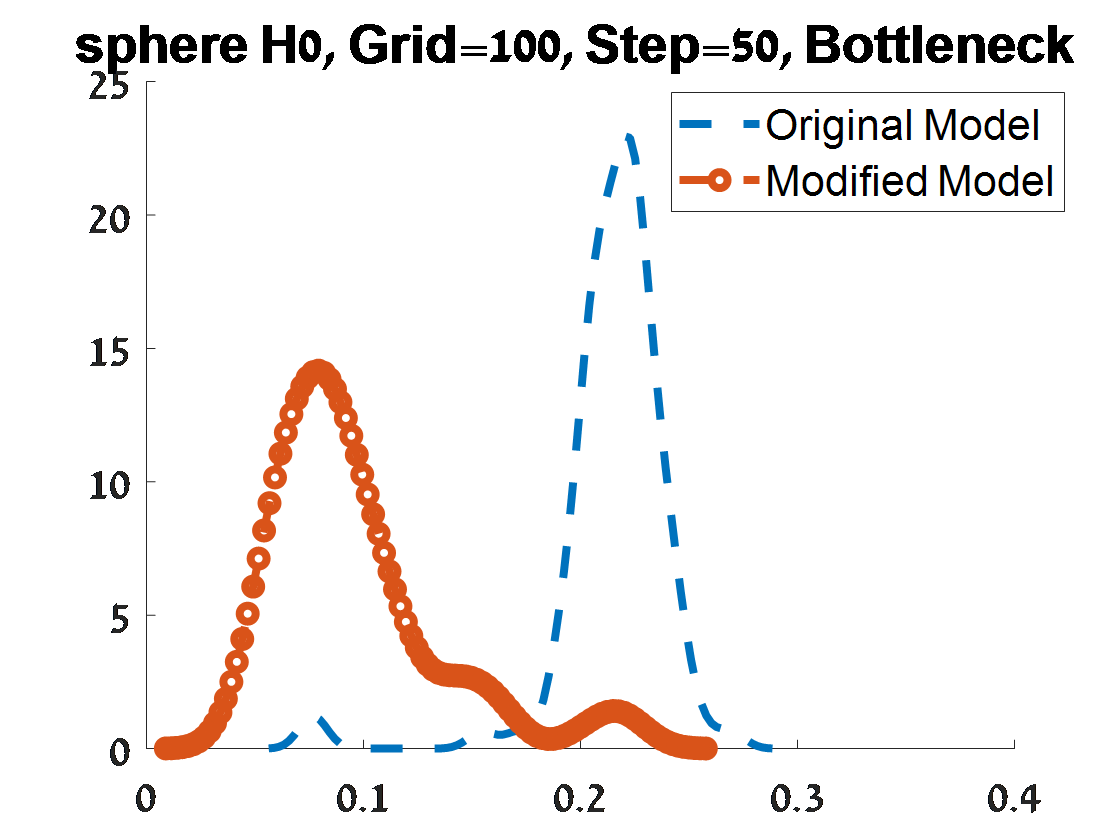}
\includegraphics[width=1.2in, height=1.4in]{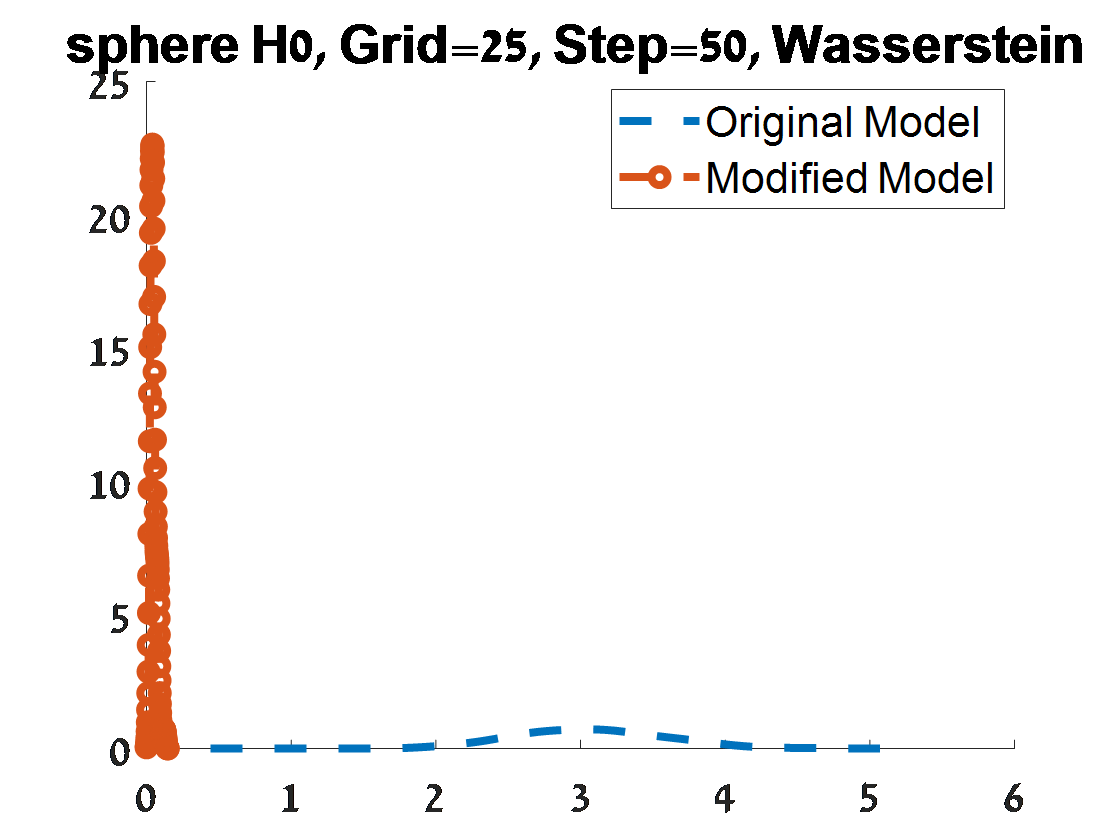}
\includegraphics[width=1.2in, height=1.4in]{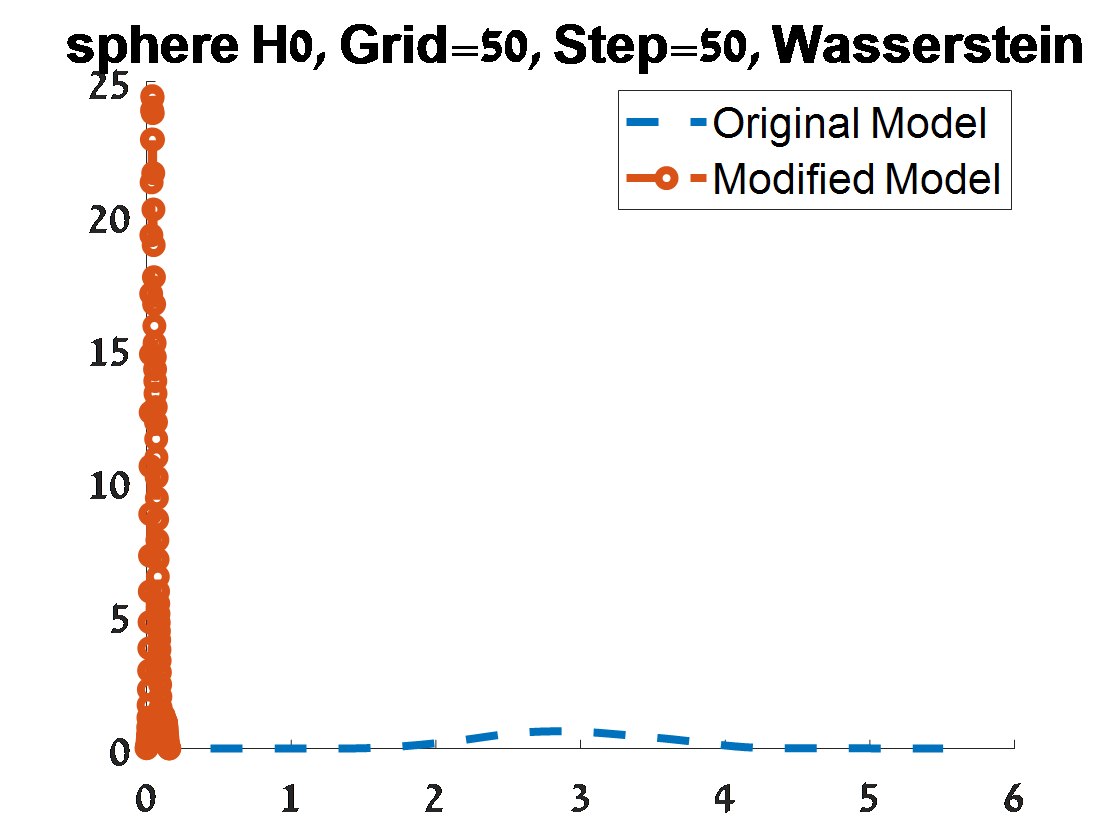}
\includegraphics[width=1.2in, height=1.4in]{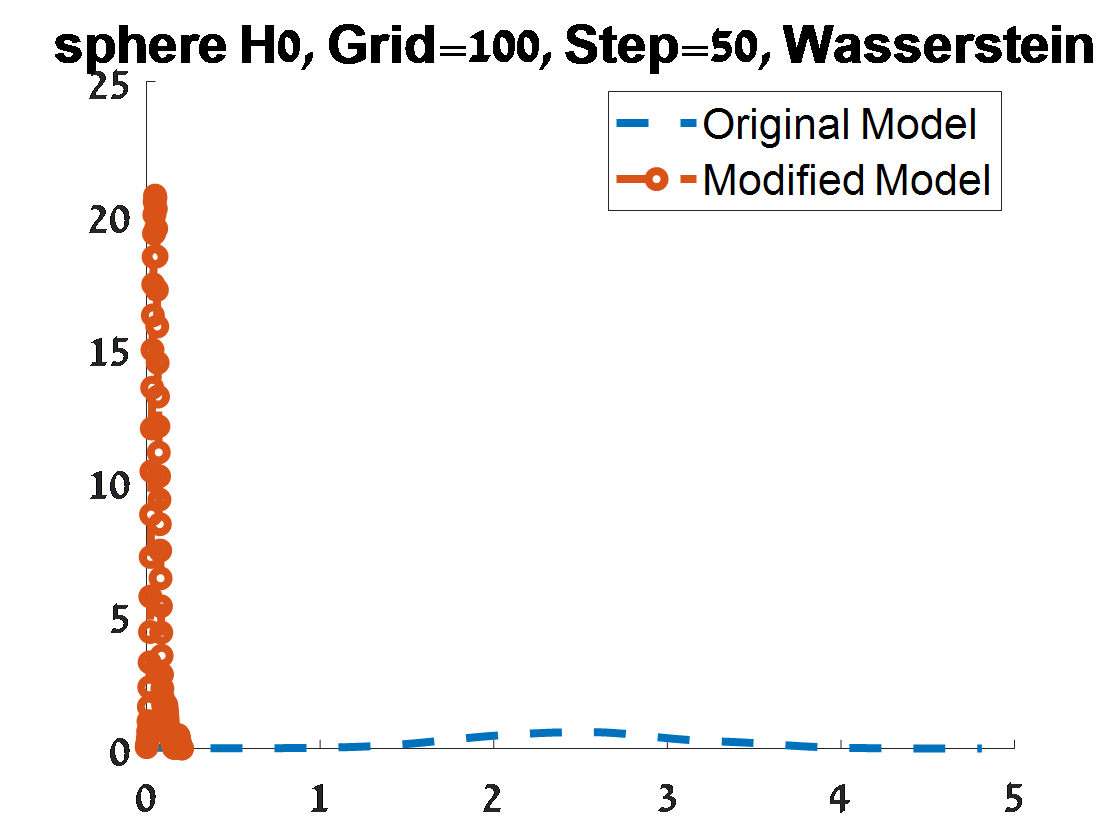}
\\
\includegraphics[width=1.2in, height=1.4in]{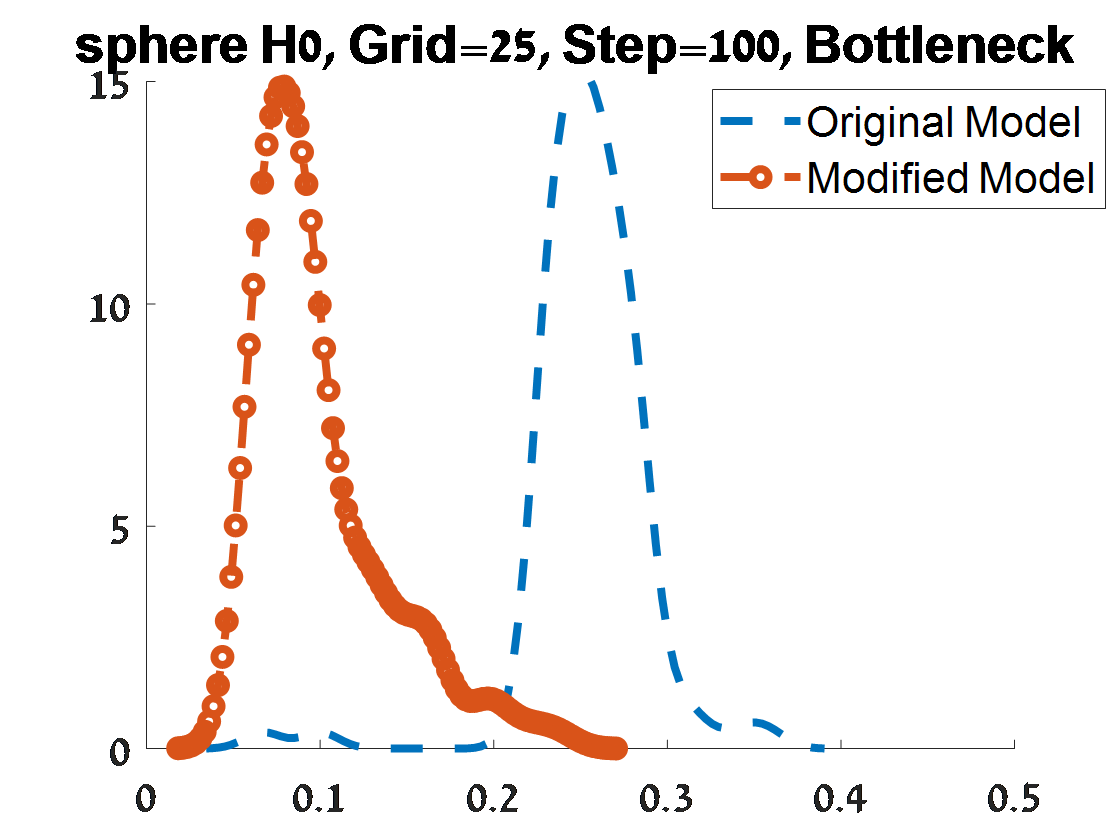}
\includegraphics[width=1.2in, height=1.4in]{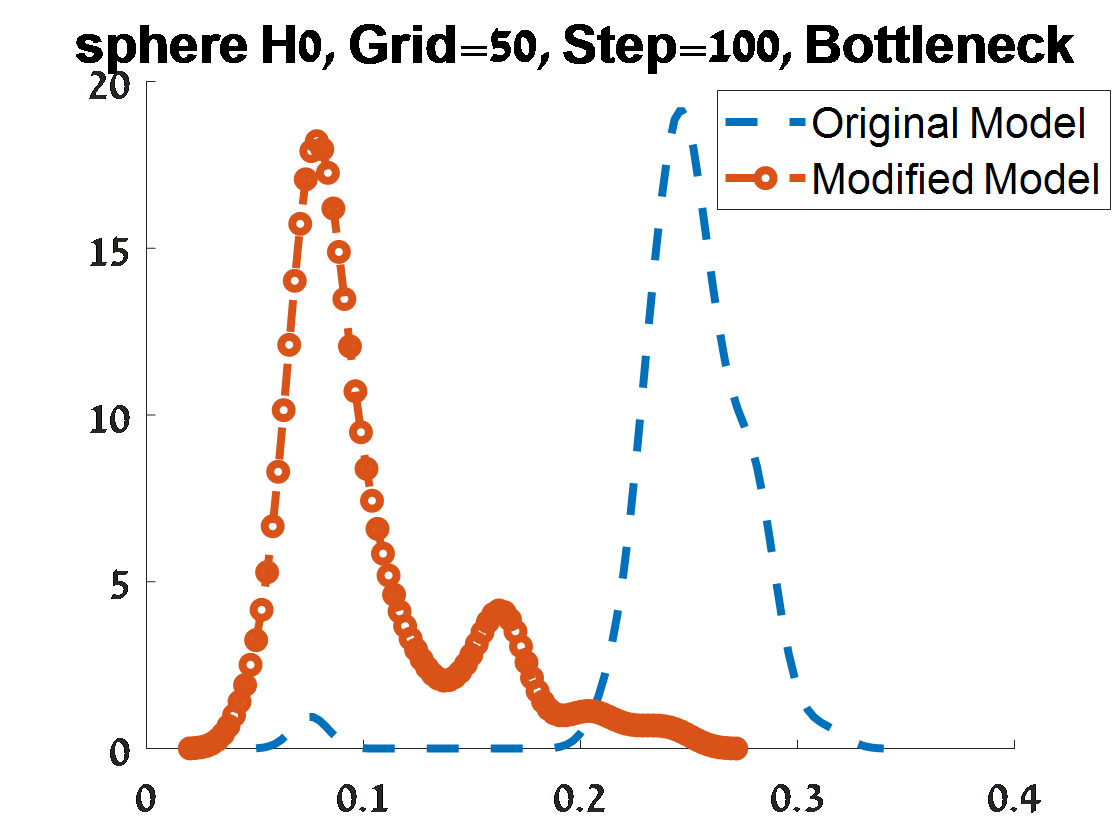}
\includegraphics[width=1.2in, height=1.4in]{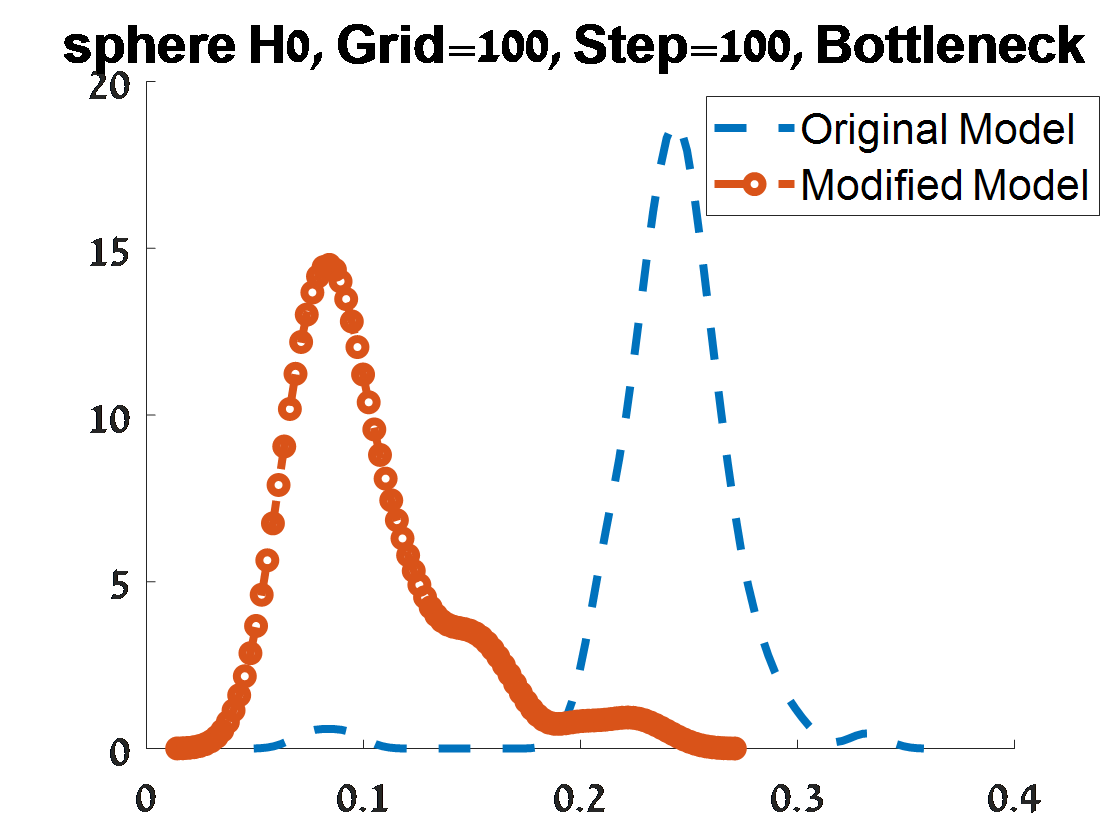}
\includegraphics[width=1.2in, height=1.4in]{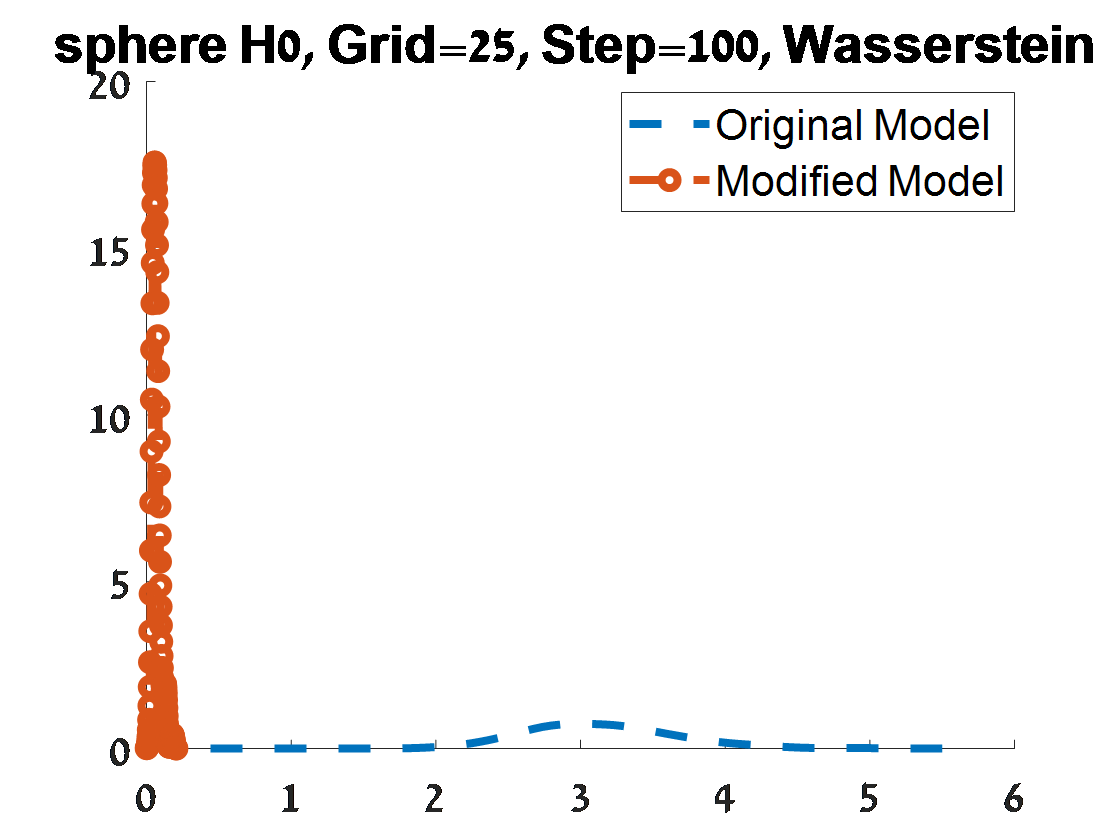}
\includegraphics[width=1.2in, height=1.4in]{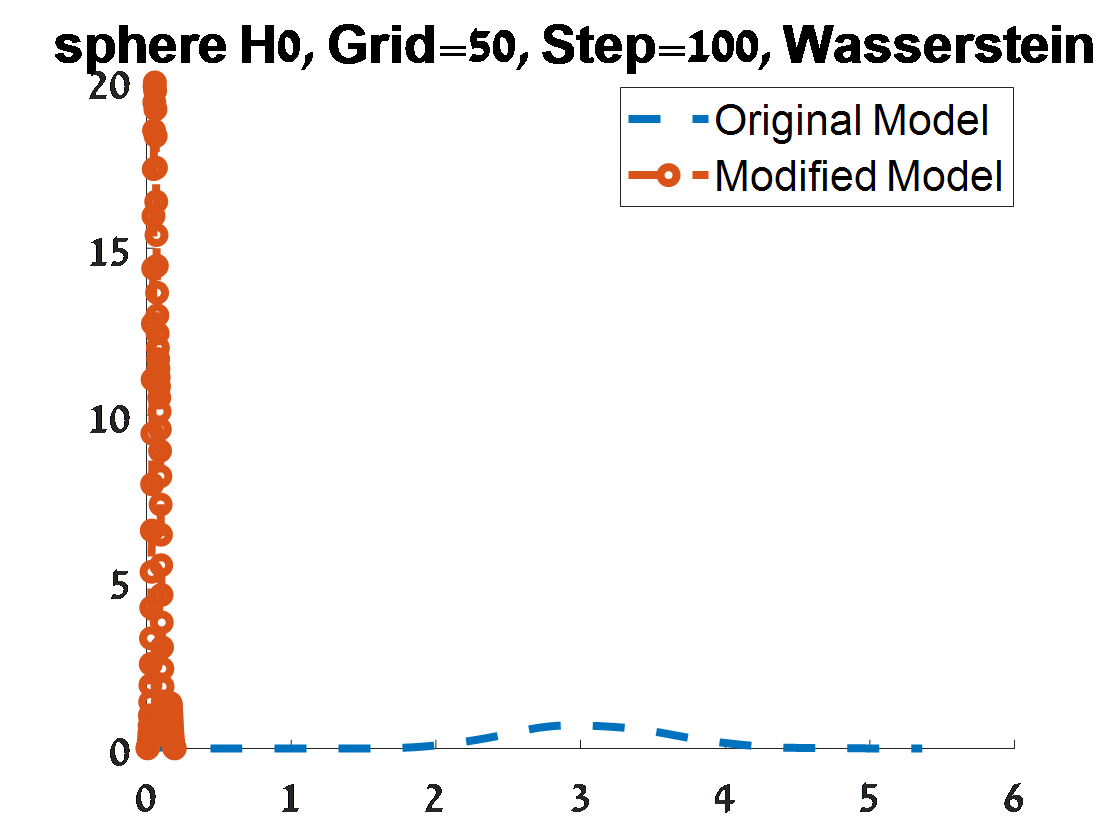}
\includegraphics[width=1.2in, height=1.4in]{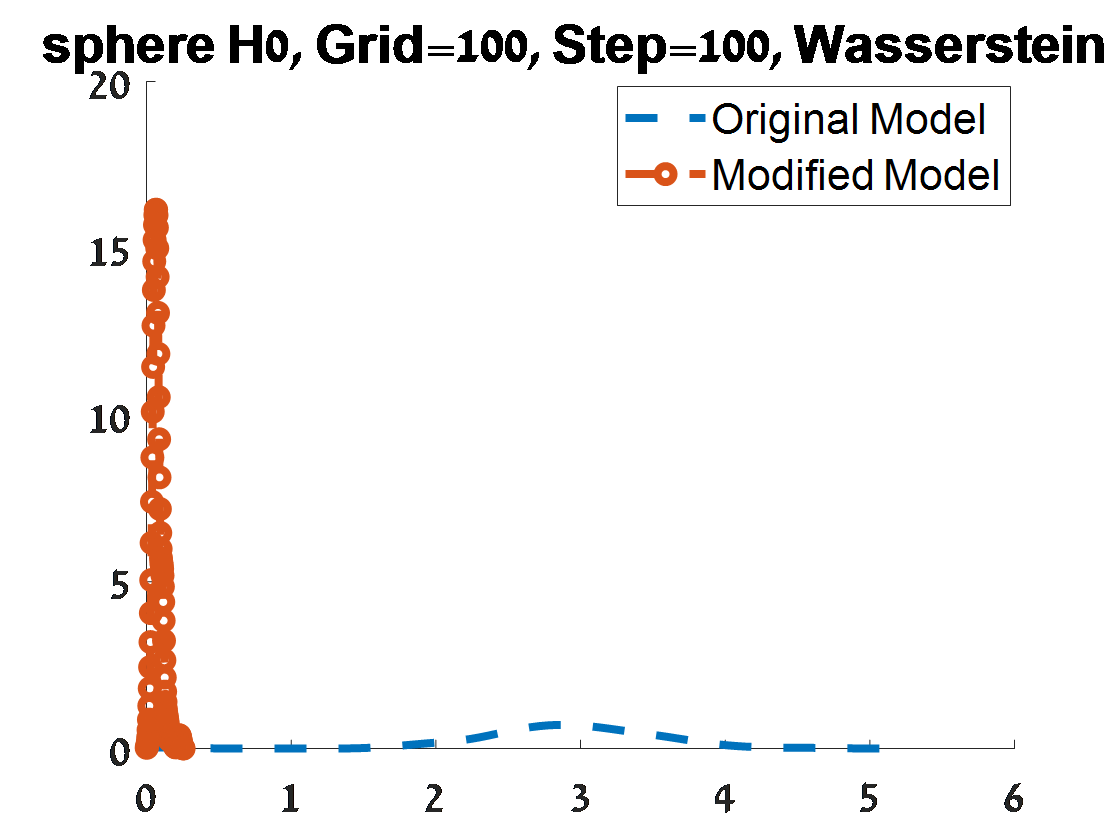}
\ec
%\caption{\footnotesize
% A random sample from two circles, 500 points from the larger circle and 300 from the smaller one,  with a kernel density
\caption{\footnotesize
 Criterion 1 of goodness of fit for 100 $H_0$ PDs corresponded to 100 samples from a unit $S^2$. The figures depend on the grid of the proposal distribution ("Grid"), and the burn-in ("Step") of the MCMC algorithm.}
\label{fig:s2_H0_a}
\end{figure}
\end{landscape}

Figure\ \ref{fig:s2_H0_a} describes the distributions over the 100 $H_0$-PDs of the first criterion of goodness of fit, and Figures 15-16 describe the distributions of the second criterion of goodness of fit.
Based on the results of criterion 1, the goodness of fit of the modified model is better relative to the original model, for both distances. That is, smaller distances between the modified model and real PDs relative to these distances between the original model and the real PDs. This result is highly prominent relative to the result of the previous examples. The Bottleneck distance behave similar, in terms of the distance values distribution, over all considered grid sizes and burn-in. For the Wasserstein distance, this distance decreases as the grid size increases for a given burn-in.
For criterion 2, here relative to the previous examples there is a larger variability between the distributions of the model's properties and those of the real PDs. But this variability is minimized under the modified model, particularly the best fitting is under the modified relative to the original model, for in burn-in of 25 and grid size of 100x100.
%VOOS Figure\ \ref{fig:s2_H0_d} presents two examples of real $H_0$ PD and its simulated $H_0$ PD based on the two model versions, only for the best scenarios we found, that is grid=50,100, and step=25. Here, in the contrary to the previous examples, the original model succeeds in creating the far points on the PD, and the modified model creates these far points more times. VOOS

\begin{landscape}
\begin{figure}[h!]
\bc
\includegraphics[width=1.2in, height=1.25in]{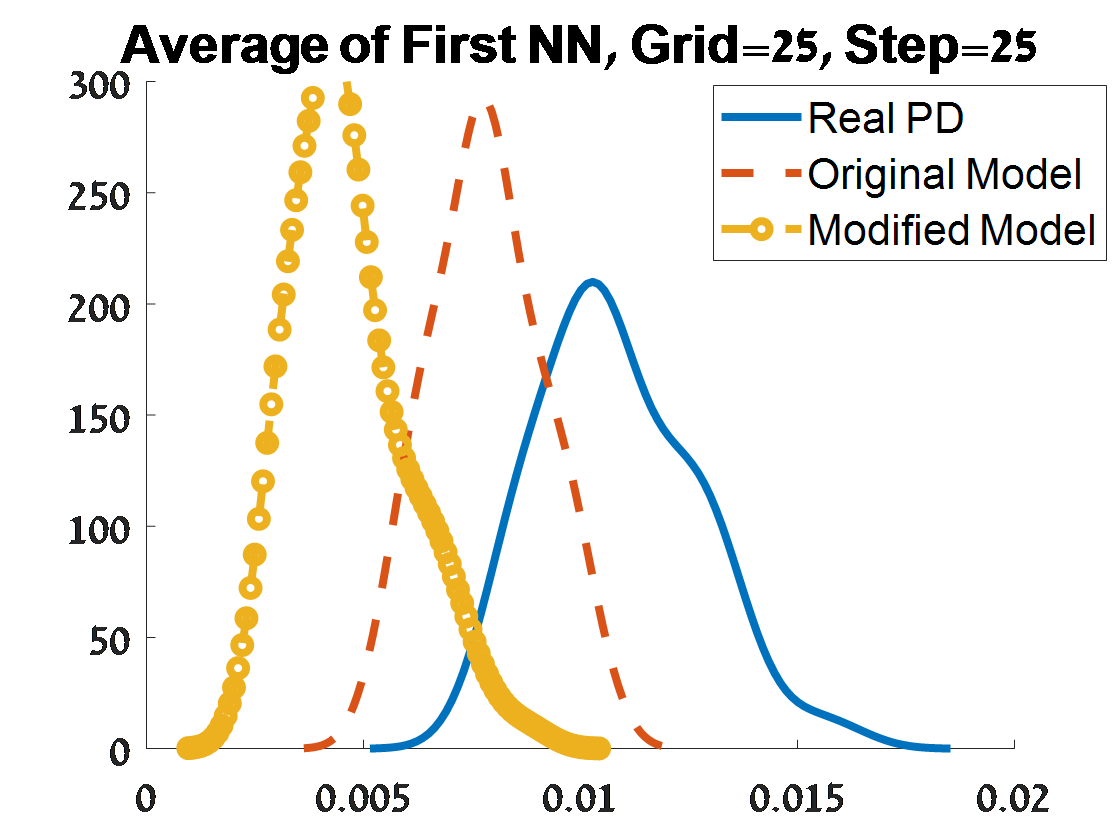}
\includegraphics[width=1.2in, height=1.25in]{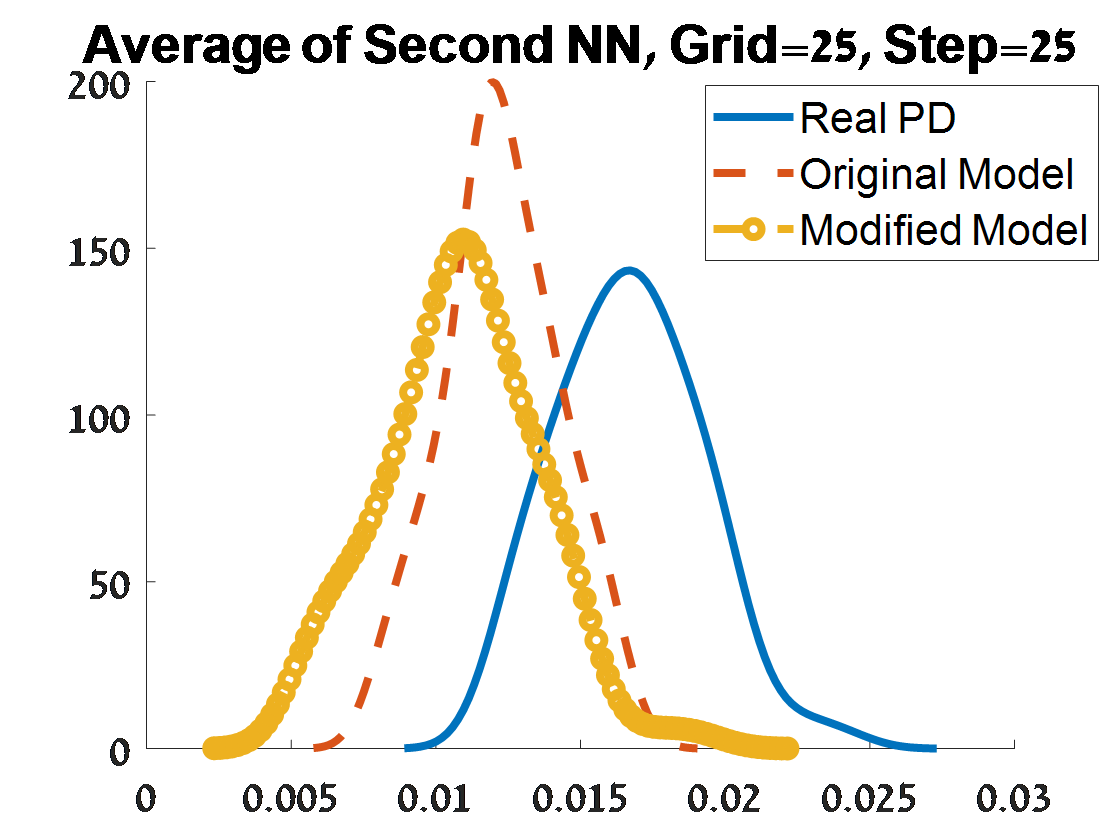}
\includegraphics[width=1.2in, height=1.25in]{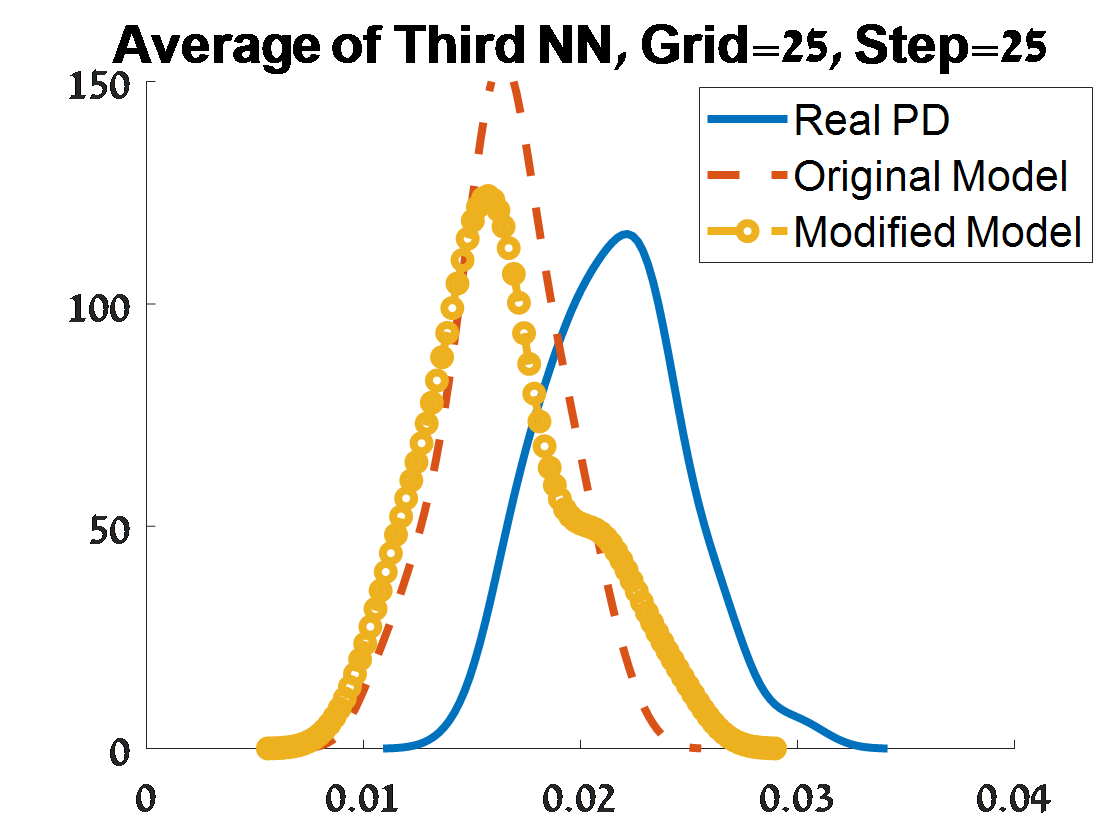}
\includegraphics[width=1.2in, height=1.25in]{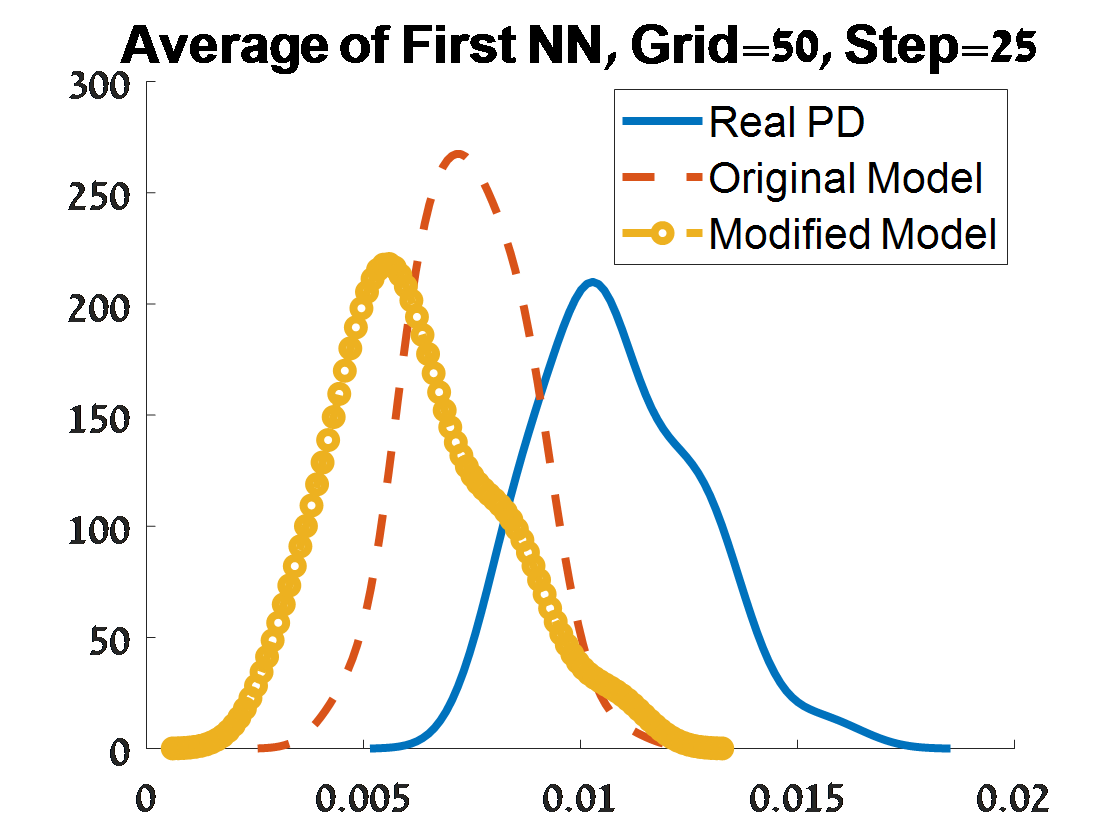}
\includegraphics[width=1.2in, height=1.25in]{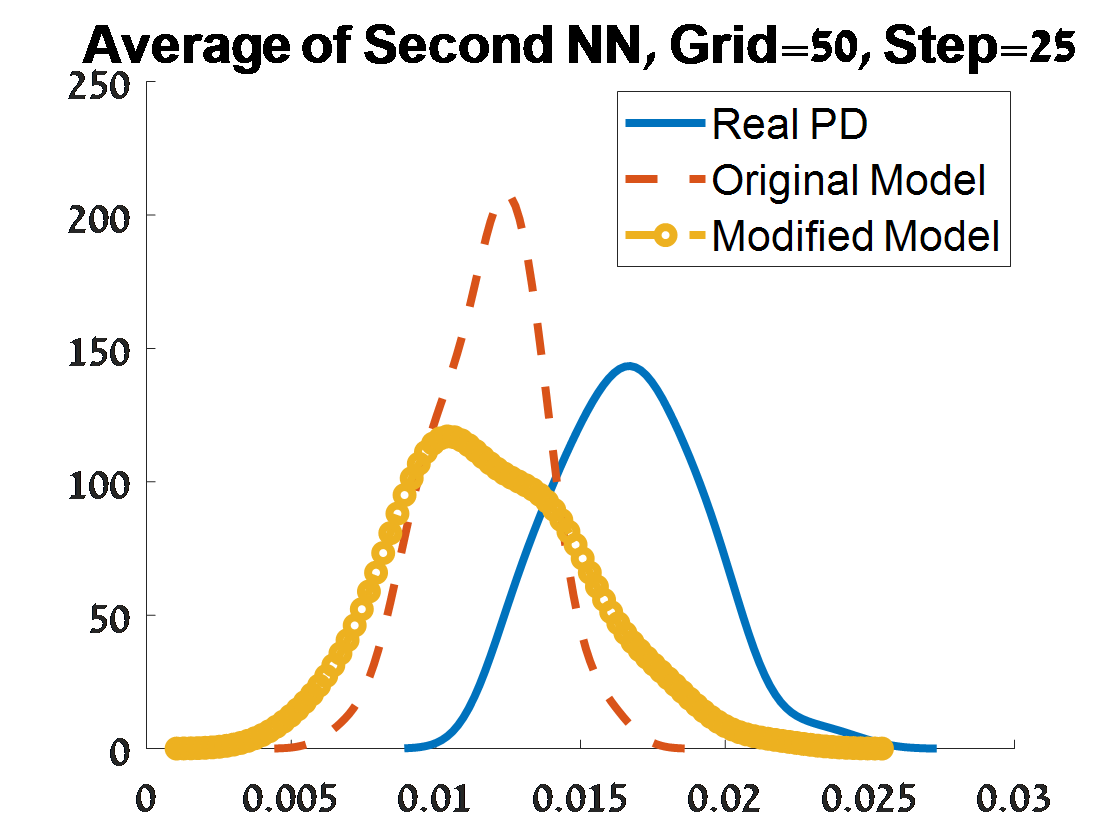}
\includegraphics[width=1.2in, height=1.25in]{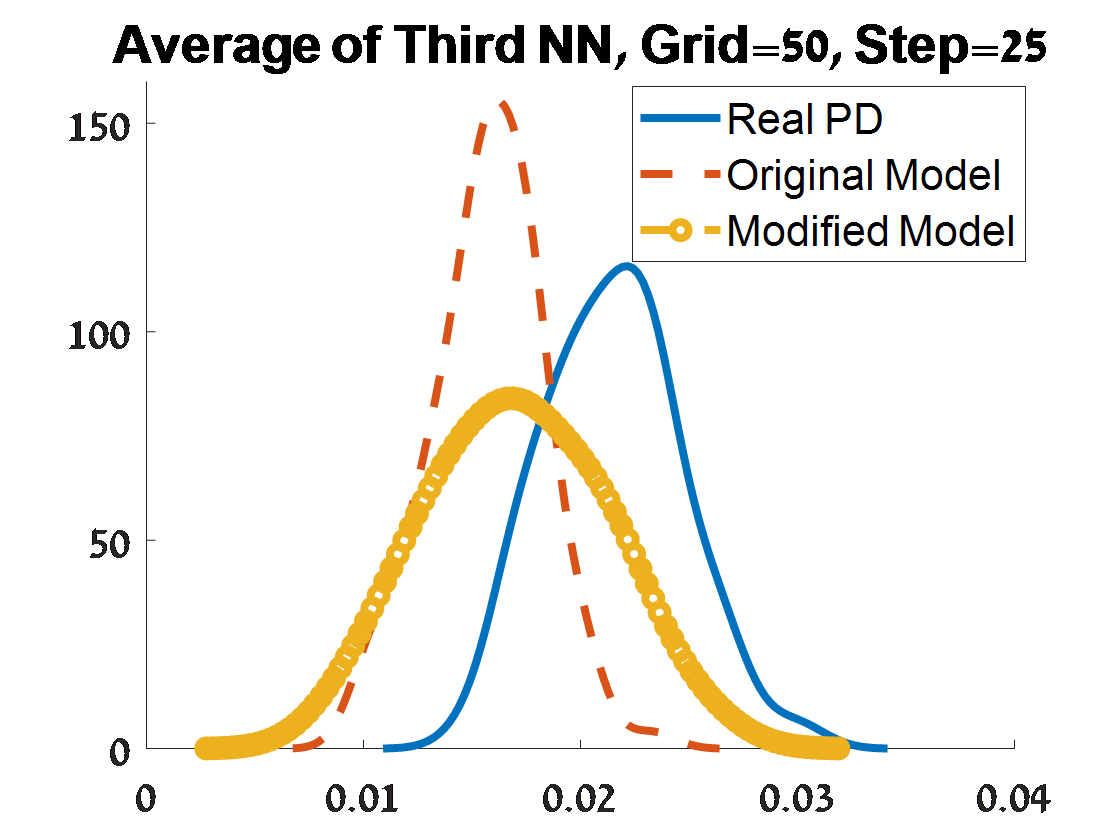}
\includegraphics[width=1.2in, height=1.25in]{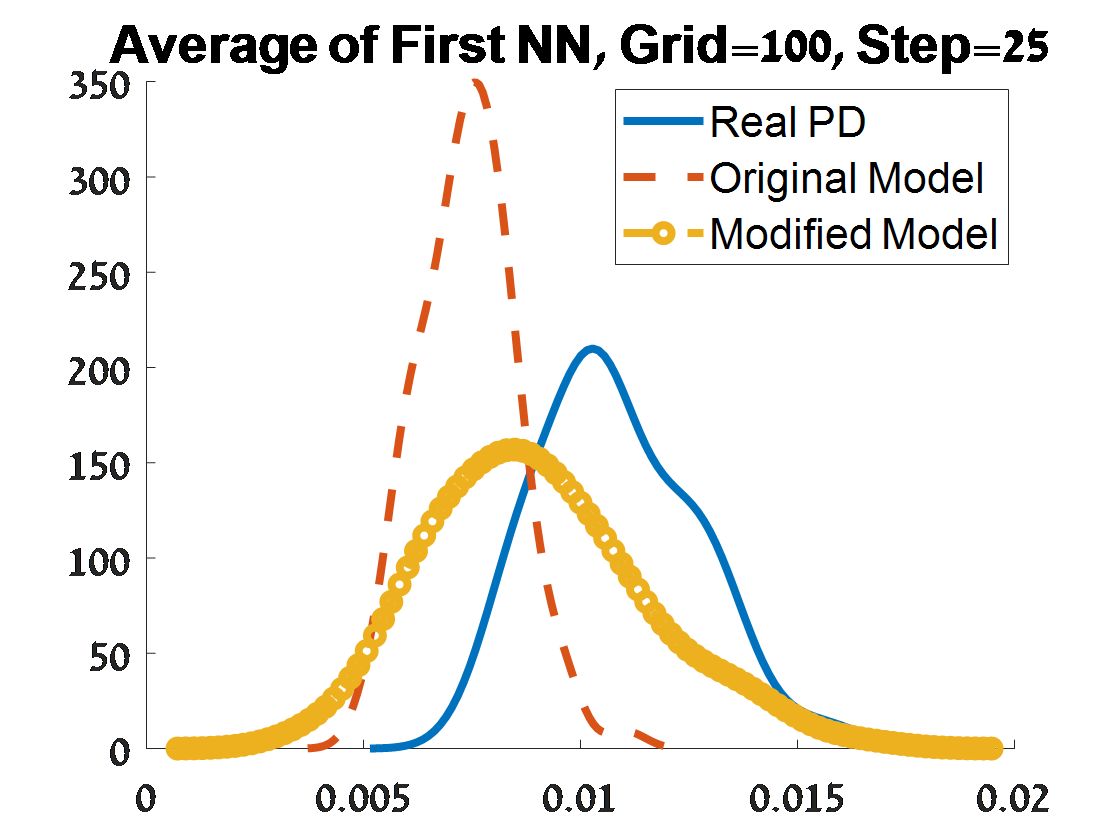}
\includegraphics[width=1.2in, height=1.25in]{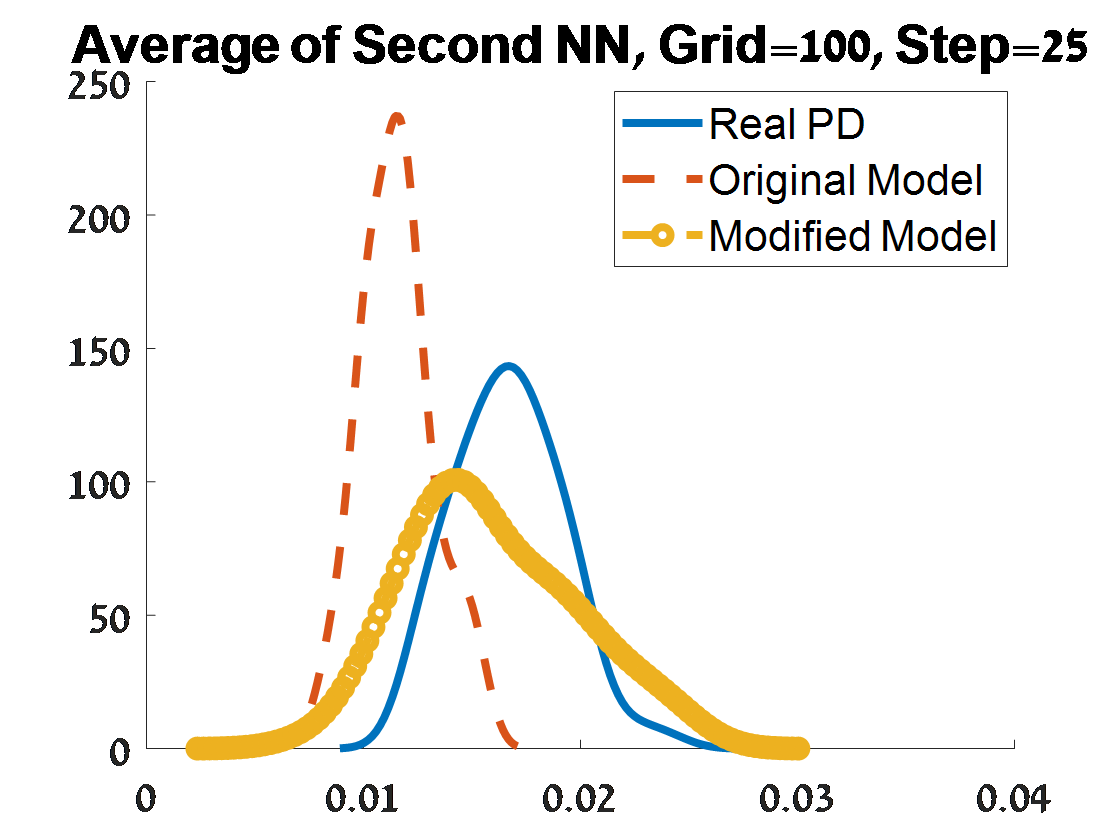}
\includegraphics[width=1.2in, height=1.25in]{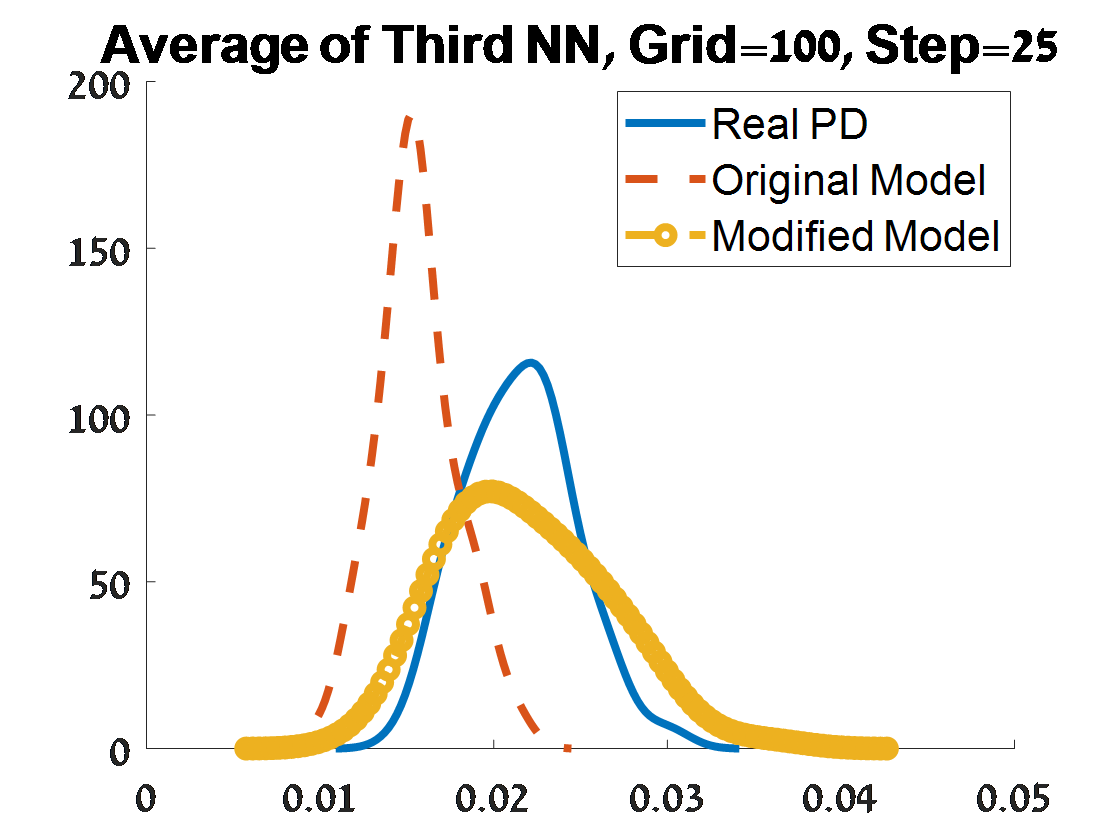}
\\
\includegraphics[width=1.2in, height=1.25in]{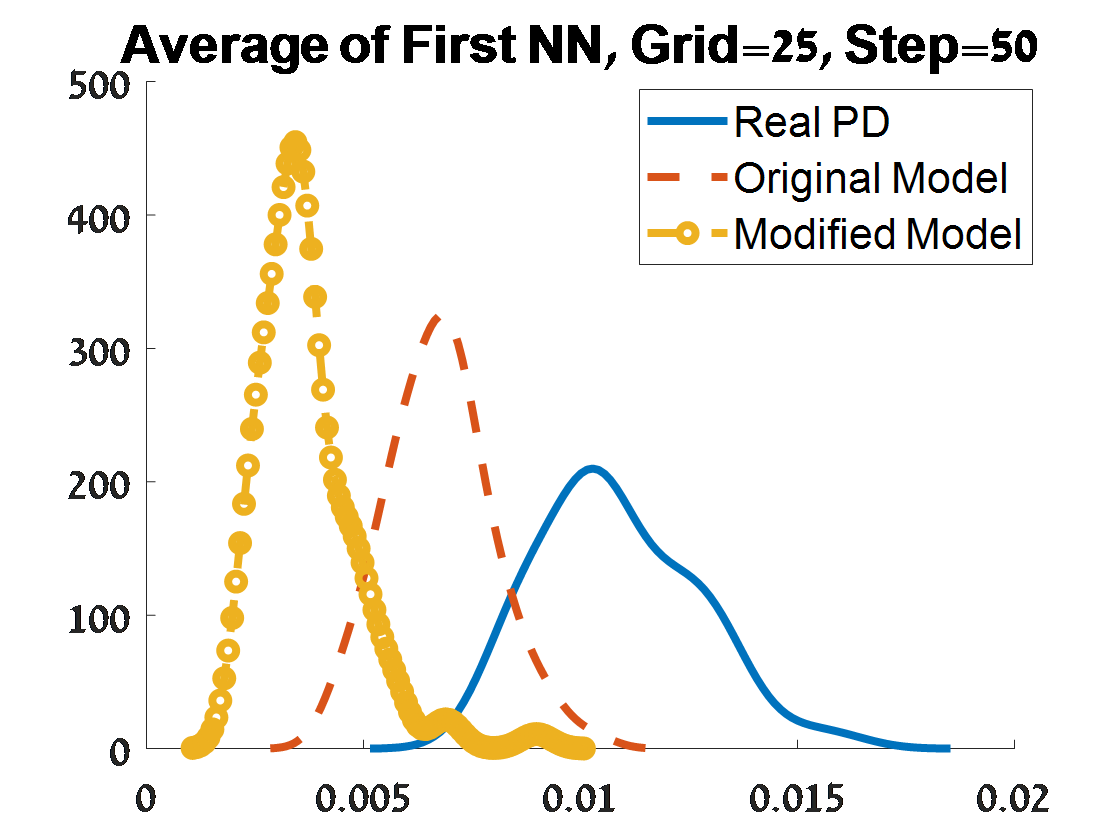}
\includegraphics[width=1.2in, height=1.25in]{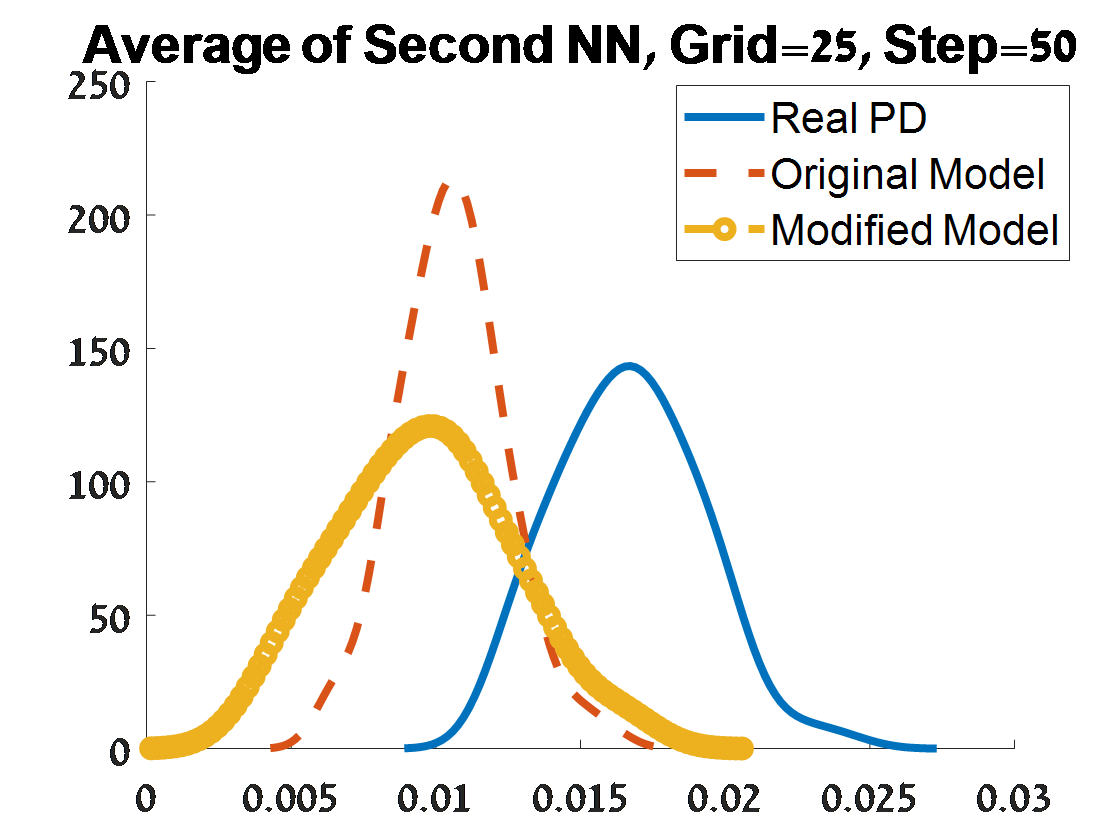}
\includegraphics[width=1.2in, height=1.25in]{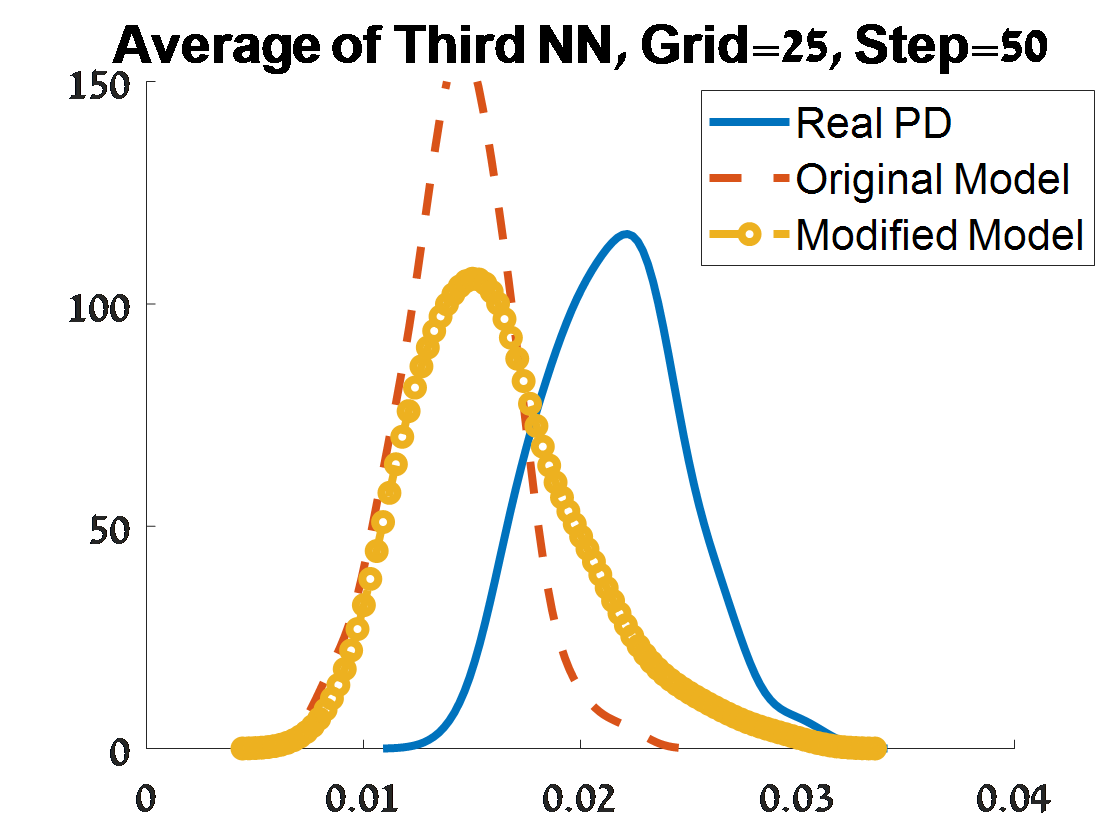}
\includegraphics[width=1.2in, height=1.25in]{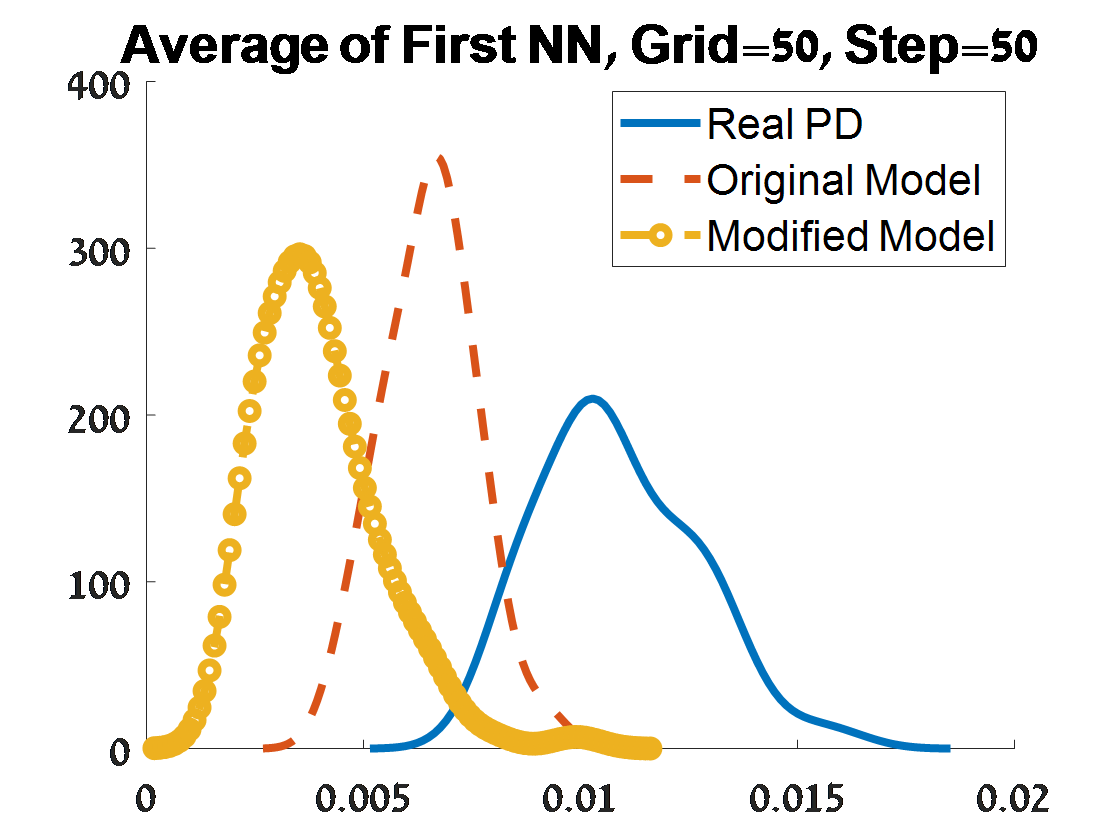}
\includegraphics[width=1.2in, height=1.25in]{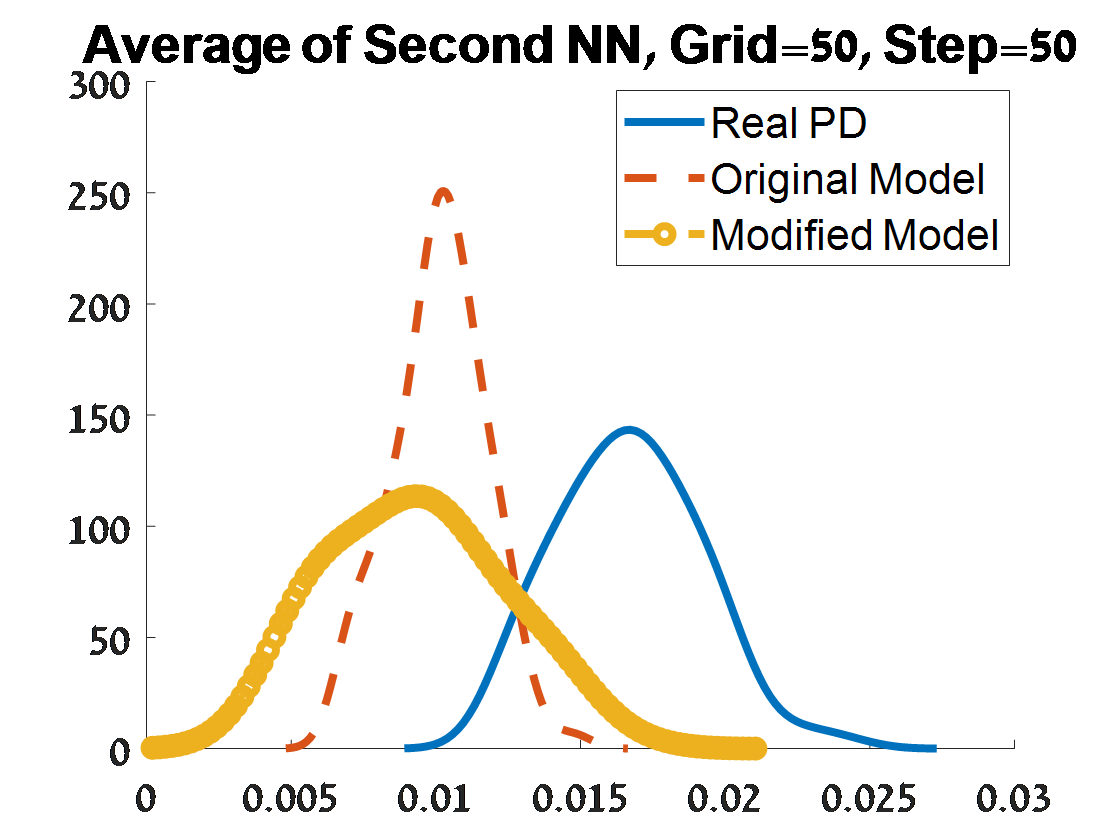}
\includegraphics[width=1.2in, height=1.25in]{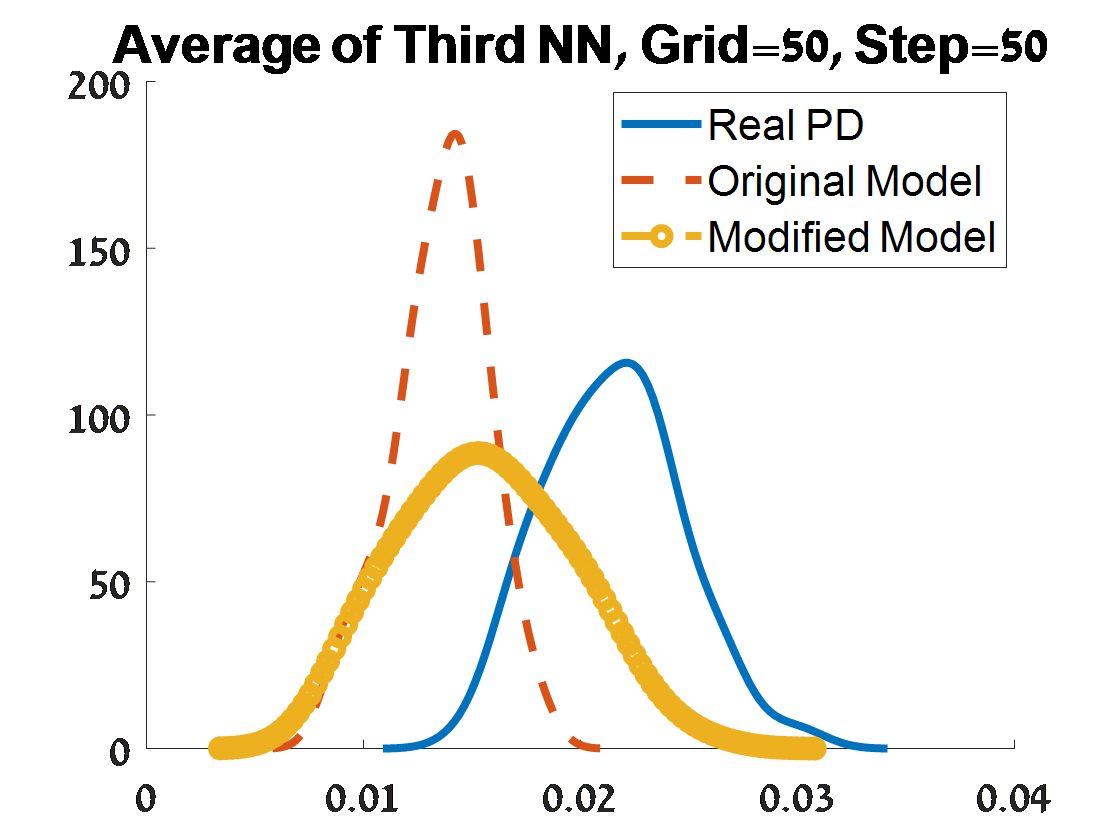}
\includegraphics[width=1.2in, height=1.25in]{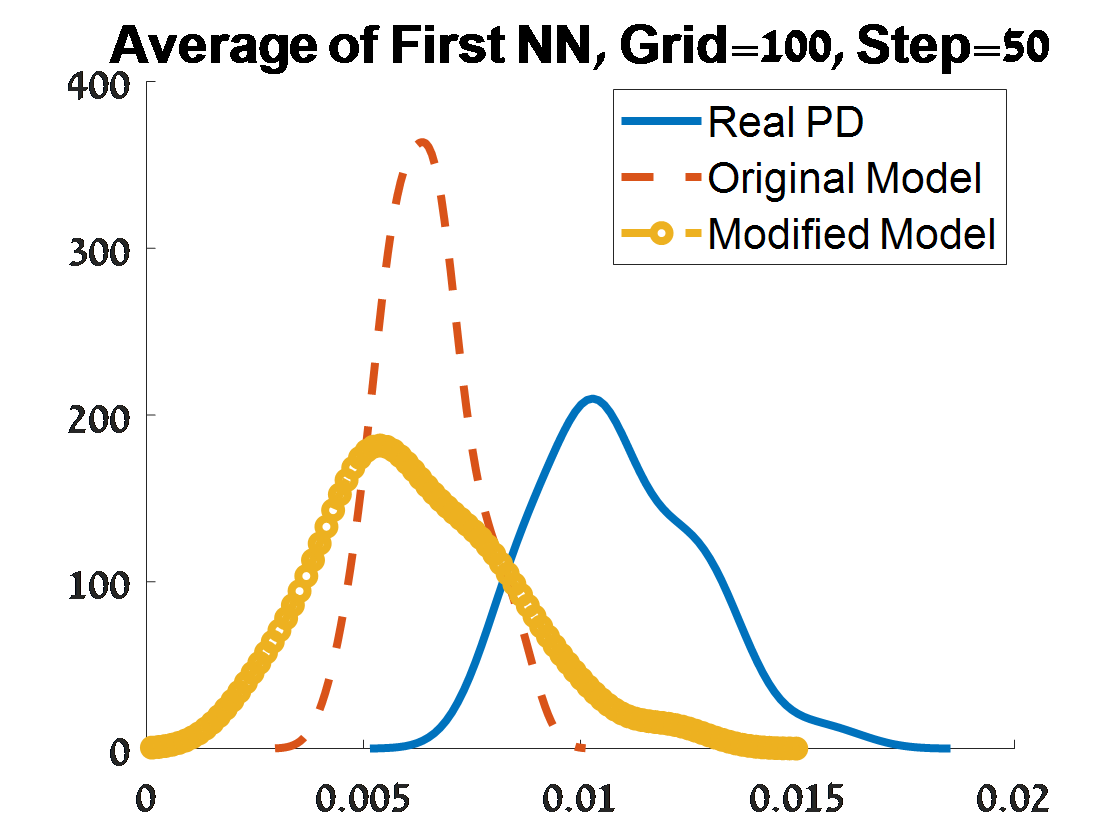}
\includegraphics[width=1.2in, height=1.25in]{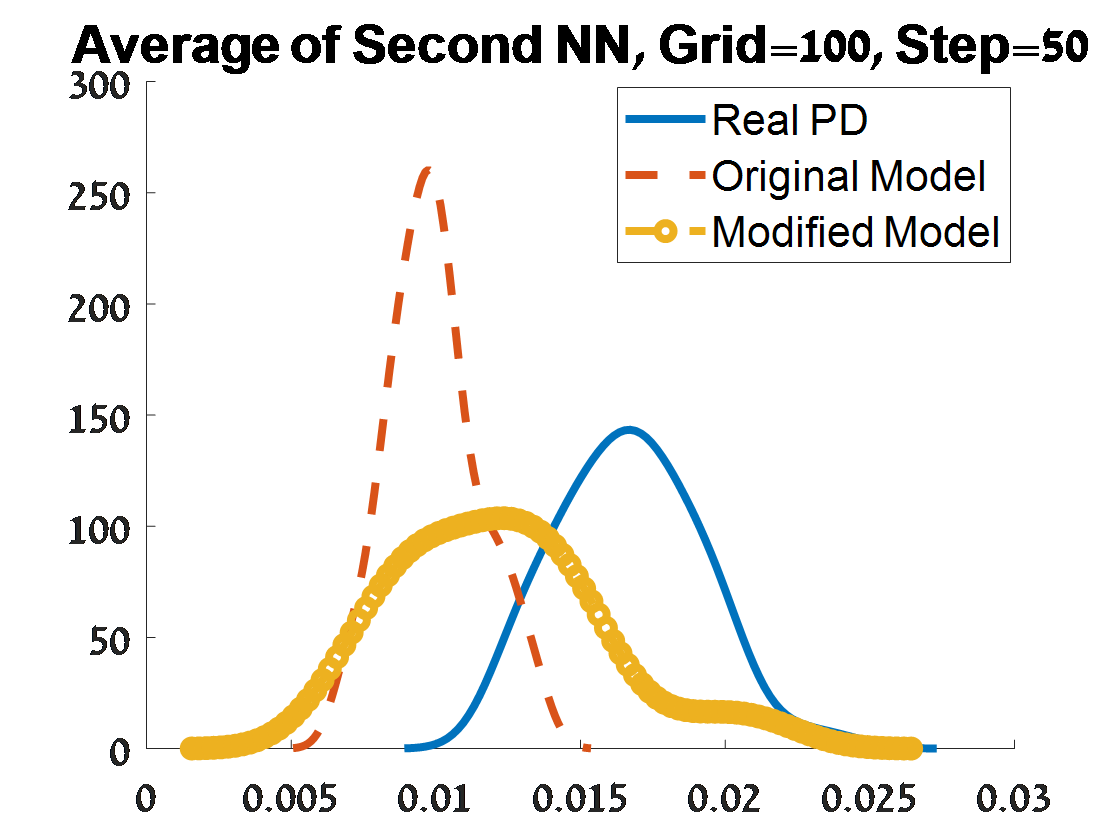}
\includegraphics[width=1.2in, height=1.25in]{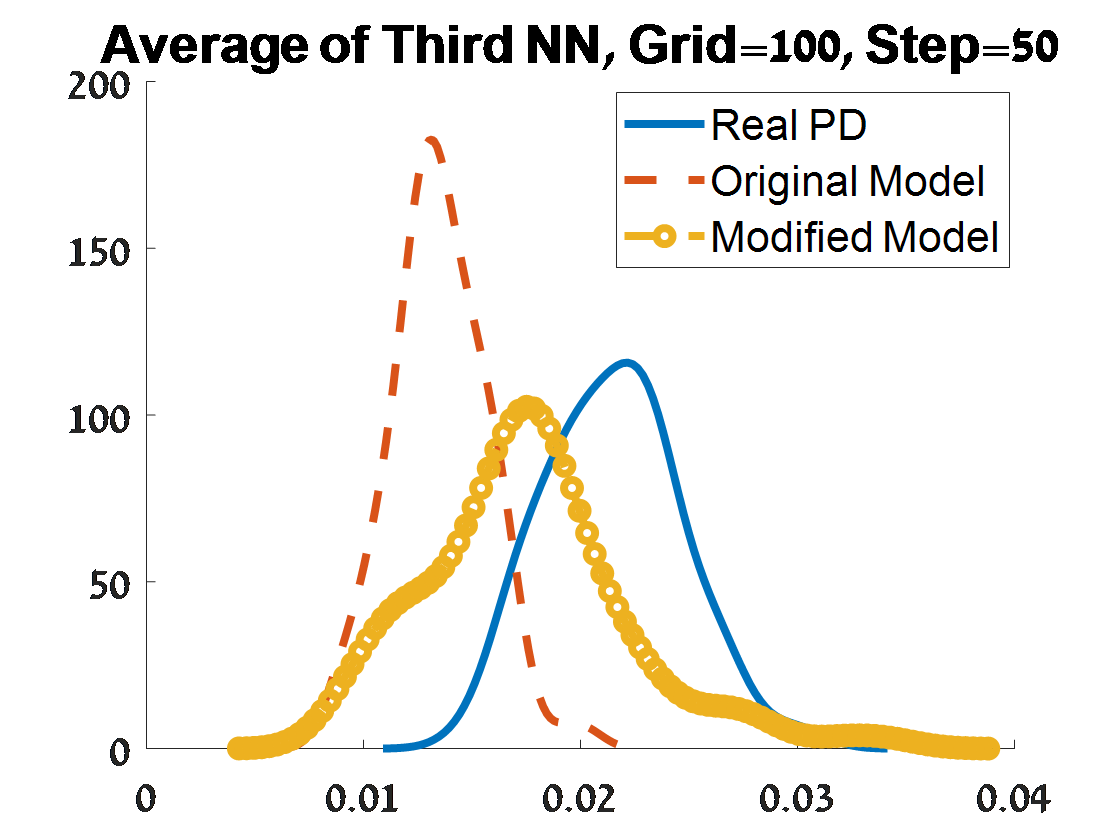}
\ec
%\caption{\footnotesize
% A random sample from two circles, 500 points from the larger circle and 300 from the smaller one,  with a kernel density
\caption{\footnotesize
 Criterion 2 of goodness of fit for 100 $H_0$ PDs corresponded to 100 samples from a unit $S^2$. The figures depend on the grid of the proposal distribution ("Grid"), and the burn-in ("Step") of the MCMC algorithm.}
\label{fig:s2_H0_b}
\end{figure}
\end{landscape}

\begin{landscape}
\begin{figure}[h!]
\bc
\includegraphics[width=1.2in, height=1.25in]{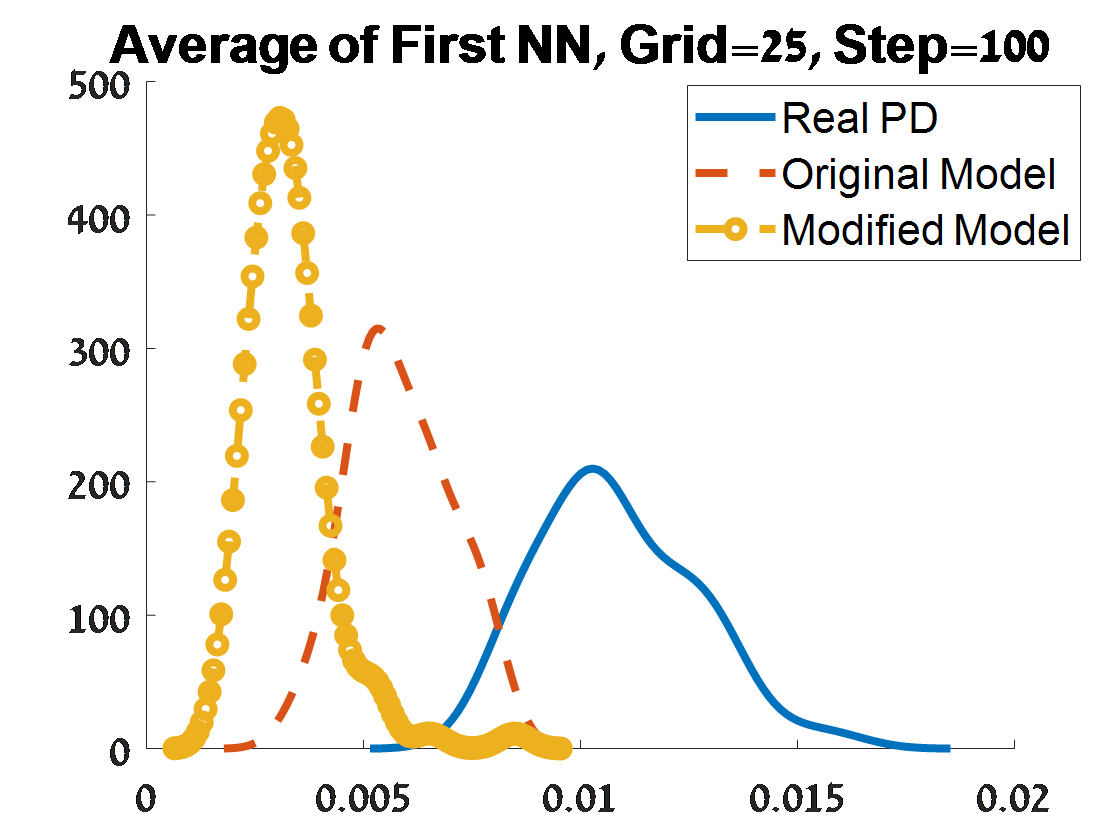}
\includegraphics[width=1.2in, height=1.25in]{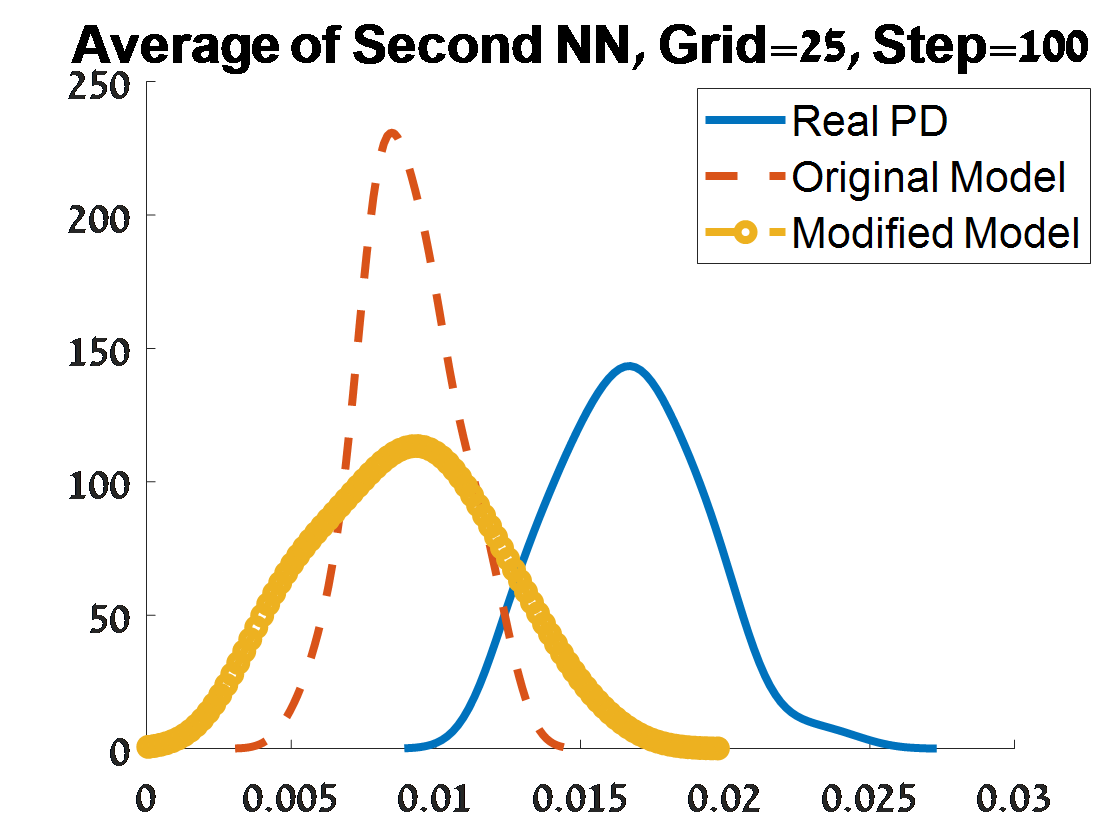}
\includegraphics[width=1.2in, height=1.25in]{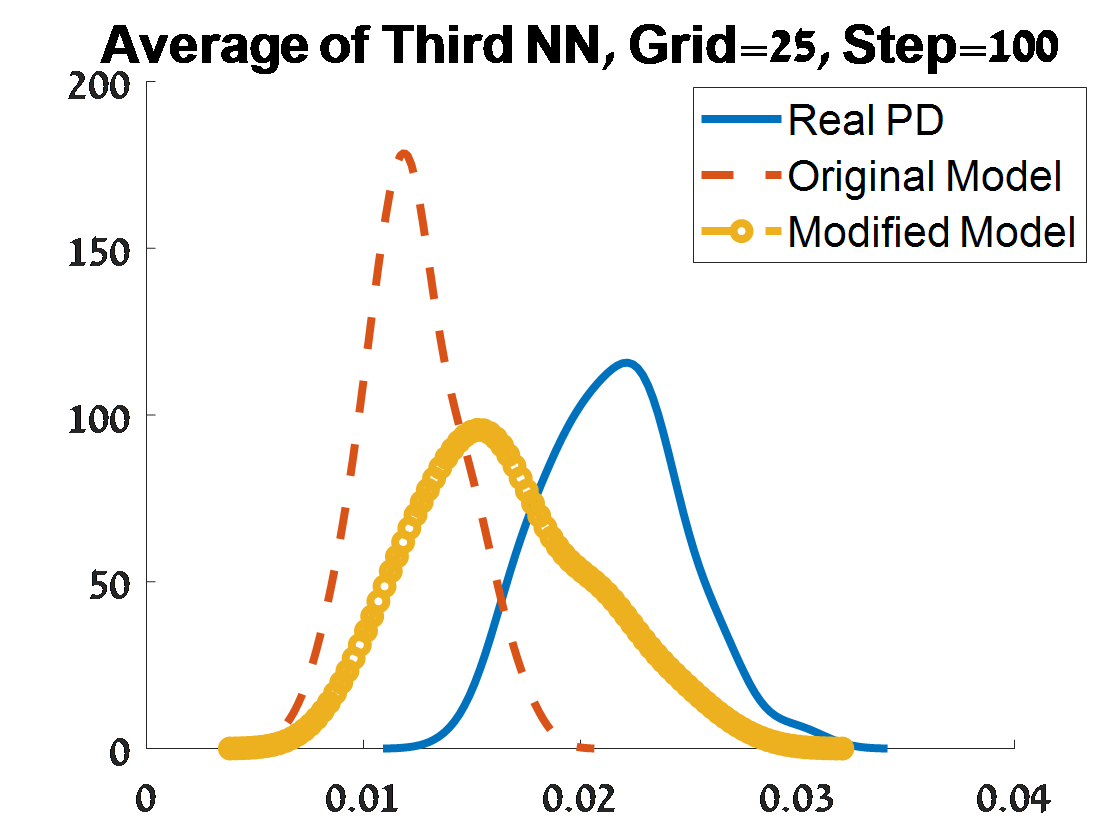}
\includegraphics[width=1.2in, height=1.25in]{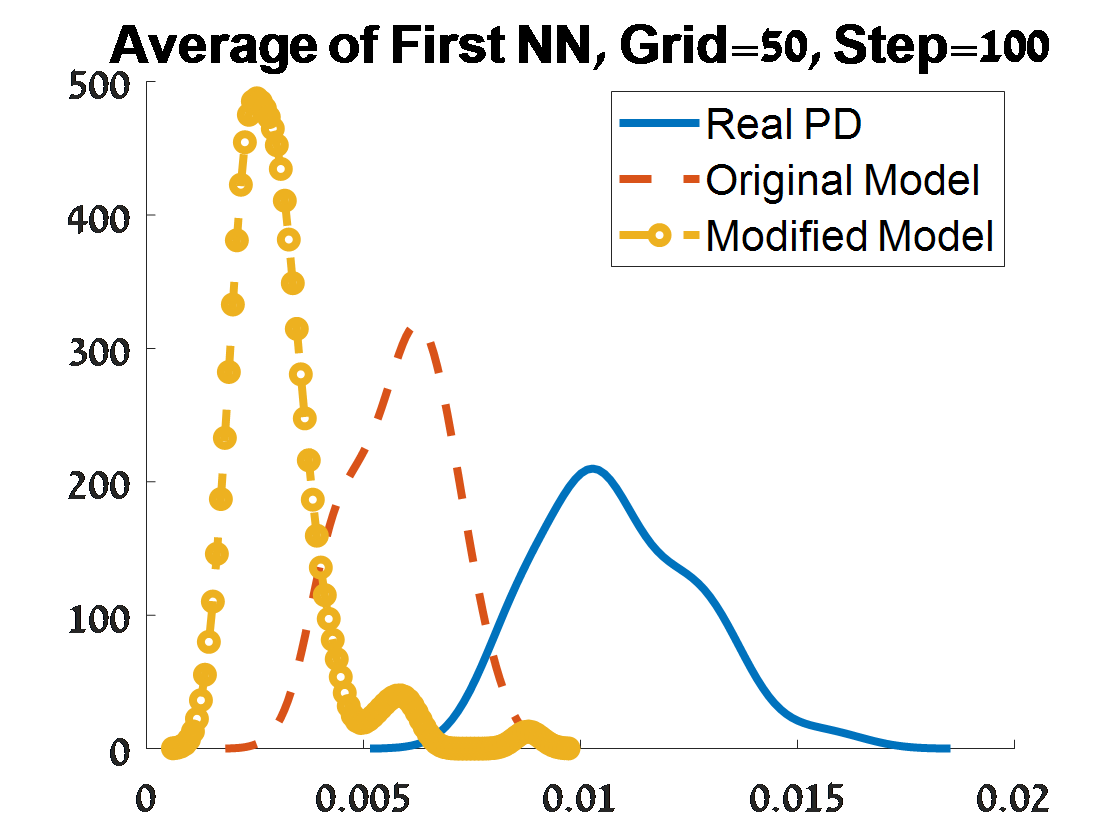}
\includegraphics[width=1.2in, height=1.25in]{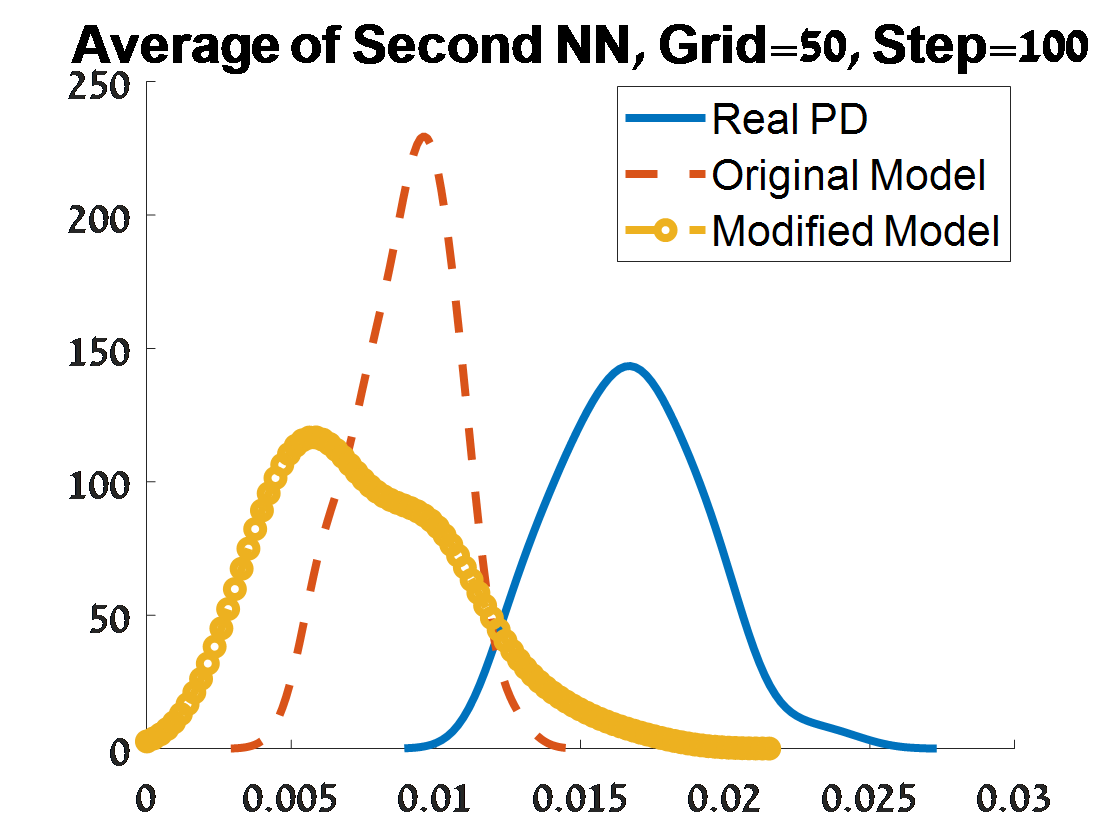}
\includegraphics[width=1.2in, height=1.25in]{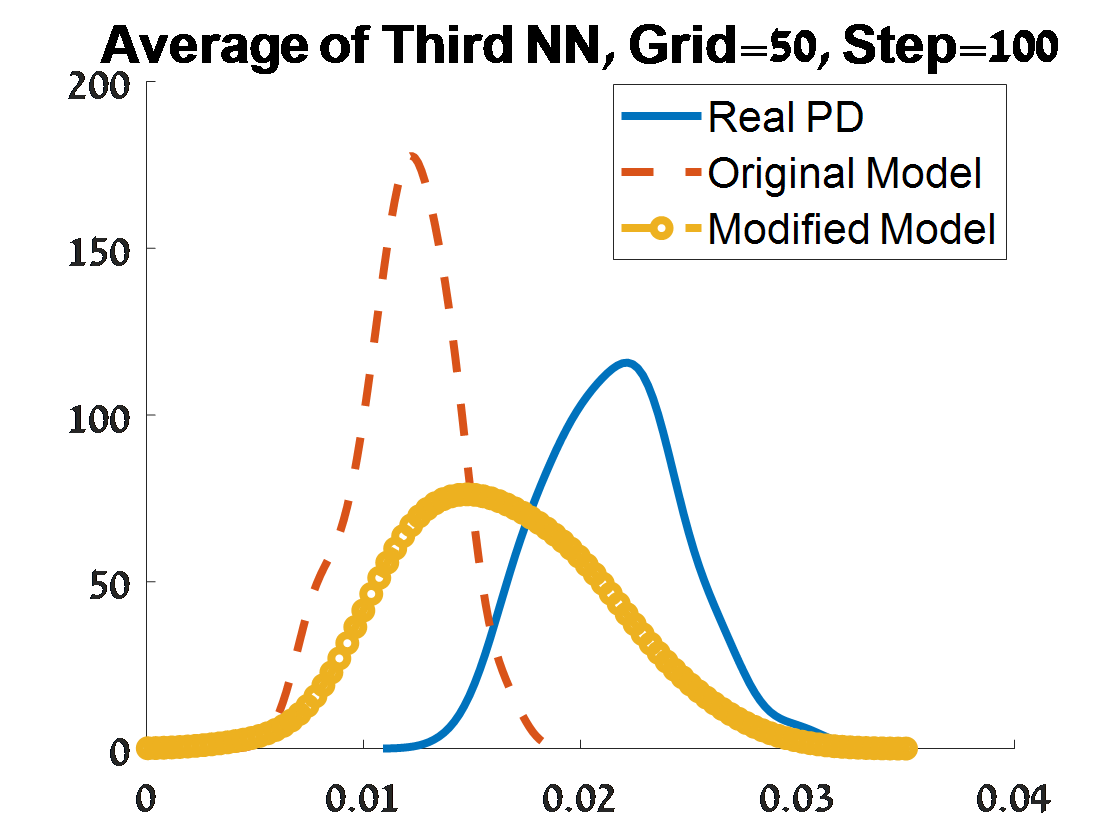}
\includegraphics[width=1.2in, height=1.25in]{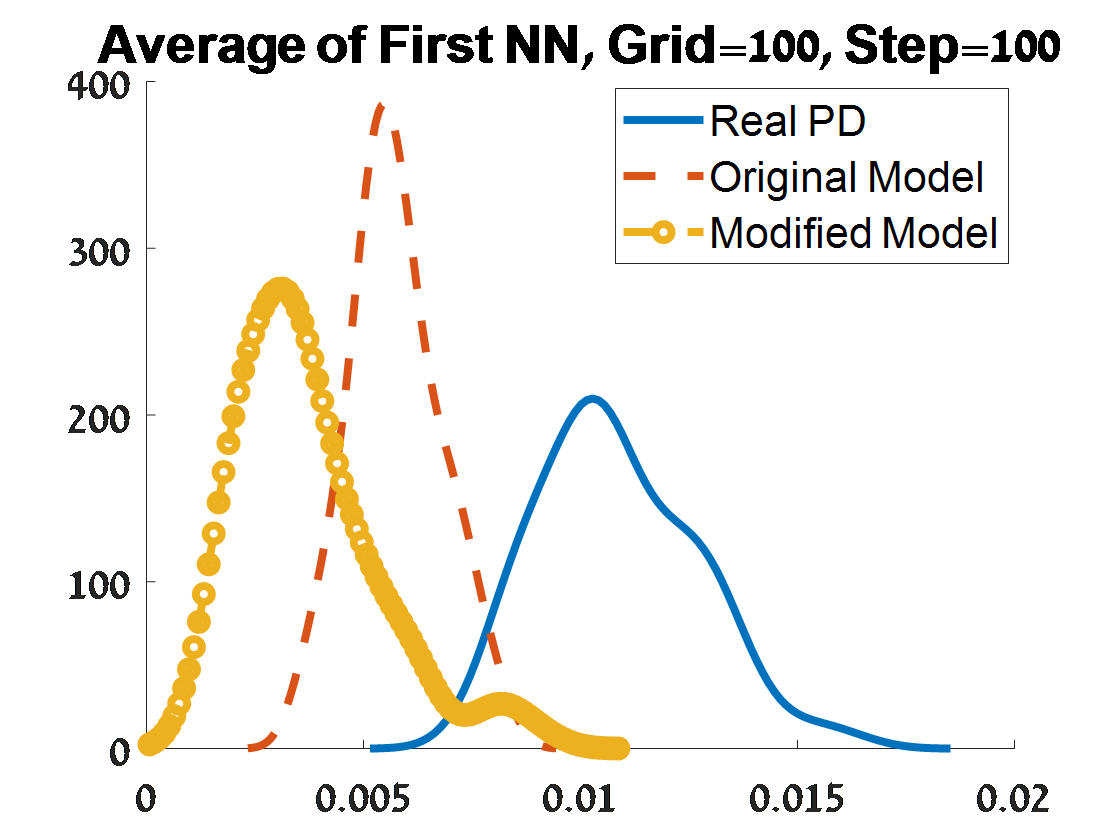}
\includegraphics[width=1.2in, height=1.25in]{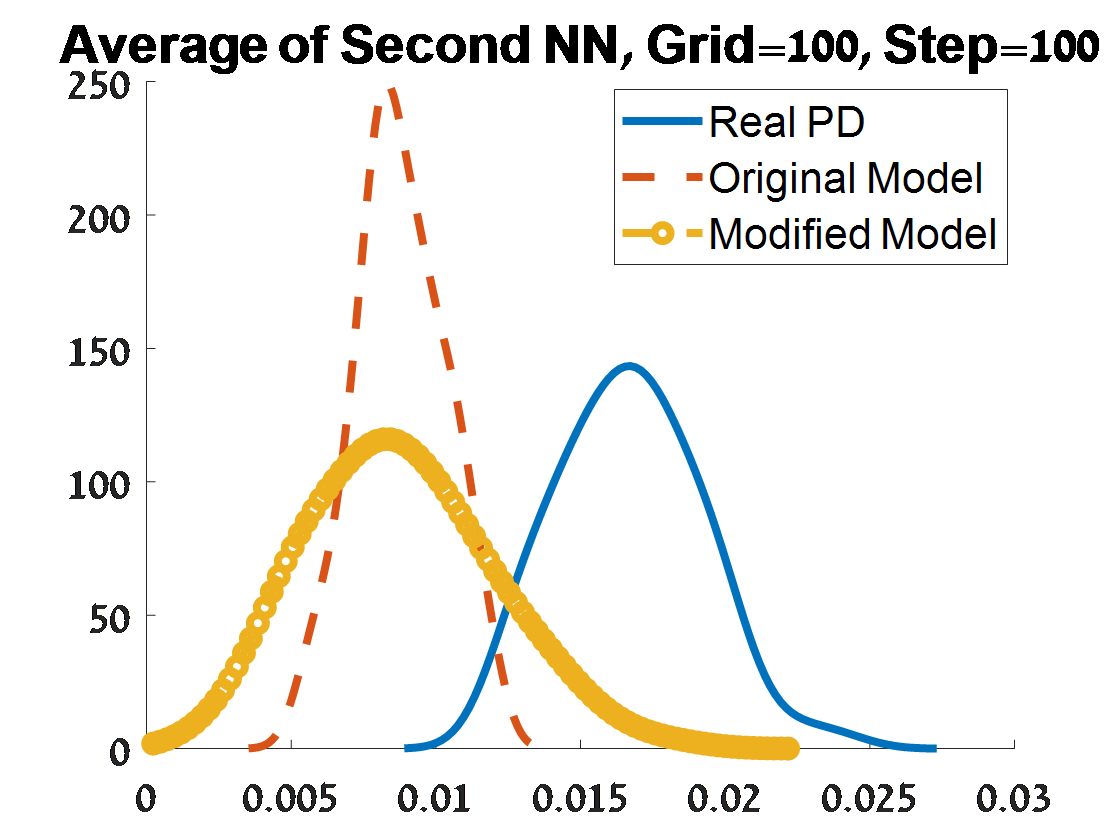}
\includegraphics[width=1.2in, height=1.25in]{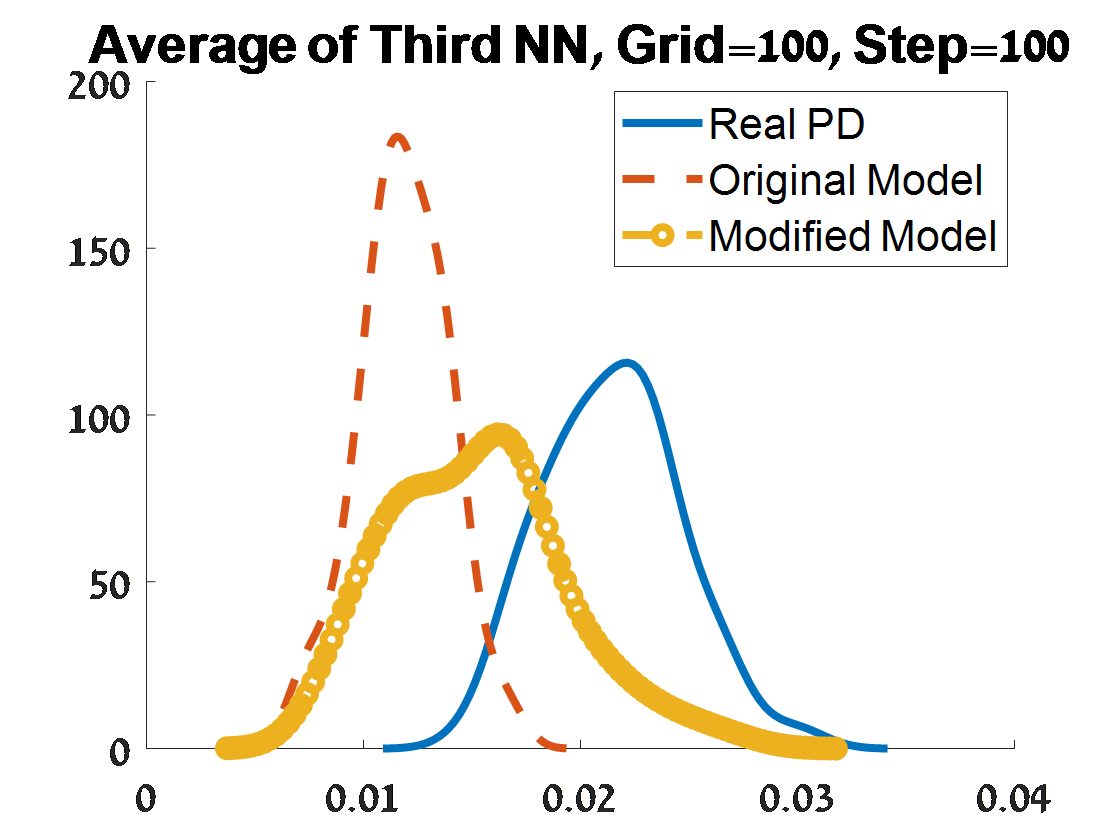}
\ec
%\caption{\footnotesize
% A random sample from two circles, 500 points from the larger circle and 300 from the smaller one,  with a kernel density
\caption{\footnotesize
 Continue of Criterion 2 of goodness of fit for 100 $H_0$ PDs corresponded to 100 samples from a unit $S^2$. The figures depend on the grid of the proposal distribution ("Grid"), and the burn-in ("Step") of the MCMC algorithm.}
\label{fig:s2_H0_c}
\end{figure}
\end{landscape}

%\begin{landscape}
%\begin{figure}[h!]
%\bc
%\includegraphics[width=1.8in, height=1.8in]{SphereH0_pd30_grid50_step25}
%\includegraphics[width=1.8in, height=1.8in]{SphereH0_pd30_grid100_step25}
%\includegraphics[width=1.8in, height=1.8in]{SphereH0_pd60_grid50_step25}
%\includegraphics[width=1.8in, height=1.8in]{SphereH0_pd60_grid100_step25}
%\ec
%\caption{\footnotesize
% A random sample from two circles, 500 points from the larger circle and 300 from the smaller one,  with a kernel density
%\caption{\footnotesize
%Examples of two $H_0$ PDs, each one is corresponded to a sample from a unit $S^2$. For each PD, the simulated PD based on the two model versions is described. The figures depend on the grid of the proposal distribution ("Grid"), and the burn-in ("Step") of the MCMC algorithm.}
%\label{fig:s2_H0_d}
%\end{figure}
%\end{landscape}

\subsubsection{The fitted model for $H_1$}
\begin{landscape}
\begin{figure}[h!]
\bc
\includegraphics[width=1.2in, height=1.4in]{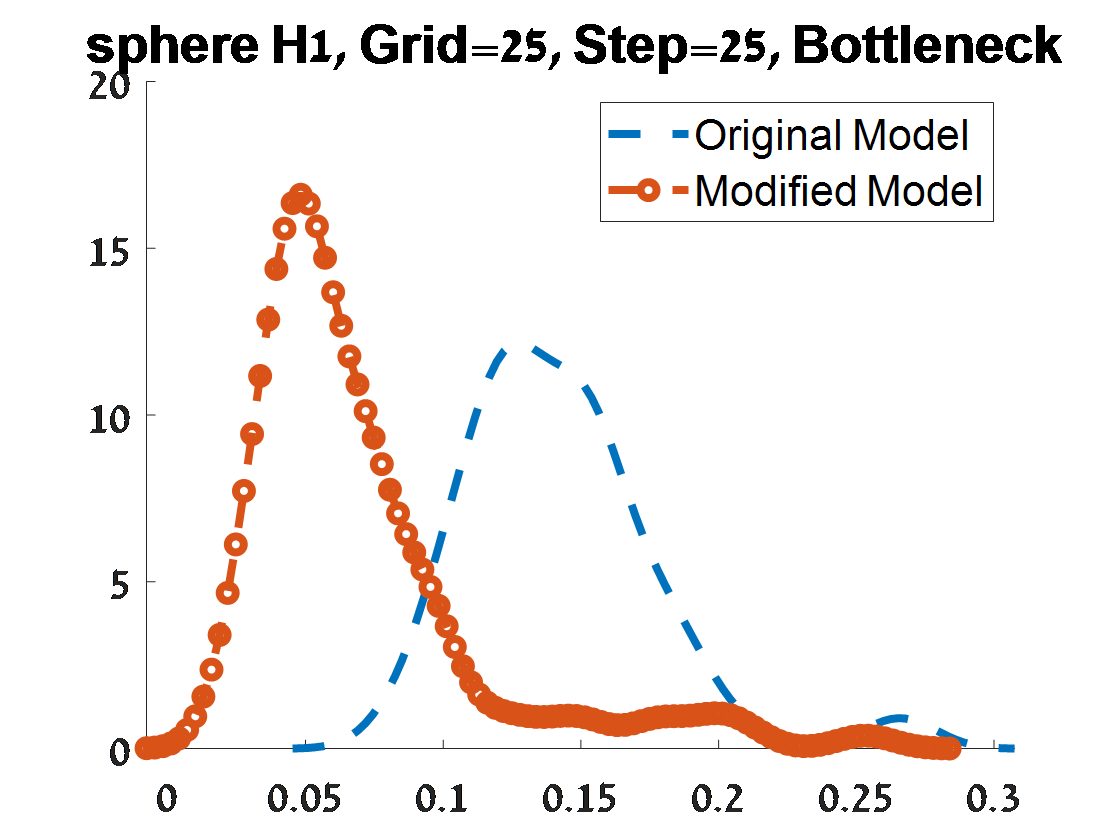}
\includegraphics[width=1.2in, height=1.4in]{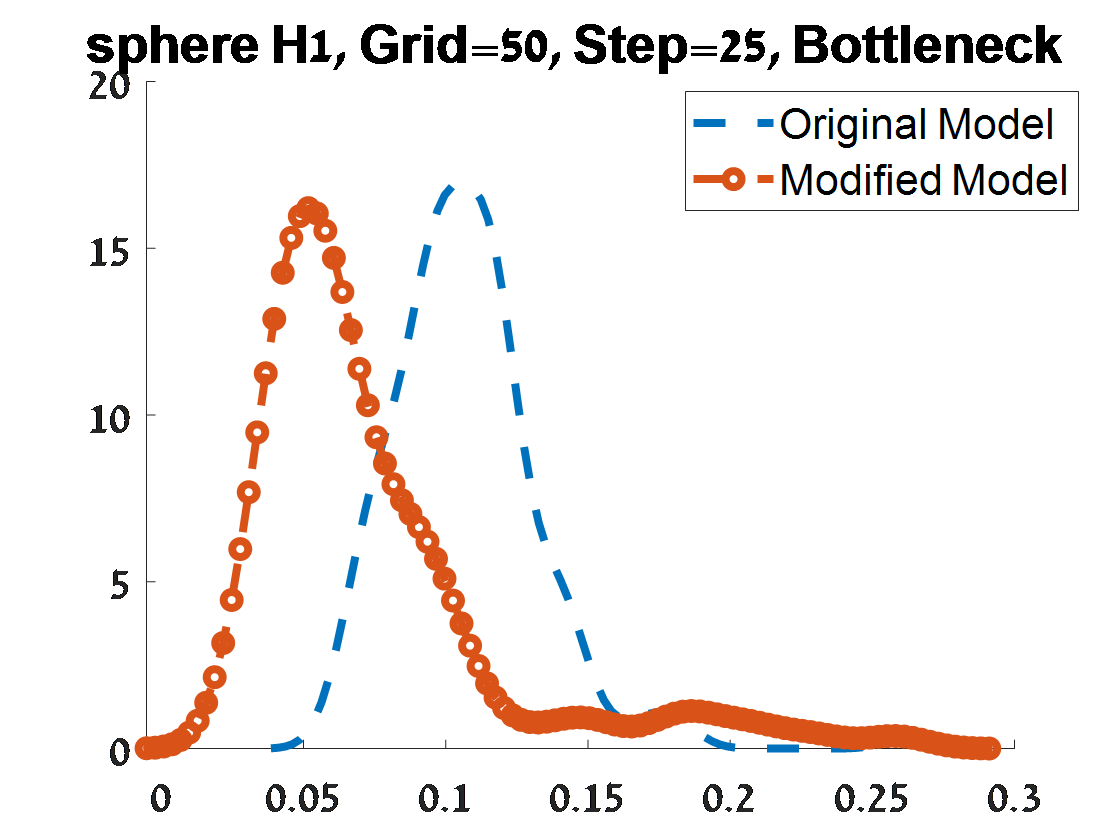}
\includegraphics[width=1.2in, height=1.4in]{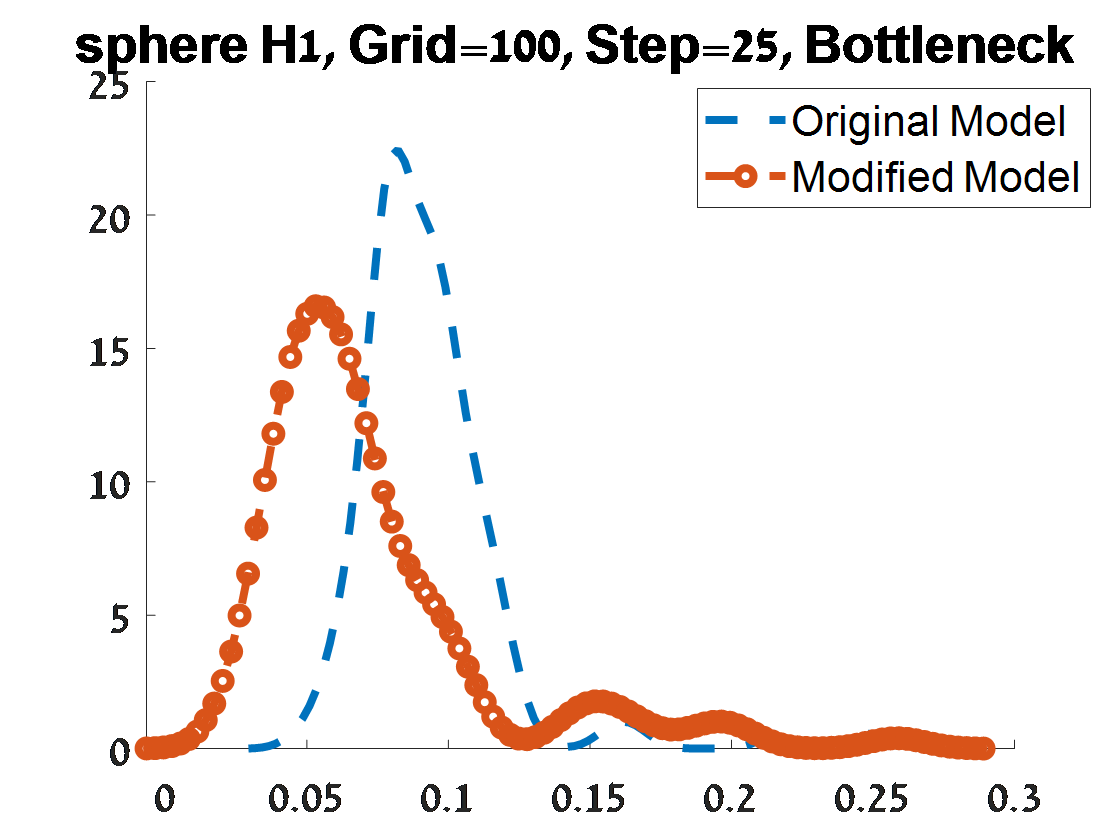}
\includegraphics[width=1.2in, height=1.4in]{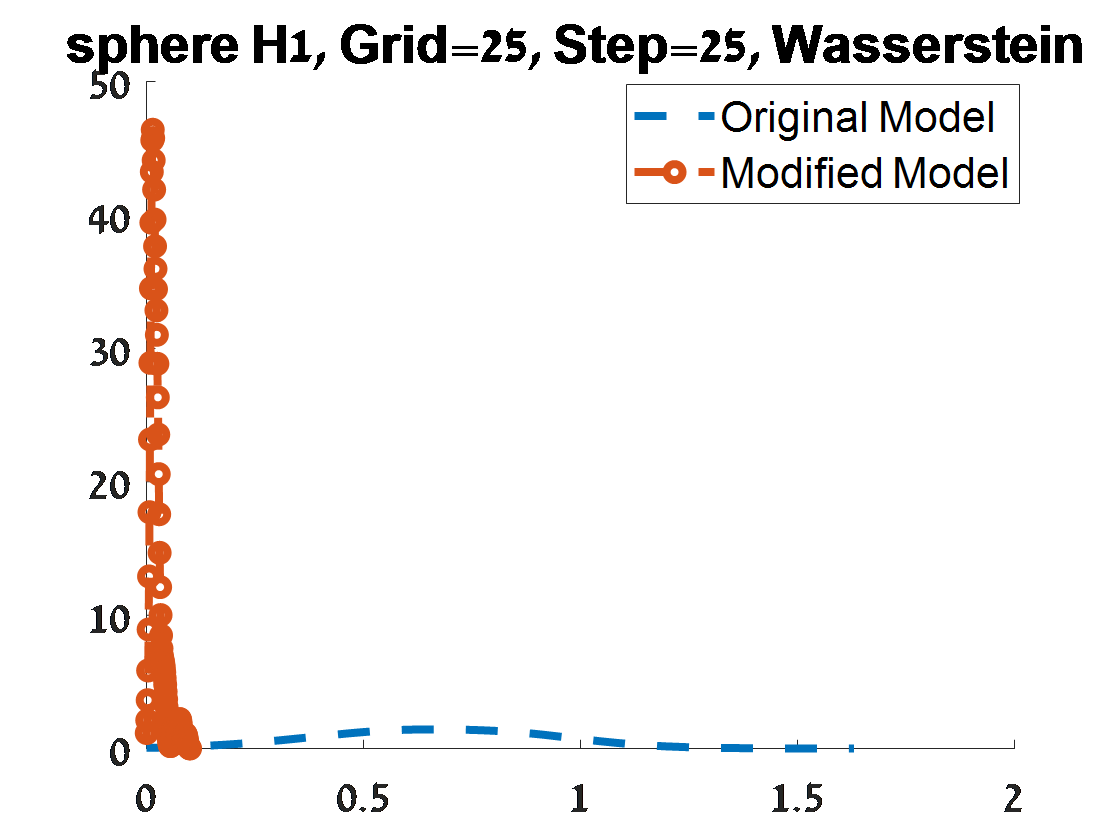}
\includegraphics[width=1.2in, height=1.4in]{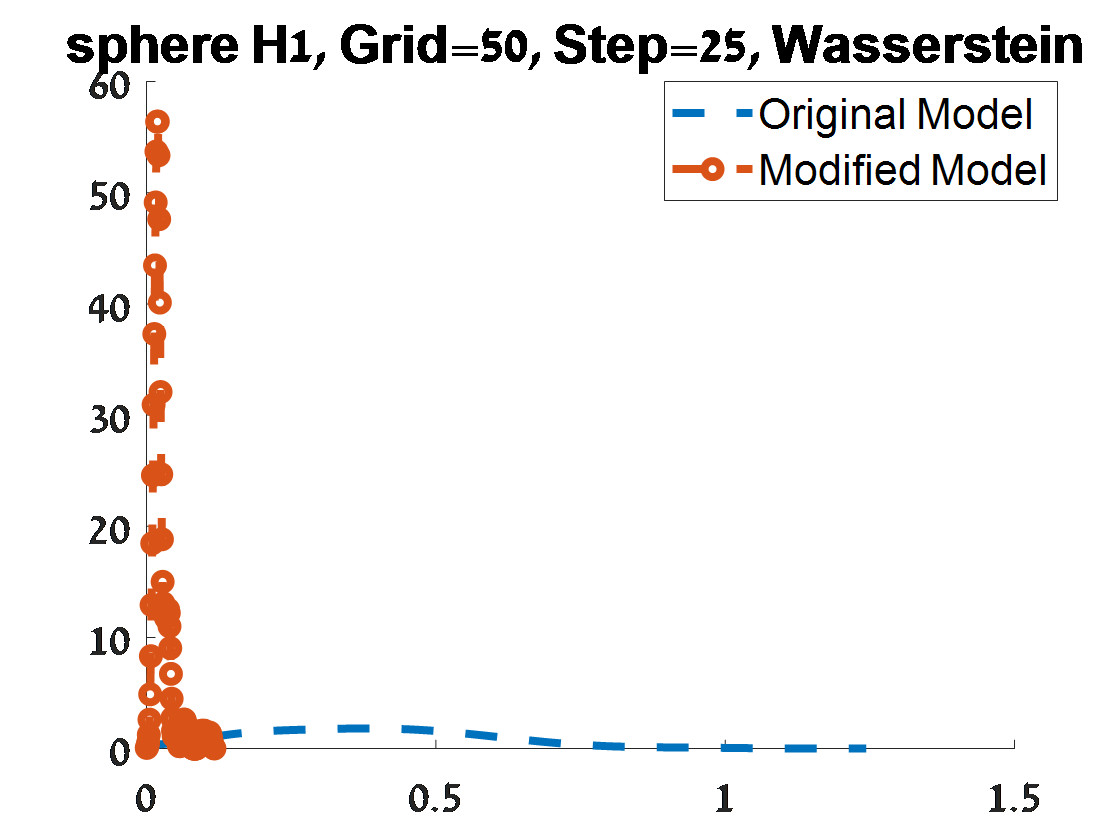}
\includegraphics[width=1.2in, height=1.4in]{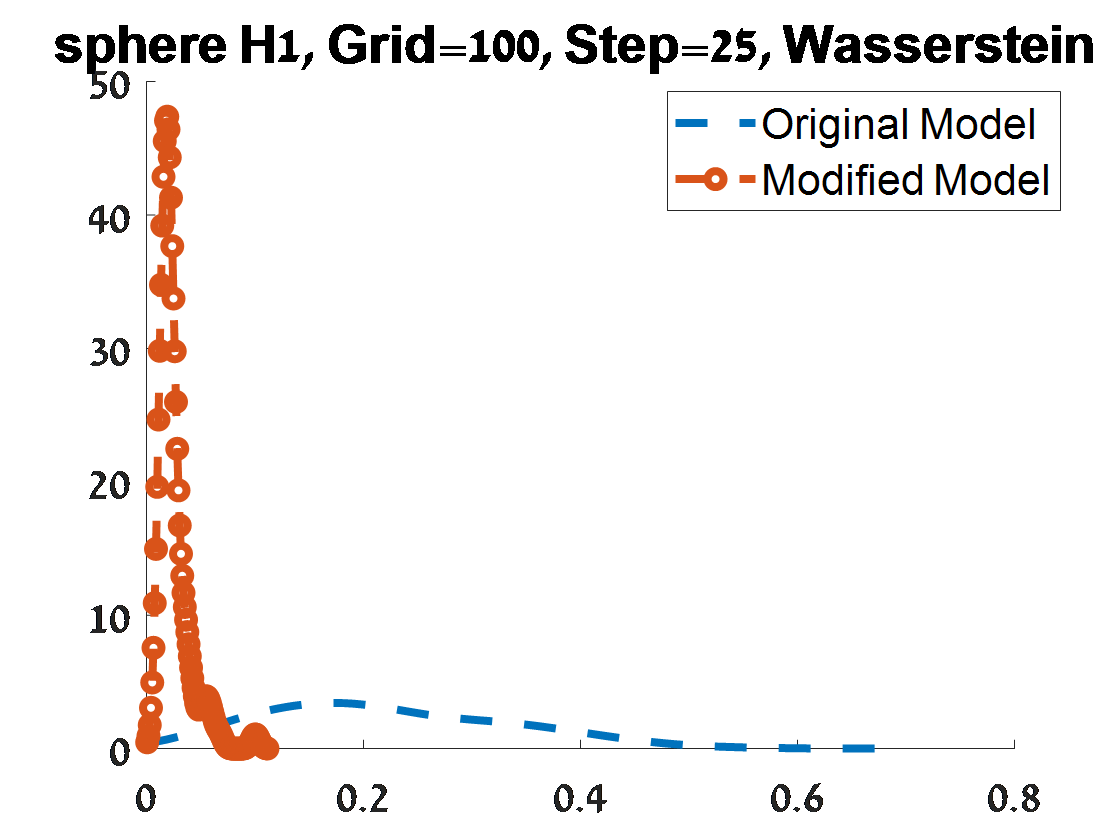}
\\
\includegraphics[width=1.2in, height=1.4in]{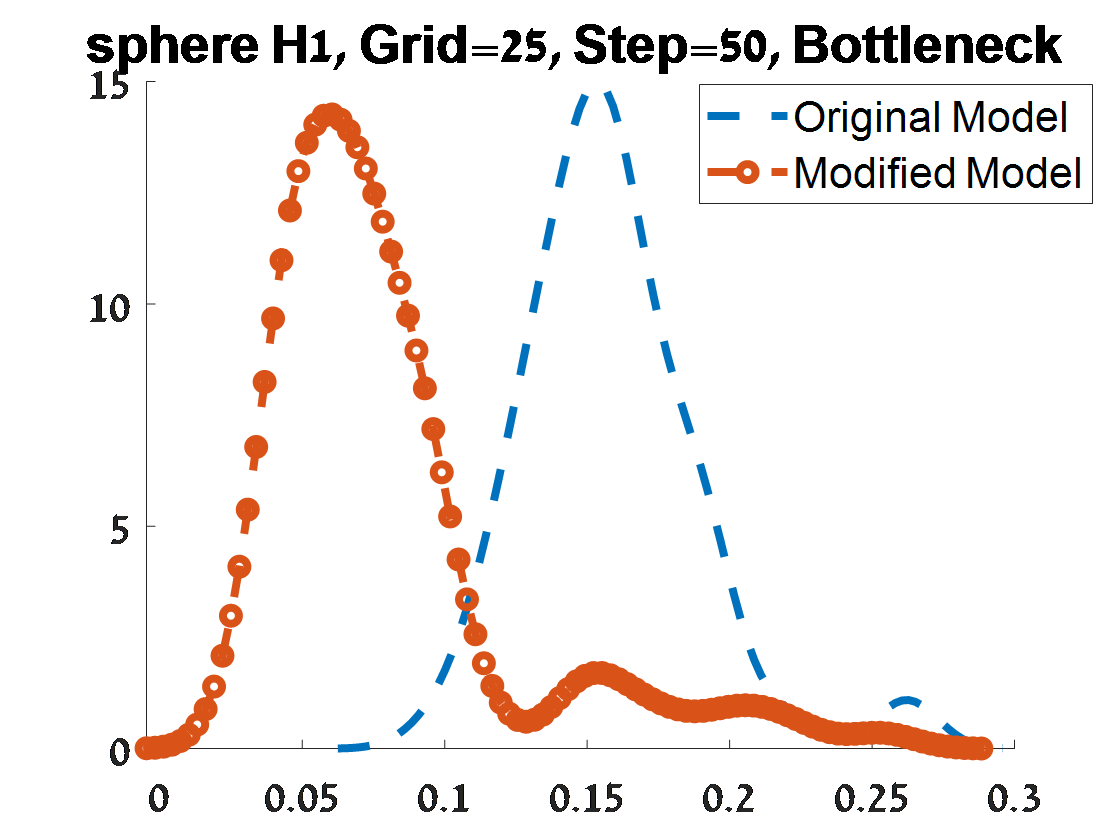}
\includegraphics[width=1.2in, height=1.4in]{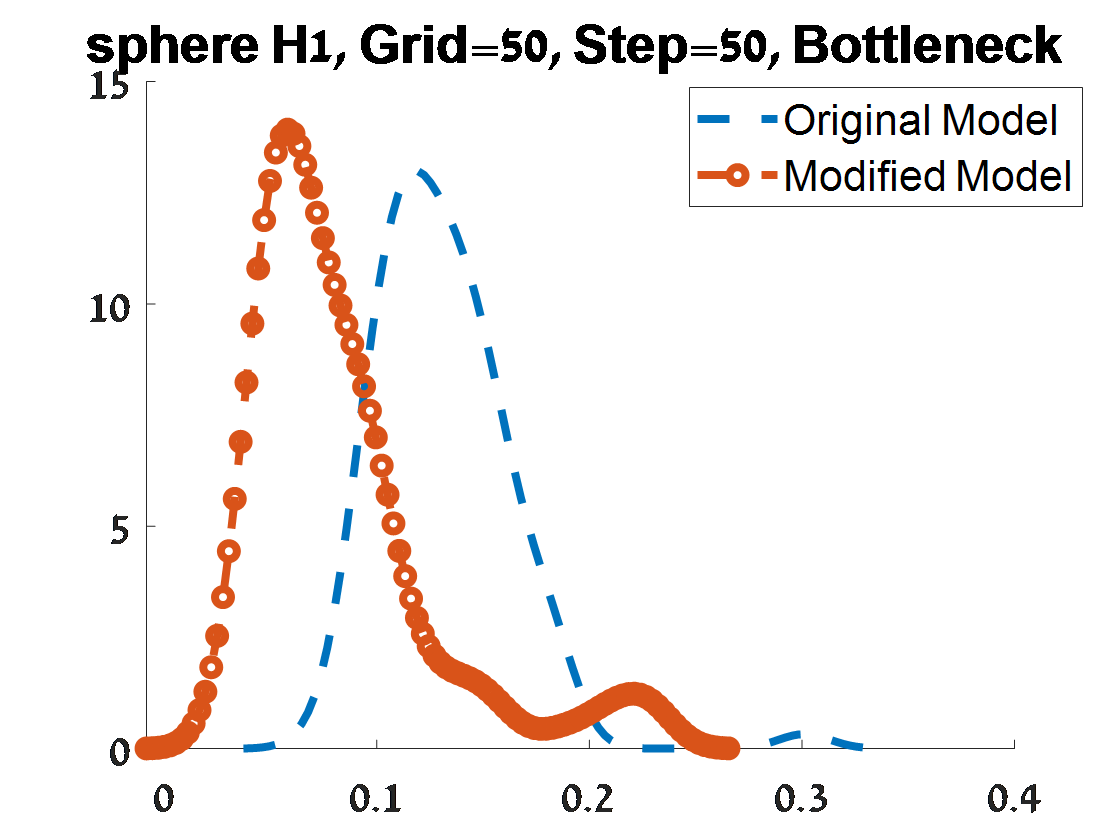}
\includegraphics[width=1.2in, height=1.4in]{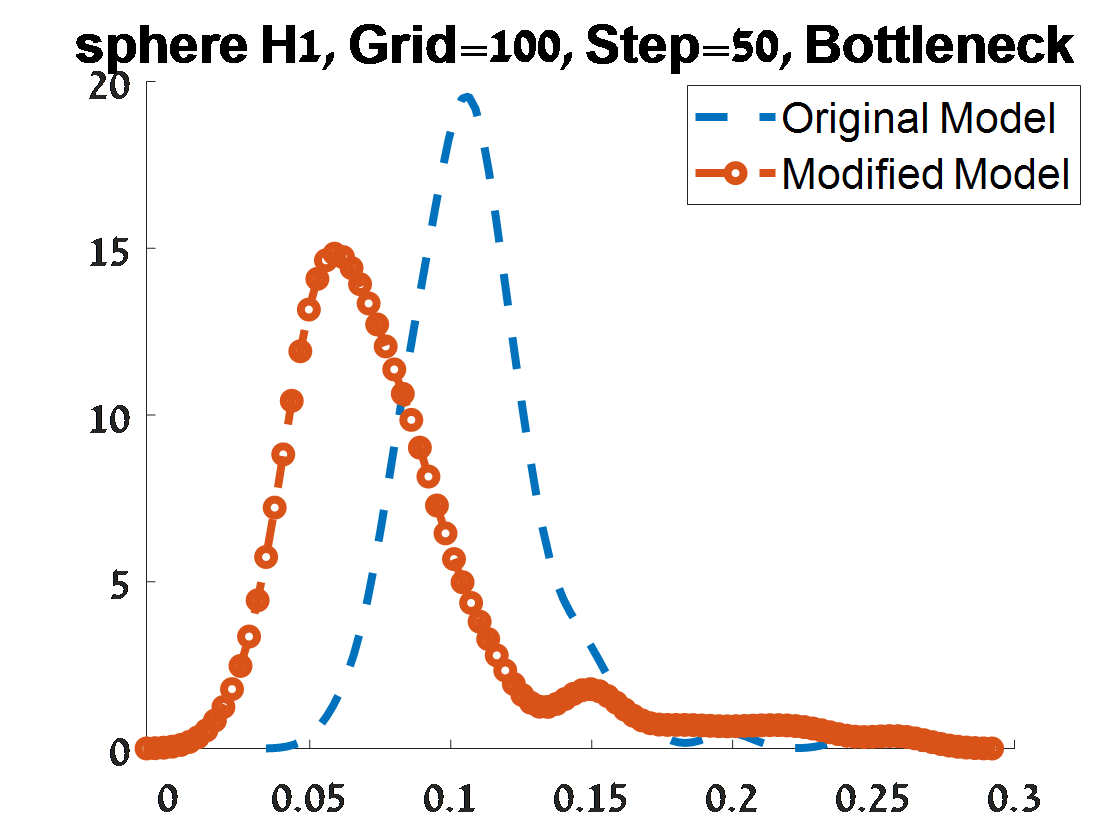}
\includegraphics[width=1.2in, height=1.4in]{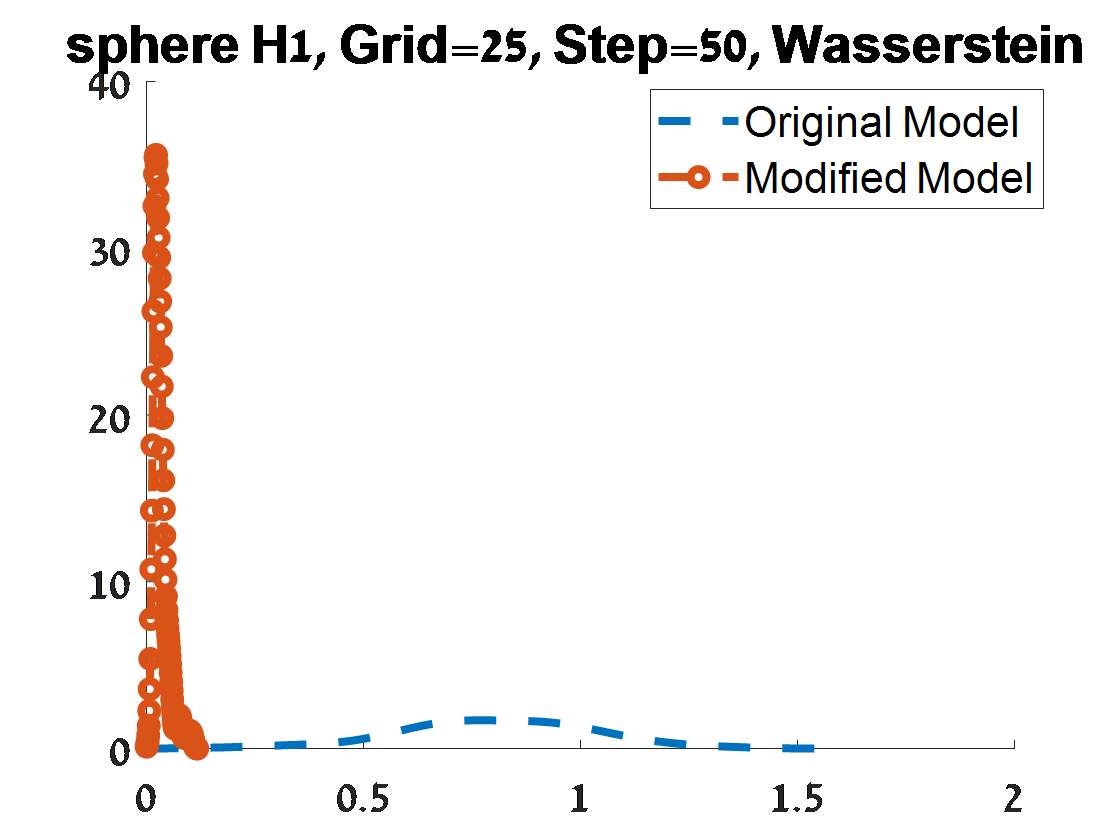}
\includegraphics[width=1.2in, height=1.4in]{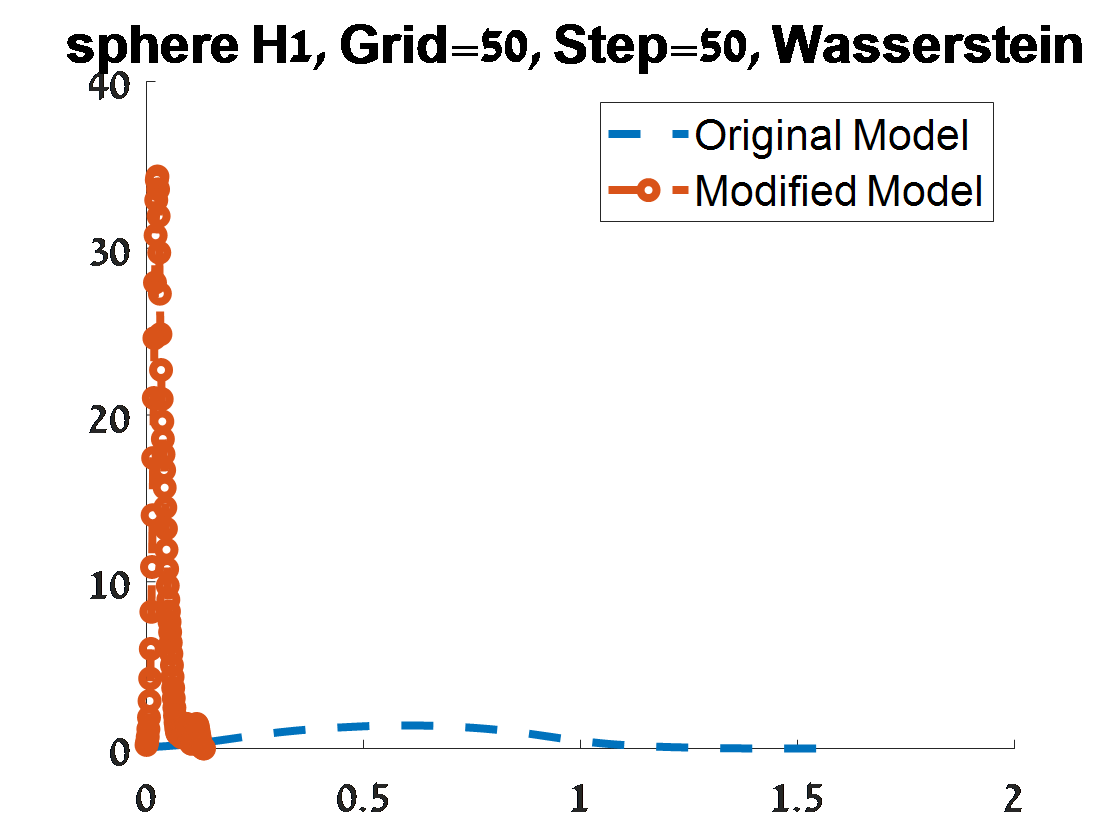}
\includegraphics[width=1.2in, height=1.4in]{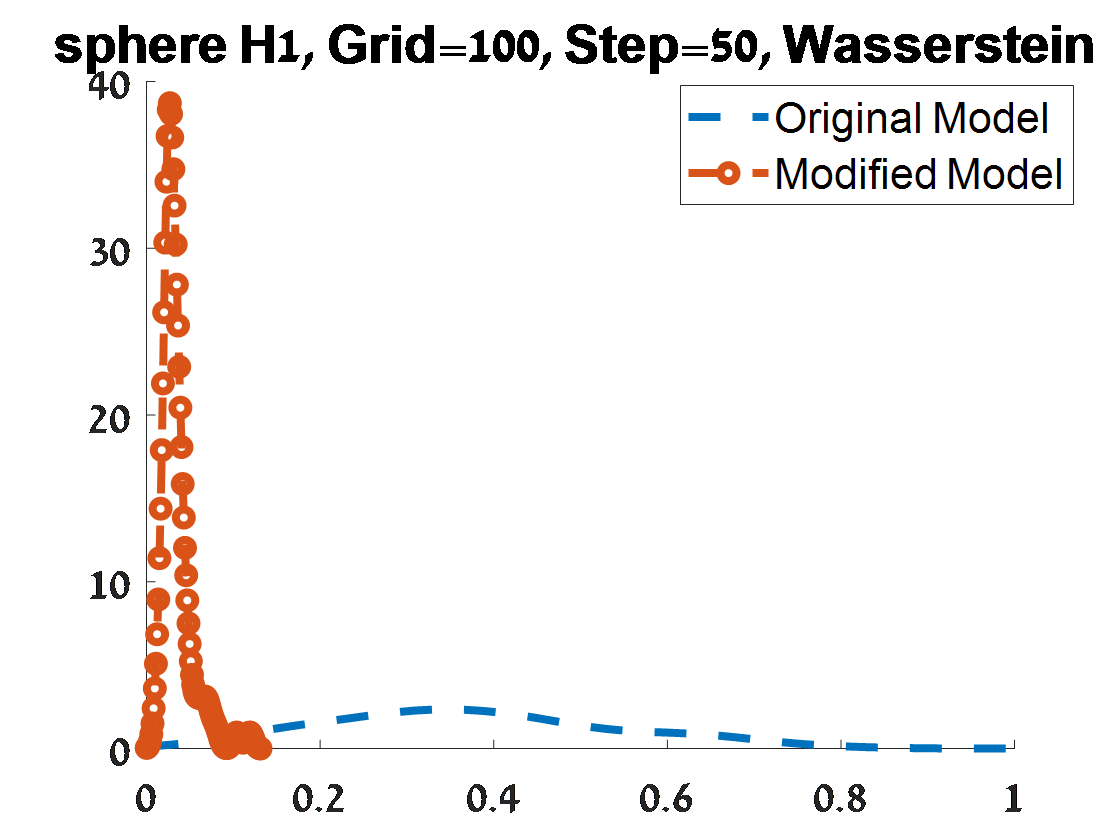}
\\
\includegraphics[width=1.2in, height=1.4in]{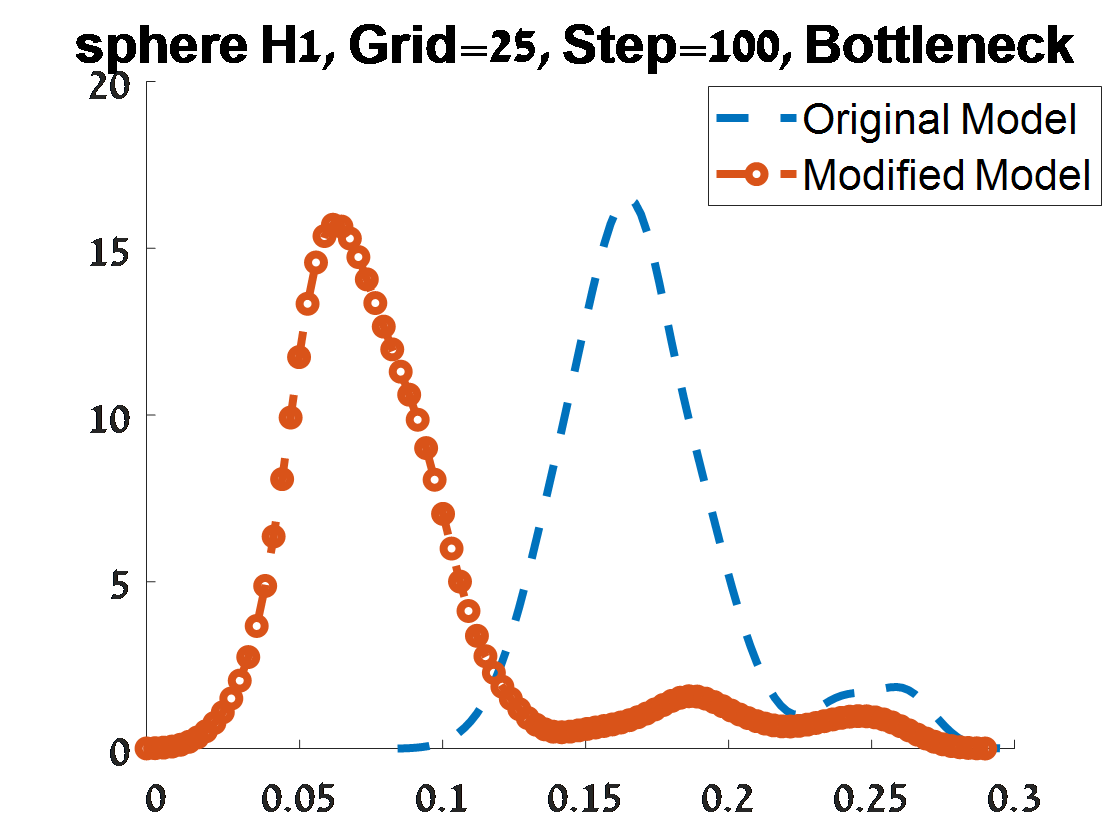}
\includegraphics[width=1.2in, height=1.4in]{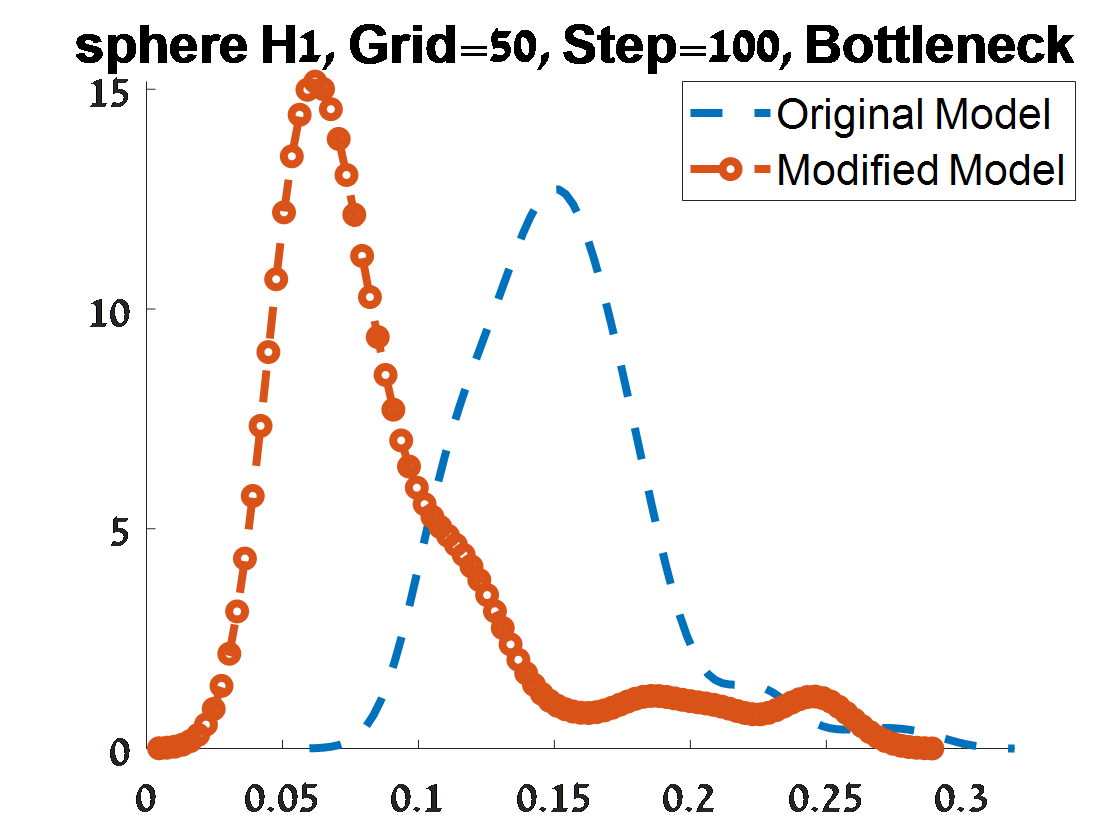}
\includegraphics[width=1.2in, height=1.4in]{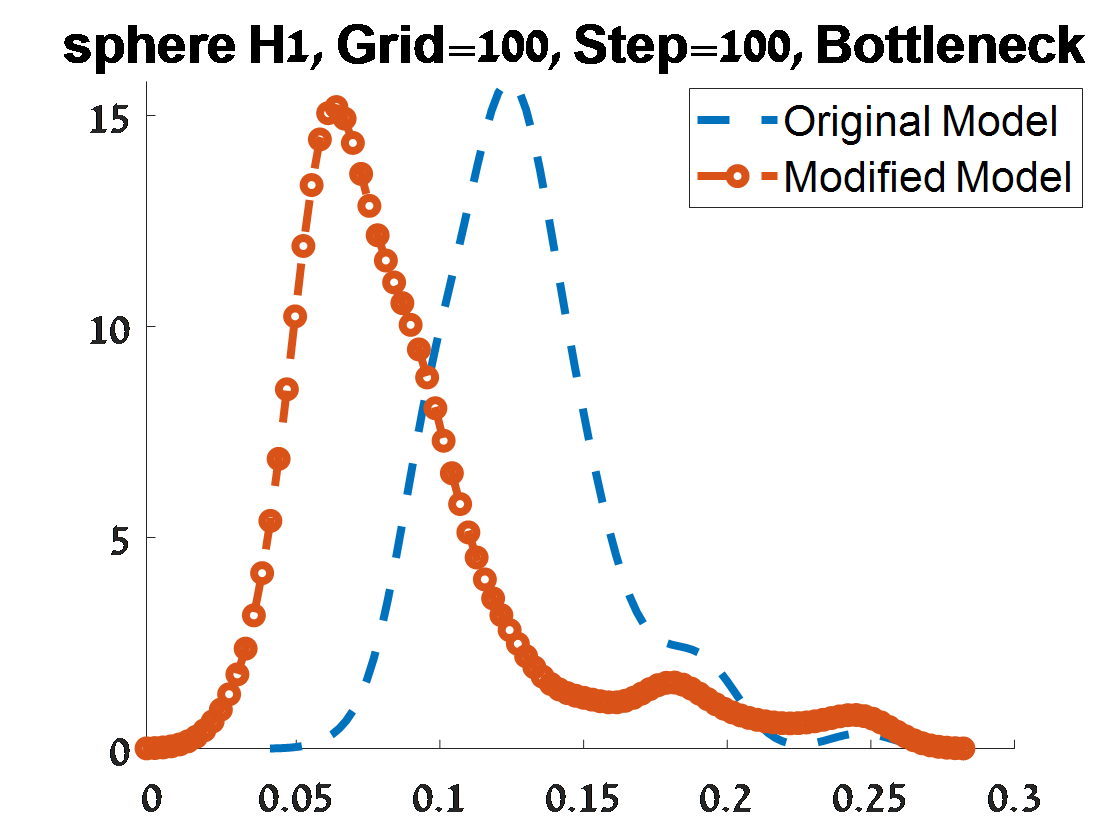}
\includegraphics[width=1.2in, height=1.4in]{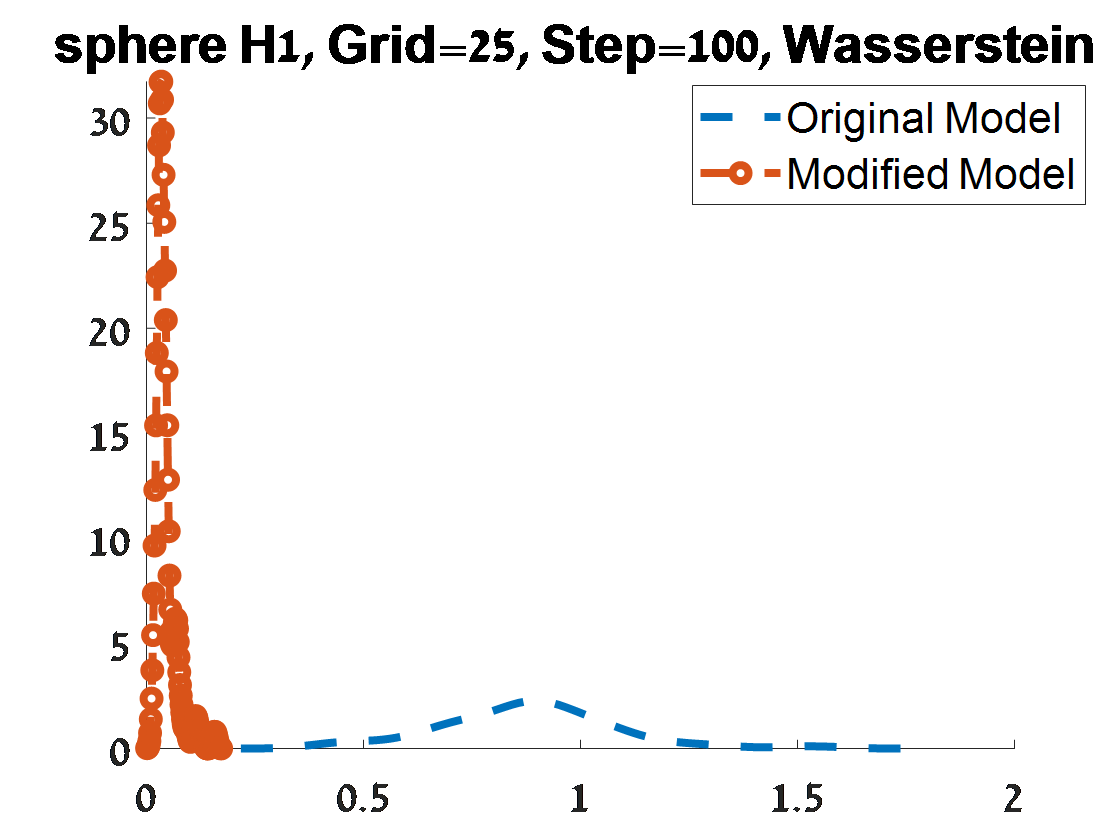}
\includegraphics[width=1.2in, height=1.4in]{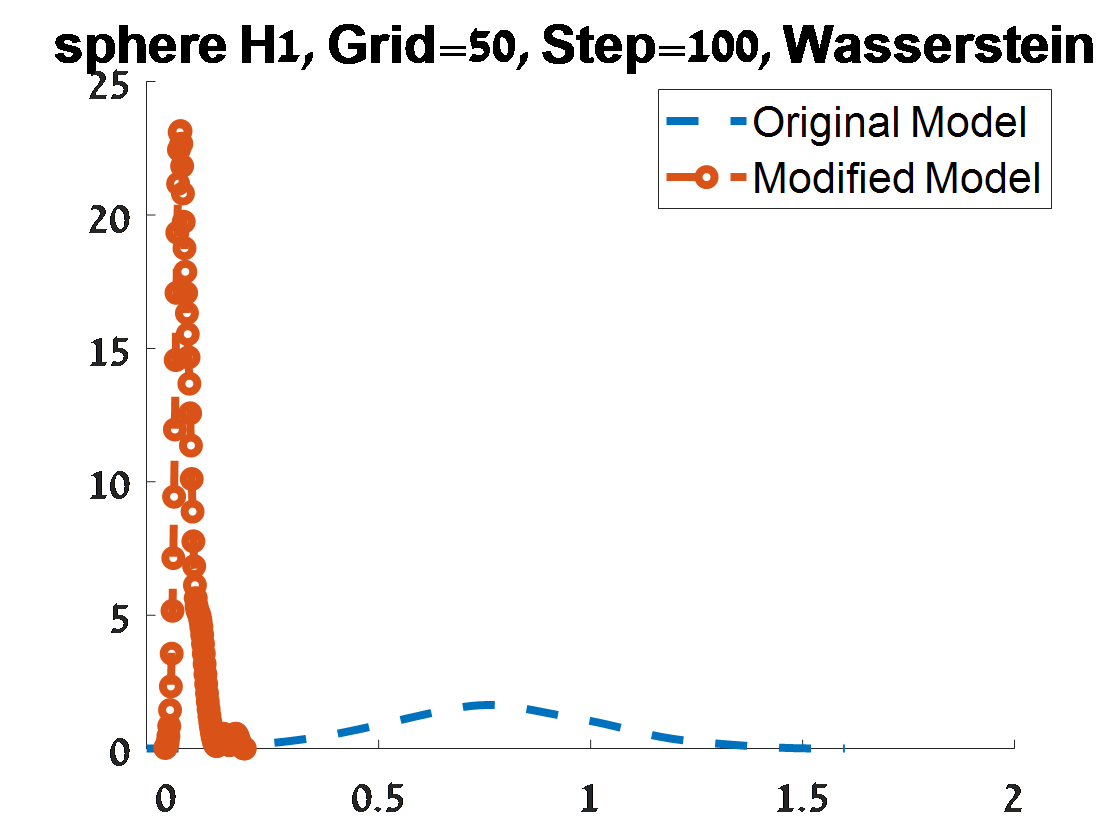}
\includegraphics[width=1.2in, height=1.4in]{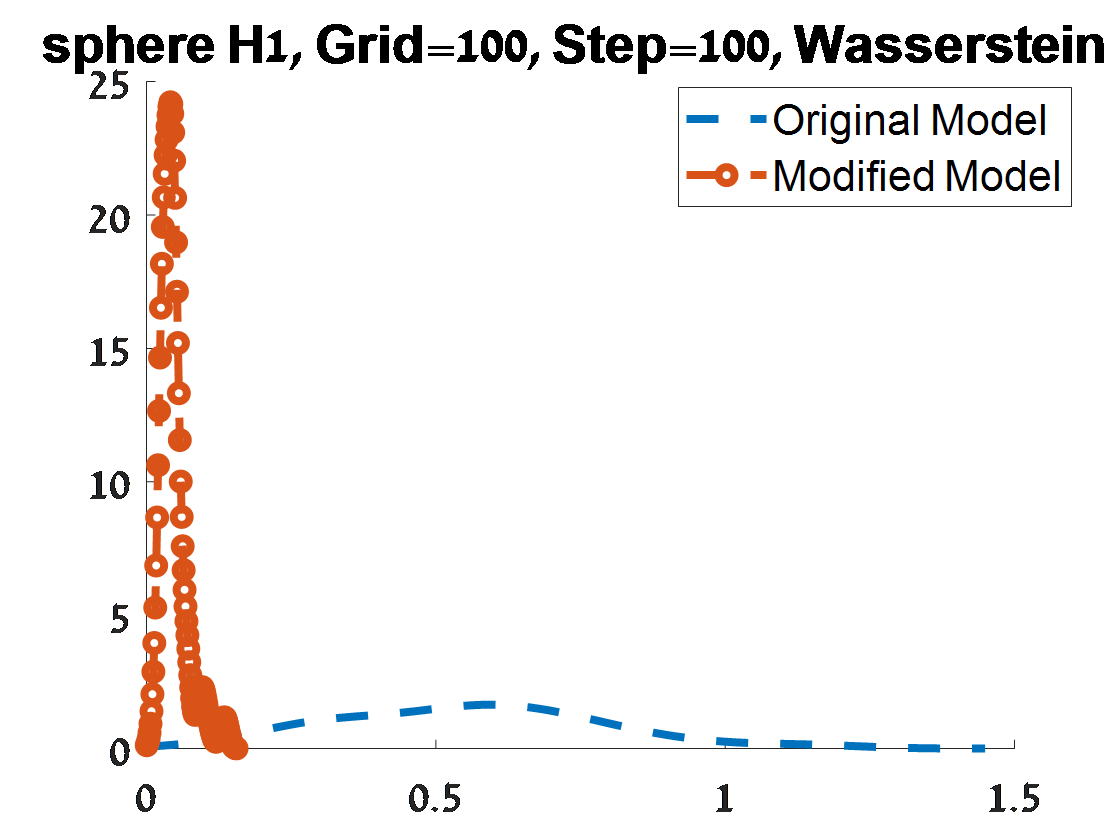}
\ec
%\caption{\footnotesize
% A random sample from two circles, 500 points from the larger circle and 300 from the smaller one,  with a kernel density
\caption{\footnotesize
Criterion 1 of goodness of fit for 100 $H_1$ PDs corresponded to 100 samples from a unit $S^2$. The figures depend on the grid of the proposal distribution ("Grid"), and the burn-in ("Step") of the MCMC algorithm.
 }
\label{fig:s2_H1_a}
\end{figure}
\end{landscape}

Figure\ \ref{fig:s2_H1_a} describes the distributions over the 100 $H_1$ PDs of the first criterion of goodness of fit, and Figures 18-19 describe the distributions of the second criterion of goodness of fit.
For criterion 1, we have, as in setting of $H_0$ PDs, that the modified model is better than the original (that is, smaller distance of the PDs under the modified model from the real PDs relative to that distance based on the original model). But, for the $H_1$ PDs, the advantage of the modified model is less extreme relative to the setting of $H_0$ PDs. That is, the distances distributions of the modified simulated PDs and those of the original simulated PDs are relative close in the setting of $H_1$ PDs comparing to these distributions in the setting of $H_0$ PDs.
In the same way, we have in criterion 2 that the distributions of the first, second, and third distances, in the real, original model, and modified model, are much similar comparing with these distributions under $H_0$ PDs. Accordingly, the closeness of these distributions under the modified model to these under the real PDs is better for all considered values of burn-in and grid size in $H_1$ than in $H_0$.
The reason for these different results in $H_1$ comparing to $H_0$ is the larger variability of the $H_0$ PD points relative to that of $H_1$ PD points, see for example in Figure\ \ref{fig:sphere_s2}.
Note that according to criterion 2, the modified model for a given grid size  is better under burn-in of 25, and the fitting is better in grid size of 100x100 for a given burn-in.
%Figure\ \ref{fig:s2_H1_d} presents two examples of real PD and its simulated PD based on the two model versions, only for the best scenarios we found, that is grid=50,100, and step=25. The both model succeeds in creating the original PD, but the modified model can also generate the outliers, i.e. extreme values from the most points.

\begin{landscape}
\begin{figure}[h!]
\bc
\includegraphics[width=1.2in, height=1.25in]{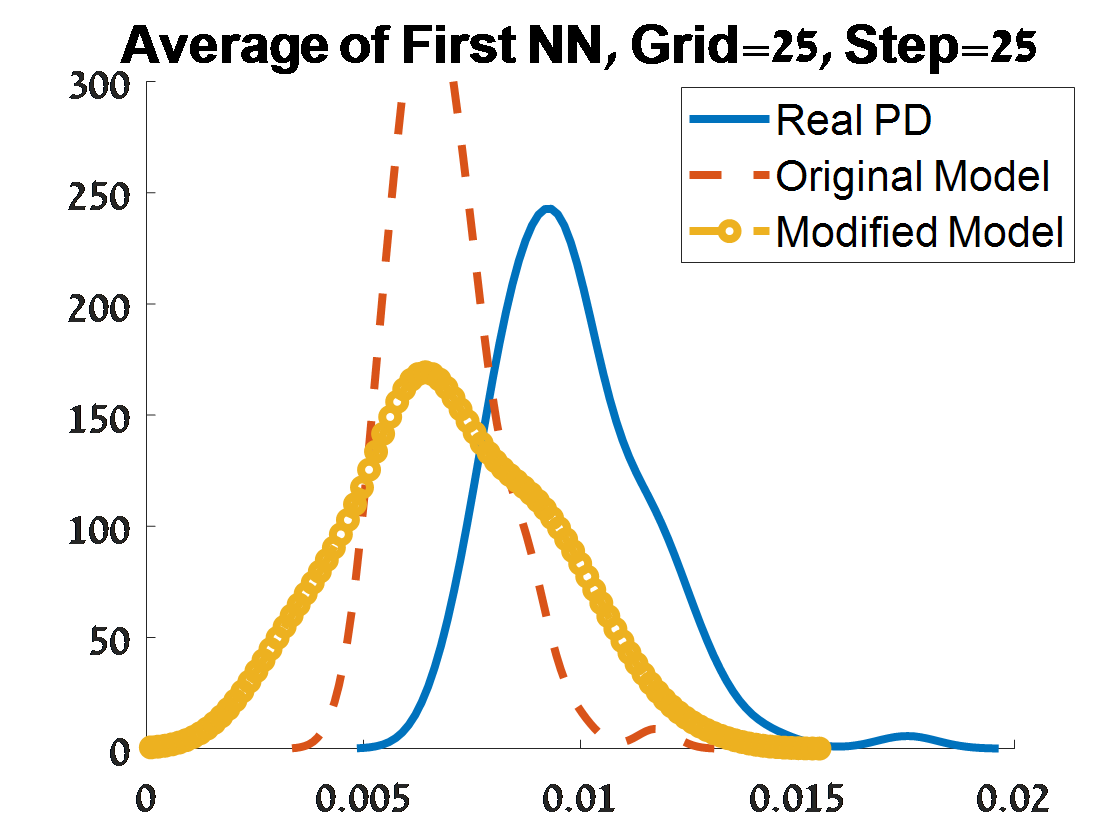}
\includegraphics[width=1.2in, height=1.25in]{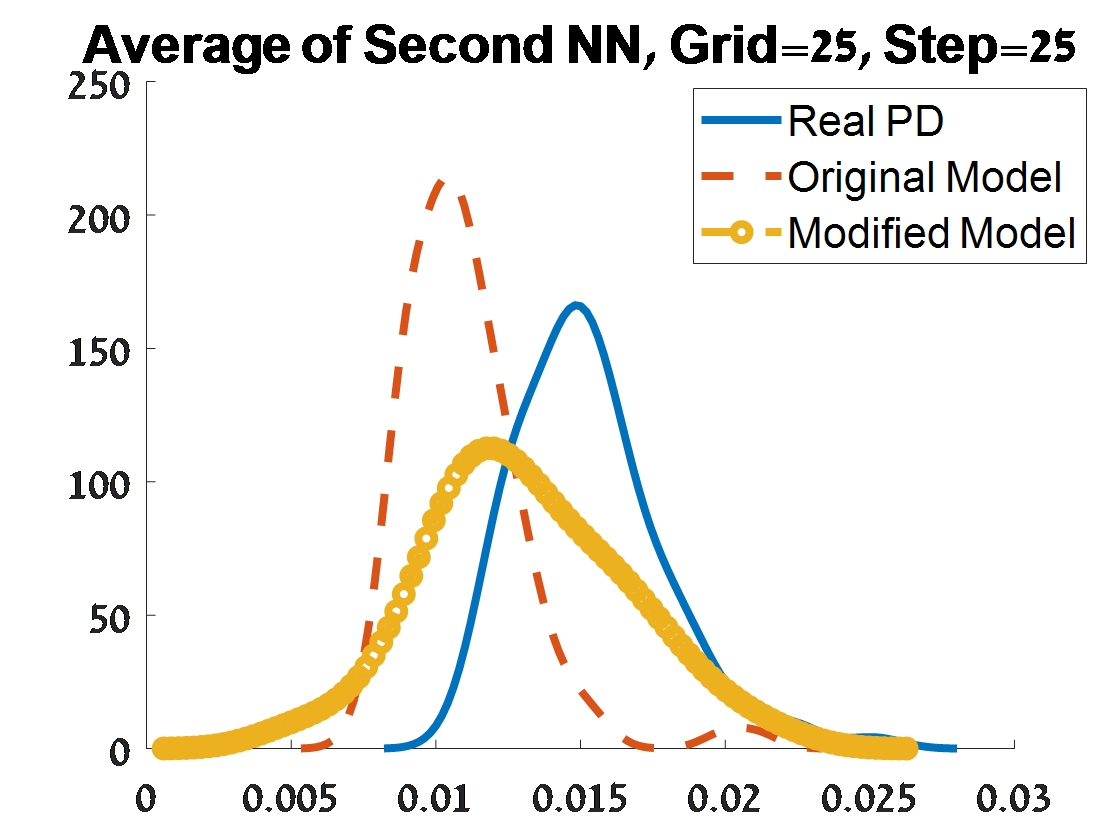}
\includegraphics[width=1.2in, height=1.25in]{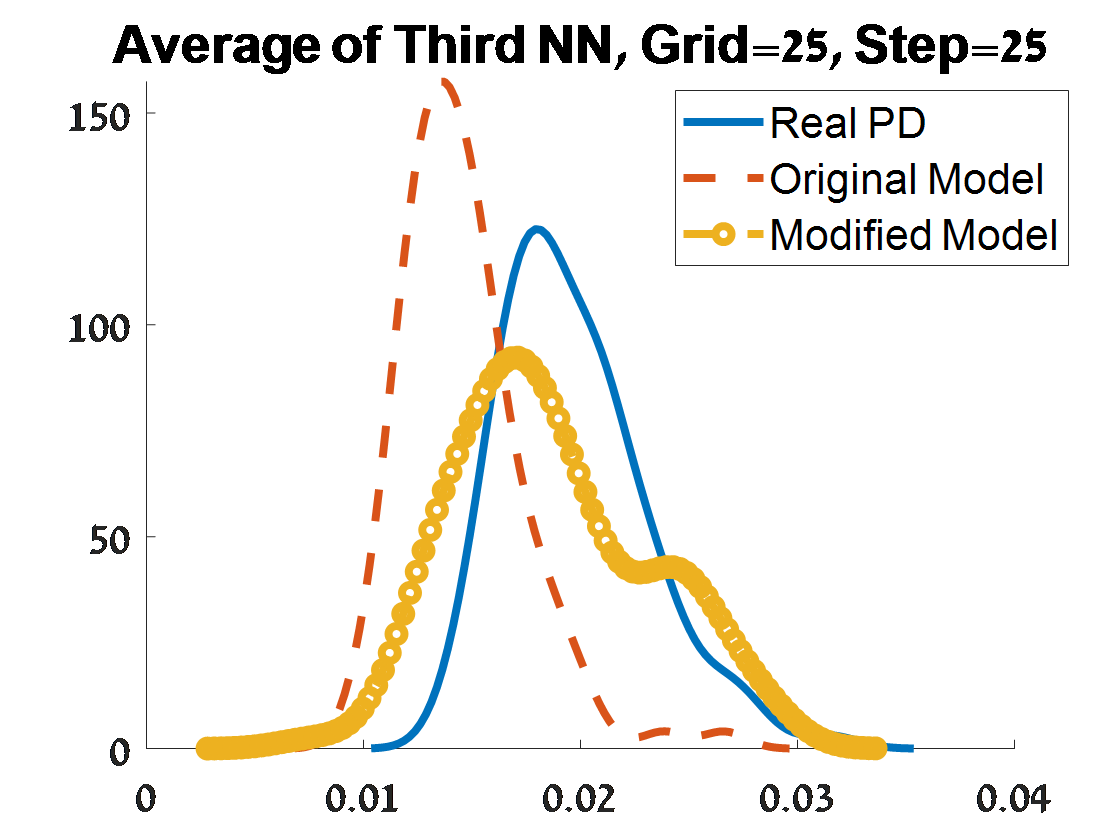}
\includegraphics[width=1.2in, height=1.25in]{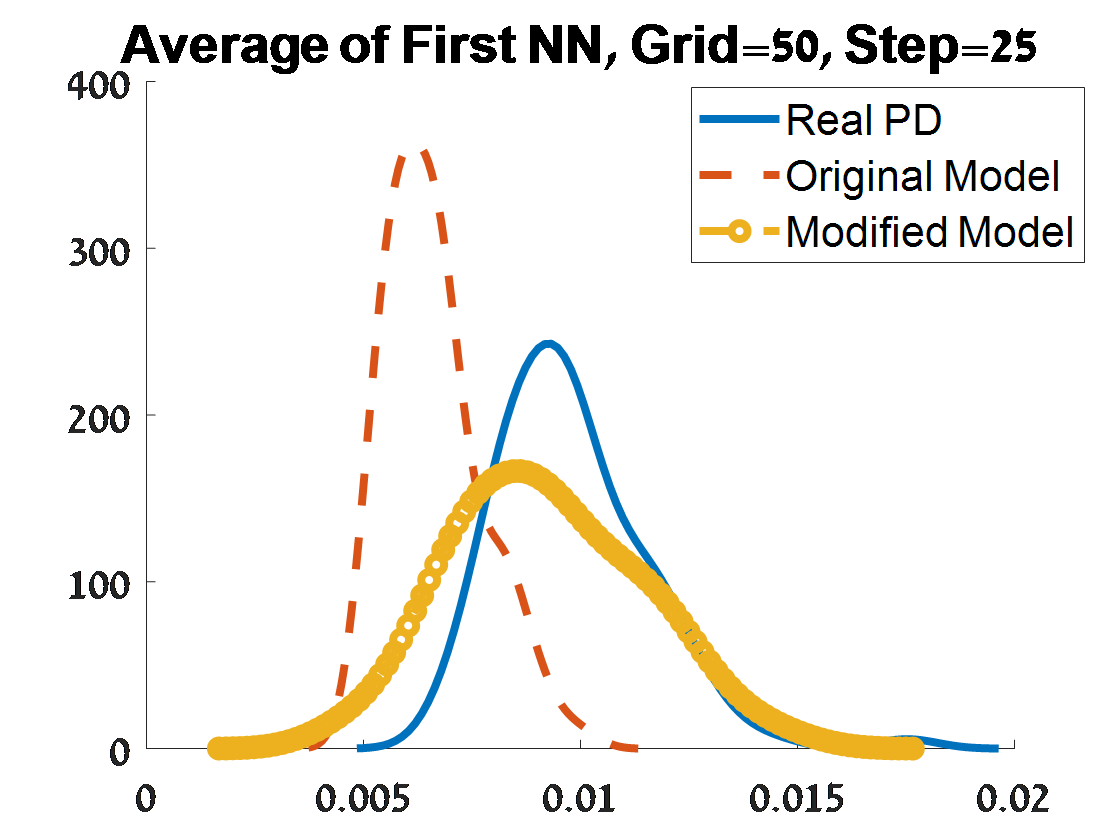}
\includegraphics[width=1.2in, height=1.25in]{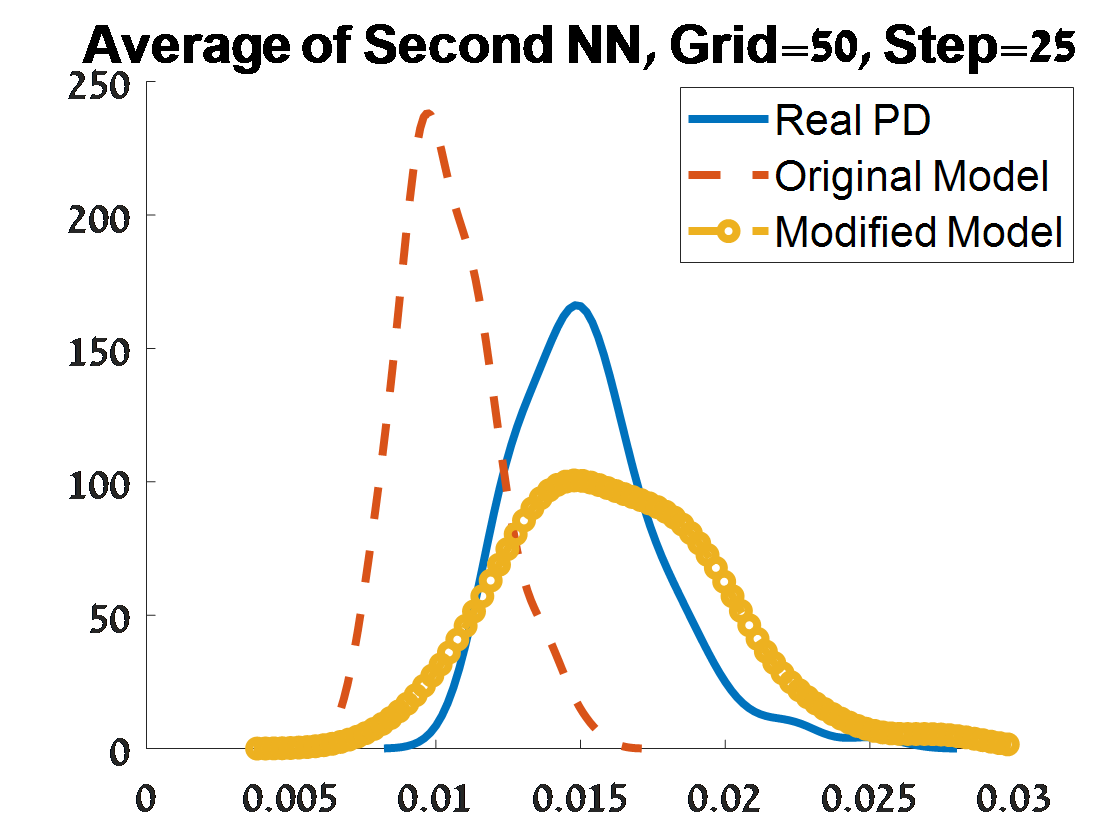}
\includegraphics[width=1.2in, height=1.25in]{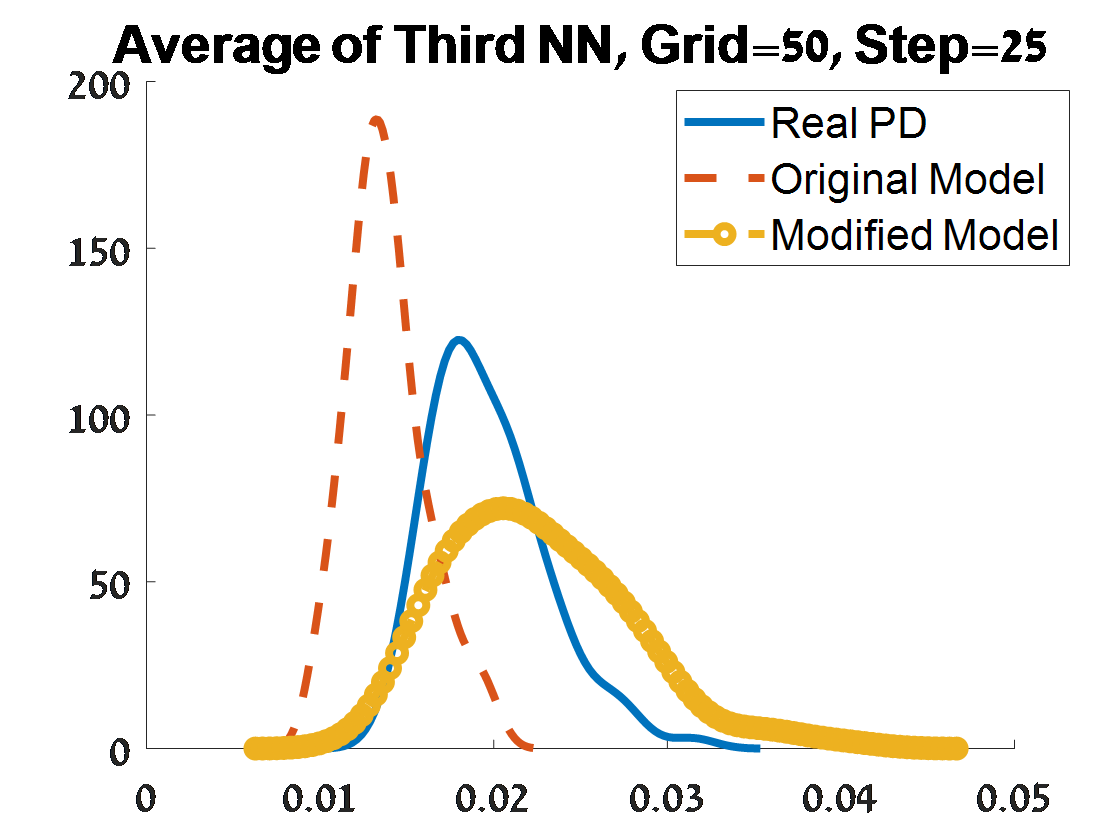}
\includegraphics[width=1.2in, height=1.25in]{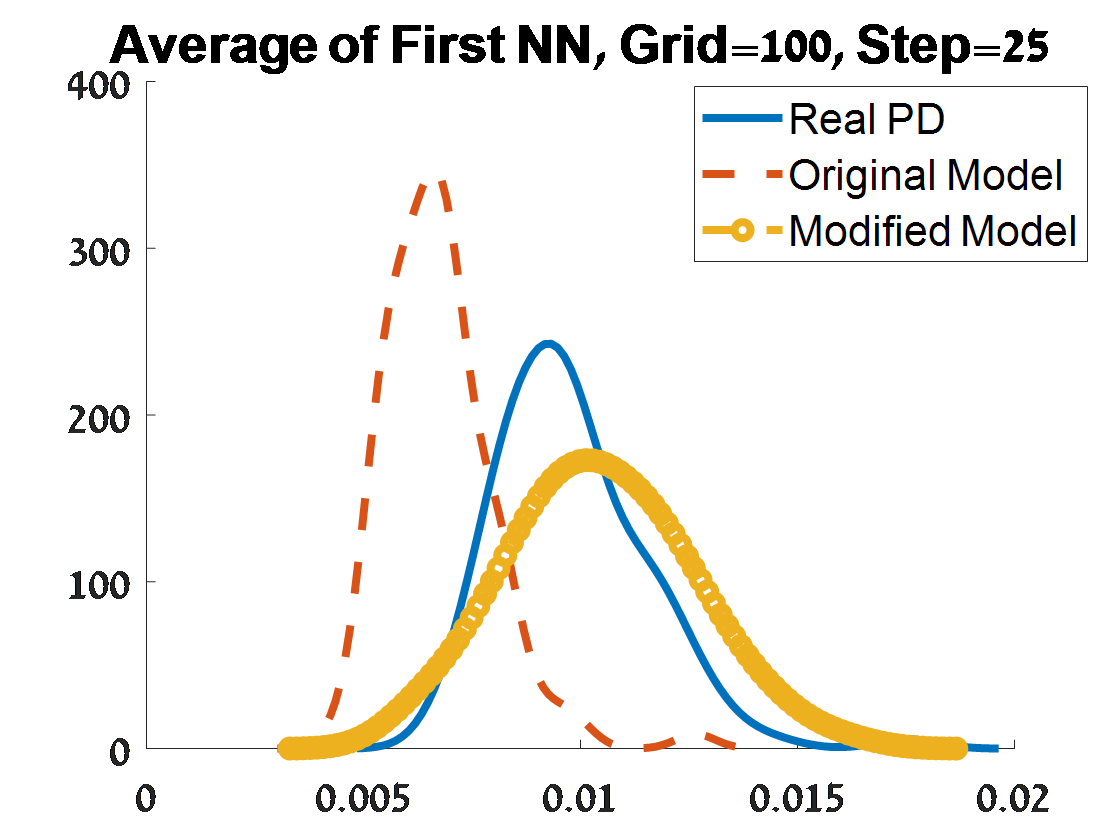}
\includegraphics[width=1.2in, height=1.25in]{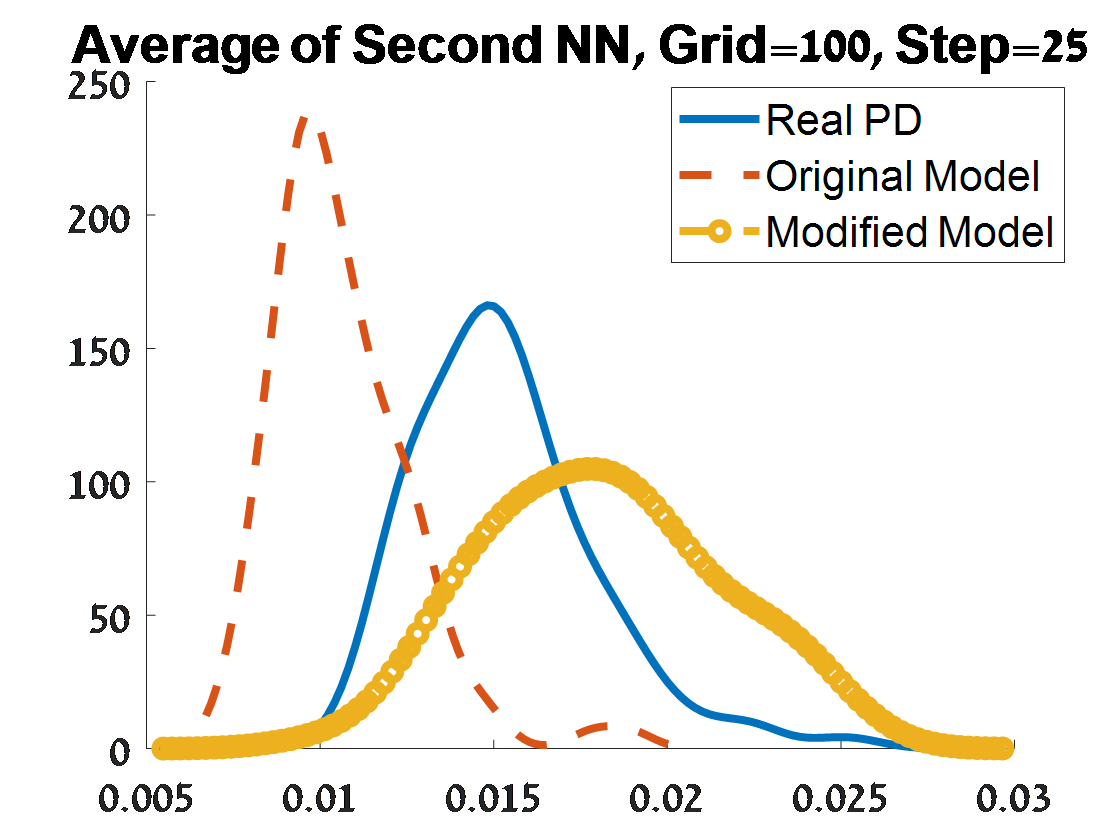}
\includegraphics[width=1.2in, height=1.25in]{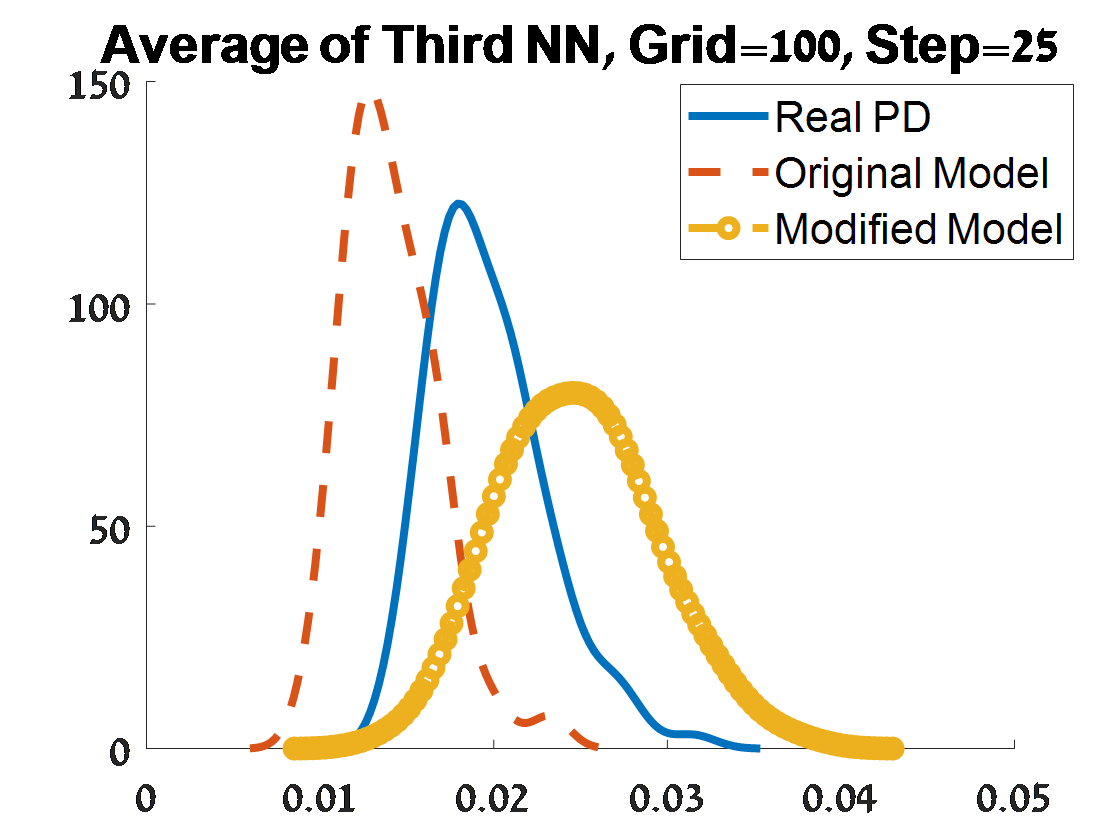}
\\
\includegraphics[width=1.2in, height=1.25in]{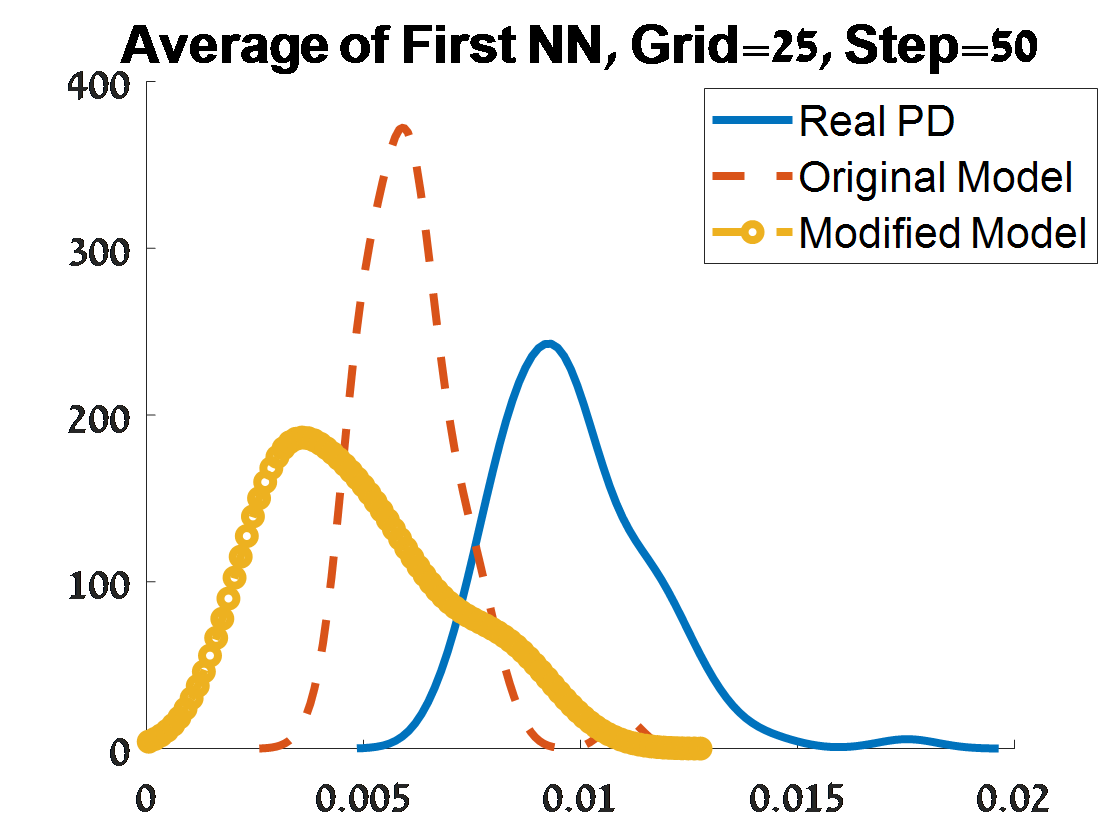}
\includegraphics[width=1.2in, height=1.25in]{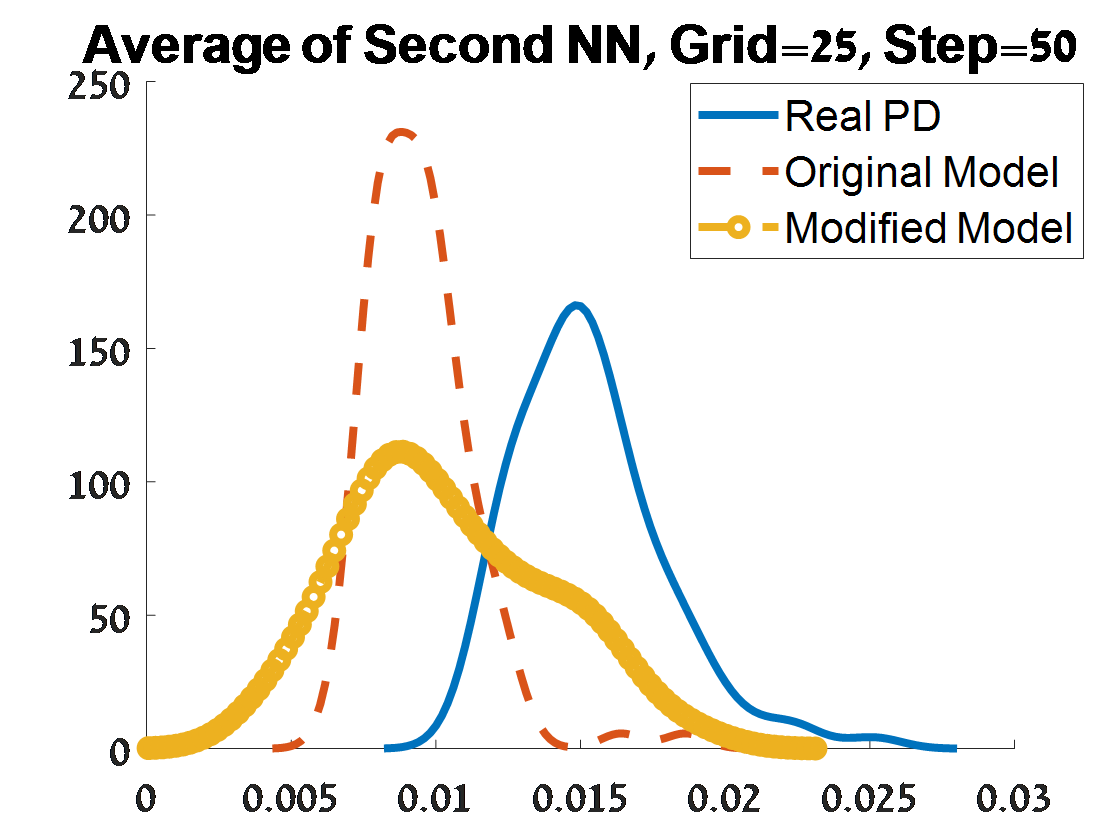}
\includegraphics[width=1.2in, height=1.25in]{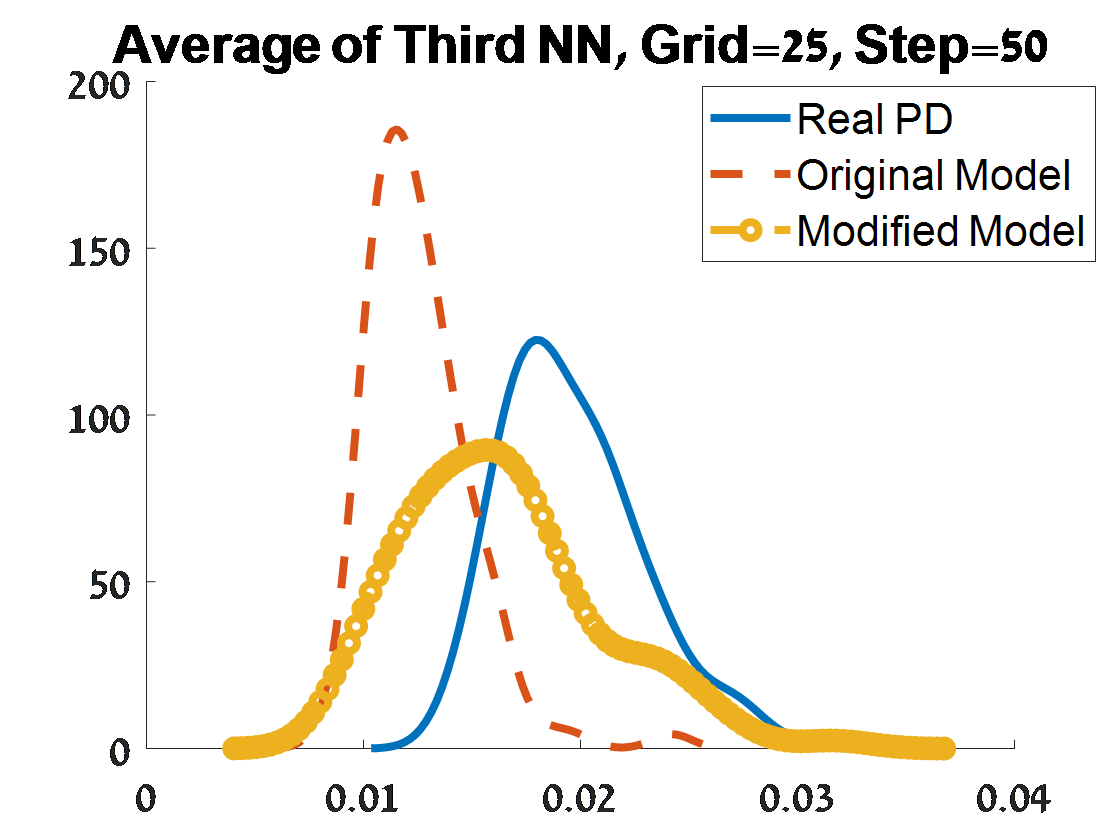}
\includegraphics[width=1.2in, height=1.25in]{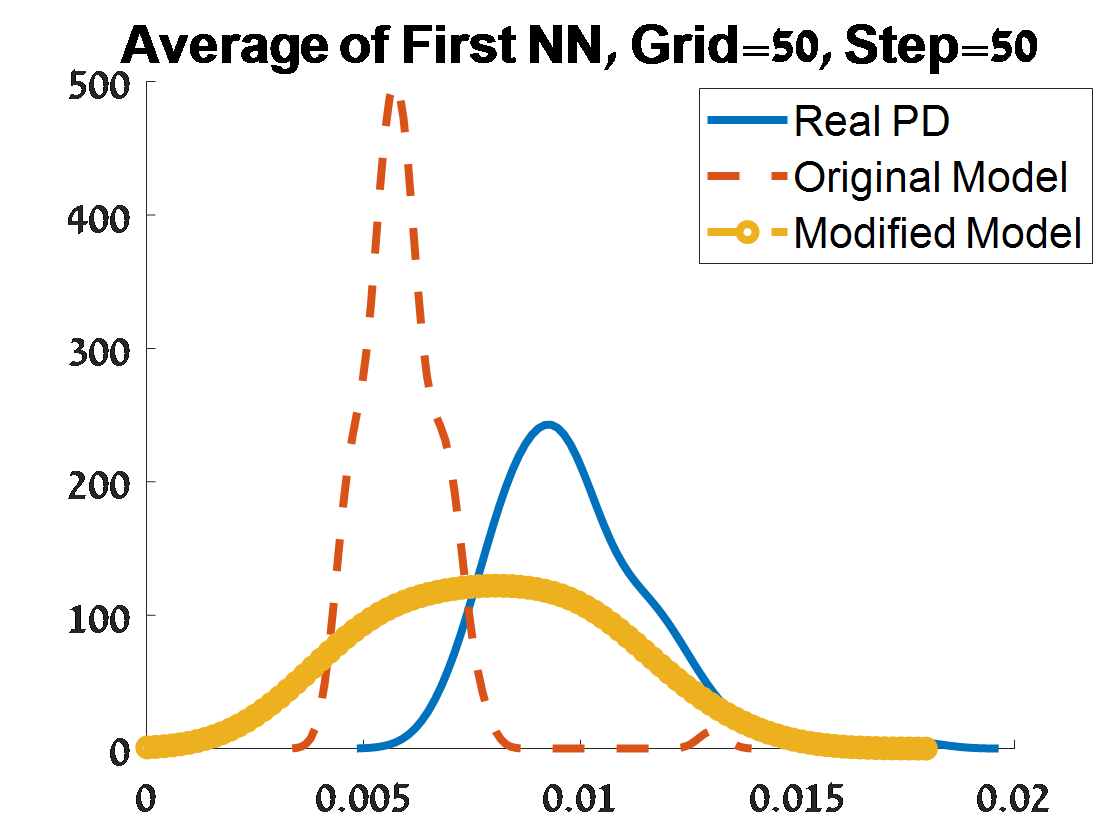}
\includegraphics[width=1.2in, height=1.25in]{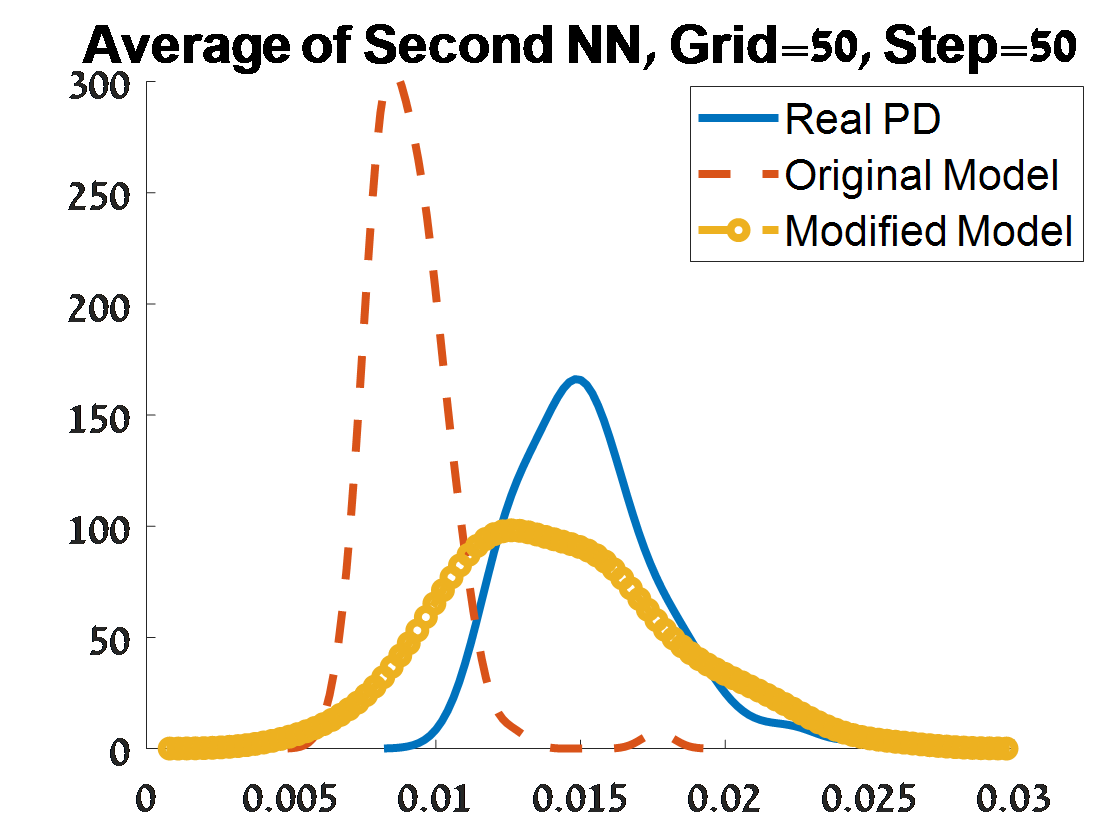}
\includegraphics[width=1.2in, height=1.25in]{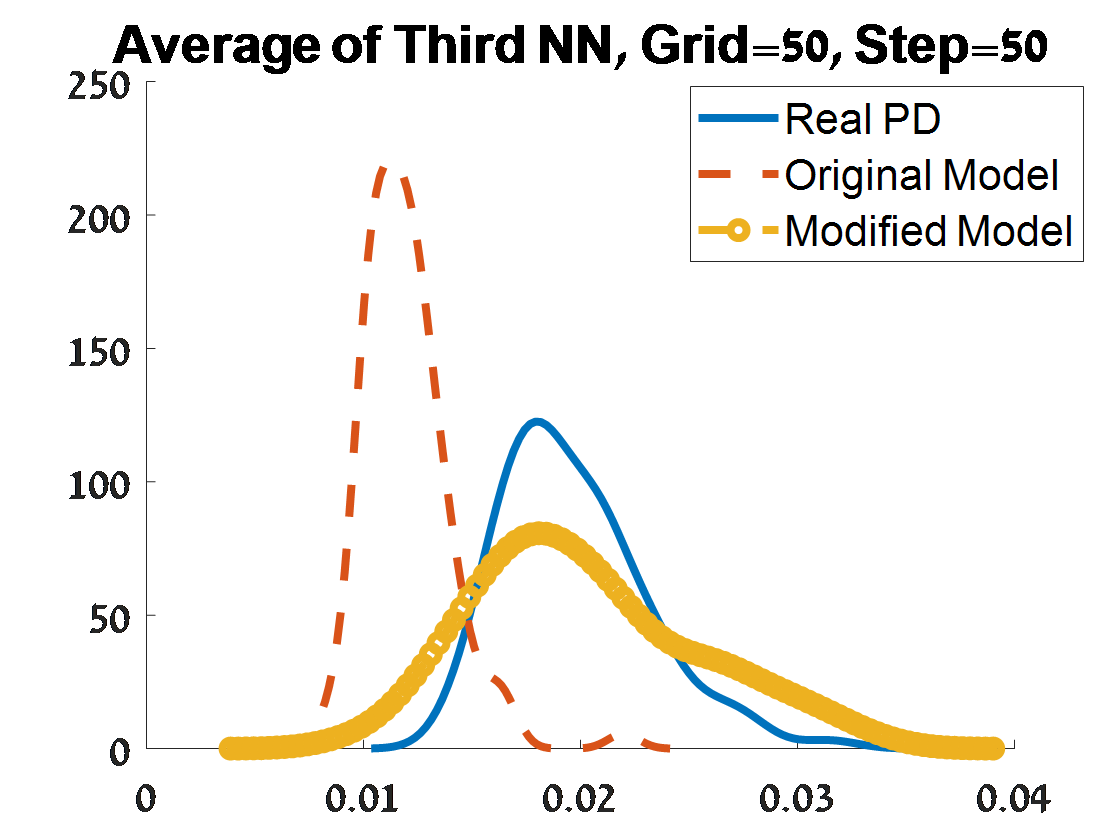}
\includegraphics[width=1.2in, height=1.25in]{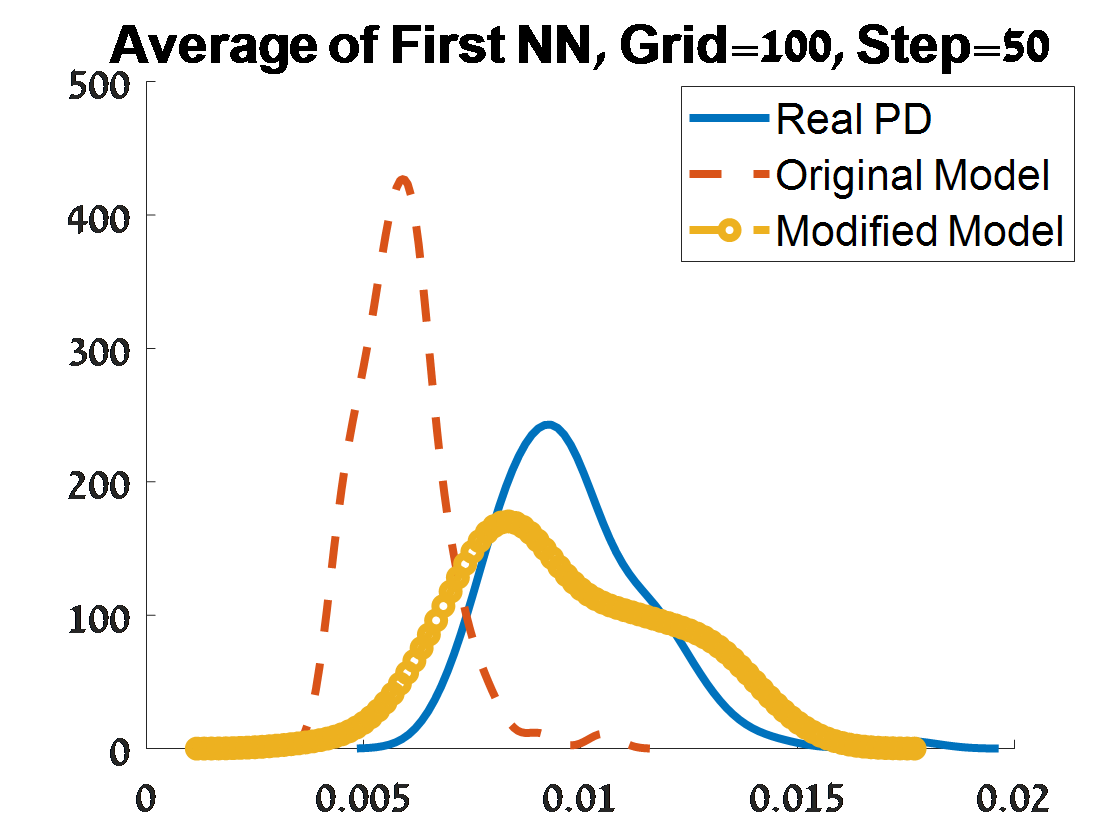}
\includegraphics[width=1.2in, height=1.25in]{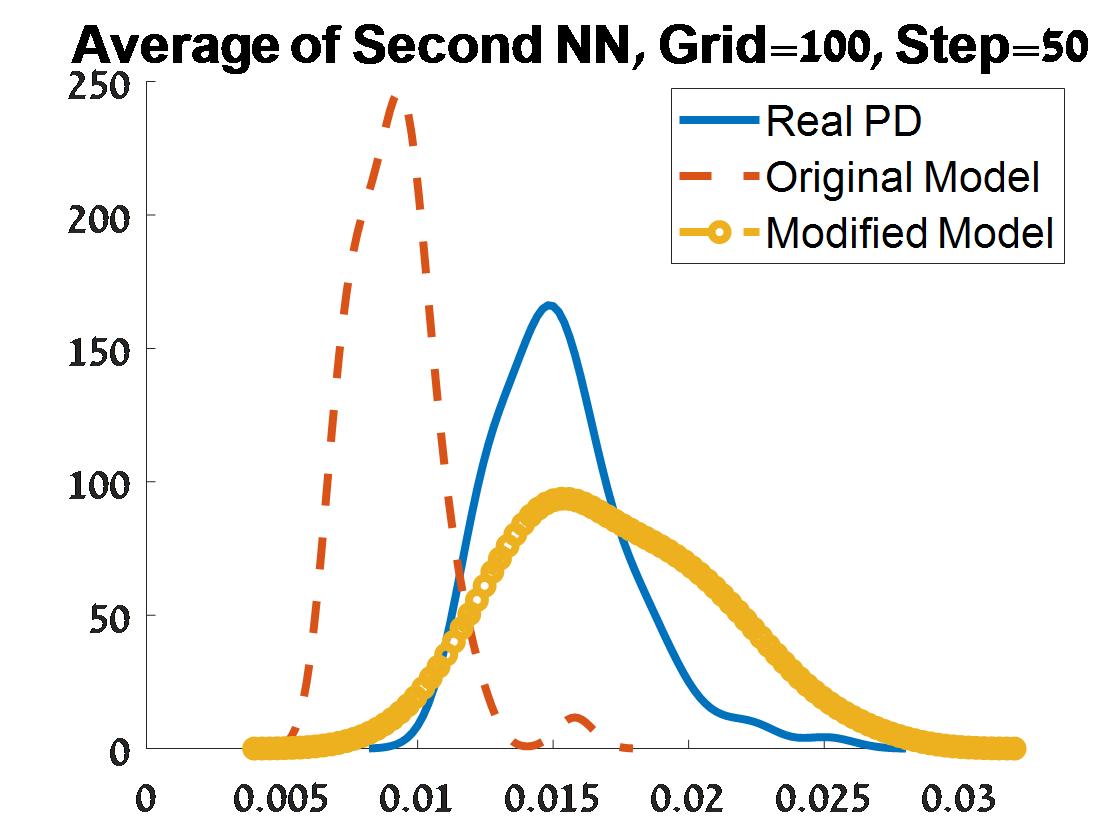}
\includegraphics[width=1.2in, height=1.25in]{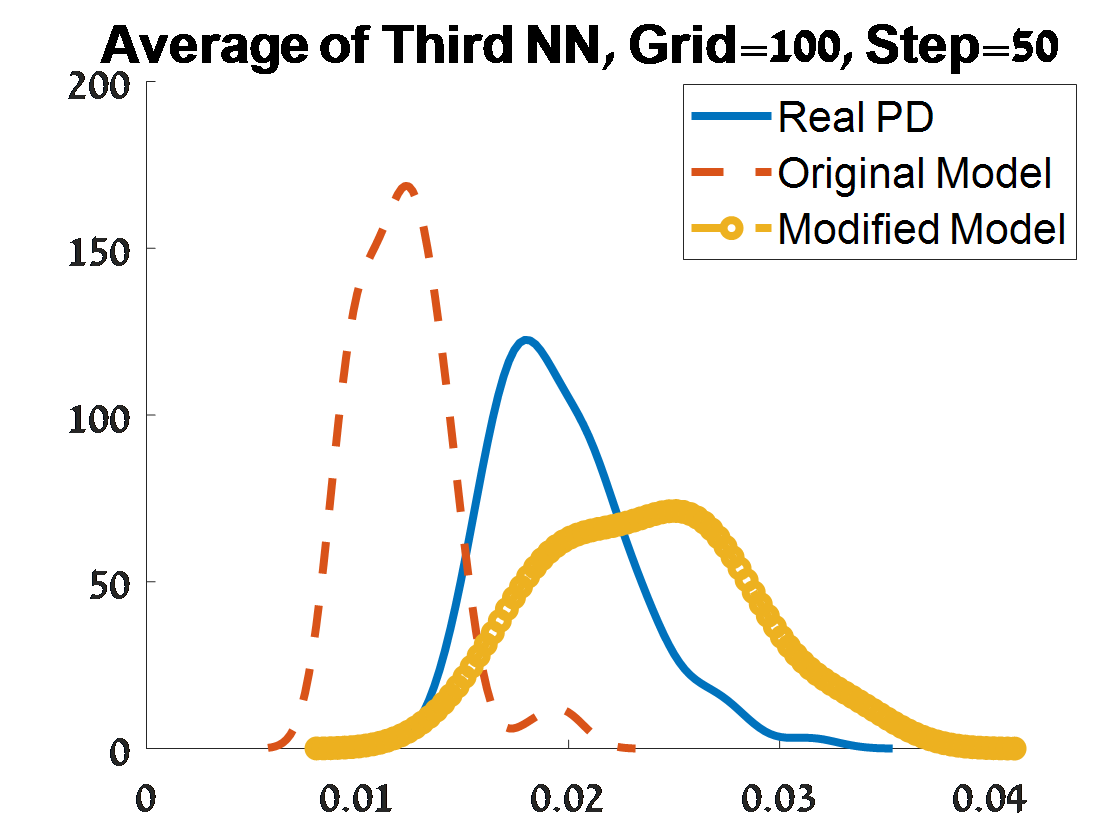}

\ec
%\caption{\footnotesize
% A random sample from two circles, 500 points from the larger circle and 300 from the smaller one,  with a kernel density
\caption{\footnotesize
 Criterion 2 of goodness of fit for 100 $H_1$ PDs corresponded to 100 samples from a unit $S^2$. The figures depend on the grid of the proposal distribution ("Grid"), and the burn-in ("Step") of the MCMC algorithm.}
\label{fig:s2_H1_b}
\end{figure}
\end{landscape}

\begin{landscape}
\begin{figure}[h!]
\bc
\includegraphics[width=1.2in, height=1.25in]{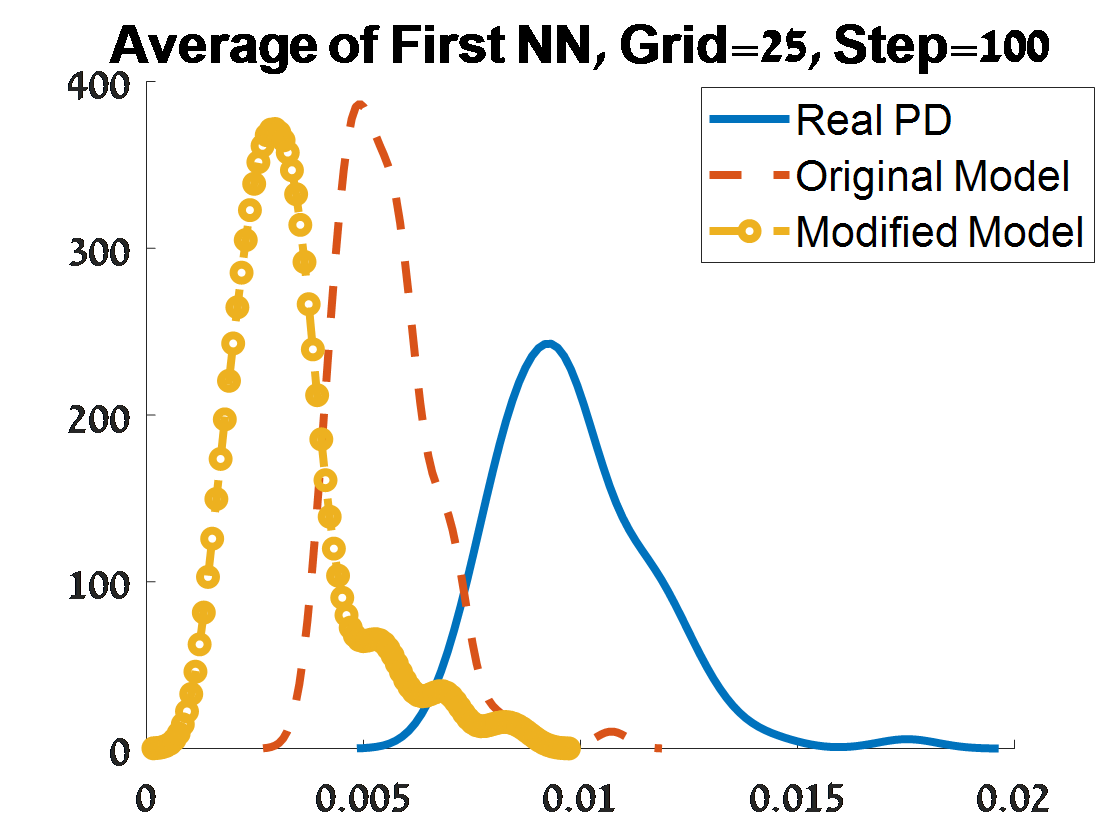}
\includegraphics[width=1.2in, height=1.25in]{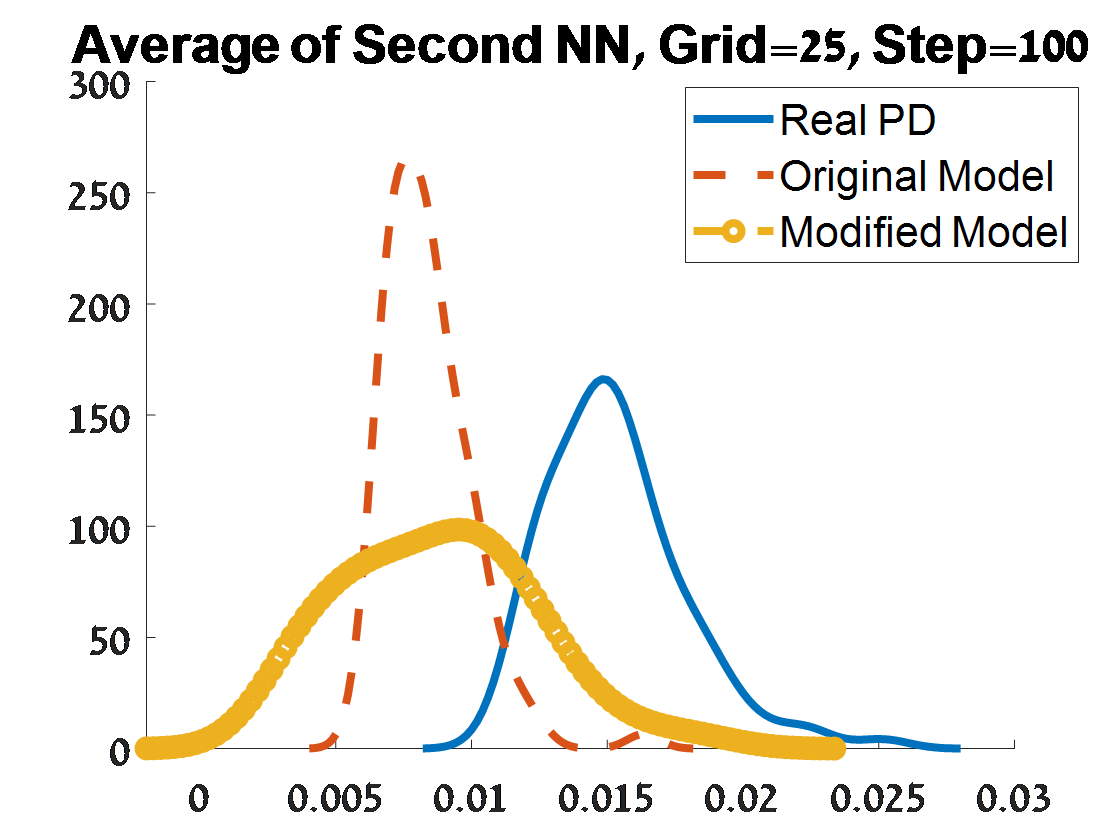}
\includegraphics[width=1.2in, height=1.25in]{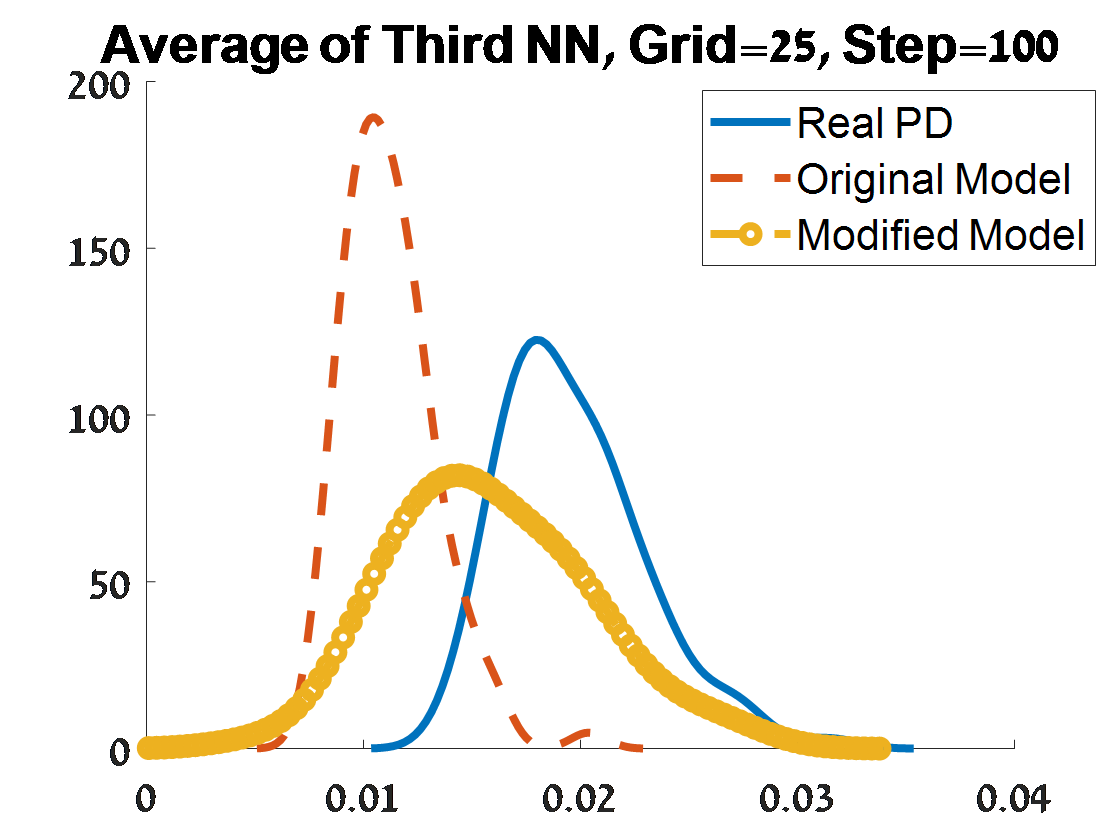}
\includegraphics[width=1.2in, height=1.25in]{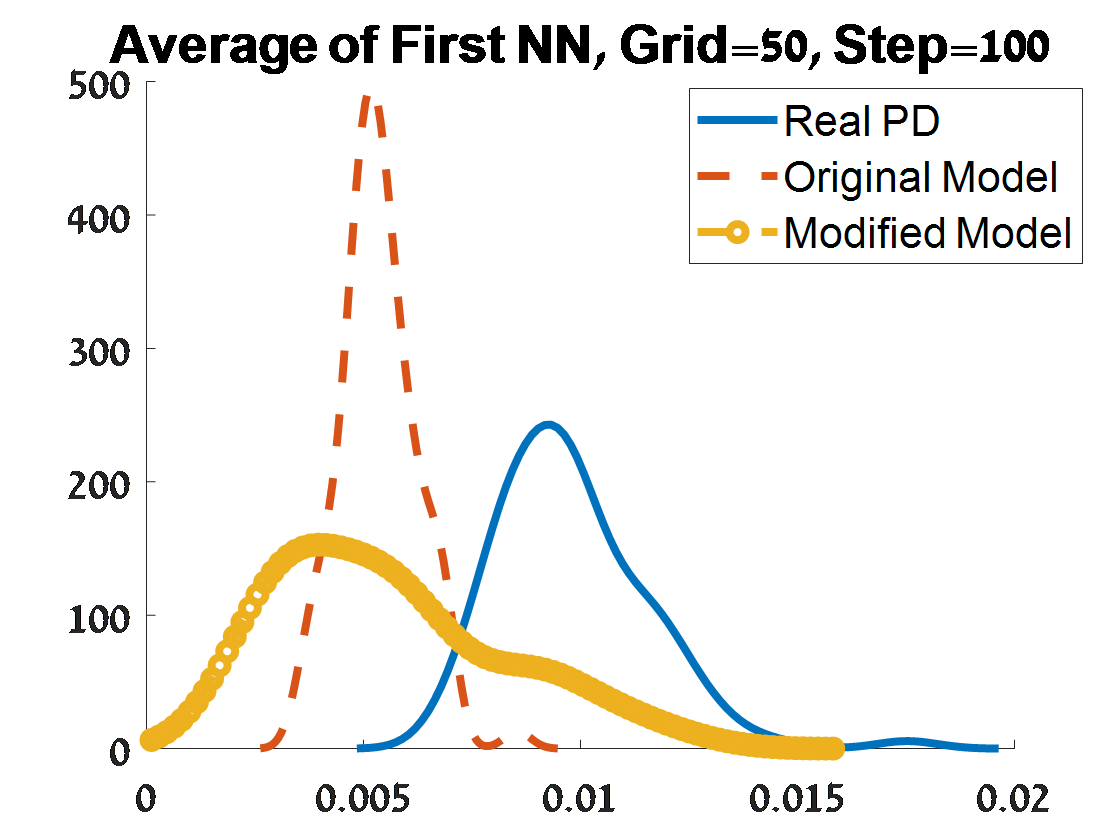}
\includegraphics[width=1.2in, height=1.25in]{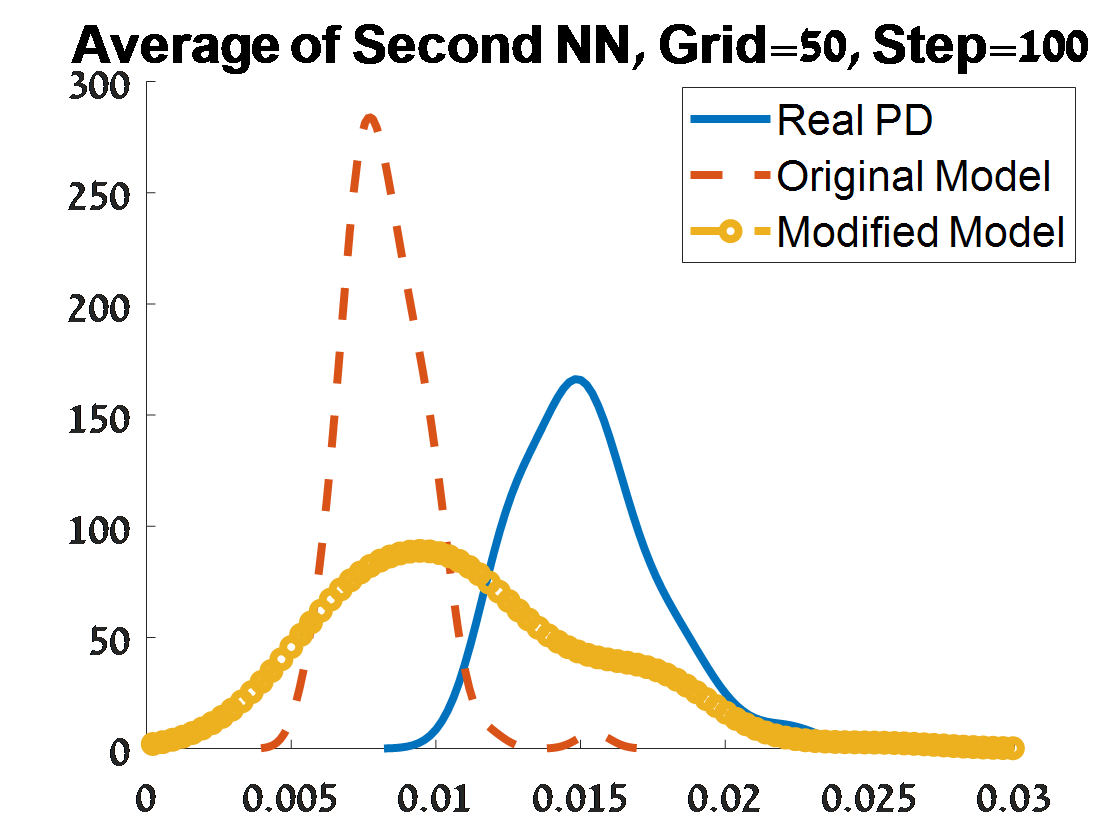}
\includegraphics[width=1.2in, height=1.25in]{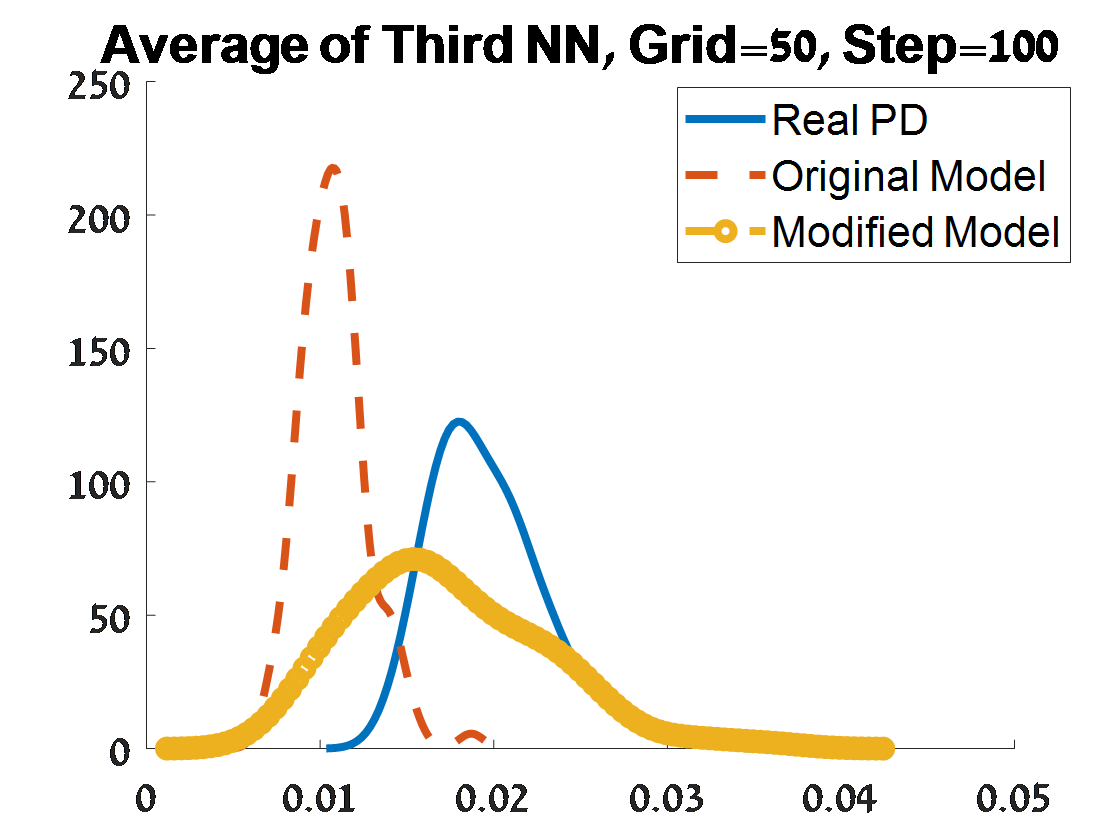}
\includegraphics[width=1.2in, height=1.25in]{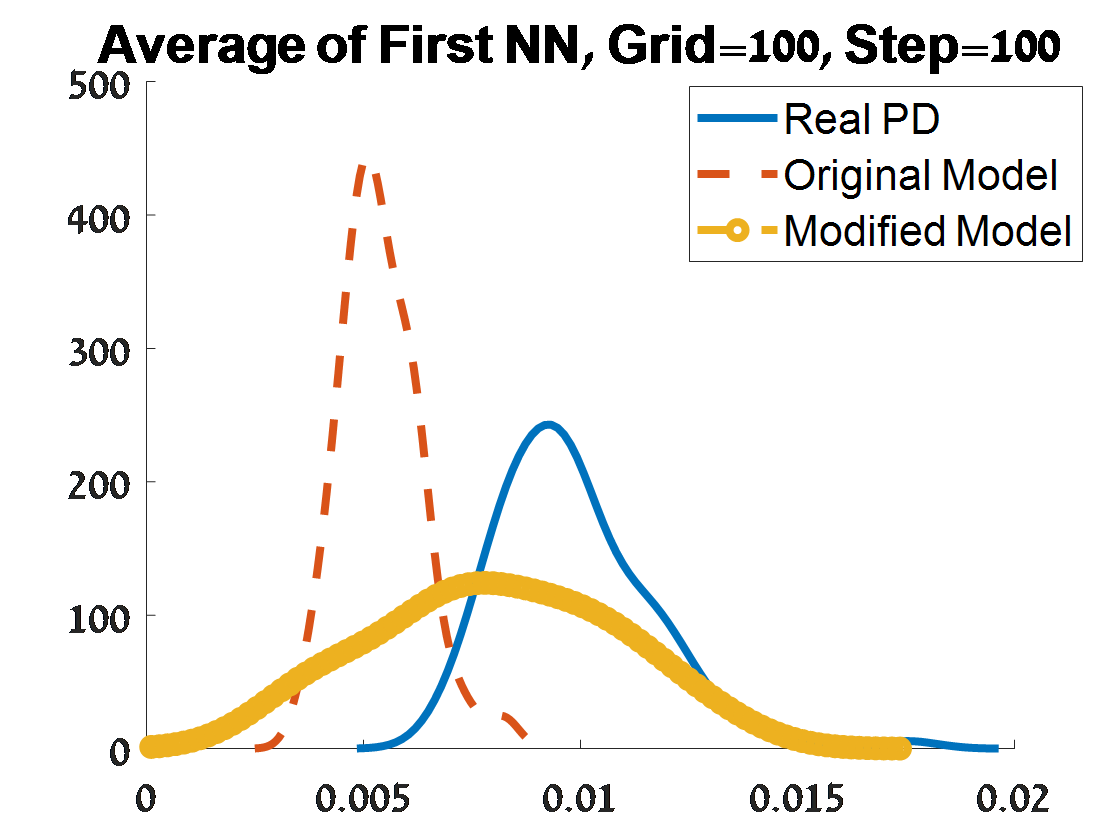}
\includegraphics[width=1.2in, height=1.25in]{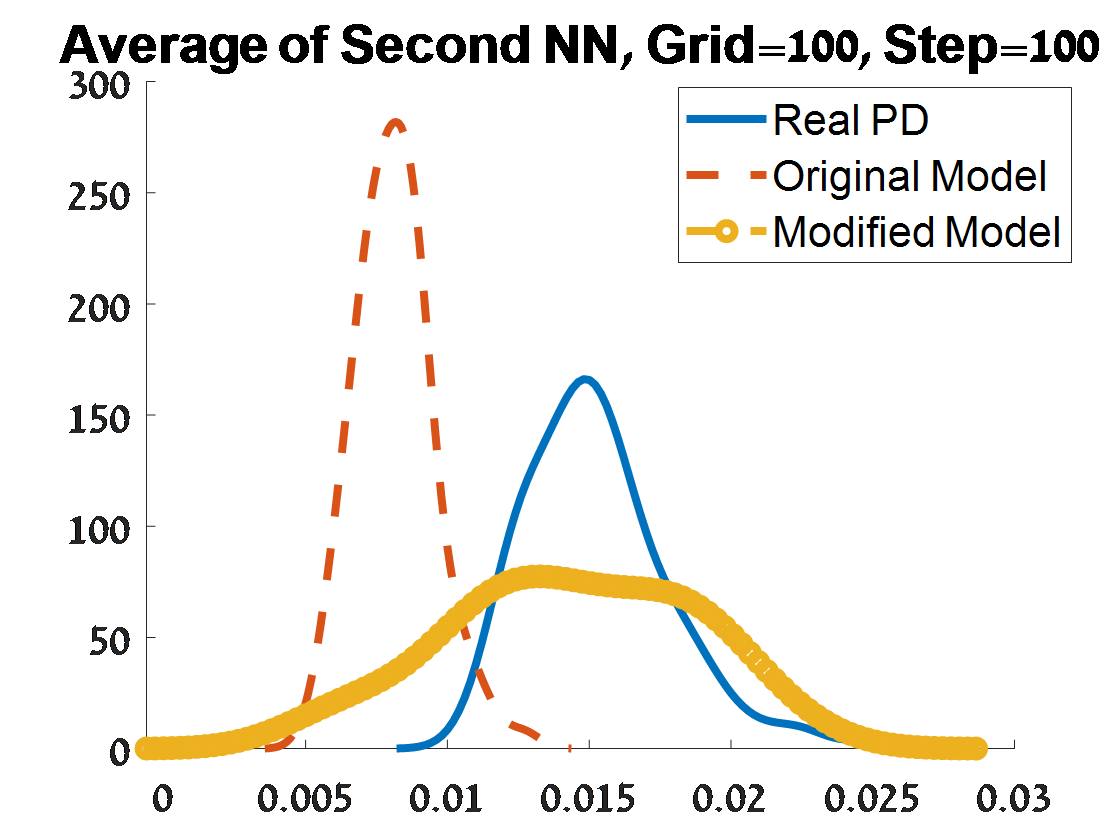}
\includegraphics[width=1.2in, height=1.25in]{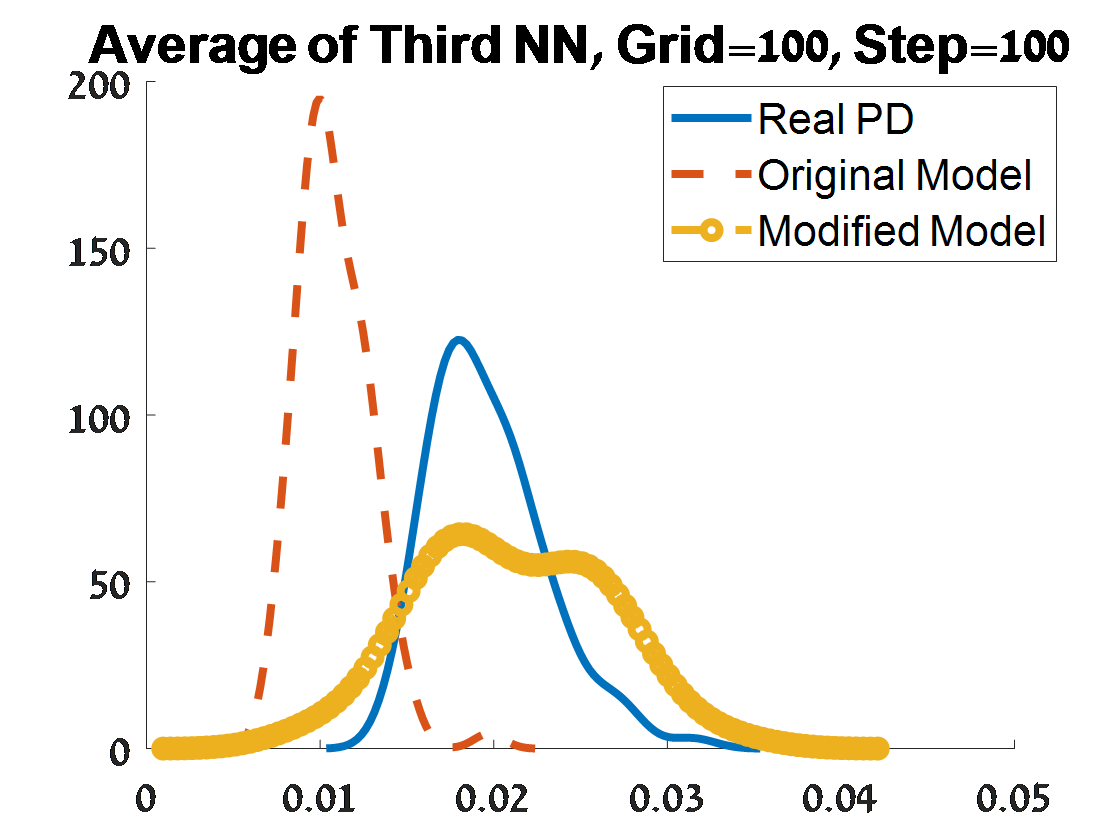}
\ec
%\caption{\footnotesize
% A random sample from two circles, 500 points from the larger circle and 300 from the smaller one,  with a kernel density
\caption{\footnotesize
Continue of Criterion 2 of goodness of fit for $H_1$ 100 PDs corresponded to 100 samples from a unit $S^2$. The figures depend on the grid of the proposal distribution ("Grid"), and the burn-in ("Step") of the MCMC algorithm.}
\label{fig:s2_H1_c}
\end{figure}
\end{landscape}

%\begin{landscape}
%\begin{figure}[h!]
%\bc
%\includegraphics[width=1.8in, height=1.8in]{SphereH1_pd30_grid50_step25}
%\includegraphics[width=1.8in, height=1.8in]{SphereH1_pd30_grid100_step25}
%\includegraphics[width=1.8in, height=1.8in]{SphereH1_pd60_grid50_step25}
%\includegraphics[width=1.8in, height=1.8in]{SphereH1_pd60_grid100_step25}
%\ec
%%\caption{\footnotesize
%% A random sample from two circles, 500 points from the larger circle and 300 from the smaller one,  with a kernel density
%\caption{\footnotesize
%Examples of two PDs, each one is corresponded to a sample from a unit $S^2$. For each PD, the simulated PD based on the two model versions is described. The figures depend on the grid of the proposal distribution ("Grid"), and the burn-in ("Step") of the MCMC algorithm.}
%\label{fig:s2_H1_d}
%\end{figure}
%\end{landscape}
\subsection{3-Sphere ($S^3$)}
This example includes a random sample of $n=1,000$ points from the uniform distribution on the sphere $S^3$ in $R^4$ with radius $r=1$.
%The black circles indicating connected components ($H_0$ persistence), the red triangles corresponding to holes ($H_1$), the blue diamonds corresponding to voids ($H_2$), and the green points to $H_3$.

\begin{landscape}
\begin{figure}[h!]
\bc
\includegraphics[width=2.1in, height=2.1in]{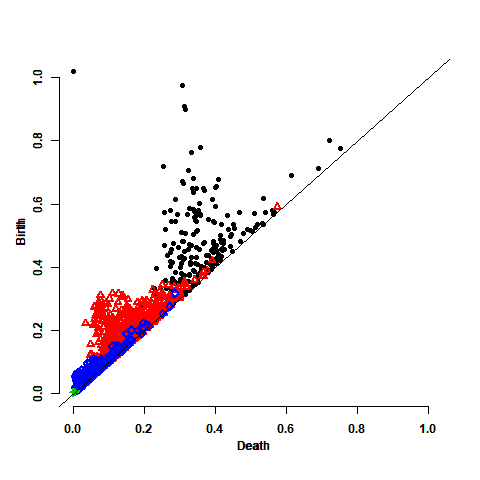}
\\
\includegraphics[width=2.1in, height=2.1in]{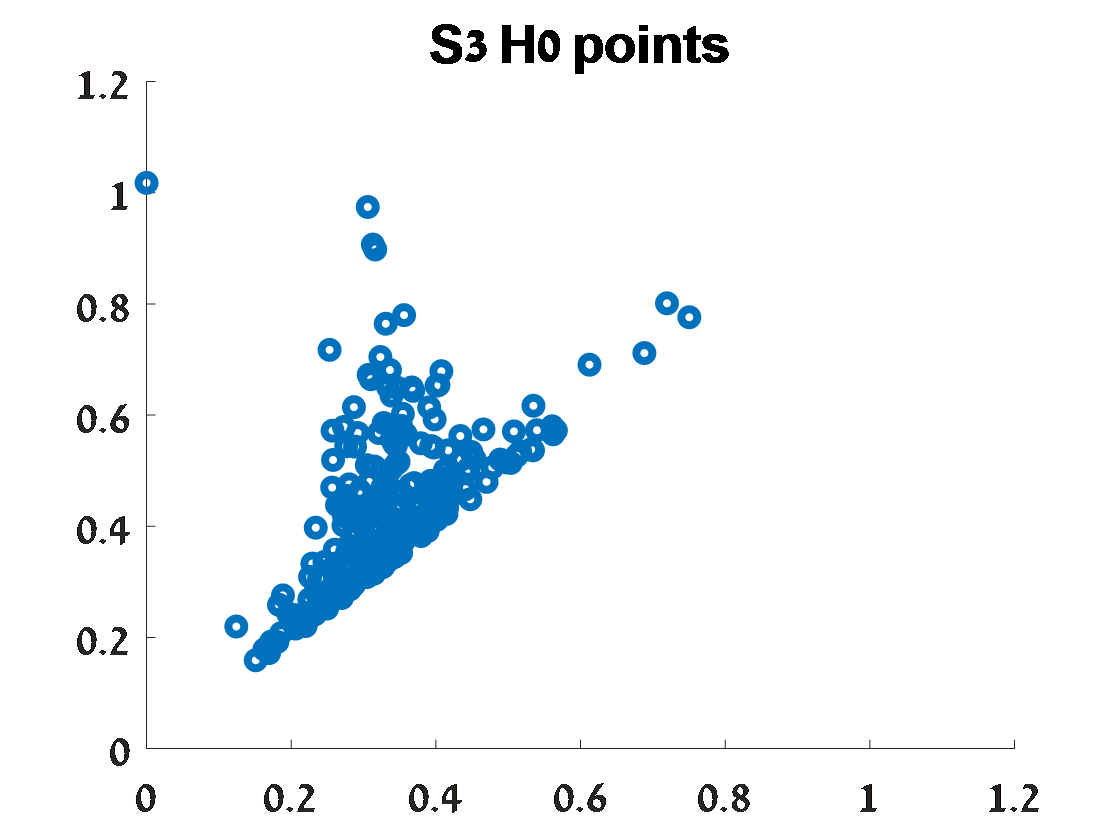}
\includegraphics[width=2.1in, height=2.1in]{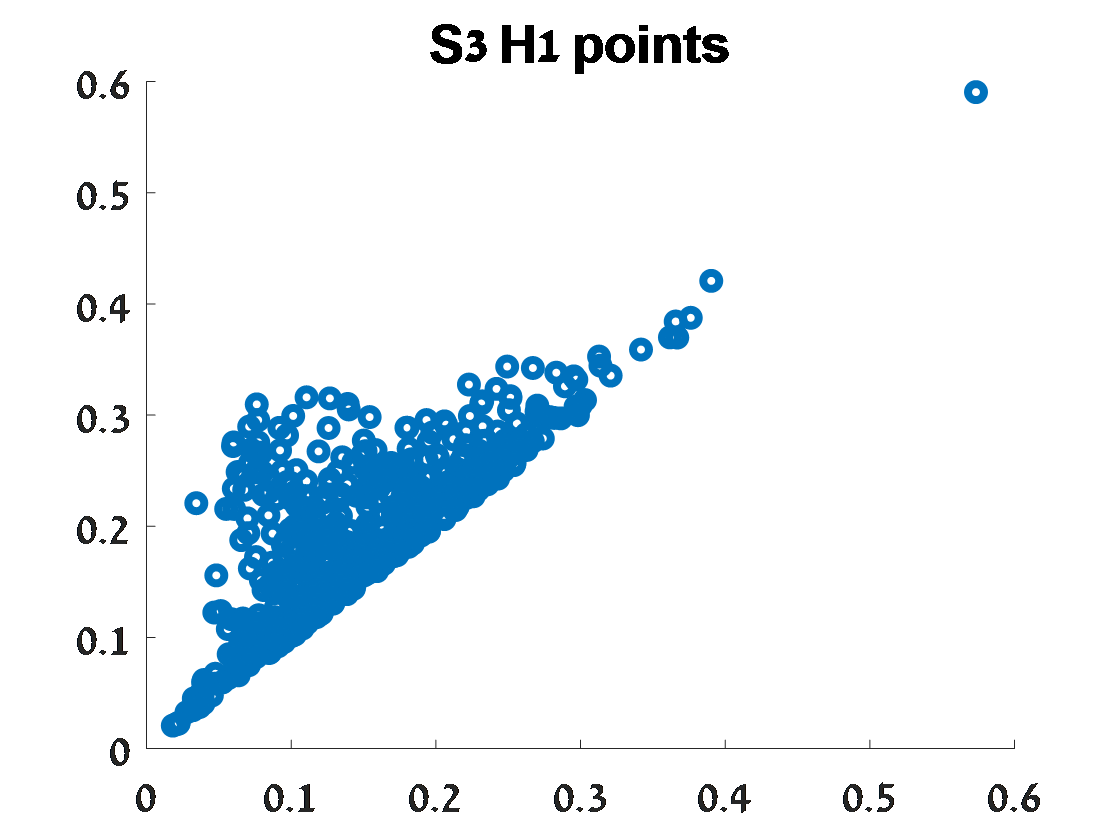}
\includegraphics[width=2.1in, height=2.1in]{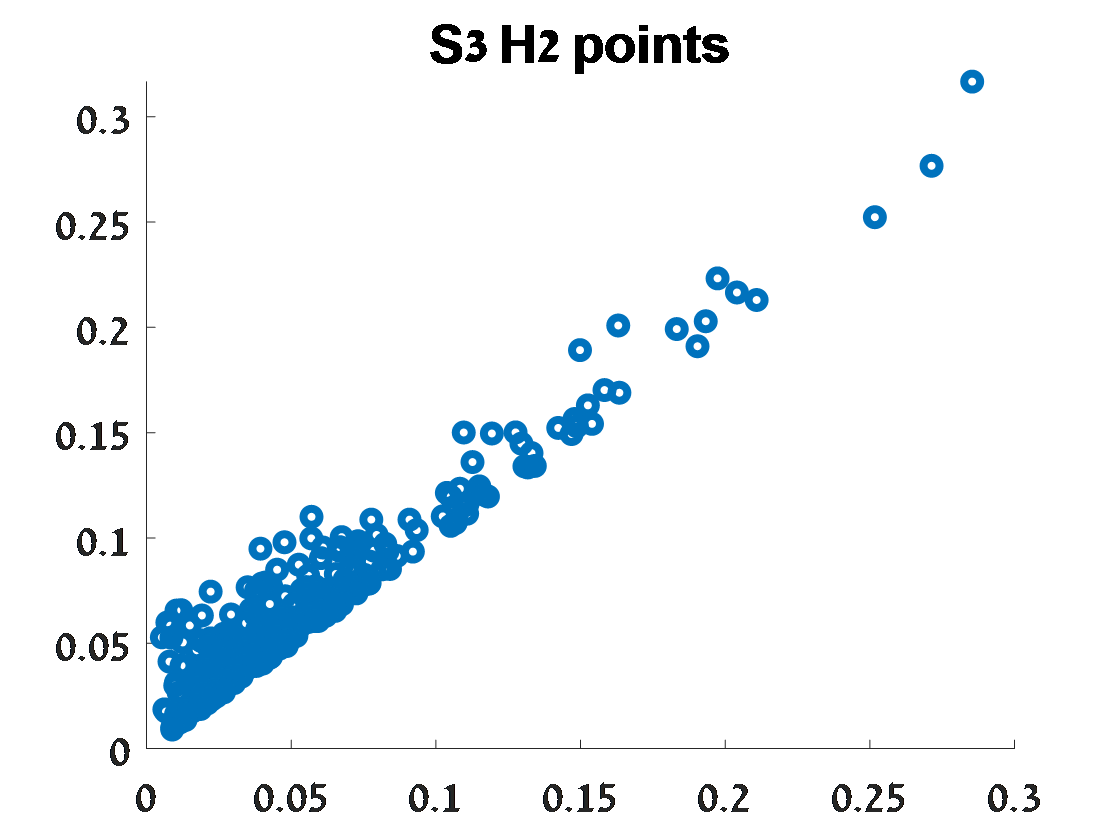}
\ec
%\caption{\footnotesize
% A random sample from two circles, 500 points from the larger circle and 300 from the smaller one,  with a kernel density
\caption{\footnotesize Top: The persistence diagram of a sample of $n=1,000$ points from the unit $S^3$, for its upper level sets. Black circles are connected components ($H_0$ persistence points), red triangles are holes ($H_1$ points), blue diamonds are voids ($H_2$), and the green points are $H_3$. Birth times are on the vertical axis. Bottom: The corresponded persistence diagram separately for each homology, except $H_3$.}
\label{fig:sphere_s3}
\end{figure}
\end{landscape}

The typical corresponded persistence diagram is presented in the left side Figure\ \ref{fig:sphere_s3}. The black circles indicating connected components ($H_0$ persistence), the red triangles corresponding to holes ($H_1$), the blue diamonds corresponding to voids ($H_2$), and the green points to $H_3$. The next plots to its right present the persistence diagram for each homology separately, except $H_3$ which has only 3 points.

In this example, there are enough points in $H_1$ and in $H_2$, so we could fitted the both original and modified models for each of these two homologies points in addition to the model's fitting for $H_0$ points.

\subsubsection{The fitted model for $H_0$}
\begin{landscape}
\begin{figure}[h!]
\bc
\includegraphics[width=1.2in, height=1.4in]{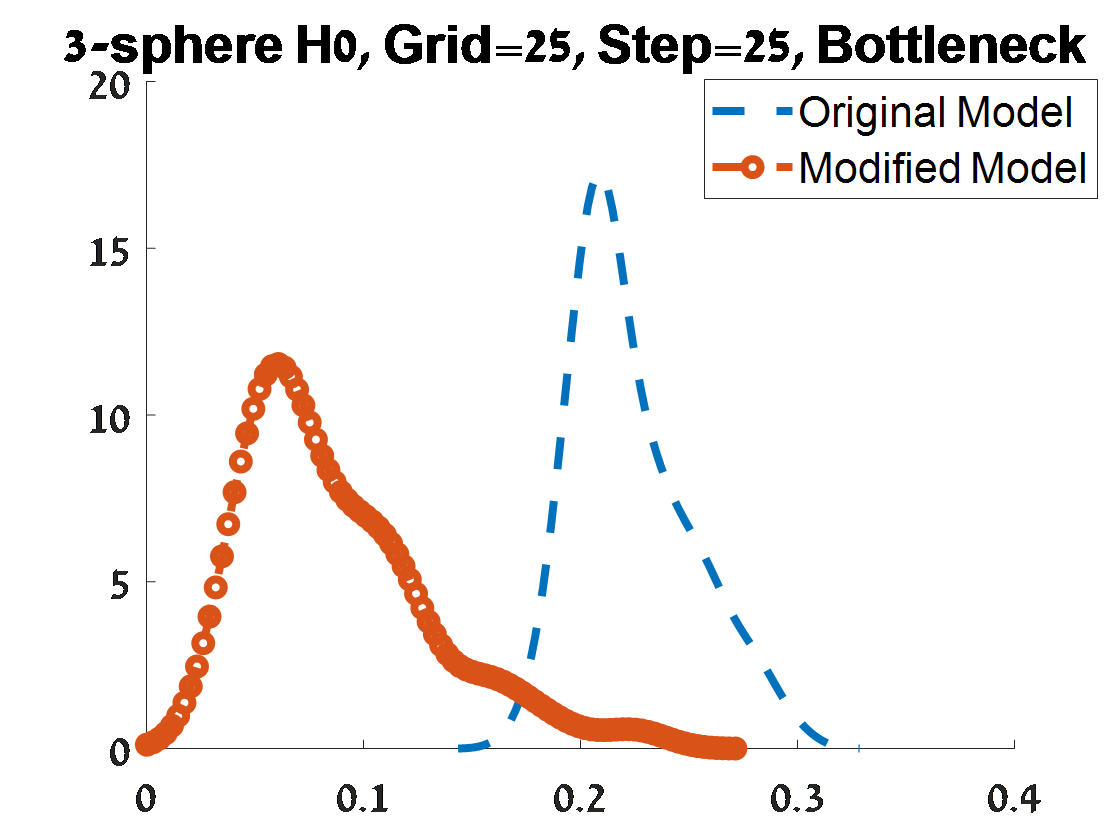}
\includegraphics[width=1.2in, height=1.4in]{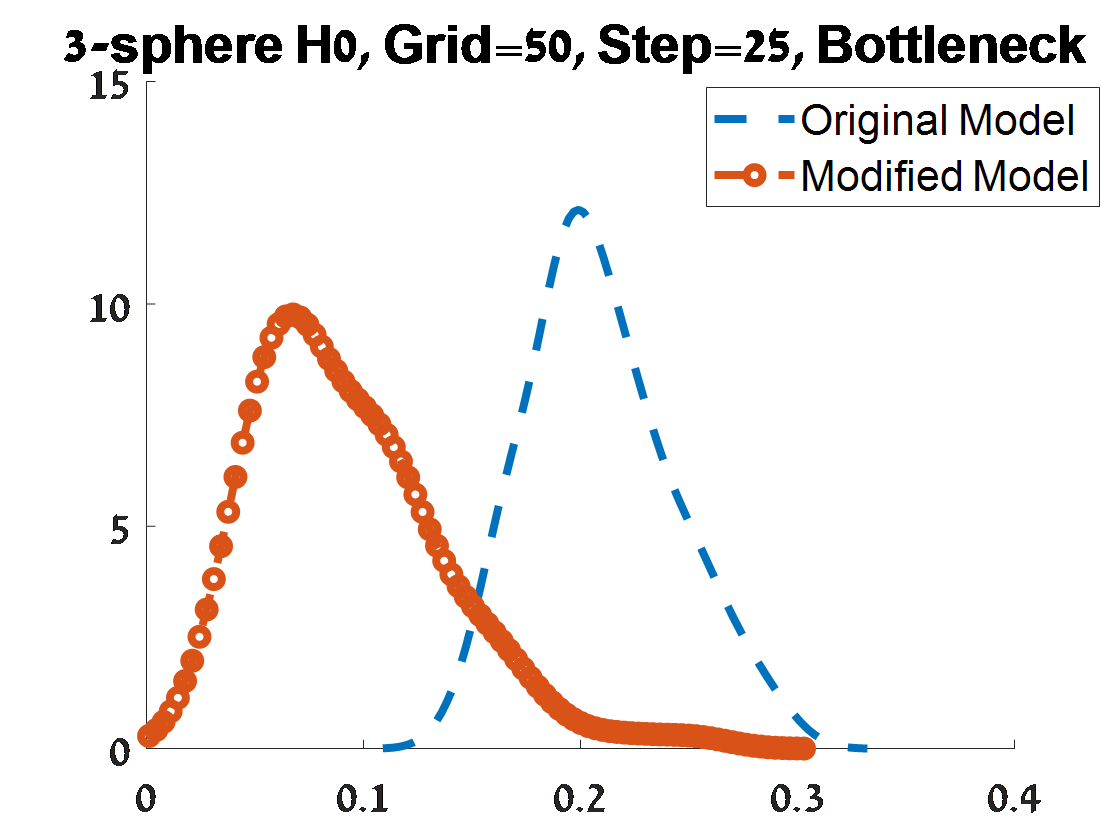}
\includegraphics[width=1.2in, height=1.4in]{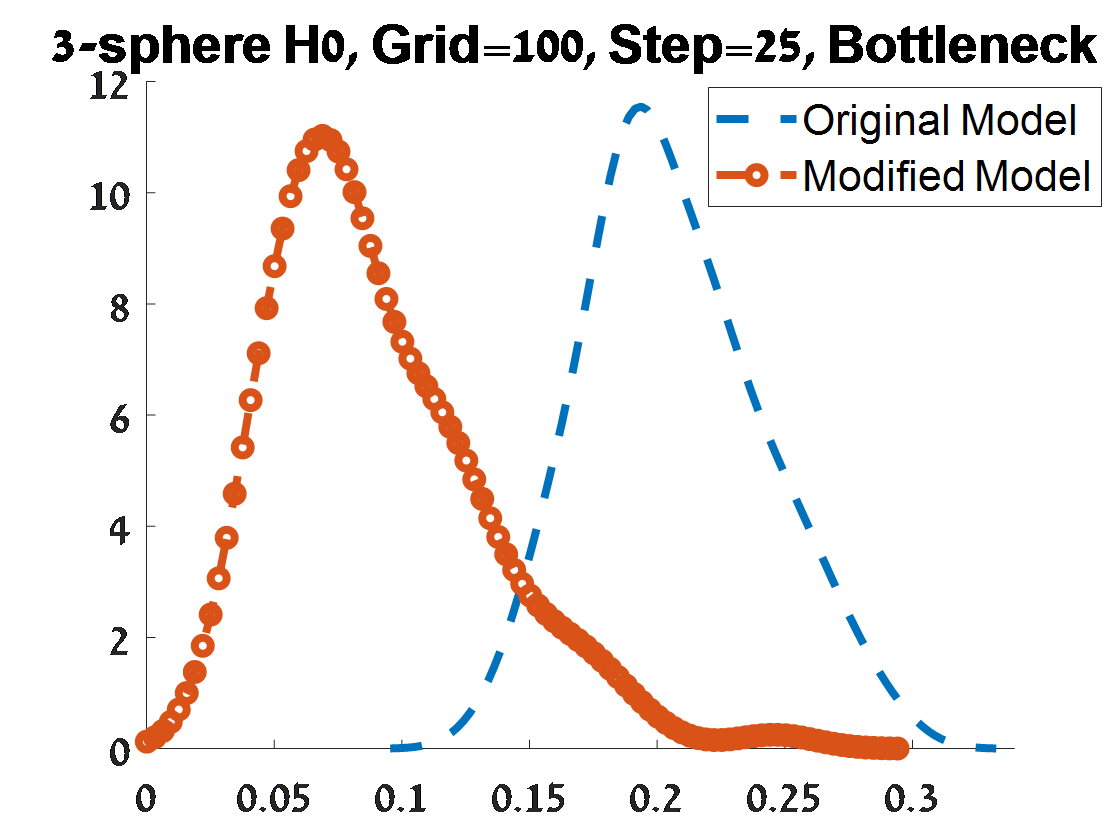}
\includegraphics[width=1.2in, height=1.4in]{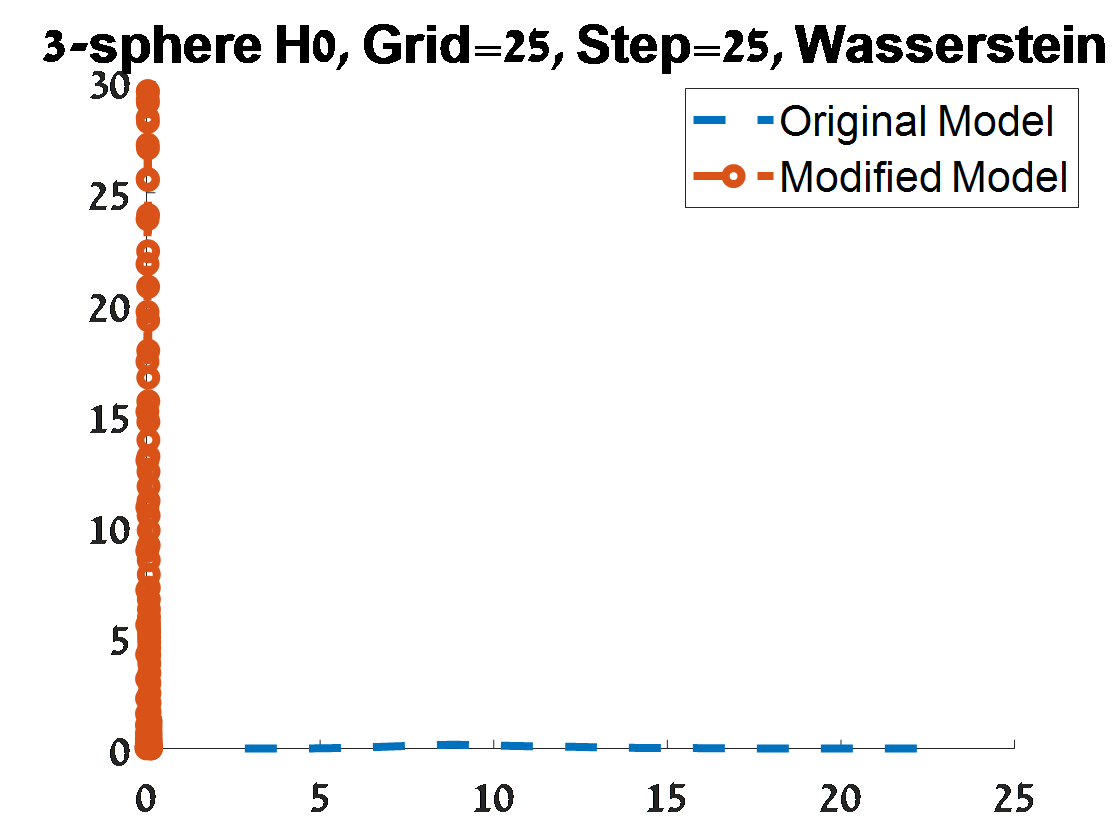}
\includegraphics[width=1.2in, height=1.4in]{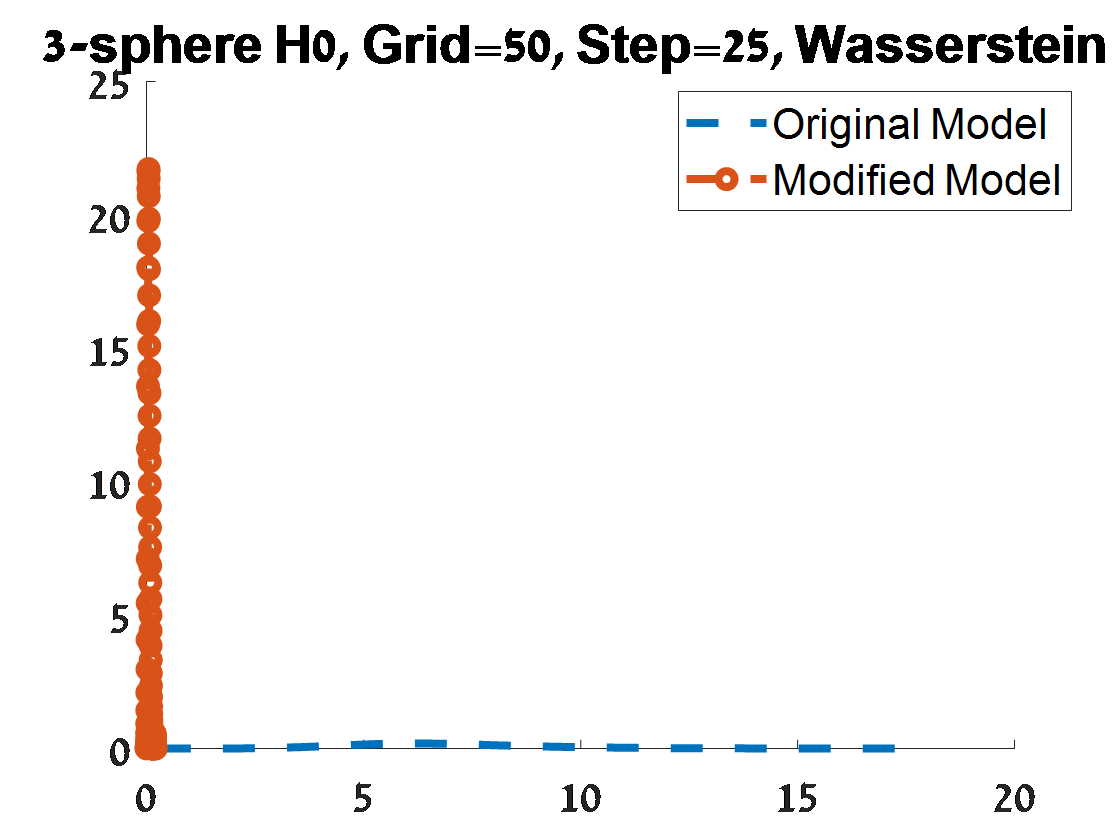}
\includegraphics[width=1.2in, height=1.4in]{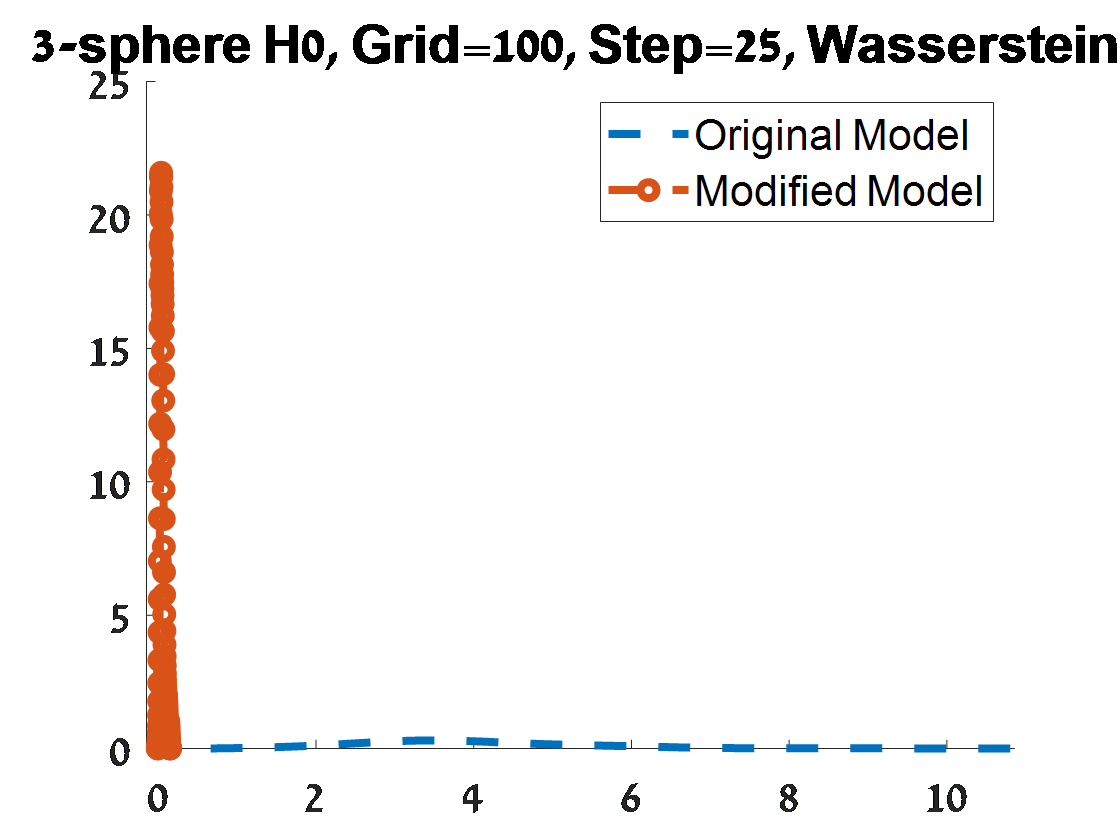}
\\
\includegraphics[width=1.2in, height=1.4in]{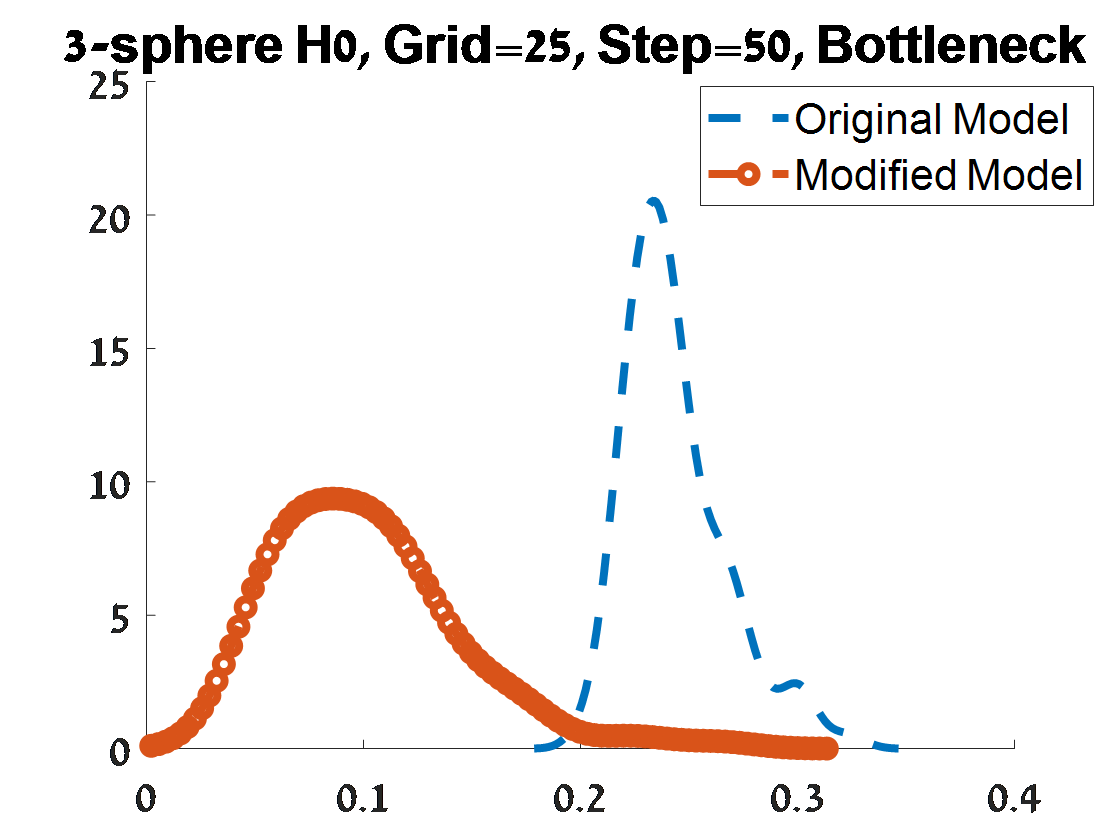}
\includegraphics[width=1.2in, height=1.4in]{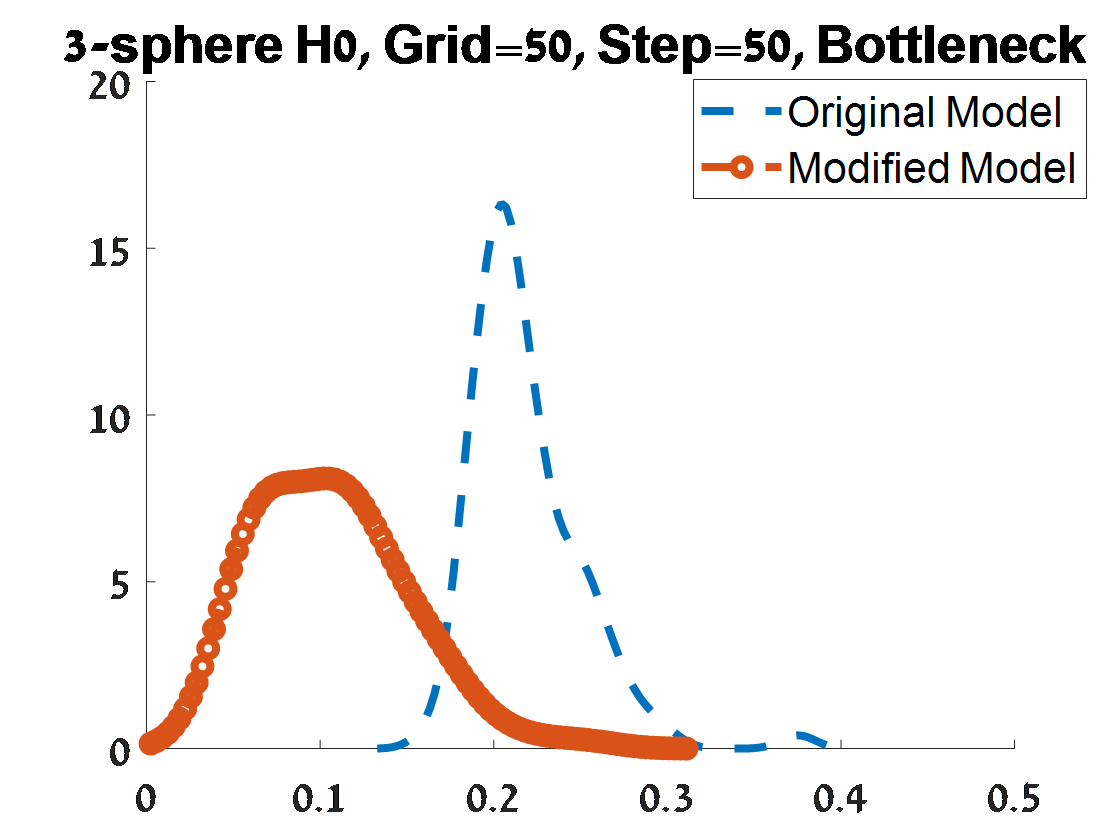}
\includegraphics[width=1.2in, height=1.4in]{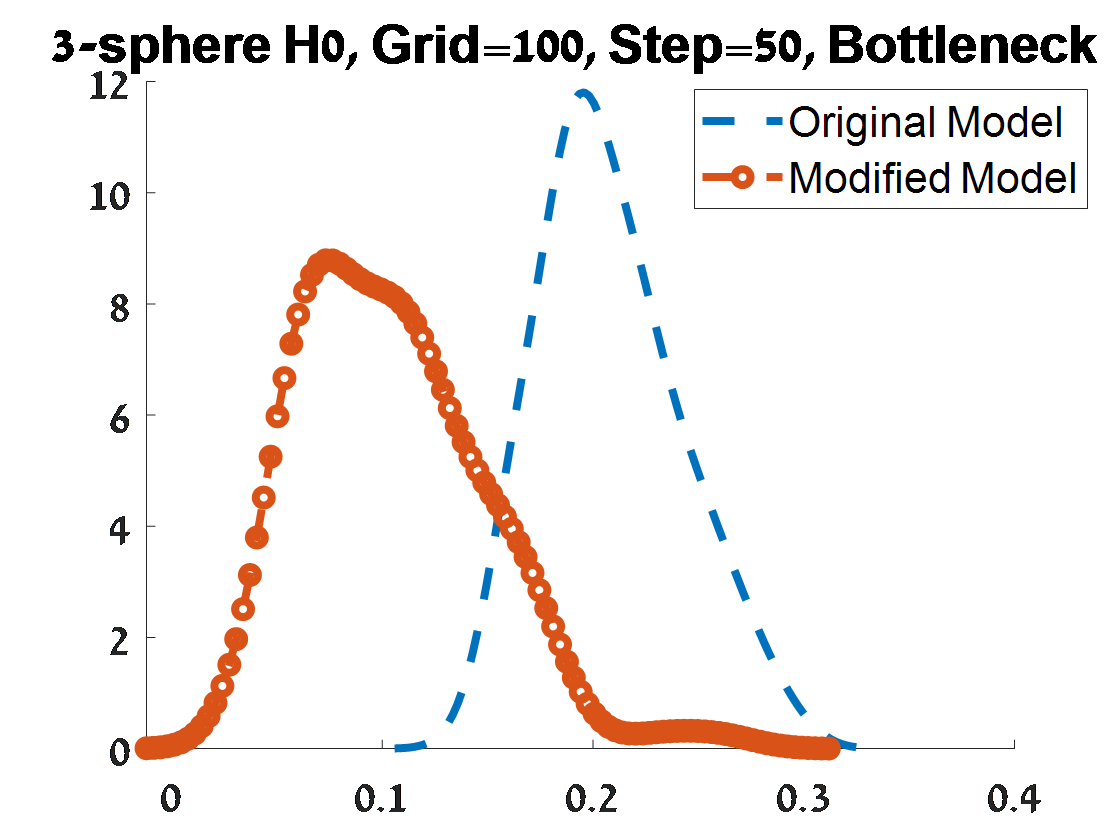}
\includegraphics[width=1.2in, height=1.4in]{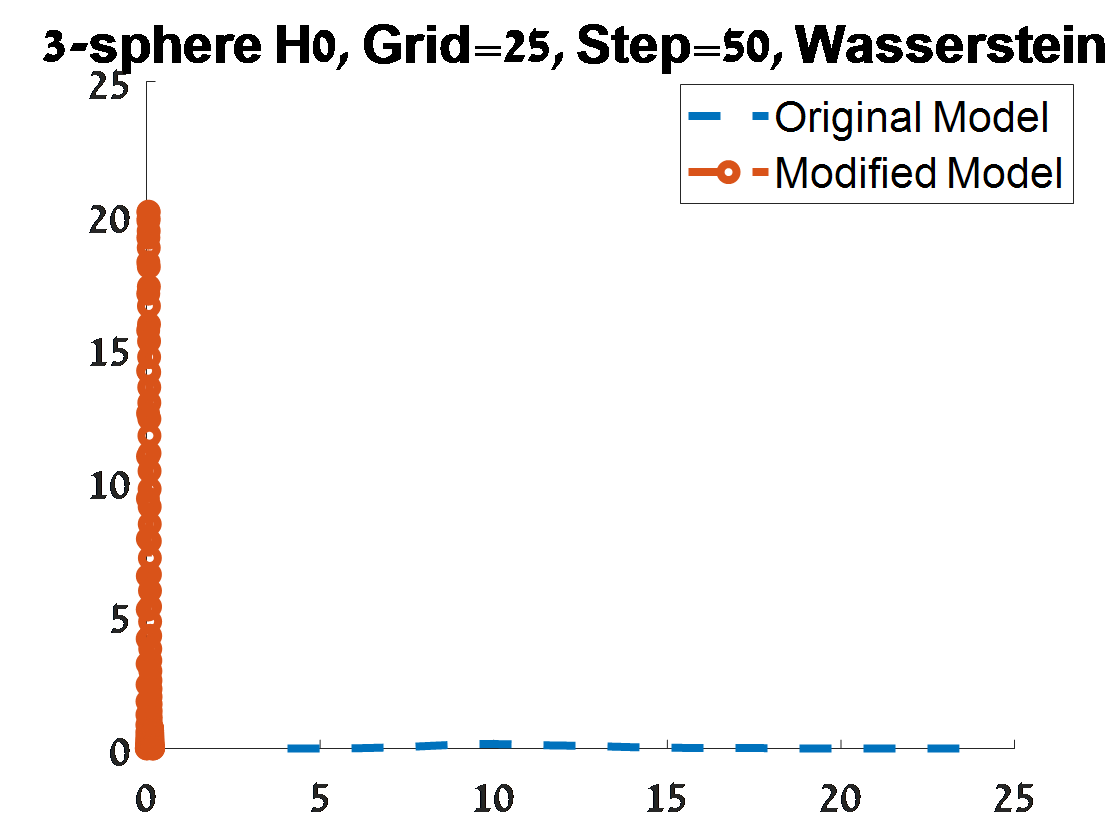}
\includegraphics[width=1.2in, height=1.4in]{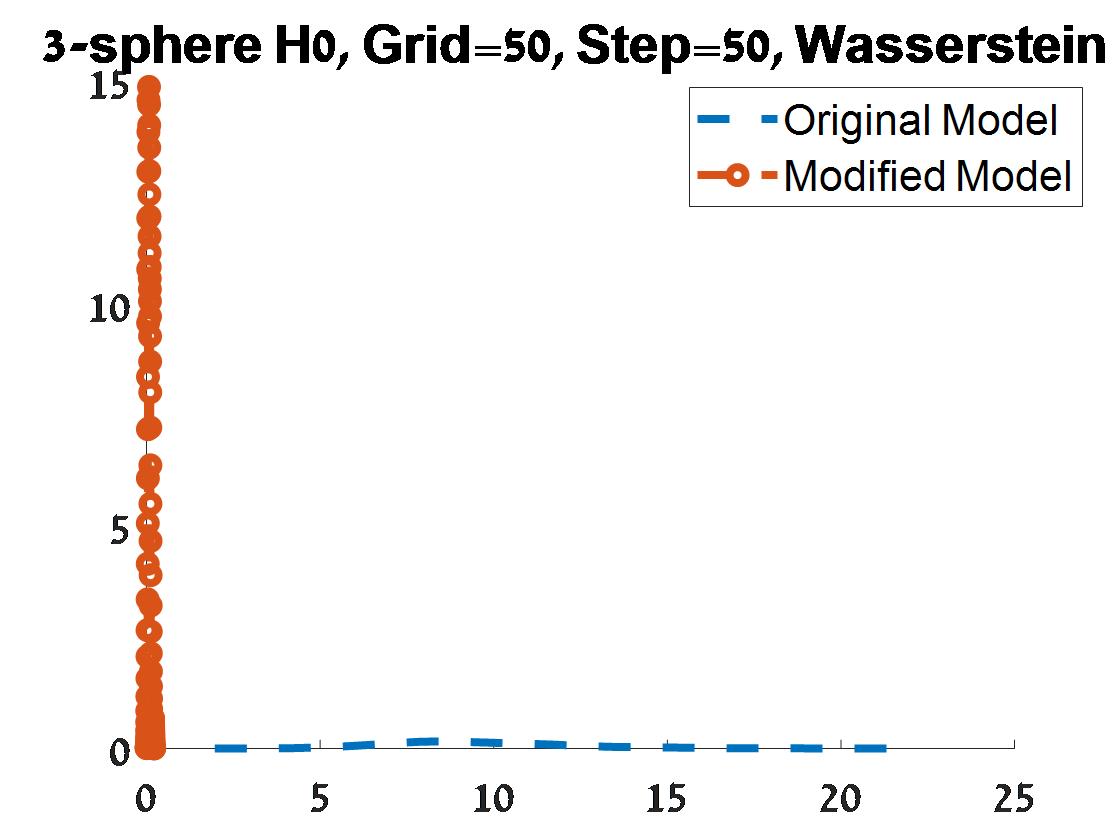}
\includegraphics[width=1.2in, height=1.4in]{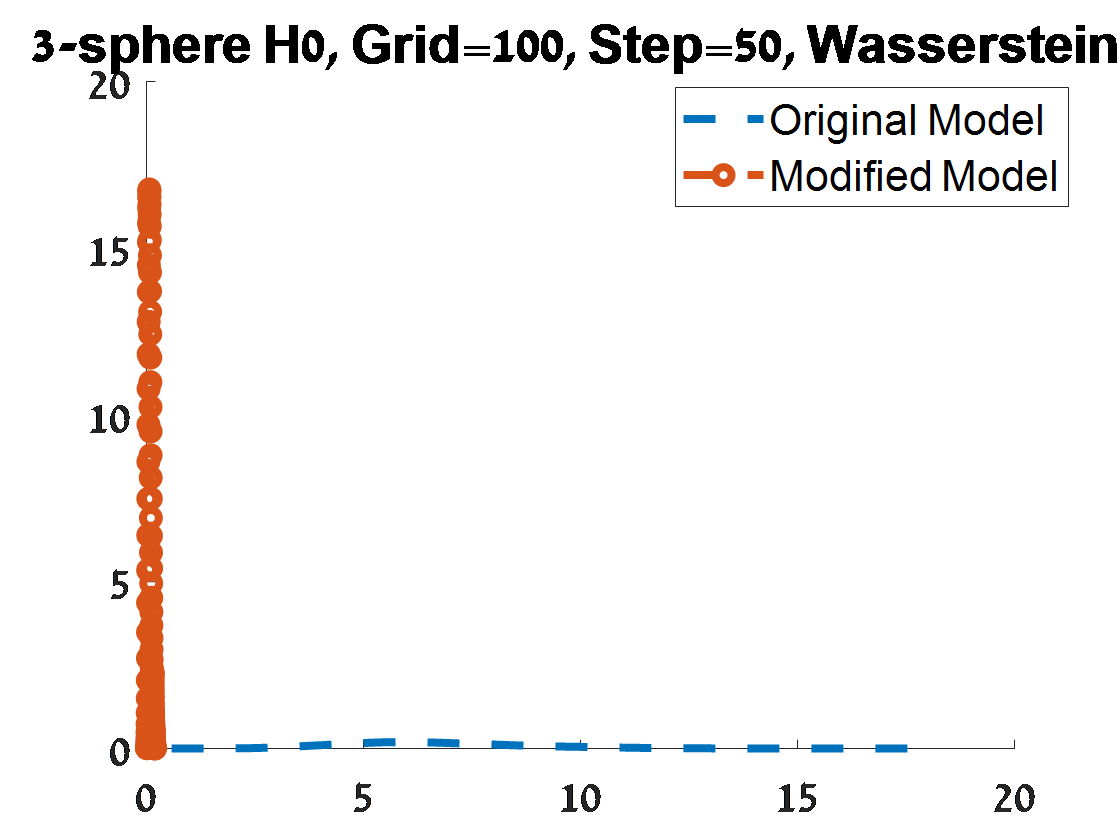}
\\
\includegraphics[width=1.2in, height=1.4in]{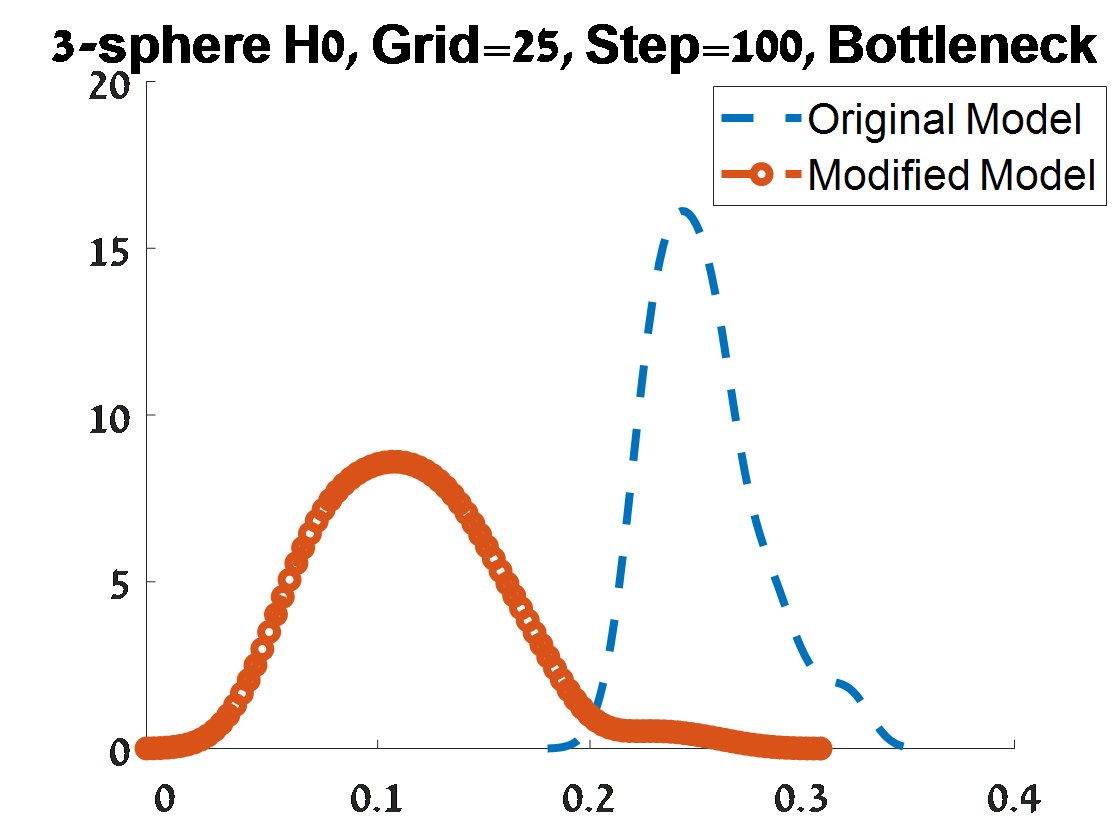}
\includegraphics[width=1.2in, height=1.4in]{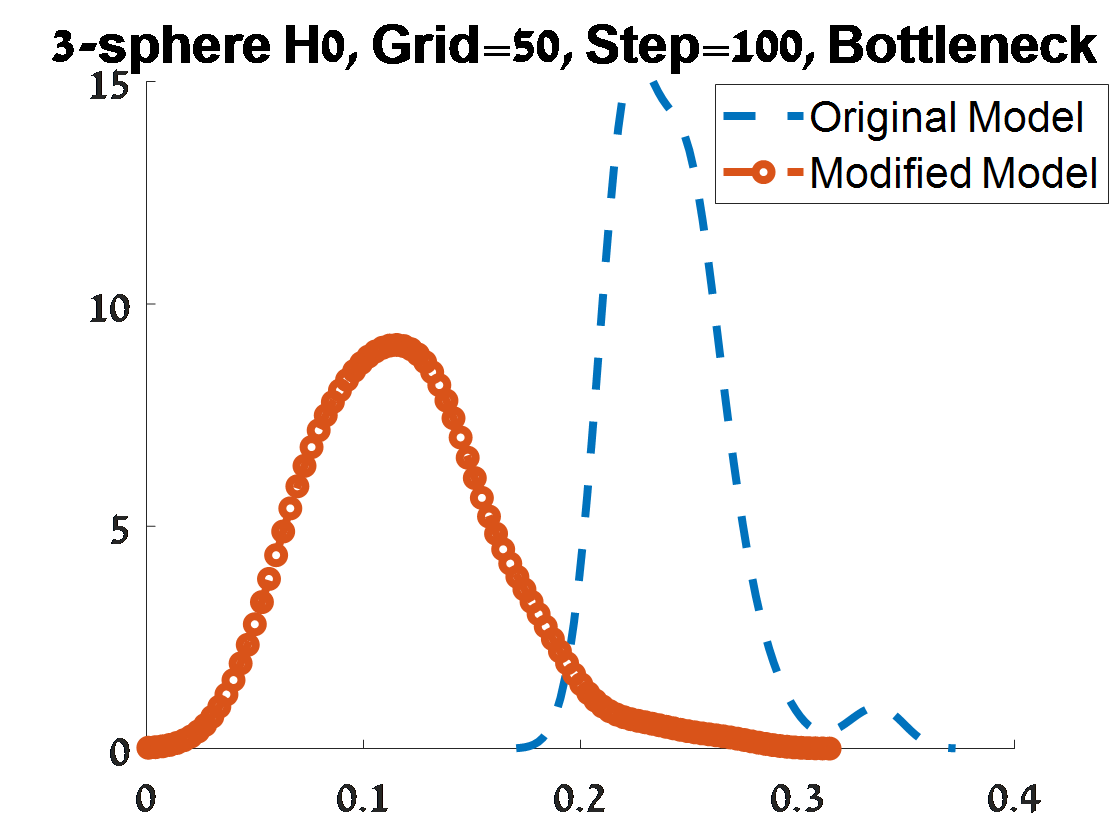}
\includegraphics[width=1.2in, height=1.4in]{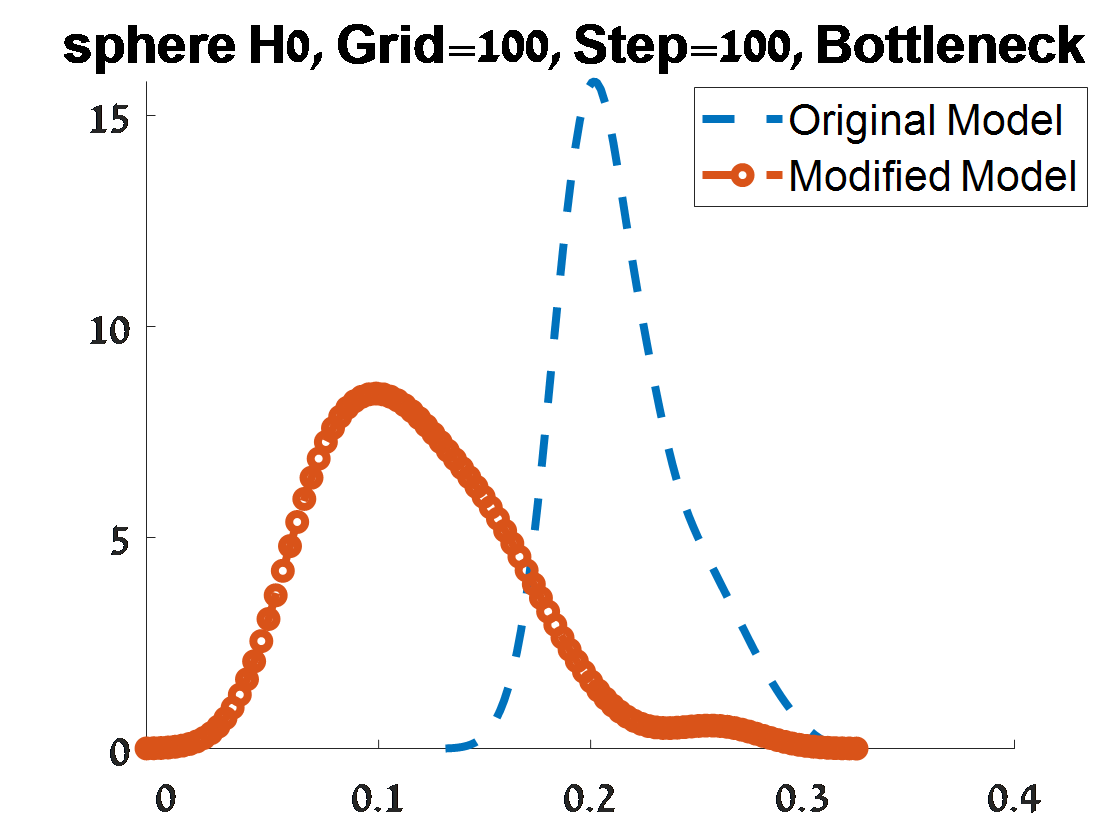}
\includegraphics[width=1.2in, height=1.4in]{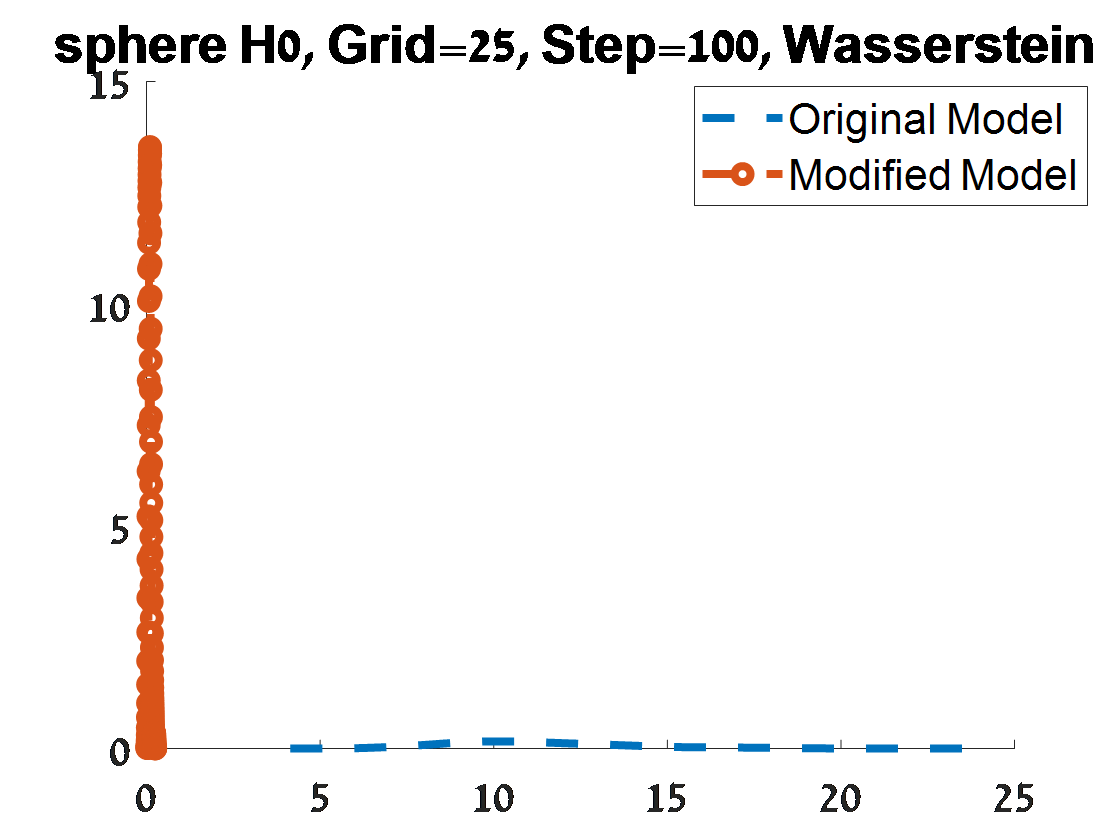}
\includegraphics[width=1.2in, height=1.4in]{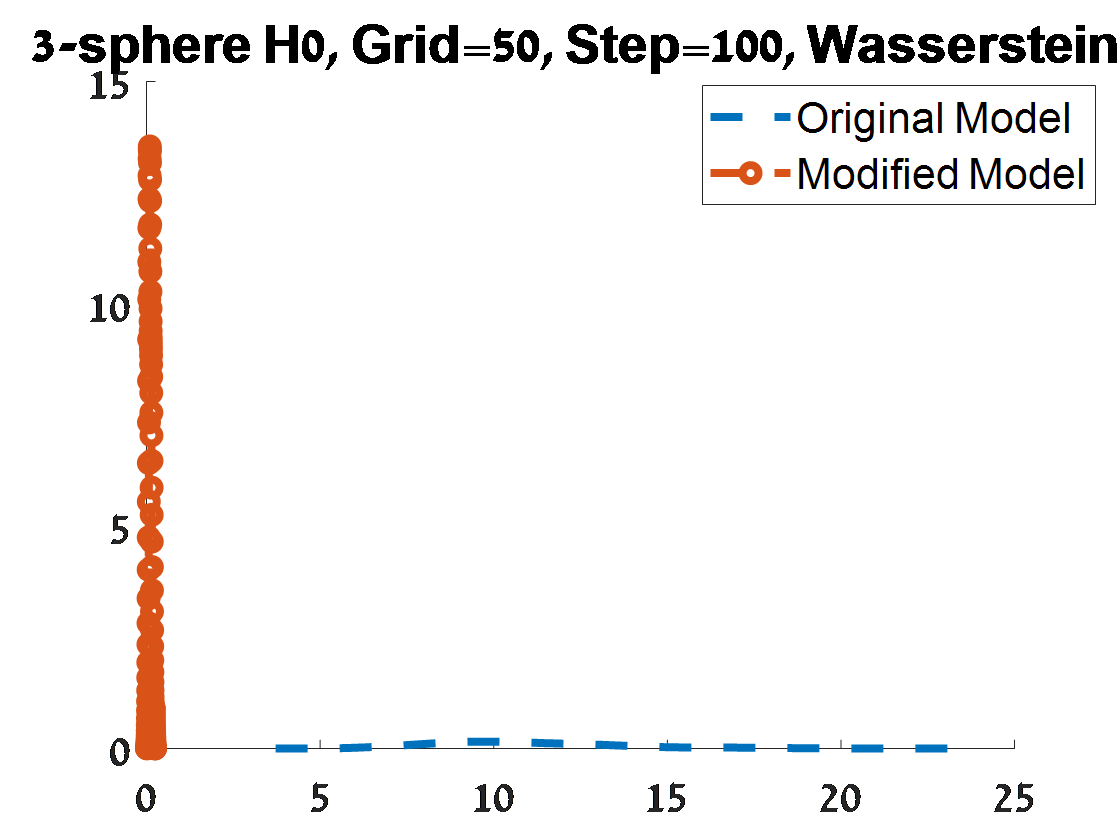}
\includegraphics[width=1.2in, height=1.4in]{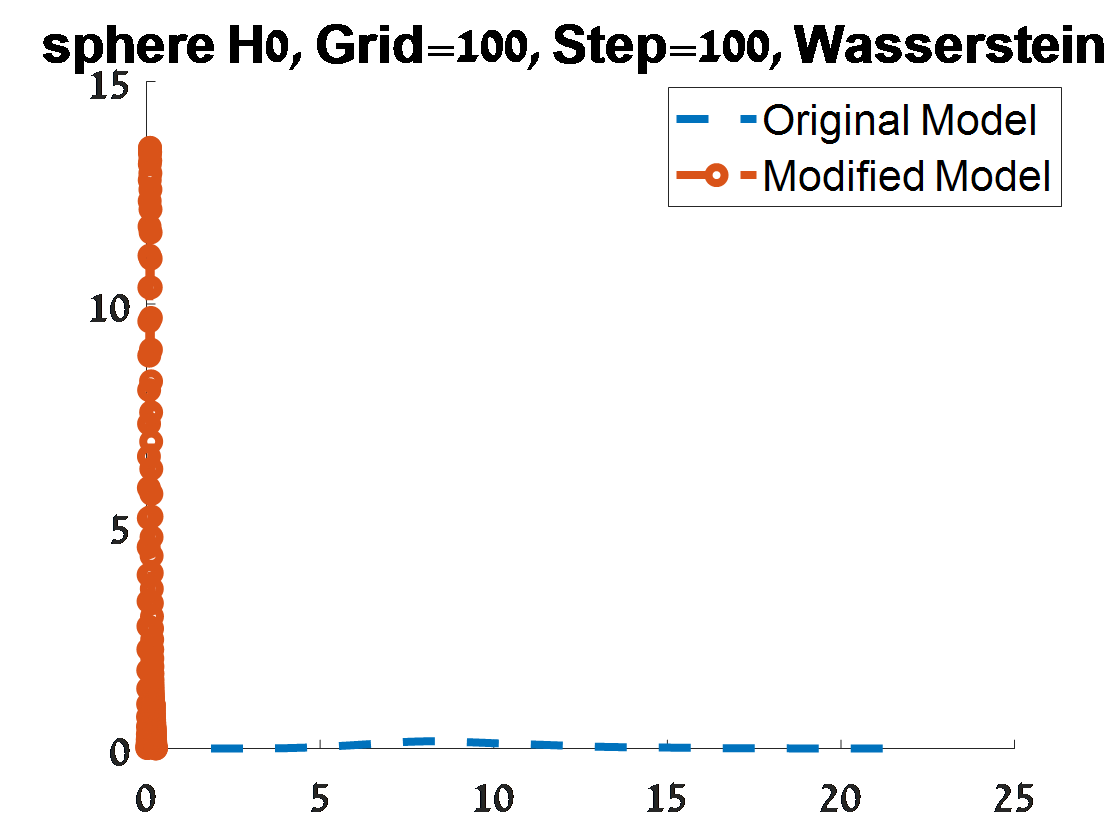}
\ec
%\caption{\footnotesize
% A random sample from two circles, 500 points from the larger circle and 300 from the smaller one,  with a kernel density
\caption{\footnotesize
 Criterion 1 of goodness of fit for 100 $H_0$ PDs corresponded to 100 samples from a unit $S^3$. The figures depend on the grid of the proposal distribution ("Grid"), and the burn-in ("Step") of the MCMC algorithm.}
\label{fig:s3_H0_a}
\end{figure}
\end{landscape}

Figure\ \ref{fig:s3_H0_a} describes the distributions over the 100 $H_0$ PDs of the first criterion of goodness of fit, and Figures 22-23 describe the distributions of the second criterion of goodness of fit.
Based on criterion 1, as in $S^2$, the smallest distance of the modified model relative to the original model is prominent in both distances, whereas in the Wasserstein distance it is even more prominent. The distances distributions for the modified model is similar over the considered grid sizes and burn-in.
For criterion 2, as in $S^2$, the variability is larger in the distributions relative to examples 1-3, where the best fitting of the modified model relative to the real PDs is under burn-in of 25 and grid size of 100.

%%Figure\ \ref{fig:s3_H0_d} presents two examples of real PD and its simulated PD based on the two model versions, only for the best scenarios we found, that is grid=50,100, and step=25. the modified model is better and succeed in creating the far points from the diagonal, where the original model is concentrating in creating more the points that are close to the diagonal and almost cannot succeed in generating the far points from the diagonal.

\begin{landscape}
\begin{figure}[h!]
\bc
\includegraphics[width=1.2in, height=1.25in]{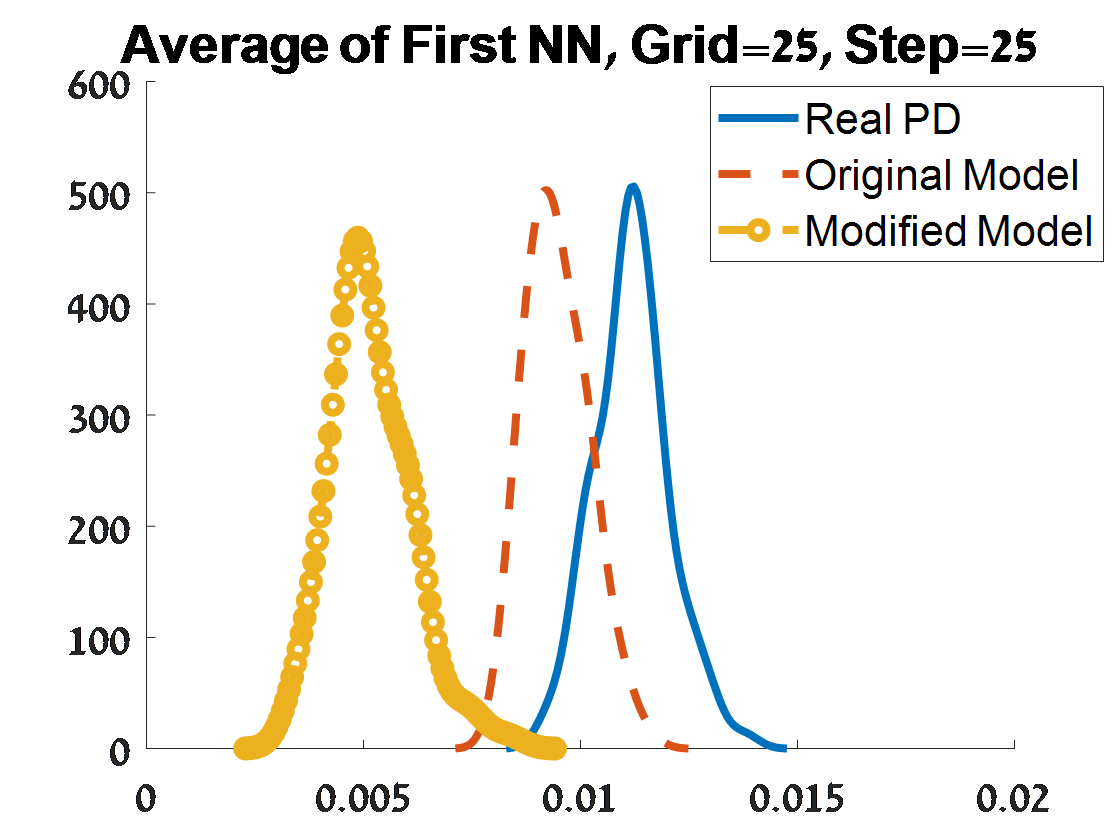}
\includegraphics[width=1.2in, height=1.25in]{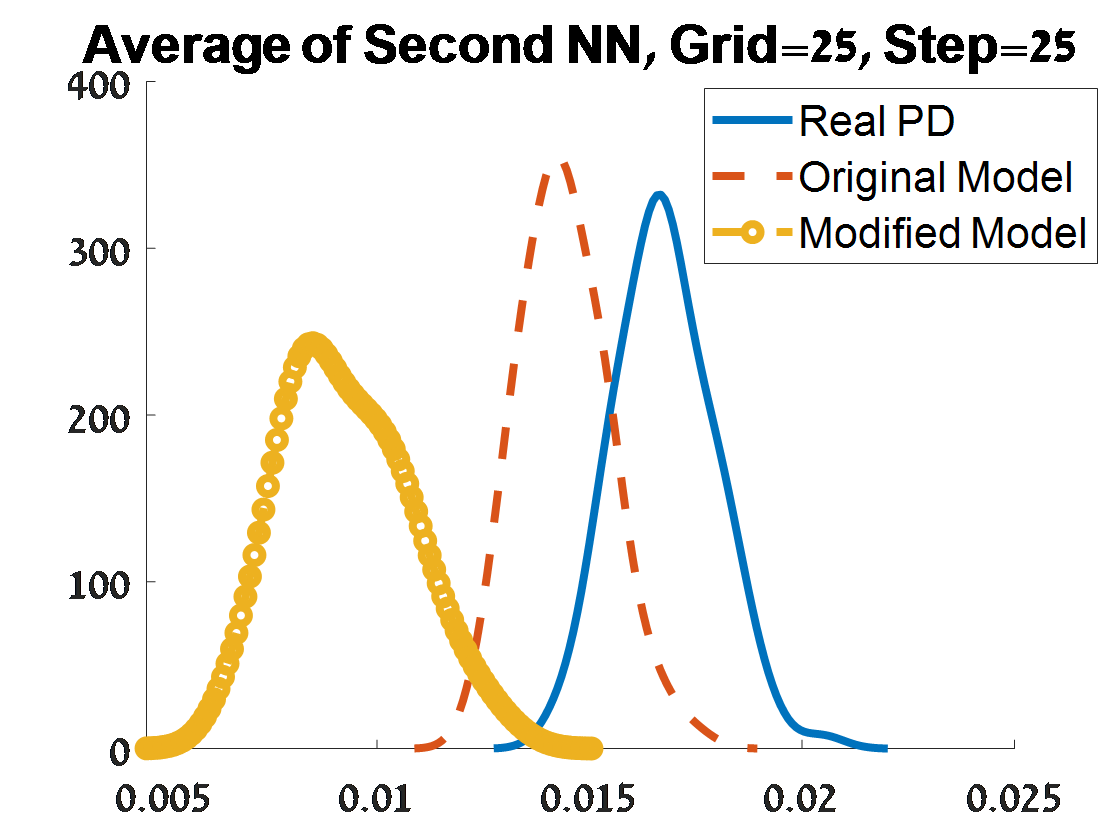}
\includegraphics[width=1.2in, height=1.25in]{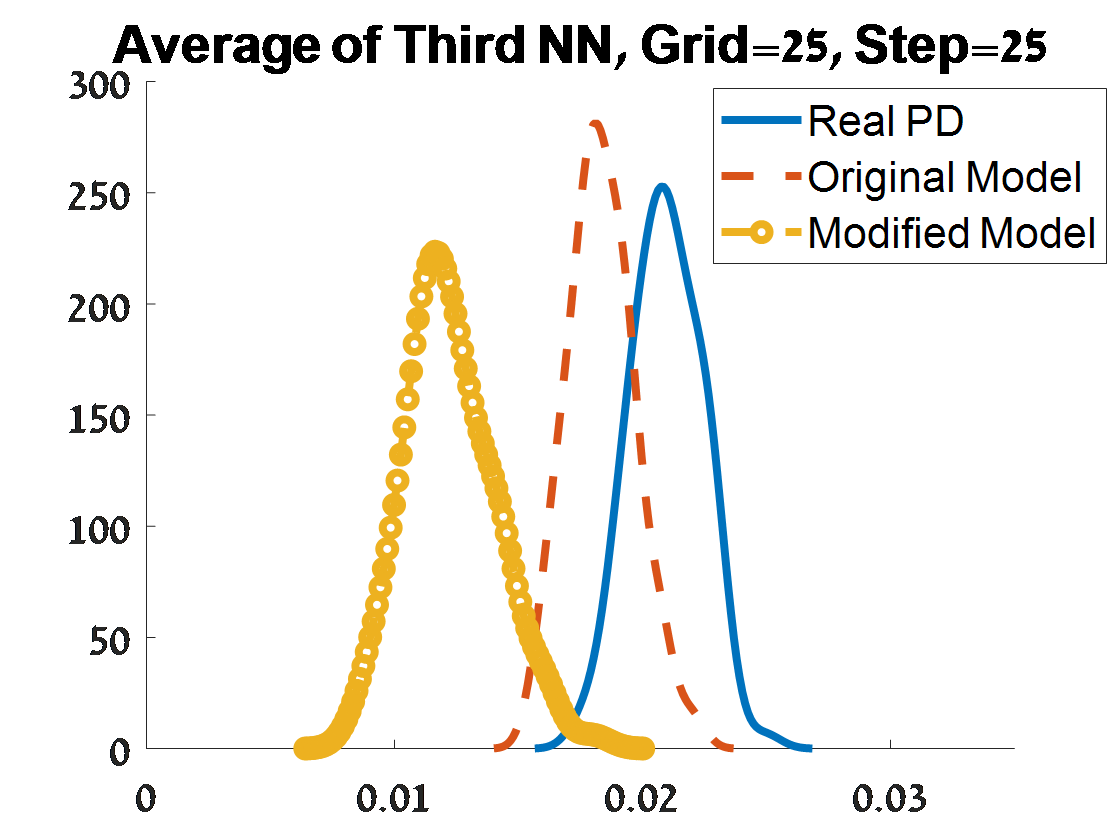}
\includegraphics[width=1.2in, height=1.25in]{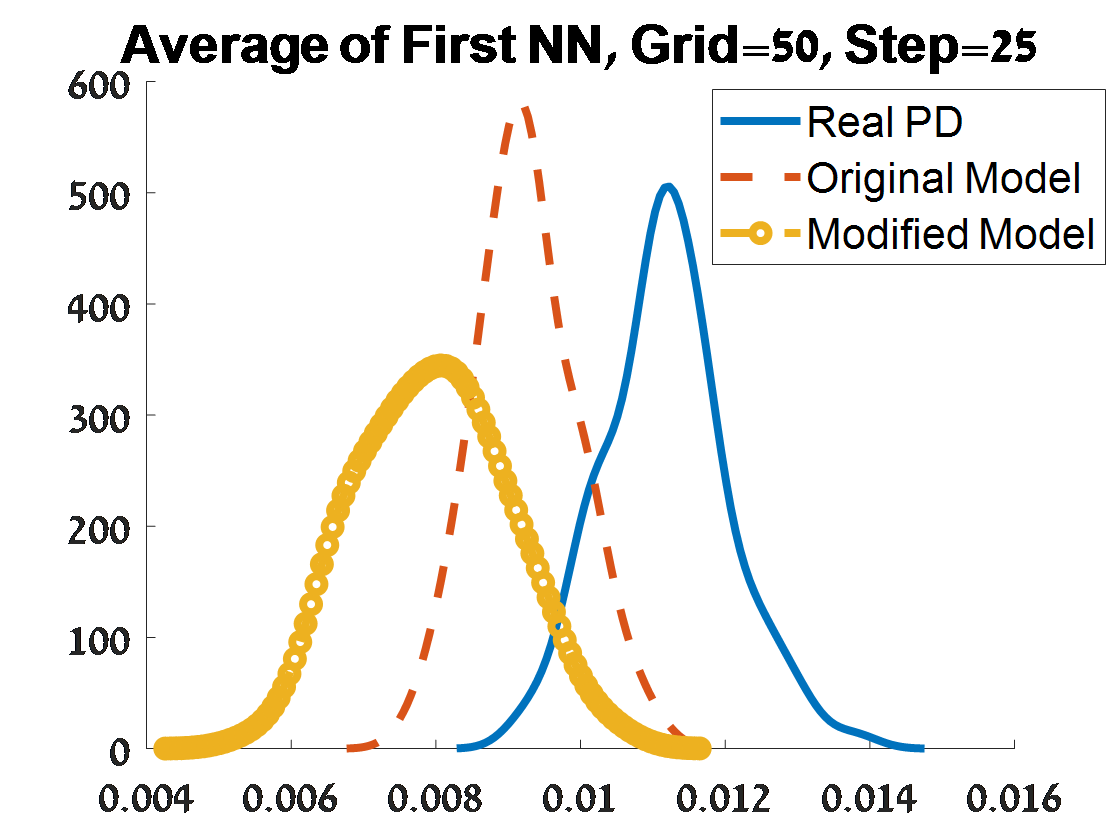}
\includegraphics[width=1.2in, height=1.25in]{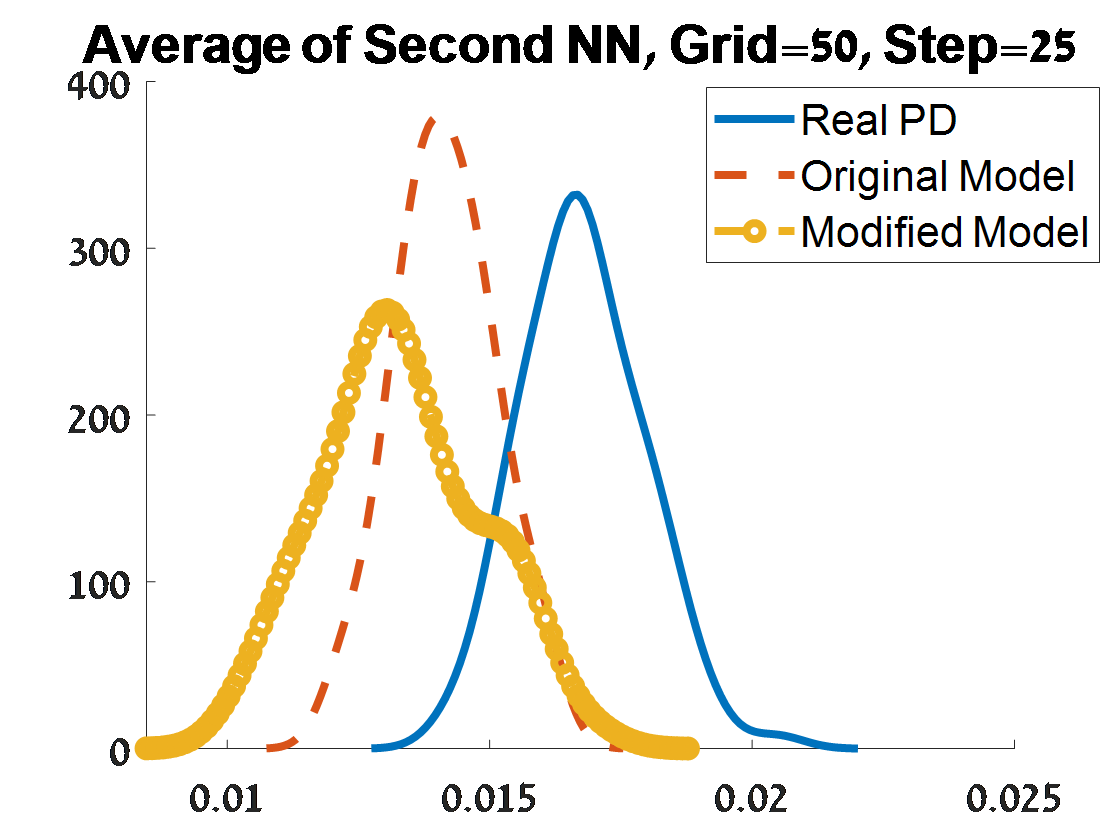}
\includegraphics[width=1.2in, height=1.25in]{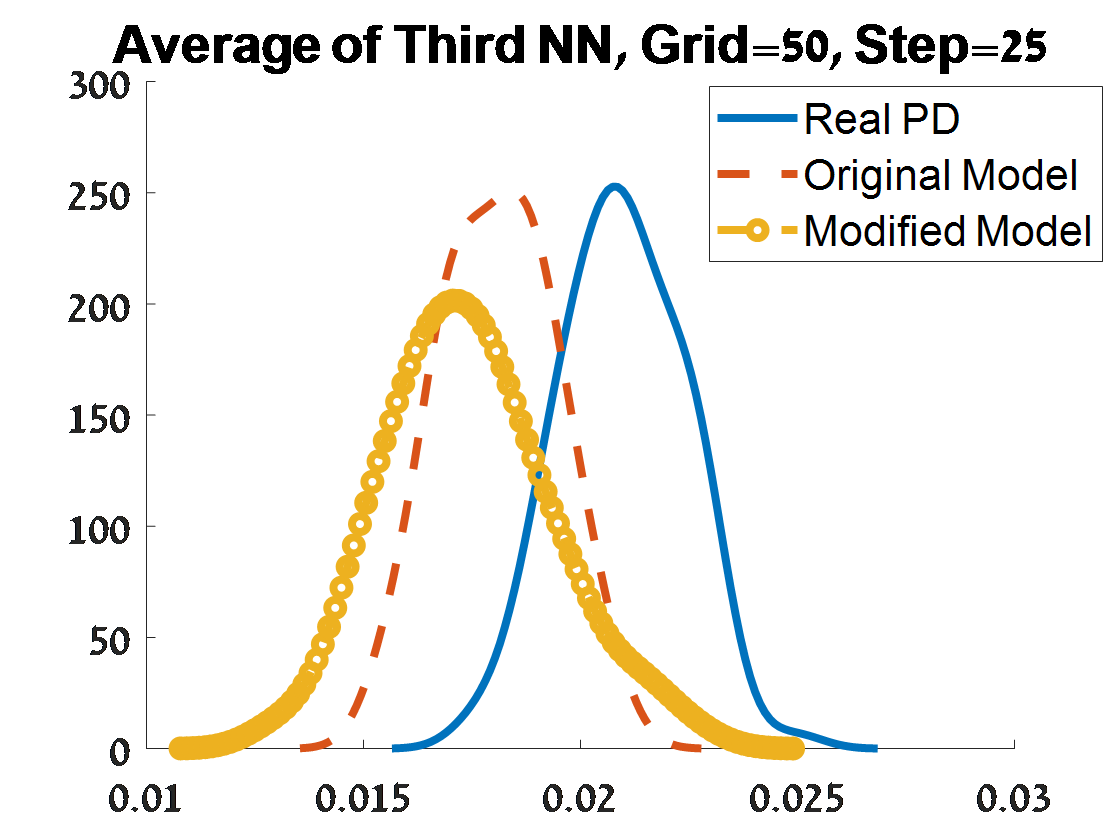}
\includegraphics[width=1.2in, height=1.25in]{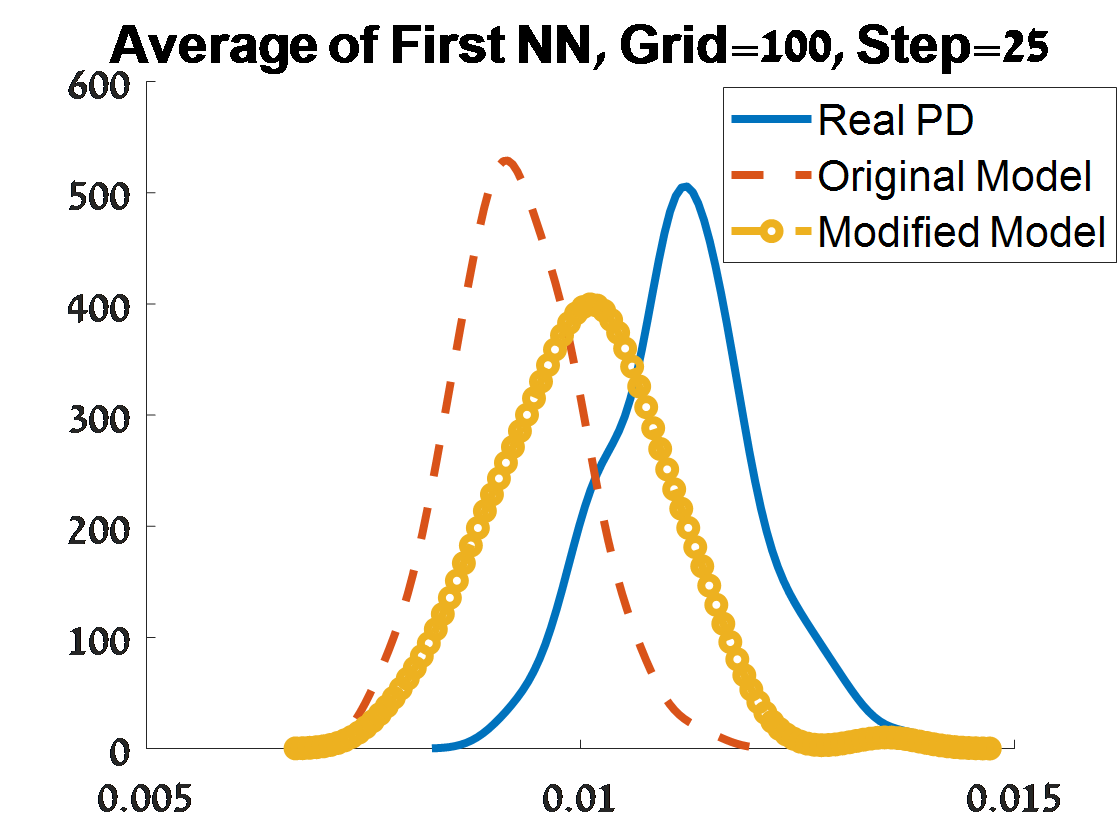}
\includegraphics[width=1.2in, height=1.25in]{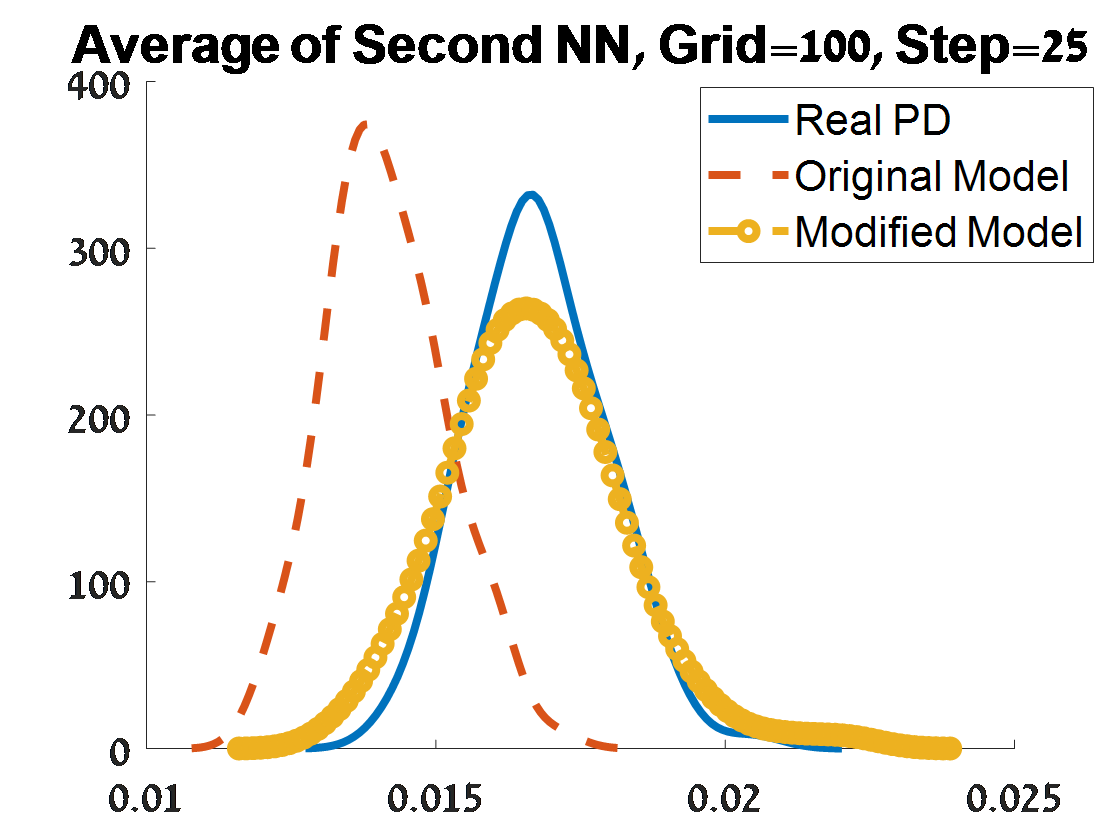}
\includegraphics[width=1.2in, height=1.25in]{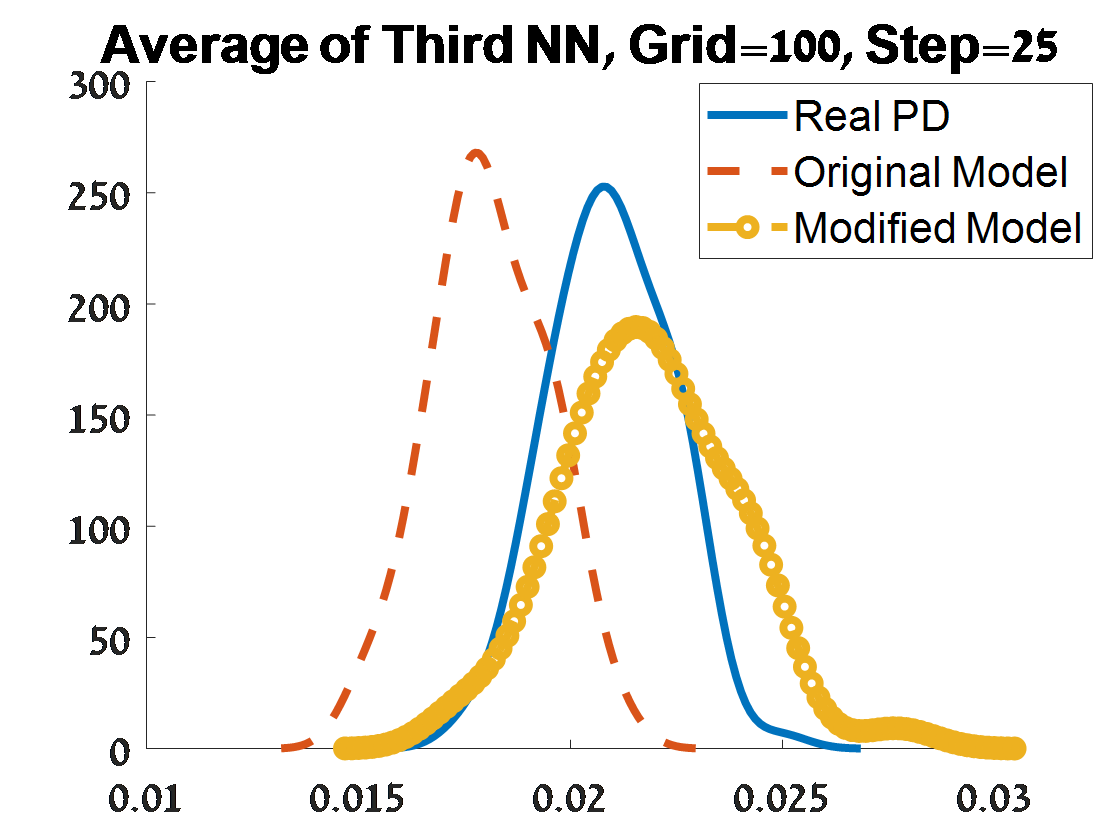}
\\
\includegraphics[width=1.2in, height=1.25in]{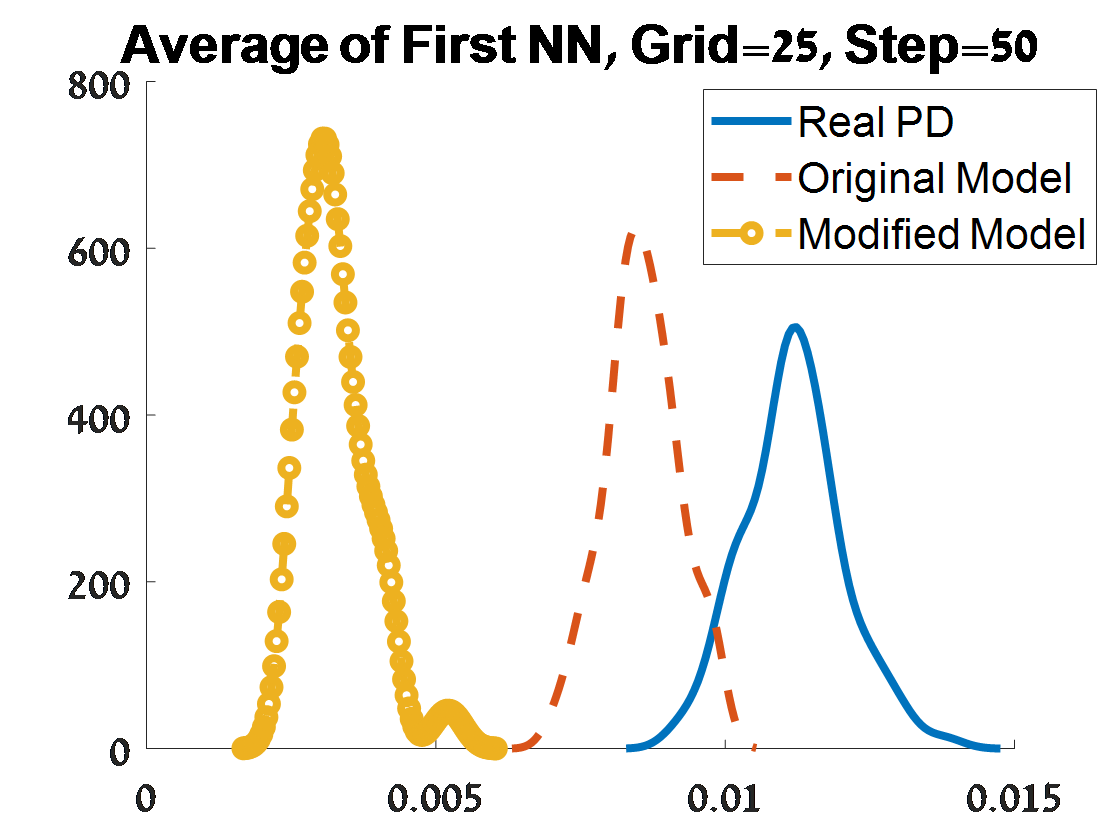}
\includegraphics[width=1.2in, height=1.25in]{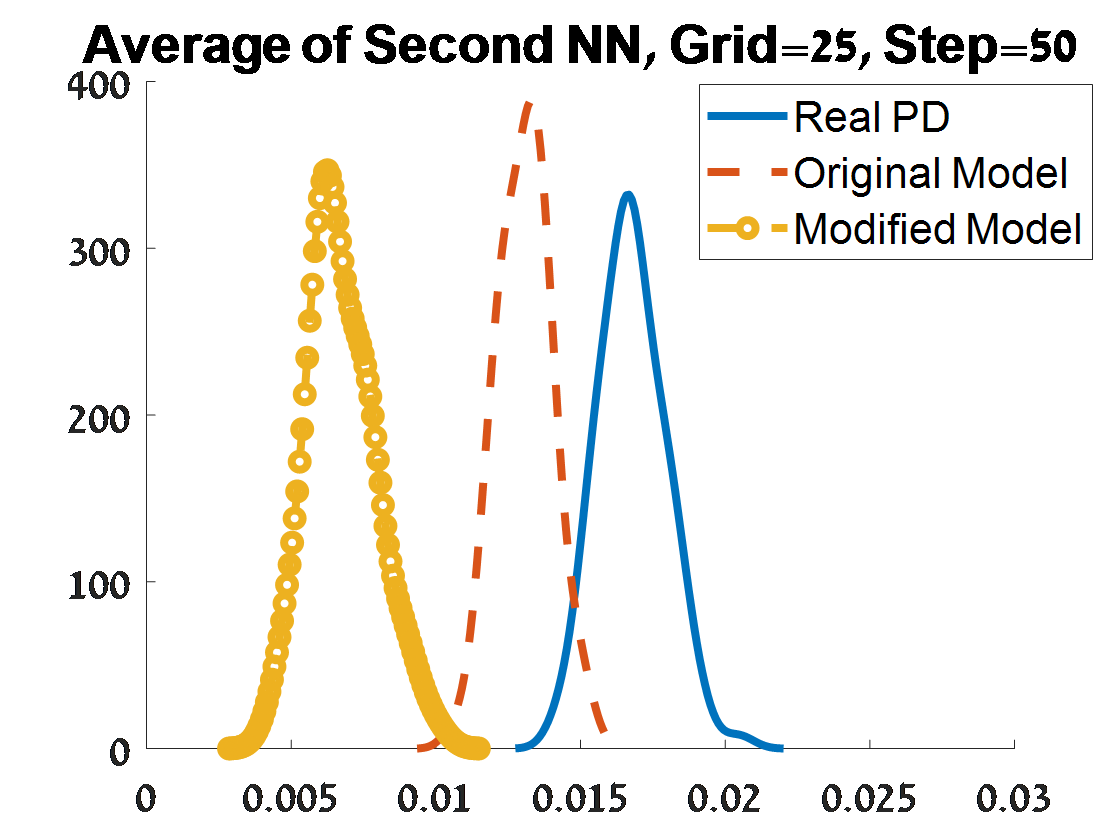}
\includegraphics[width=1.2in, height=1.25in]{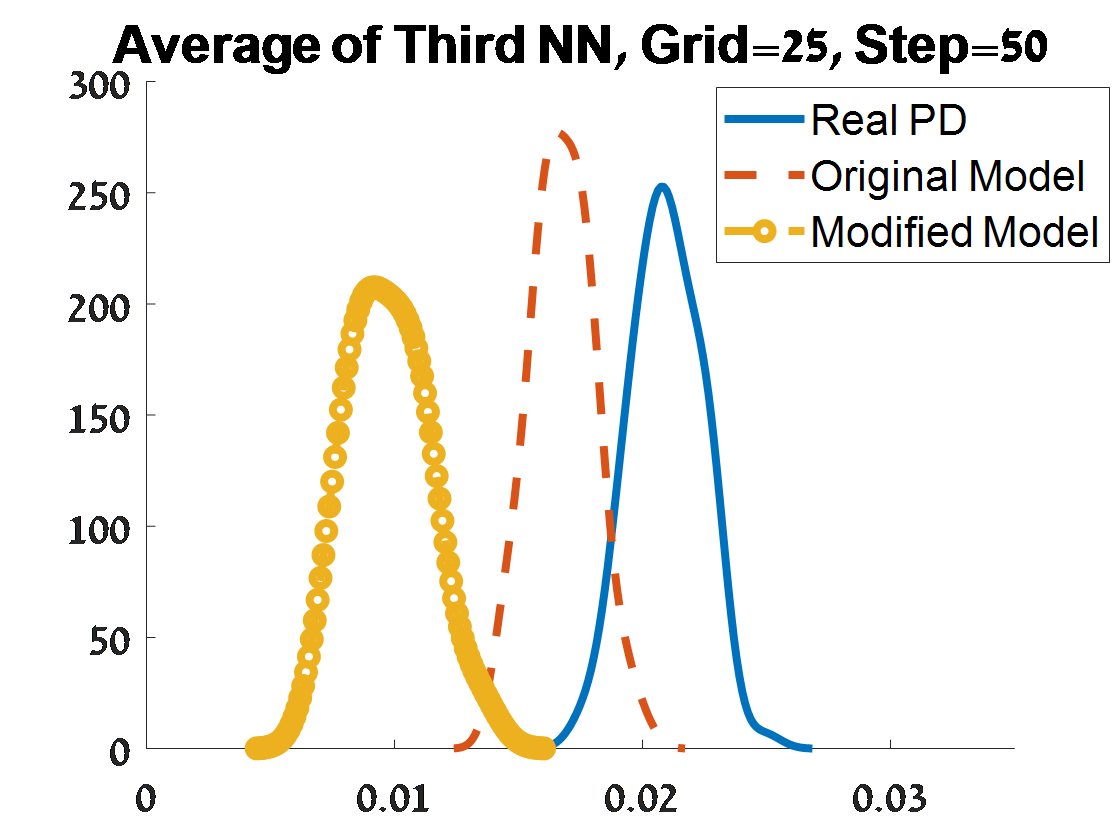}
\includegraphics[width=1.2in, height=1.25in]{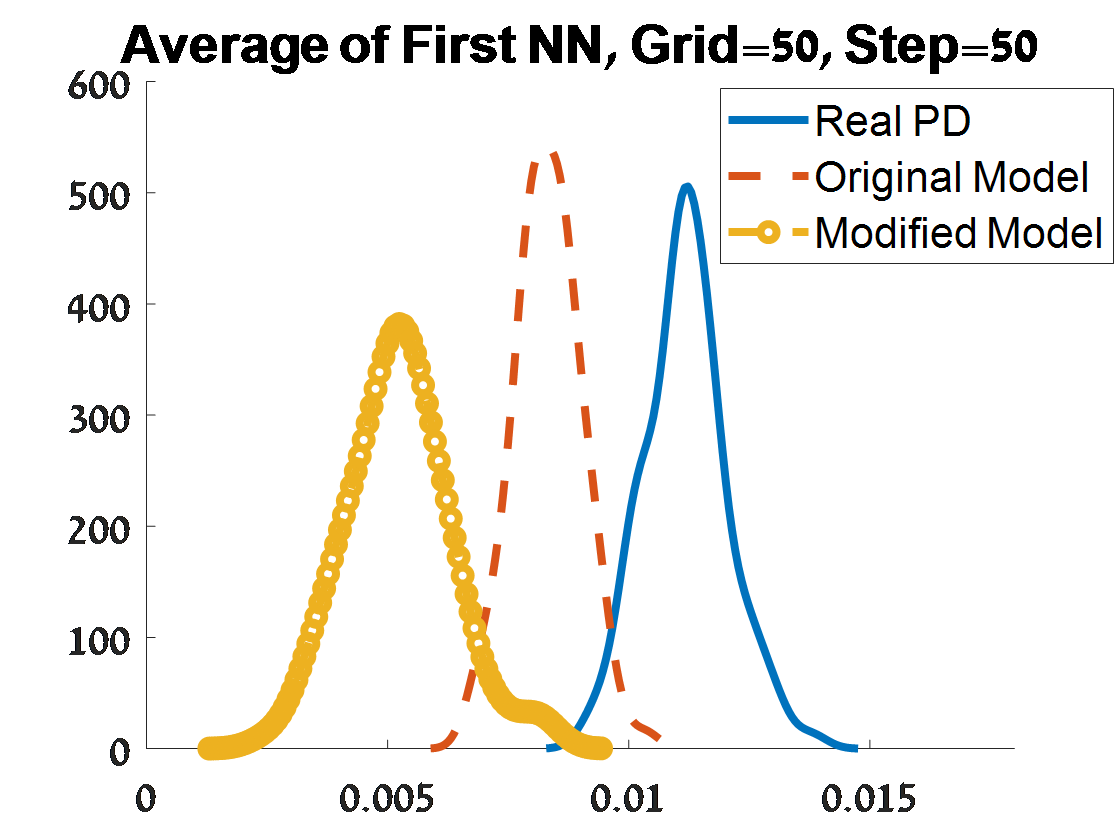}
\includegraphics[width=1.2in, height=1.25in]{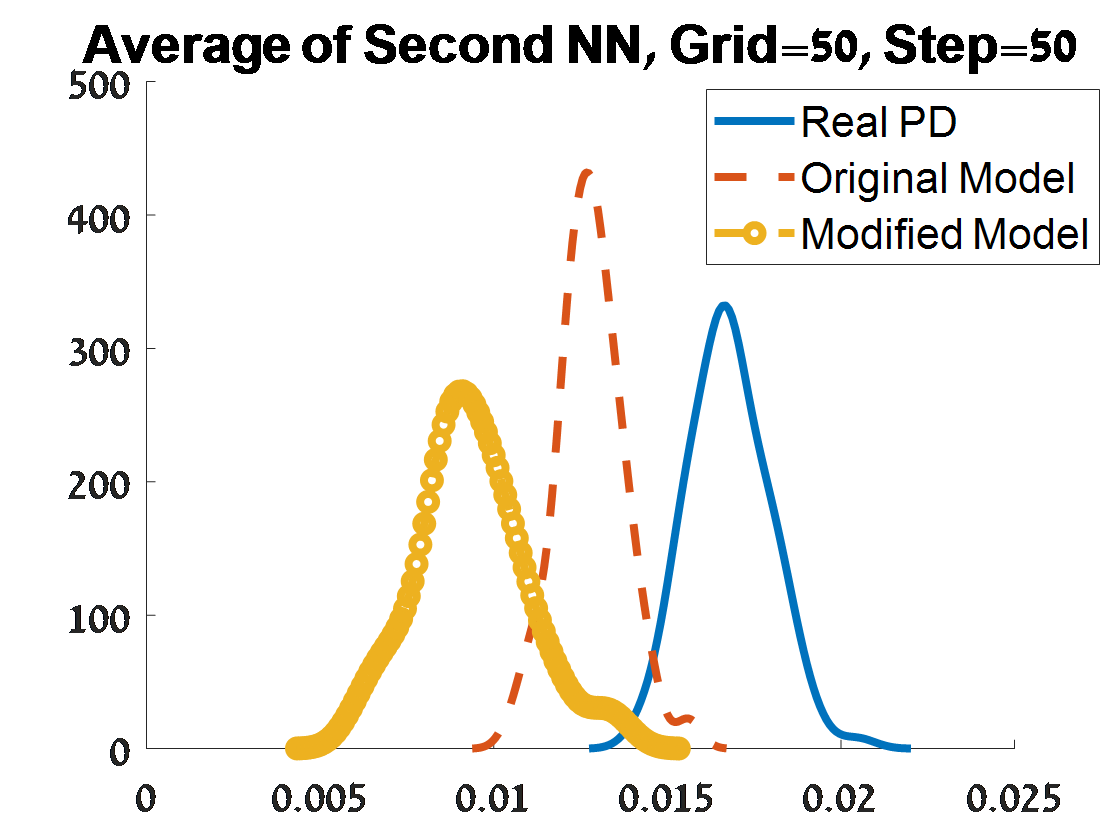}
\includegraphics[width=1.2in, height=1.25in]{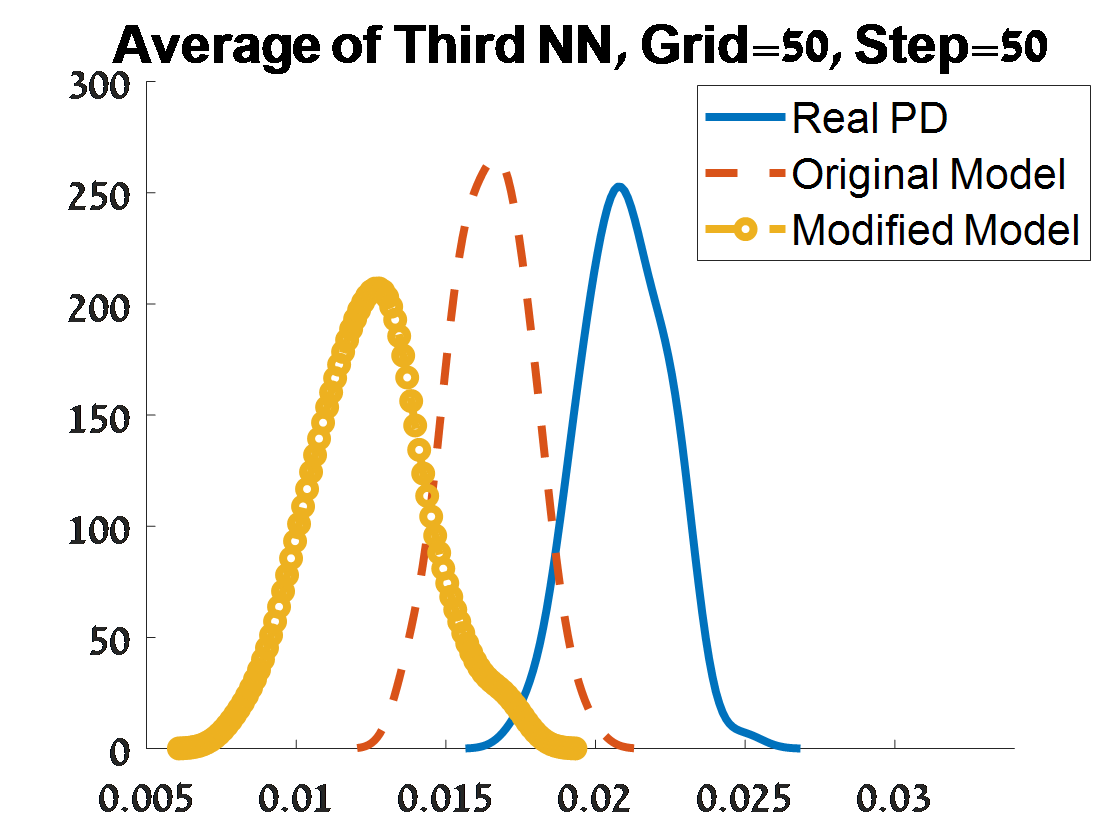}
\includegraphics[width=1.2in, height=1.25in]{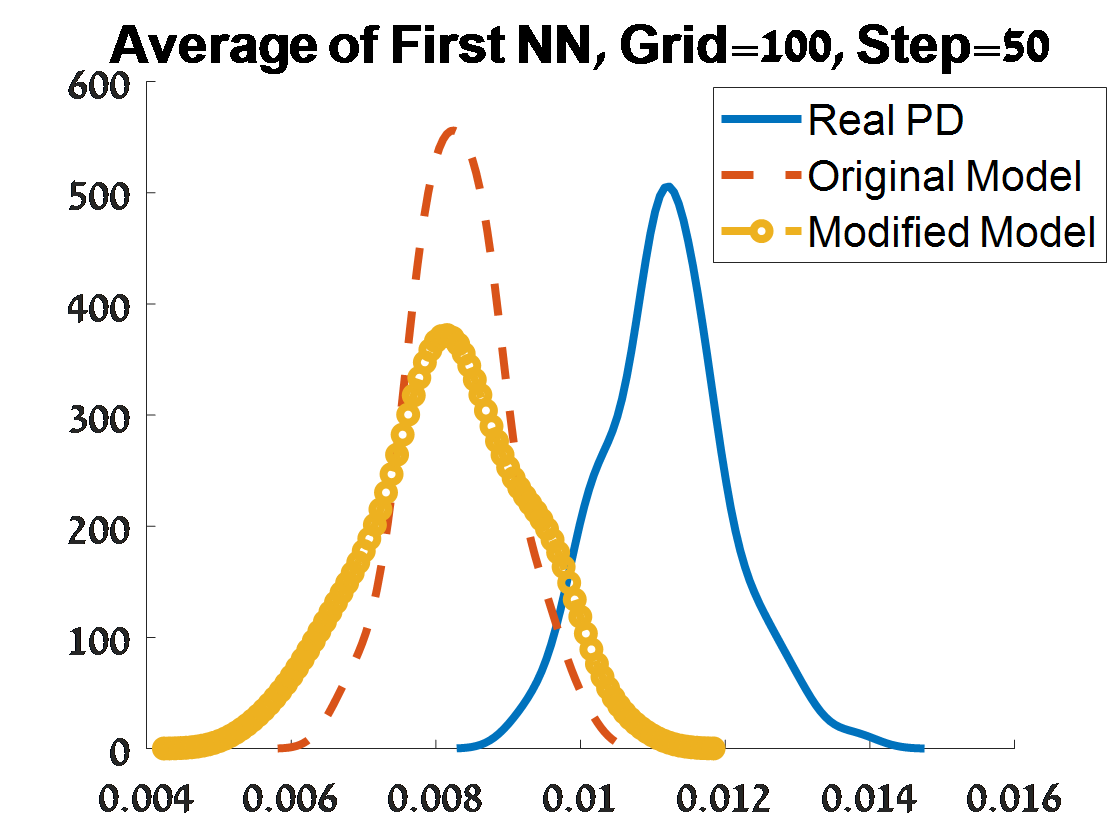}
\includegraphics[width=1.2in, height=1.25in]{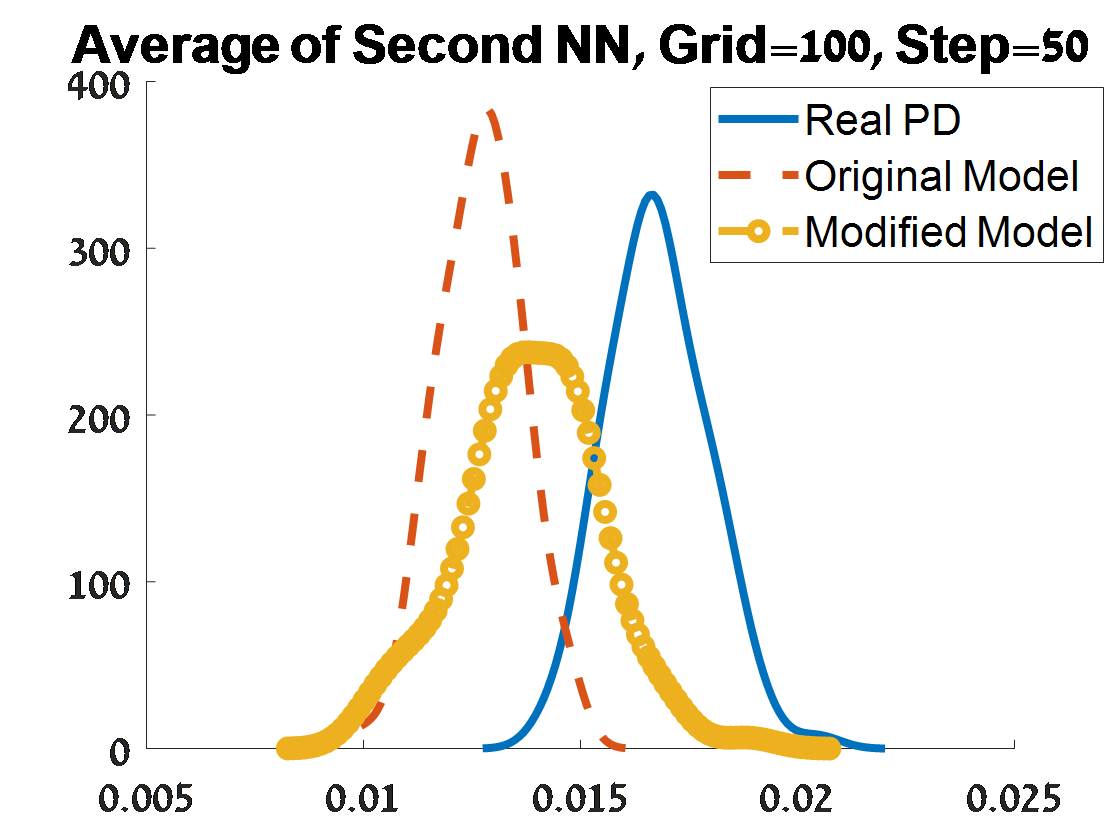}
\includegraphics[width=1.2in, height=1.25in]{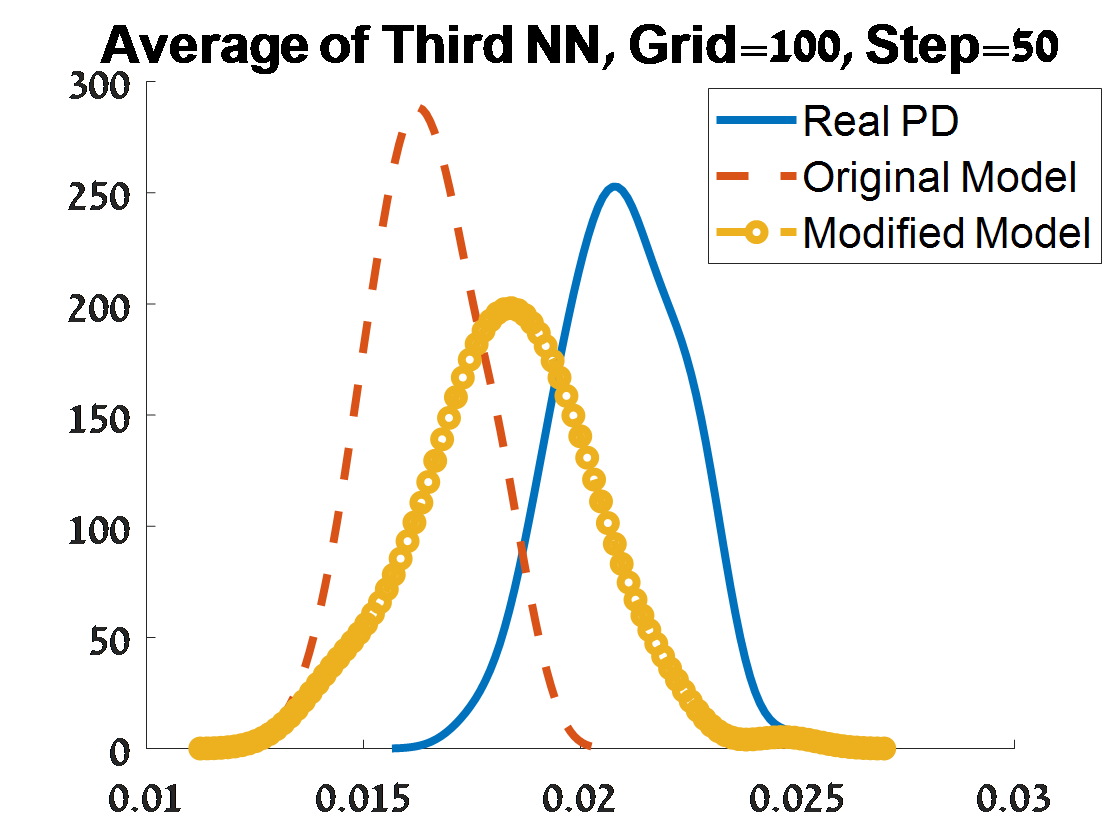}
\ec
%\caption{\footnotesize
% A random sample from two circles, 500 points from the larger circle and 300 from the smaller one,  with a kernel density
\caption{\footnotesize
Criterion 2 of goodness of fit for 100 $H_0$ PDs corresponded to 100 samples from a unit $S^3$. The figures depend on the grid of the proposal distribution ("Grid"), and the burn-in ("Step") of the MCMC algorithm.}
\label{fig:s3_H0_b}
\end{figure}
\end{landscape}

\begin{landscape}
\begin{figure}[h!]
\bc
\includegraphics[width=1.2in, height=1.25in]{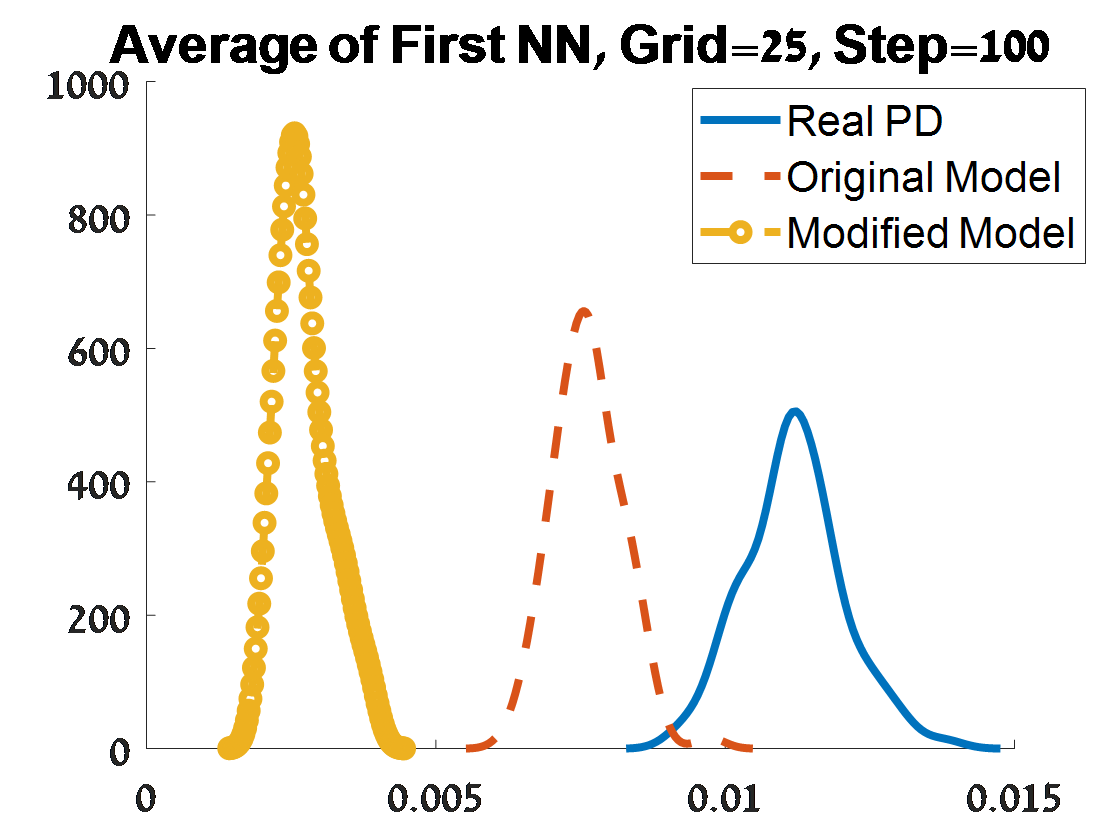}
\includegraphics[width=1.2in, height=1.25in]{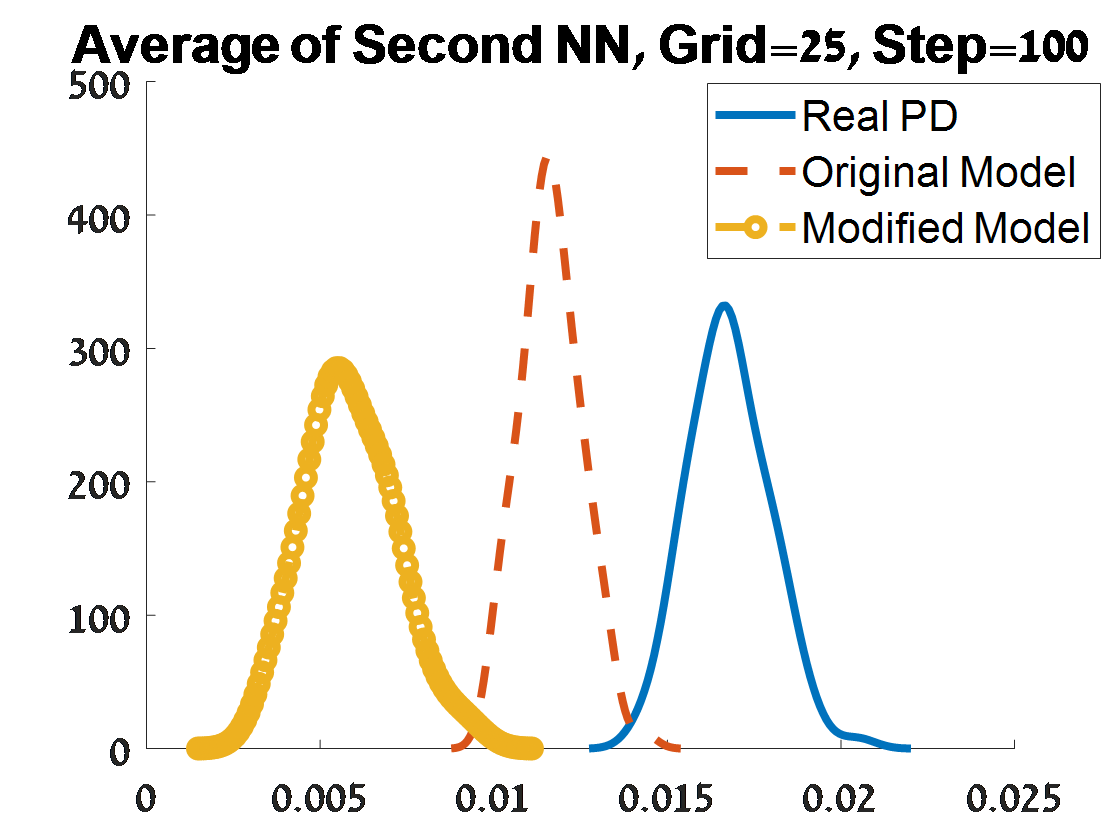}
\includegraphics[width=1.2in, height=1.25in]{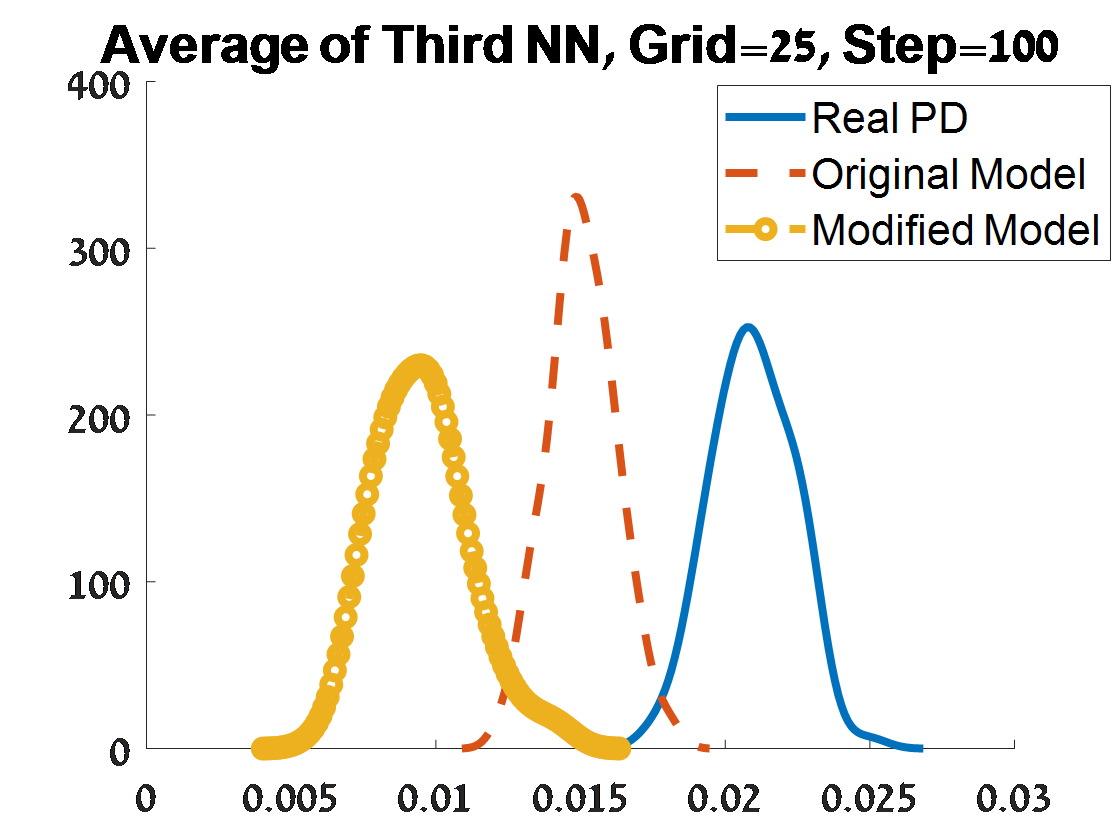}
\includegraphics[width=1.2in, height=1.25in]{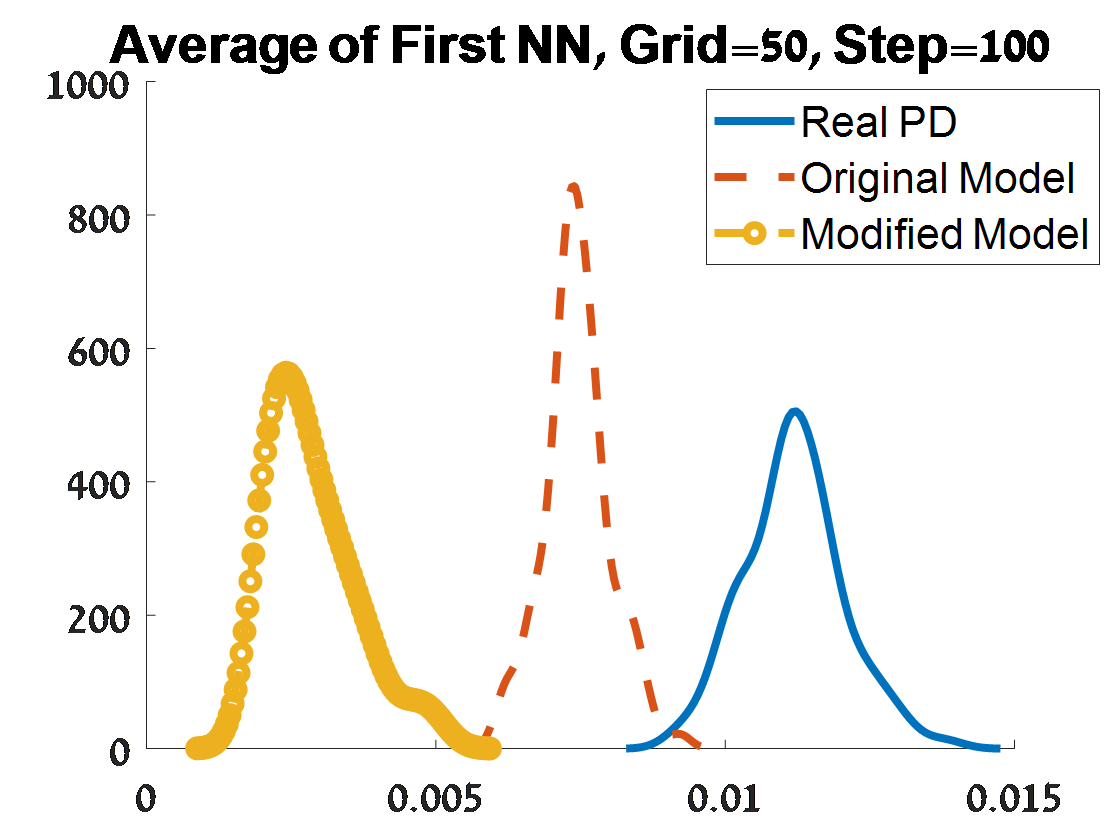}
\includegraphics[width=1.2in, height=1.25in]{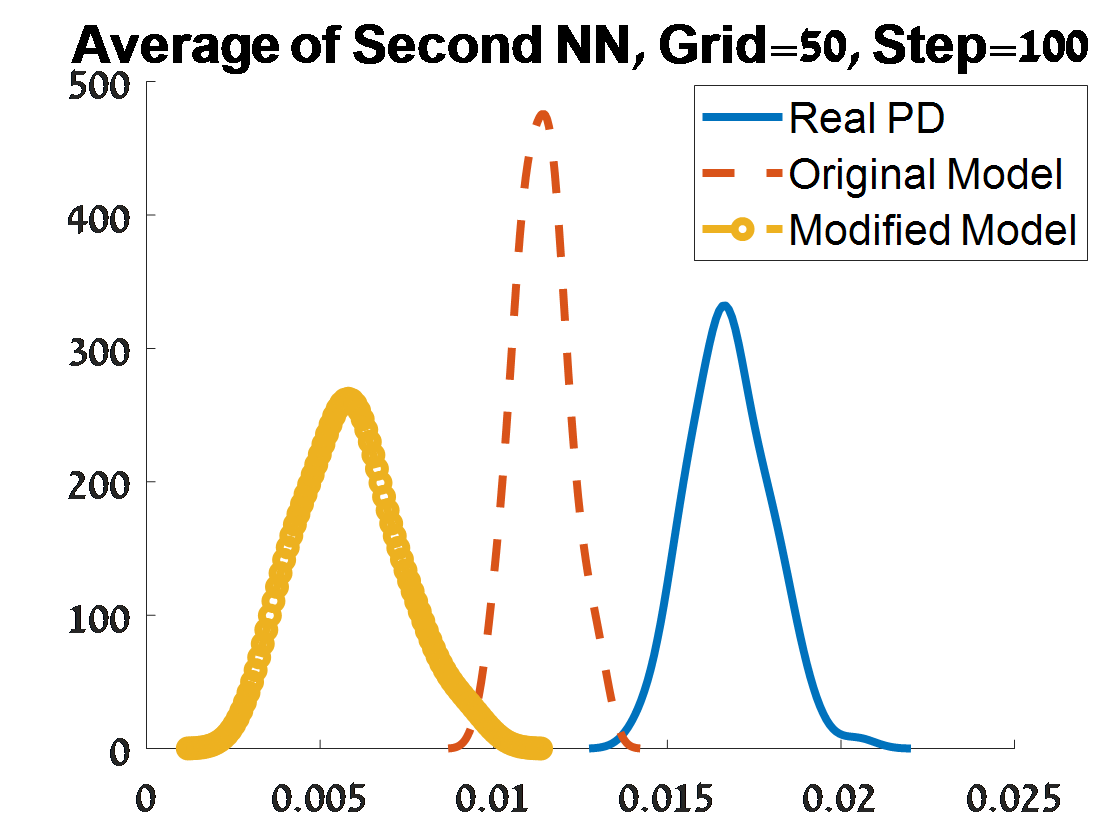}
\includegraphics[width=1.2in, height=1.25in]{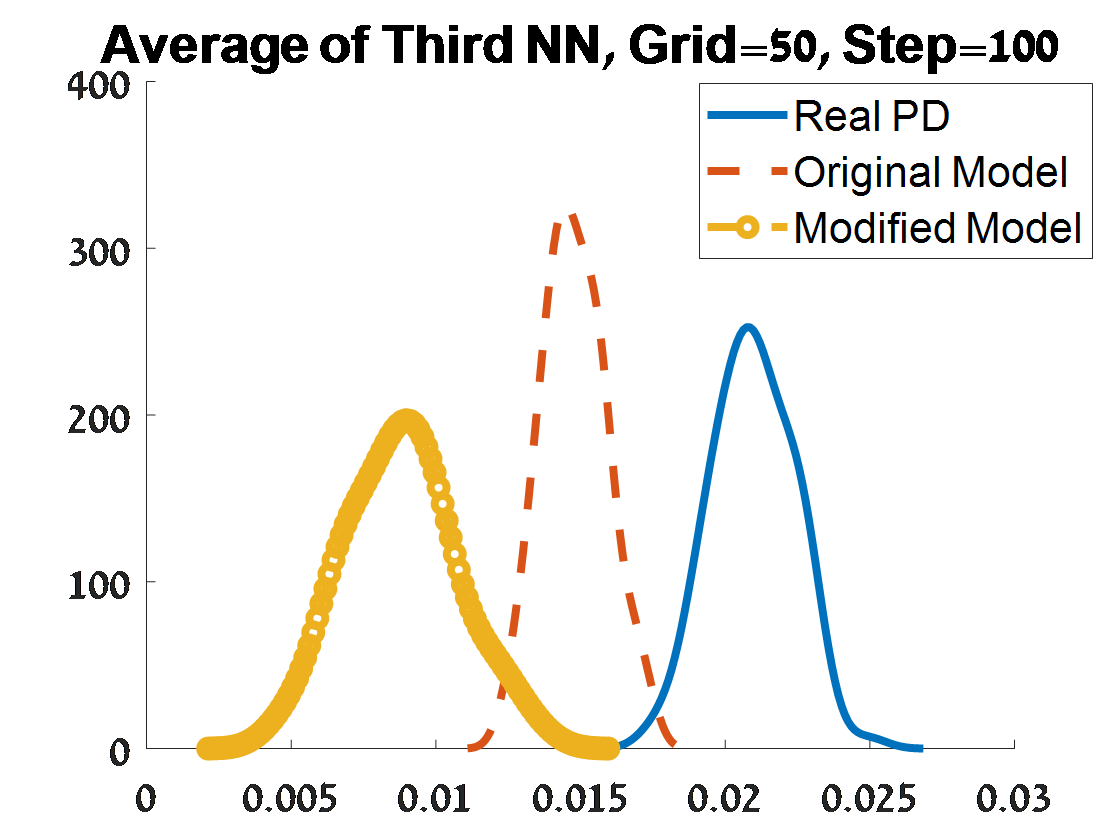}
\includegraphics[width=1.2in, height=1.25in]{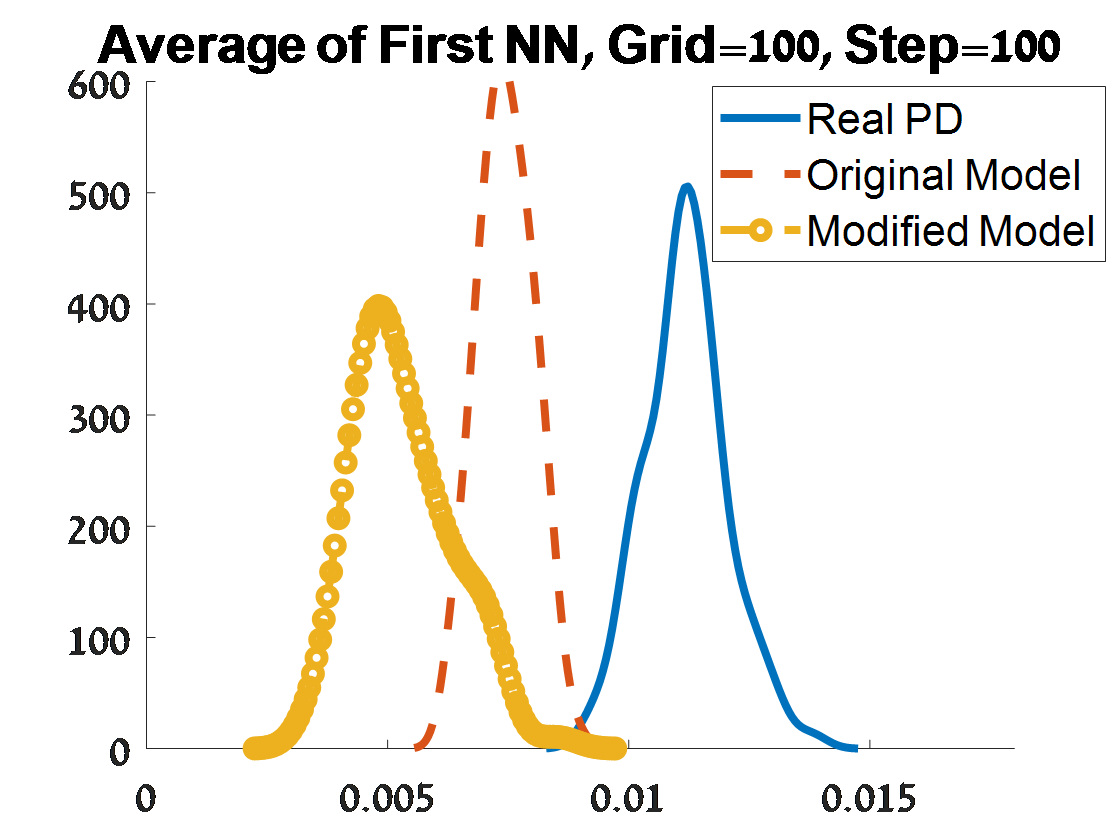}
\includegraphics[width=1.2in, height=1.25in]{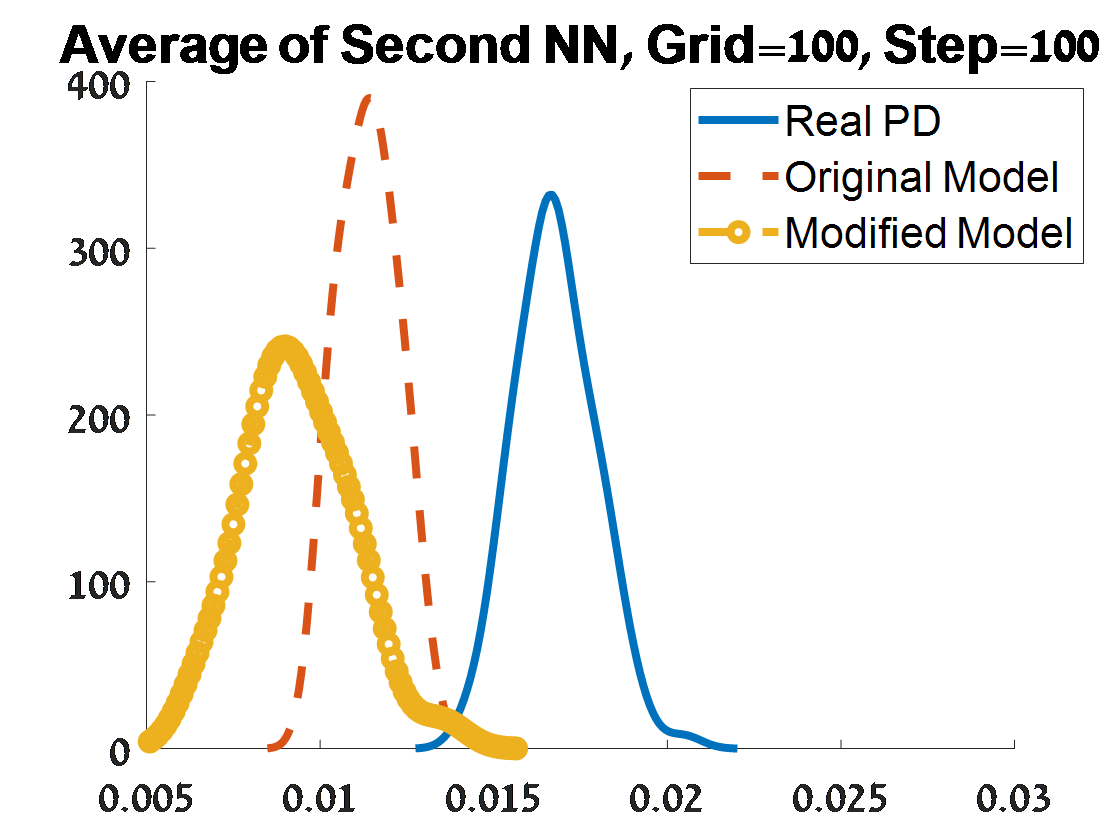}
\includegraphics[width=1.2in, height=1.25in]{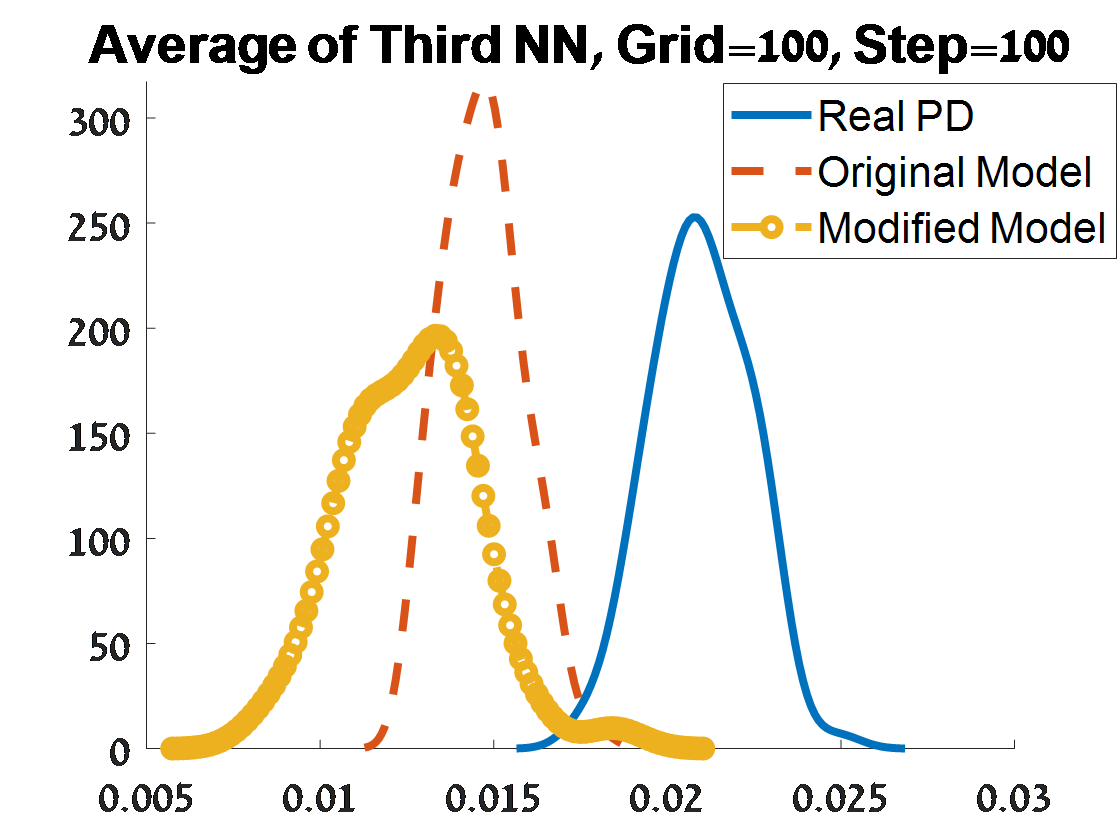}
\ec
%\caption{\footnotesize
% A random sample from two circles, 500 points from the larger circle and 300 from the smaller one,  with a kernel density
\caption{\footnotesize
 Continue of Criterion 2 of goodness of fit for 100 $H_0$ PDs corresponded to 100 samples from a unit $S^3$. The figures depend on the grid of the proposal distribution ("Grid"), and the burn-in ("Step") of the MCMC algorithm.}
\label{fig:s3_H0_c}
\end{figure}
\end{landscape}

%\begin{landscape}
%\begin{figure}[h!]
%\bc
%\includegraphics[width=1.8in, height=1.8in]{Sphere_s3_H0_pd30_grid50_step25}
%\includegraphics[width=1.8in, height=1.8in]{Sphere_s3_H0_pd30_grid100_step25}
%\includegraphics[width=1.8in, height=1.8in]{Sphere_s3_H0_pd60_grid50_step25}
%\includegraphics[width=1.8in, height=1.8in]{Sphere_s3_H0_pd60_grid100_step25}
%\ec
%%\caption{\footnotesize
%% A random sample from two circles, 500 points from the larger circle and 300 from the smaller one,  with a kernel density
%\caption{\footnotesize
%Examples of two PDs, each one is corresponded to a sample from a unit $S^3$. For each PD, the simulated PD based on the two model versions is described. The figures depend on the grid of the proposal distribution ("Grid"), and the burn-in ("Step") of the MCMC algorithm.}
%\label{fig:s3_H0_d}
%\end{figure}
%\end{landscape}

\subsubsection{The fitted model for $H_1$}
\begin{landscape}
\begin{figure}[h!]
\bc
\includegraphics[width=1.2in, height=1.4in]{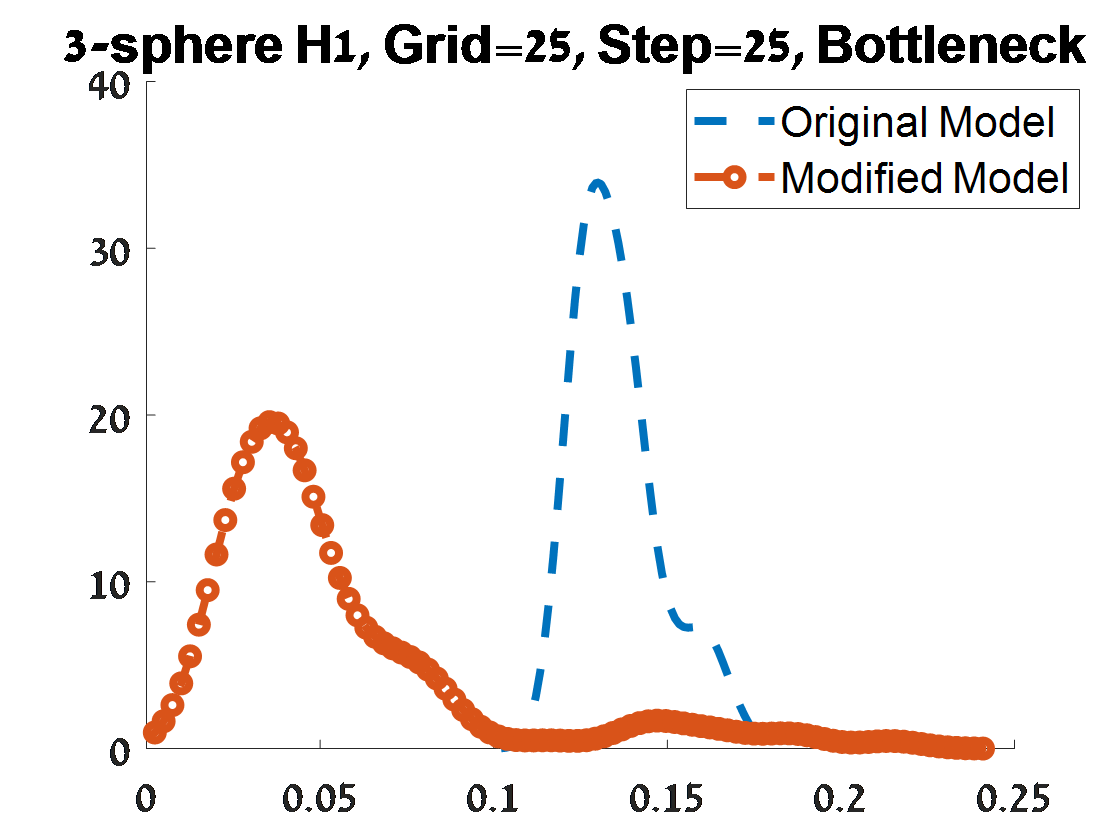}
\includegraphics[width=1.2in, height=1.4in]{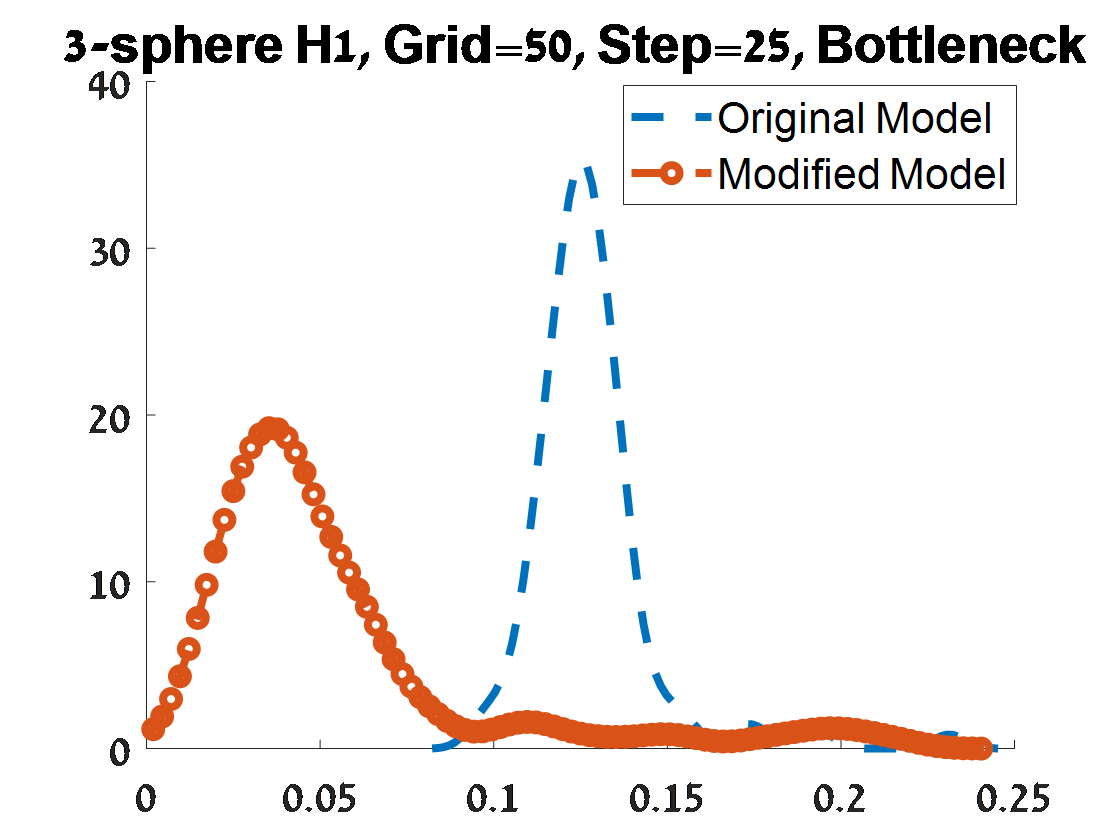}
\includegraphics[width=1.2in, height=1.4in]{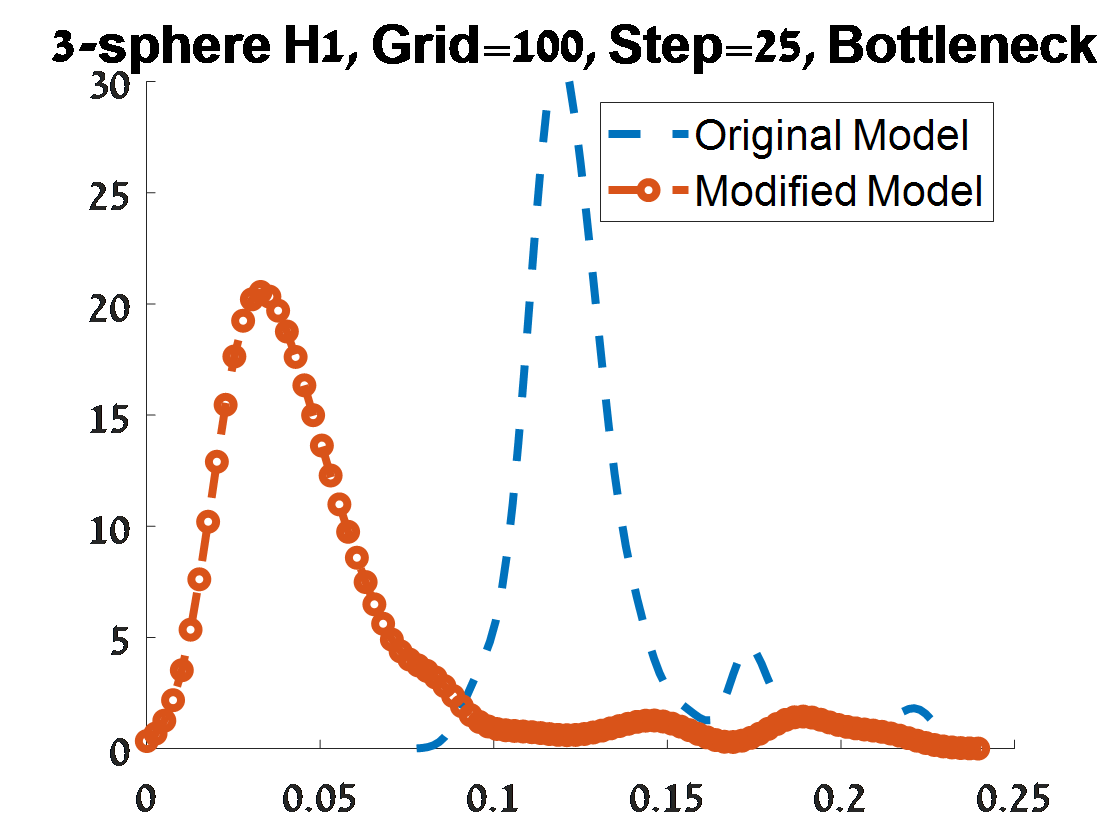}
\includegraphics[width=1.2in, height=1.4in]{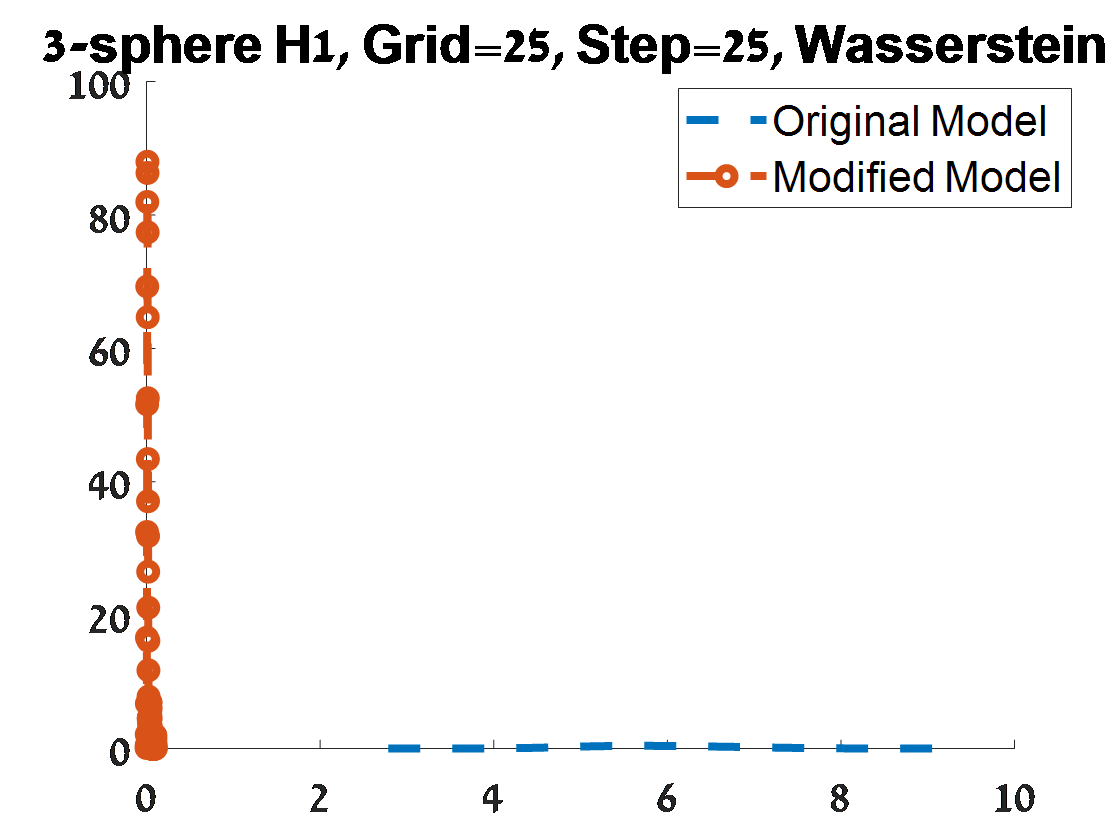}
\includegraphics[width=1.2in, height=1.4in]{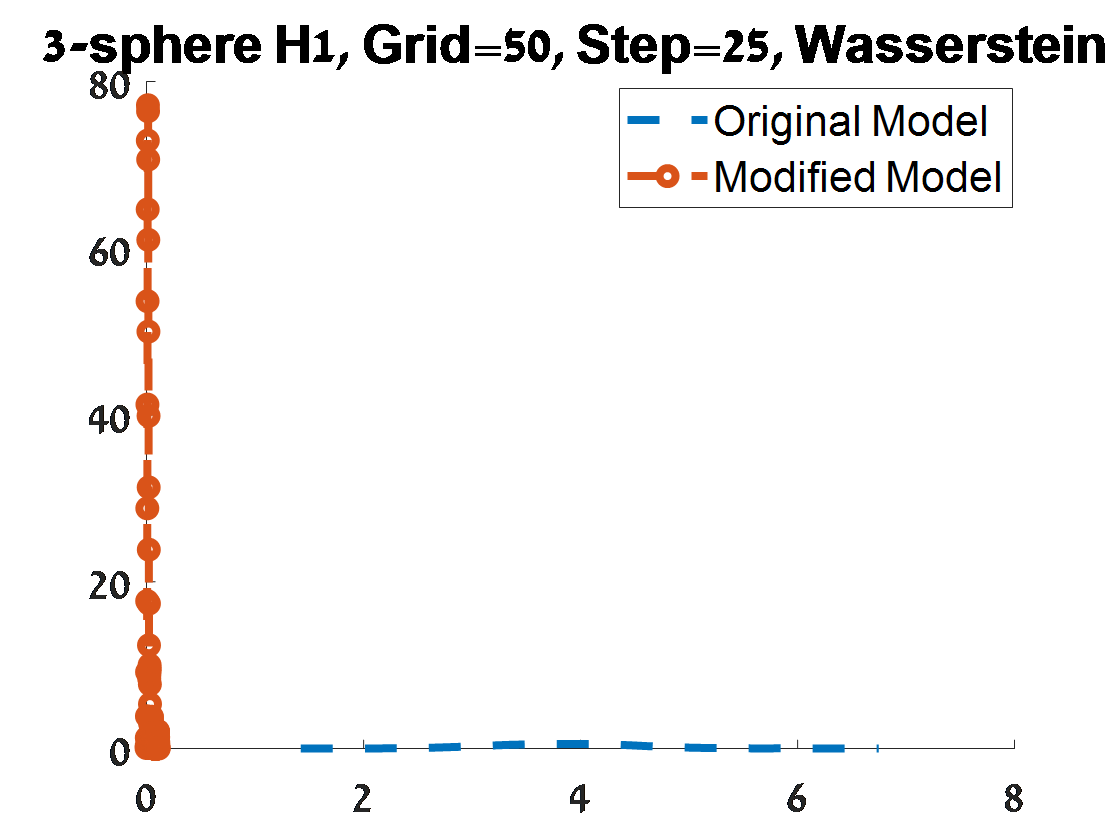}
\includegraphics[width=1.2in, height=1.4in]{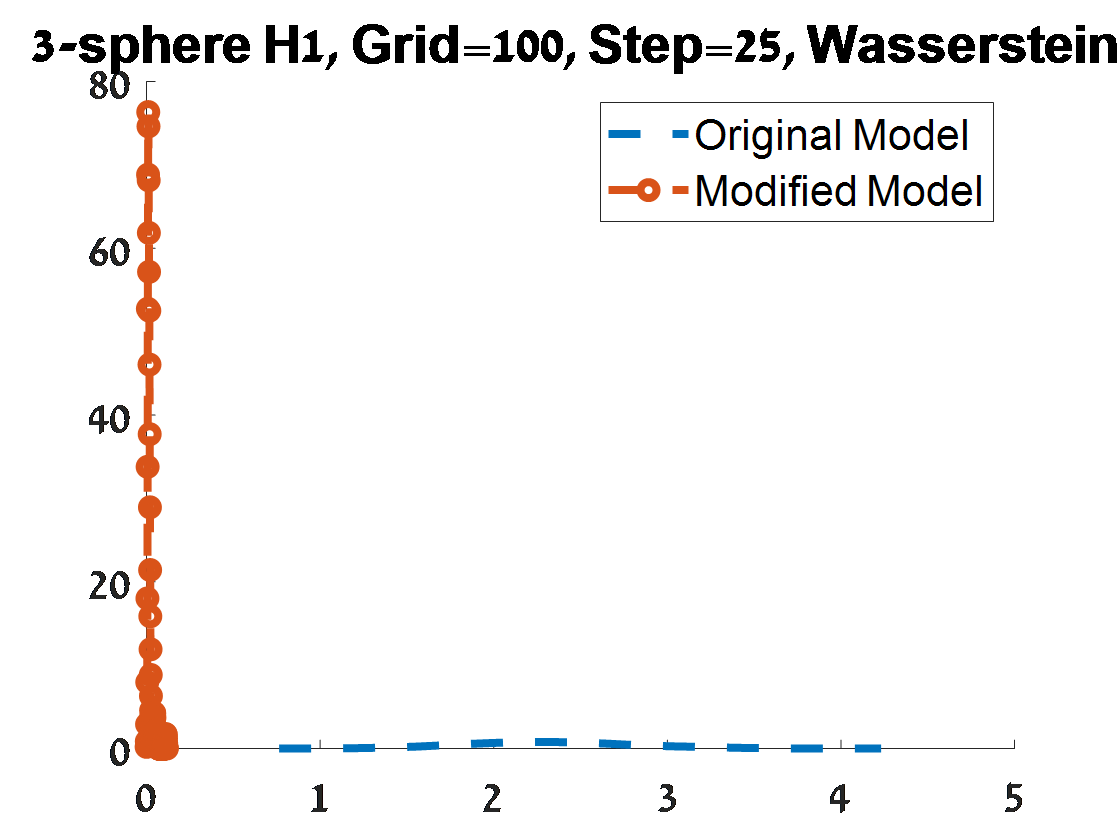}
%\\
\includegraphics[width=1.2in, height=1.4in]{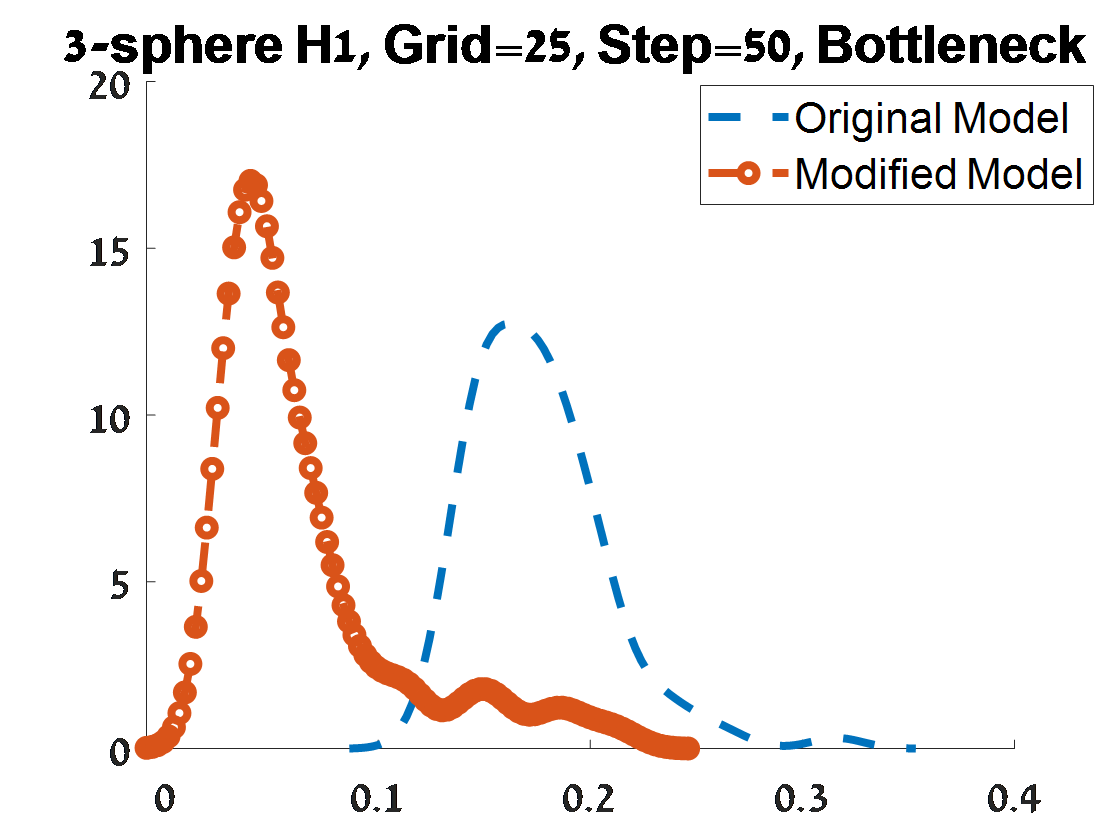}
\includegraphics[width=1.2in, height=1.4in]{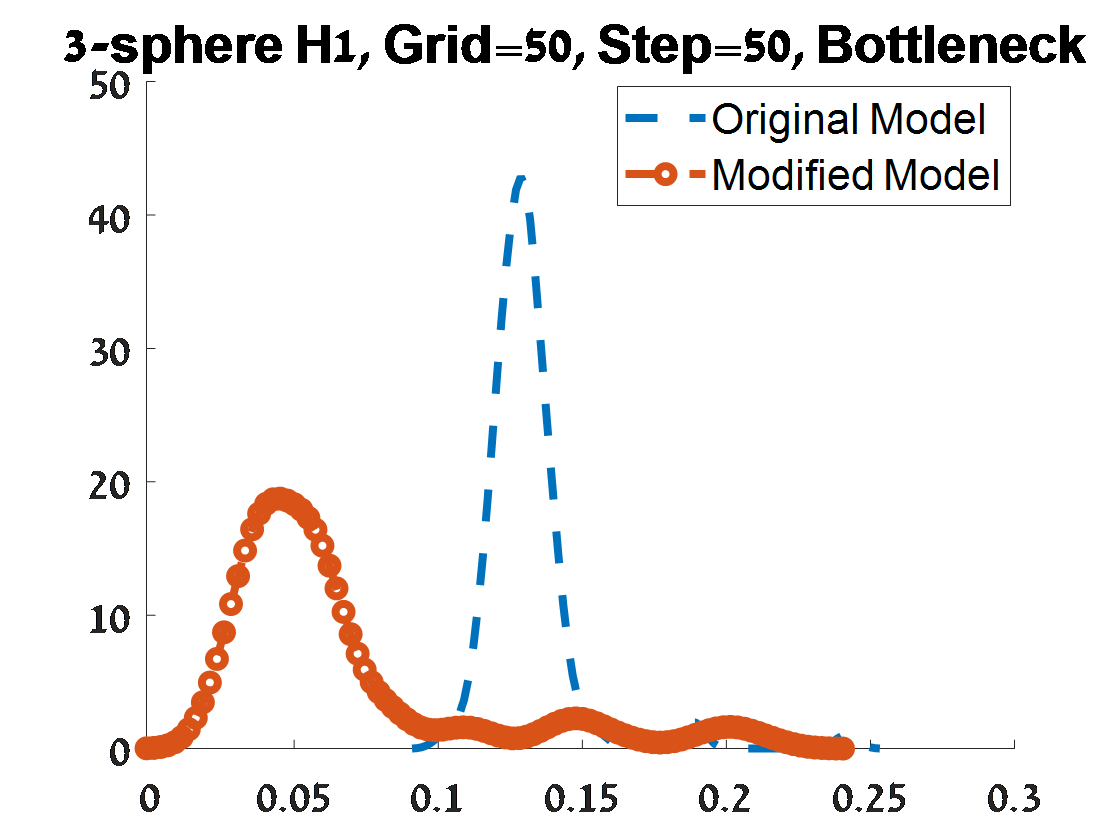}
\includegraphics[width=1.2in, height=1.4in]{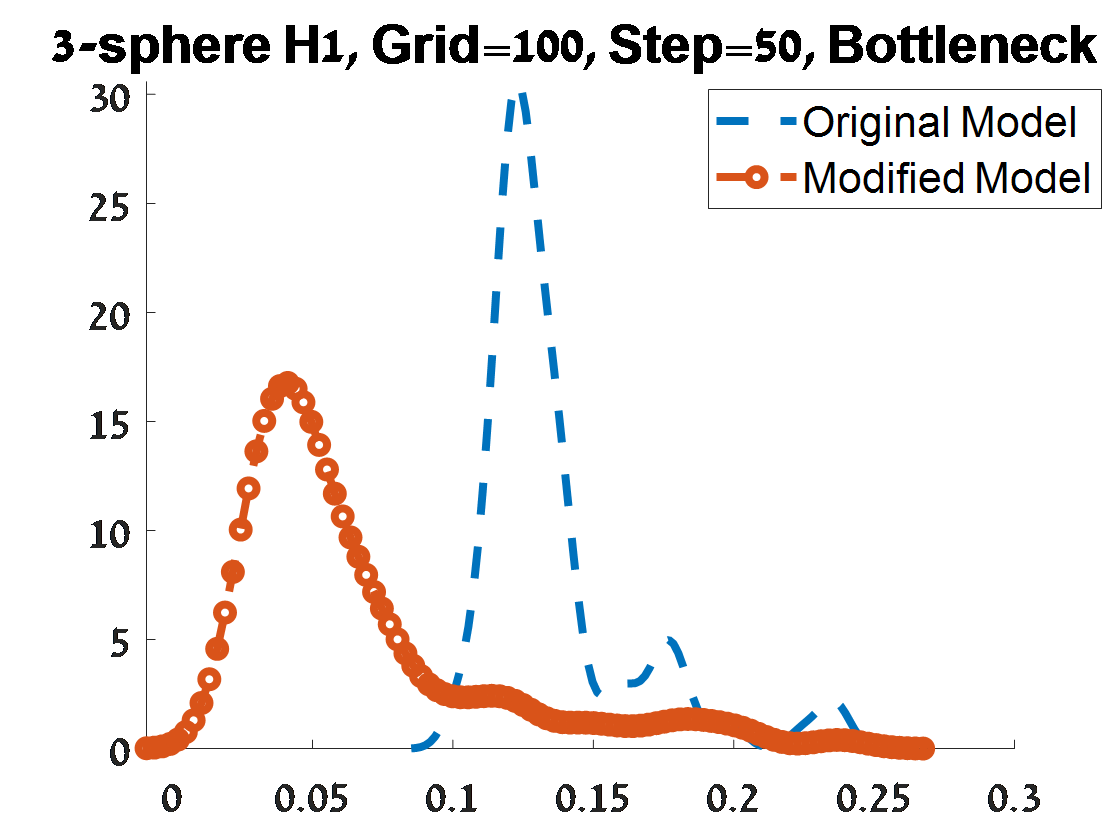}
\includegraphics[width=1.2in, height=1.4in]{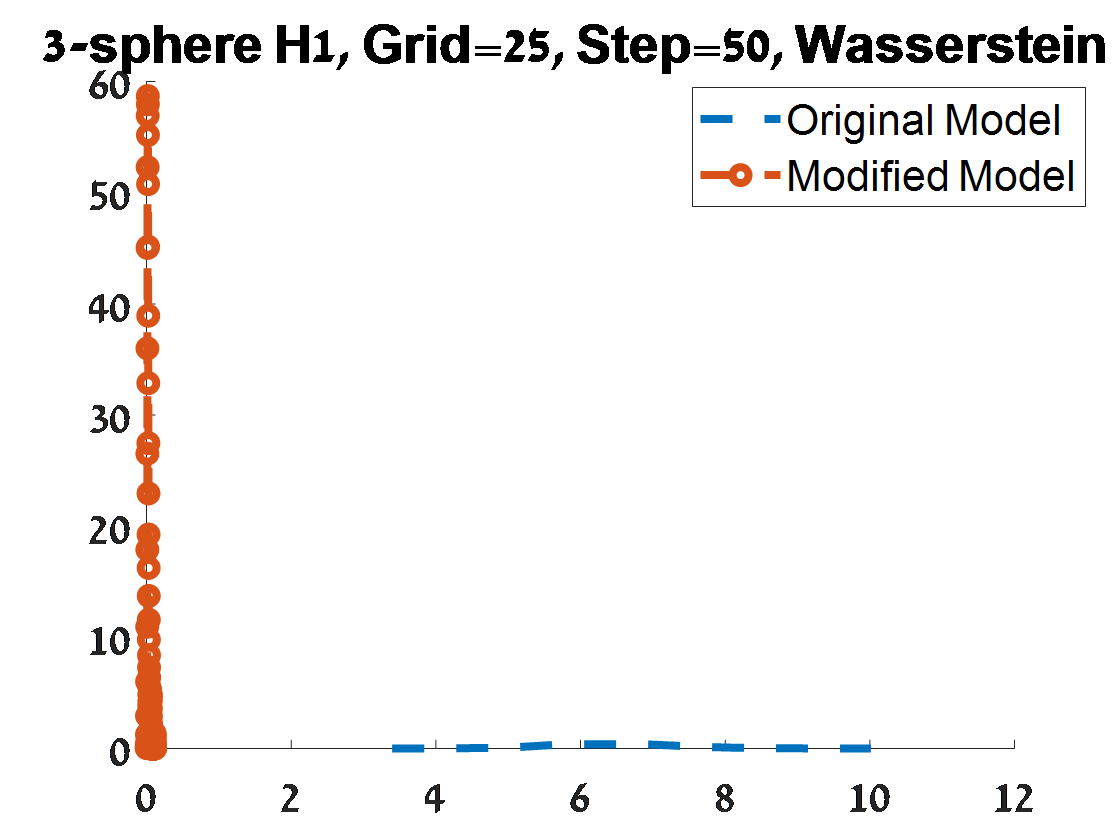}
\includegraphics[width=1.2in, height=1.4in]{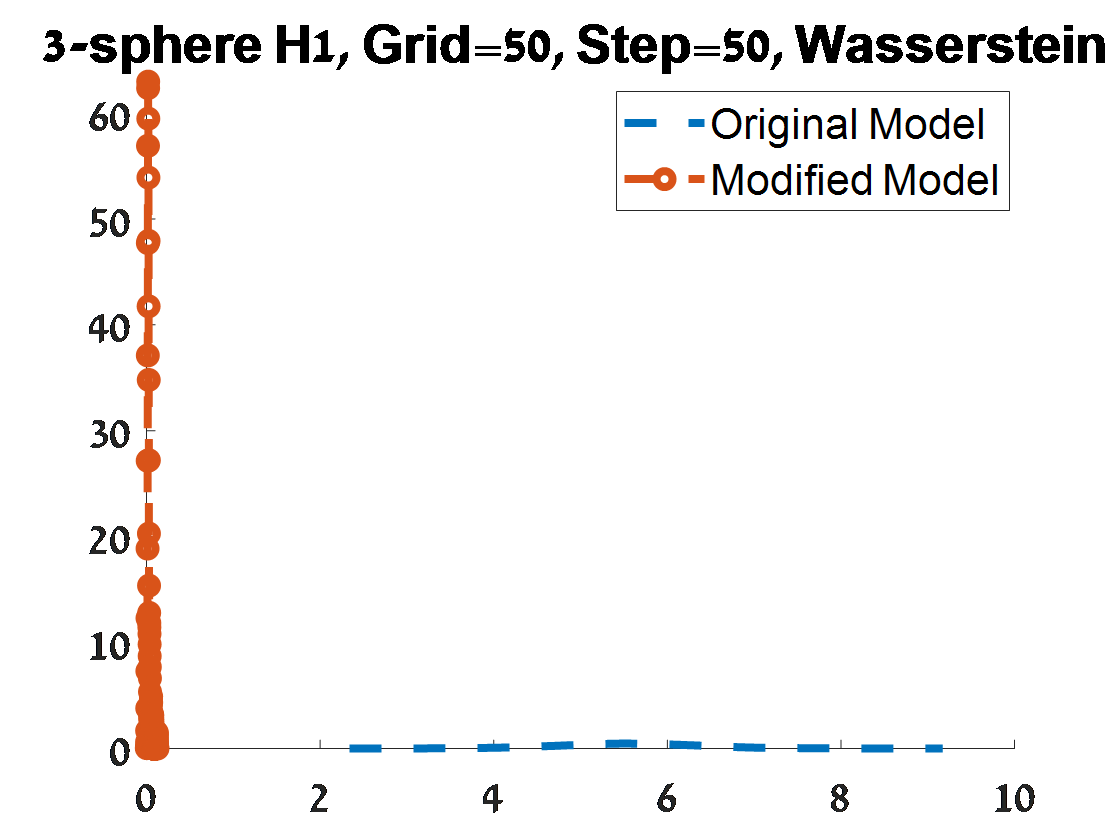}
\includegraphics[width=1.2in, height=1.4in]{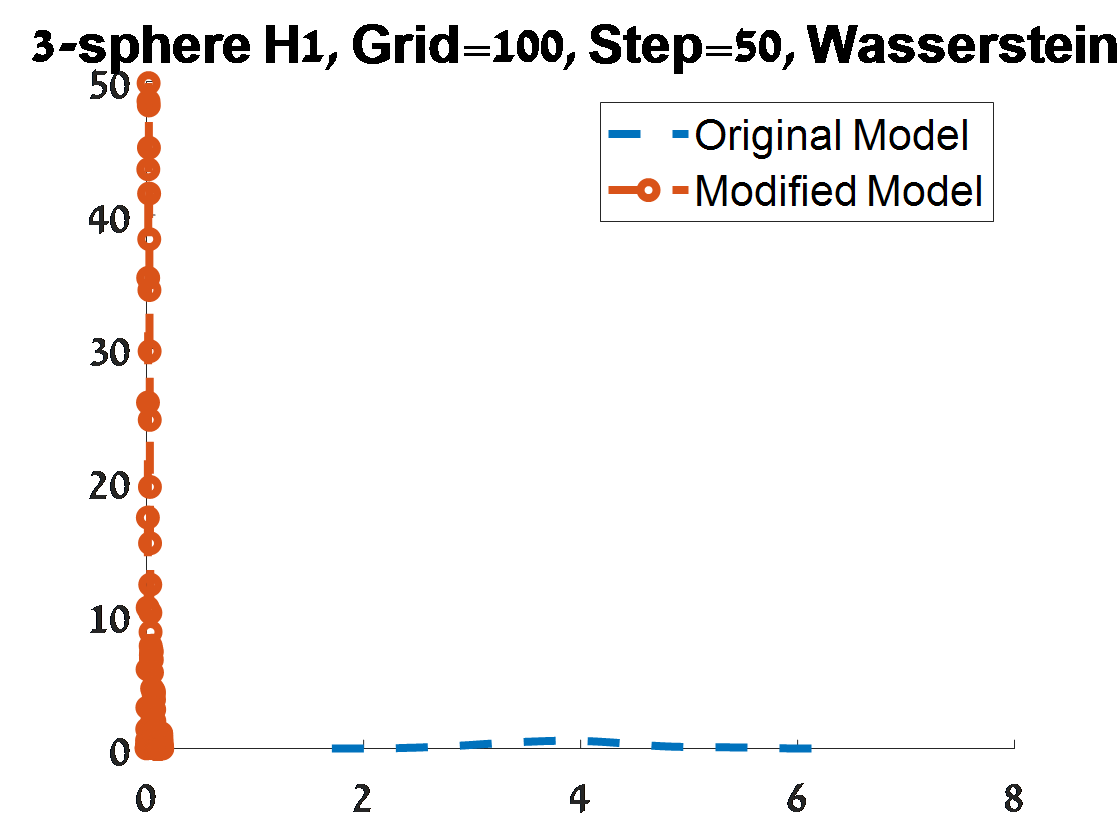}
%\\
\includegraphics[width=1.2in, height=1.4in]{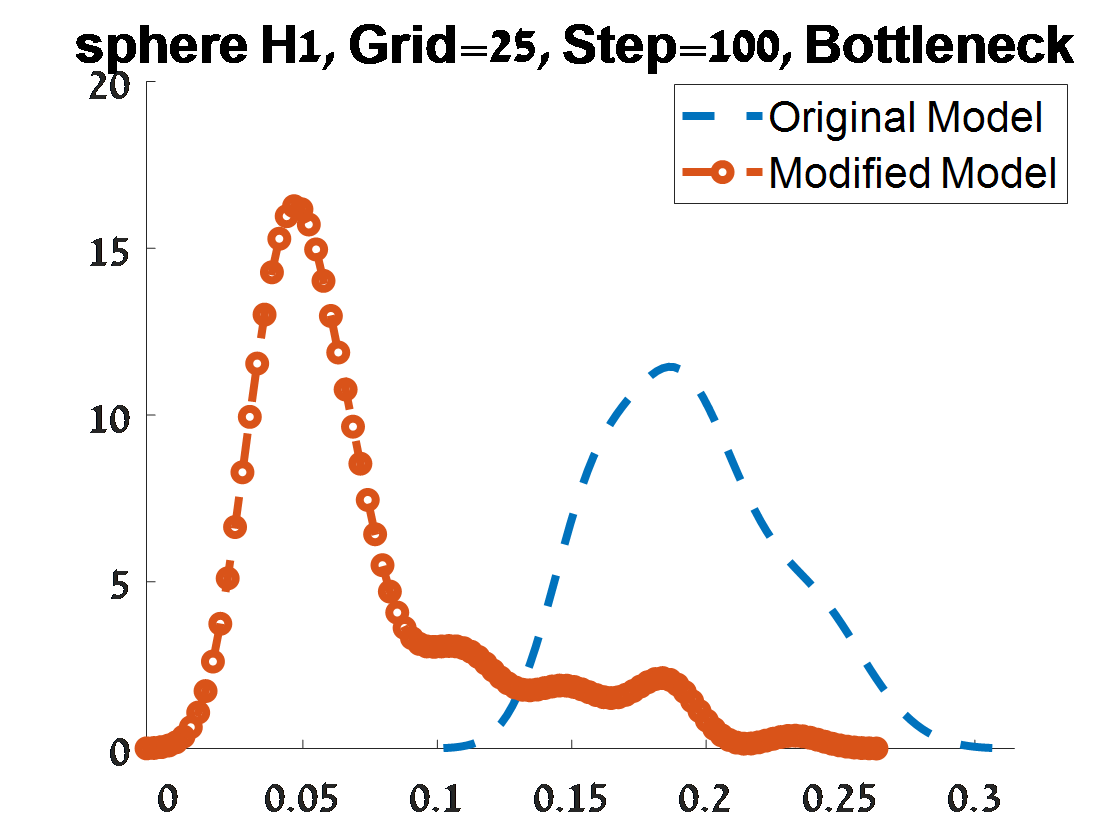}
\includegraphics[width=1.2in, height=1.4in]{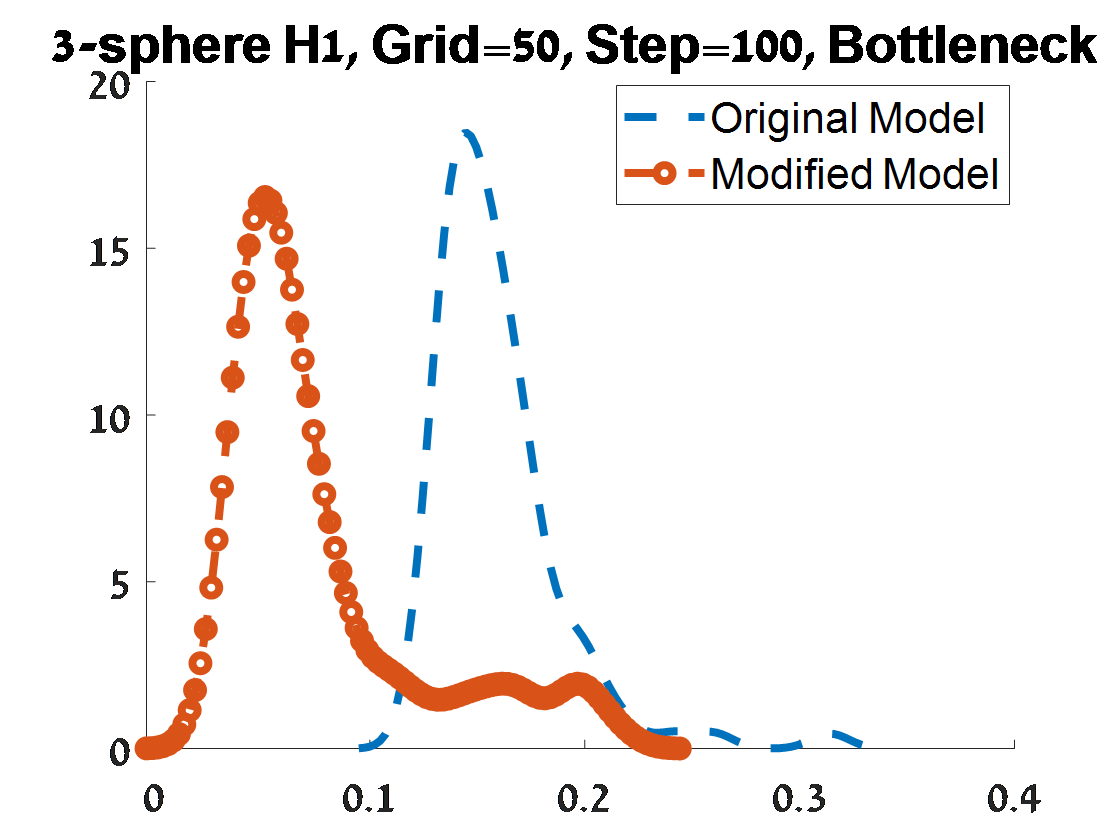}
\includegraphics[width=1.2in, height=1.4in]{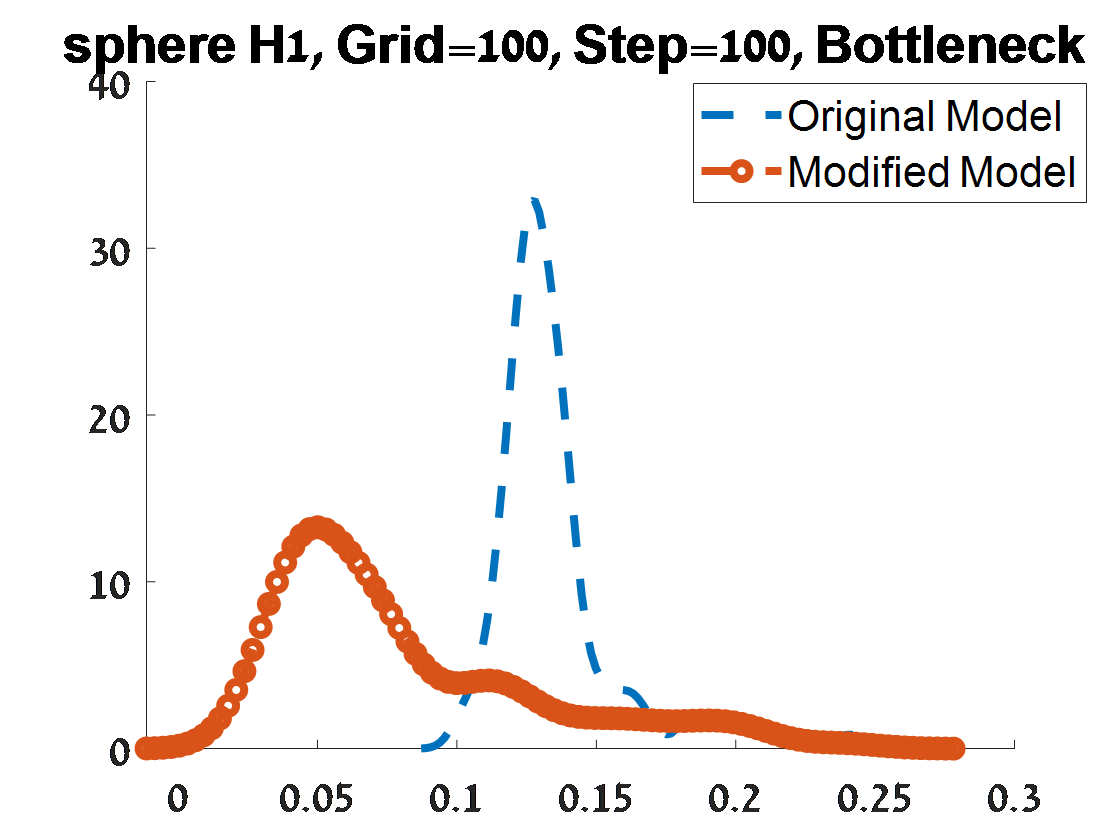}
\includegraphics[width=1.2in, height=1.4in]{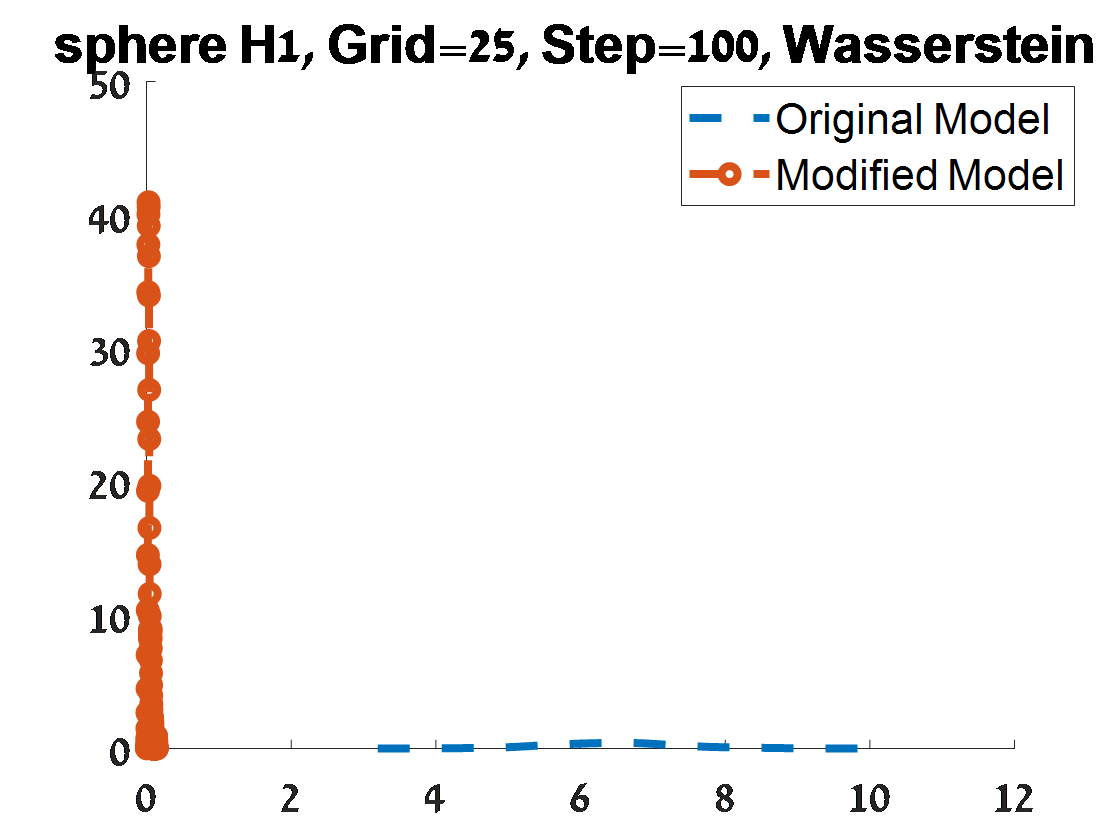}
\includegraphics[width=1.2in, height=1.4in]{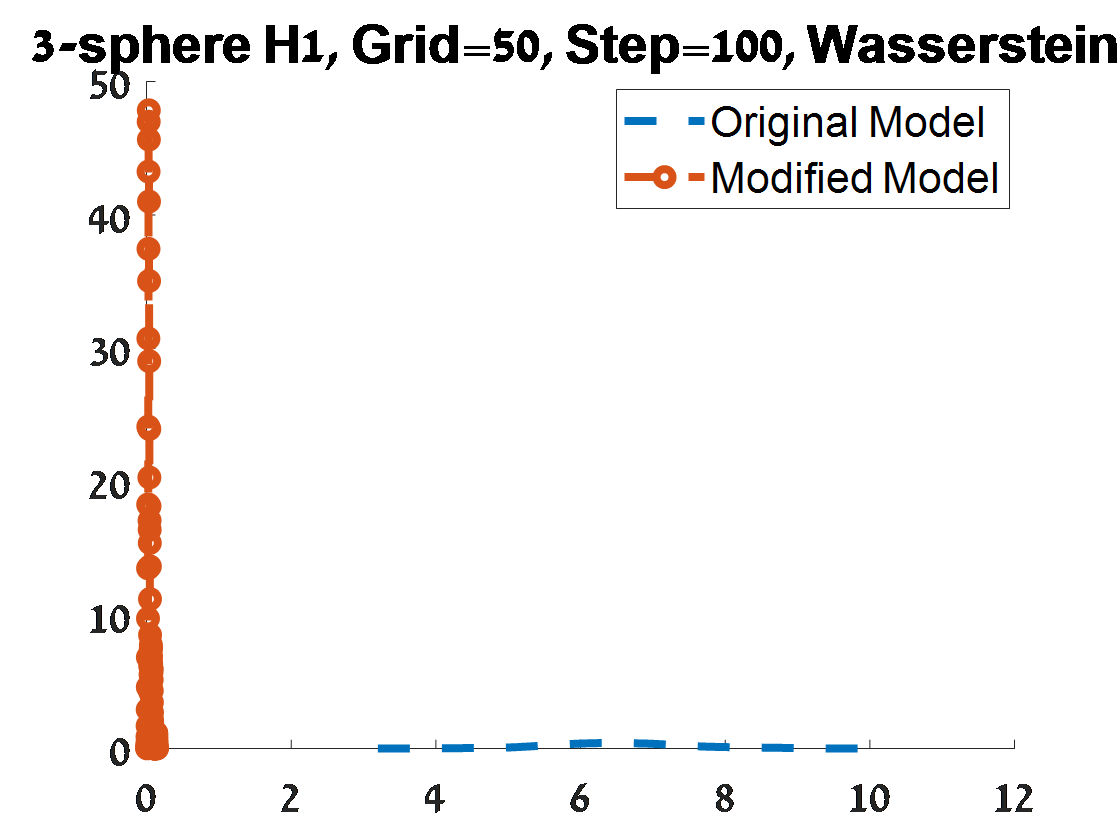}
\includegraphics[width=1.2in, height=1.4in]{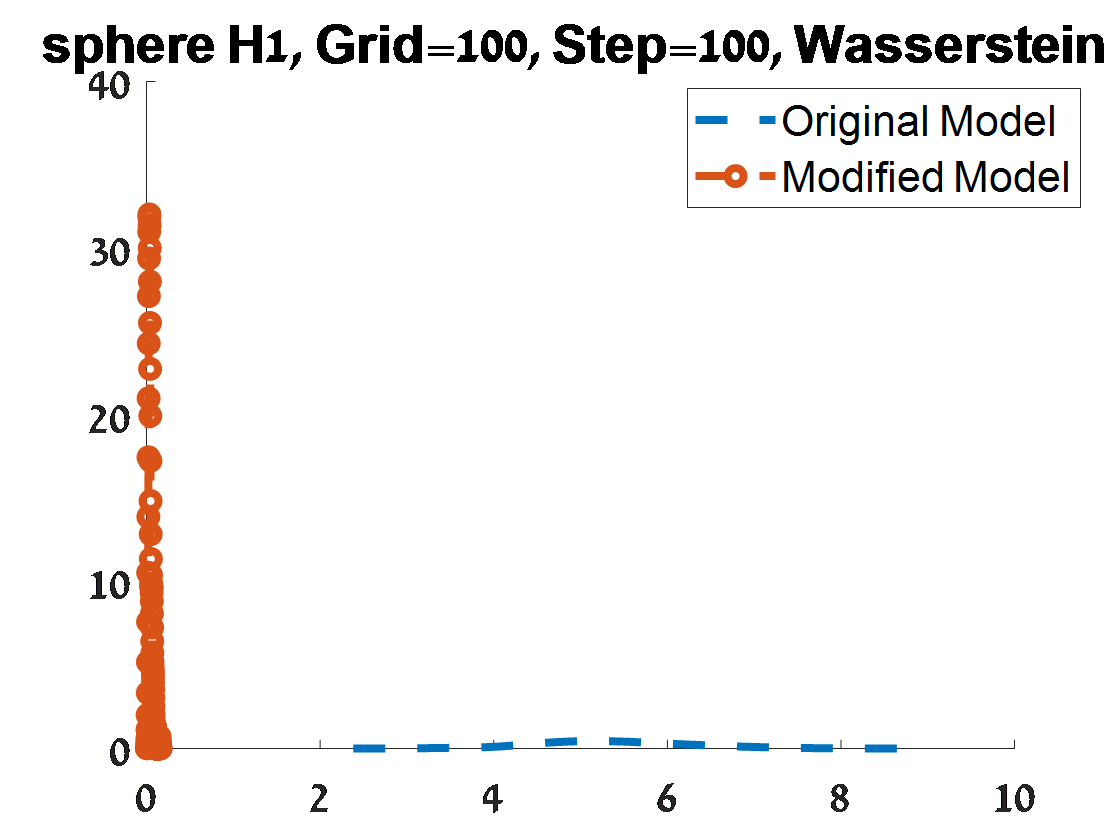}
\ec
\caption{\footnotesize
 Criterion 1 of goodness of fit for 100 $H_1$ PDs with  points corresponded to 100 samples from a unit $S^3$. The figures depend on the grid of the proposal distribution ("Grid"), and the burn-in ("Step") of the MCMC algorithm.}
\label{fig:s3_H1_a}
\end{figure}
\end{landscape}

Figure\ \ref{fig:s3_H1_a} describes the distributions over the 100 $H_1$ PDs of the first criterion of goodness of fit, and Figures 25-26 describe the distributions of the second criterion of goodness of fit.
Based on criterion 1, the modified model seems better under the both distance measures, that is, the distance of simulated based on the modified model is smaller than that under the original model. This is especially prominent in the Wasserstein distance.
As in the $S^2$ example, the advantage of the modified model under the $H_1$ PDs is less extreme relative to the setting of $H_0$ PDs. The Wasserstein distance has higher values in $H_0$ than in $H_1$. The Bottleneck distance is a little higher in $H_0$ than in $H_1$, but still has moderate values relative to the values of Wasserstein distance. The reason for these results is the large variability in the points on the persistence diagrams H0 relative to H1.

Bases on criterion 2, as in $H_0$, the distributional properties for small grid sizes under the original model are close to those of the real PDs than the the distributional properties under the modified model. But the contrary for grid size of 100x100 and burn-in of 25, where properties of the modified model are close to these properties of the real PDs. The same variability of these distributions that had observed in $H_0$ also can be seen here for $H_1$.

%Figure\ \ref{fig:s3_H1_d} presents two examples of real PD and its simulated PD based on the two model versions, only for the best scenarios we found, that is grid=50,100, and step=25. the modified model is better and succeed in creating the far points from the diagonal, where the original model is concentrating in creating more the points that are close to the diagonal and almost cannot succeed in generating the far points from the diagonal.

 %A comparison of two samples drawn from a circle with radius $r=1$ (the unit circle).
%
%
%Each example is described in Figure\ \ref{fig:circle} based on a sample of $n=1000$. The first plot of each example describes the object sample, and to its right we present the corresponding persistence diagram. The black circles indicating connected components ($H_0$ persistence), the red triangles corresponding to holes ($H_1$), and the blue diamonds corresponding to voids ($H_2$).
\begin{landscape}
\begin{figure}[h!]
\bc
\includegraphics[width=1.2in, height=1.25in]{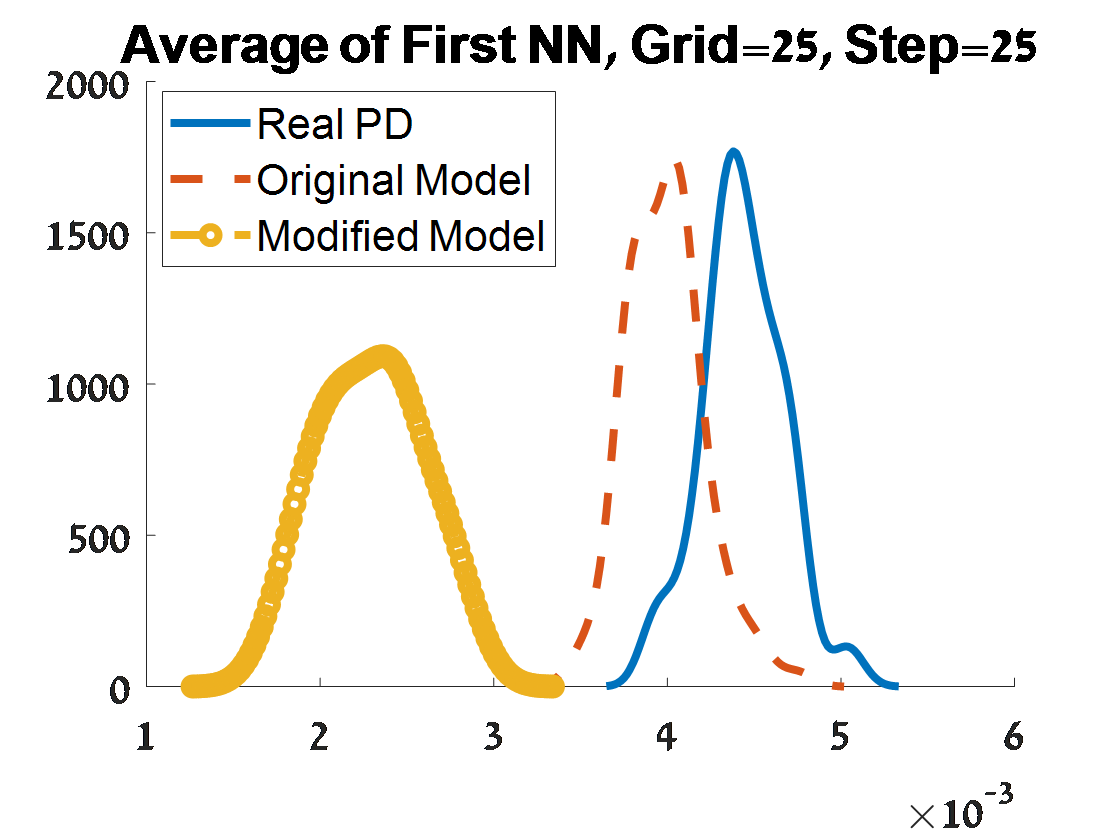}
\includegraphics[width=1.2in, height=1.25in]{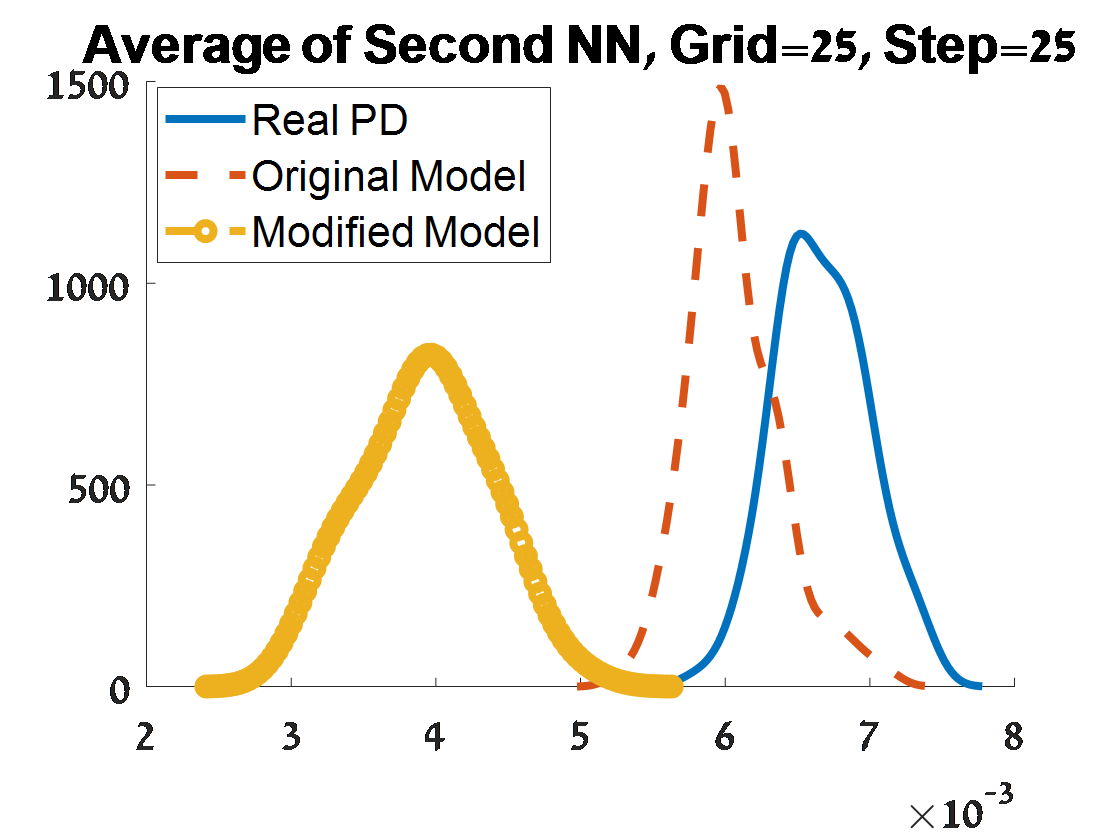}
\includegraphics[width=1.2in, height=1.25in]{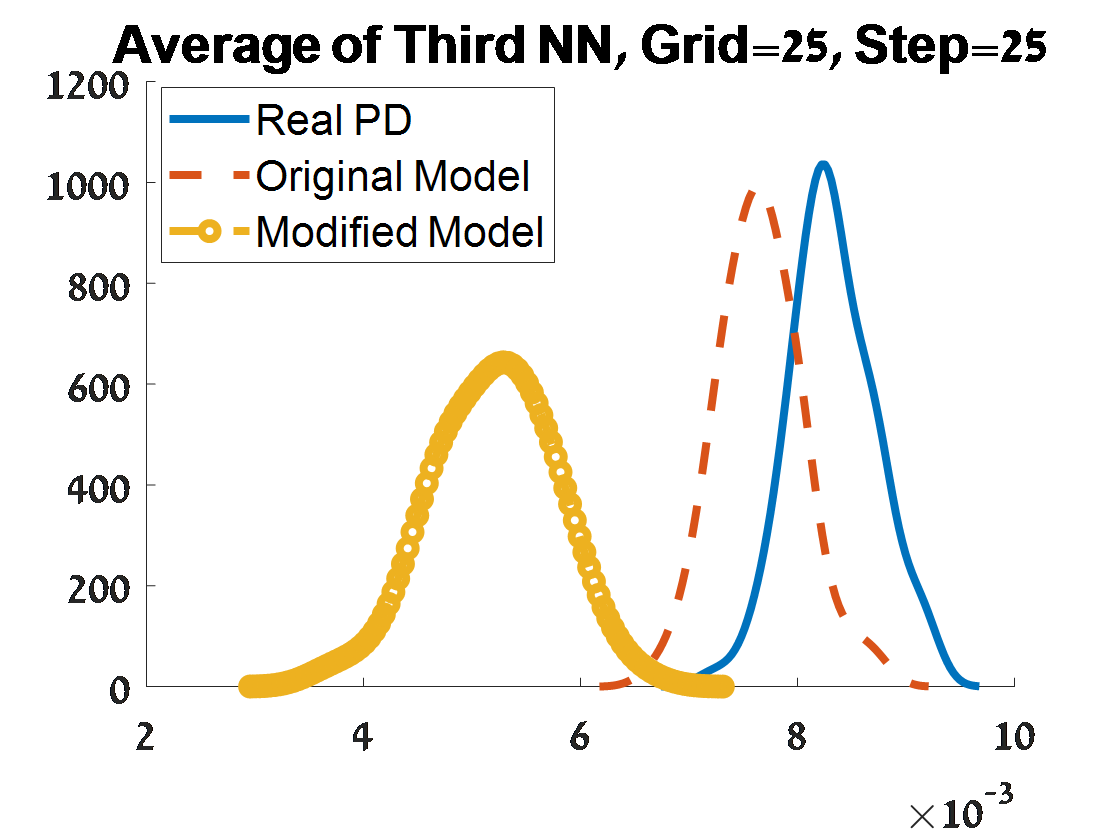}
\includegraphics[width=1.2in, height=1.25in]{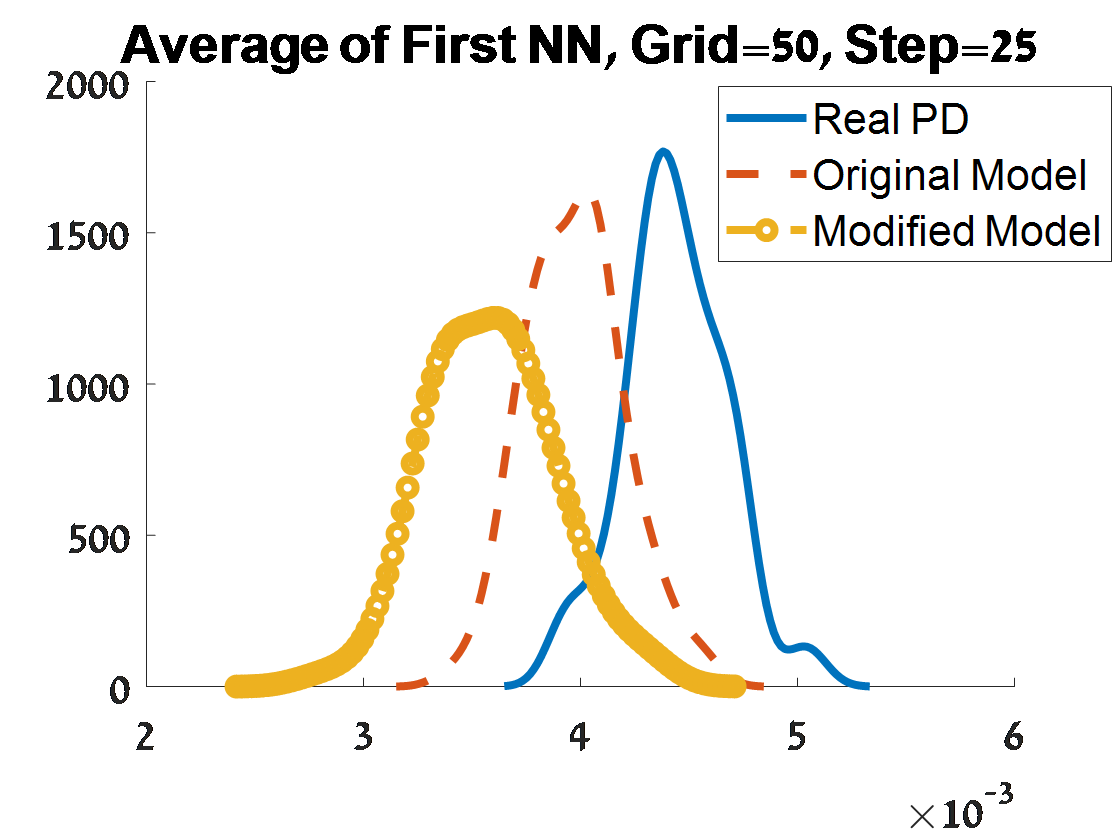}
\includegraphics[width=1.2in, height=1.25in]{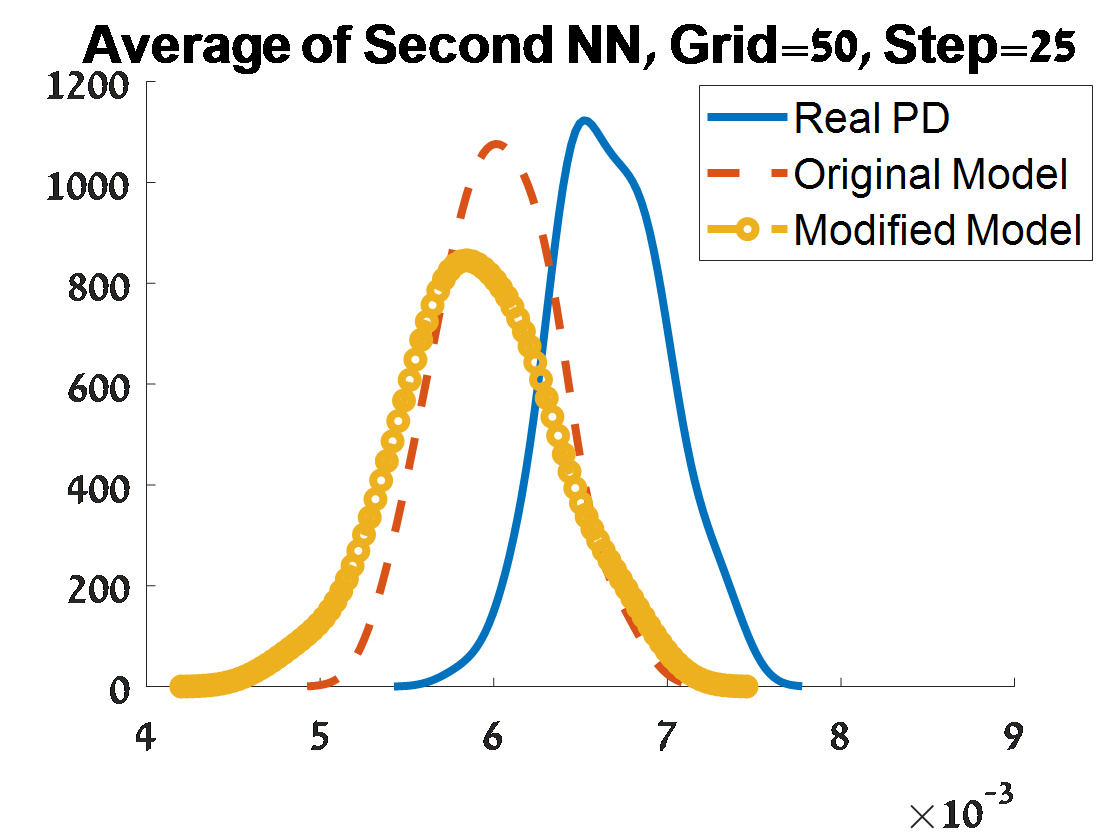}
\includegraphics[width=1.2in, height=1.25in]{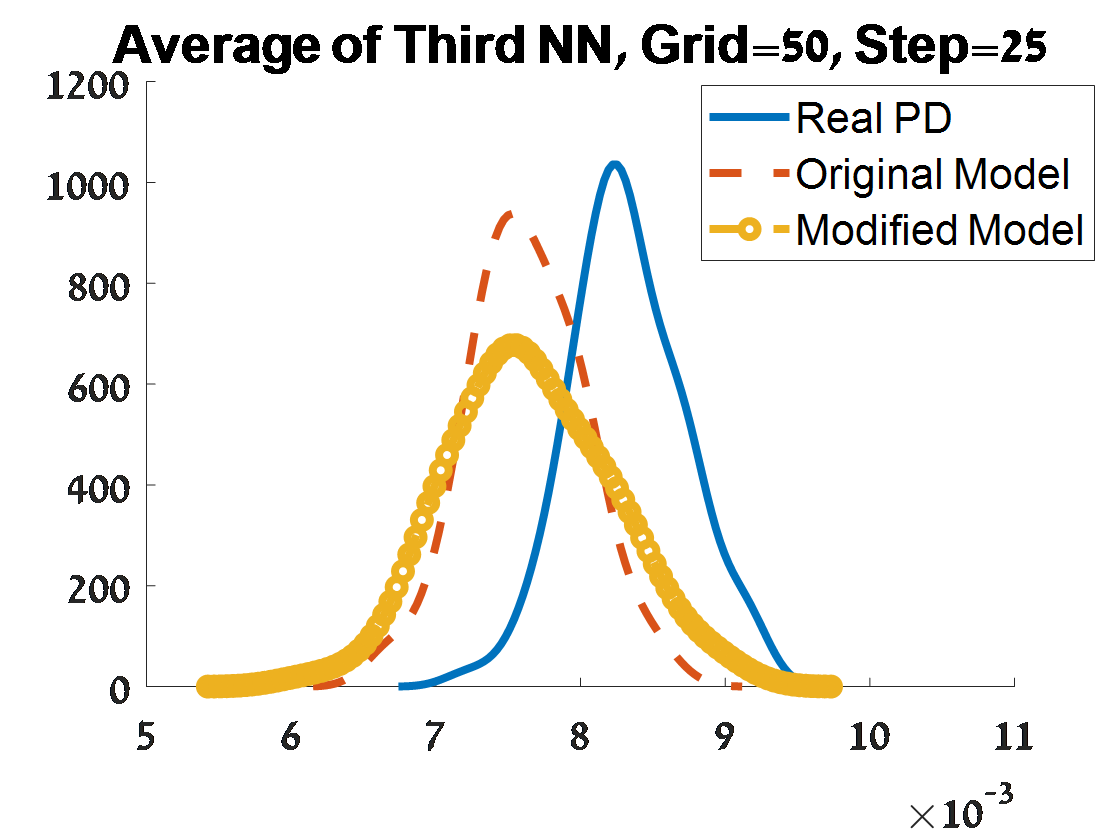}
\includegraphics[width=1.2in, height=1.25in]{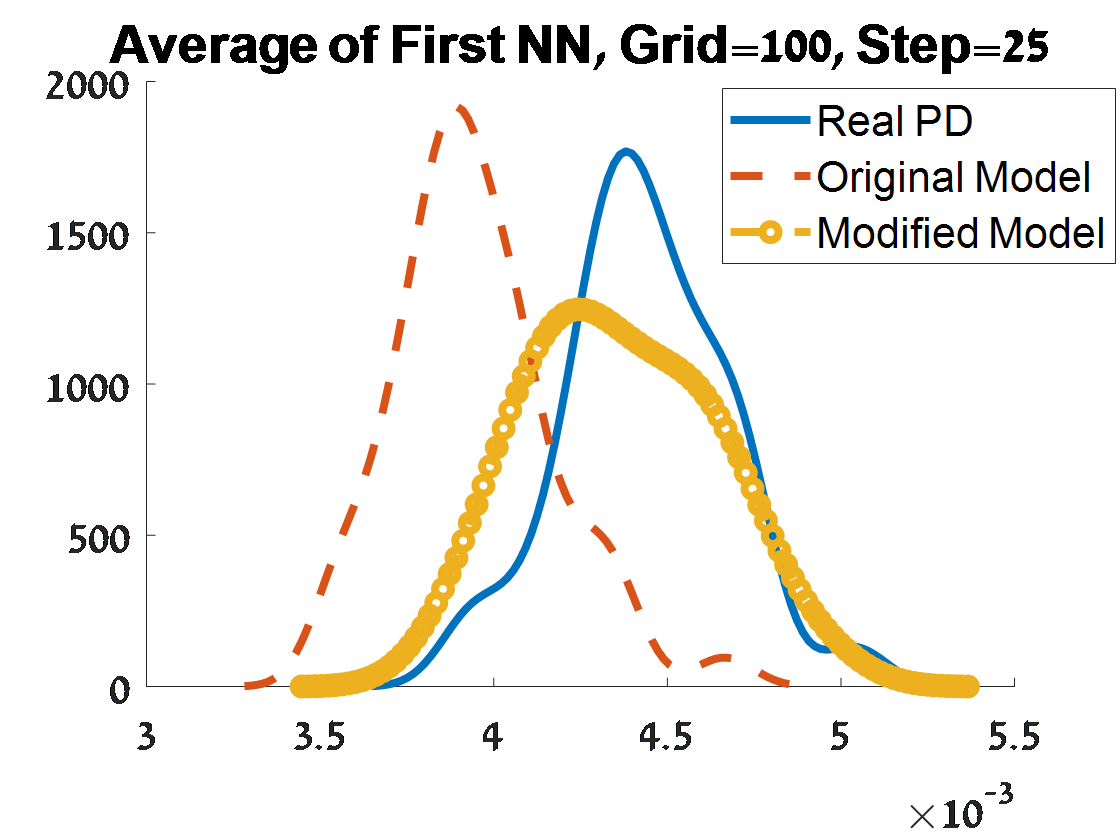}
\includegraphics[width=1.2in, height=1.25in]{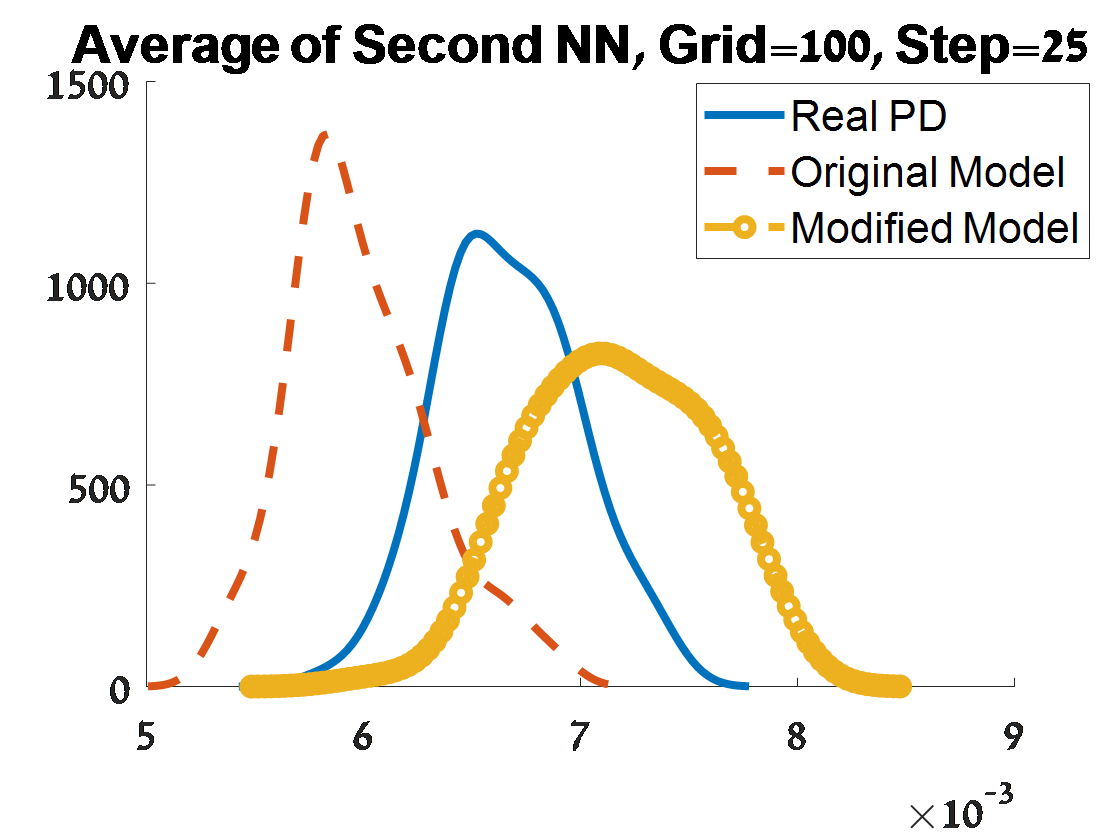}
\includegraphics[width=1.2in, height=1.25in]{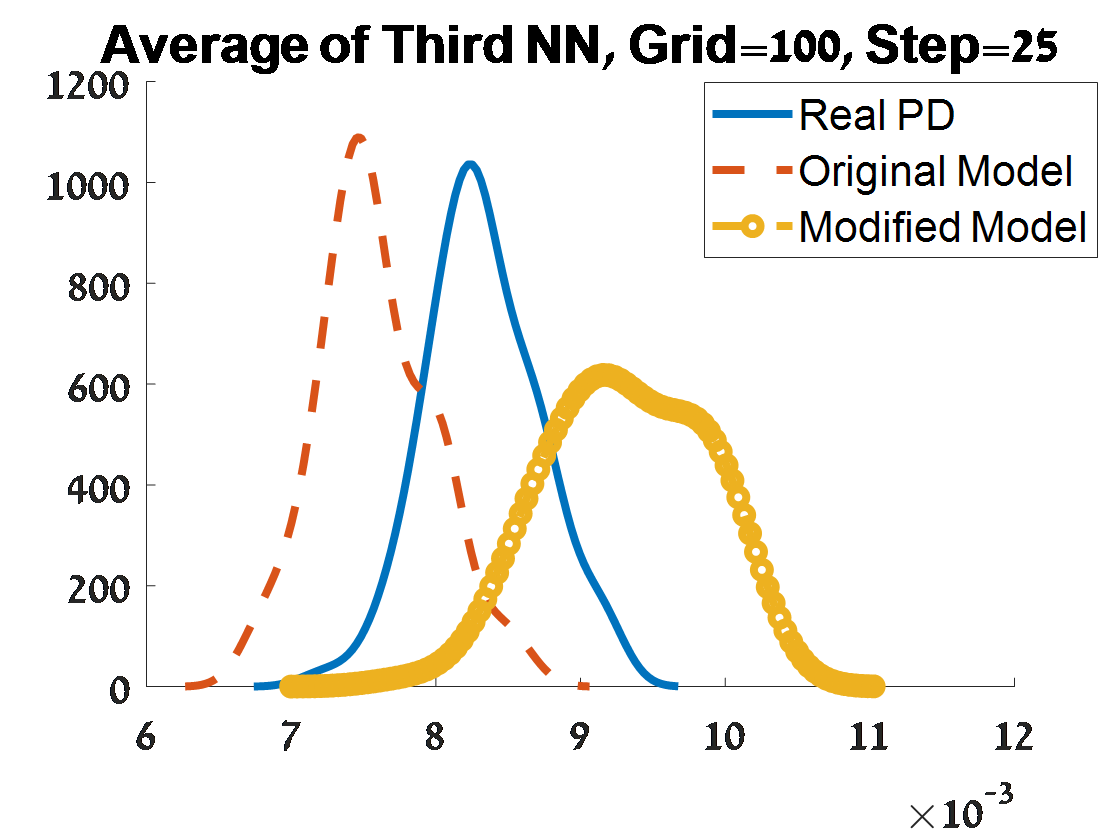}
\\
\includegraphics[width=1.2in, height=1.25in]{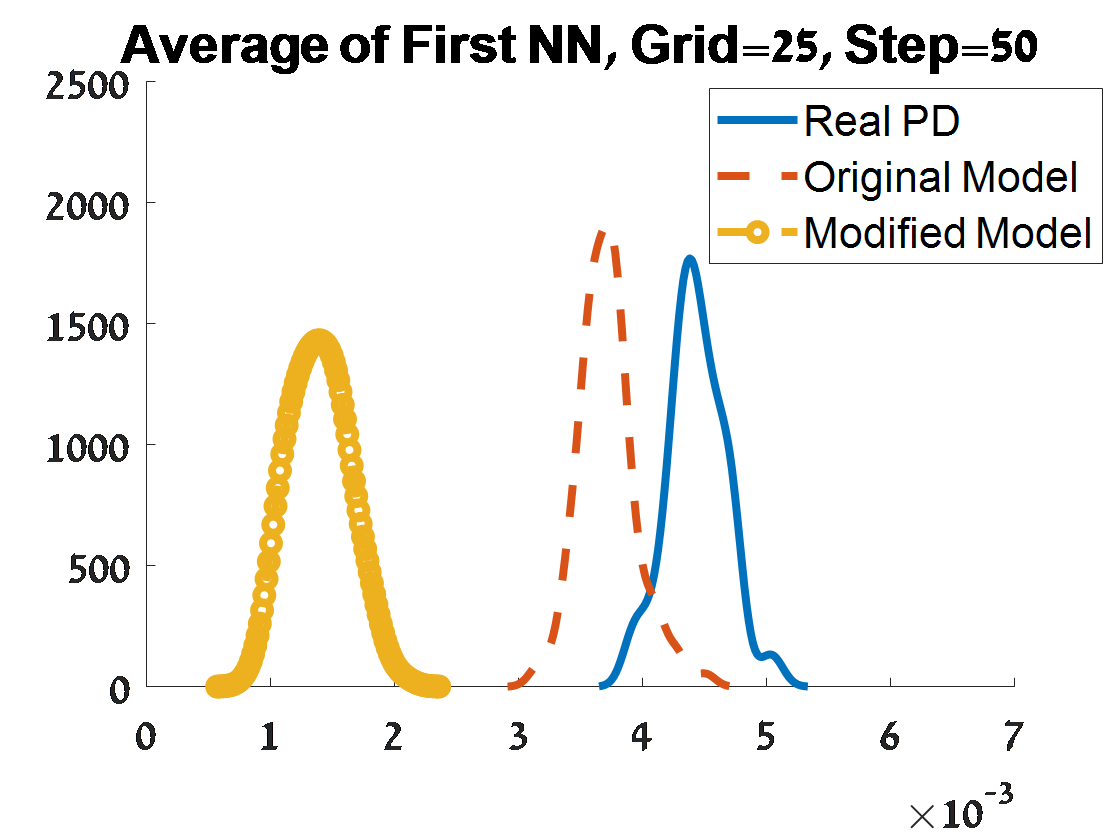}
\includegraphics[width=1.2in, height=1.25in]{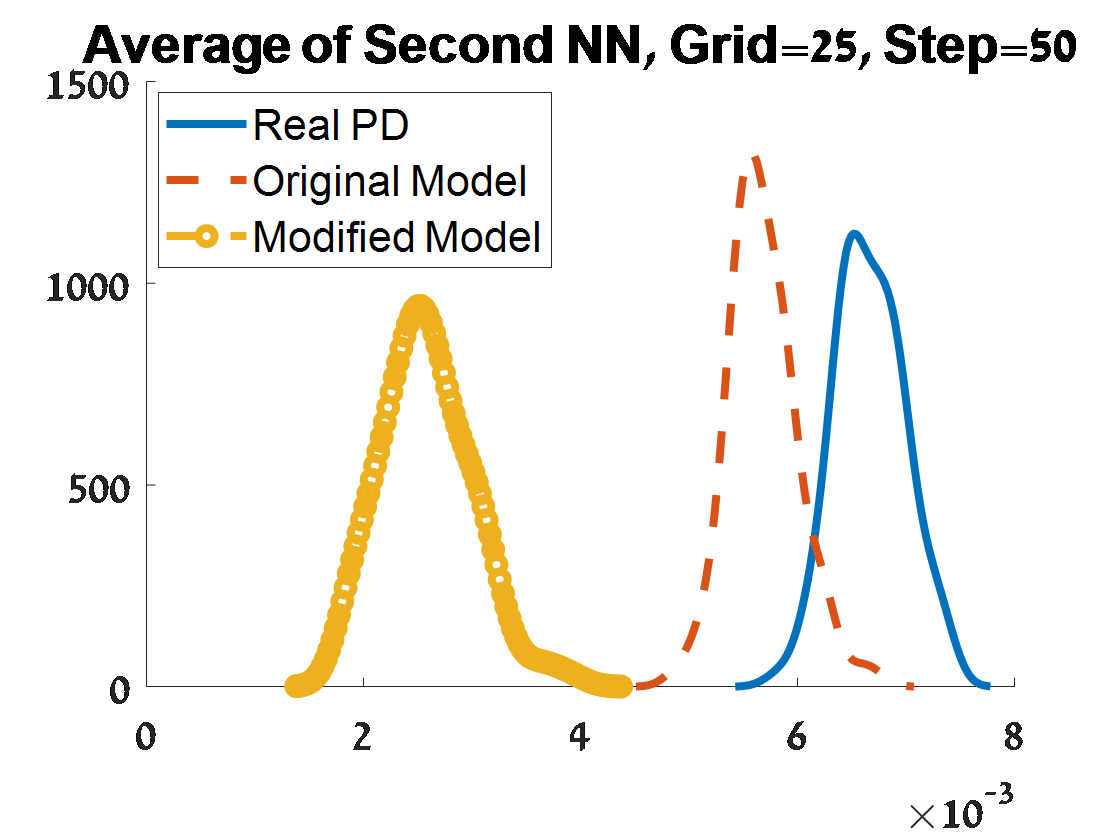}
\includegraphics[width=1.2in, height=1.25in]{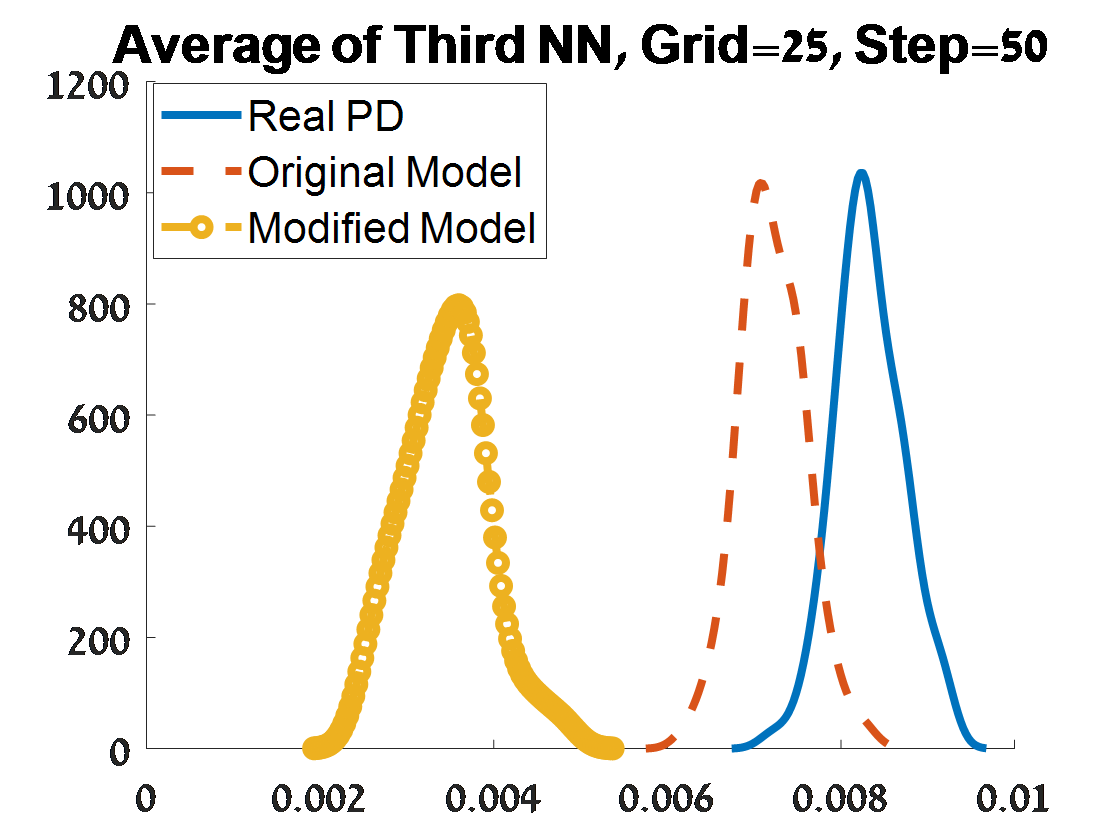}
\includegraphics[width=1.2in, height=1.25in]{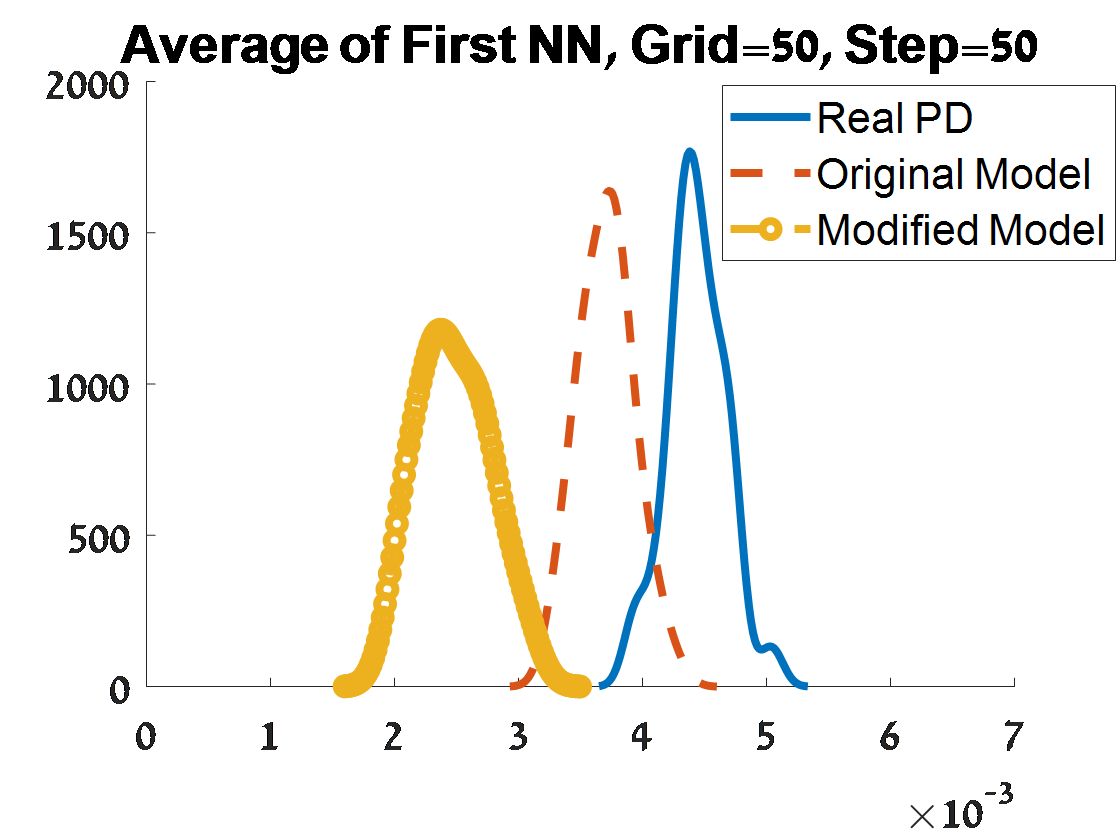}
\includegraphics[width=1.2in, height=1.25in]{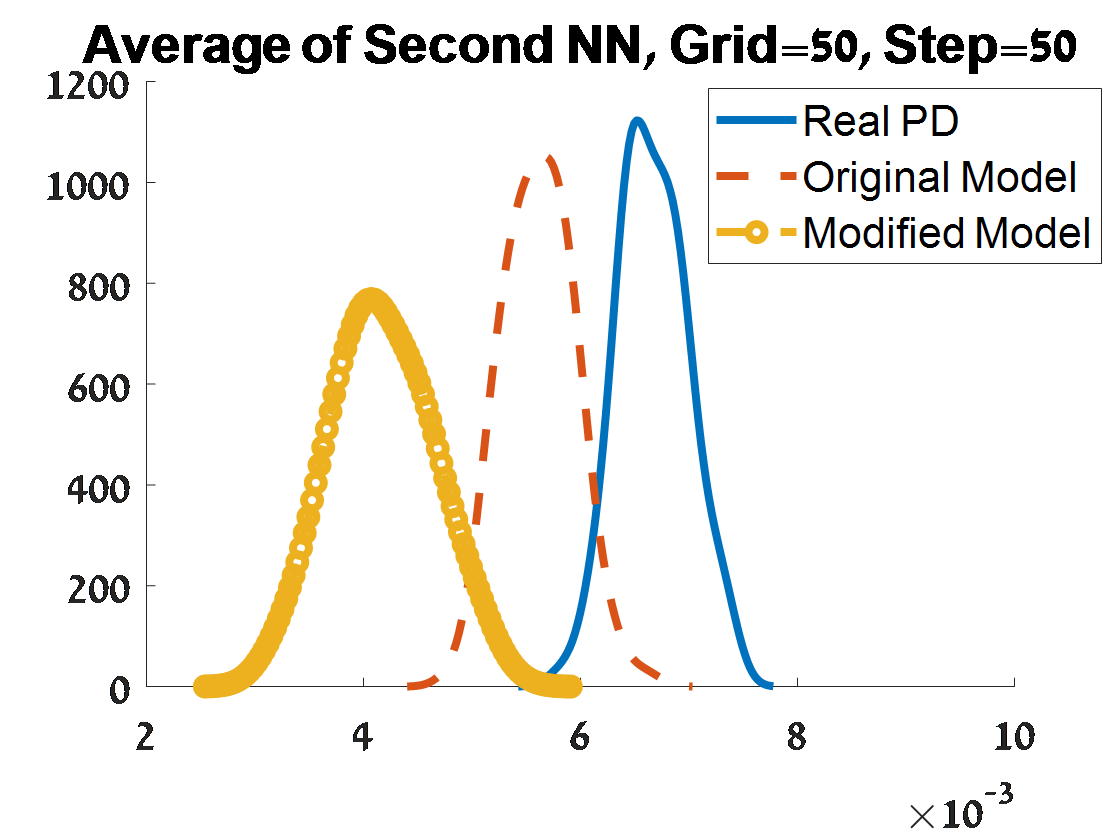}
\includegraphics[width=1.2in, height=1.25in]{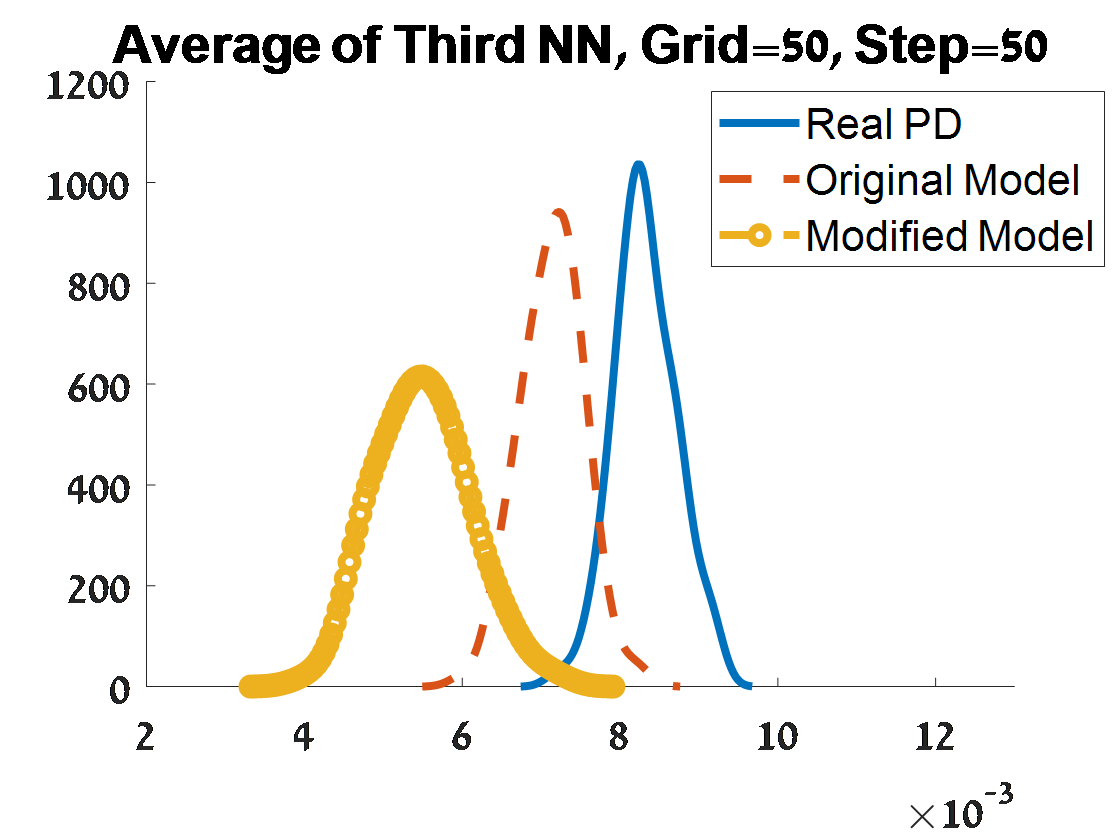}
\includegraphics[width=1.2in, height=1.25in]{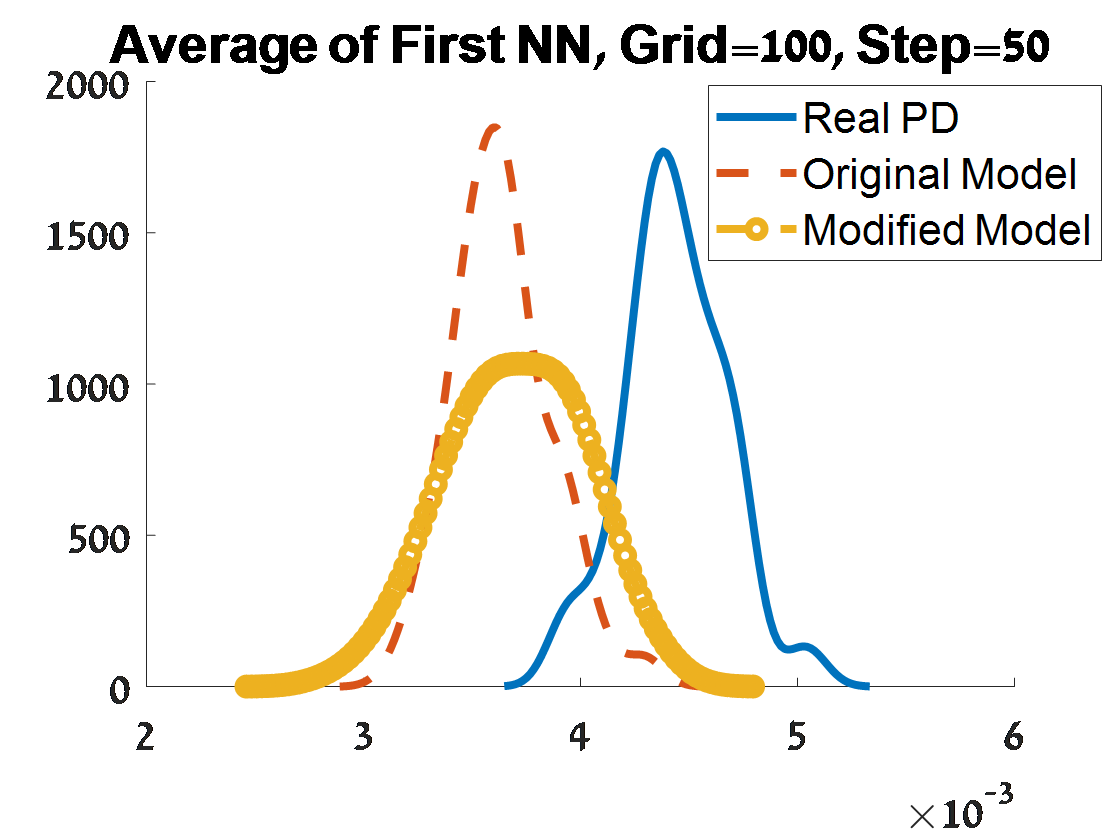}
\includegraphics[width=1.2in, height=1.25in]{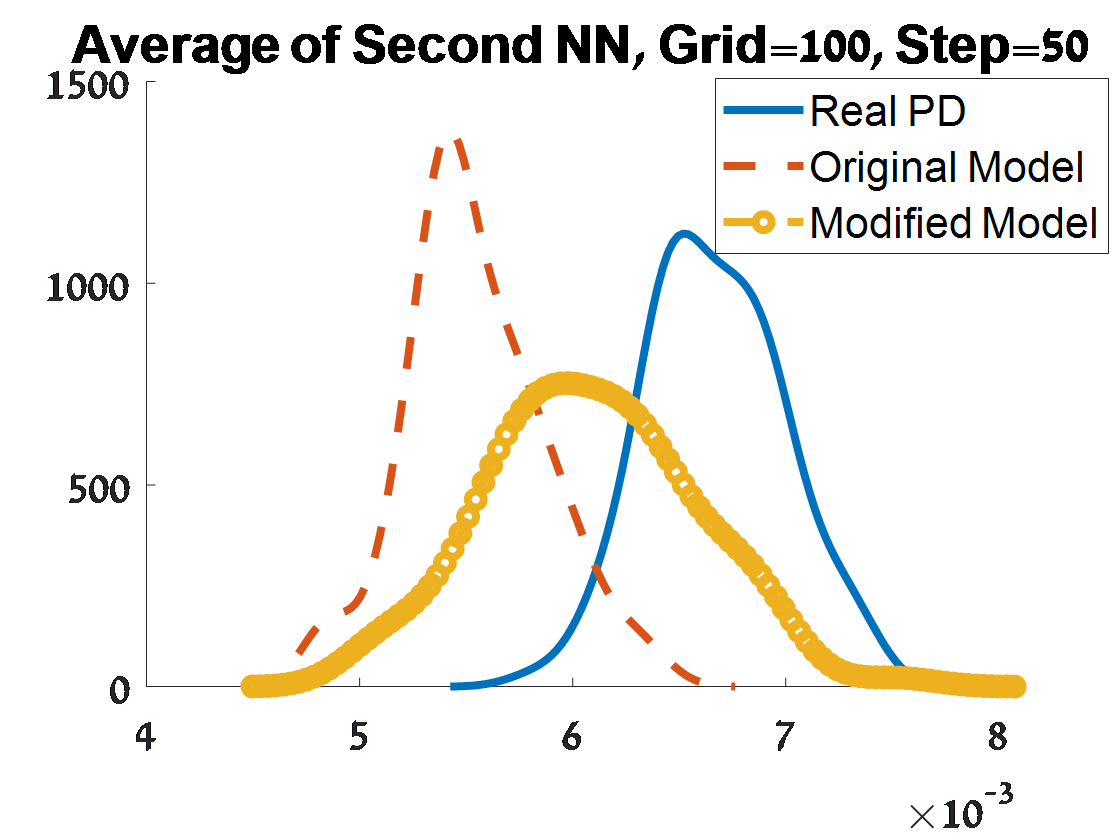}
\includegraphics[width=1.2in, height=1.25in]{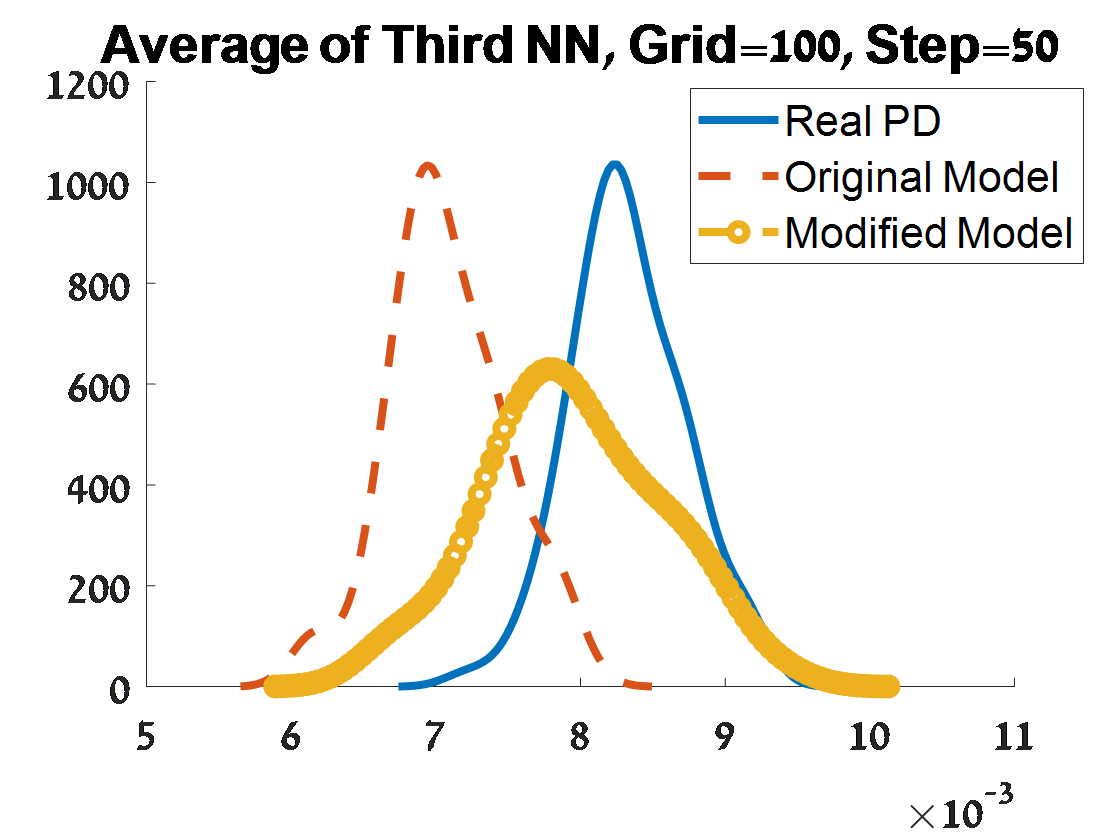}
\ec
%\caption{\footnotesize
% A random sample from two circles, 500 points from the larger circle and 300 from the smaller one,  with a kernel density
\caption{\footnotesize
 Criterion 2 of goodness of fit for 100 $H_1$ PDs corresponded to 100 samples from a unit $S^3$. The figures depend on the grid of the proposal distribution ("Grid"), and the burn-in ("Step") of the MCMC algorithm.}
\label{fig:s3_H1_b}
\end{figure}
\end{landscape}

\begin{landscape}
\begin{figure}[h!]
\bc
\includegraphics[width=1.2in, height=1.25in]{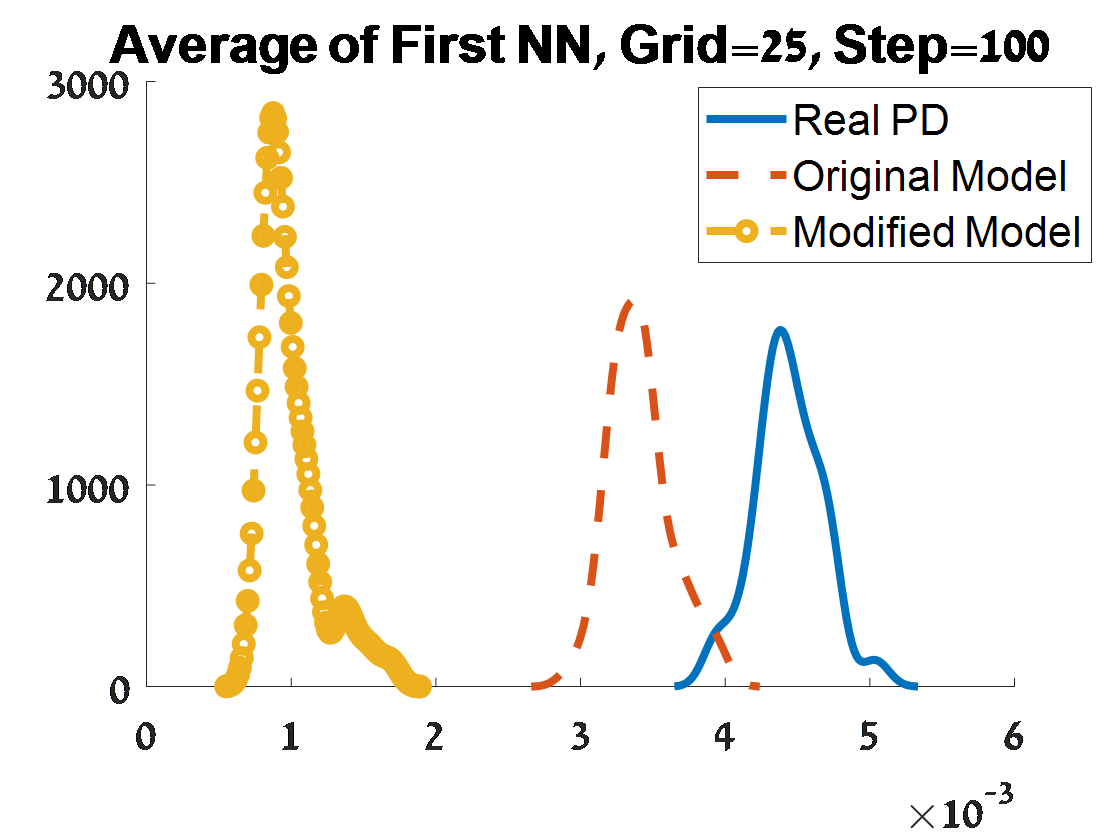}
\includegraphics[width=1.2in, height=1.25in]{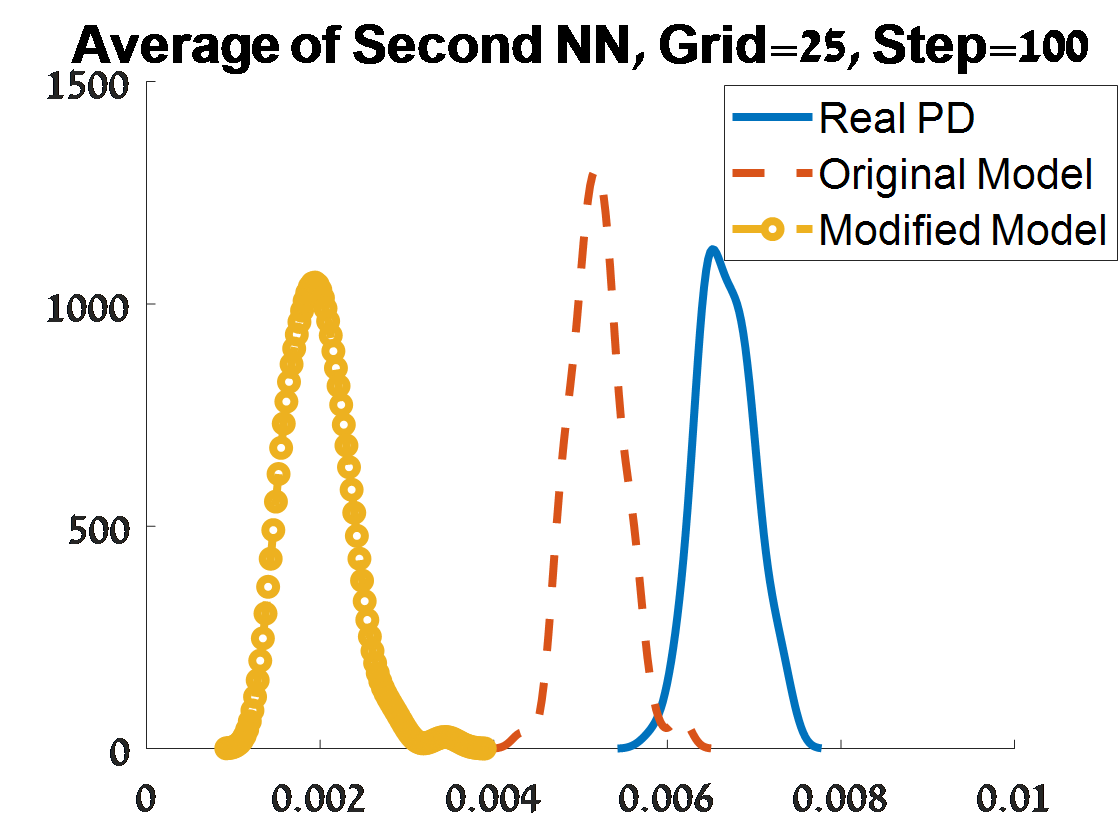}
\includegraphics[width=1.2in, height=1.25in]{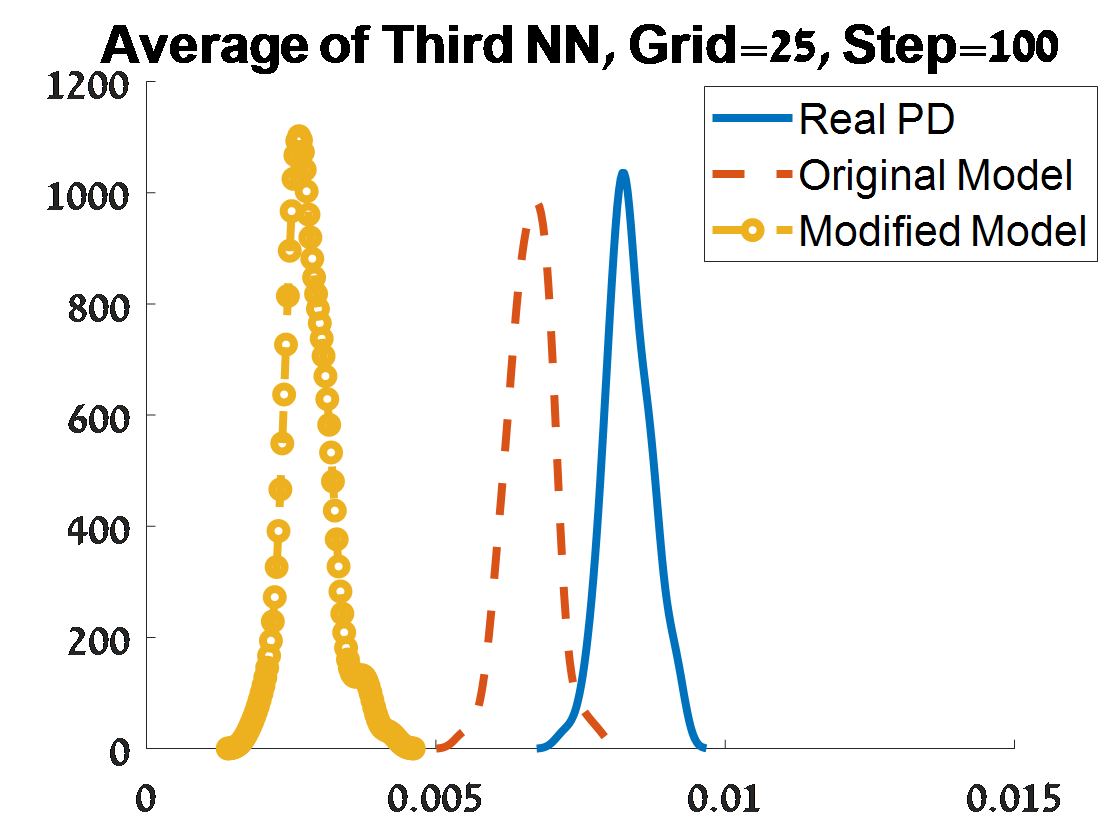}
\includegraphics[width=1.2in, height=1.25in]{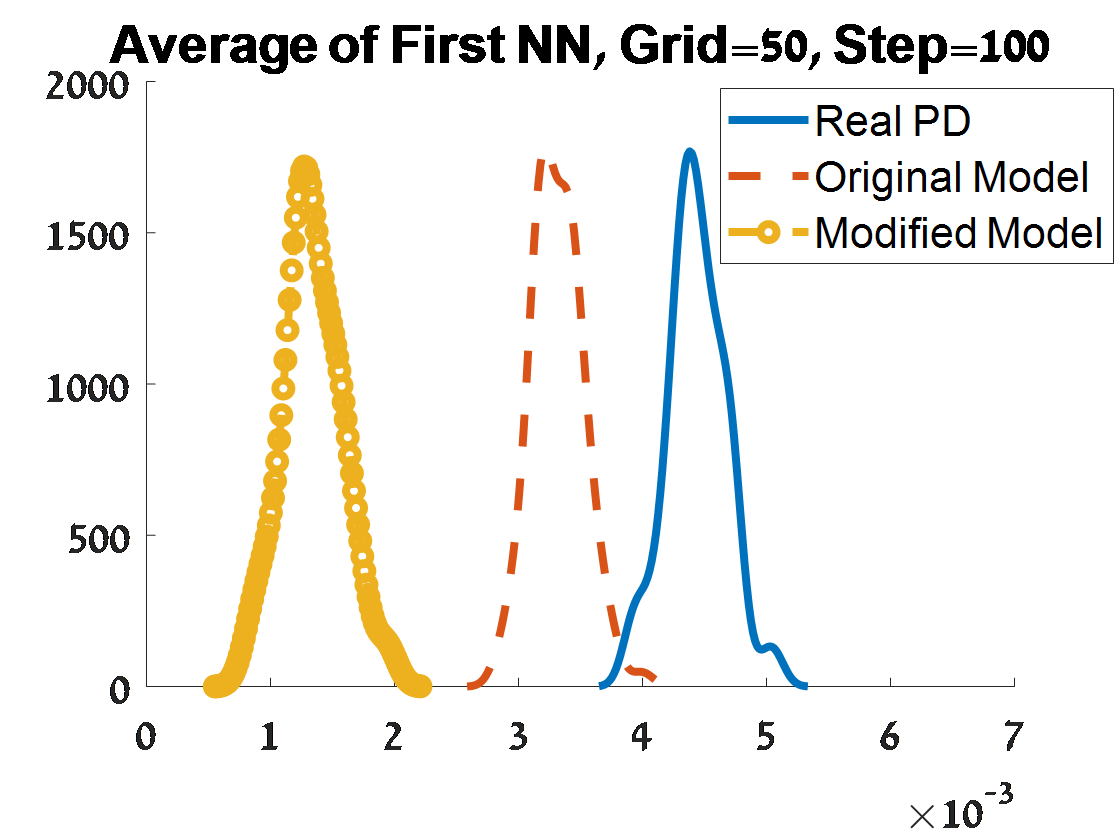}
\includegraphics[width=1.2in, height=1.25in]{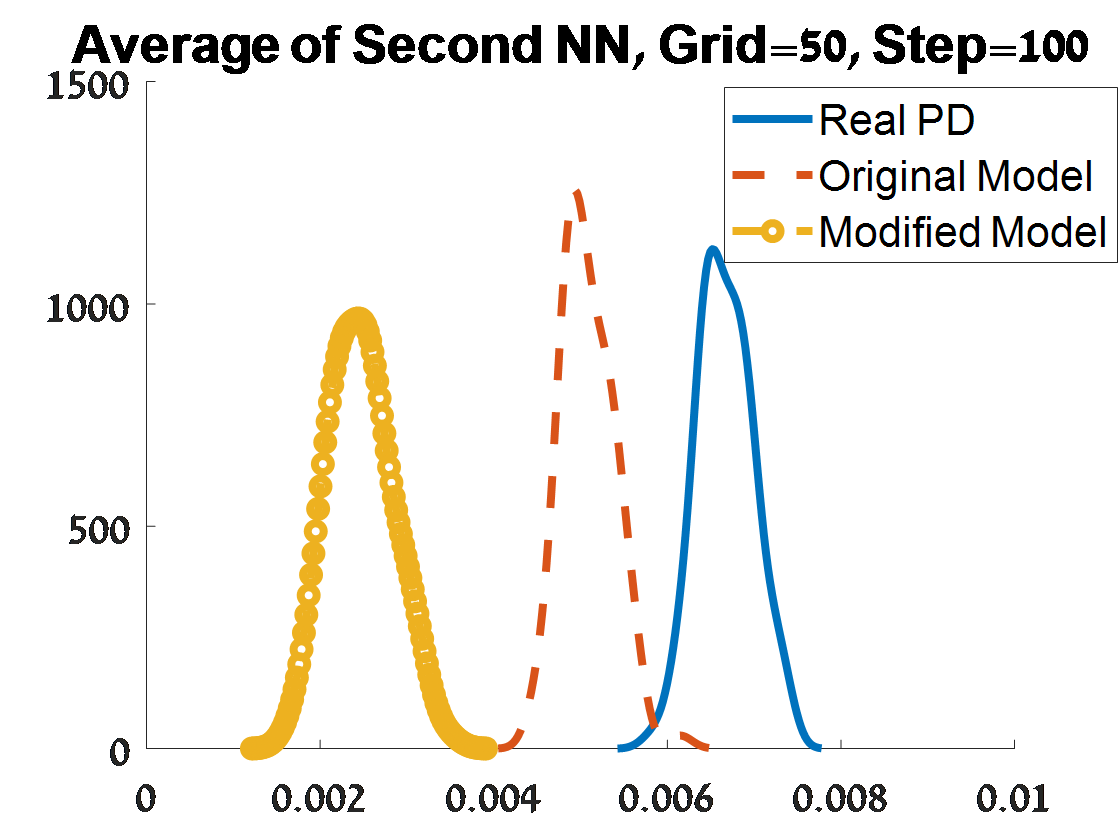}
\includegraphics[width=1.2in, height=1.25in]{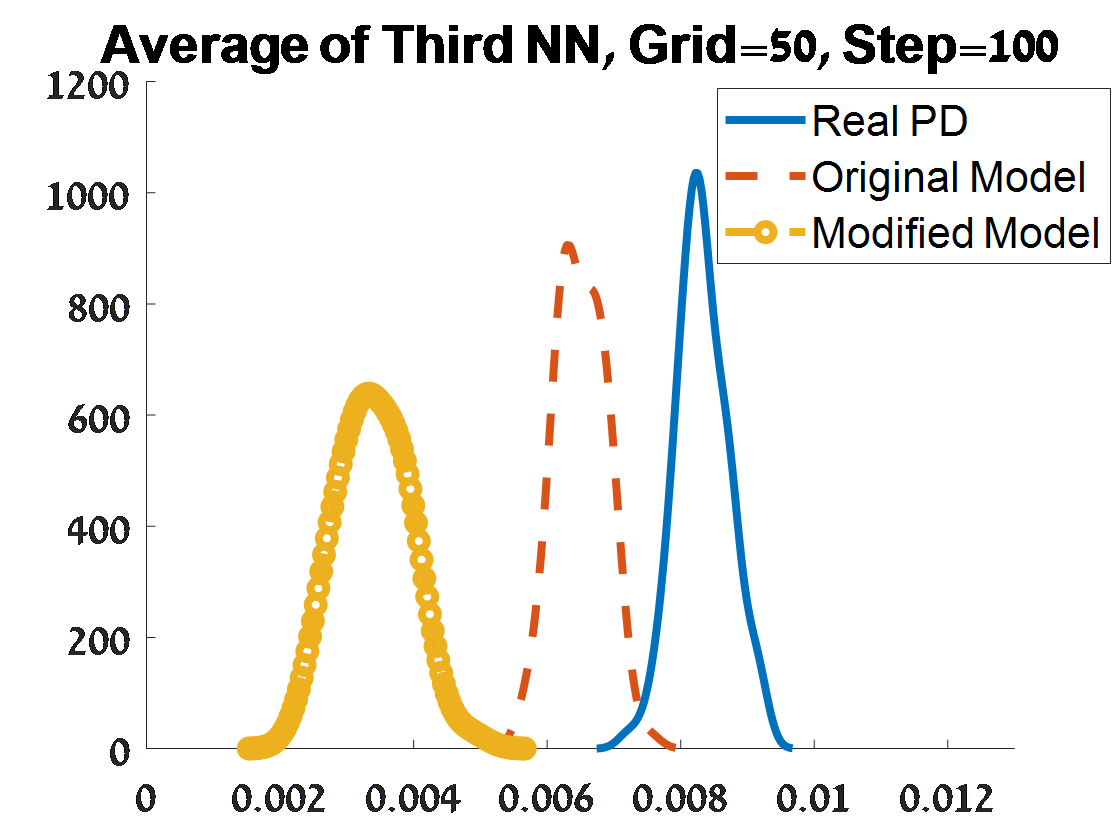}
\includegraphics[width=1.2in, height=1.25in]{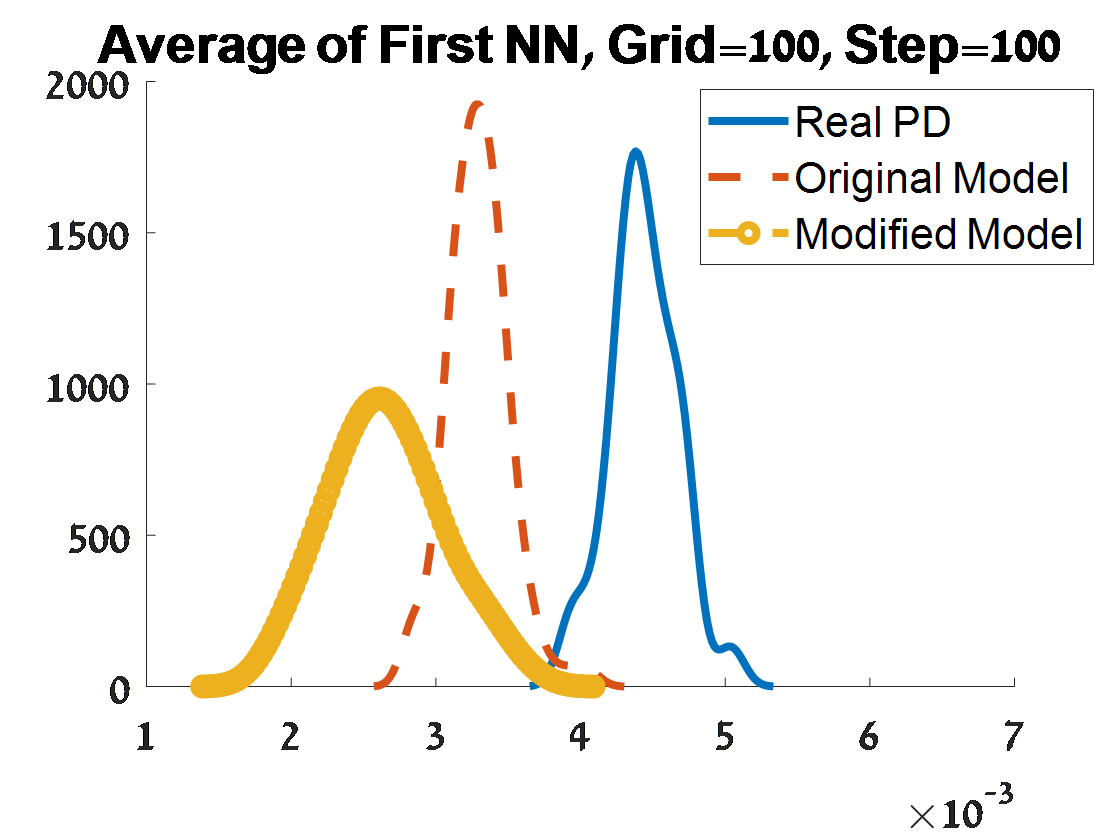}
\includegraphics[width=1.2in, height=1.25in]{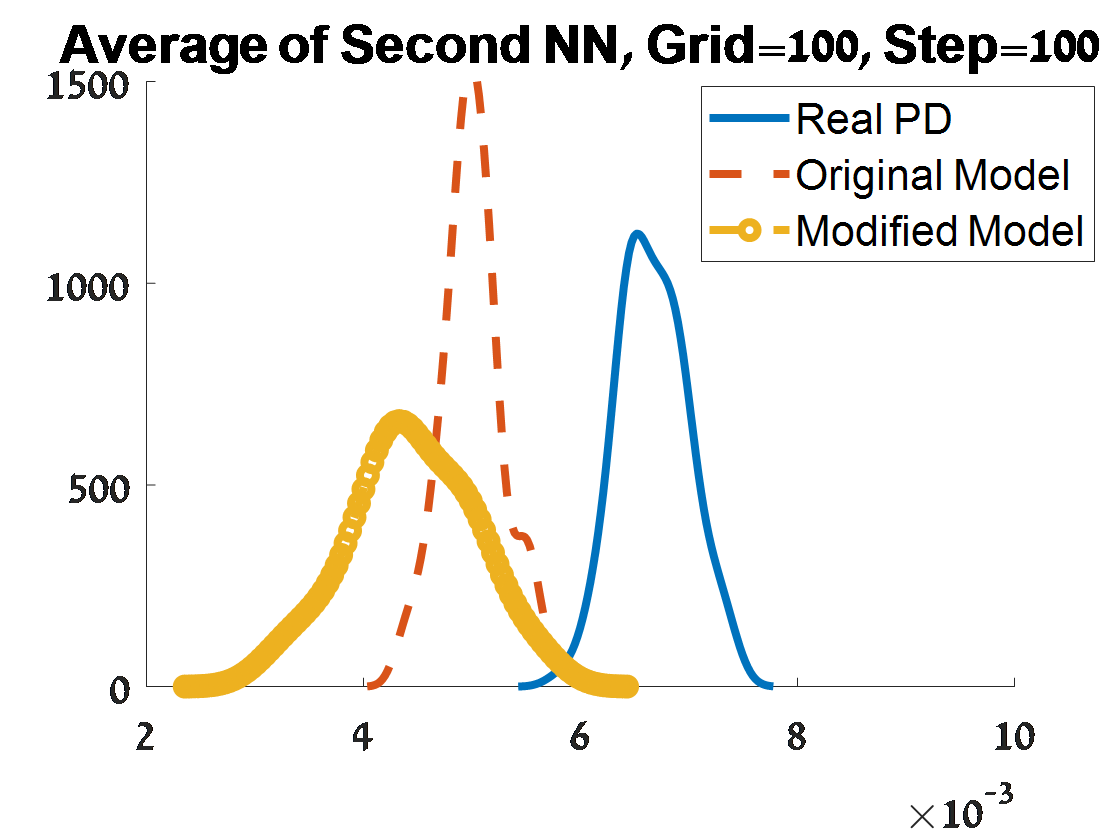}
\includegraphics[width=1.2in, height=1.25in]{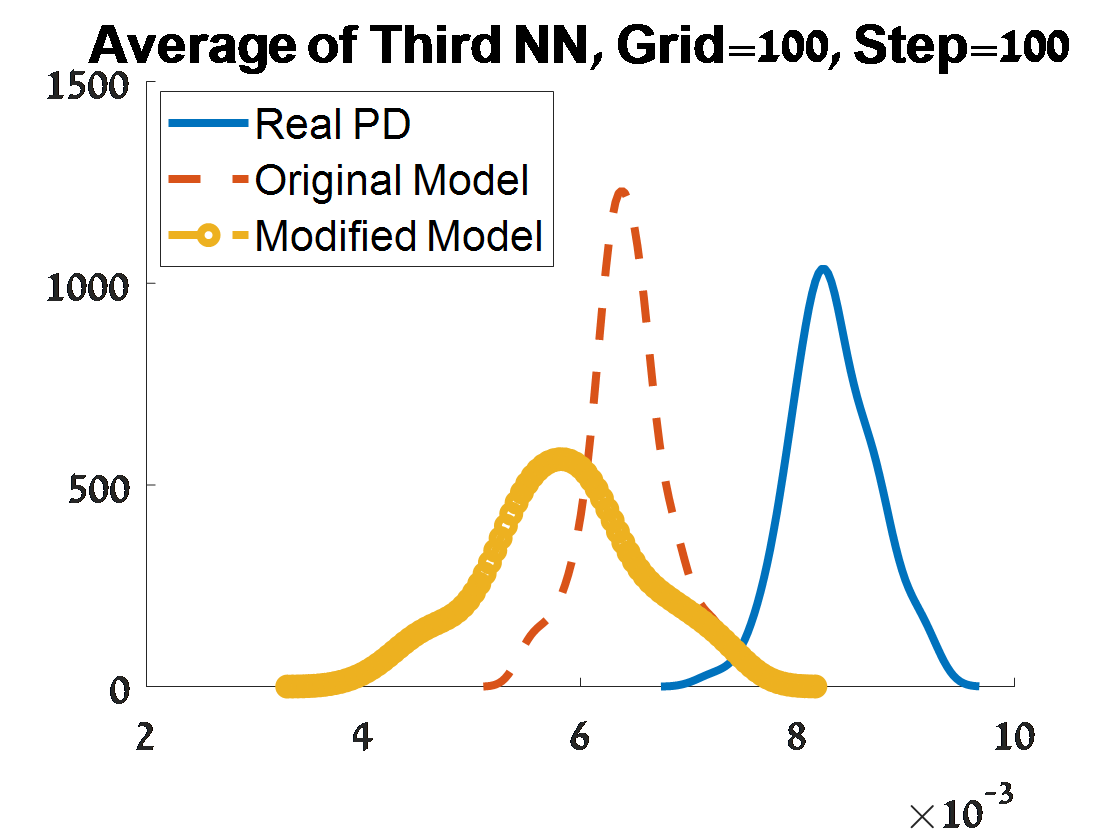}
\ec
\caption{\footnotesize
 Continue of Criterion 2 of goodness of fit for $H_1$ PDs corresponded to 100 samples from a unit $S^3$. The figures depend on the grid of the proposal distribution ("Grid"), and the burn-in ("Step") of the MCMC algorithm.}
\label{fig:s3_H1_c}
\end{figure}
\end{landscape}

%\begin{landscape}
%\begin{figure}[h!]
%\bc
%\includegraphics[width=1.8in, height=1.8in]{Sphere_s3_H1_pd30_grid50_step25}
%\includegraphics[width=1.8in, height=1.8in]{Sphere_s3_H1_pd30_grid100_step25}
%\includegraphics[width=1.8in, height=1.8in]{Sphere_s3_H1_pd60_grid50_step25}
%\includegraphics[width=1.8in, height=1.8in]{Sphere_s3_H1_pd60_grid100_step25}
%\ec
%%\caption{\footnotesize
%% A random sample from two circles, 500 points from the larger circle and 300 from the smaller one,  with a kernel density
%\caption{\footnotesize
%Examples of two PDs, each one is corresponded to a sample from a unit $S^3$. For each PD, the simulated PD based on the two model versions is described. The figures depend on the grid of the proposal distribution ("Grid"), and the burn-in ("Step") of the MCMC algorithm.}
%\label{fig:s3_H1_d}
%\end{figure}
%\end{landscape}

%---------------------------------------------------------------------------------------------------%
\subsubsection{The fitted model for $H_2$}
As noted above, we used the search range of [0,4] for estimating the parameter $\alpha$. But here for the $H_2$ points, $50\%$ of the PDs (over the 100 PDs) had a problem in the likelihood once using this search range. Limiting the search range to [0,1] solved this problem. This later search range is reasonable since the estimates of $\alpha$ that were obtained under the search range [0,4] were smaller than 1.
Specifically, the distribution of $\alpha$ estimates under the constrained range of [0,1] had ranged in [0.016,0.756] with median of 0.070, while this distribution under the search range of [0,4] had ranged in [0.576,0.951] with median of 0.755. That is, a smaller estimate of $\alpha$ once constraining it to the smaller search range.

\begin{landscape}
\begin{figure}[h!]
\bc
\includegraphics[width=1.2in, height=1.4in]{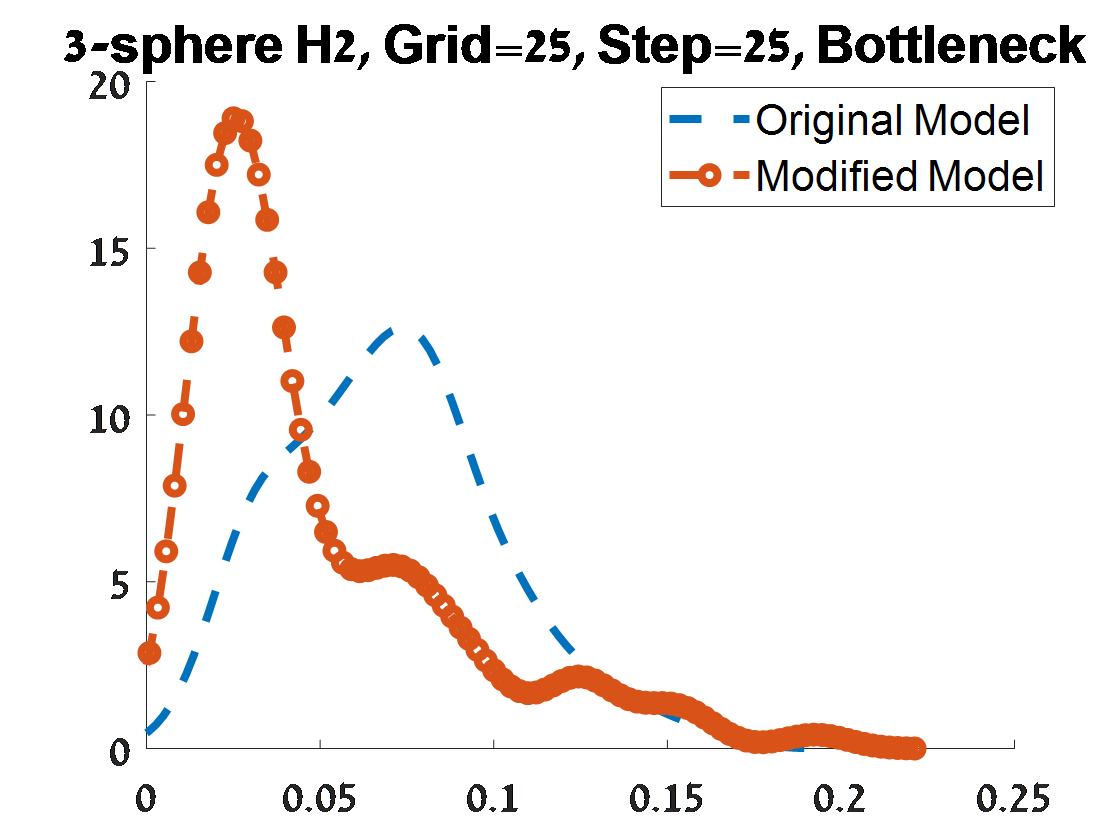}
\includegraphics[width=1.2in, height=1.4in]{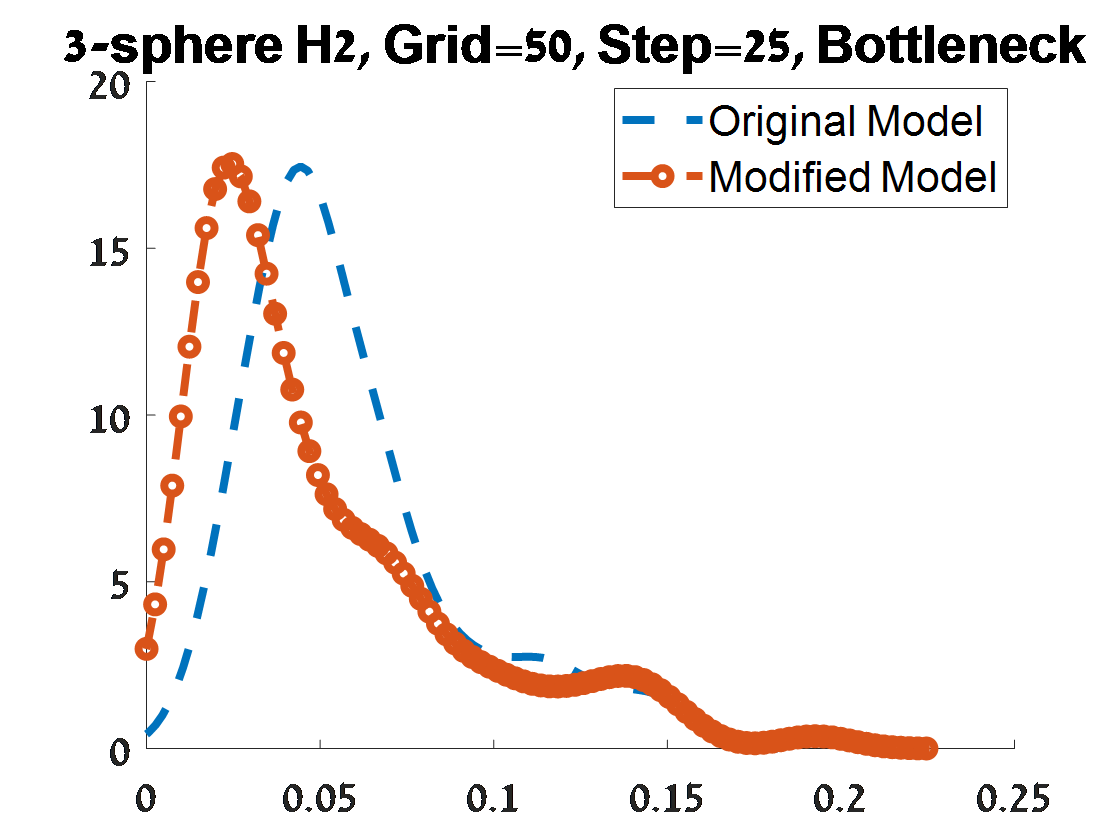}
\includegraphics[width=1.2in, height=1.4in]{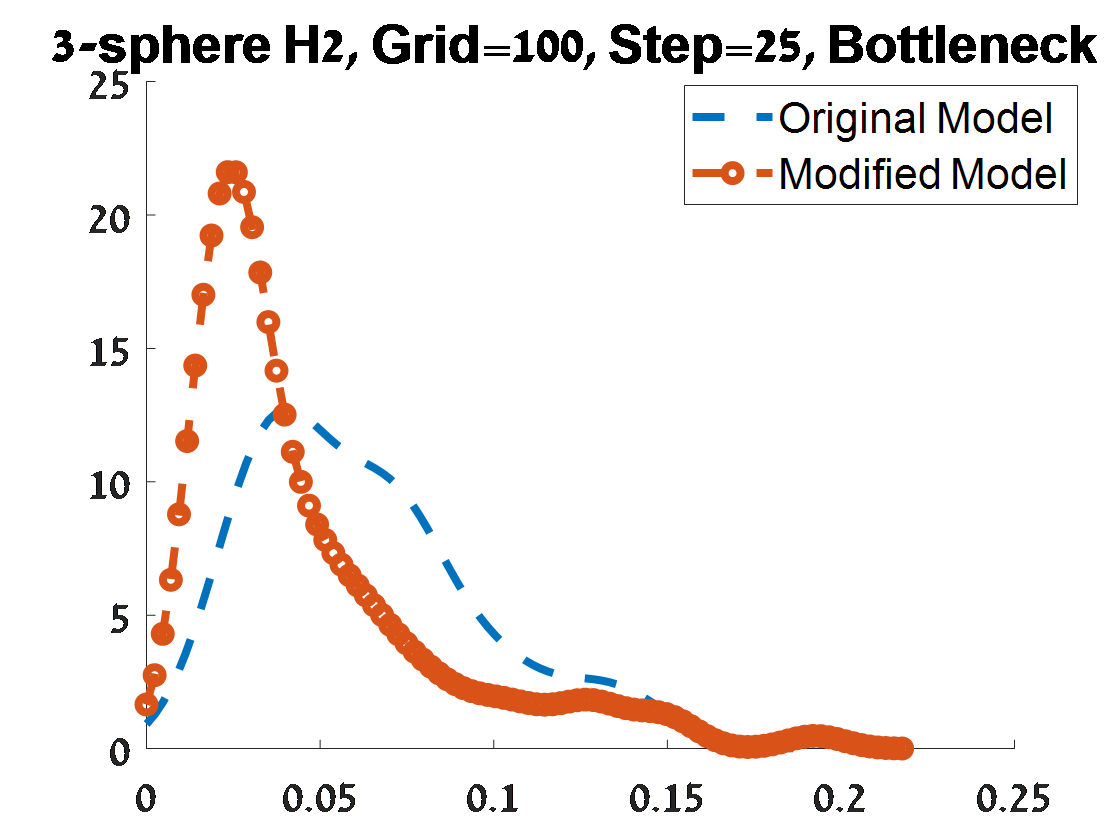}
\includegraphics[width=1.2in, height=1.4in]{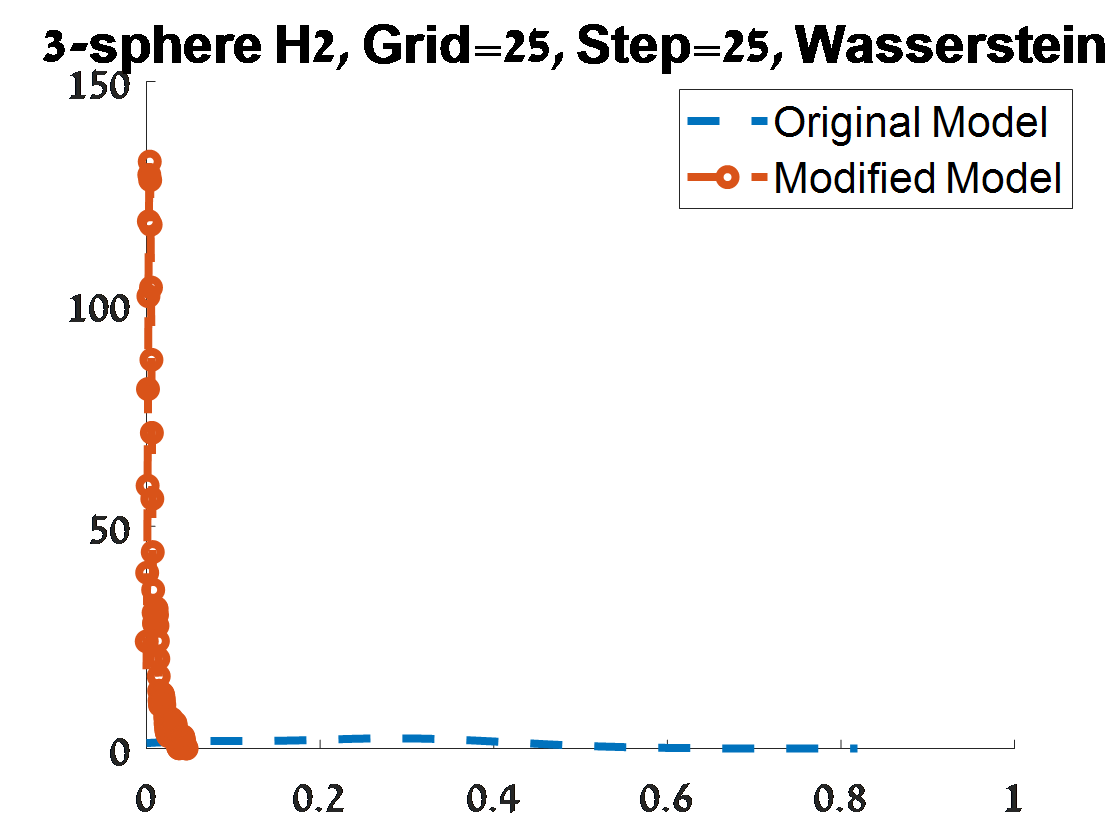}
\includegraphics[width=1.2in, height=1.4in]{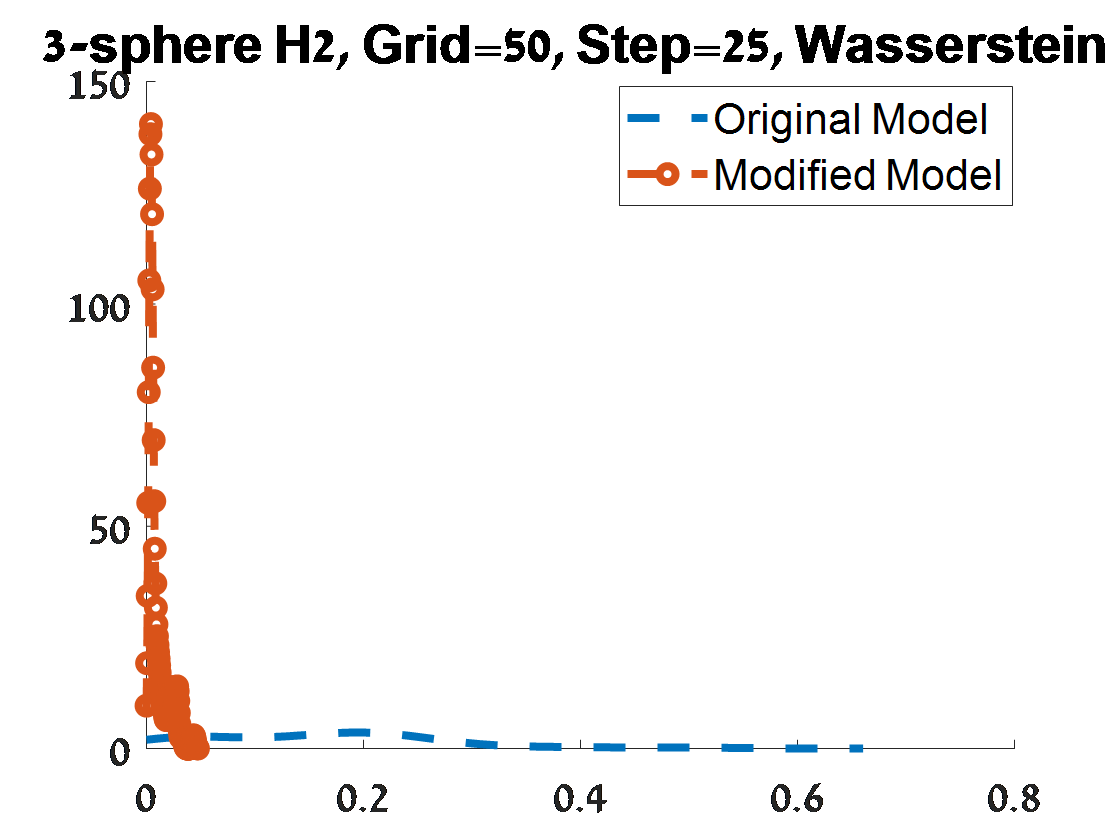}
\includegraphics[width=1.2in, height=1.4in]{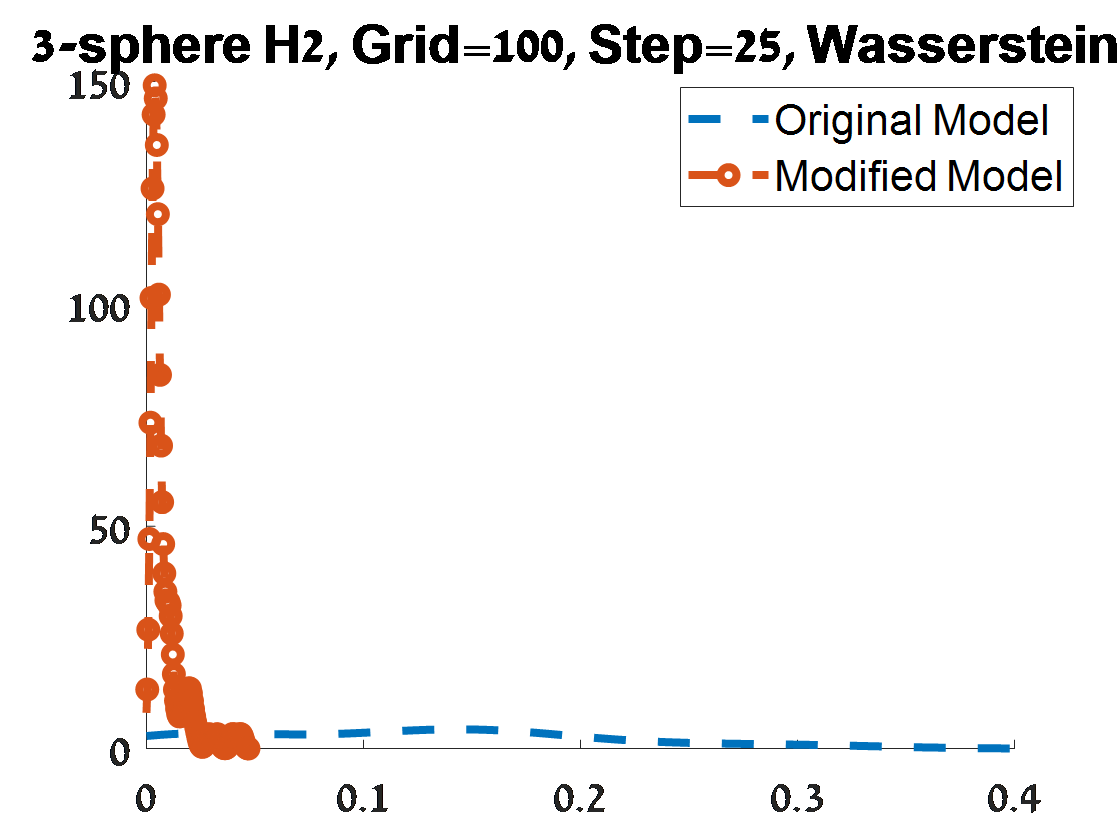}
%%\\
\includegraphics[width=1.2in, height=1.4in]{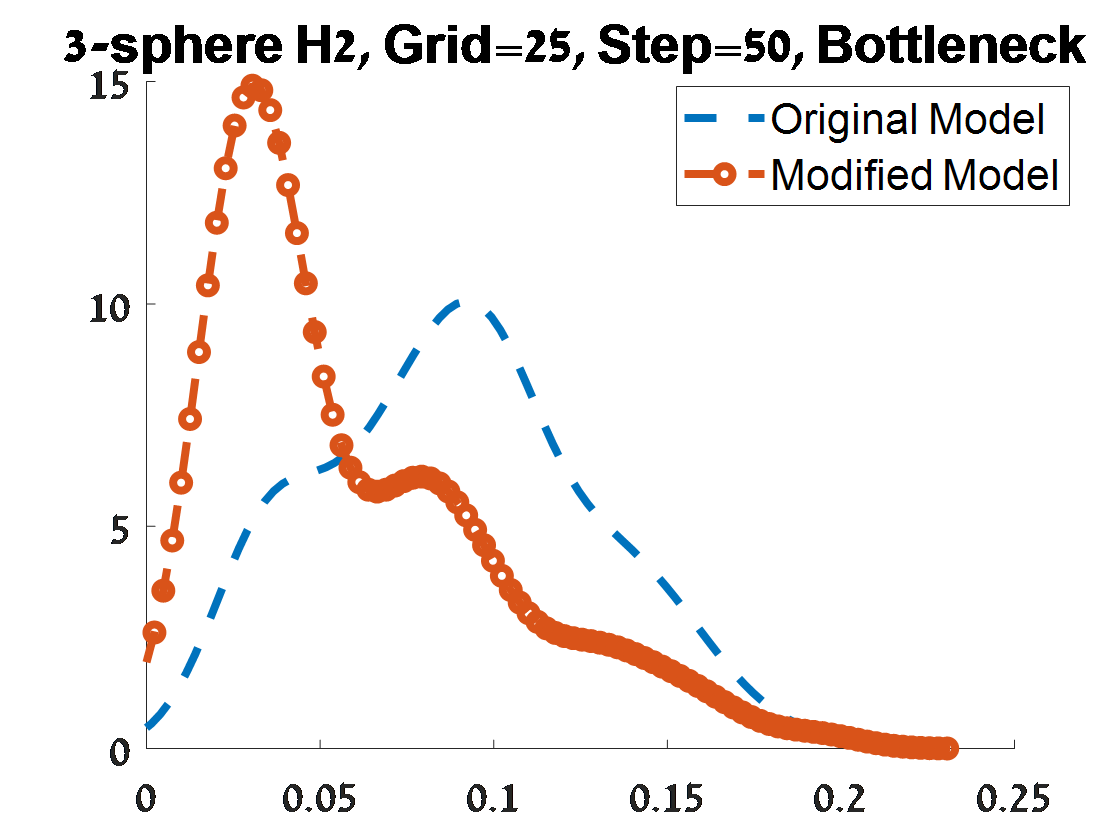}
\includegraphics[width=1.2in, height=1.4in]{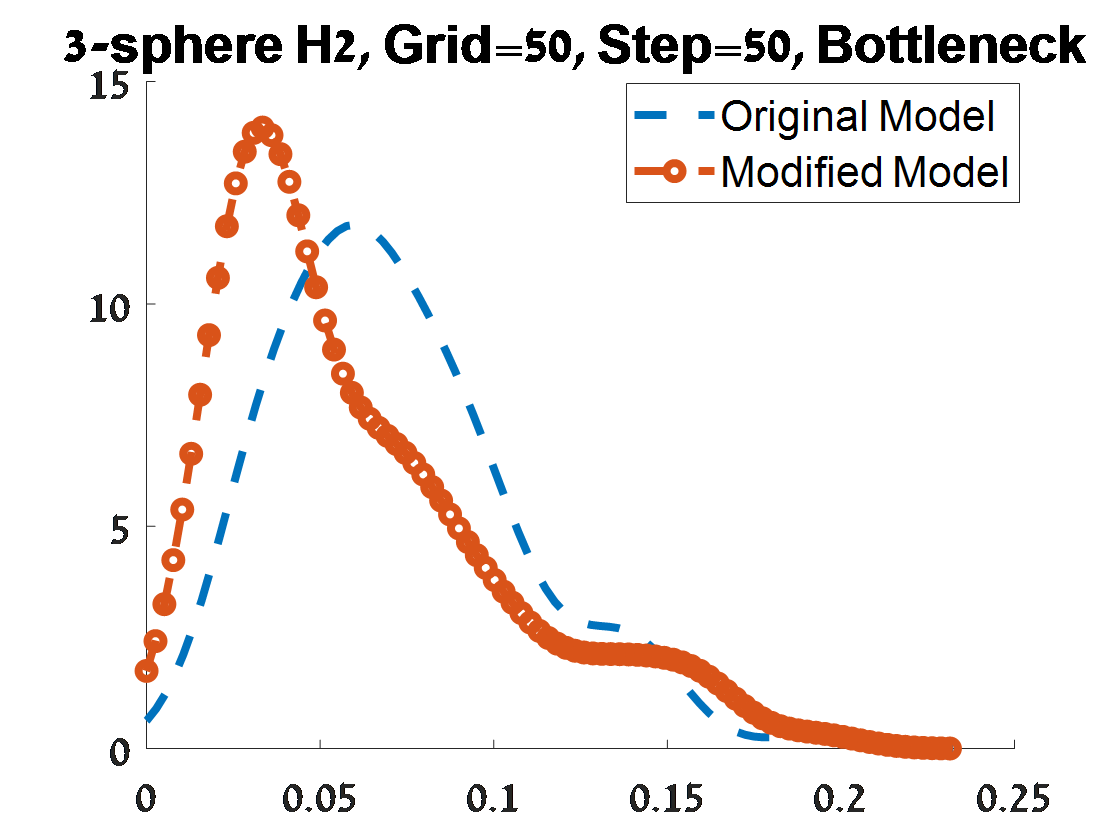}
\includegraphics[width=1.2in, height=1.4in]{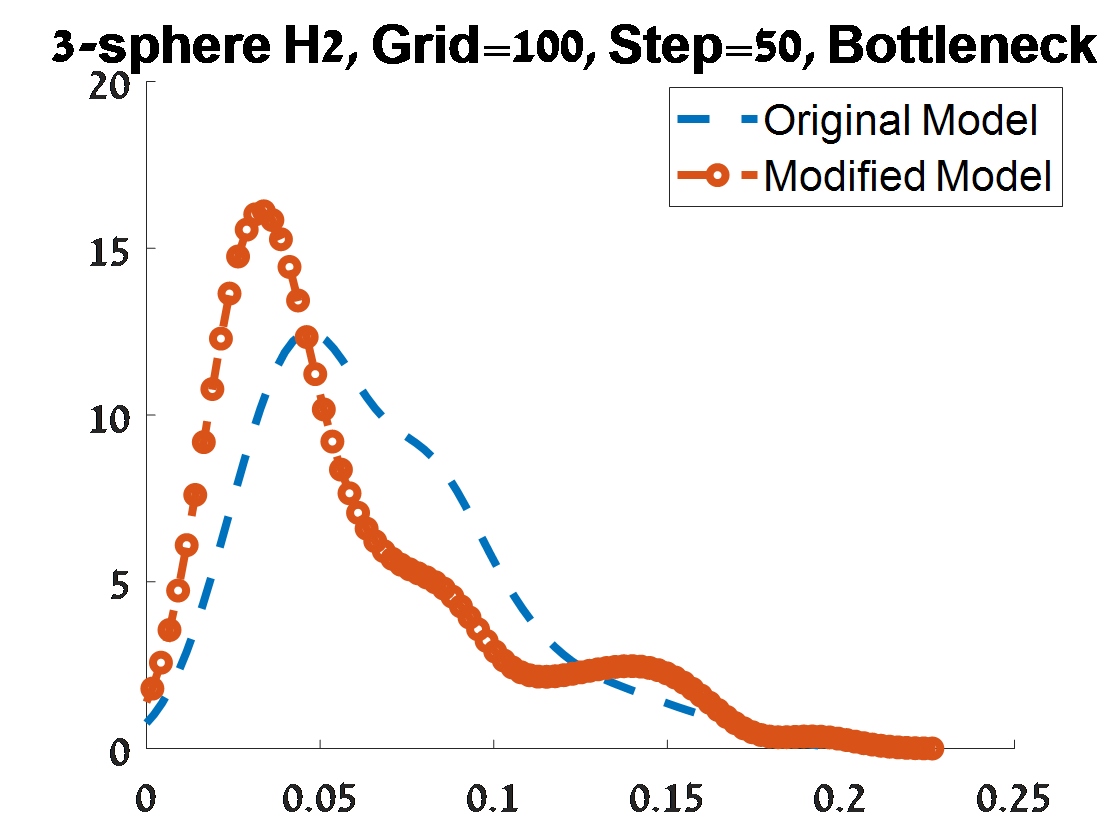}
\includegraphics[width=1.2in, height=1.4in]{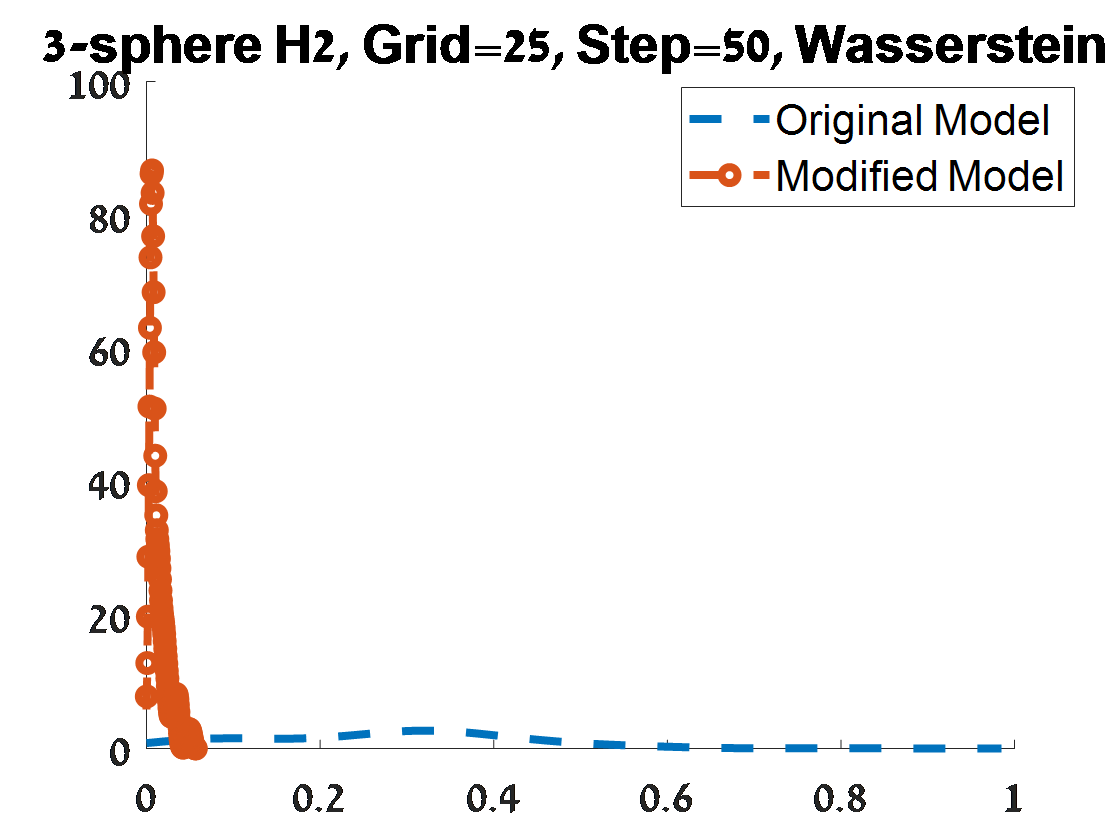}
\includegraphics[width=1.2in, height=1.4in]{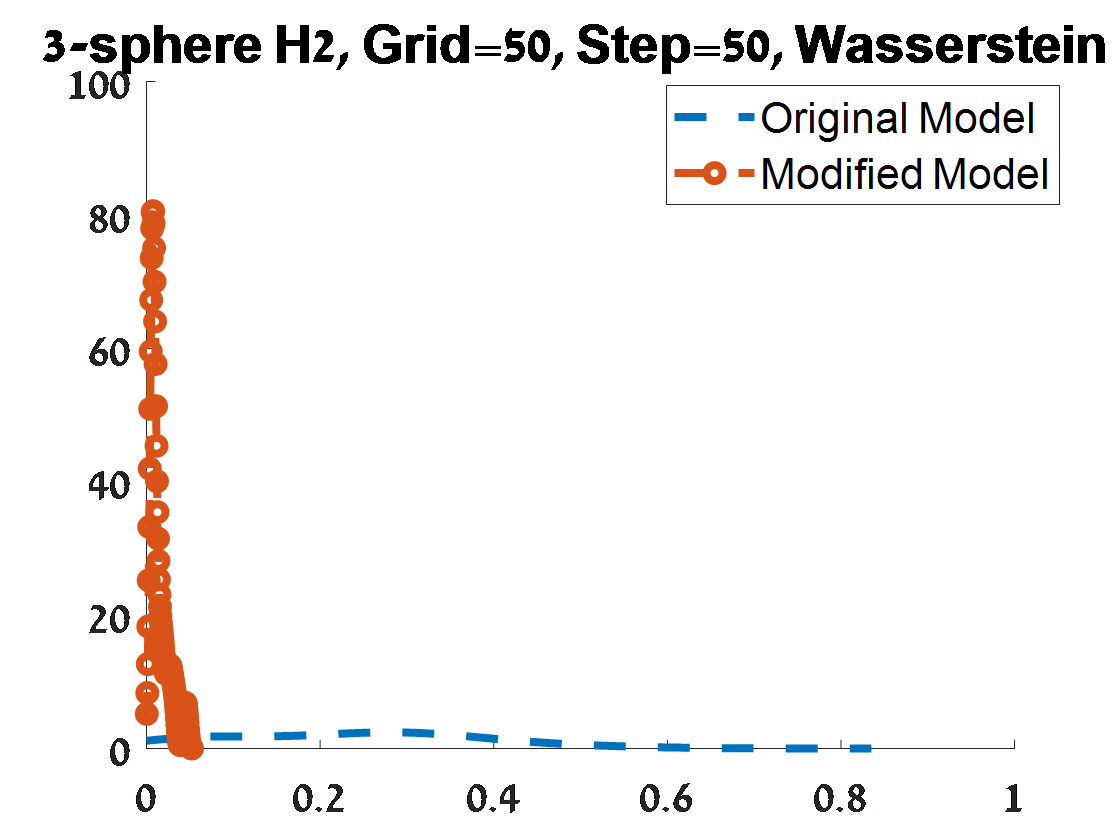}
\includegraphics[width=1.2in, height=1.4in]{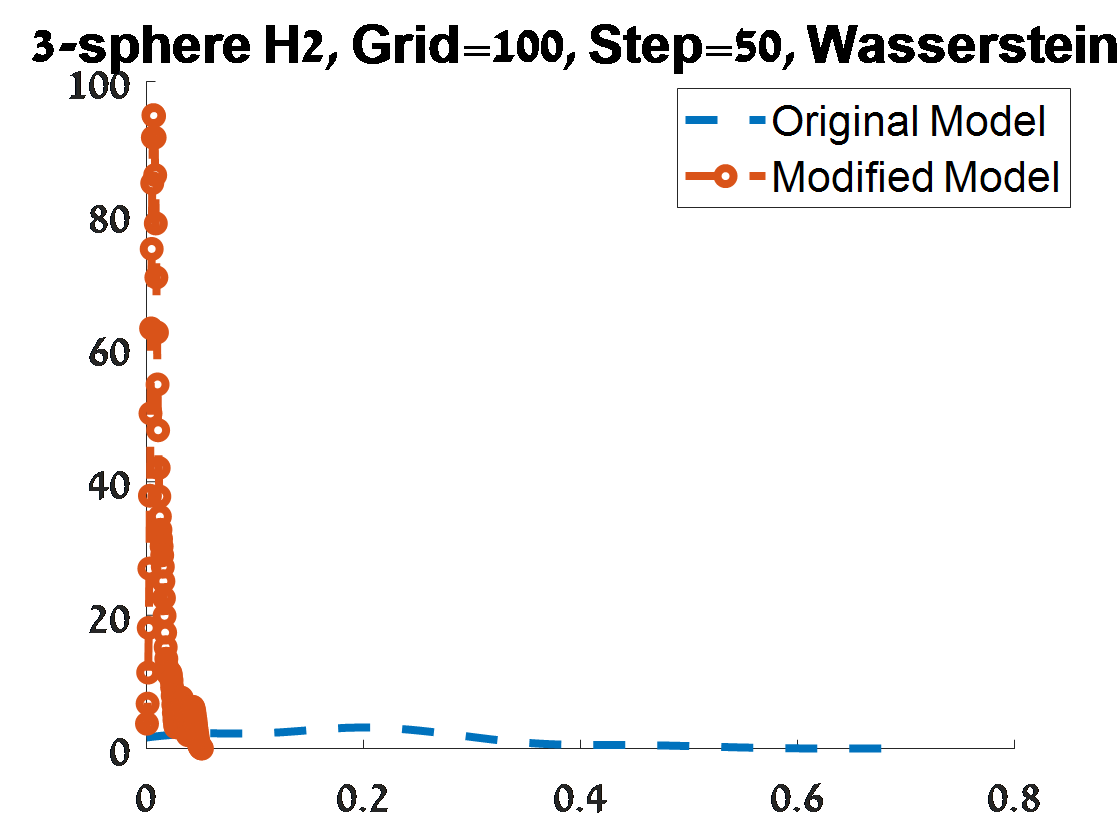}
%%\\
\includegraphics[width=1.2in, height=1.4in]{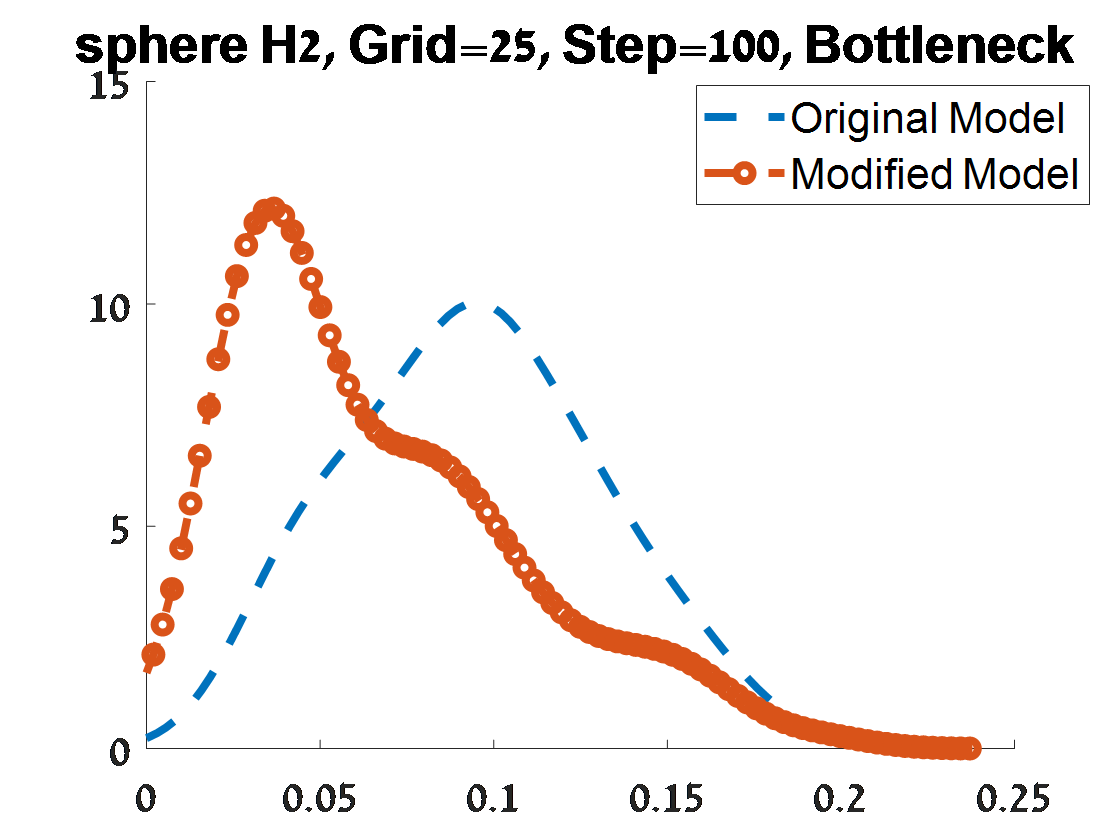}
\includegraphics[width=1.2in, height=1.4in]{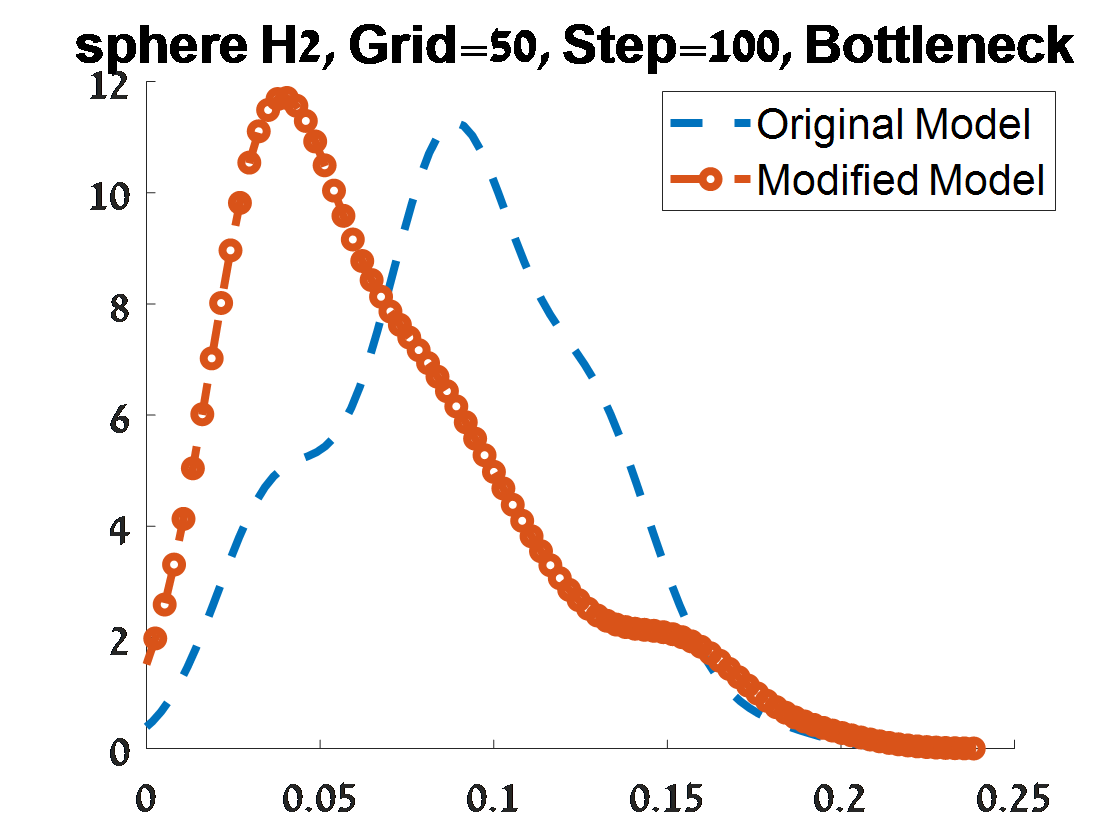}
\includegraphics[width=1.2in, height=1.4in]{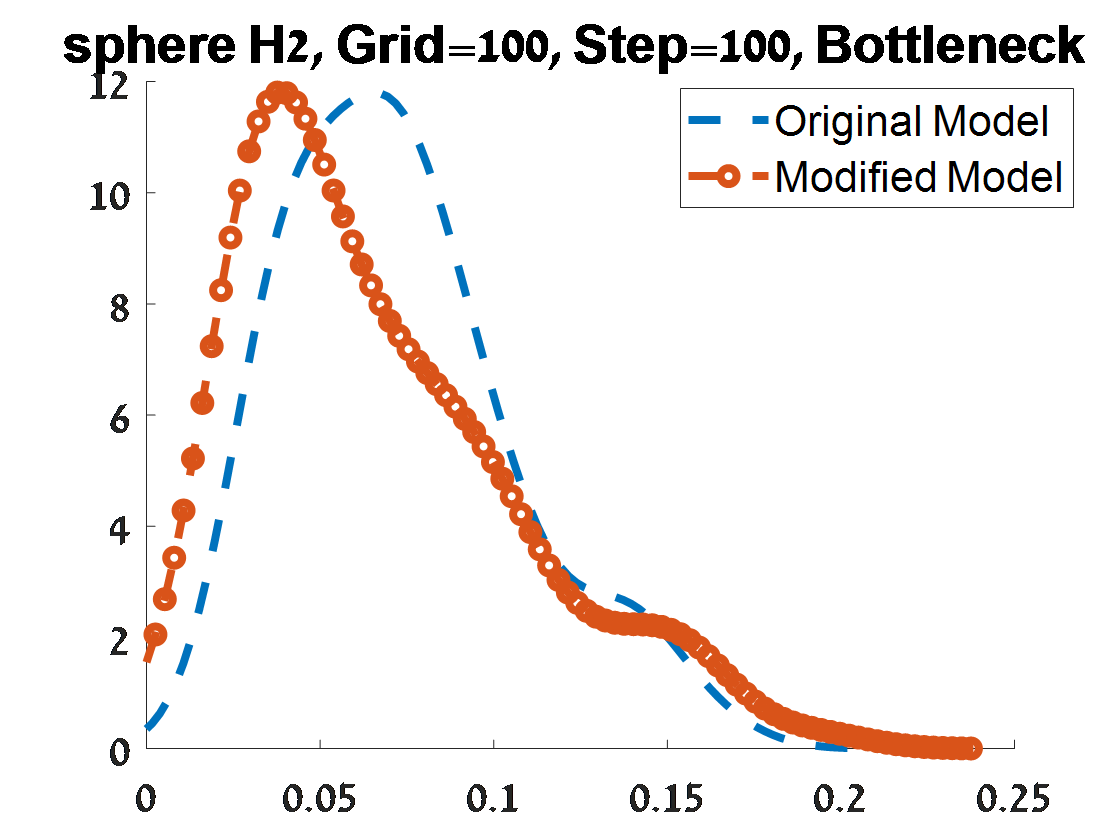}
\includegraphics[width=1.2in, height=1.4in]{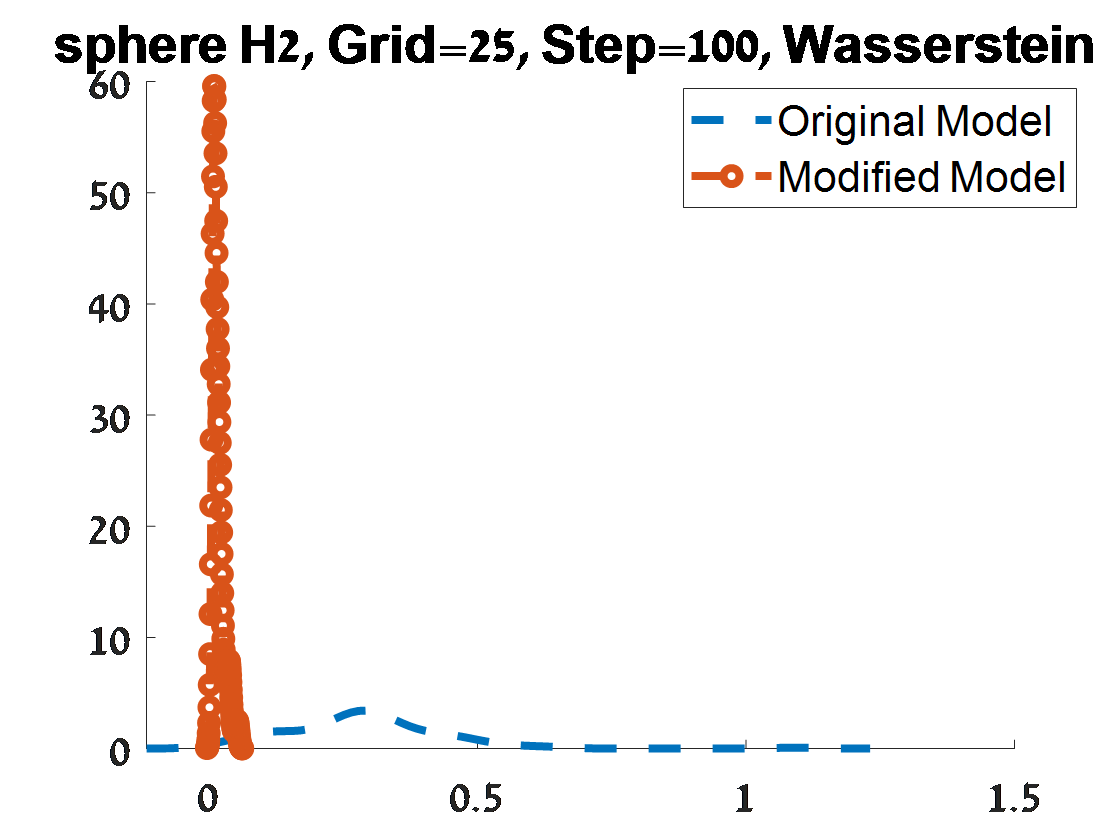}
\includegraphics[width=1.2in, height=1.4in]{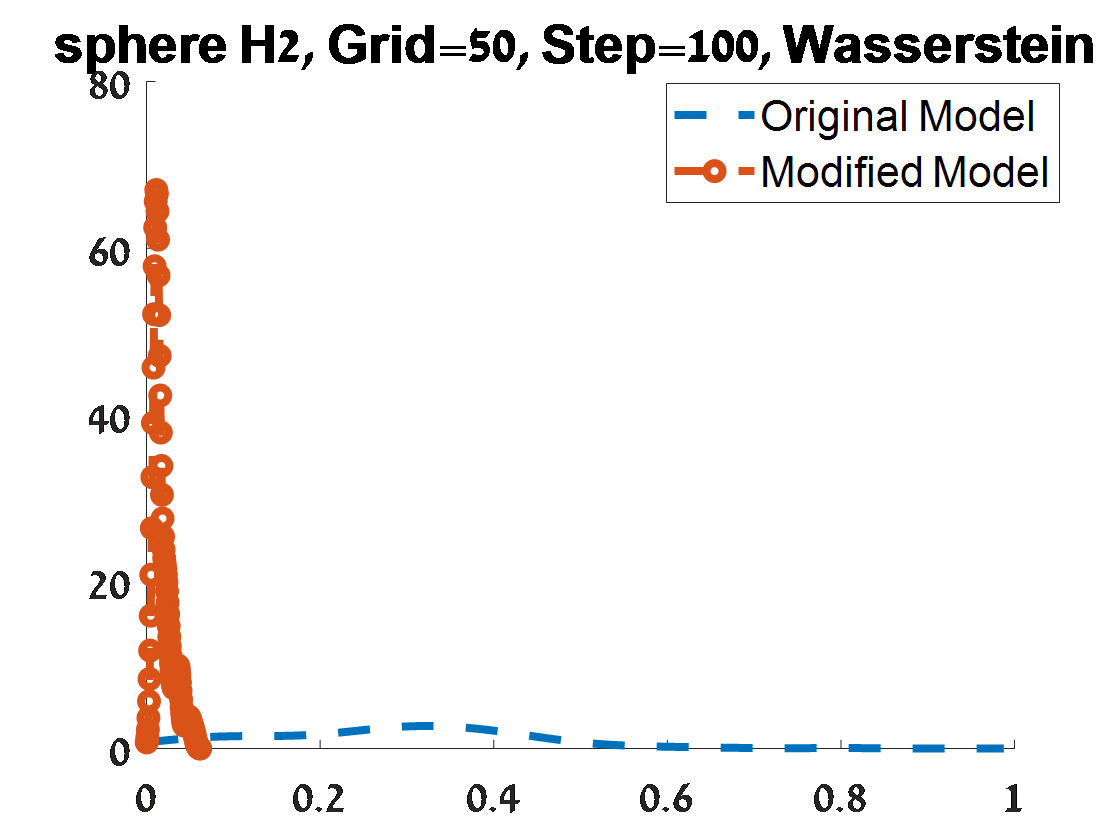}
\includegraphics[width=1.2in, height=1.4in]{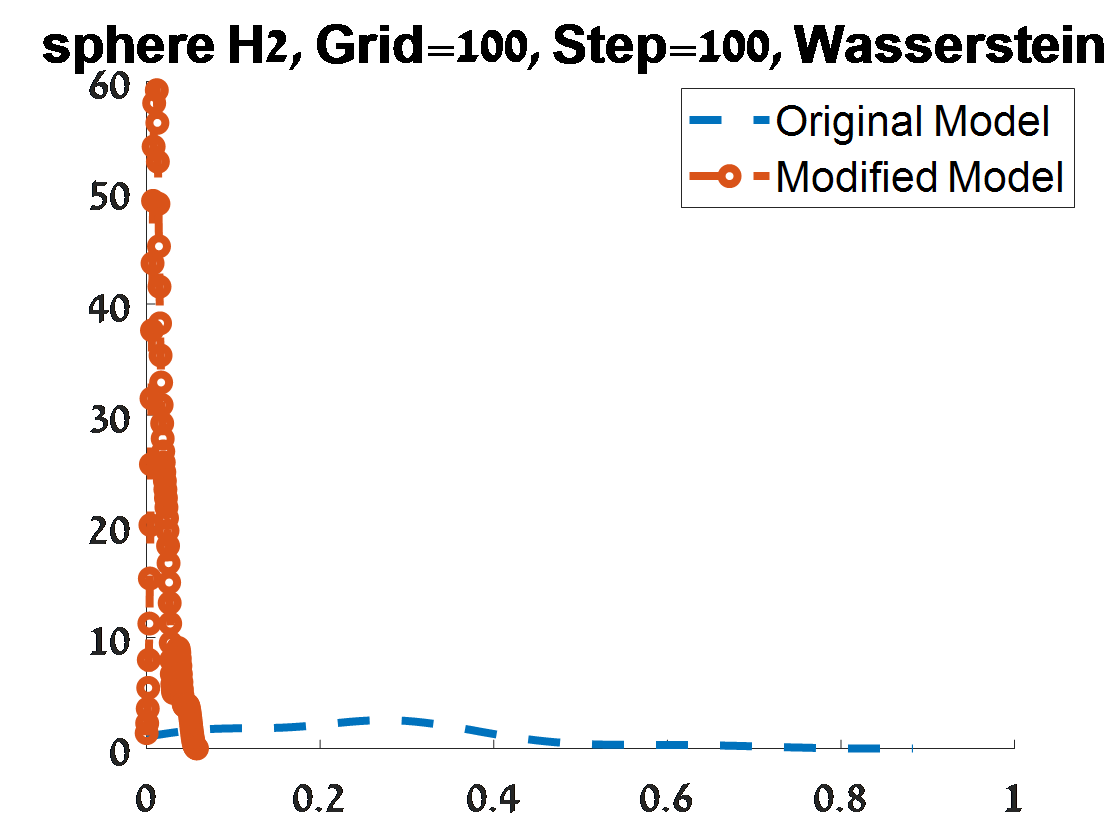}
\ec
\caption{\footnotesize
 Criterion 1 of goodness of fit for 100 $H_2$ PDs corresponded to 100 samples from a unit $S^3$. The figures depend on the grid of the proposal distribution ("Grid"), and the burn-in ("Step") of the MCMC algorithm.}
\label{fig:s3_H2_a}
\end{figure}
\end{landscape}

Figure\ \ref{fig:s3_H2_a} describes the distributions over the 100 PD of the first goodness of fit criterion, and Figures 28-29
%\ \ref{fig:s3_H2_b} and Figure\ \ref{fig:s3_H2_c}
describe the distributions of the second goodness of fit criterion.

Here we see a different pattern at the distribution shape of the properties under criterion 2 relative to this shape in $H_0$ and $H_1$: the shape of the distributions is different for the the modified model than these shape for the original model and the real PDs. The reason is the different shape of the points on the $H_2$ PD relative to this shape of the $H_0$ and $H_1$ points.

Based on criterion 1, the modified model seems better under the both distance measures, but still behave close under the Bottleneck distance whereas the advantage of the modified model is prominent in the Wasserstein distance. Relative to these distances in $H_0$ and $H_1$ points, the Bottleneck distances of the modified and the original models are similar in $H_2$ relative to these distances in $H_0$ and $H_1$. In addition, the values of the Bottleneck and Wasserstein distances are smaller for the both models in $H_2$ than in $H_0$ and $H_1$.
The reason for these results is the large variability in the points on the persistence diagrams H0 and H1 relative to H2.

\begin{landscape}
\begin{figure}[h!]
\bc
\includegraphics[width=1.2in, height=1.25in]{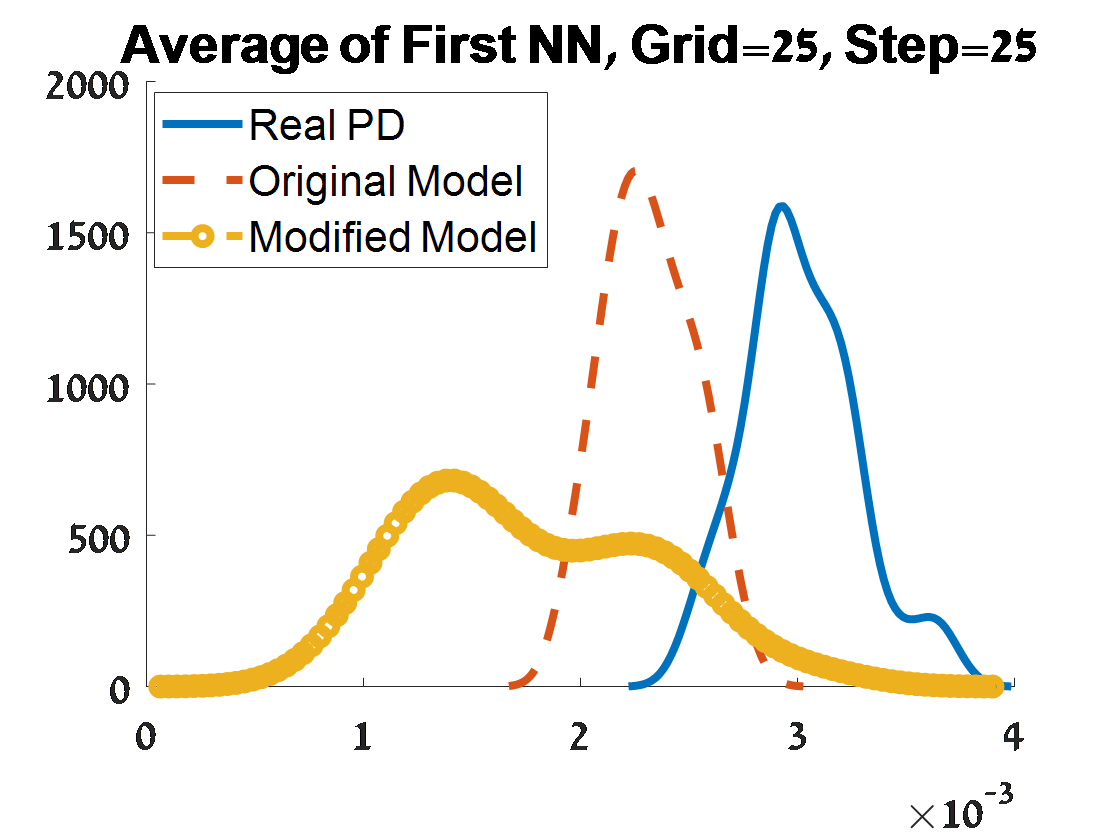}
\includegraphics[width=1.2in, height=1.25in]{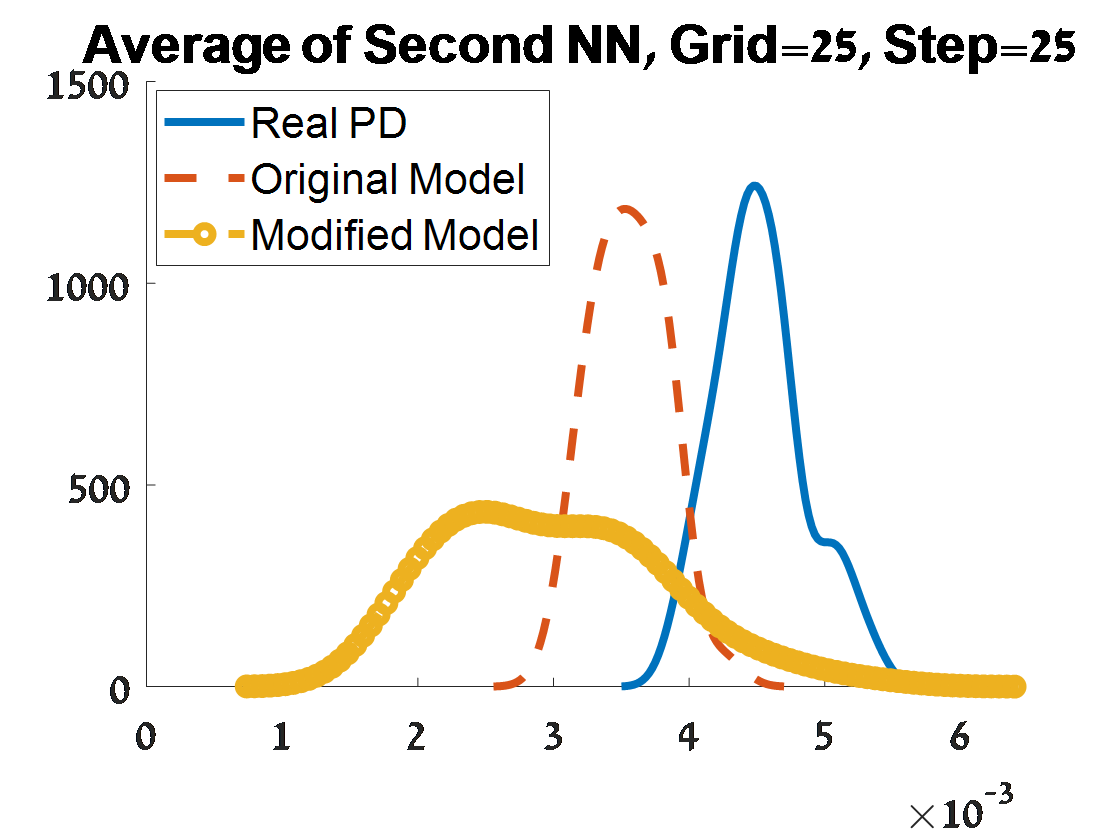}
\includegraphics[width=1.2in, height=1.25in]{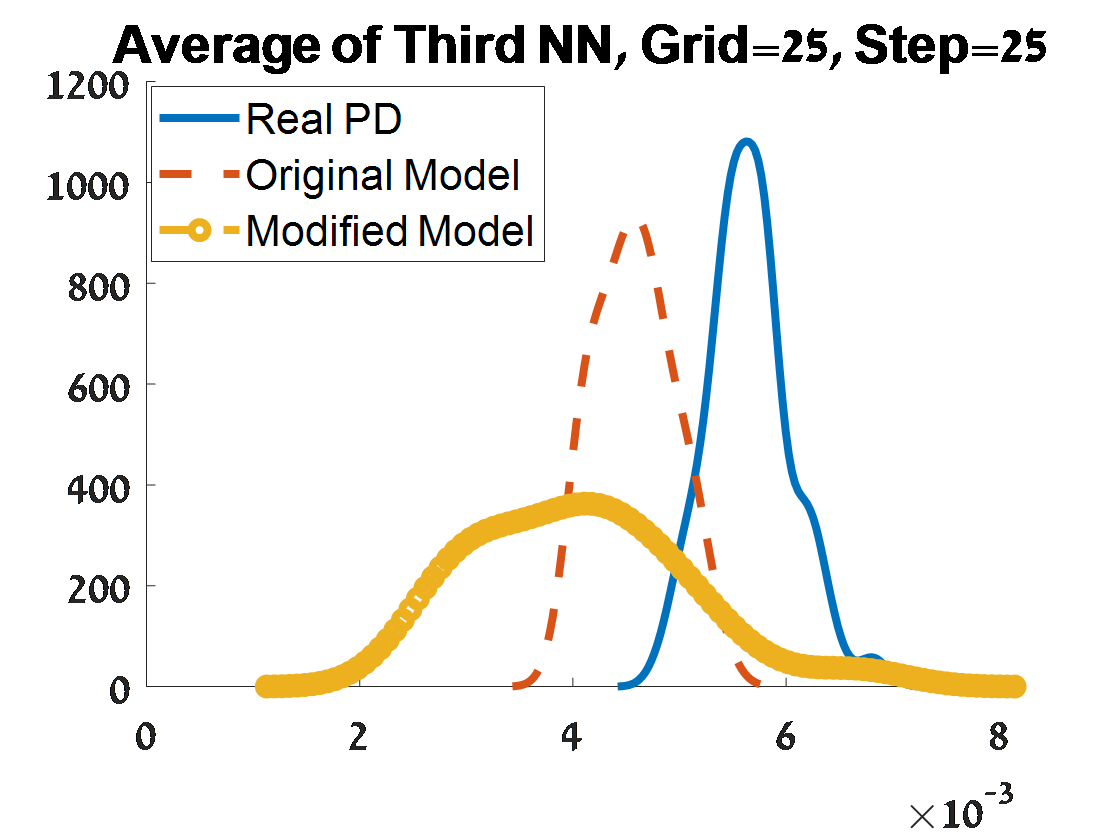}
\includegraphics[width=1.2in, height=1.25in]{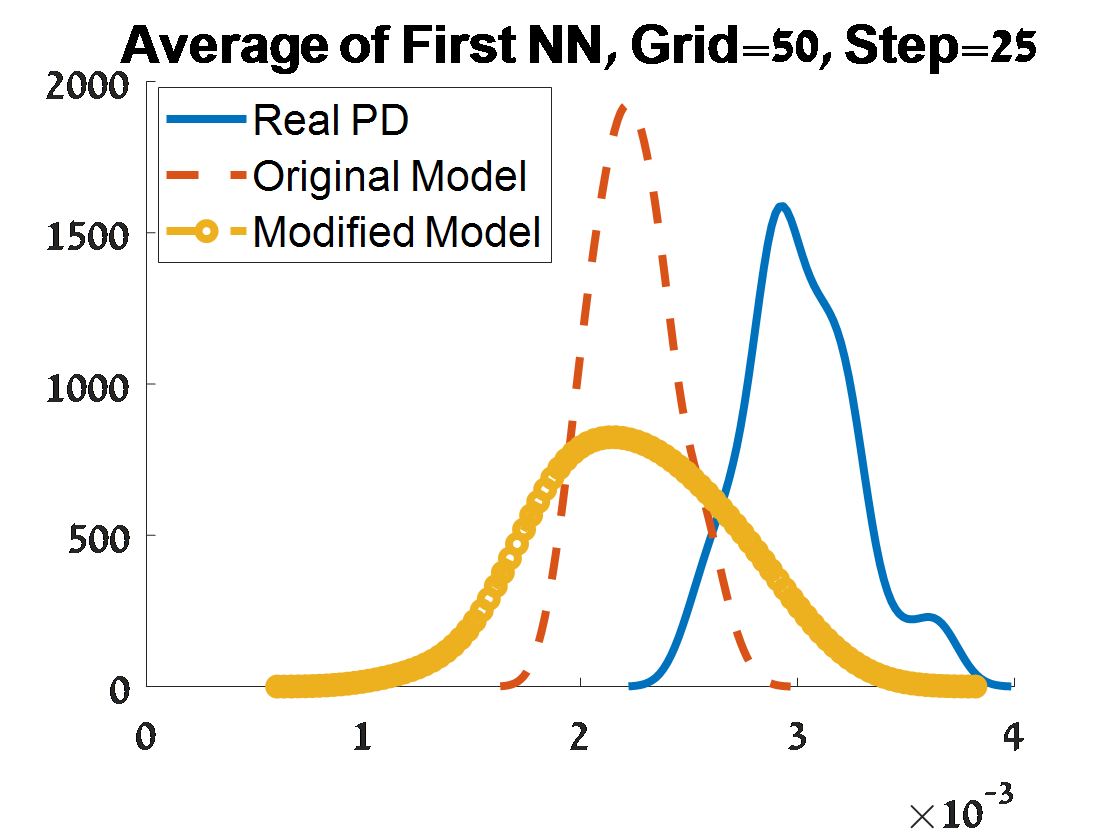}
\includegraphics[width=1.2in, height=1.25in]{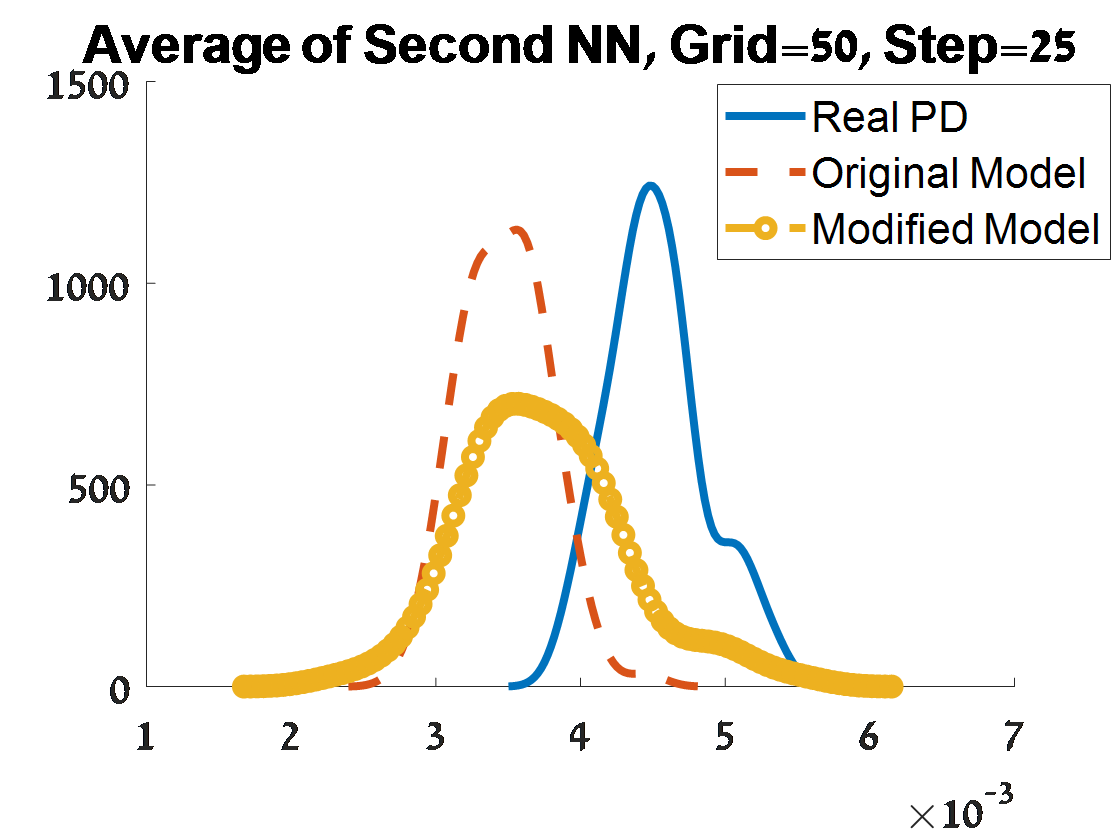}
\includegraphics[width=1.2in, height=1.25in]{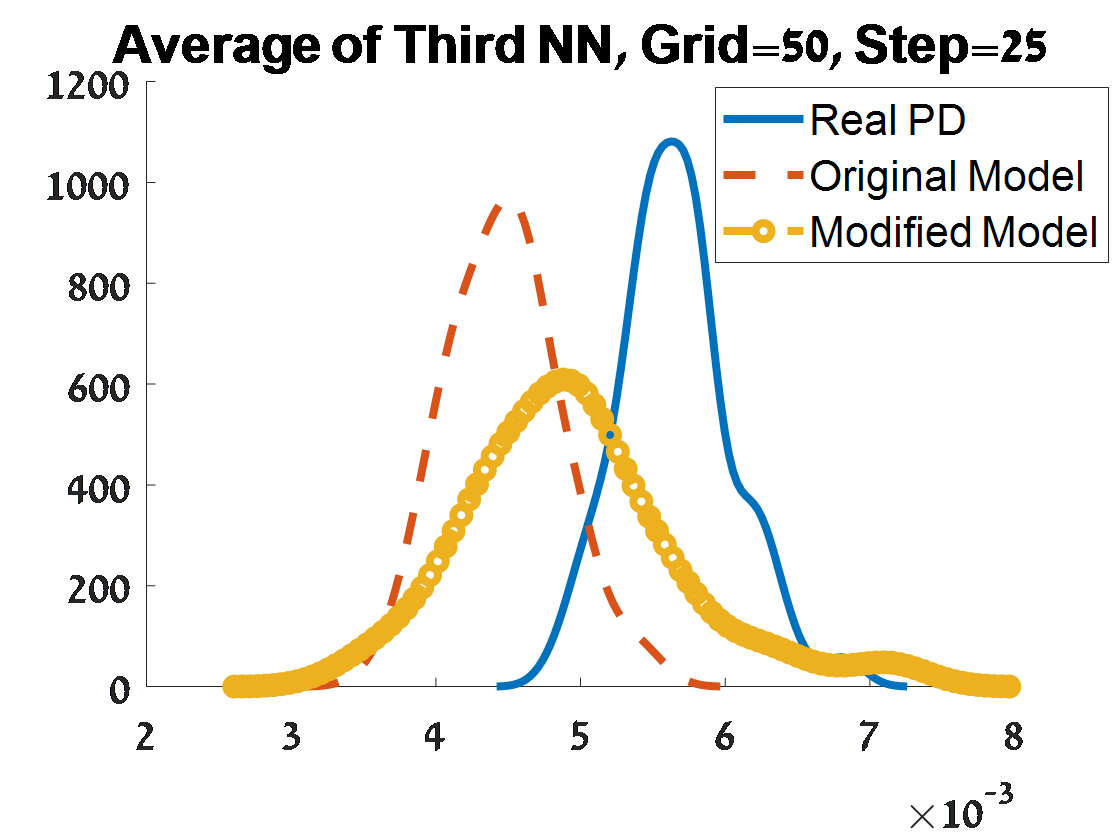}
\includegraphics[width=1.2in, height=1.25in]{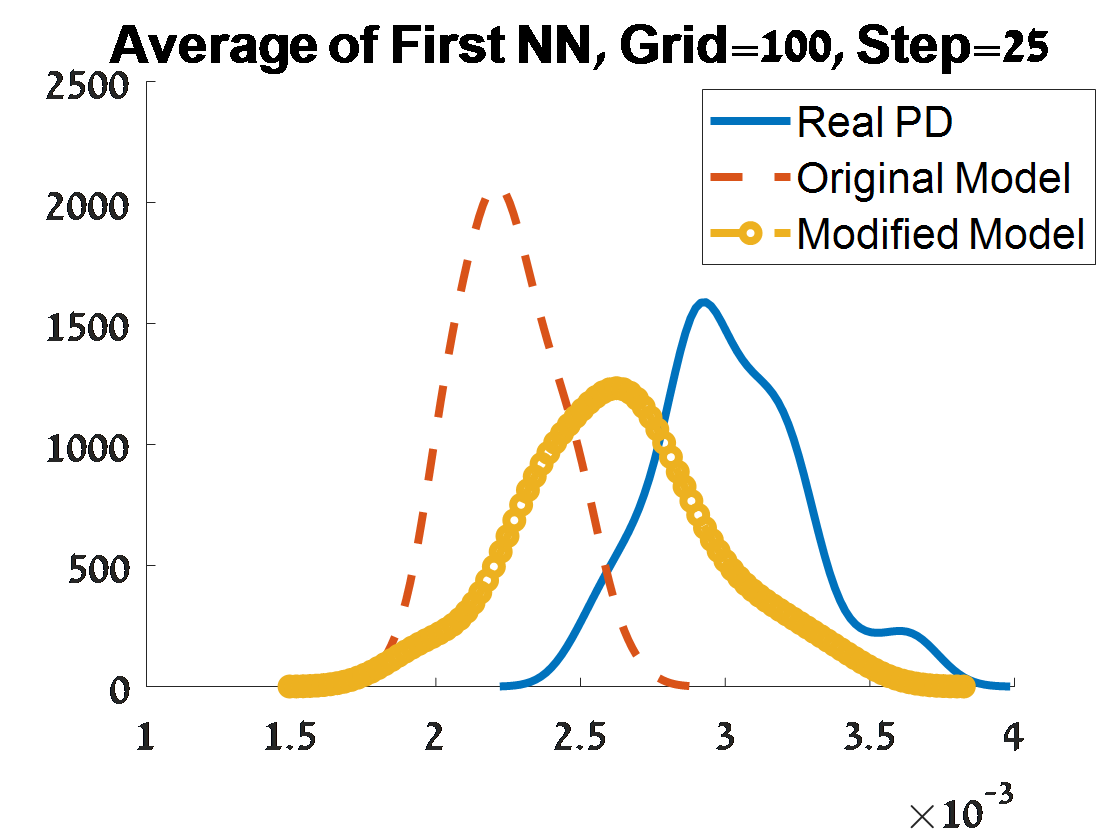}
\includegraphics[width=1.2in, height=1.25in]{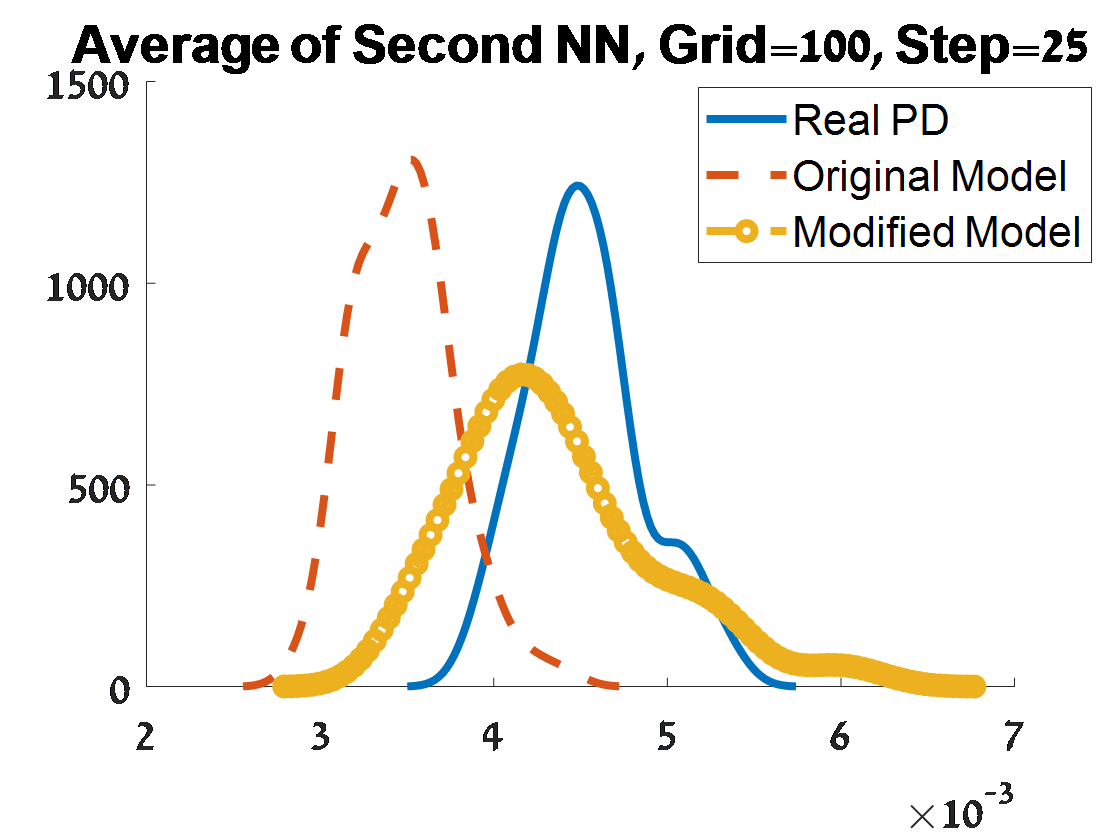}
\includegraphics[width=1.2in, height=1.25in]{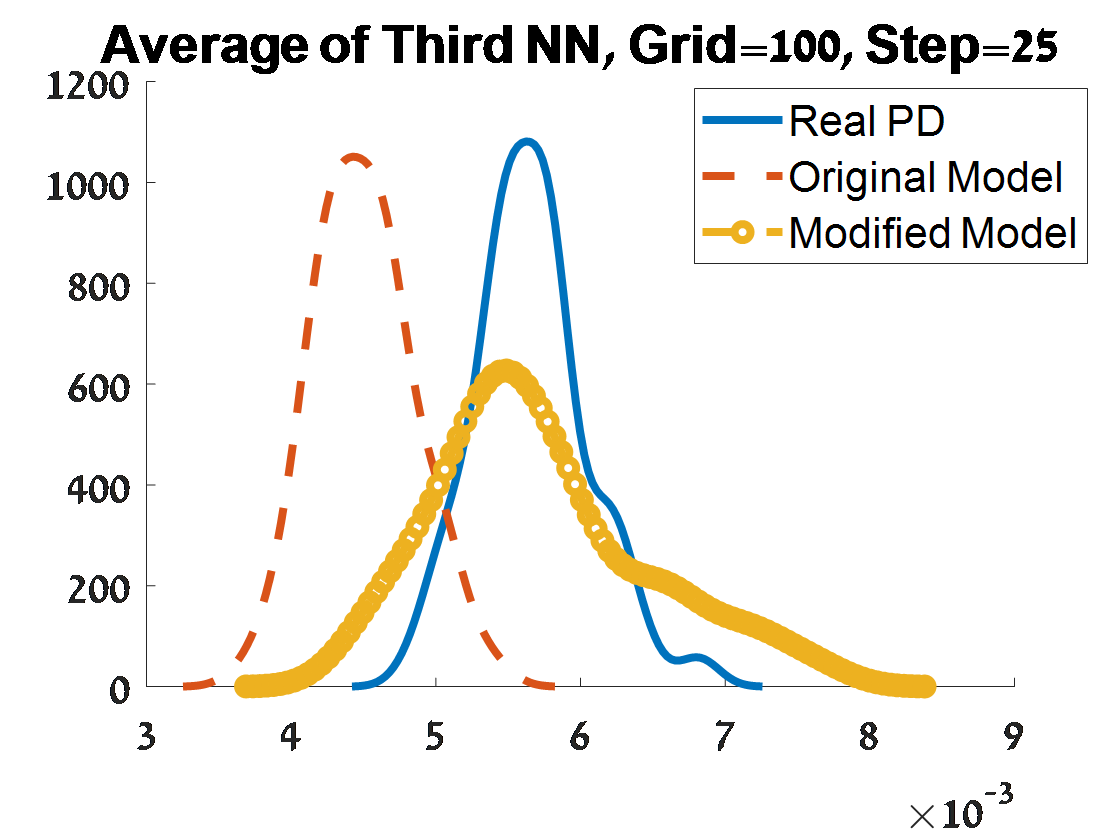}
\\
\includegraphics[width=1.2in, height=1.25in]{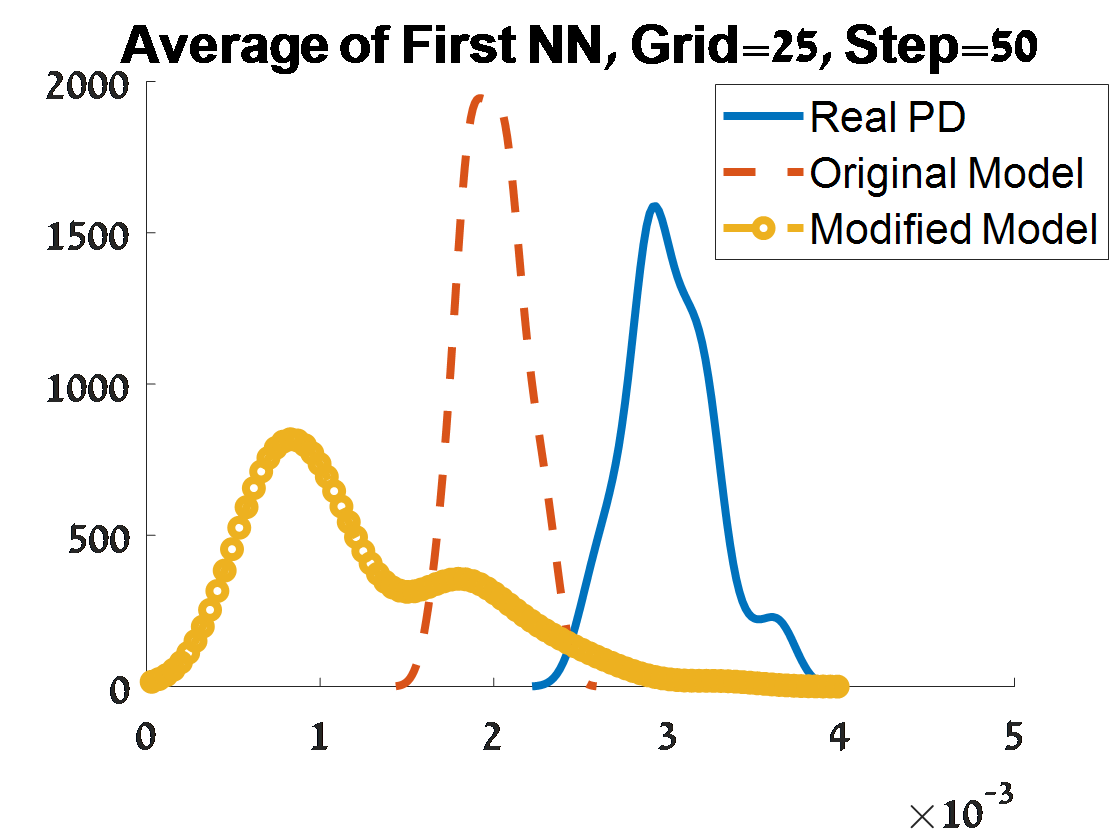}
\includegraphics[width=1.2in, height=1.25in]{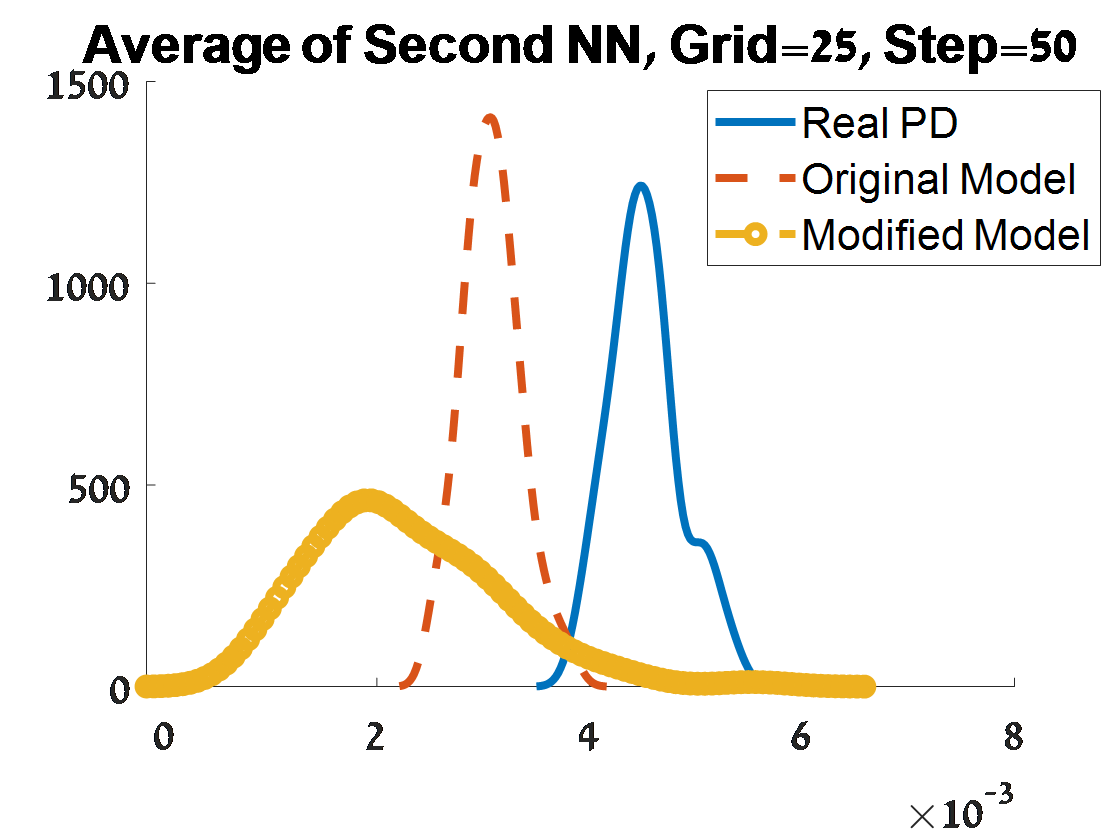}
\includegraphics[width=1.2in, height=1.25in]{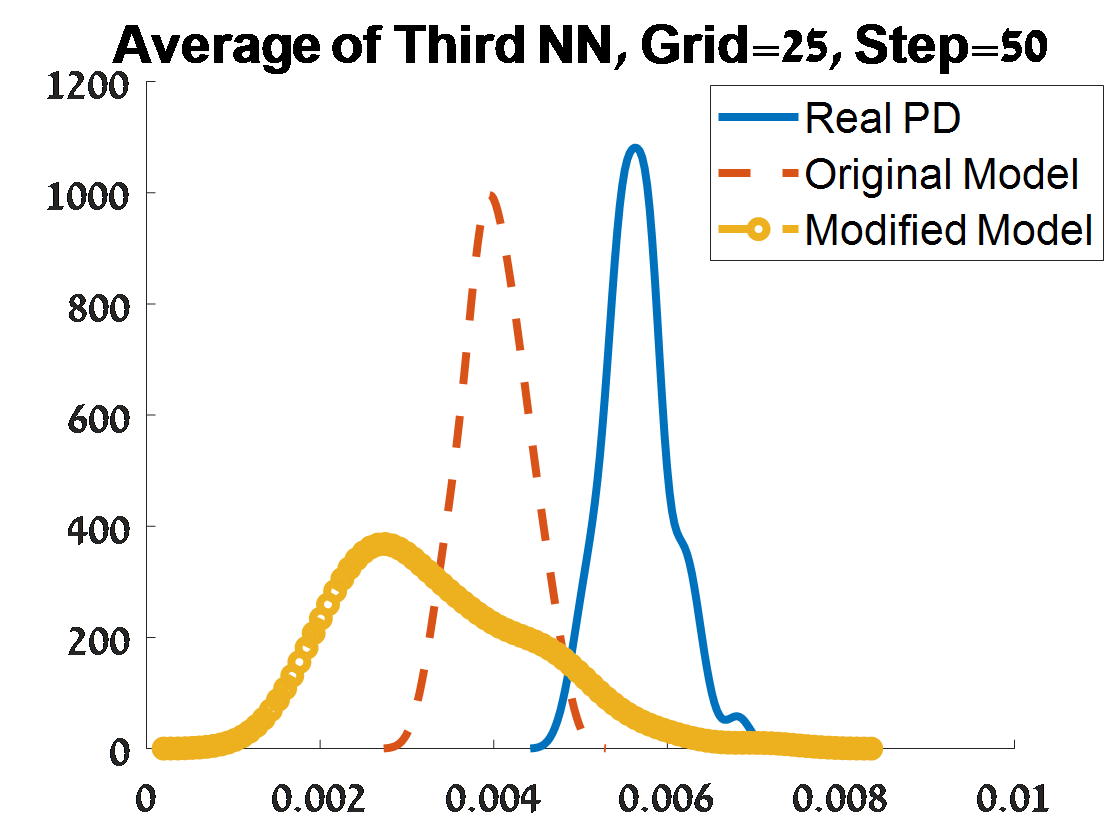}
\includegraphics[width=1.2in, height=1.25in]{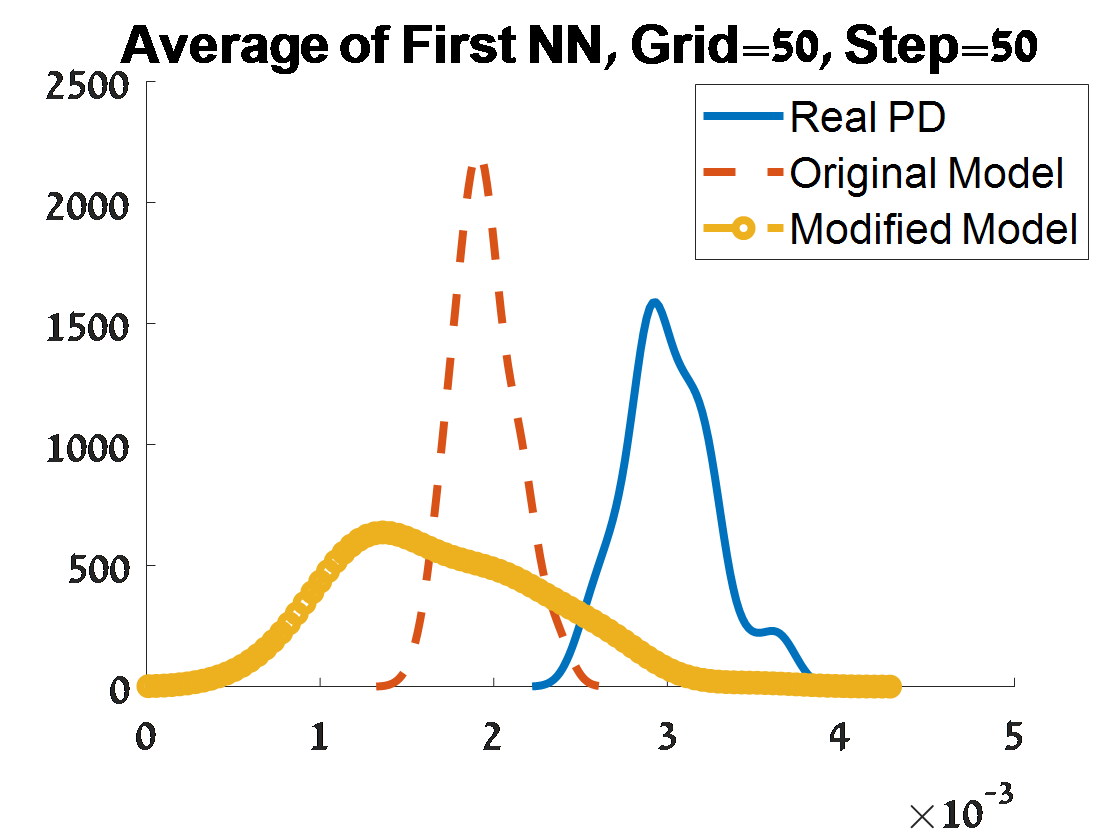}
\includegraphics[width=1.2in, height=1.25in]{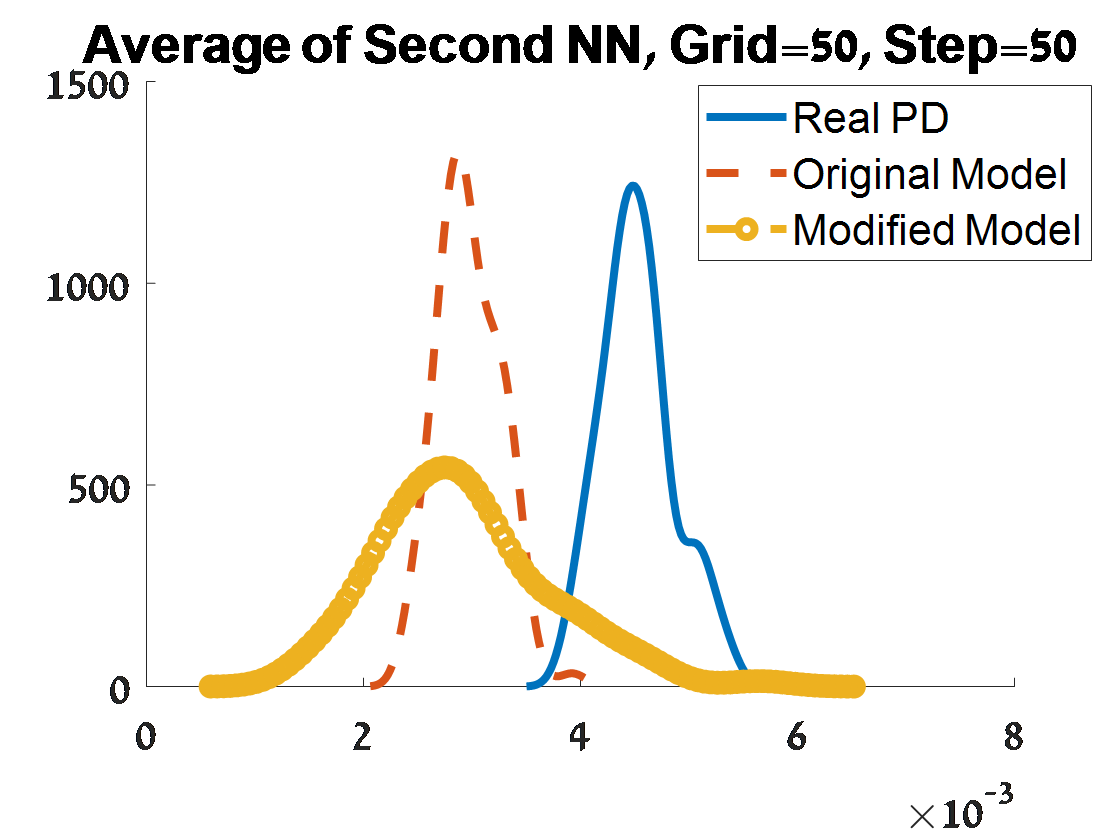}
\includegraphics[width=1.2in, height=1.25in]{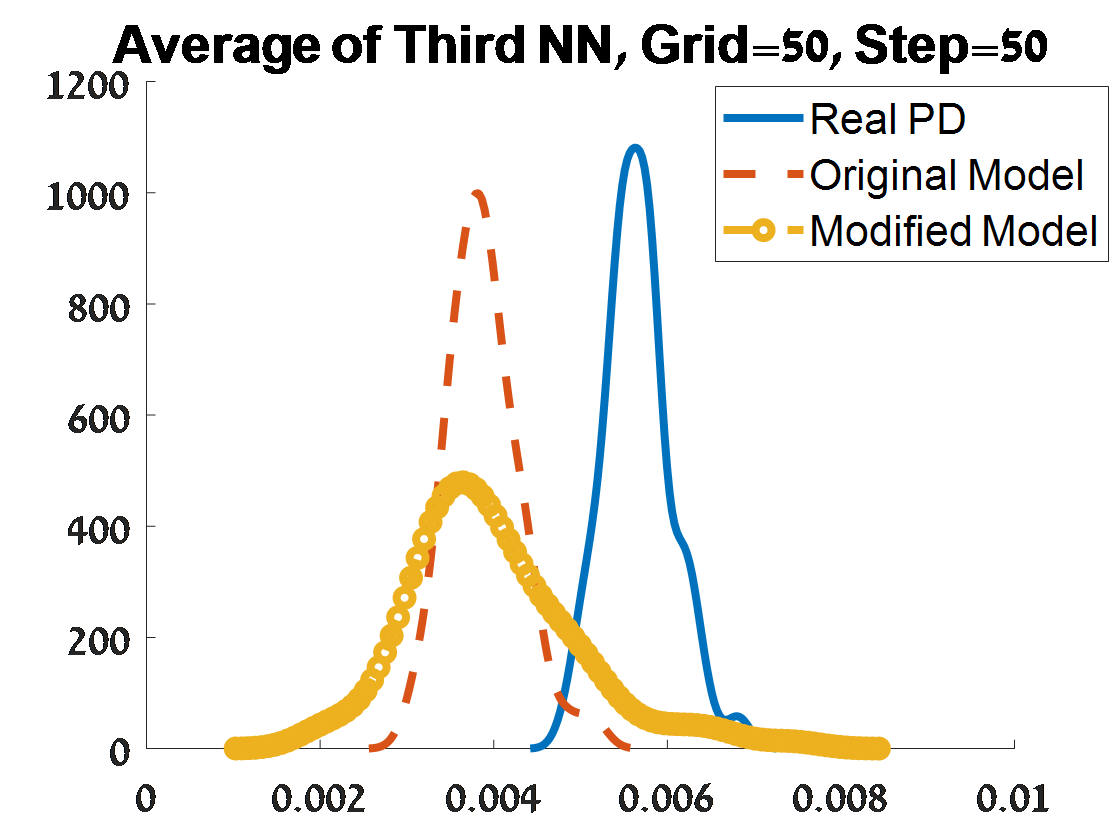}
\includegraphics[width=1.2in, height=1.25in]{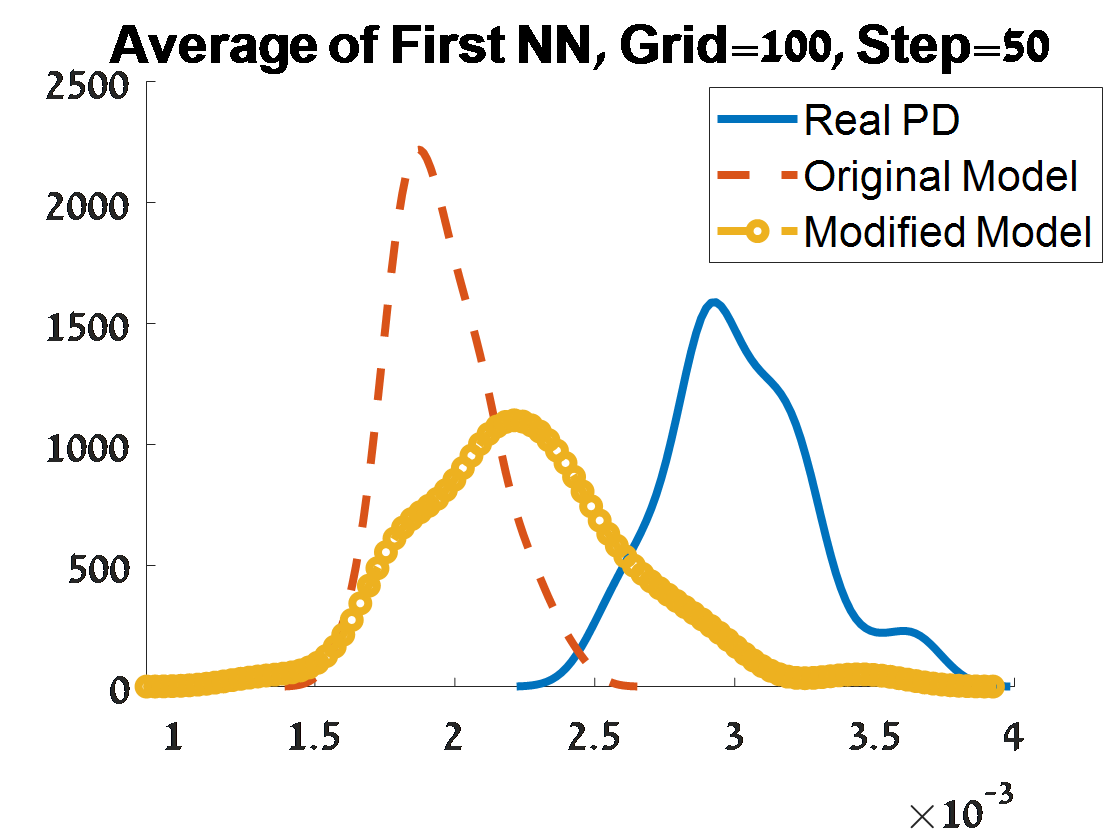}
\includegraphics[width=1.2in, height=1.25in]{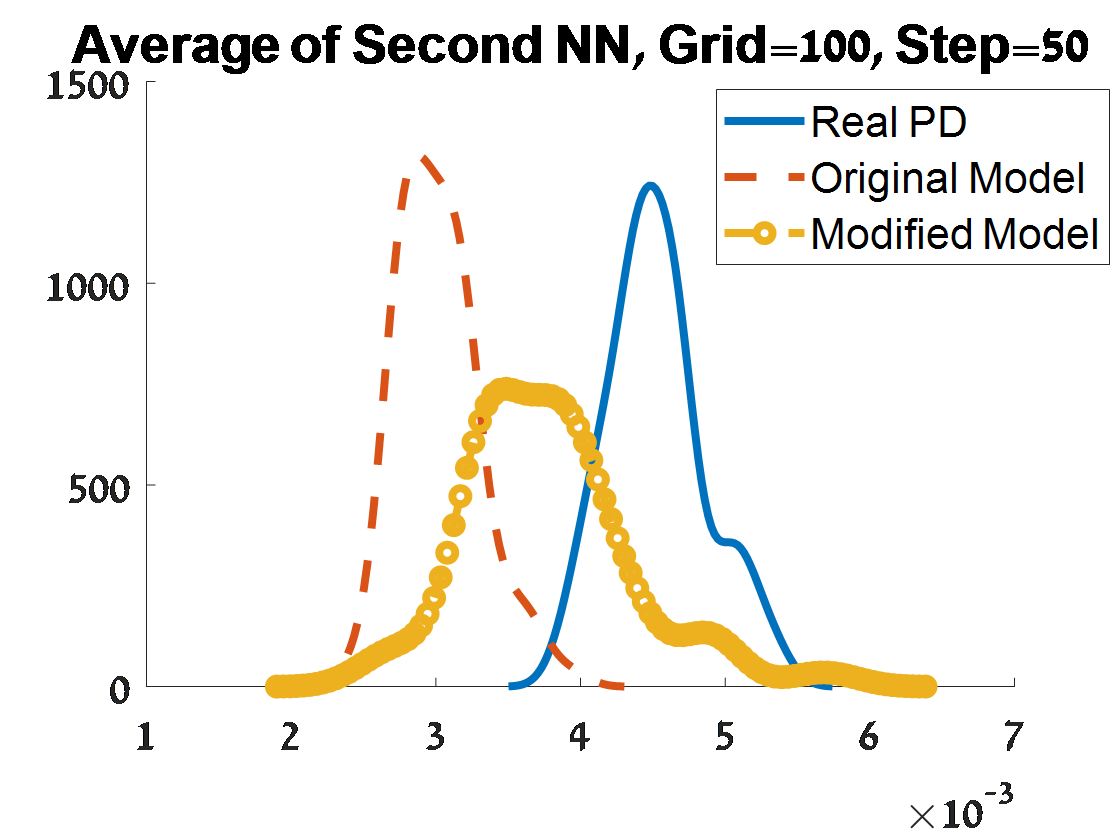}
\includegraphics[width=1.2in, height=1.25in]{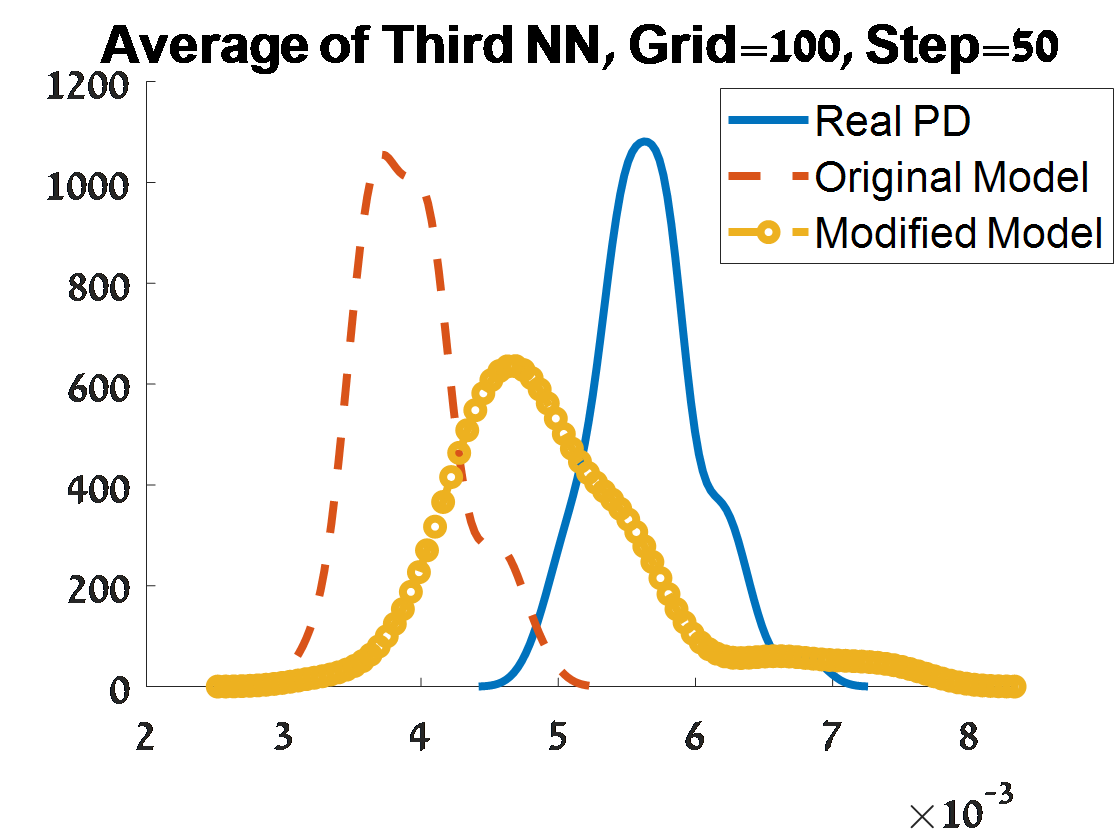}
\ec
%\caption{\footnotesize
% A random sample from two circles, 500 points from the larger circle and 300 from the smaller one,  with a kernel density
\caption{\footnotesize
 Criterion 2 of goodness of fit for 100 $H_2$ PDs corresponded to 100 samples from a unit $S^3$. The figures depend on the grid of the proposal distribution ("Grid"), and the burn-in ("Step") of the MCMC algorithm.}
\label{fig:s3_H2_b}
\end{figure}
\end{landscape}

\begin{landscape}
\begin{figure}[h!]
\bc
\includegraphics[width=1.2in, height=1.25in]{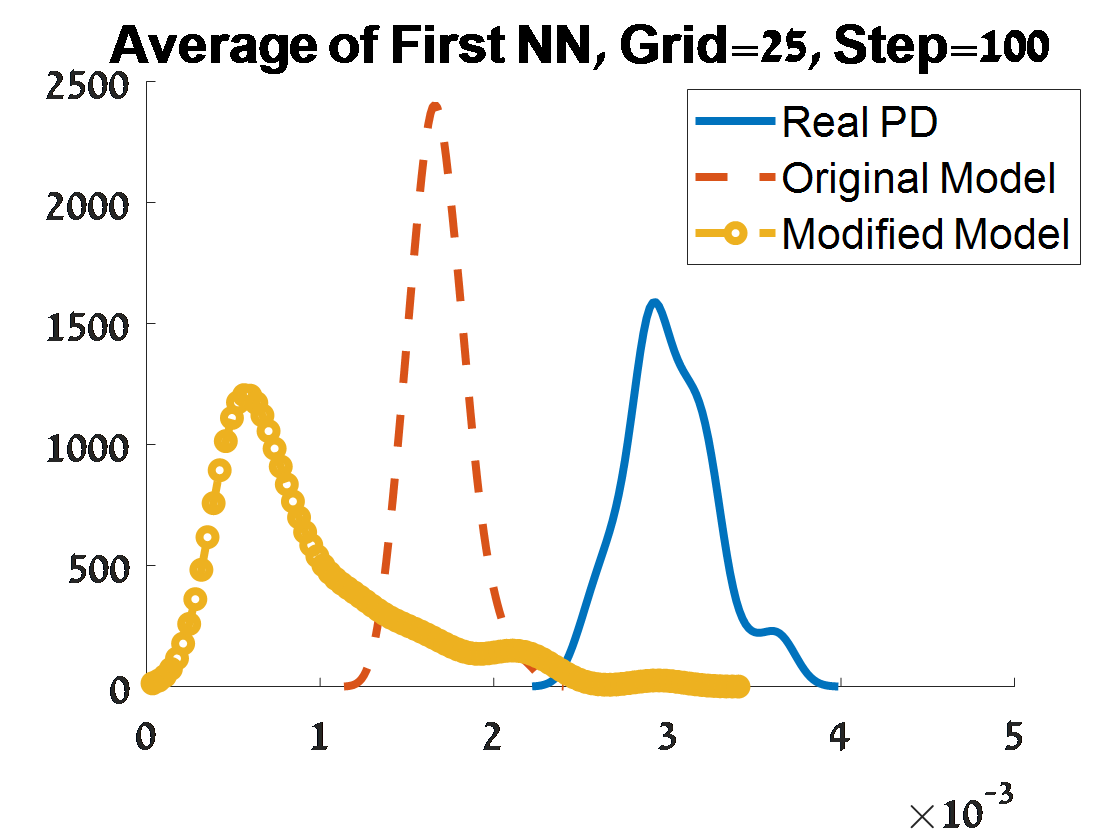}
\includegraphics[width=1.2in, height=1.25in]{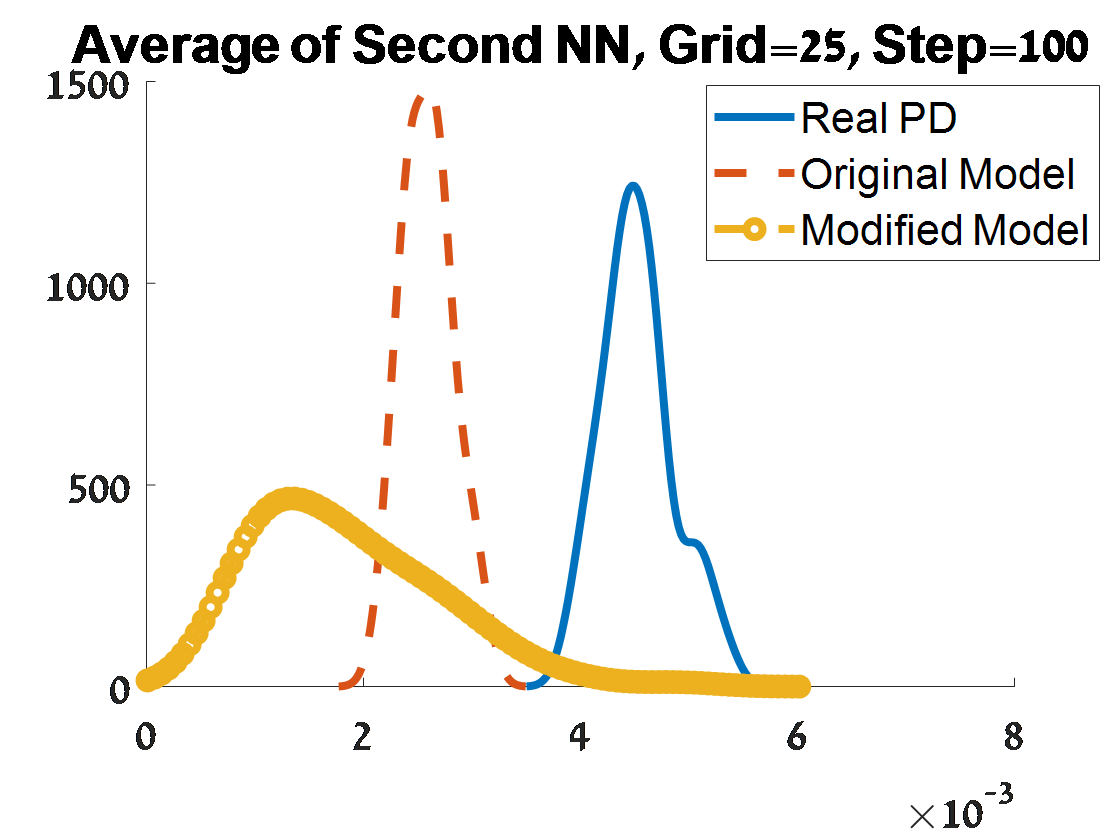}
\includegraphics[width=1.2in, height=1.25in]{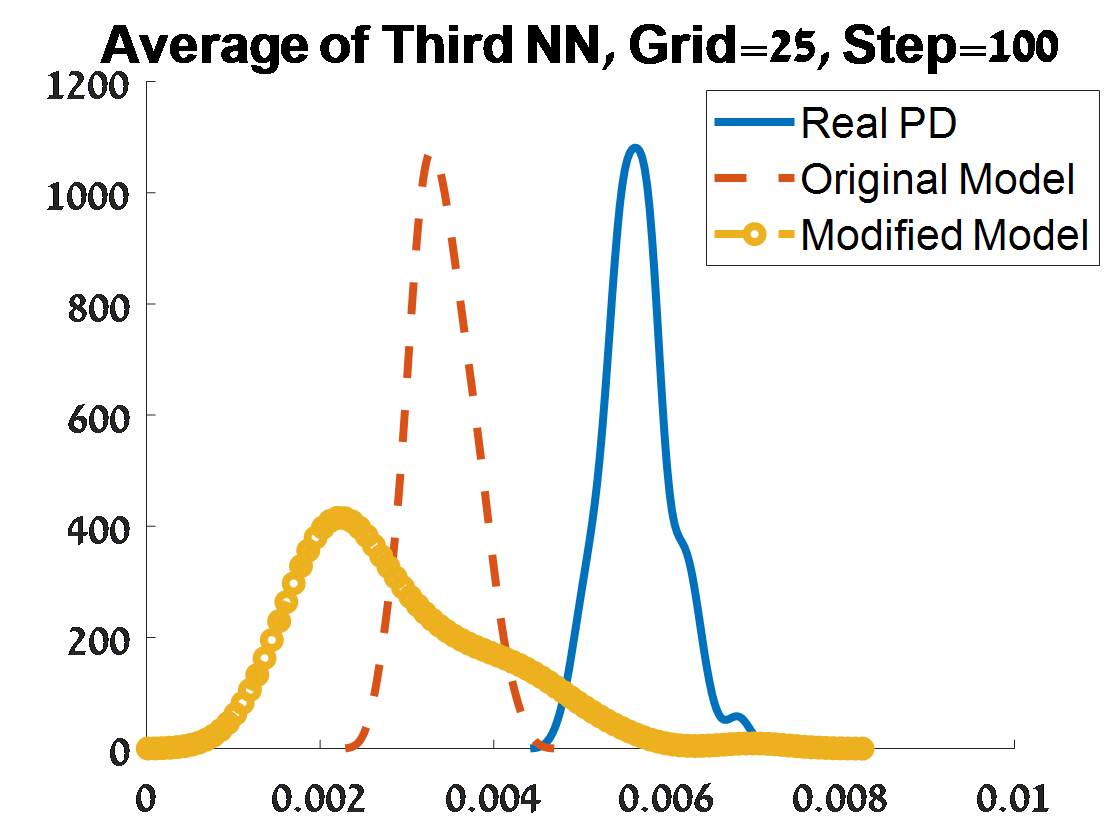}
\includegraphics[width=1.2in, height=1.25in]{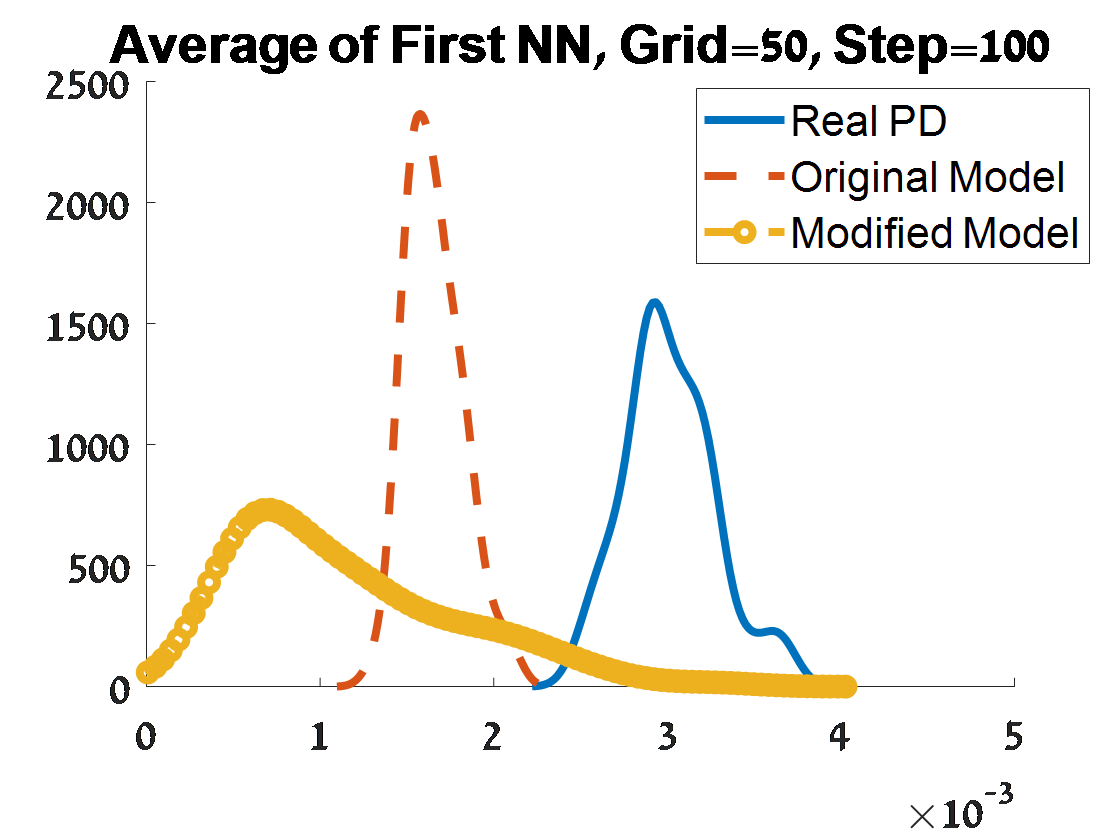}
\includegraphics[width=1.2in, height=1.25in]{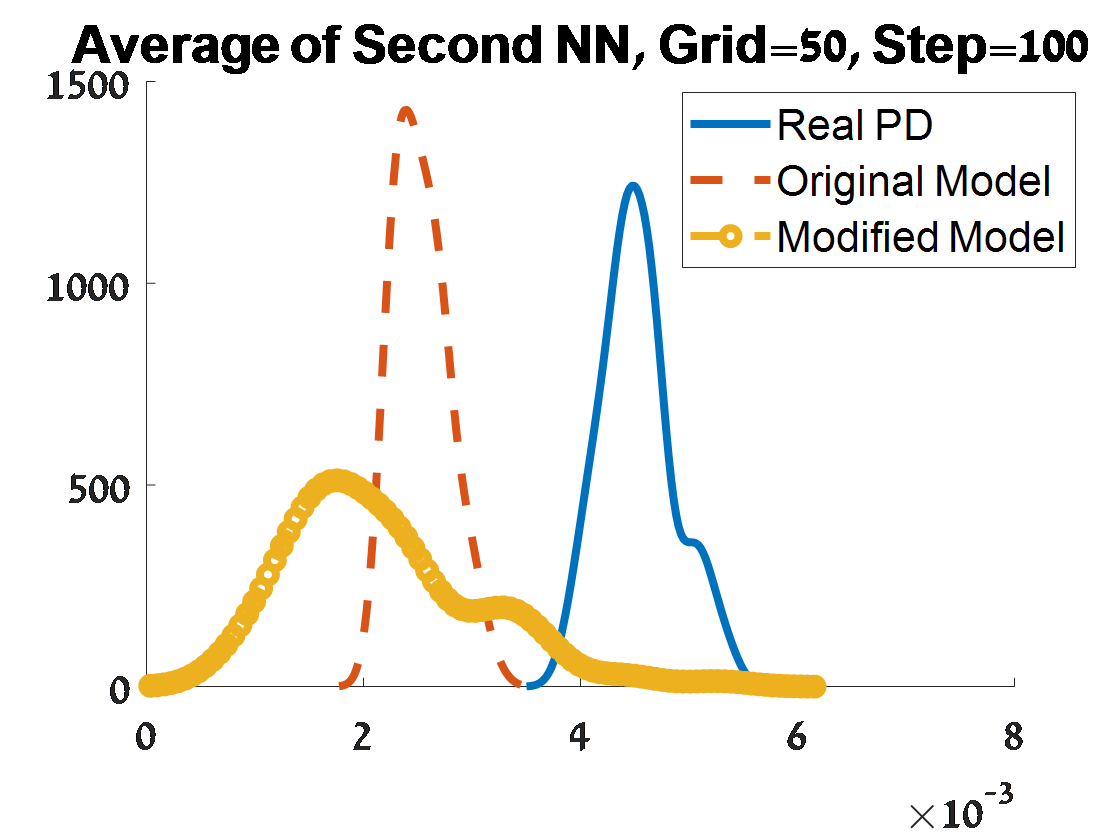}
\includegraphics[width=1.2in, height=1.25in]{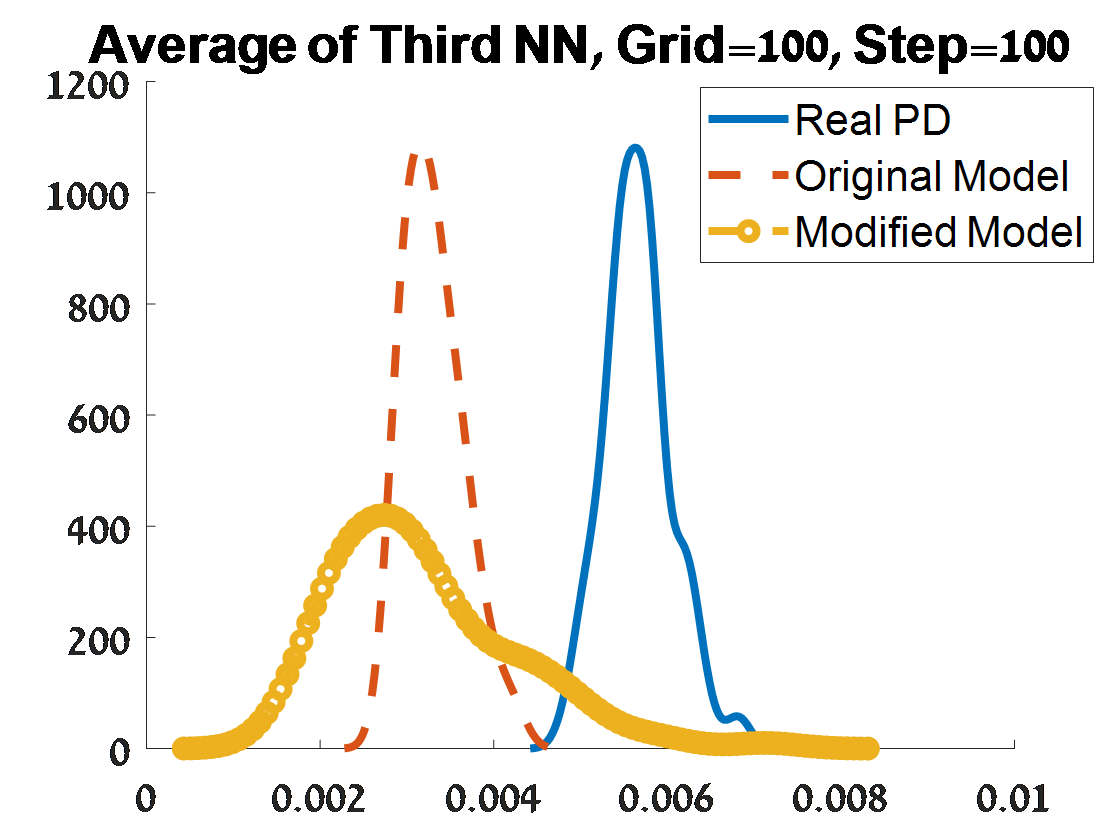}
\includegraphics[width=1.2in, height=1.25in]{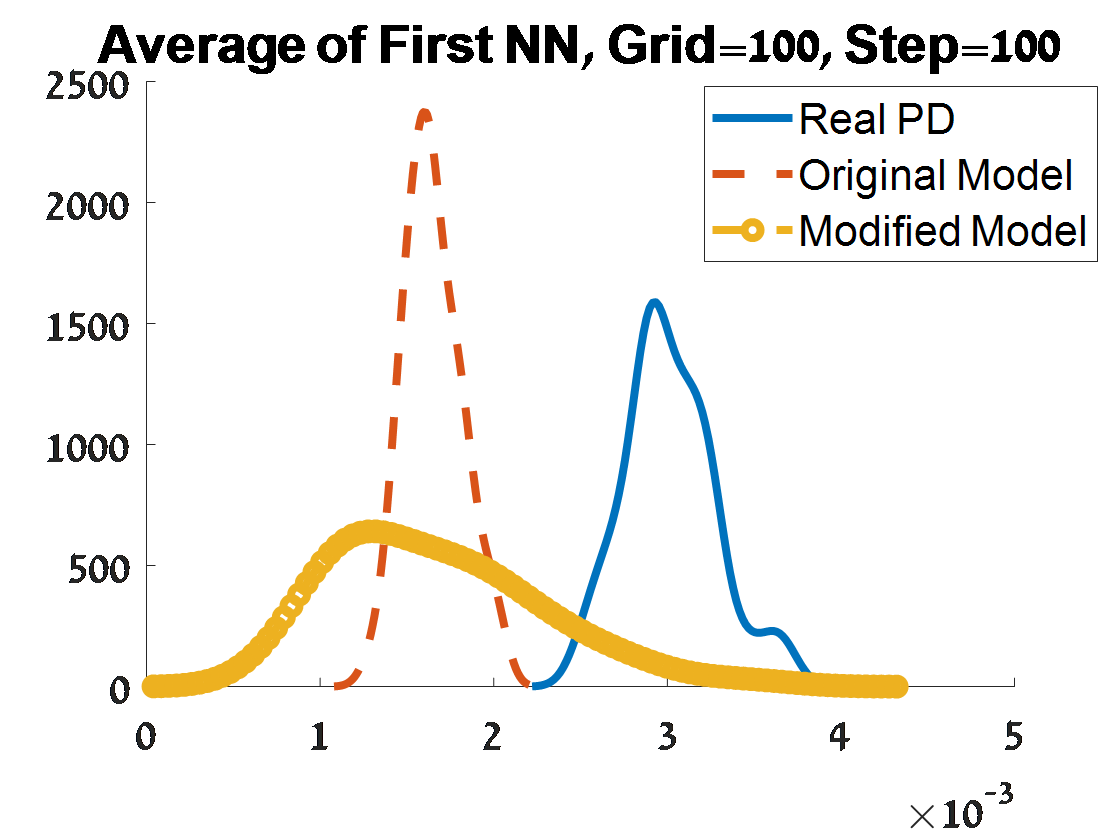}
\includegraphics[width=1.2in, height=1.25in]{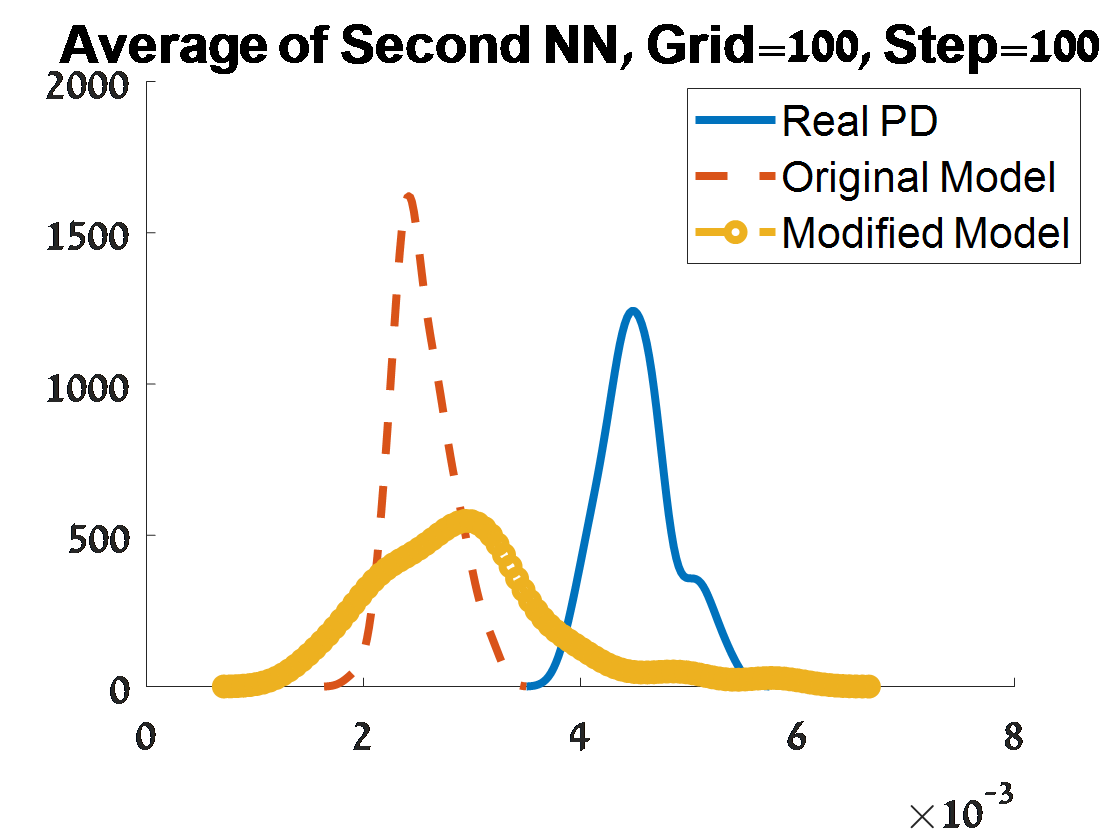}
\includegraphics[width=1.2in, height=1.25in]{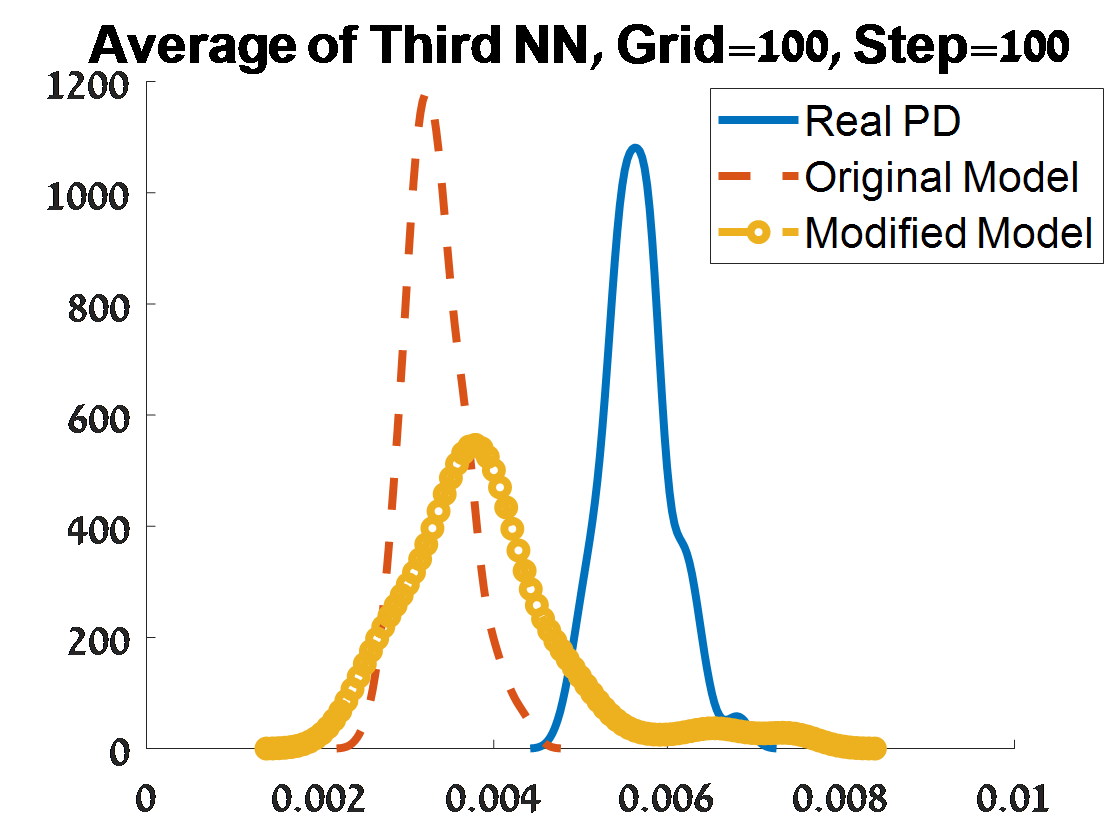}
\ec
\caption{\footnotesize
 Continue of Criterion 2 of goodness of fit for 100 $H_2$ PDs corresponded to 100 samples from a unit $S^3$. The figures depend on the grid of the proposal distribution ("Grid"), and the burn-in ("Step") of the MCMC algorithm.}
\label{fig:s3_H2_c}
\end{figure}
\end{landscape}

%\begin{landscape}
%\begin{figure}[h!]
%\bc
%\includegraphics[width=1.8in, height=1.8in]{Sphere_s3_H1_pd30_grid50_step25}
%\includegraphics[width=1.8in, height=1.8in]{Sphere_s3_H1_pd30_grid100_step25}
%\includegraphics[width=1.8in, height=1.8in]{Sphere_s3_H1_pd60_grid50_step25}
%\includegraphics[width=1.8in, height=1.8in]{Sphere_s3_H1_pd60_grid100_step25}
%\ec
%%\caption{\footnotesize
%% A random sample from two circles, 500 points from the larger circle and 300 from the smaller one,  with a kernel density
%\caption{\footnotesize
%Examples of two PDs, each one is corresponded to a sample from a unit $S^3$. For each PD, the simulated PD based on the two model versions is described. The figures depend on the grid of the proposal distribution ("Grid"), and the burn-in ("Step") of the MCMC algorithm.}
%\label{fig:s3_H1_d}
%\end{figure}
%\end{landscape}
\subsection{Distribution of the estimates}
Constraining the search range for the estimation of $\alpha$ as we saw in $S^3$ for $H_2$ points gives a motivation to compare the range of $\alpha$ estimates over the examined settings. In addition, by this, it is interesting to see if and how the distribution of the estimates of $\alpha$ (denoted by $\hat \alpha$) influent the distribution of $\Theta$ estimates (denoted by  $\hat \Theta$). For this purpose, we distinguished between the different combinations of signs of $\theta_1, \theta_2, \theta_3$ estimates, and examined the range of $\alpha$ estimates for each such combination. Table 1 summarizes the results. We see that the range of $\hat \alpha$ is pretty similar for the different combinations of $\hat \Theta$ signs, except the case of $\hat \Theta>0$ which has small values toward zero of $\hat \alpha$. The later case is generally for all the examined examples except for $S^3$ in $H_2$ points which has a weight of $29\%$. The common case is of $\hat \theta_1>0$, $\hat \theta_2<0$, $\hat \theta_3<0$.

\begin{center}
%\small
\fontsize{7.4}{1.16}\selectfont
\begin{tabular}{c|c|c|c|c}

%\mc{15}{c}{Table 4}       \\
\mc{5}{c}{\bf{Table 1.} Distribution of the modified model's estimates} \\
\\
\\
\\
\\
\\
\\
\\
\\
\\
\\
\\
\\
\\
\\
%\mc{11}{c} {\bf Two Samples of a Unit Circle $^a$} \\
%\\
%\\
%\\
%\\
%\\
%\boldmath{Homology} && \textbf{Bottleneck$^b$}&&& \textbf{ Wasserstein$^c$} &&\textbf{ Ratio$^d$}&&&\textbf{ RST $^e$}\\
\textbf{Geometrical object} & \textbf{Homology}& \boldmath{$\hat \alpha$} \textbf{Range} &\boldmath{ $\hat \Theta$} \textbf{Sign}&\boldmath{$\%$} \textbf{Cases}\\
\\
\\
\\
\\
\\
\\
\\
\\

\textbf{One circle} &$H_0$	&	[0.085,2.876]&$\hat \theta_1>0$, $\hat \theta_2<0$, $\hat \theta_3<0$ &68\\
\\
\\
\\
 &&	[0.028,2.733]&$\hat \theta_1>0$, $\hat \theta_2>0$, $\hat \theta_3<0$ &28\\
\\
\\
\\
 &&	[0.015,0.024]&$\hat \theta_1>0$, $\hat \theta_2>0$, $\hat \theta_3>0$ &2\\
\\
\\
\\
 &&	[2.572,2.628]&$\hat \theta_1<0$, $\hat \theta_2>0$, $\hat \theta_3<0$ &2\\
\\
\\
\\
\\
\\
\\
\textbf{Concentric circles} &$H_0$	&	[0.091,3.112]&$\hat \theta_1>0$, $\hat \theta_2<0$, $\hat \theta_3<0$ &72\\
\\
\\
\\
 &&	[0.060,2.900]&$\hat \theta_1>0$, $\hat \theta_2>0$, $\hat \theta_3<0$ &25\\
\\
\\
\\
 &&	[0.016, 0.022]&$\hat \theta_1>0$, $\hat \theta_2>0$, $\hat \theta_3>0$ &3\\
\\
\\
\\
\\
\\
\\
\\
\\
\\
\textbf{Distinct circles} &$H_0$	&	[0.129,1.203]&$\hat \theta_1>0$, $\hat \theta_2<0$, $\hat \theta_3<0$ &65\\
\\
\\
\\
 &&	[0.101,1.371]&$\hat \theta_1>0$, $\hat \theta_2>0$, $\hat \theta_3<0$ &35\\
\\
\\
\\
\\
\\
\\
\\
\\
\\
\textbf{2-Sphere } &$H_0$	&	[0.201,0.903]&$\hat \theta_1>0$, $\hat \theta_2<0$, $\hat \theta_3<0$ &41\\
\\
\\
\\
 &&	[0.208,1.145]&$\hat \theta_1>0$, $\hat \theta_2>0$, $\hat \theta_3<0$ &57\\
\\
\\
\\
&&	[0.037,0.047]&$\hat \theta_1>0$, $\hat \theta_2>0$, $\hat \theta_3>0$ &2\\
\\
\\
\\
\\
\\
\\
\\
\\
&$H_1$	&	[0.159,1.125]&$\hat \theta_1>0$, $\hat \theta_2<0$, $\hat \theta_3<0$ &61\\
\\
\\
\\
 &&	[0.105,1.074]&$\hat \theta_1>0$, $\hat \theta_2>0$, $\hat \theta_3<0$ &39\\
\\
\\
\\
\\
\\
\\
\\
\\
\\
\textbf{3-Sphere } &$H_0$	&	[0.453,0.863]&$\hat \theta_1>0$, $\hat \theta_2<0$, $\hat \theta_3<0$ &74\\
\\
\\
\\
 &&	[0.495,0.864]&$\hat \theta_1>0$, $\hat \theta_2>0$, $\hat \theta_3<0$ &26\\
\\
\\
\\
\\
\\
\\
\\
\\
&$H_1$	&	[0.720,1.320]&$\hat \theta_1>0$, $\hat \theta_2<0$, $\hat \theta_3<0$ &93\\
\\
\\
\\
 &&	[0.859,1.084]&$\hat \theta_1>0$, $\hat \theta_2>0$, $\hat \theta_3<0$ &7\\
\\
\\
\\
\\
\\
\\
\\
\\
&$H_2$	&	[0.403,0.946]&$\hat \theta_1>0$, $\hat \theta_2<0$, $\hat \theta_3<0$ &52\\
\\
\\
\\
 &&	[0.584,0.951]&$\hat \theta_1>0$, $\hat \theta_2>0$, $\hat \theta_3<0$ &19\\
\\
\\
\\
 &&	[0.016,0.127]&$\hat \theta_1>0$, $\hat \theta_2>0$, $\hat \theta_3>0$ &29\\
\\
\\
\\
\end{tabular}
\end{center}
\footnotesize{Behaviour of estimates of $\alpha$ ( $\hat \Theta$) and $\Theta$  ($\hat \Theta$) over 100 PDs of each geometrical objects.}
\normalsize
\section{Results}
Generally, in all the considered examples we have that, based on the examined criteria, the modified RST behave better than the original one in fitting the distribution of the points on the PD (for a given homology). Specifically, for the two-dimensional examples, which include one or two geometrical objects, the modified RST is better for all considered grid sizes and burn-in values, but the best fitting is in grid sizes of 50 and 100, and burn-in of 25. But for the three and four dimensional examples, the modified RST is better only in grid size of 100 and burn-in of 25. The reason for the need in large grid in the later examples is to capture the large variability of the points on the PD due to the high dimensionality.

\section{Summary}
In this paper suggest a modified RST for modeling the points on persistence diagram. We examined the performance of this modified version relative to the original one by using two criteria. We have found that the modified RST fits the distribution of the points on the persistence diagram better than the original RST. Particularly, the best fitting is achieved usually in a larger grid for which the proposal distribution of the MCMC algorithm is calculated (in our simulations, grid sizes of 50 or 100), and in a smaller burn-in of the MCMC algorithm (in our simulations, burn-in of 25). Therefore we recommend to use the modified RST with considering these values of the MCMC parameters.

\end{document}